\documentclass[a4paper,12pt]{article}
\pdfoutput=1
\usepackage{epsfig,amsmath,amsfonts,amsthm}
\usepackage[figuresright]{rotating}
\usepackage[left=28mm,right=20mm,top=30mm,bottom=20mm]{geometry}
\usepackage{subfig}
\usepackage{graphicx}
\usepackage{rotating}
\usepackage{booktabs}
\usepackage{moreverb}

\DeclareGraphicsExtensions{.pdf,.png,.jpg}
\usepackage{natbib}
\numberwithin{equation}{section}

\def\e{\hbox{E}}

\def\var{\hbox{Var}}

\def\max{\hbox{max}}

\providecommand{\keywords}[1]{\textbf{{Keywords}} #1}
\usepackage[nokeyprefix]{refstyle}
\usepackage{varioref}
\usepackage{xr}
\usepackage{hyperref}

\begin{document}

\title{Simulations for estimation of heterogeneity variance and overall effect with constant and inverse-variance weights in meta-analysis of difference in standardized means (DSM)}

\author{Elena Kulinskaya and  David C. Hoaglin }

\date{\today}

\maketitle

\begin{abstract}
When the individual studies assembled for a meta-analysis report means ($\mu_C$, $\mu_T$) for their treatment (T) and control (C) arms, but those data are on different scales or come from different instruments, the customary measure of effect is the standardized mean difference (SMD).
The SMD is defined as the difference between the means in the treatment and control arms, standardized by the assumed common standard deviation, $\sigma$. However, if the variances in the two arms differ, there is no consensus on a definition of SMD. Thus, we propose a new effect measure, the difference of standardized means (DSM), defined as $\Delta = \mu_T/\sigma_T - \mu_C/\sigma_C$. The estimated DSM can easily be  used as an effect measure in standard meta-analysis.

For random-effects meta-analysis of DSM, we introduce new point and interval estimators of the between-studies variance ($\tau^2$) based on the $Q$ statistic with effective-sample-size weights, $Q_F$. We study, by simulation, bias and coverage of these new estimators of $\tau^2$ and related estimators of $\Delta$. For comparison, we also study bias and coverage of well-known estimators based on the $Q$ statistic with inverse-variance weights, $Q_{IV}$, such as the Mandel-Paule, DerSimonian-Laird, and restricted-maximum-likelihood estimators.

\end{abstract}


\section{Introduction}

When the individual studies assembled for a meta-analysis report means for their treatment and control arms, but those data are on different scales or come from different instruments, the customary measure of effect is the standardized mean difference (SMD).   The SMD is considered to be
the most appropriate effect-size index in psychological research \citep{sanches-2010}, and was also found to be  more generalizable than the mean difference \citep{Takeshima2014}.  The SMD is extensively used in social and health sciences, in ecology and conservation, and in numerous other areas.

The SMD is defined as the standardized difference of the means in the treatment (T) and control (C) arms, $\delta = (\mu_T - \mu_C)/\sigma$, assuming equal variances ($\sigma^2_T = \sigma^2_C = \sigma^2$. An extensive literature discusses its estimation, properties, and use in meta-analysis. However, the strong assumption of homogeneity of variances is often not satisfied in practice. Unfortunately, when the variances differ, there is no consensus on a definition of SMD.

To avoid assuming homogeneity, we propose a new effect measure, the difference of standardized means (DSM), defined as $\Delta = \mu_T/\sigma_T - \mu_C/\sigma_C$. The estimated DSM can easily be  used as an effect measure in standard meta-analyses. Alternatively, the standardized means, $\mu_j/\sigma_j$ for $j =$ T or C, can be used directly in a meta-regression with treatment/control status as one of the predictors.

In studying estimation of the overall effect in random-effects meta-analyses of the mean difference (MD),  the standardized mean difference (SMD), and the log-odds-ratio (LOR), we found that SSW, a weighted mean whose weights involve only the studies' arm-level sample sizes, performed well, avoiding shortcomings associated with estimators that use inverse-variance weights based on estimated variances (\cite{BHK2018SMD, BHK2020LOR}).

We also previously studied $Q_F$, a version of Cochran's $Q$ statistic \citep{cochran1954combination} for assessment of heterogeneity that uses those effective-sample-size-based weights. That work produced favorable results when the measure of effect is the mean difference (MD) (\cite{Qfixed}) or the standardized mean difference (SMD) (\cite{Qfixed}). Here, for DSM, we study by simulation an approximation to the distribution of $Q_F$.

We also introduce and investigate two new estimators of $\tau^2$ for DSM, based on $Q_F$: an estimator based on the first moment of $Q_F$ and a novel median-unbiased estimator. We study, by simulation, the bias of point estimators of $\tau^2$ and the coverage of confidence intervals for $\tau^2$.  We also study the bias and coverage of SSW-based point and interval estimators of $\Delta$. For comparison we include the usual version of $Q$ ($Q_{IV}$), which uses inverse-variance weights, and familiar point and interval estimators of $\tau^2$ and $\Delta$.

In Section~\ref{sec:DSM}, we review estimation of DSM and its variance. In section~\ref{sec:Approx} we describe approximations to the distributions of $Q_F$ and $Q_{IV}$ for DSM. Section~\ref{sec:tau} discusses approaches for estimating $\tau^2$ and $\Delta$ in the random-effects model (REM) for DSM.  In Section~\ref{simul} we describe our simulation and report results for performance of these methods. The user-friendly R programs
implementing our methods are available in \cite{DSMprogs}.

\section{Sample properties of the DSM} \label{sec:DSM}

Consider a study with $n_C$  observations $X_{iC}\sim N(  \mu_C,\sigma_C^2)$ in the control arm and $n_T$  observations $X_{iT}\sim N(  \mu_T,\sigma_T^2)$ in the treatment arm. We denote the total sample size by $n = n_C + n_T$ and the fraction in the control arm by $q = n_C/N$.
The standardized sample means are  $d_j = \bar{X}_j/s_j$ for $j = $ T or C, where $s_j$ is the square root of the sample variance $s_j^2$.  $d_j$ is a biased estimate of $\delta_j = \mu_j / \sigma_j$.  Using the correction of \cite{hedges1983random}, $g_j = J(n_j - 1)d_j$ is an unbiased estimate of  $\delta_j$, where $J(m) = \Gamma(m/2) / (\sqrt{m/2} \Gamma((m - 1)/2))$. It has a scaled noncentral t-distribution, $$ {n_j^{1/2}} g_j /J(n_j - 1) \sim t(n_j^{1/2} \delta_j, n_j - 1).$$ By a derivation similar to that in \cite{hedges1983random}, its variance is
\begin{equation} \label{eq:var_g}
\var(g_j) = v_j^2 = \frac{(n_j - 1) J^2(n_j - 1)} {n_j (n_j - 3)} \left(1 + n_j \delta^2_j \right) - \delta_j^2
\end{equation}
and  an unbiased estimate of its variance is
\begin{equation} \label{eq:g_var}
\widehat{\var(g_j)} = \hat v_j^2  = \frac{1}{n_j} + \left(1 - \frac{(n_{j} - 3)}
{(n_{j} - 1) J^2(n_{j} - 1)} \right) g^2_{j}.
\end{equation}

The DSM $\Delta = \delta_T - \delta_C$  is estimated by $\hat\Delta = g_T - g_C$; and its variance $v^2 = \var(g_T) + \var(g_C)$, assuming independence of T and C, is estimated by $\hat v^2 = \widehat{\var(g_T)} +  \widehat{\var(g_C)}$.  We denote the effective sample size by $\tilde n = n_T n_C / n$. Then
\begin{equation} \label{eq:g_varful}
\hat v^2 = \frac{1}{\tilde n} + \left(1 - \frac{(n_{T} - 3)}
{(n_{T} - 1) J^2(n_{T} - 1)} \right) g^2_{T} + \left(1 - \frac{(n_{C} - 3)}
{(n_{C} - 1) J^2(n_{C} - 1)} \right) g^2_{C}.
\end{equation}

\cite{Bentkus2007} investigated the large-sample properties of a t-statistic $T_n = \sqrt{n} \bar{X} / s_n$ for a general underlying distribution.
From their results it follows that, when the observations in the T and C arms arise from the same family of distributions with finite second and fourth moments, the DSM is asymptotically normally distributed:
\begin{equation}\label{eq:as_var}
\hat\Delta \xrightarrow{\text{{D}}} N(\Delta, \frac{1} {\tilde{n}} - \left(\frac{\delta_T} {n_T} + \frac{\delta_C} {n_C}\right) \alpha_3 + \left(\frac{\delta^2_T} {n_T} + \frac{\delta_C^2} {n_C} \right) (\alpha_4 - 1)/4),
\end{equation}
where $\alpha_k = \e((X - \mu)^k) / \sigma^k$ for $k = 3, \;4$.

However, our simulations use only the underlying normal distribution.

\section{Approximations to the distributions of $Q_F$ and $Q_{IV}$} \label{sec:Approx}

Cochran's $Q$ statistic is a weighted sum of the squared deviations of the estimated effects $\hat\Delta_k$ from their weighted mean $\bar\Delta_w = \sum w_i \hat\Delta_k / \sum w_k$:
\begin{equation} \label{Q}
Q = \sum w_k (\hat{\Delta}_k - \bar{\Delta}_w)^2.
\end{equation}
In \cite{cochran1954combination}  $w_k$ is the reciprocal of the \textit{estimated} variance of $\hat{\Delta}_k$. We denote this traditional version of $Q$ with inverse-variance weights by $Q_{IV}$. In meta-analysis those $w_k$ come from the fixed-effect model.
\cite{dersimonian2007random} discussed a version of $Q$ in which the $w_k$ are arbitrary positive constants. In what follows, we examine $Q_F$, previously studied by \cite{Qfixed}, in which $w_k = \tilde{n}_k$.

The standard approximation to the distribution of $Q_{IV}$ is the chi-square distribution with $K - 1$ degrees of freedom. A few estimators of $\tau^2$, such as \cite{dersimonian1986meta} (DL) and \cite{mandel1970interlaboratory} (MP), are based on the chi-square moments. However, it is well known that this approximation is usually not satisfactory for small to moderate sample sizes.

For DSM, $Q_F$ is a quadratic form in differences of noncentral $t$ variates.  Because of their asymptotic normality, the algorithm of \cite{Farebrother1984} may provide a satisfactory approximation, especially for larger sample sizes. To apply it, we again plug in estimated variances. We investigate the quality of that approximation, which we denote by F SSW.

 For comparison, our simulations include the generic chi-square approximation to the distribution of $Q_{IV}$.

 \section{Point and interval estimators of $\tau^2$ and $\Delta$} \label{sec:tau}

\subsection{Point estimators of $\tau^2$ for DSM}

Under the random-effects model, it is straightforward to obtain the first moment of $Q_F$ as
\begin{equation} \label{M1Q}
\e(Q_F) =  W \left[ \sum p_k (1 - p_k) (\e(v_k^2) + \tau^2) \right],
\end{equation}
where  $W = \sum w_k$ and   $p_k = w_k / W$.\\
This expression, derived in \cite{Qfixed}, is similar to Equation (4) in \cite{dersimonian2007random};  they use the conditional variance  $\hat{v}_k^2$ instead of its unconditional mean $\e(v_k^2)$. 

From this equation,  \cite{dersimonian2007random} 
derived the conditional moment estimator
\begin{equation} \label{tau_DSK}
\hat\tau^2_{MC} = \max \left(0, \;\frac{Q / W - \sum p_k (1 - p_k) \hat v_k^2} {\sum p_k (1 - p_k)} \right).
\end{equation}
 For DSM we study this estimator, denoted by SSC, with the effective-sample-size weights, $\tilde{n}_k$.

The estimator  $\hat{\tau}^2_{MC}$
arose from setting the observed value of $Q$ equal to its expected value and solving for $\tau^2$. Instead of the expected value, one could use the median of the non-null distribution of $Q$ given $\tau^2$. If the true (or approximate) cumulative distribution function is $F(\cdot | \tau^2)$, a point estimator of $\tau^2$ can be found as
$$\hat{\tau}^2_{med} = \max(0, \; \{\tau^2: F(Q | \tau^2) = 0.5\} ).$$
When the Farebrother approximation to the distribution of $Q$ (Section~\ref{sec:Approx}) is used, with the conditional estimated variances, we denote the resulting estimator by SMC.

For comparison, our simulations (Section~\ref{simul}) include three estimators that use inverse-variance weights: DL, REML, and MP.

\subsection{Interval estimators of $\tau^2$ for DSM}

Straightforward use of the cumulative distribution function $F(\cdot | \tau^2)$ also yields a $ 100(1 - \alpha)\% $ confidence interval for $\tau^2$: $$\{\tau^2\geq 0: F(Q | \tau^2) \in [\alpha/2, 1 - \alpha/2] \}.$$
We use  the conditional estimated variances in the Farebrother approximation to $F$; we refer to the resulting profile estimator as the FPC  (Farebrother Profile Conditional) interval. \cite{Jackson2013CI} introduced a similar approach. The FPC interval can be obtained from the \textit{confint} procedure in \textit{metafor} for GENQ or GENQM objects that used $nbar$ weights (\cite{metafor}).

Our simulations (Section~\ref{simul}) also include the Q-profile (QP) interval (\cite{viechtbauer2007confidence}) and the profile-likelihood (PL) interval (\cite{Hardy_1996_StatMed_619}).

\subsection{Point and interval estimators of $\Delta$}

We consider a point estimator of $\Delta$ with the effective-sample-size weights $\tilde n_k$, denoted by SSW.  For comparison, our simulations (Section~\ref{simul}) include three estimators that use inverse-variance weights $w_k = \hat v_k^2 + \hat \tau^2$ with $\tau^2$ estimated by the DL, REML, and MP methods.

For interval estimation of $\Delta$, we consider two confidence intervals centered at SSW and using SMC or SSC estimators of $\tau^2$ in estimating the standard error of SSW in combination with percentiles from the normal distribution. We refer to these intervals also as SMC and SSC. We also consider the standard DL-, REML-, and MP-based intervals and, additionally,  the HKSJ interval \citep{HA&KN&SI11, sidik2002simple}, which is based on the DL estimate of $\tau^2$ and uses t-quantiles.

\section{Simulation design  and results} \label{simul}

\subsection{Simulation design}

Our simulation design follows that described in \cite{BHK2018SMD}. Briefly,
we varied five parameters: the standardized mean in the control arm ($\delta_C$),  the overall true effect ($\Delta$), the between-studies variance ($\tau^2$), the number of studies ($K$), and the studies' total sample size ($n$ or $\bar{n}$). The proportion of observations in the control arm ($f$) was fixed at $1/2$.   Table~\ref{tab:design} lists the values of each parameter.

The values of $\Delta$ (= $-2$, $-1$, $-0.5$, 0, 0.5, 1, 2) and $\delta_C$ (= $-2.5$, $-1$, 0, 1, 2.5) aim to represent the range containing most values encountered in practice, as tabulated by us from  8628 studies using the SMD from issue 4 of the Cochrane database (2004, compact disk edition).

The values of $\tau^2$ (0, 0.1, 0.5, 1, 1.5) align well  with the results of \cite{Rubio-Aparicio_2018_BehavResMeth_2057}.

The numbers of studies ($K$ = 5, 10, and 30) reflect the sizes of many meta-analyses and have yielded valuable insights in previous work.

Many studies allocate subjects equally to the two groups ($f = 1/2$), and rough equality holds more widely (as in the studies analyzed by Rubio-Aparicio et al.).

In practice, many studies' total sample sizes fall in the ranges covered by our choices ($n$ = 20, 40, 100, and 250 when all studies have the same $n$, and $\bar{n}$ =  30, 60, 100, and 160 when sample sizes vary among studies). The choices of unequal individual sample sizes follow a suggestion of \cite{sanches-2000}, who constructed the studies' sample sizes to have skewness 1.464, which they regarded as typical in behavioral and health sciences.

We generated the studies' true effects from a normal distribution:  $\Delta_{k} \sim N(\Delta, \tau^2)$. The  values of $\delta_C$ and $\Delta_k$ defined the values of $\delta_{kT}$, and the standardized means $g_{kC}$ and $g_{kT}$ were generated from the respective scaled noncentral $t$ distributions.  We used a total of $10,000$ repetitions for each combination of parameters.

Overall, we considered 2100 combinations of parameters for the case of equal sample sizes and 2100 combinations for unequal sample sizes.. R statistical software \citep{rrr} was used for simulations. The user-friendly R programs implementing our methods are available at https://osf.io/3gytv .


\begin{table}[ht]
	\caption{ \label{tab:design} \emph{Values of parameters in the simulations}}
	\begin{footnotesize}
		\begin{center}
			\begin{tabular}
				{|l|l|l|}
				\hline
				Parameter & Equal study sizes & Unequal study sizes \\
                                	& & \\
				\hline
				$K$ (number of studies) & 5, 10, 30 & 5, 10, 30 \\
				\hline
				$n$ or $\bar{n}$ (average (individual) study size ---  & 20, 40, 100, 250 & 30 (12,16,18,20,84), \\
				total of the two arms) &  & 60 (24,32,36,40,168), \\
				For  $K = 10$ and $K = 30$, the same set of unequal & & 100 (64,72,76,80,208), \\
				 study sizes is used twice or six times, respectively. & &160 (124,132,136,140,268) \\
                			\hline
				$f$ (proportion of observations in the control arm) & 1/2 &1/2  \\
				\hline
                			$\delta_{C}$ (standardized mean in the control arm) & $-2.5$, $-1$, 0, 1, 2.5 & $-2.5$, $-1$, 0, 1, 2.5\\
                			\hline
				$\Delta$ (true value of DSM) & $-2$, $-1$, $-0.5$, 0, 0.5, 1, 2 & $-2$, $-1$, $-0.5$, 0, 0.5, 1, 2 \\
                			$\tau^{2}$ (variance of random effects) & 0, 0.1, 0.5, 1, 1.5 & 0, 0.1, 0.5, 1, 1.5\\
                			\hline                			
			\end{tabular}
		\end{center}
	\end{footnotesize}
\end{table}

\subsection{Summary of simulation results}

\subsubsection{  Relative error in the level of the test for heterogeneity of DSM (Appendix A)}

Typically, F SSW overestimates the levels and the $\chi^2$ approximation underestimates them, with relative errors consideraby higher at the lowest levels.  F SSW appears to be considerably better than Chisq outside the square $\{ |\delta_T|<1,\; |\delta_C|<1 \}$, considerably worse within this square, and often equally bad (though with the opposite sign) on the boundary.  The reason may be that the noncentral t distribution approaches asymptotic normality comparatively quickly, as opposed to much slower approach of (near)-central t to normal.

\subsubsection{ Empirical level at $\alpha = .05$ of the test for heterogeneity of DSM (Appendix B)}

Results of this section provide more detail on the results in Section A. When $\delta_C = -2.5$ or $2.5$, the F SSW approximation is considerably better than the ChiSq. ChiSq values are much too low, especially when $K = 30$. Both approximations work well by $n = 250$.

For $\delta_C = -1$, F SSW is better when $K = 5$ or when $\Delta < 0$ (i.e., for the more-extreme negative values of $\delta_T$). And for positive $\Delta$, the ChiSq approximation is better. Similarly, for  $\delta_C = 1$, the F SSW approximation is better for $\Delta > 0$ (i.e., for the more-extreme positive values of $\delta_T$), and ChiSq is better for $\Delta < 0$.

Finally, the results for $\delta_C = 0$ demonstrate striking differences between the two approximations, depending on the distance of both $\delta_C$ and $\delta_T$ from 0. F SSW grossly overestimates the .05 level at $\Delta = 0$, but is nearly perfect at the edges of $\Delta = \pm 2$.

\subsubsection{ Bias of point estimators of $\tau^2$ for DSM (Appendix C)}

The bias of all estimators of $\tau^2$ is highest at $\tau^2 = 0$. For MP, REML, and DL, it then slowly decreases in $\tau^2$, crossing zero at some stage and becoming negative. SMC has the highest positive bias, and REML and especially  DL  have considerable negative bias when $\tau^2 > 0.4$.  SSC and MP are the best estimators of $\tau^2$. Both typically perform reasonably well when $n \geq 40$ or $\bar n \geq 60$. Estimation improves for larger $K$; $K=5$ is the most challenging, requiring larger sample sizes.  In general, MP gives somewhat lower estimates than SSC. Therefore, it has less bias for small $\tau^2$, though it may underestimate large $\tau^2$. For $\delta_C$ and $\delta_T$ within $\pm 1$, SSC often provides very good estimation when $n \geq 20$, especially for $K \geq 10$. When $n \geq 100$, SSC, MP, and (in most scenarios) REML perform reasonably well.

\subsubsection{ Coverage of interval estimators of $\tau^2$ for DSM (Appendix D)}

We considered three interval estimators of $\tau^2$: QP, FPC, and PL. The performance of PL was not satisfactory, especially for small sample sizes and $K = 30$. Its coverage was too high for small values of $\tau^2$, and it demonstrated erratic behavior for large values. Coverage of QP and FPC did not differ too much, though coverage of QP typically was higher (by 1 to 2 percentage points).  Near zero, all intervals had too high coverage. For $\tau^2 \geq 0.1$, FPC provided almost nominal coverage when $n \geq 20$ or $\bar n \geq 30$ for $K = 5$; but for $K \geq 10$, coverage of QP tended to be somewhat closer to nominal, whereas that of FPC hovered at about 93-94\% for $n \geq 30$.

\subsubsection{ Bias of point estimators of $\Delta$  (Appendix E)}

The SSW estimator of $\Delta$ was practically unbiased in all scenarios. All other estimators had very similar and rather substantial bias.   The bias can be both positive and negative. It was highly  positive for large negative values of $\Delta$, and it decreased for less-negative values. As $\Delta$ increased, starting from $\Delta = 0.5$, and sometimes from $\Delta = 0$ (for  $\delta_C \geq 1$), the bias became increasingly negative. The bias typically increased with $\tau^2$ and $K$,  decreased with $n$, and was usually at or below  0.1 for $n \geq 100$.

\subsubsection{ Coverage of interval estimators of $\Delta$  (Appendix F)}

The best interval estimator of $\Delta$ for $K = 5$ is HKSJ, and for $K \geq 10$, SMC, closely followed by SSC. These three estimators provide coverage close to nominal when $n \geq 40$, and often even for $n = 20$.  The rest of the estimators (DL, REML and MP) often result in considerable under-coverage.  In general, coverage is better for smaller absolute values of $\Delta$, and it  improves with increasing $n$, so the standard estimators usually, but not always, provide acceptable coverage when $n \geq 100$. The large $K$ are the most problematic for the standard estimators, but SMC and SSC perform well.

\section*{Acknowledgements}
The work by E. Kulinskaya was supported by the Economic and Social Research Council
[grant number ES/L011859/1].

\clearpage
\bibliography{refs.bib}
\clearpage


\section*{Appendices}
\begin{itemize}
\item Appendix A: Relative error in the level of the test for heterogeneity of DSM
\item Appendix B: Empirical level at $\alpha = .05$ of the test for heterogeneity of DSM
\item Appendix C: Bias in point estimators of $\tau^2$
\item Appendix D: Coverage of 95\% confidence intervals for $\tau^2$
\item Appendix E: Bias in point estimators of $\Delta$
\item Appendix F: Coverage of 95\% confidence intervals for $\Delta$
\end{itemize}

\setcounter{figure}{0}
\setcounter{section}{0}
\clearpage

\section*{Appendix A: Plots of relative error in the level of the test for heterogeneity of DSM for two approximations for the null distribution of $Q$}

Each figure corresponds to a value of the standardized mean in the Control arm $\delta_{C}$  (= $-2.5$, $-1$, 0, 1, 2.5)  and a value of the overall DSM $\Delta$ (= $-2$, $-1$, $-0.5$, 0, 0.5, 1, 2) . \\
The fraction of each study's sample size in the Control arm ($f$) is held constant at 0.5.

For each combination of a value of $n$ (= 20, 40, 100, 250) or  $\bar{n}$ (= 30, 60, 100, 160) and a value of $K$ (= 5, 10, 30), a panel plots, versus the nominal upper tail area (= .005, .01, .05, .01, 0.25, .5), the relative error between the achieved level and the nominal level for two approximations to the null distribution of $Q$:
\begin{itemize}
\item ChiSq (Chi-square approximation with $K - 1$ df, inverse-variance weights)
\item F SSW  (Farebrother approximation, effective-sample-size weights)
\end{itemize}

\clearpage

\setcounter{figure}{0}
\setcounter{section}{0}
\renewcommand{\thefigure}{A.\arabic{figure}}


\begin{figure}[ht]
	\centering
	\includegraphics[scale=0.33]{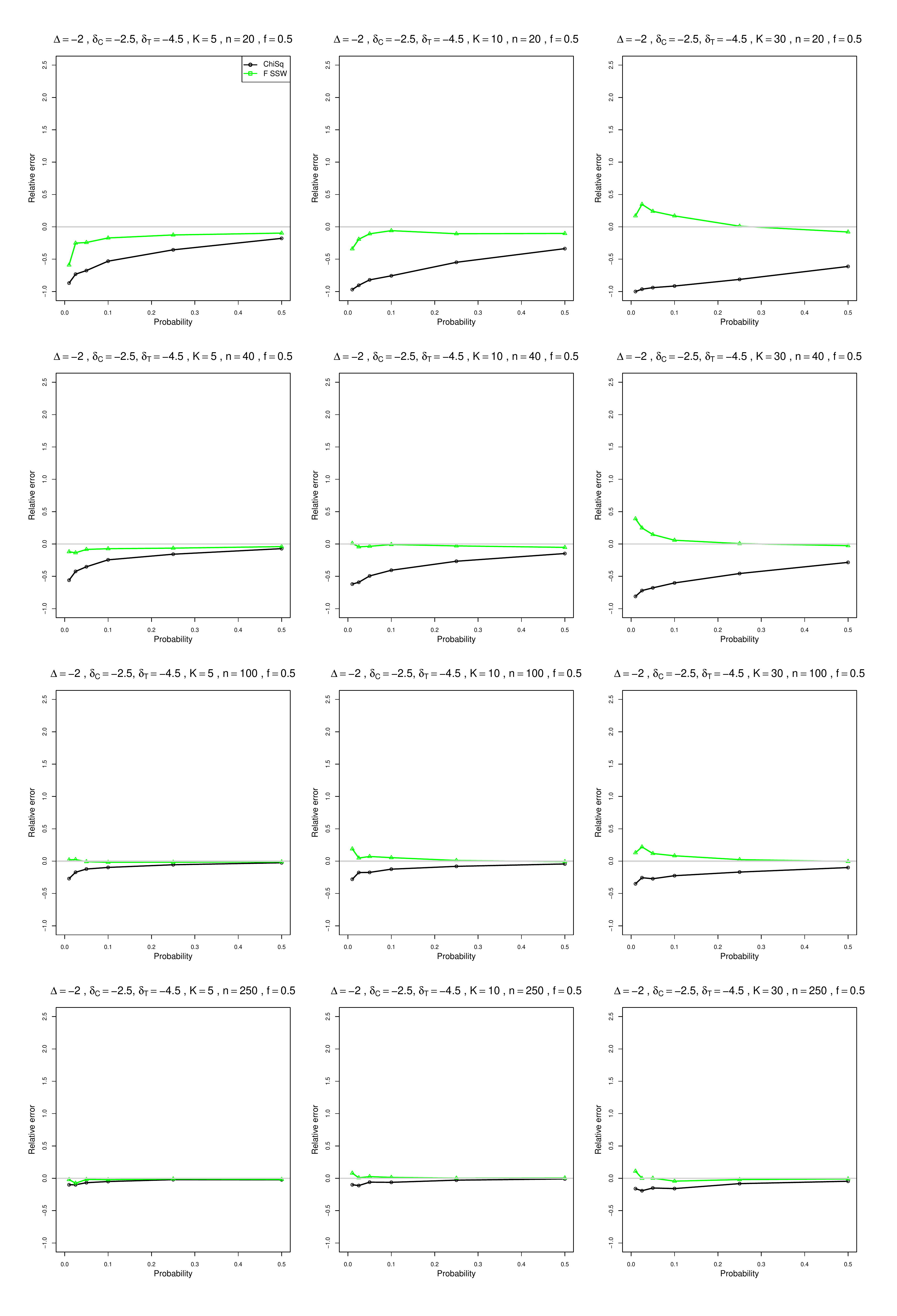}
	\caption{Relative error between the achieved level and the nominal level for two approximations to the null distribution of Q for DSM (Chisq and F SSW) vs upper tail probability, for equal sample sizes $n=20,\;40,\;100$ and $250$, $\delta_{iC} = -2.5$, $\Delta=-2$ and  $f = 0.5$.   }
	\label{Pplot_relative_truncated_deltaC_-25deltaT=-4.5_DSM_equal_sample_sizes.pdf}
\end{figure}

\begin{figure}[ht]
	\centering
	\includegraphics[scale=0.33]
    {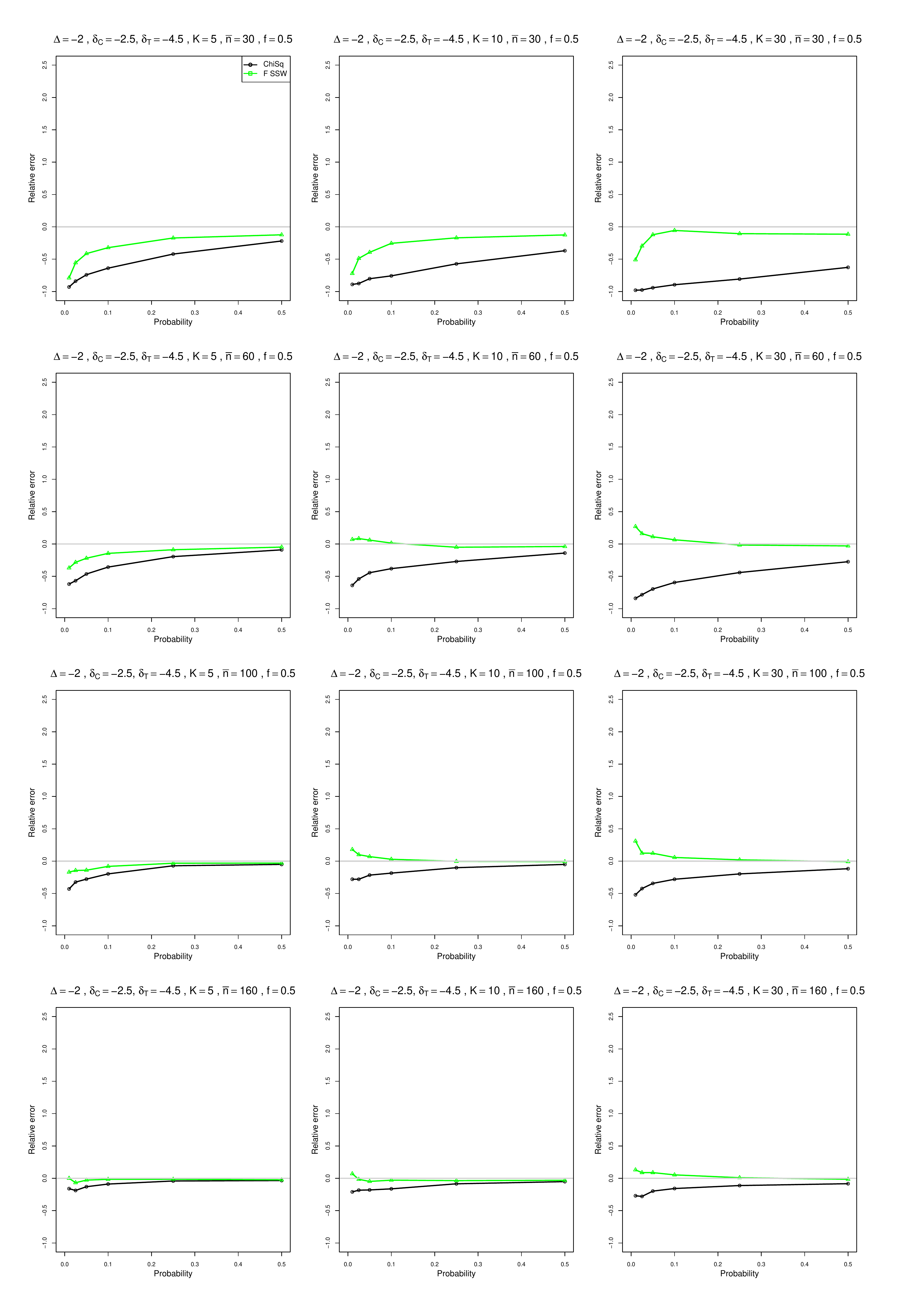}
	\caption{Relative error between the achieved level and the nominal level for two approximations to the null distribution of Q for DSM (Chisq and F SSW) vs upper tail probability, for unequal sample sizes $\bar{n}=30,\;60,\;100$ and $160$, $\delta_{iC} = -2.5$, $\Delta=-2$ and  $f = 0.5$.   }
	\label{Pplot_relative_truncated_deltaC_-25deltaT=-4.5_DSM_unequal_sample_sizes.pdf}
\end{figure}

\begin{figure}[ht]
	\centering
	\includegraphics[scale=0.33]{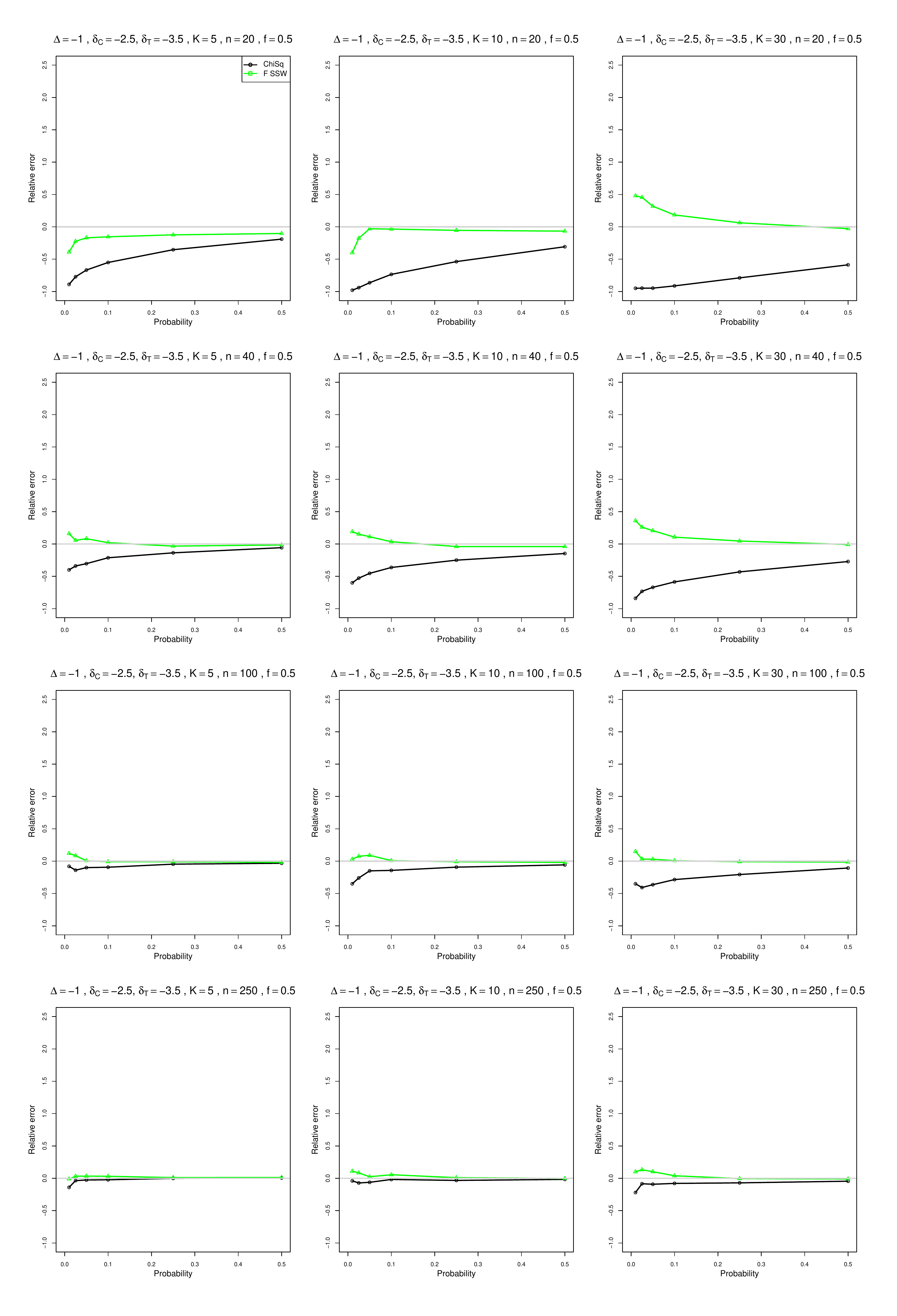}
	\caption{Relative error between the achieved level and the nominal level for two approximations to the null distribution of Q for DSM (Chisq and F SSW) vs upper tail probability, for equal sample sizes $n=20,\;40,\;100$ and $250$, $\delta_{iC} = -2.5$, $\Delta=-1$ and  $f = 0.5$.   }
	\label{Pplot_relative_truncated_deltaC_-25deltaT=-3.5_DSM_equal_sample_sizes.pdf}
\end{figure}

\begin{figure}[ht]
	\centering
	\includegraphics[scale=0.33]{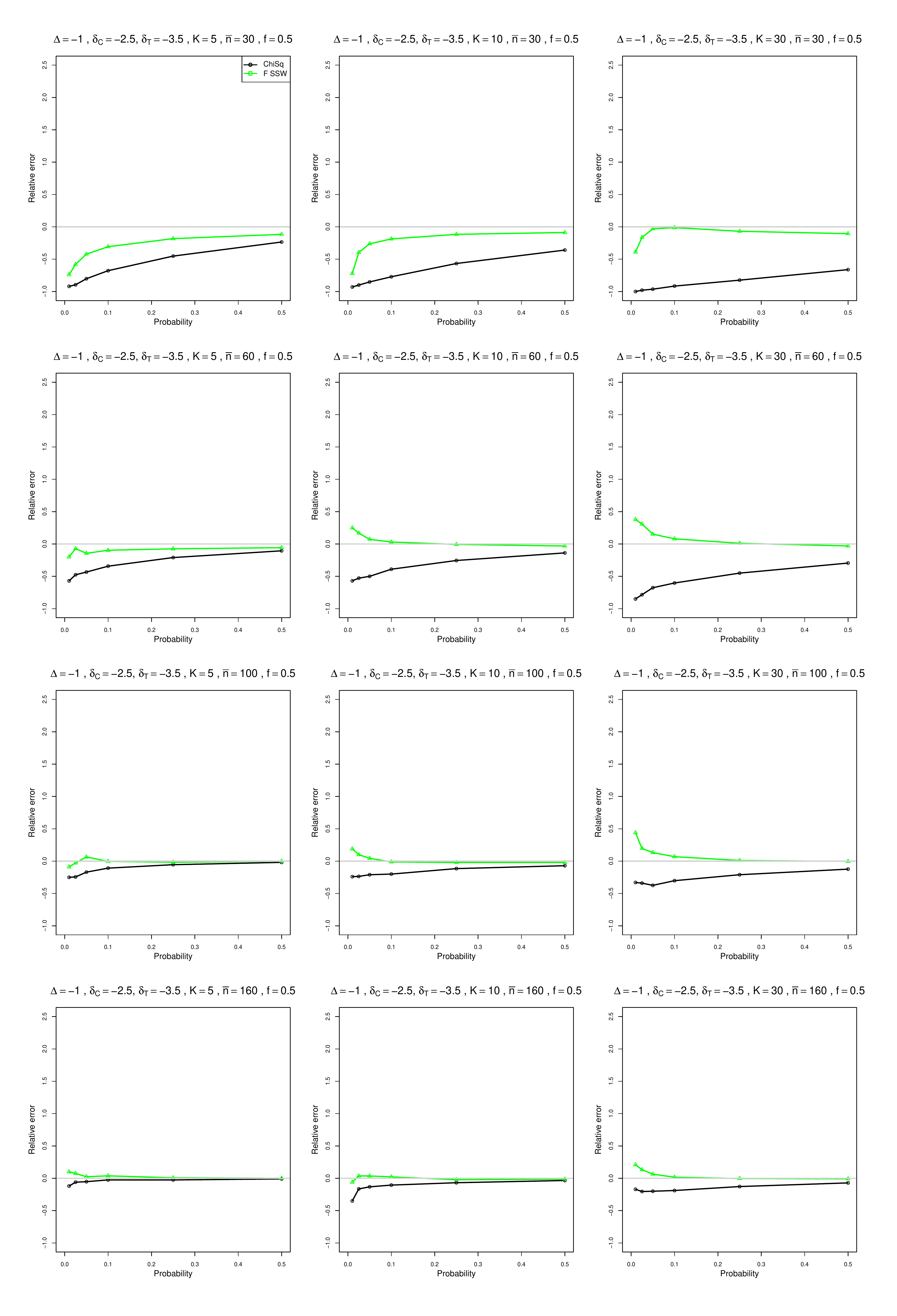}
	\caption{Relative error between the achieved level and the nominal level for two approximations to the null distribution of Q for DSM (Chisq and F SSW) vs upper tail probability, for unequal sample sizes $\bar{n}=30,\;60,\;100$ and $160$, $\delta_{iC} = -2.5$, $\Delta=-1$ and  $f = 0.5$.   }
	\label{Pplot_relative_truncated_deltaC_-25deltaT=-3.5_DSM_unequal_sample_sizes.pdf}
\end{figure}

\begin{figure}[ht]
	\centering
	\includegraphics[scale=0.33]{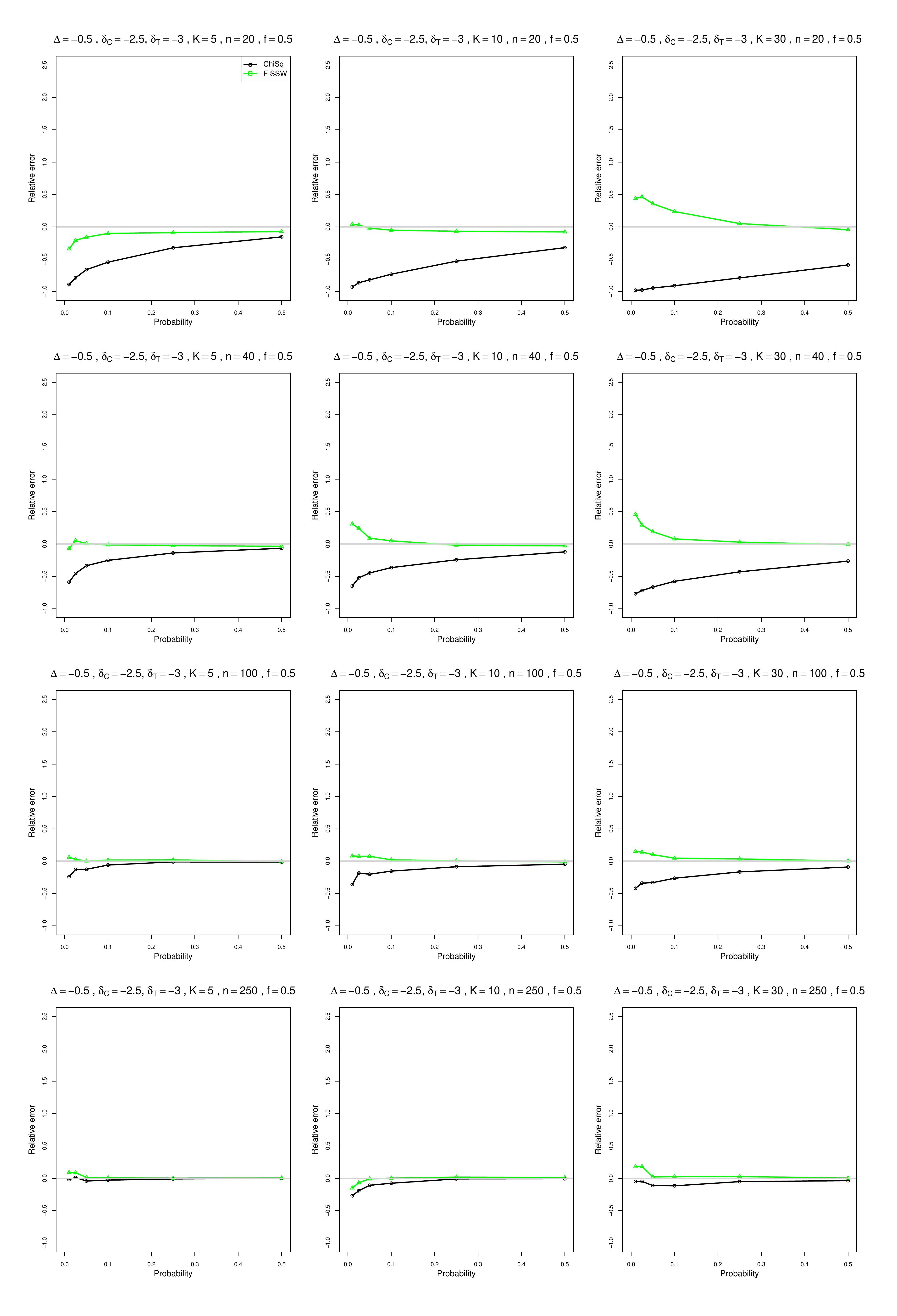}
	\caption{Relative error between the achieved level and the nominal level for two approximations to the null distribution of Q for DSM (Chisq and F SSW) vs upper tail probability, for equal sample sizes $n=20,\;40,\;100$ and $250$, $\delta_{iC} = -2.5$, $\Delta=-0.5$ and  $f = 0.5$.   }
	\label{Pplot_relative_truncated_deltaC_-25deltaT=-3_DSM_equal_sample_sizes.pdf}
\end{figure}

\begin{figure}[ht]
	\centering
	\includegraphics[scale=0.33]{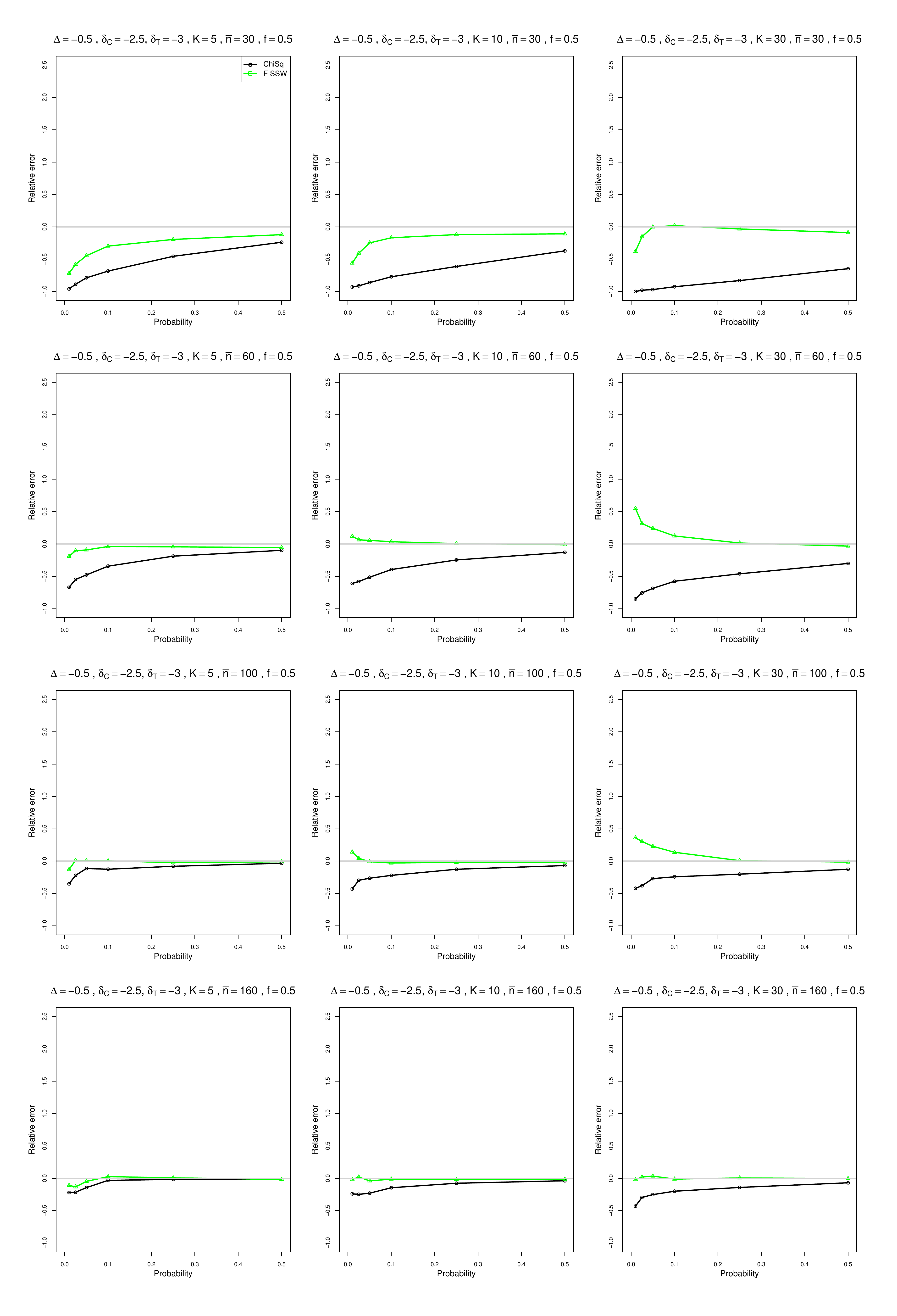}
	\caption{Relative error between the achieved level and the nominal level for two approximations to the null distribution of Q for DSM (Chisq and F SSW) vs upper tail probability, for unequal sample sizes $\bar{n}=30,\;60,\;100$ and $160$, $\delta_{iC} = -2.5$, $\Delta=-0.5$ and  $f = 0.5$.   }
	\label{Pplot_relative_truncated_deltaC_-25deltaT=-3_DSM_unequal_sample_sizes.pdf}
\end{figure}

\begin{figure}[ht]
	\centering
	\includegraphics[scale=0.33]{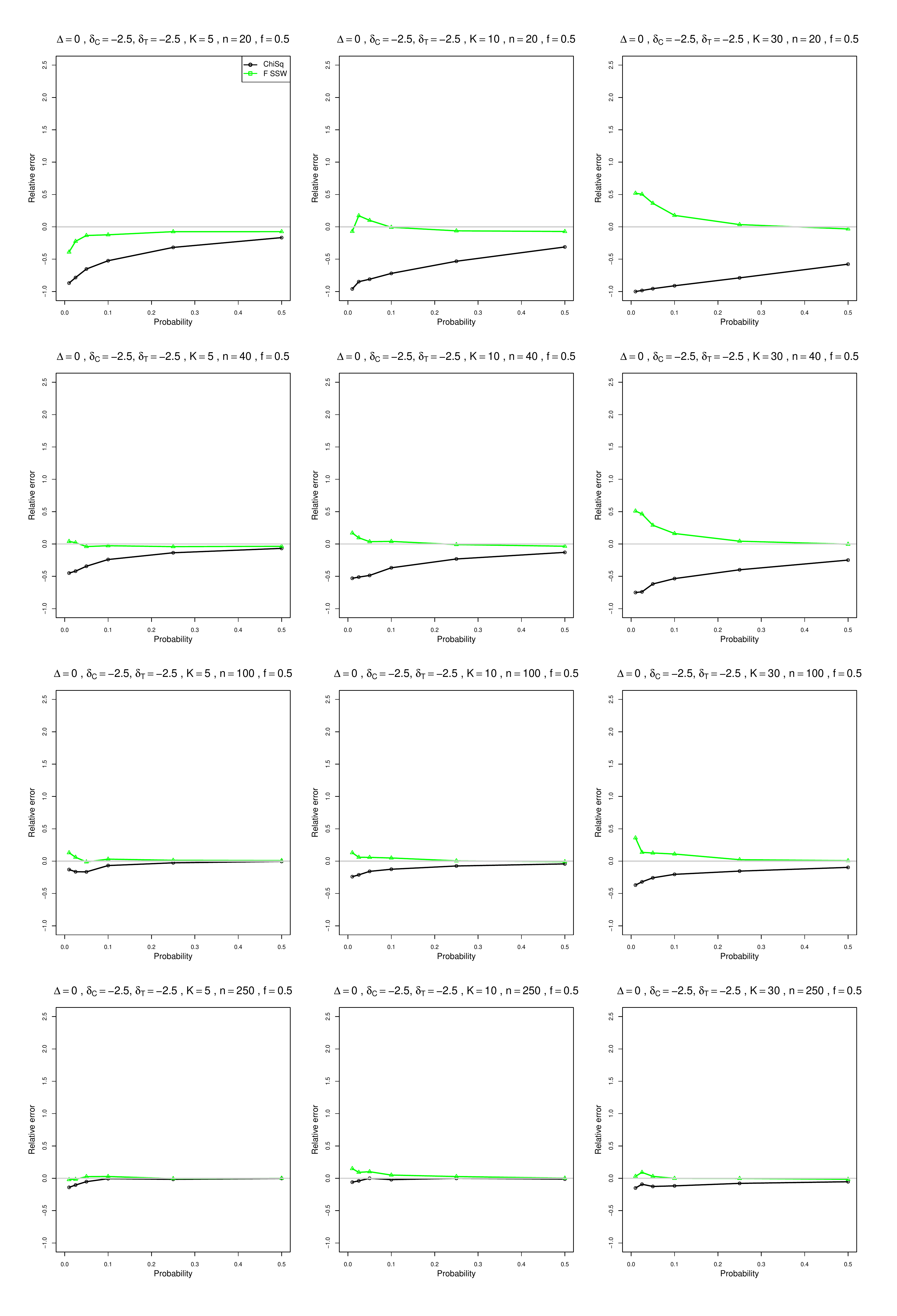}
	\caption{Relative error between the achieved level and the nominal level for two approximations to the null distribution of Q for DSM (Chisq and F SSW) vs upper tail probability, for equal sample sizes $n=20,\;40,\;100$ and $250$, $\delta_{iC} = -2.5$, $\Delta=0$ and  $f = 0.5$.   }
	\label{Pplot_relative_truncated_deltaC_-25deltaT=-2.5_DSM_equal_sample_sizes.pdf}
\end{figure}

\begin{figure}[ht]
	\centering
	\includegraphics[scale=0.33]{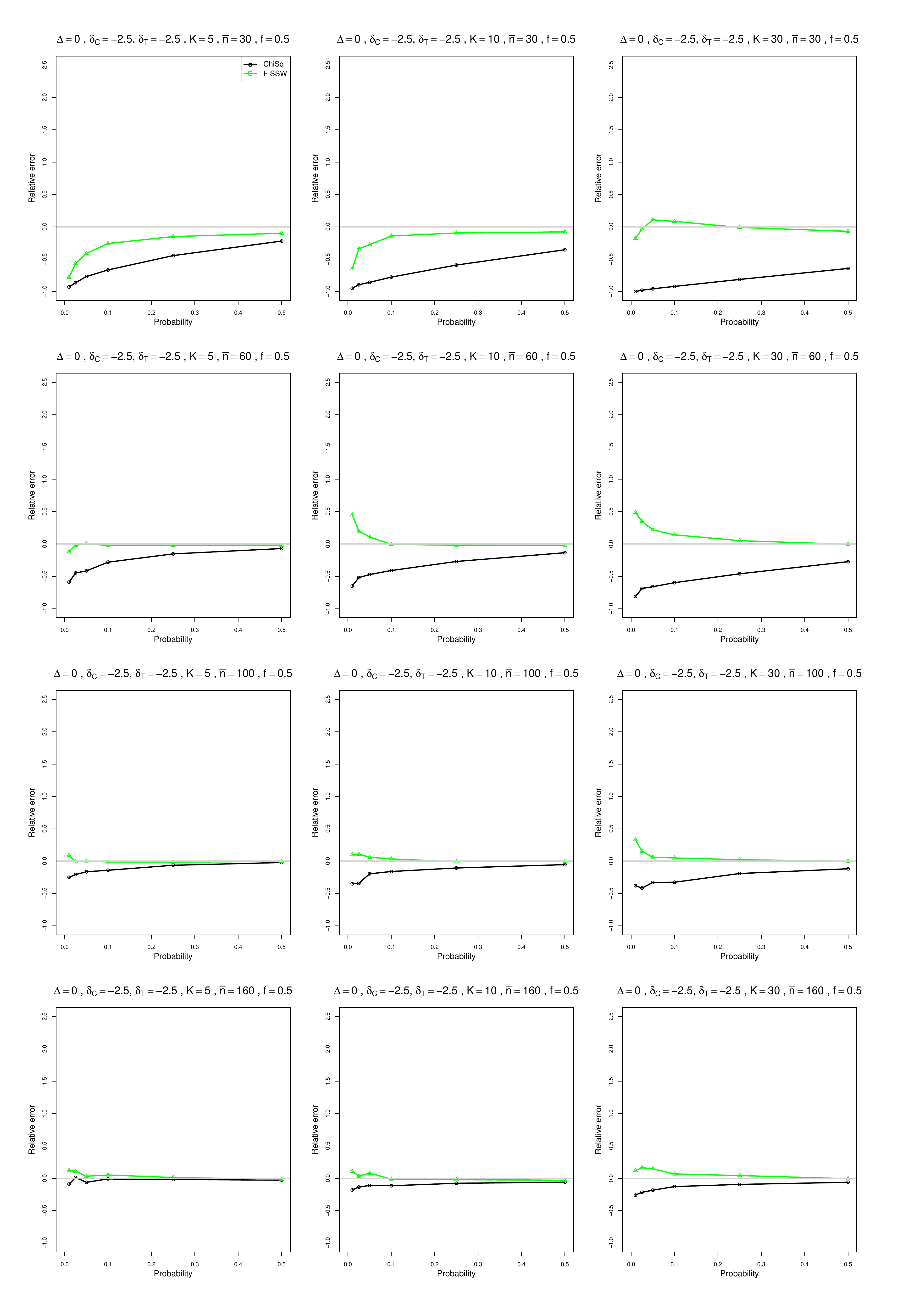}
	\caption{Relative error between the achieved level and the nominal level for two approximations to the null distribution of Q for DSM (Chisq and F SSW) vs upper tail probability, for unequal sample sizes $\bar{n}=30,\;60,\;100$ and $160$, $\delta_{iC} = -2.5$, $\Delta=0$ and  $f = 0.5$.   }
	\label{Pplot_relative_truncated_deltaC_-25deltaT=-2.5_DSM_unequal_sample_sizes.pdf}
\end{figure}

\begin{figure}[ht]
	\centering
	\includegraphics[scale=0.33]{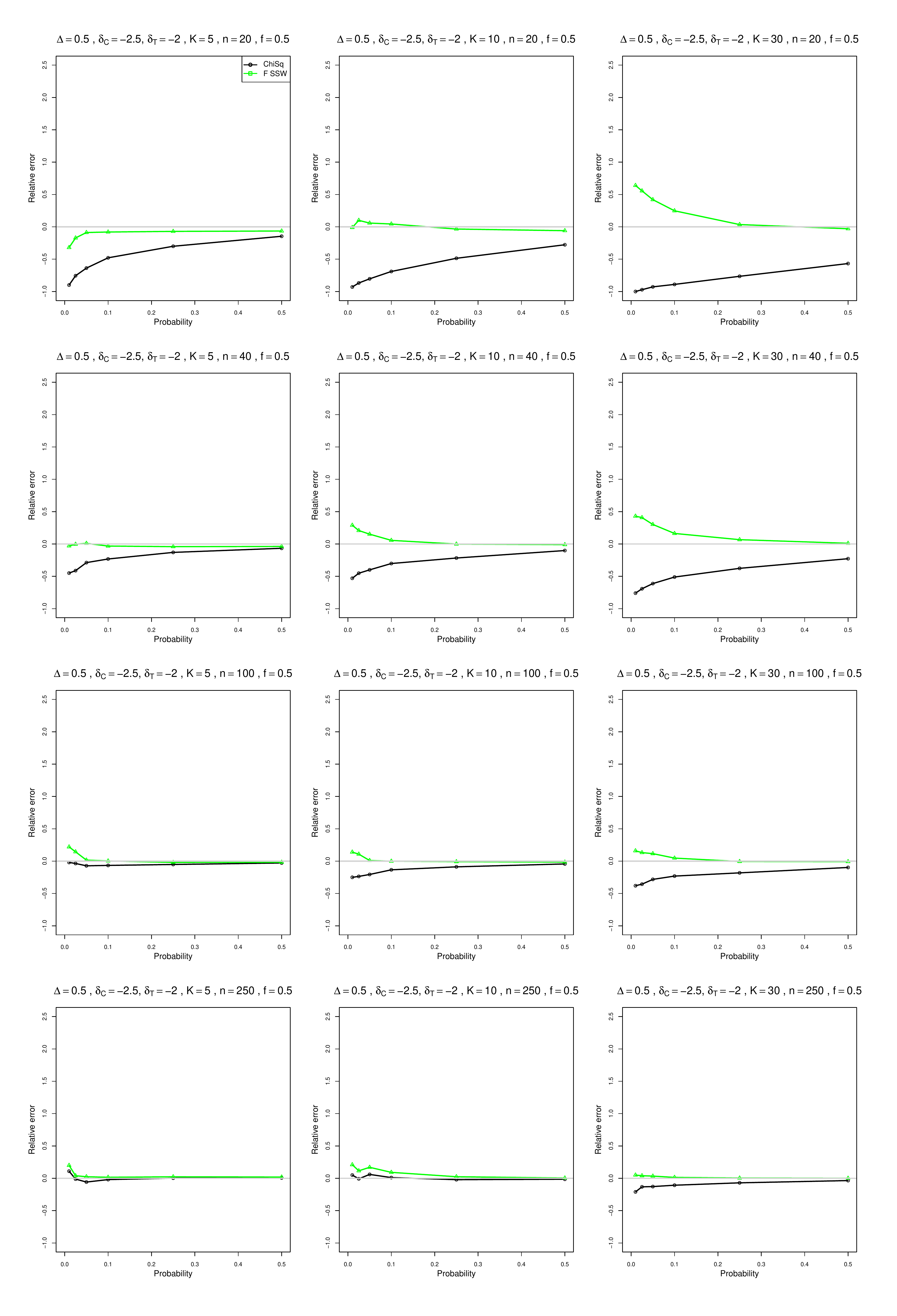}
	\caption{Relative error between the achieved level and the nominal level for two approximations to the null distribution of Q for DSM (Chisq and F SSW) vs upper tail probability, for equal sample sizes $n=20,\;40,\;100$ and $250$, $\delta_{iC} = -2.5$, $\Delta=0.5$ and  $f = 0.5$.   }
	\label{Pplot_relative_truncated_deltaC_-25deltaT=-2_DSM_equal_sample_sizes.pdf}
\end{figure}

\begin{figure}[ht]
	\centering
	\includegraphics[scale=0.33]{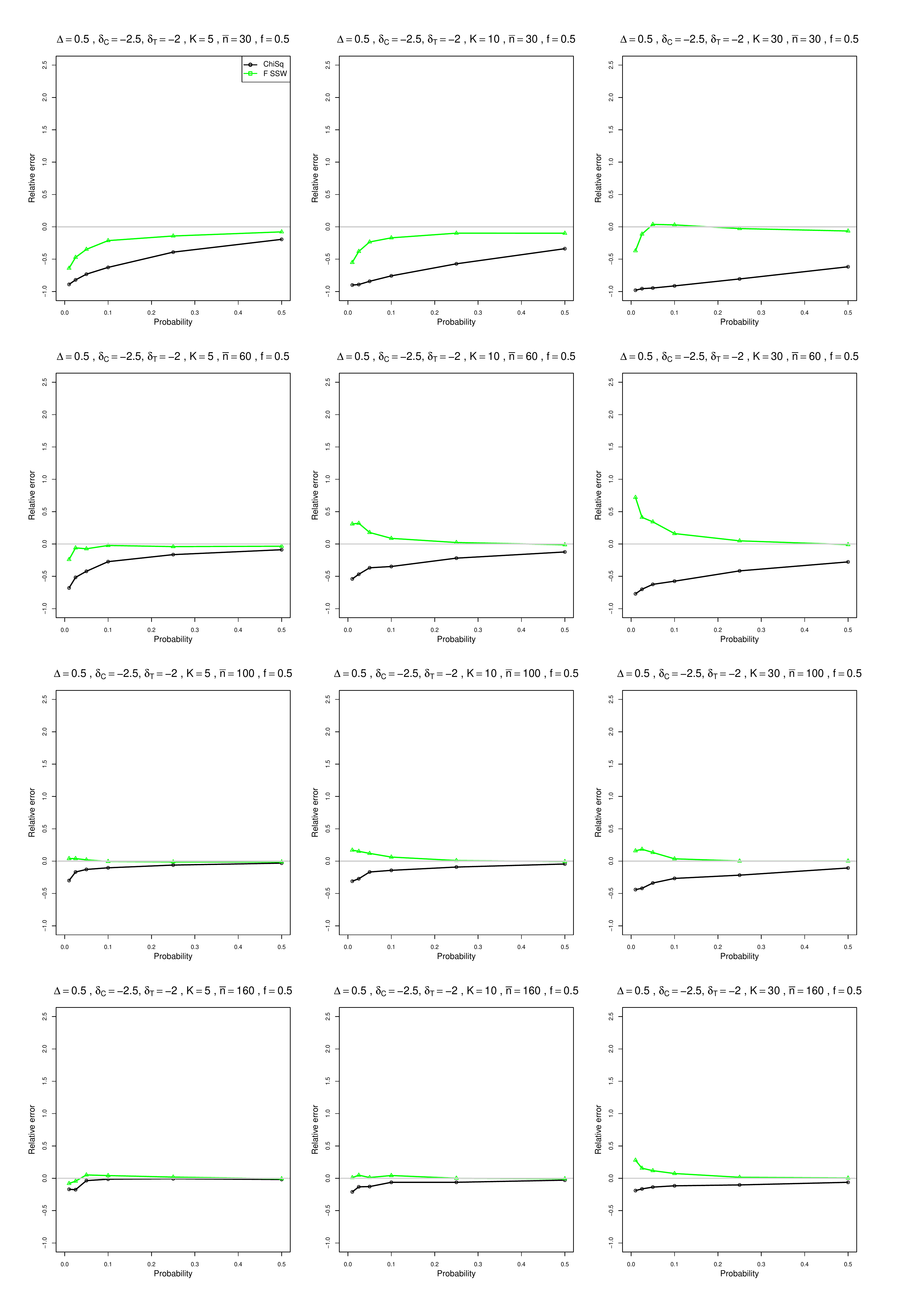}
	\caption{Relative error between the achieved level and the nominal level for two approximations to the null distribution of Q for DSM (Chisq and F SSW) vs upper tail probability, for unequal sample sizes $\bar{n}=30,\;60,\;100$ and $160$, $\delta_{iC} = -2.5$, $\Delta=0.5$ and  $f = 0.5$.   }
	\label{Pplot_relative_truncated_deltaC_-25deltaT=-2_DSM_unequal_sample_sizes.pdf}
\end{figure}

\begin{figure}[ht]
	\centering
	\includegraphics[scale=0.33]{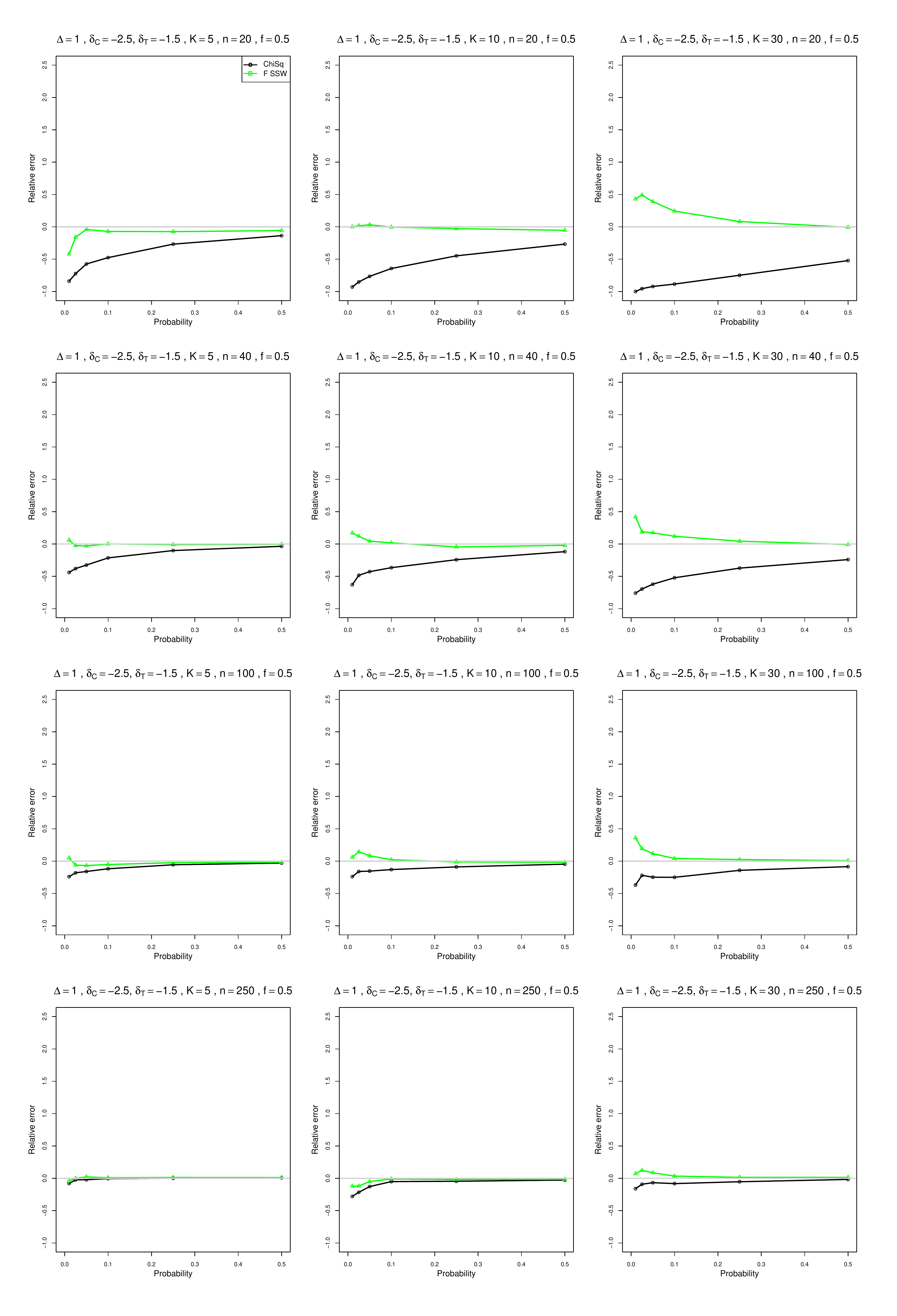}
	\caption{Relative error between the achieved level and the nominal level for two approximations to the null distribution of Q for DSM (Chisq and F SSW) vs upper tail probability, for equal sample sizes $n=20,\;40,\;100$ and $250$, $\delta_{iC} = -2.5$, $\Delta=1$ and  $f = 0.5$.   }
	\label{Pplot_relative_truncated_deltaC_-25deltaT=-1.5_DSM_equal_sample_sizes.pdf}
\end{figure}

\begin{figure}[ht]
	\centering
	\includegraphics[scale=0.33]{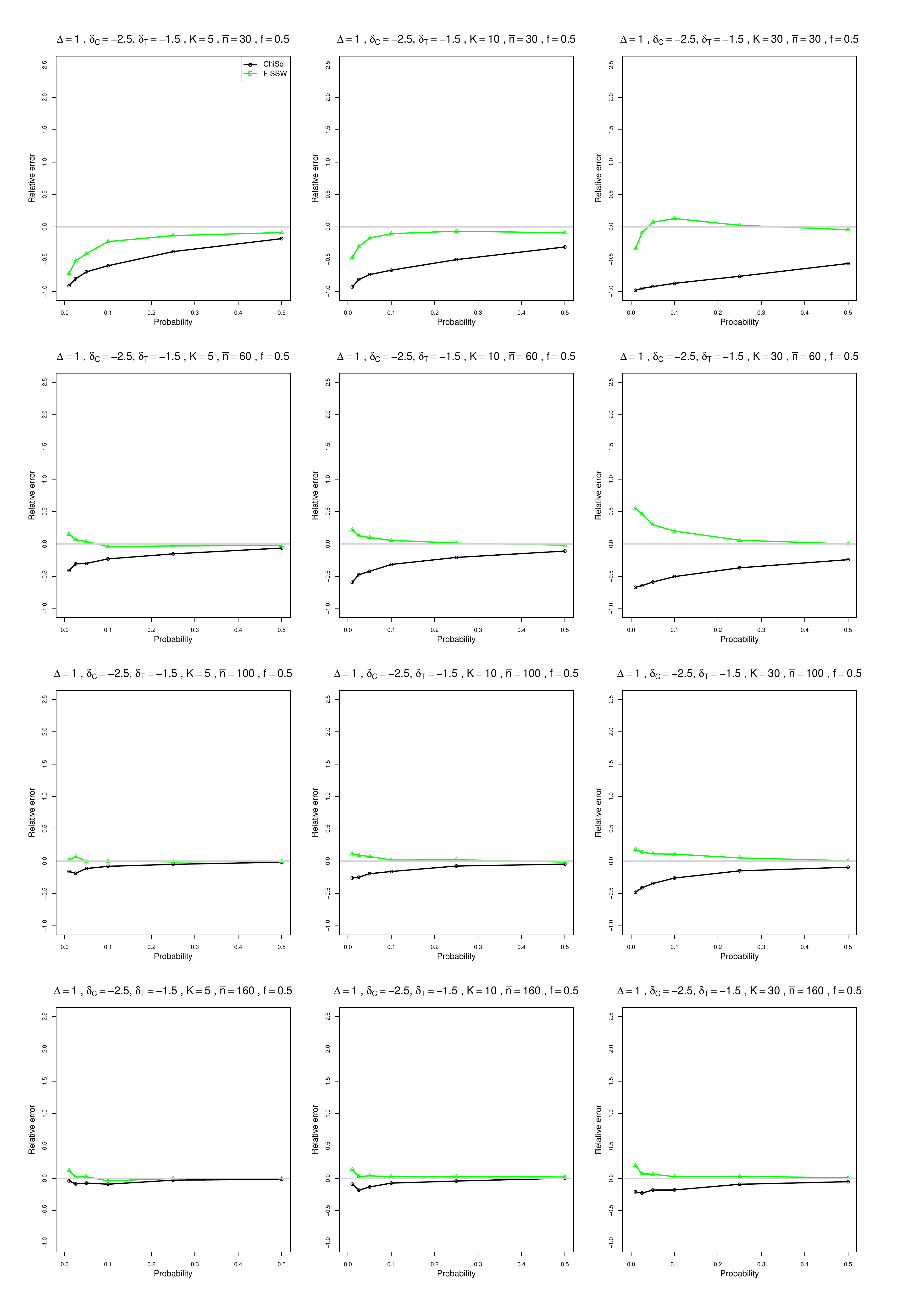}
	\caption{Relative error between the achieved level and the nominal level for two approximations to the null distribution of Q for DSM (Chisq and F SSW) vs upper tail probability, for unequal sample sizes $\bar{n}=30,\;60,\;100$ and $160$, $\delta_{iC} = -2.5$, $\Delta=1$ and  $f = 0.5$.   }
	\label{Pplot_relative_truncated_deltaC_-25deltaT=-1.5_DSM_unequal_sample_sizes.pdf}
\end{figure}

\begin{figure}[ht]
	\centering
	\includegraphics[scale=0.33]{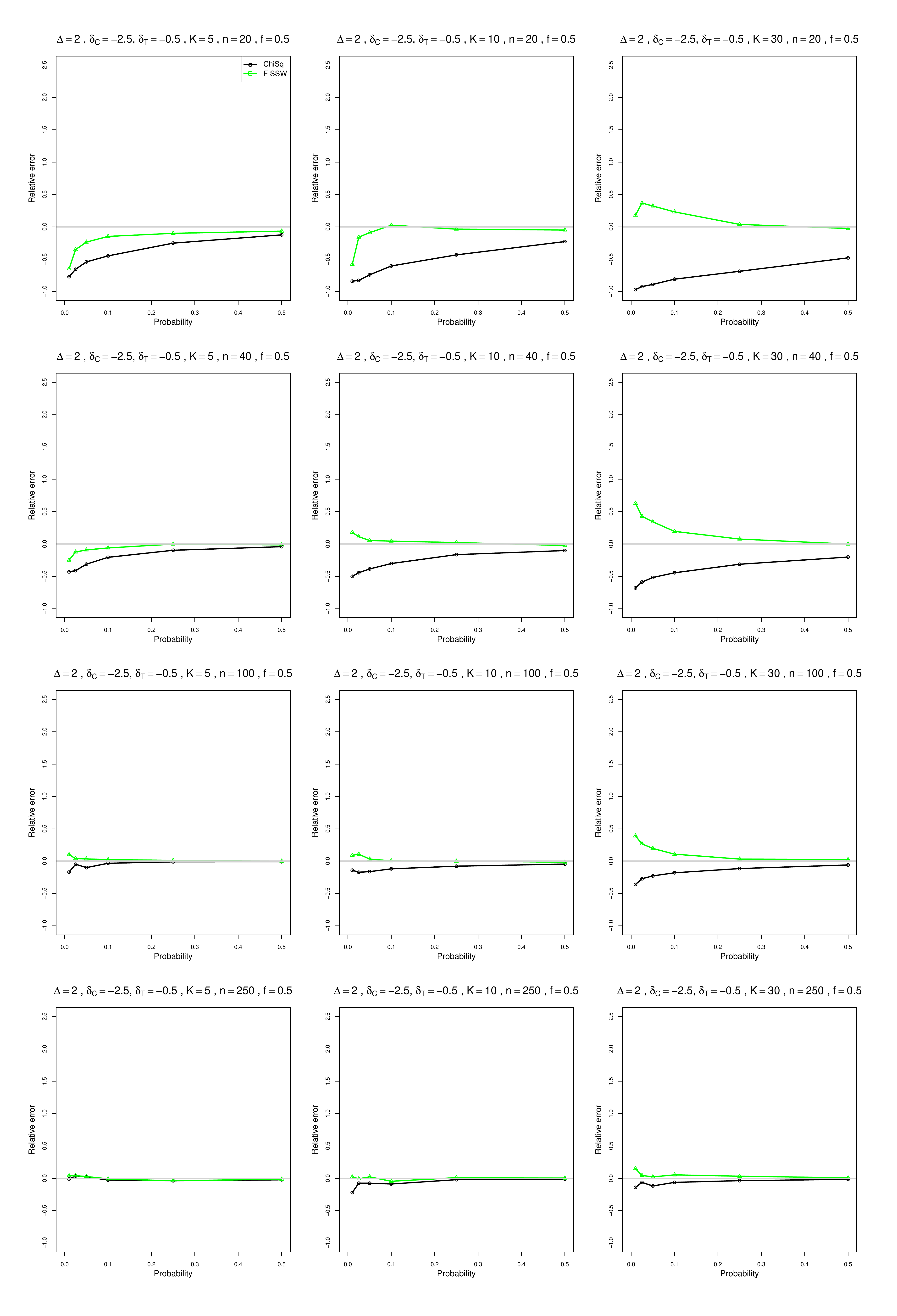}
	\caption{Relative error between the achieved level and the nominal level for two approximations to the null distribution of Q for DSM (Chisq and F SSW) vs upper tail probability, for equal sample sizes $n=20,\;40,\;100$ and $250$, $\delta_{iC} = -2.5$, $\Delta=2$ and  $f = 0.5$.   }
	\label{Pplot_relative_truncated_deltaC_-25deltaT=-0.5_DSM_equal_sample_sizes.pdf}
\end{figure}

\begin{figure}[ht]
	\centering
	\includegraphics[scale=0.33]{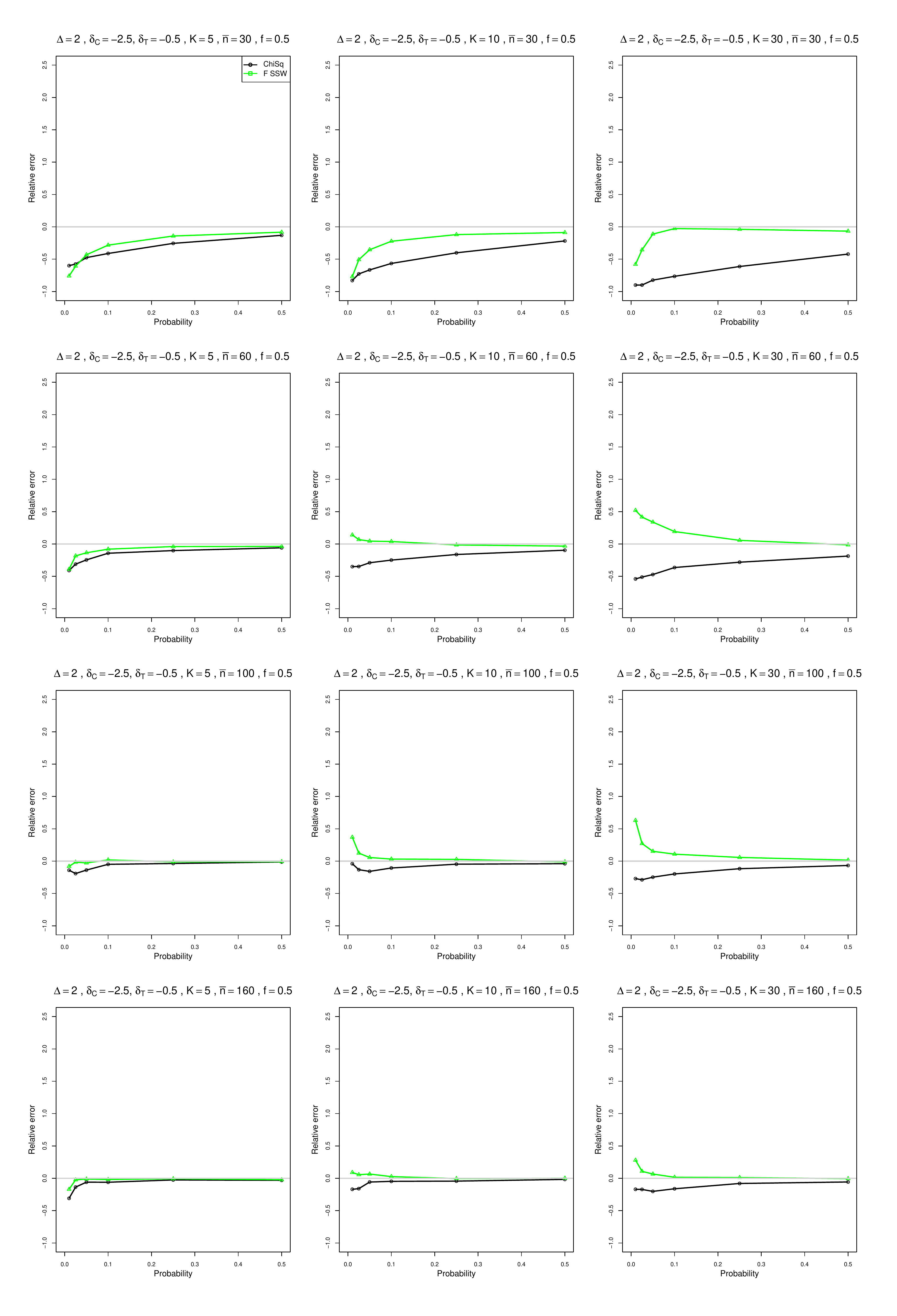}
	\caption{Relative error between the achieved level and the nominal level for two approximations to the null distribution of Q for DSM (Chisq and F SSW) vs upper tail probability, for unequal sample sizes $\bar{n}=30,\;60,\;100$ and $160$, $\delta_{iC} = -2.5$, $\Delta=2$ and  $f = 0.5$.   }
	\label{Pplot_relative_truncated_deltaC_-25deltaT=-0.5_DSM_unequal_sample_sizes.pdf}
\end{figure}
\begin{figure}[ht]
	\centering
	\includegraphics[scale=0.33]{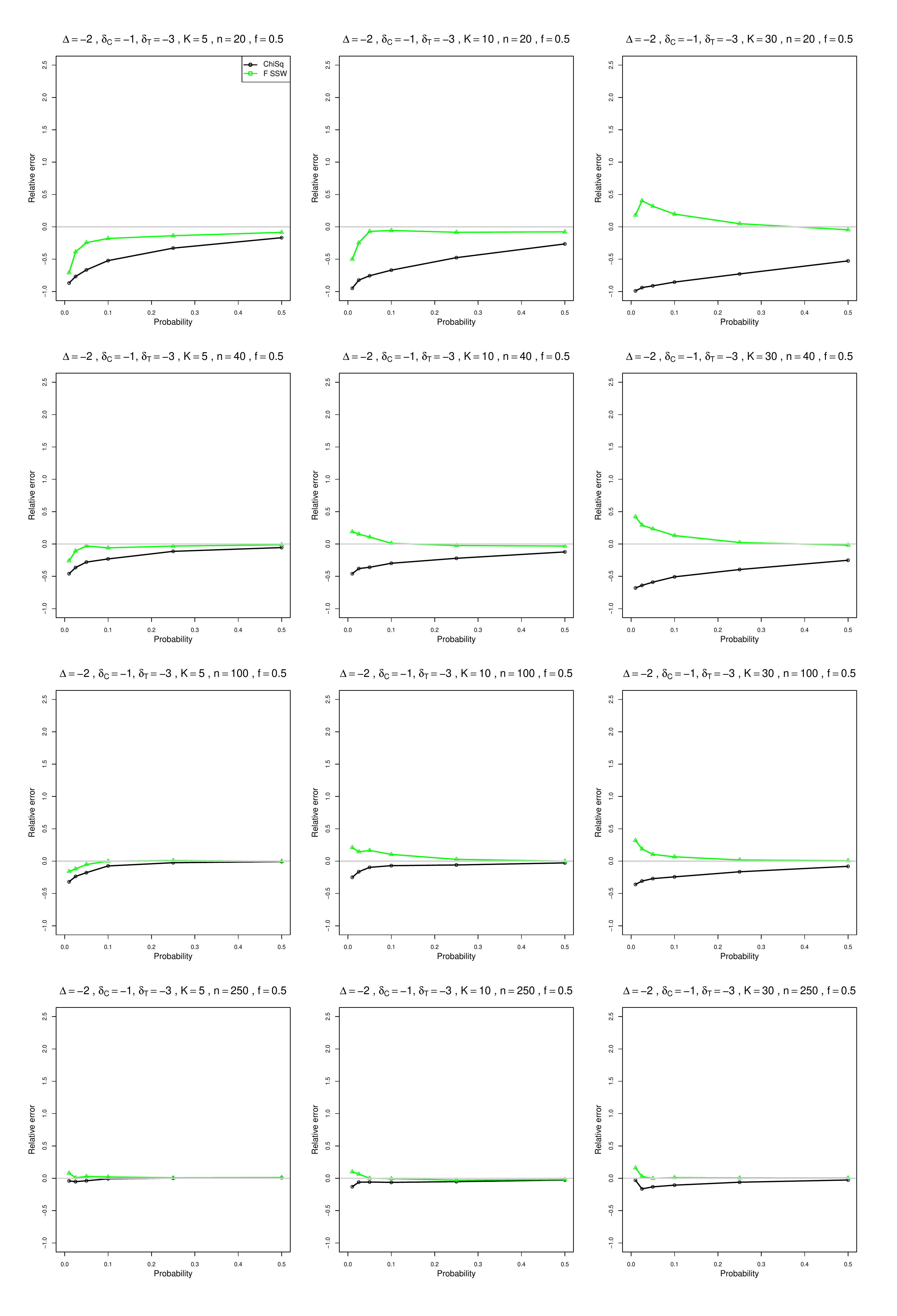}
	\caption{Relative error between the achieved level and the nominal level for two approximations to the null distribution of Q for DSM (Chisq and F SSW) vs upper tail probability, for equal sample sizes $n=20,\;40,\;100$ and $250$, $\delta_{iC} = -1$, $\Delta=-2$ and  $f = 0.5$.   }
	\label{Pplot_relative_truncated_deltaC_-1deltaT=-3_DSM_equal_sample_sizes.pdf}
\end{figure}

\begin{figure}[ht]
	\centering
	\includegraphics[scale=0.33]{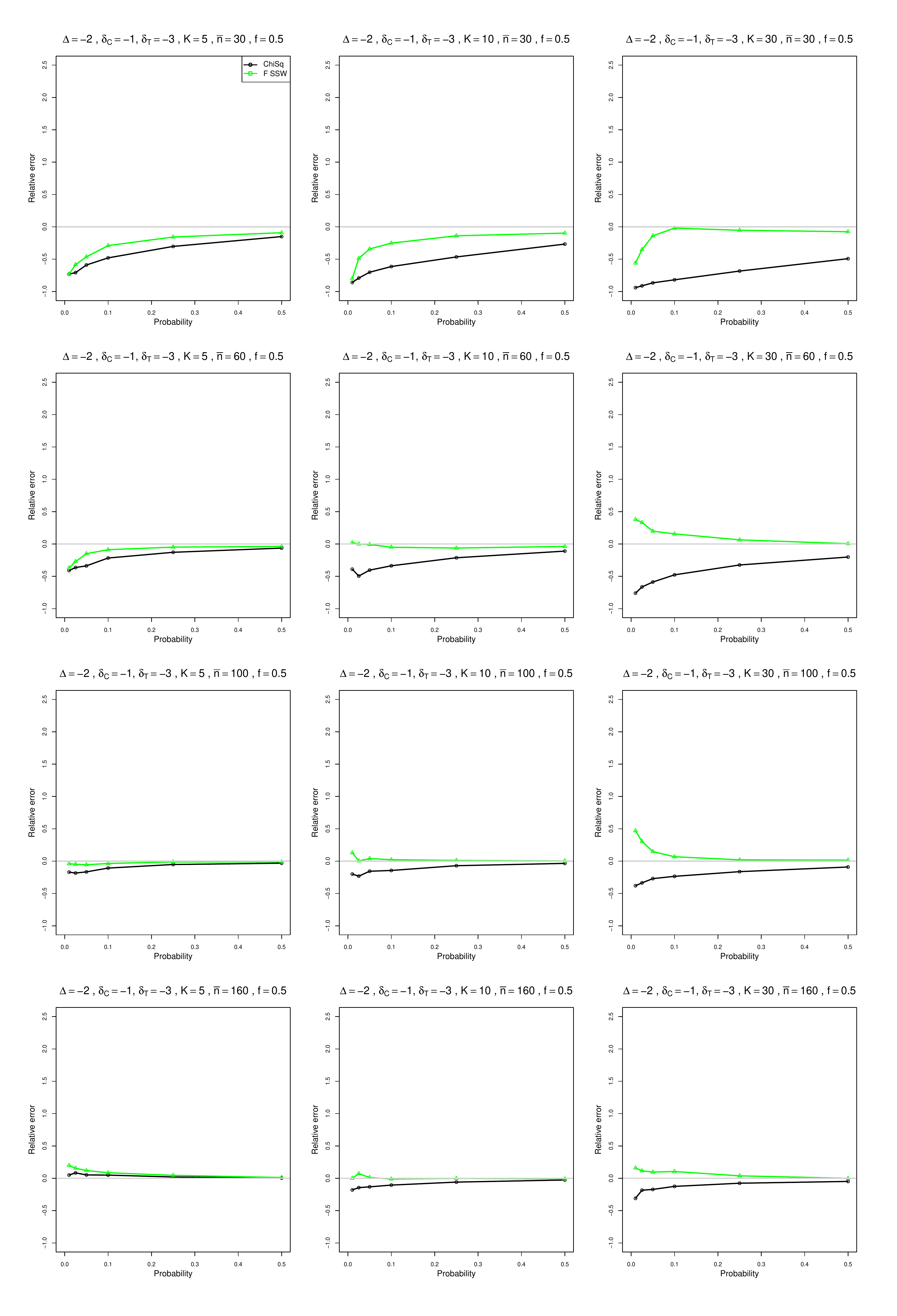}
	\caption{Relative error between the achieved level and the nominal level for two approximations to the null distribution of Q for DSM (Chisq and F SSW) vs upper tail probability, for unequal sample sizes $\bar{n}=30,\;60,\;100$ and $160$, $\delta_{iC} = -1$, $\Delta=-2$ and  $f = 0.5$.   }
	\label{Pplot_relative_truncated_deltaC_-1deltaT=-3_DSM_unequal_sample_sizes.pdf}
\end{figure}

\begin{figure}[ht]
	\centering
	\includegraphics[scale=0.33]{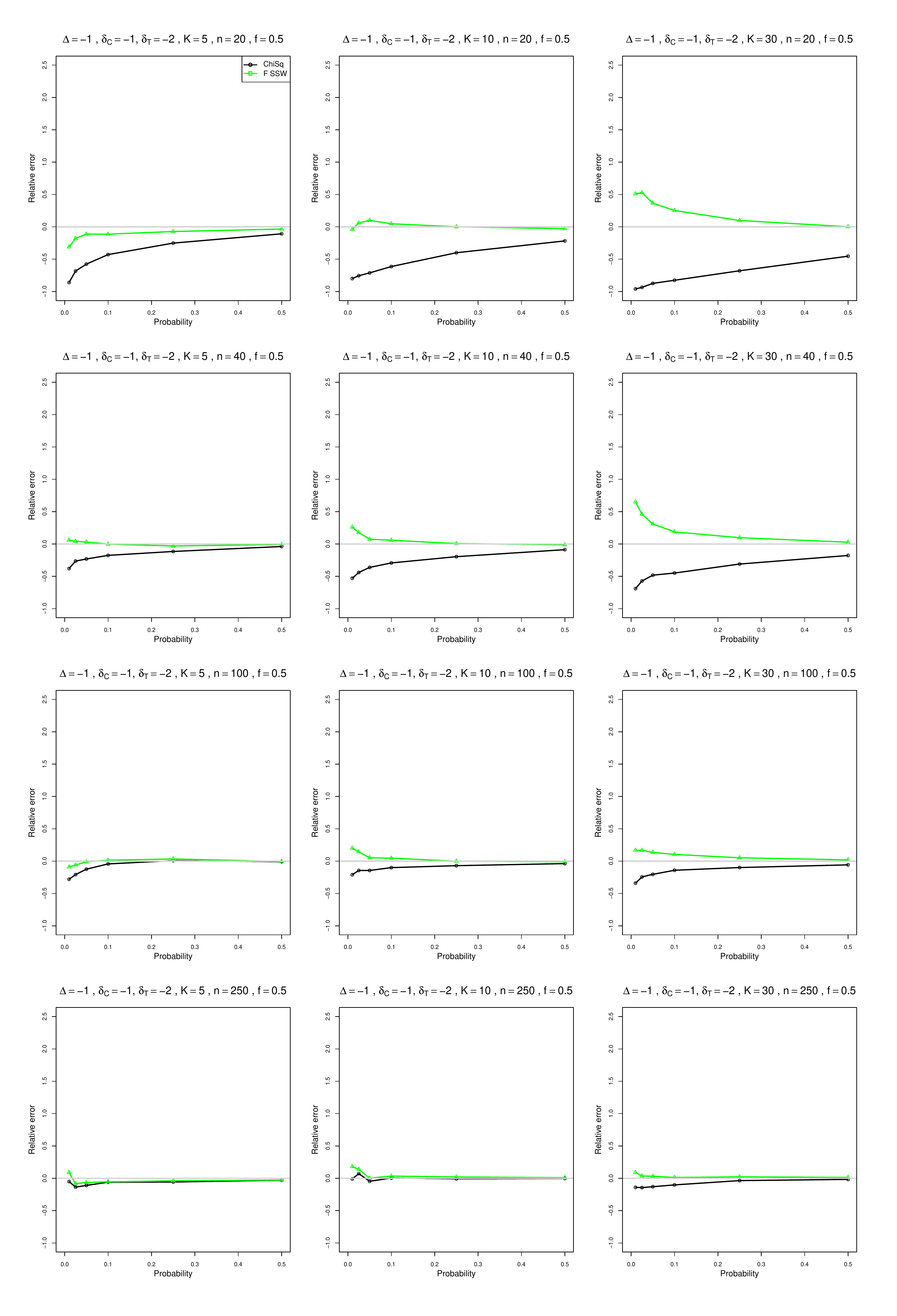}
	\caption{Relative error between the achieved level and the nominal level for two approximations to the null distribution of Q for DSM (Chisq and F SSW) vs upper tail probability, for equal sample sizes $n=20,\;40,\;100$ and $250$, $\delta_{iC} = -1$, $\Delta=-1$ and  $f = 0.5$.   }
	\label{Pplot_relative_truncated_deltaC_-1deltaT=-2_DSM_equal_sample_sizes.pdf}
\end{figure}

\begin{figure}[ht]
	\centering
	\includegraphics[scale=0.33]{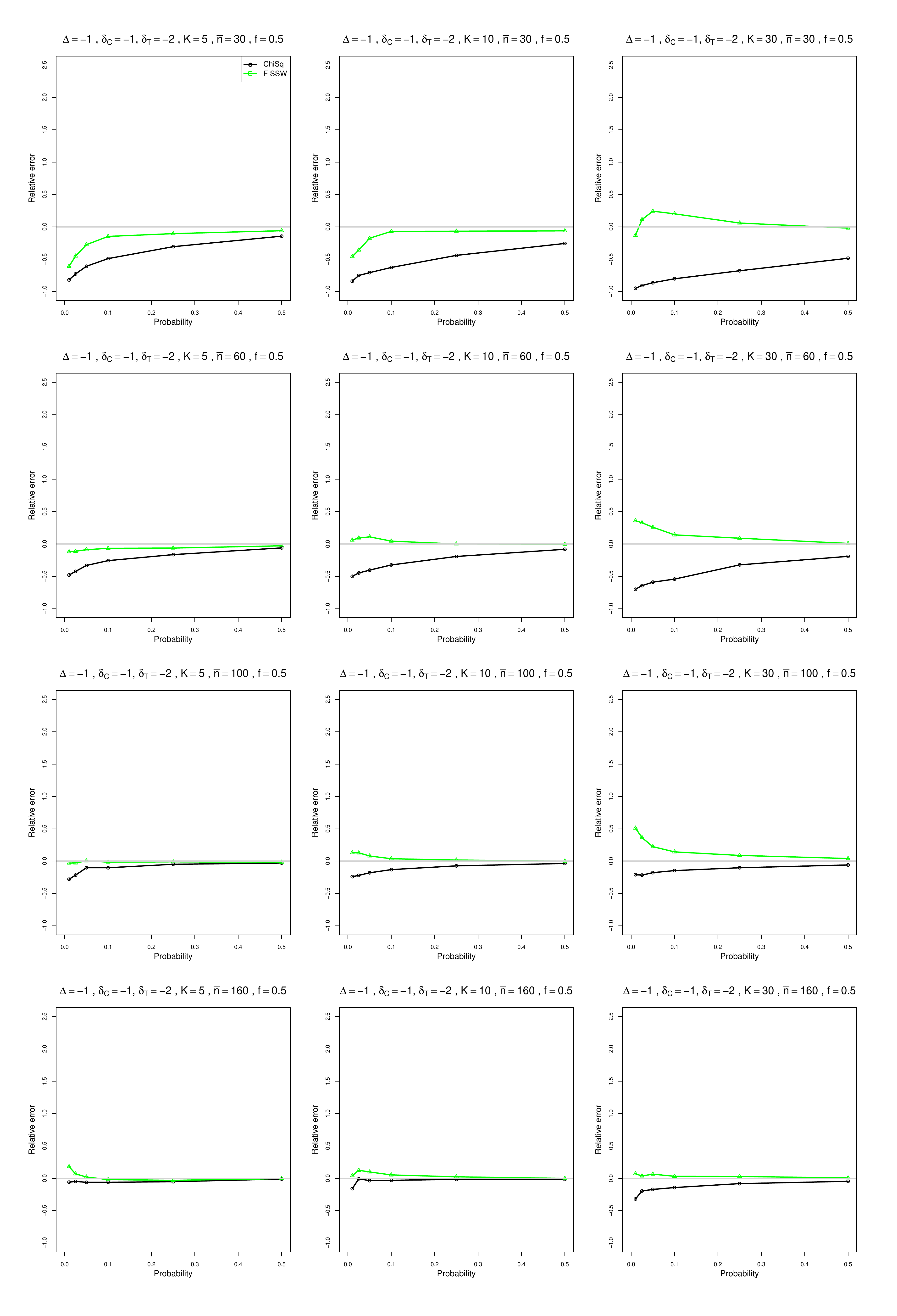}
	\caption{Relative error between the achieved level and the nominal level for two approximations to the null distribution of Q for DSM (Chisq and F SSW) vs upper tail probability, for unequal sample sizes $\bar{n}=30,\;60,\;100$ and $160$, $\delta_{iC} = -1$, $\Delta=-1$ and  $f = 0.5$.   }
	\label{Pplot_relative_truncated_deltaC_-1deltaT=-2_DSM_unequal_sample_sizes.pdf}
\end{figure}

\begin{figure}[ht]
	\centering
	\includegraphics[scale=0.33]{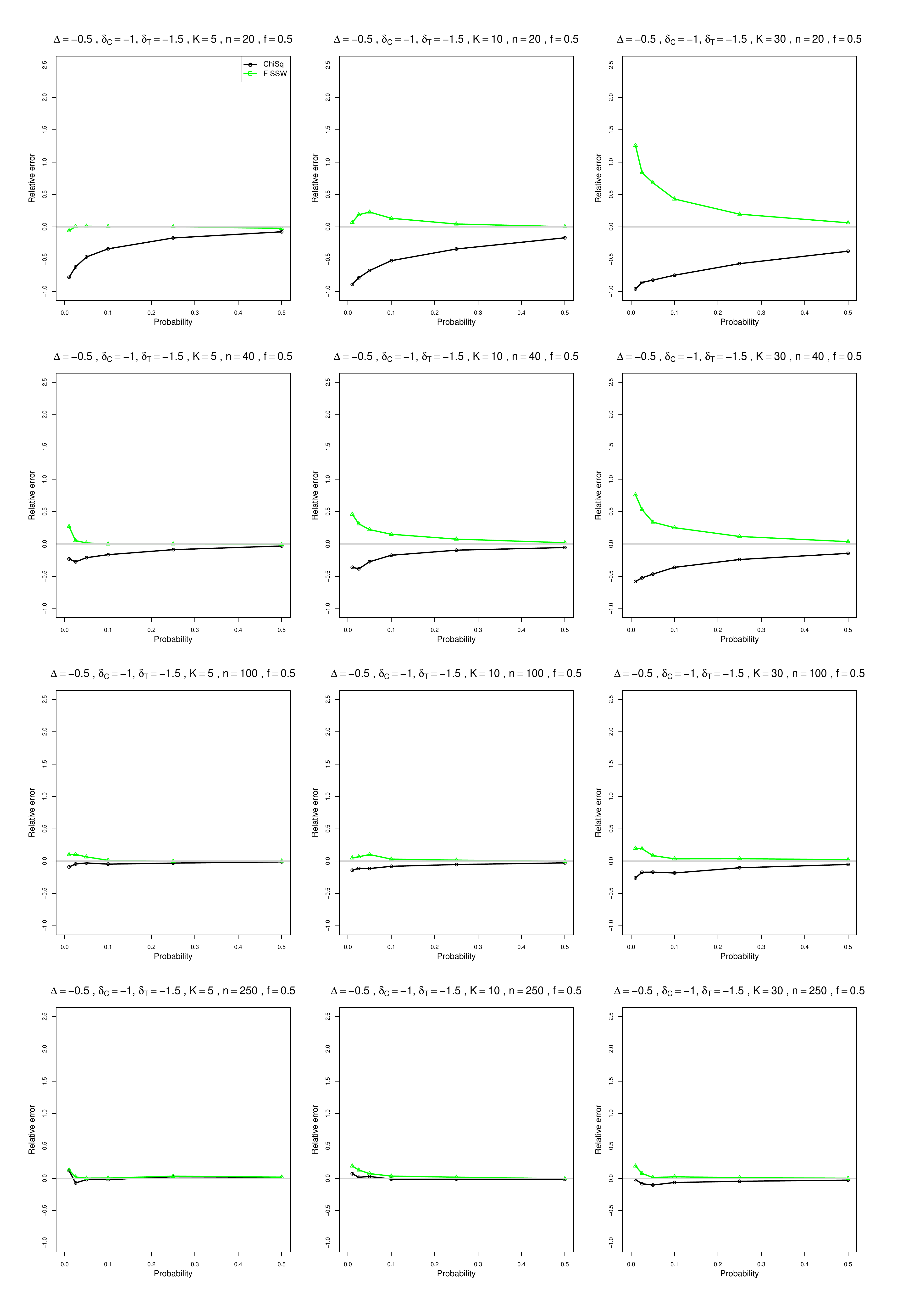}
	\caption{Relative error between the achieved level and the nominal level for two approximations to the null distribution of Q for DSM (Chisq and F SSW) vs upper tail probability, for equal sample sizes $n=20,\;40,\;100$ and $250$, $\delta_{iC} = -1$, $\Delta=-0.5$ and  $f = 0.5$.   }
	\label{Pplot_relative_truncated_deltaC_-1deltaT=-1.5_DSM_equal_sample_sizes.pdf}
\end{figure}

\begin{figure}[ht]
	\centering
	\includegraphics[scale=0.33]{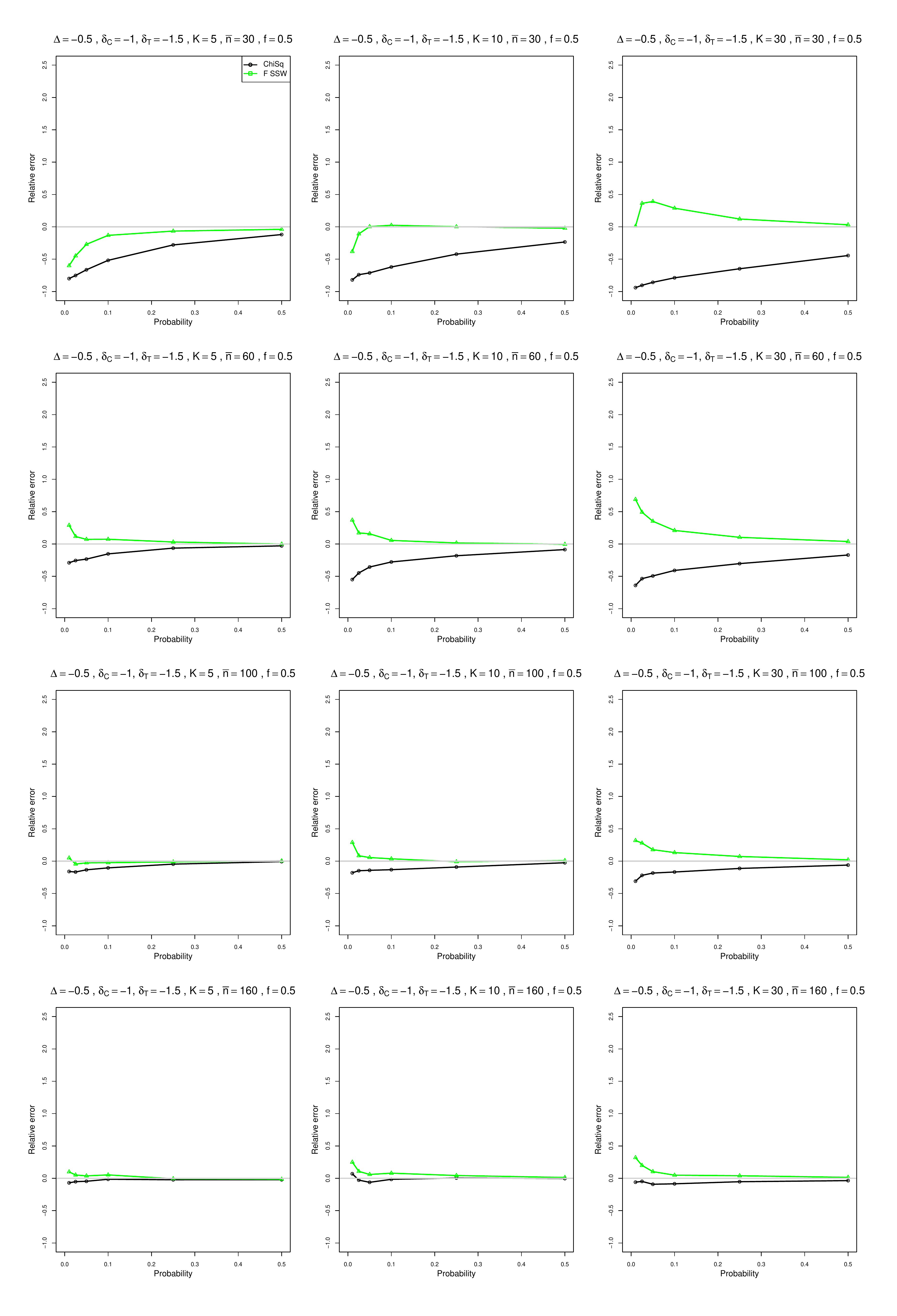}
	\caption{Relative error between the achieved level and the nominal level for two approximations to the null distribution of Q for DSM (Chisq and F SSW) vs upper tail probability, for unequal sample sizes $\bar{n}=30,\;60,\;100$ and $160$, $\delta_{iC} = -1$, $\Delta=-0.5$ and  $f = 0.5$.   }
	\label{Pplot_relative_truncated_deltaC_-1deltaT=-1,5_DSM_unequal_sample_sizes.pdf}
\end{figure}

\begin{figure}[ht]
	\centering
	\includegraphics[scale=0.33]{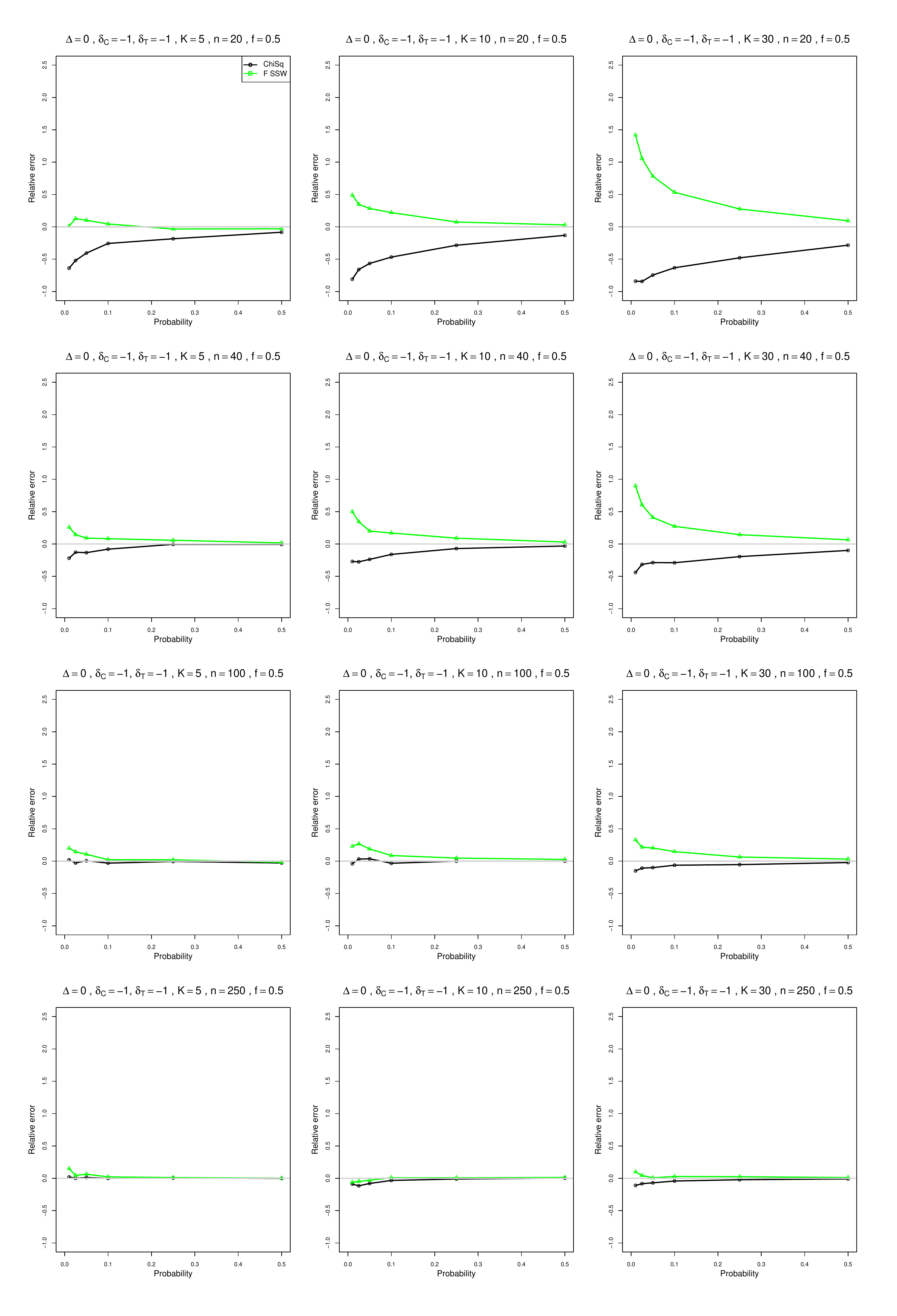}
	\caption{Relative error between the achieved level and the nominal level for two approximations to the null distribution of Q for DSM (Chisq and F SSW) vs upper tail probability, for equal sample sizes $n=20,\;40,\;100$ and $250$, $\delta_{iC} = -1$, $\Delta=0$ and  $f = 0.5$.   }
	\label{Pplot_relative_truncated_deltaC_-1deltaT=-1_DSM_equal_sample_sizes.pdf}
\end{figure}

\begin{figure}[ht]
	\centering
	\includegraphics[scale=0.33]{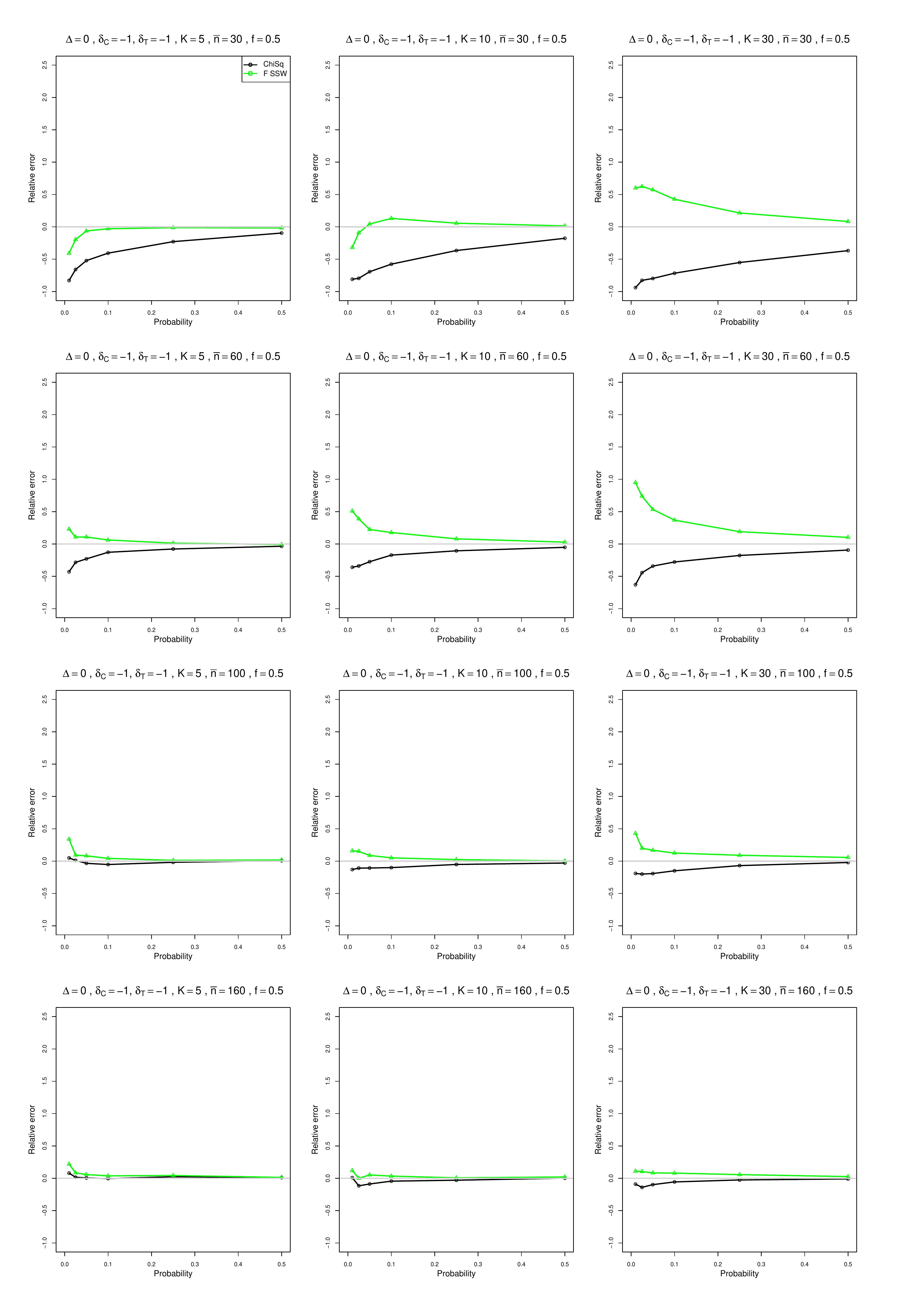}
	\caption{Relative error between the achieved level and the nominal level for two approximations to the null distribution of Q for DSM (Chisq and F SSW) vs upper tail probability, for unequal sample sizes $\bar{n}=30,\;60,\;100$ and $160$, $\delta_{iC} = -1$, $\Delta=0$ and  $f = 0.5$.   }
	\label{Pplot_relative_truncated_deltaC_-1deltaT=-1_DSM_unequal_sample_sizes.pdf}
\end{figure}

\begin{figure}[ht]
	\centering
	\includegraphics[scale=0.33]{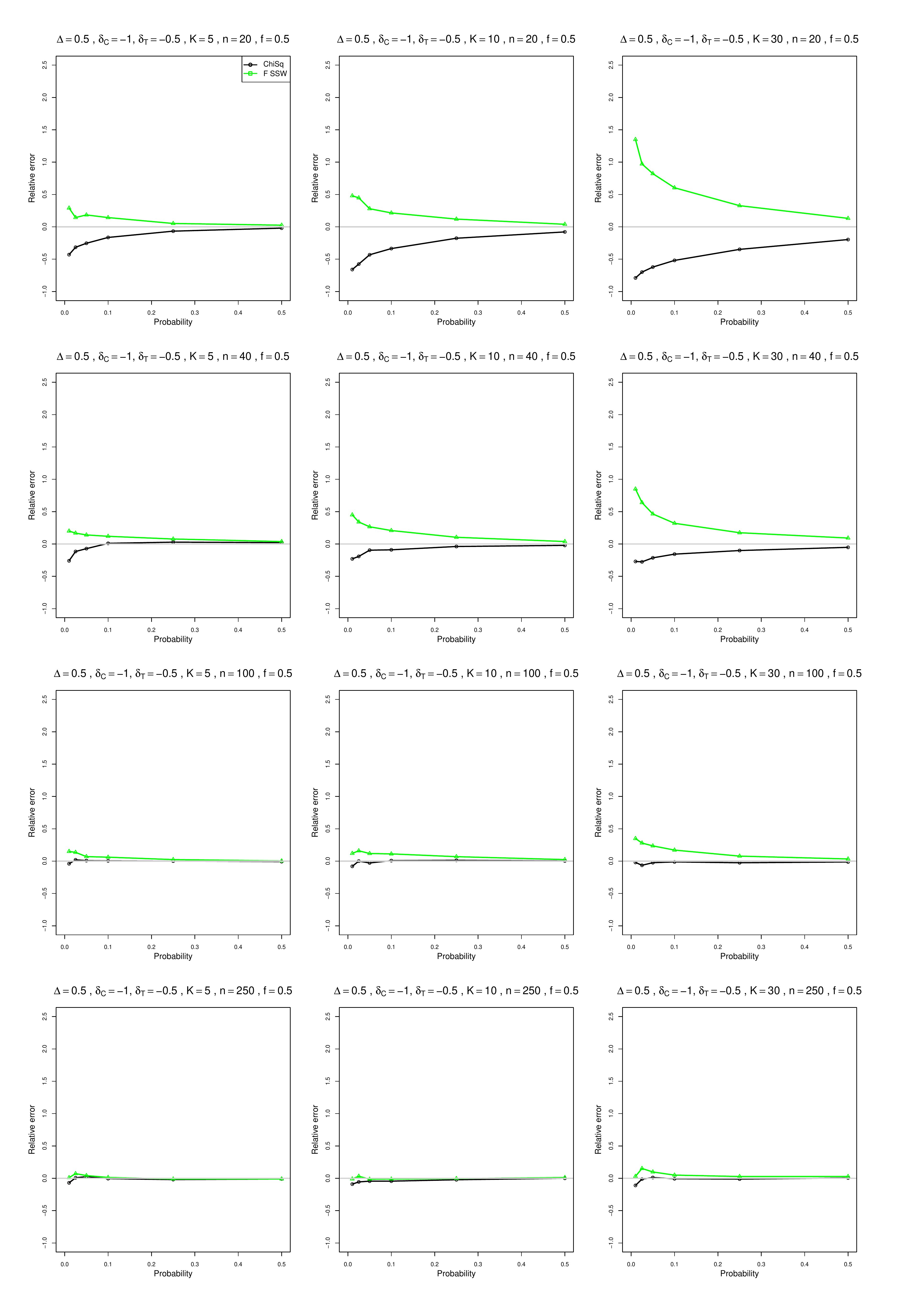}
	\caption{Relative error between the achieved level and the nominal level for two approximations to the null distribution of Q for DSM (Chisq and F SSW) vs upper tail probability, for equal sample sizes $n=20,\;40,\;100$ and $250$, $\delta_{iC} = -1$, $\Delta=0.5$ and  $f = 0.5$.   }
	\label{Pplot_relative_truncated_deltaC_-1deltaT=-0.5_DSM_equal_sample_sizes.pdf}
\end{figure}

\begin{figure}[ht]
	\centering
	\includegraphics[scale=0.33]{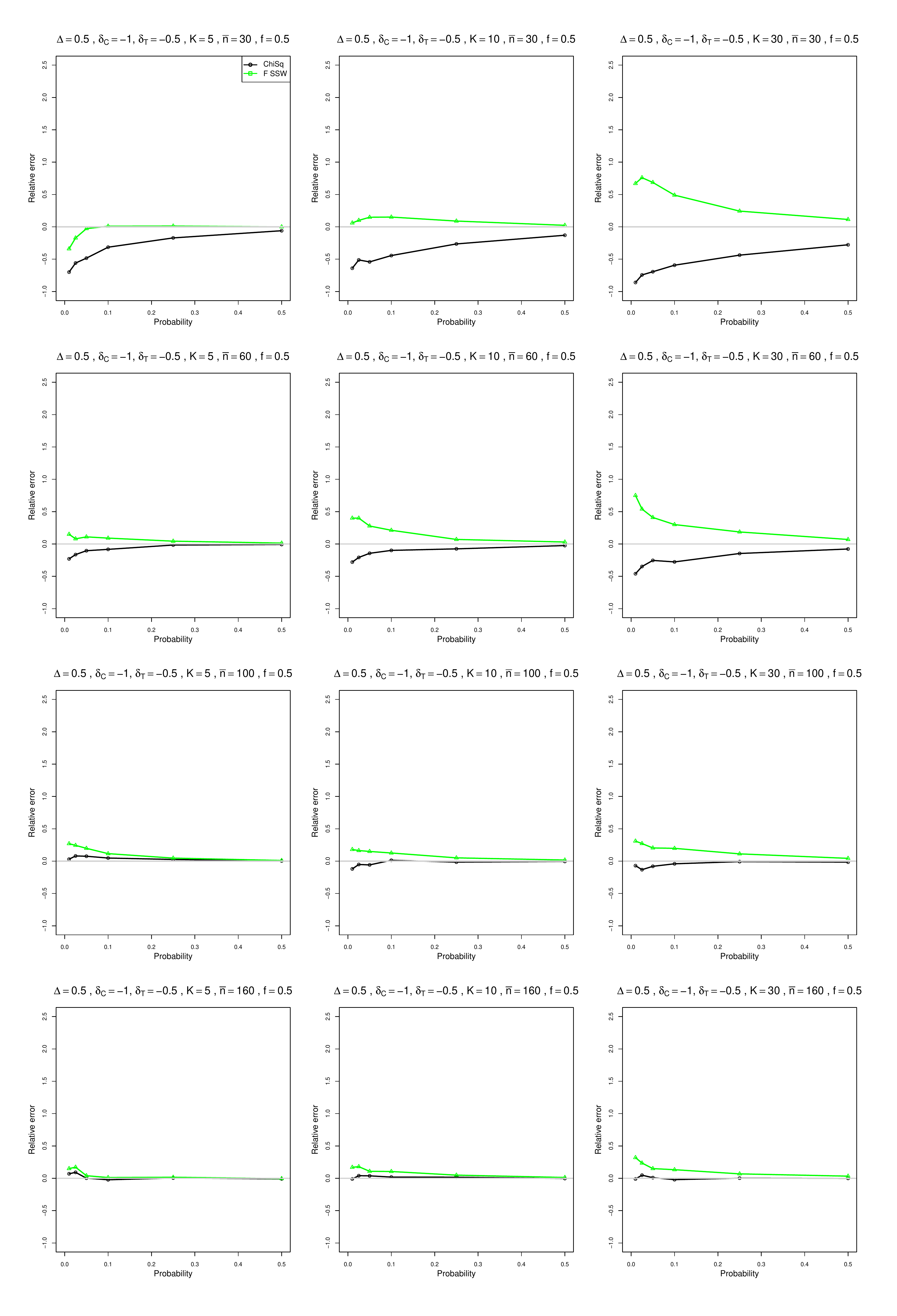}
	\caption{Relative error between the achieved level and the nominal level for two approximations to the null distribution of Q for DSM (Chisq and F SSW) vs upper tail probability, for unequal sample sizes $\bar{n}=30,\;60,\;100$ and $160$, $\delta_{iC} = -1$, $\Delta=0.5$ and  $f = 0.5$.   }
	\label{Pplot_relative_truncated_deltaC_-1deltaT=-0,5_DSM_unequal_sample_sizes.pdf}
\end{figure}

\begin{figure}[ht]
	\centering
	\includegraphics[scale=0.33]{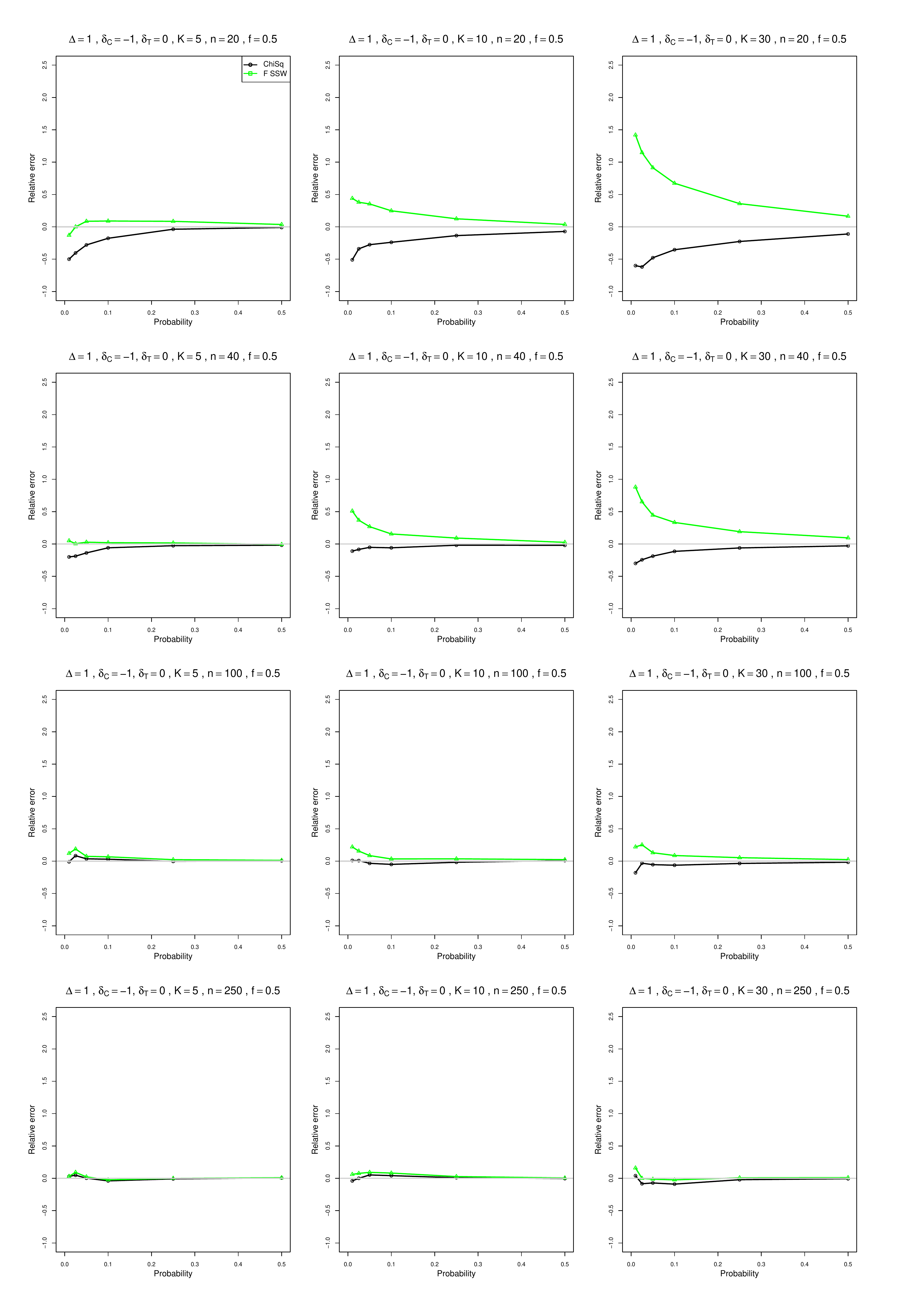}
	\caption{Relative error between the achieved level and the nominal level for two approximations to the null distribution of Q for DSM (Chisq and F SSW) vs upper tail probability, for equal sample sizes $n=20,\;40,\;100$ and $250$, $\delta_{iC} = -1$, $\Delta=1$ and  $f = 0.5$.   }
	\label{Pplot_relative_truncated_deltaC_--1deltaT=0_DSM_equal_sample_sizes.pdf}
\end{figure}

\begin{figure}[ht]
	\centering
	\includegraphics[scale=0.33]{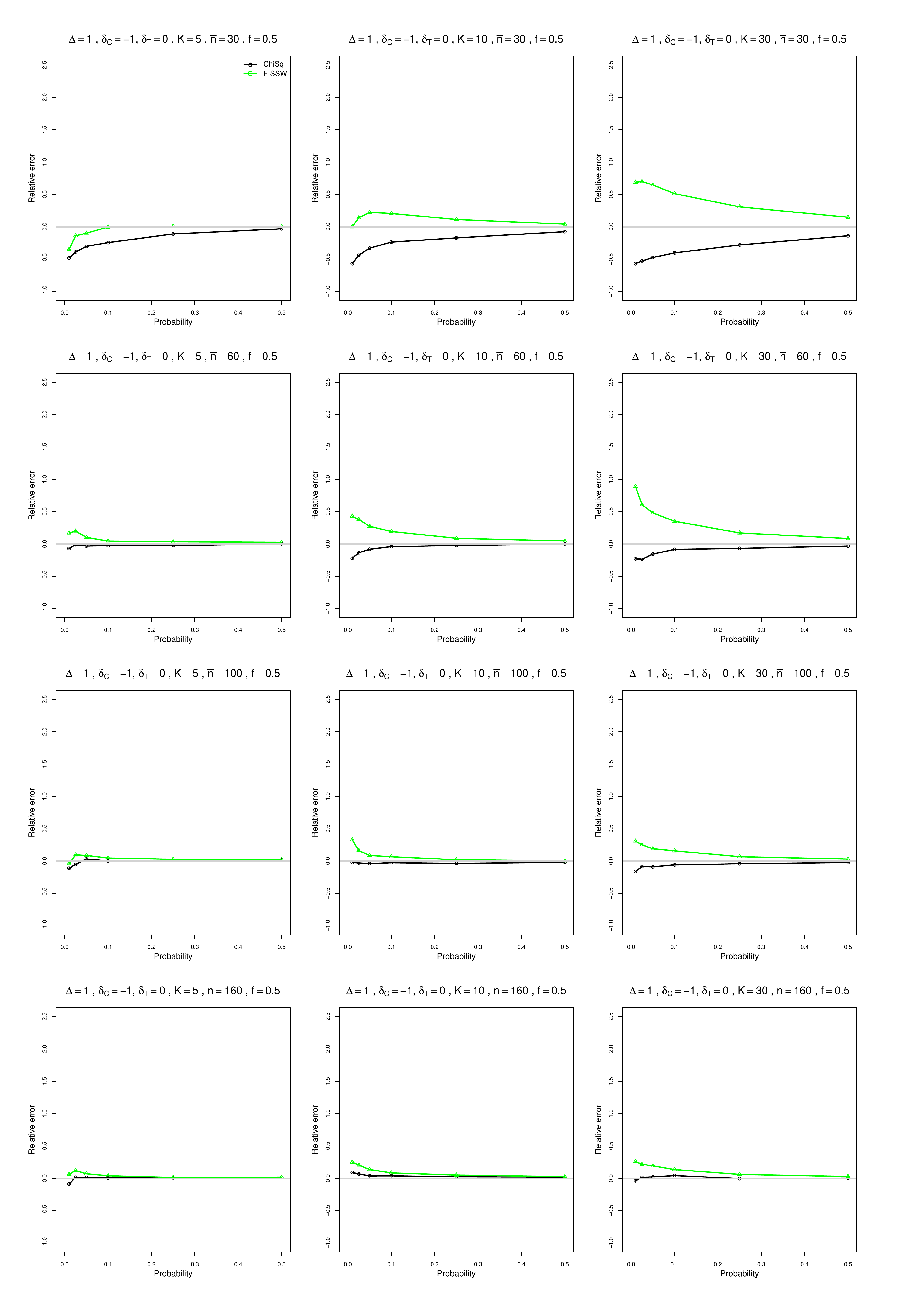}
	\caption{Relative error between the achieved level and the nominal level for two approximations to the null distribution of Q for DSM (Chisq and F SSW) vs upper tail probability, for unequal sample sizes $\bar{n}=30,\;60,\;100$ and $160$, $\delta_{iC} = -1$, $\Delta=1$ and  $f = 0.5$.   }
	\label{Pplot_relative_truncated_deltaC_-1deltaT=0_DSM_unequal_sample_sizes.pdf}
\end{figure}

\begin{figure}[ht]
	\centering
	\includegraphics[scale=0.33]{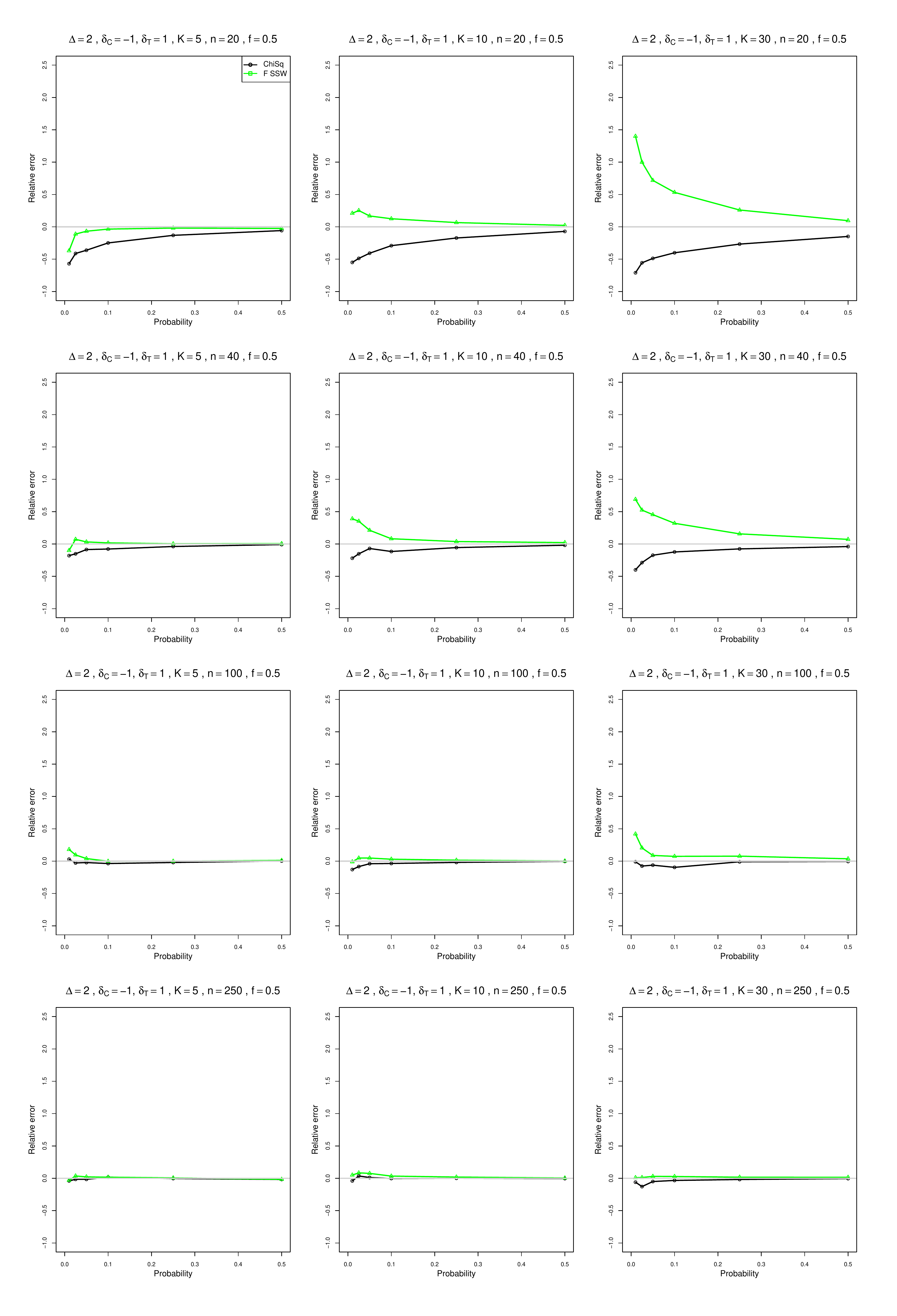}
	\caption{Relative error between the achieved level and the nominal level for two approximations to the null distribution of Q for DSM (Chisq and F SSW) vs upper tail probability, for equal sample sizes $n=20,\;40,\;100$ and $250$, $\delta_{iC} = -1$, $\Delta=2$ and  $f = 0.5$.   }
	\label{Pplot_relative_truncated_deltaC_-1deltaT=1_DSM_equal_sample_sizes.pdf}
\end{figure}

\begin{figure}[ht]
	\centering
	\includegraphics[scale=0.33]{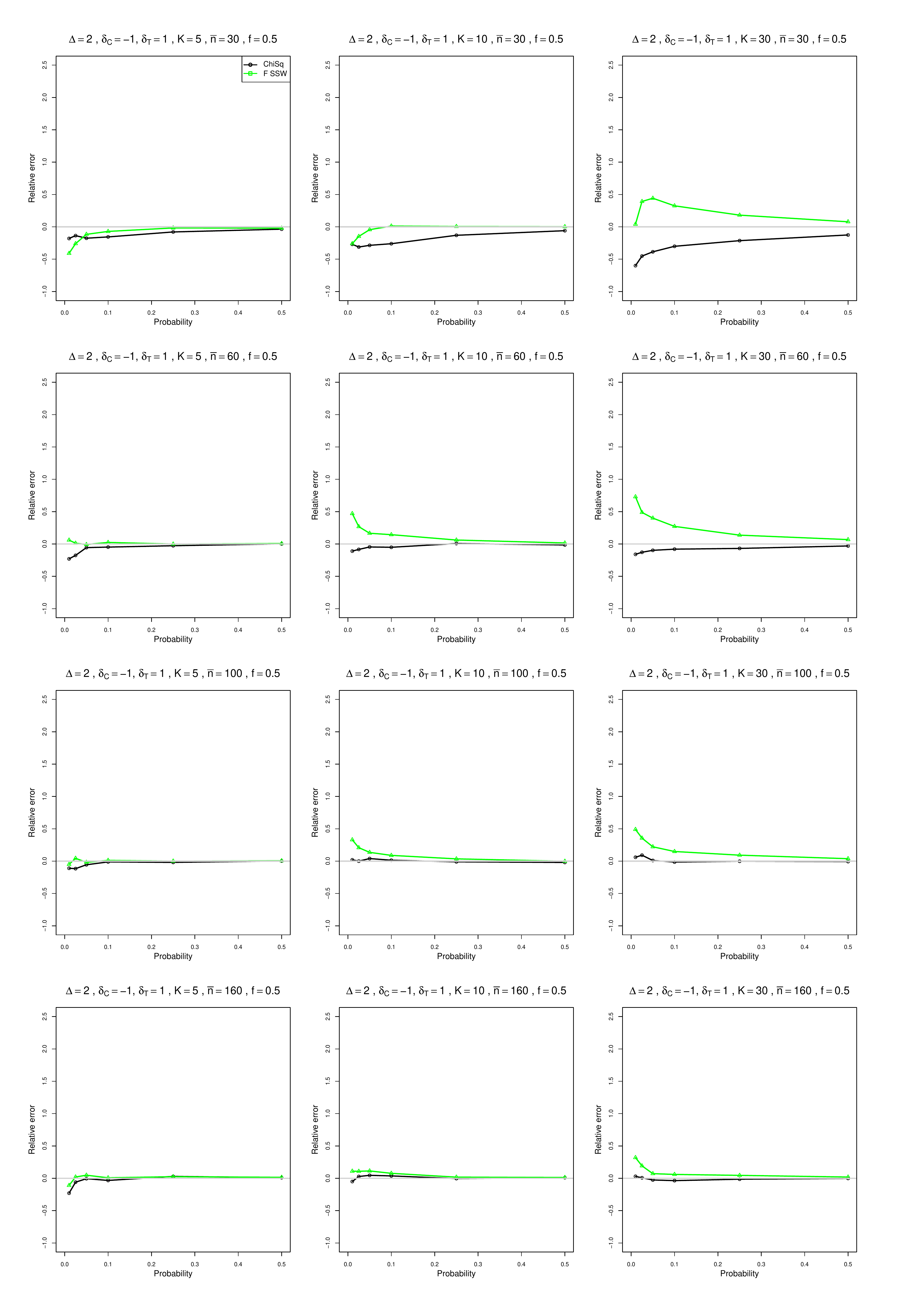}
	\caption{Relative error between the achieved level and the nominal level for two approximations to the null distribution of Q for DSM (Chisq and F SSW) vs upper tail probability, for unequal sample sizes $\bar{n}=30,\;60,\;100$ and $160$, $\delta_{iC} = -1$, $\Delta=2$ and  $f = 0.5$.   }
	\label{Pplot_relative_truncated_deltaC_-1deltaT=1_DSM_unequal_sample_sizes.pdf}
\end{figure}


\begin{figure}[ht]
	\centering
	\includegraphics[scale=0.33]{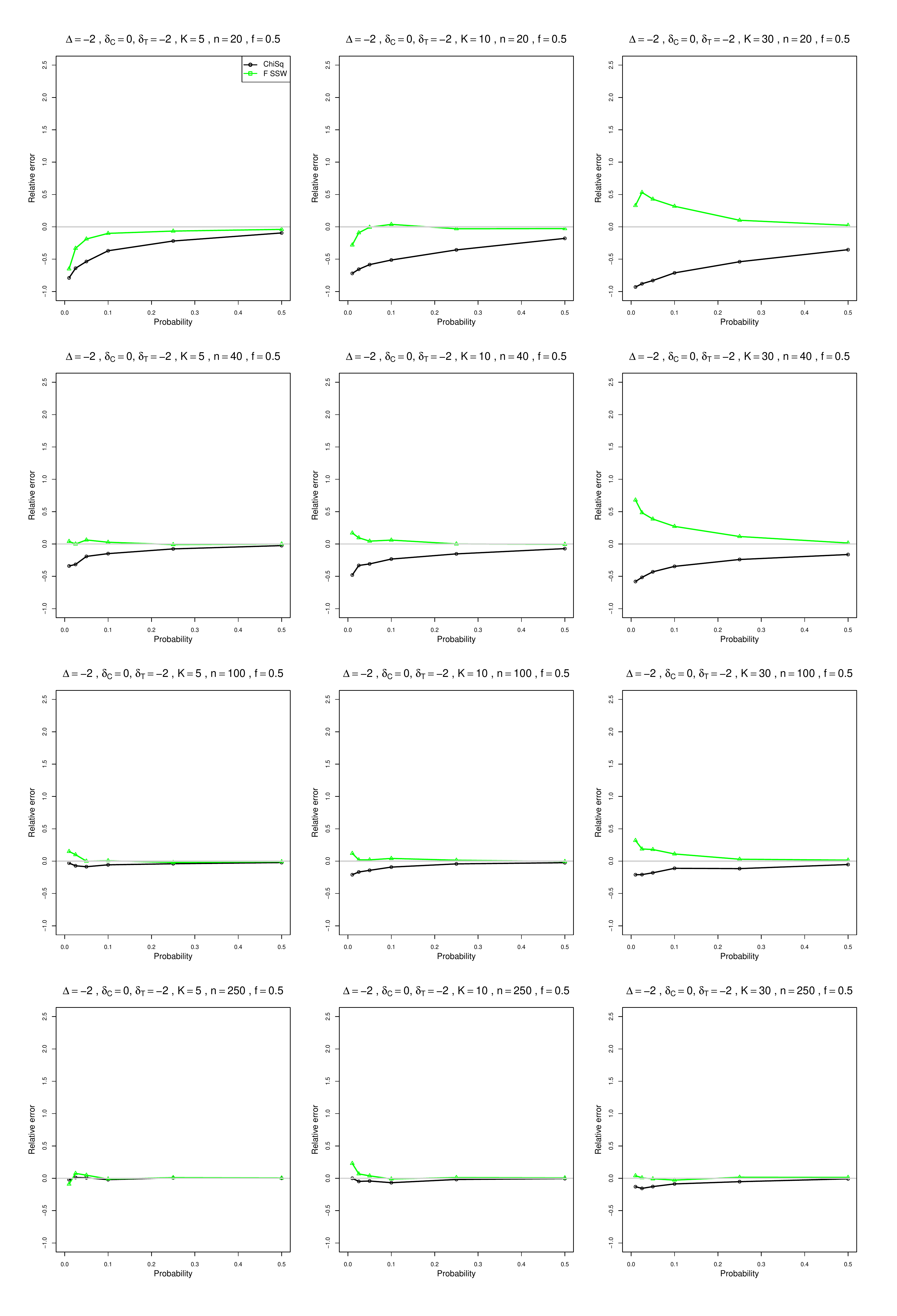}
	\caption{Relative error between the achieved level and the nominal level for two approximations to the null distribution of Q for DSM (Chisq and F SSW) vs upper tail probability, for equal sample sizes $n=20,\;40,\;100$ and $250$, $\delta_{iC} = 0$, $\Delta=-2$ and  $f = 0.5$.   }
	\label{Pplot_relative_truncated_deltaC_0deltaT=-2_DSM_equal_sample_sizes.pdf}
\end{figure}

\begin{figure}[ht]
	\centering
	\includegraphics[scale=0.33]{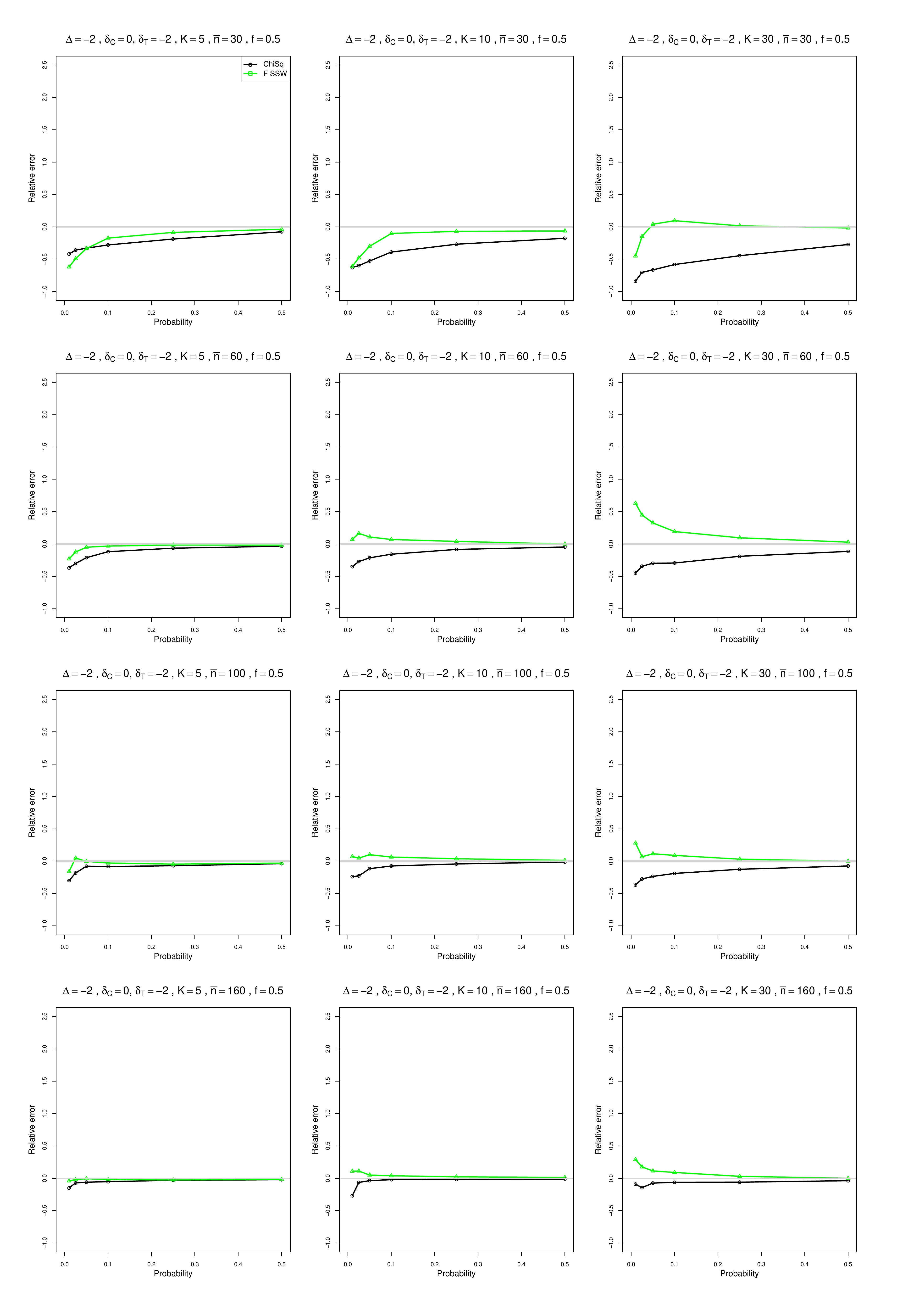}
	\caption{Relative error between the achieved level and the nominal level for two approximations to the null distribution of Q for DSM (Chisq and F SSW) vs upper tail probability, for unequal sample sizes $\bar{n}=30,\;60,\;100$ and $160$, $\delta_{iC} = 0$, $\Delta=-2$ and  $f = 0.5$.   }
	\label{Pplot_relative_truncated_deltaC_-1deltaT=-3_DSM_unequal_sample_sizes.pdf}
\end{figure}

\begin{figure}[ht]
	\centering
	\includegraphics[scale=0.33]{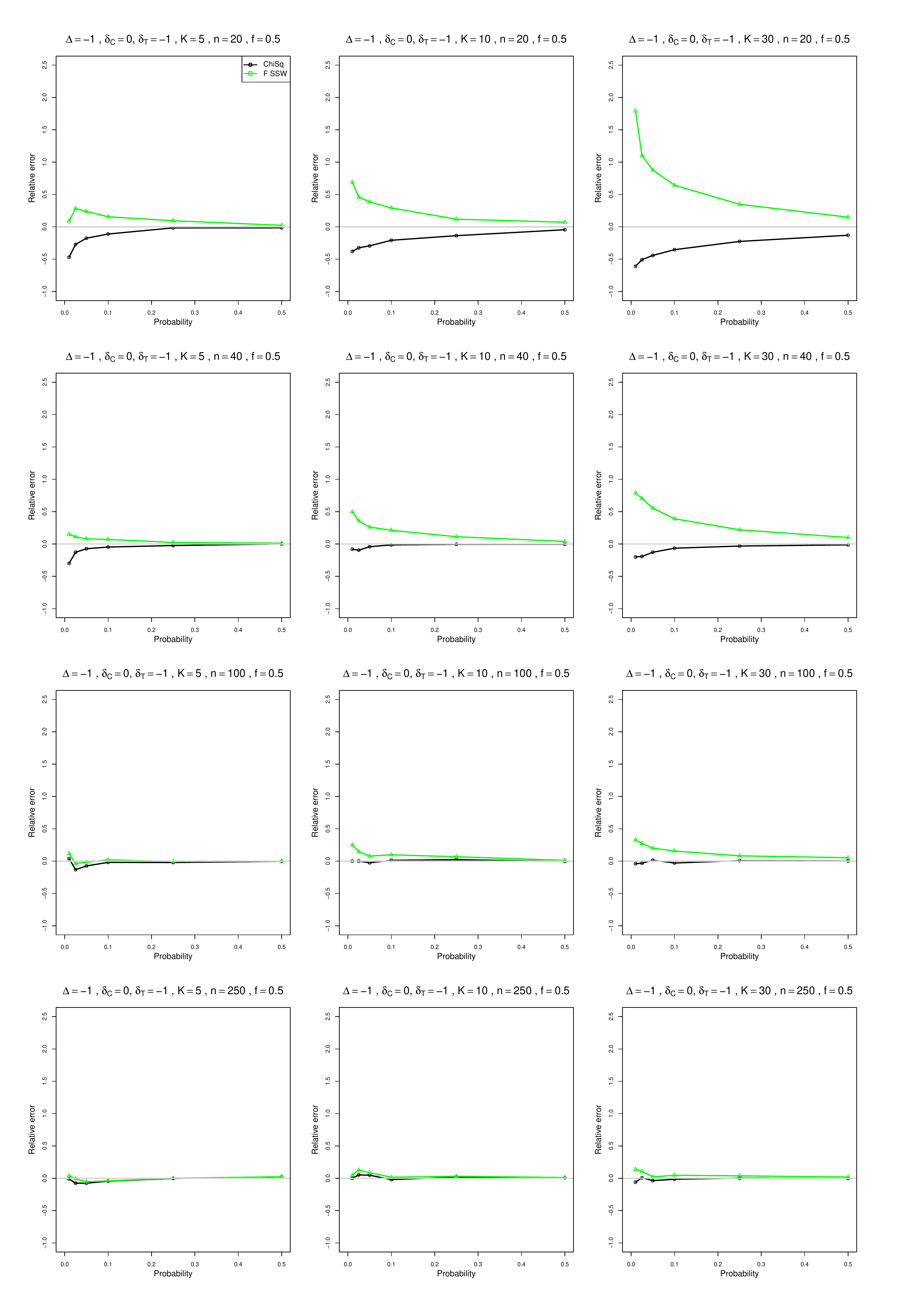}
	\caption{Relative error between the achieved level and the nominal level for two approximations to the null distribution of Q for DSM (Chisq and F SSW) vs upper tail probability, for equal sample sizes $n=20,\;40,\;100$ and $250$, $\delta_{iC} = 0$, $\Delta=-1$ and  $f = 0.5$.   }
	\label{Pplot_relative_truncated_deltaC_-0deltaT=-1_DSM_equal_sample_sizes.pdf}
\end{figure}

\begin{figure}[ht]
	\centering
	\includegraphics[scale=0.33]{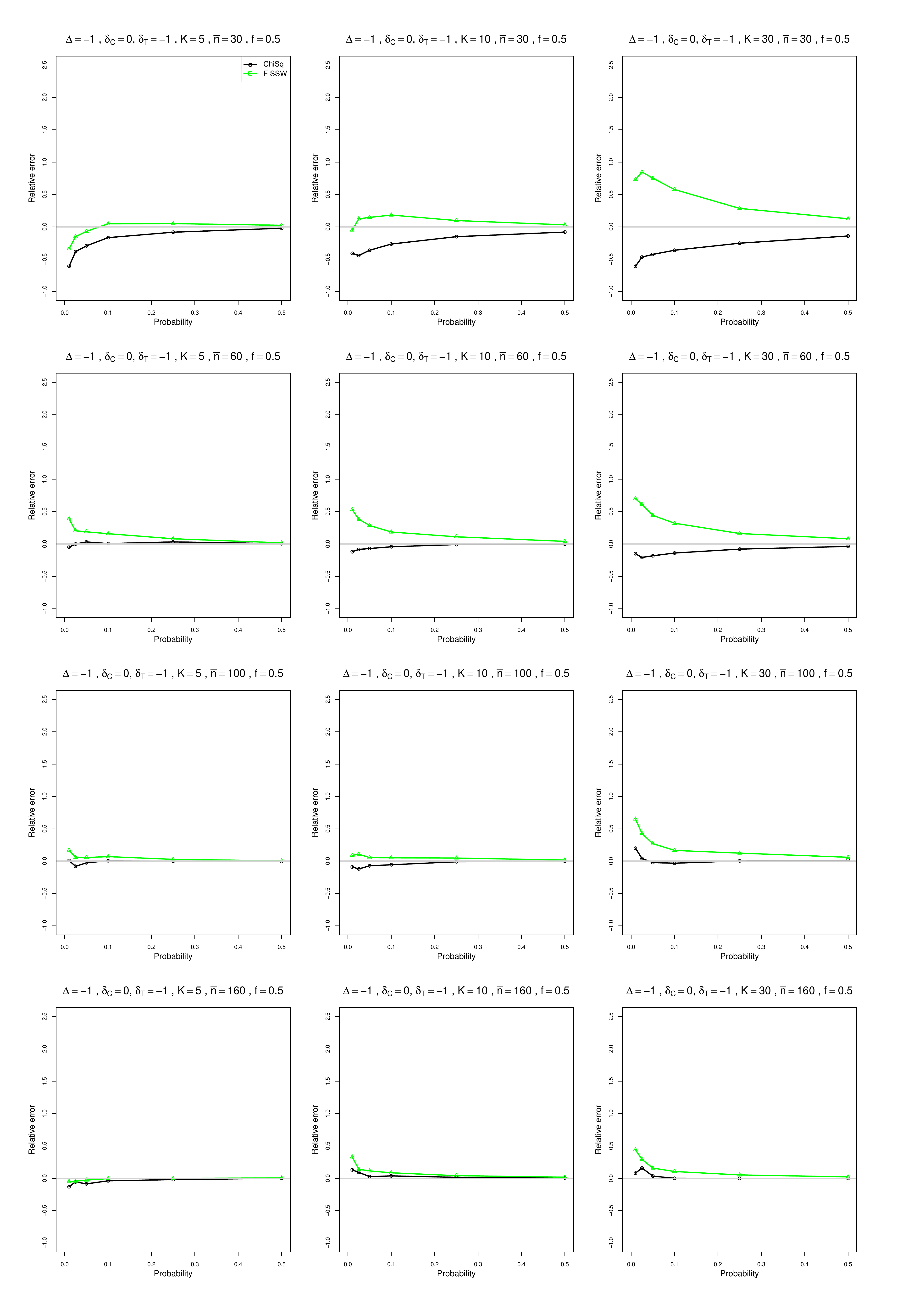}
	\caption{Relative error between the achieved level and the nominal level for two approximations to the null distribution of Q for DSM (Chisq and F SSW) vs upper tail probability, for unequal sample sizes $\bar{n}=30,\;60,\;100$ and $160$, $\delta_{iC} = 0$, $\Delta=-1$ and  $f = 0.5$.   }
	\label{Pplot_relative_truncated_deltaC_0deltaT=-1_DSM_unequal_sample_sizes.pdf}
\end{figure}

\begin{figure}[ht]
	\centering
	\includegraphics[scale=0.33]{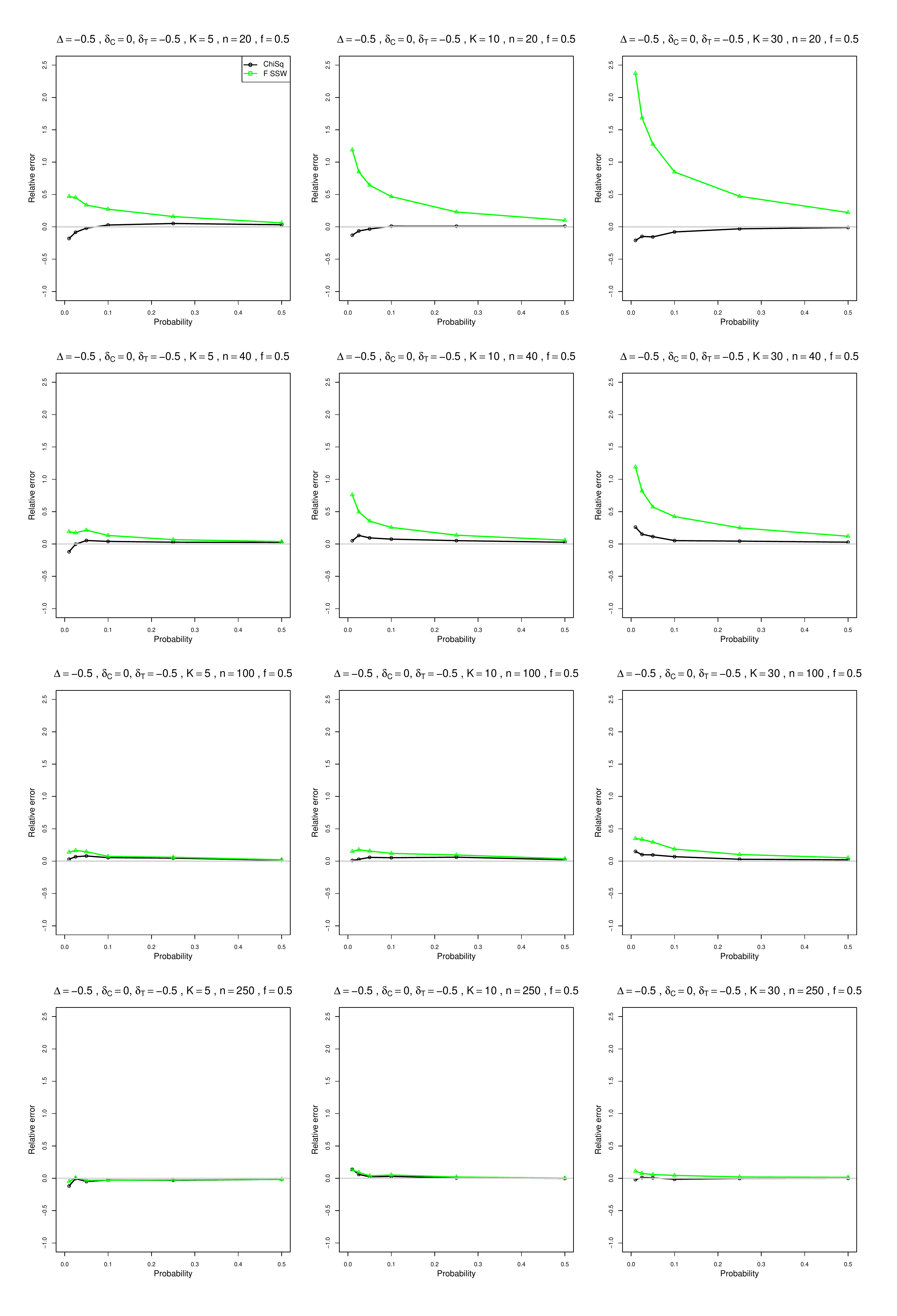}
	\caption{Relative error between the achieved level and the nominal level for two approximations to the null distribution of Q for DSM (Chisq and F SSW) vs upper tail probability, for equal sample sizes $n=20,\;40,\;100$ and $250$, $\delta_{iC} = 0$, $\Delta=-0.5$ and  $f = 0.5$.   }
	\label{Pplot_relative_truncated_deltaC_-0deltaT=-0.5_DSM_equal_sample_sizes.pdf}
\end{figure}

\begin{figure}[ht]
	\centering
	\includegraphics[scale=0.33]{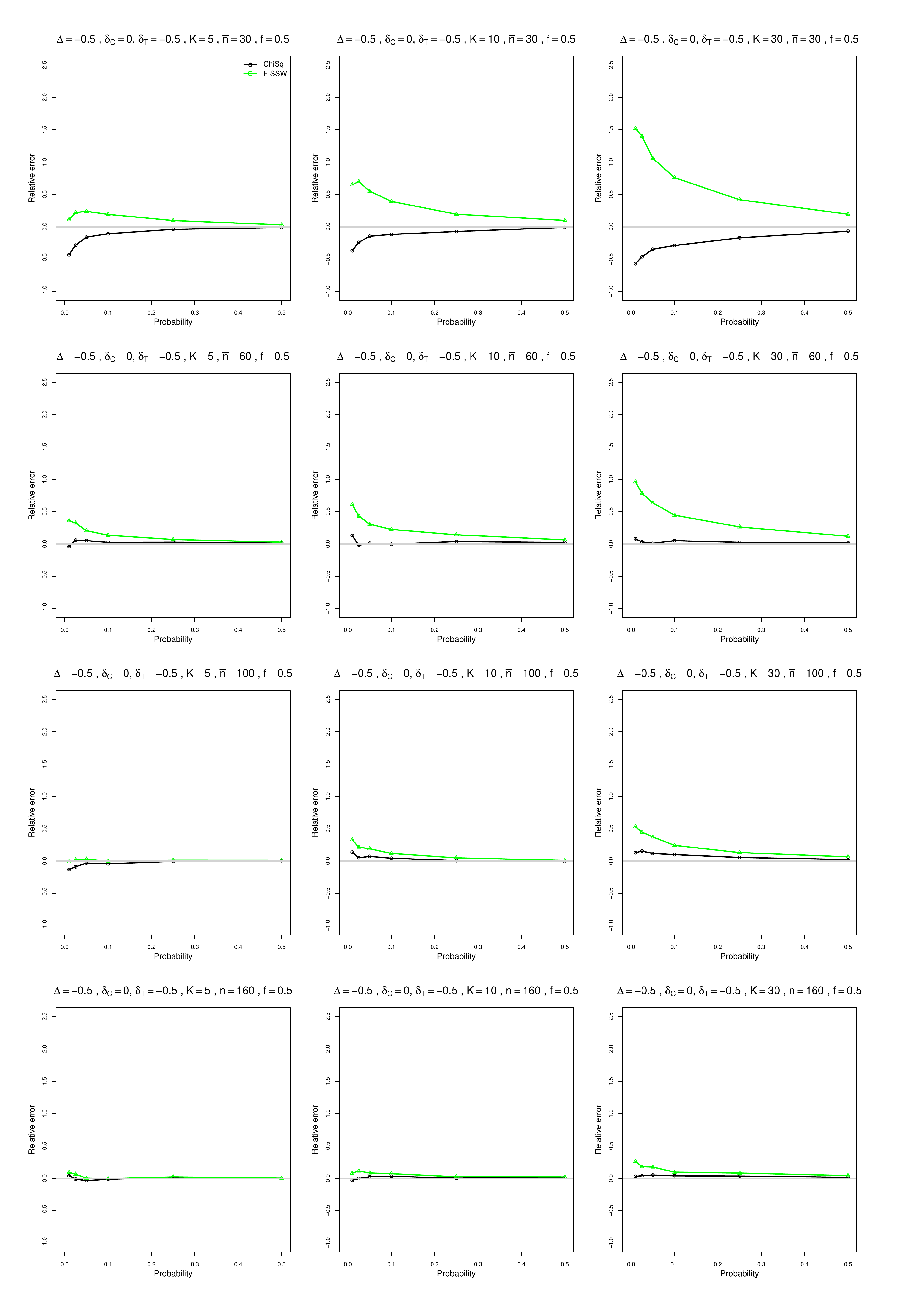}
	\caption{Relative error between the achieved level and the nominal level for two approximations to the null distribution of Q for DSM (Chisq and F SSW) vs upper tail probability, for unequal sample sizes $\bar{n}=30,\;60,\;100$ and $160$, $\delta_{iC} = 0$, $\Delta=-0.5$ and  $f = 0.5$.   }
	\label{Pplot_relative_truncated_deltaC_0deltaT=-0.5_DSM_unequal_sample_sizes.pdf}
\end{figure}

\begin{figure}[ht]
	\centering
	\includegraphics[scale=0.33]{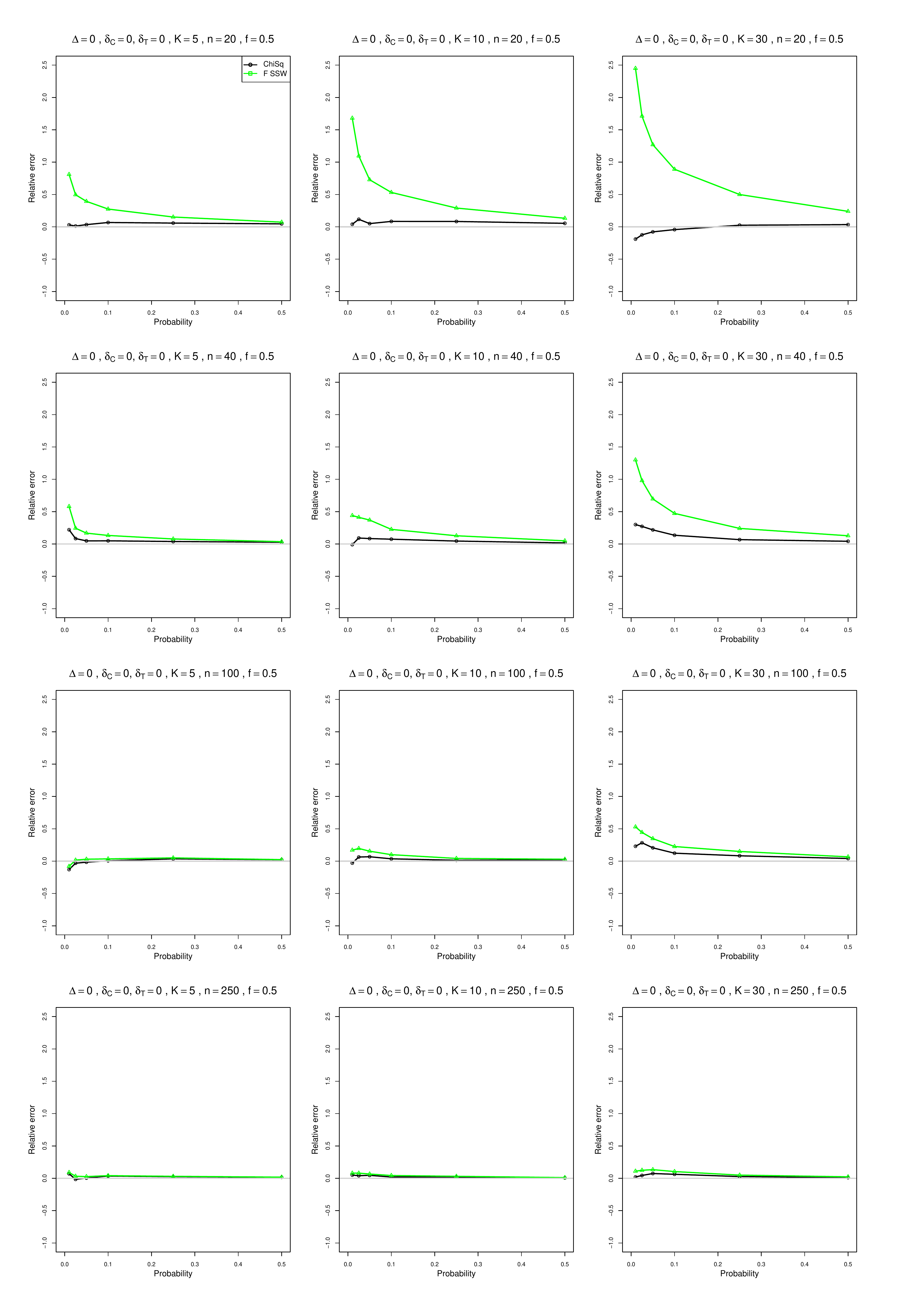}
	\caption{Relative error between the achieved level and the nominal level for two approximations to the null distribution of Q for DSM (Chisq and F SSW) vs upper tail probability, for equal sample sizes $n=20,\;40,\;100$ and $250$, $\delta_{iC} = 0$, $\Delta=0$ and  $f = 0.5$.   }
	\label{Pplot_relative_truncated_deltaC_0deltaT=0_DSM_equal_sample_sizes.pdf}
\end{figure}

\begin{figure}[ht]
	\centering
	\includegraphics[scale=0.33]{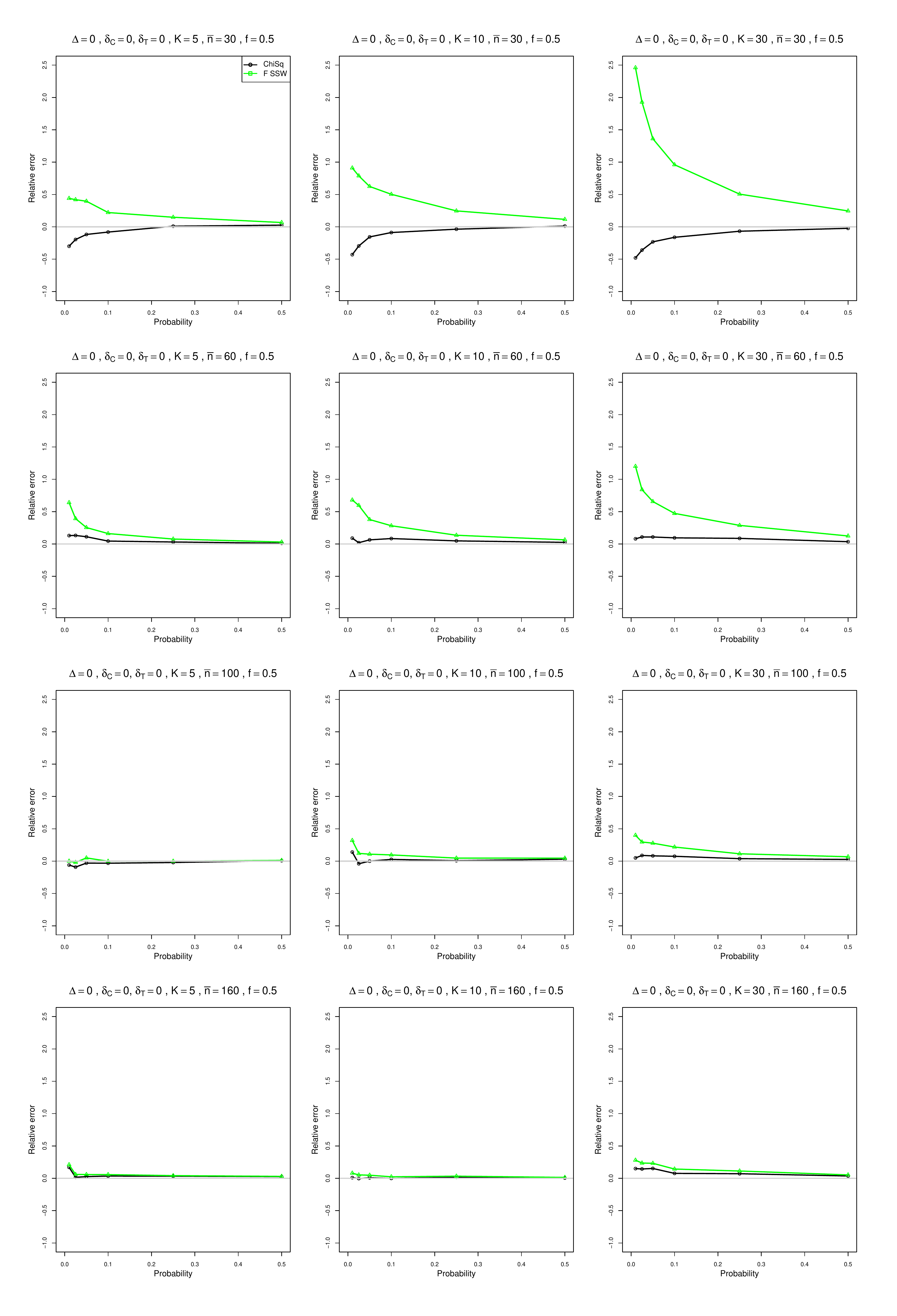}
	\caption{Relative error between the achieved level and the nominal level for two approximations to the null distribution of Q for DSM (Chisq and F SSW) vs upper tail probability, for unequal sample sizes $\bar{n}=30,\;60,\;100$ and $160$, $\delta_{iC} = 0$, $\Delta=0$ and  $f = 0.5$.   }
	\label{Pplot_relative_truncated_deltaC_0deltaT=0_DSM_unequal_sample_sizes.pdf}
\end{figure}

\begin{figure}[ht]
	\centering
	\includegraphics[scale=0.33]{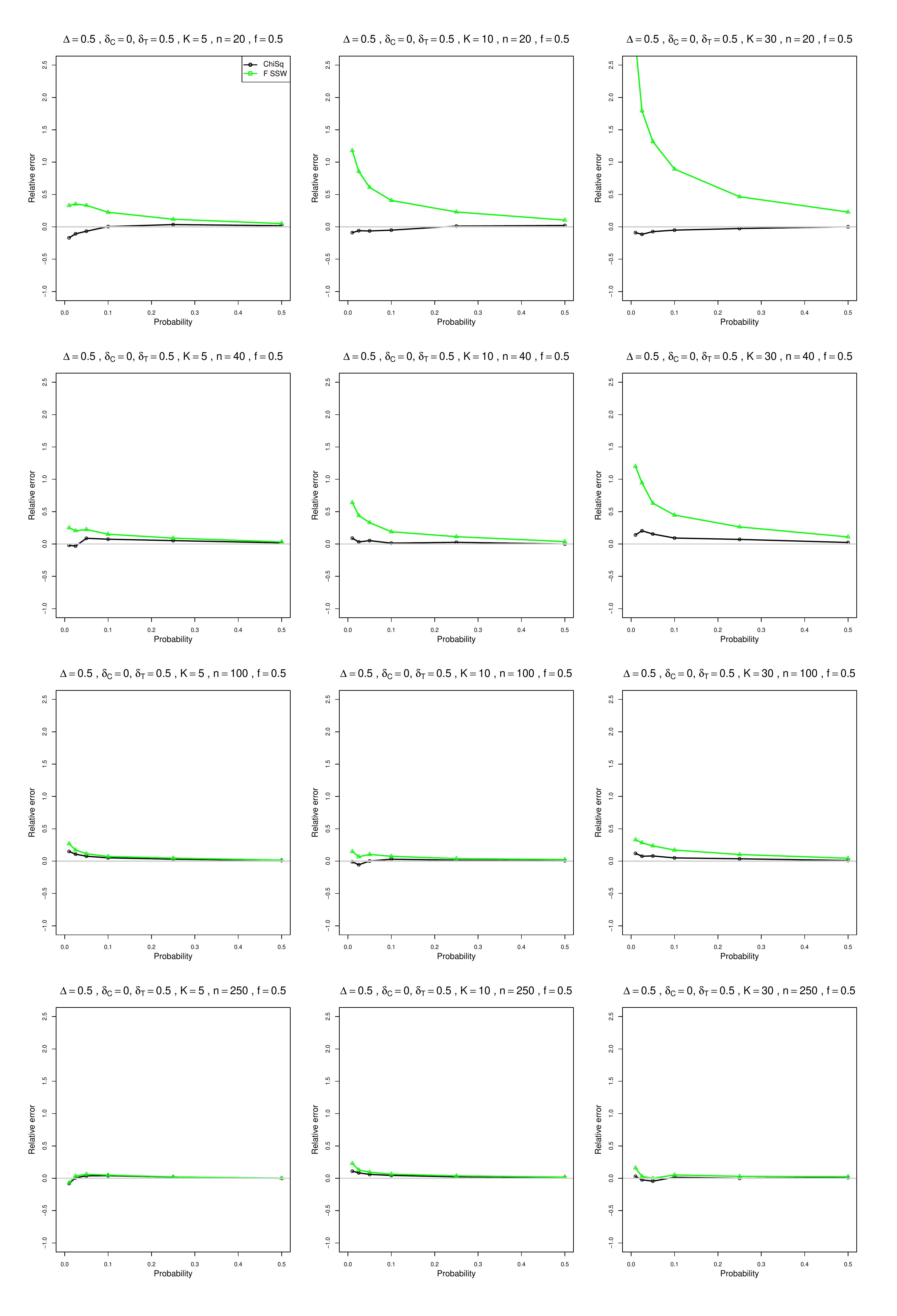}
	\caption{Relative error between the achieved level and the nominal level for two approximations to the null distribution of Q for DSM (Chisq and F SSW) vs upper tail probability, for equal sample sizes $n=20,\;40,\;100$ and $250$, $\delta_{iC} = 0$, $\Delta=0.5$ and  $f = 0.5$.   }
	\label{Pplot_relative_truncated_deltaC_0deltaT=0.5_DSM_equal_sample_sizes.pdf}
\end{figure}

\begin{figure}[ht]
	\centering
	\includegraphics[scale=0.33]{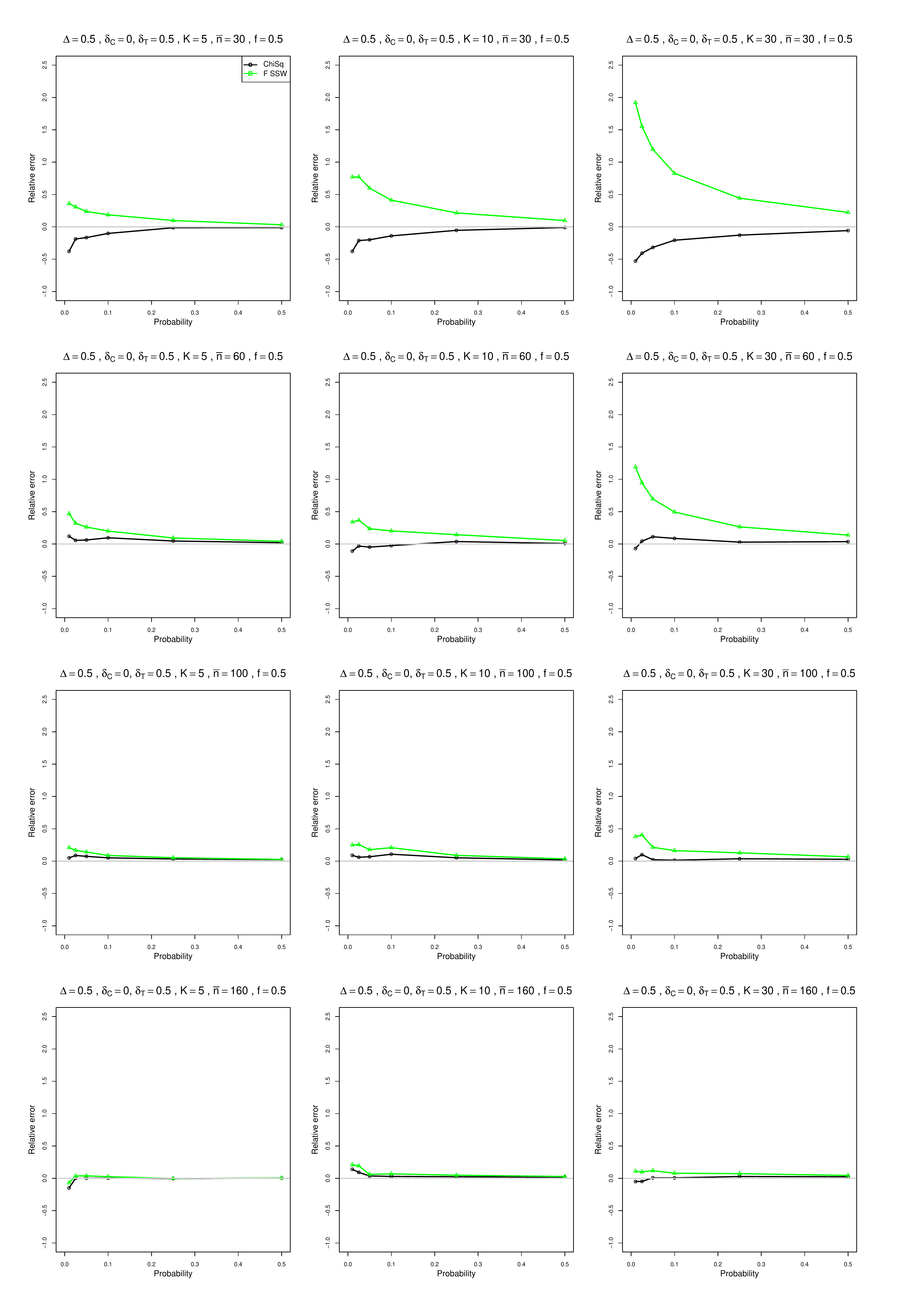}
	\caption{Relative error between the achieved level and the nominal level for two approximations to the null distribution of Q for DSM (Chisq and F SSW) vs upper tail probability, for unequal sample sizes $\bar{n}=30,\;60,\;100$ and $160$, $\delta_{iC} = 0$, $\Delta=0.5$ and  $f = 0.5$.   }
	\label{Pplot_relative_truncated_deltaC_0deltaT=-0.5_DSM_unequal_sample_sizes.pdf}
\end{figure}

\begin{figure}[ht]
	\centering
	\includegraphics[scale=0.33]{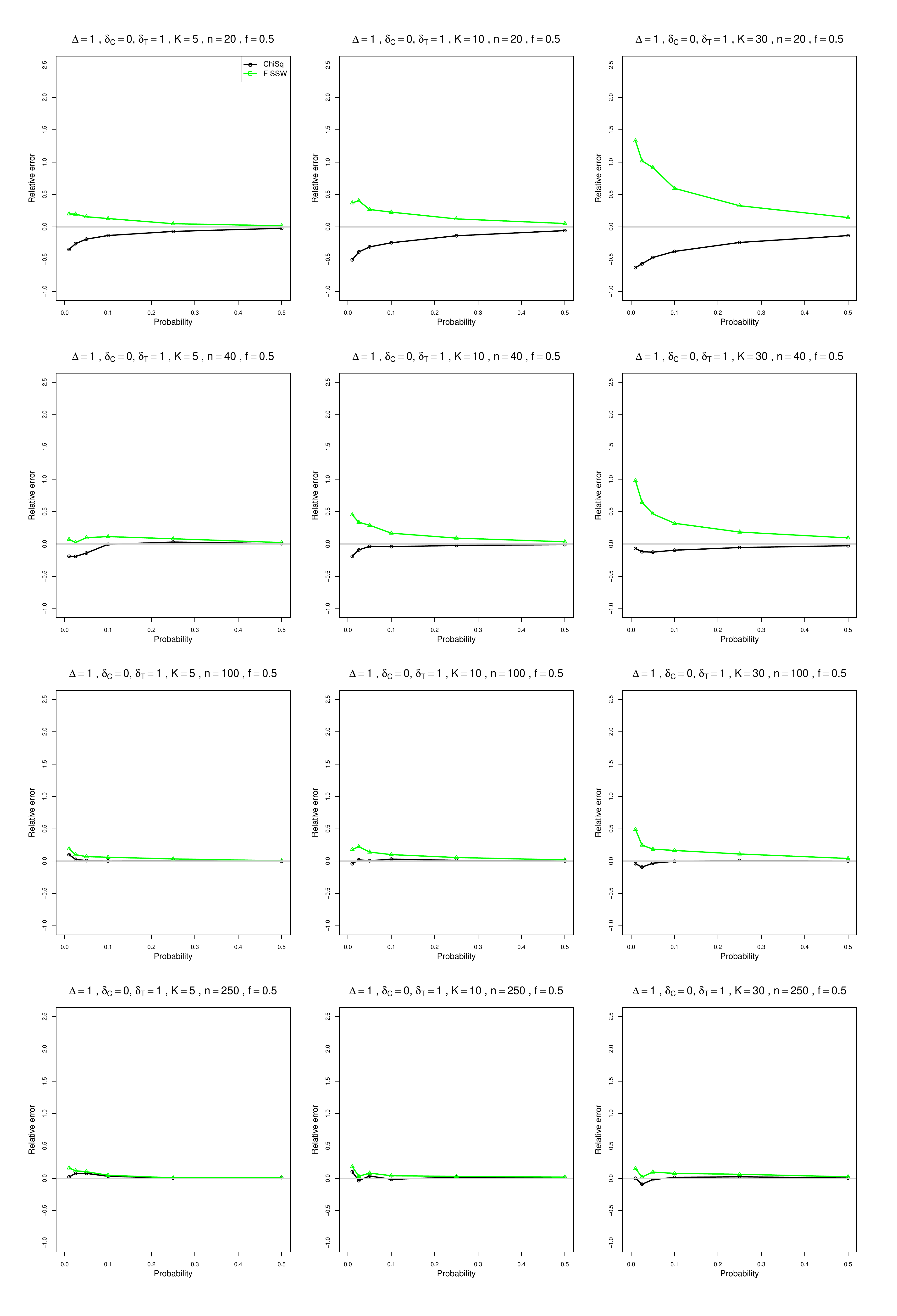}
	\caption{Relative error between the achieved level and the nominal level for two approximations to the null distribution of Q for DSM (Chisq and F SSW) vs upper tail probability, for equal sample sizes $n=20,\;40,\;100$ and $250$, $\delta_{iC} = 0$, $\Delta=1$ and  $f = 0.5$.   }
	\label{Pplot_relative_truncated_deltaC_=0deltaT=1_DSM_equal_sample_sizes.pdf}
\end{figure}

\begin{figure}[ht]
	\centering
	\includegraphics[scale=0.33]{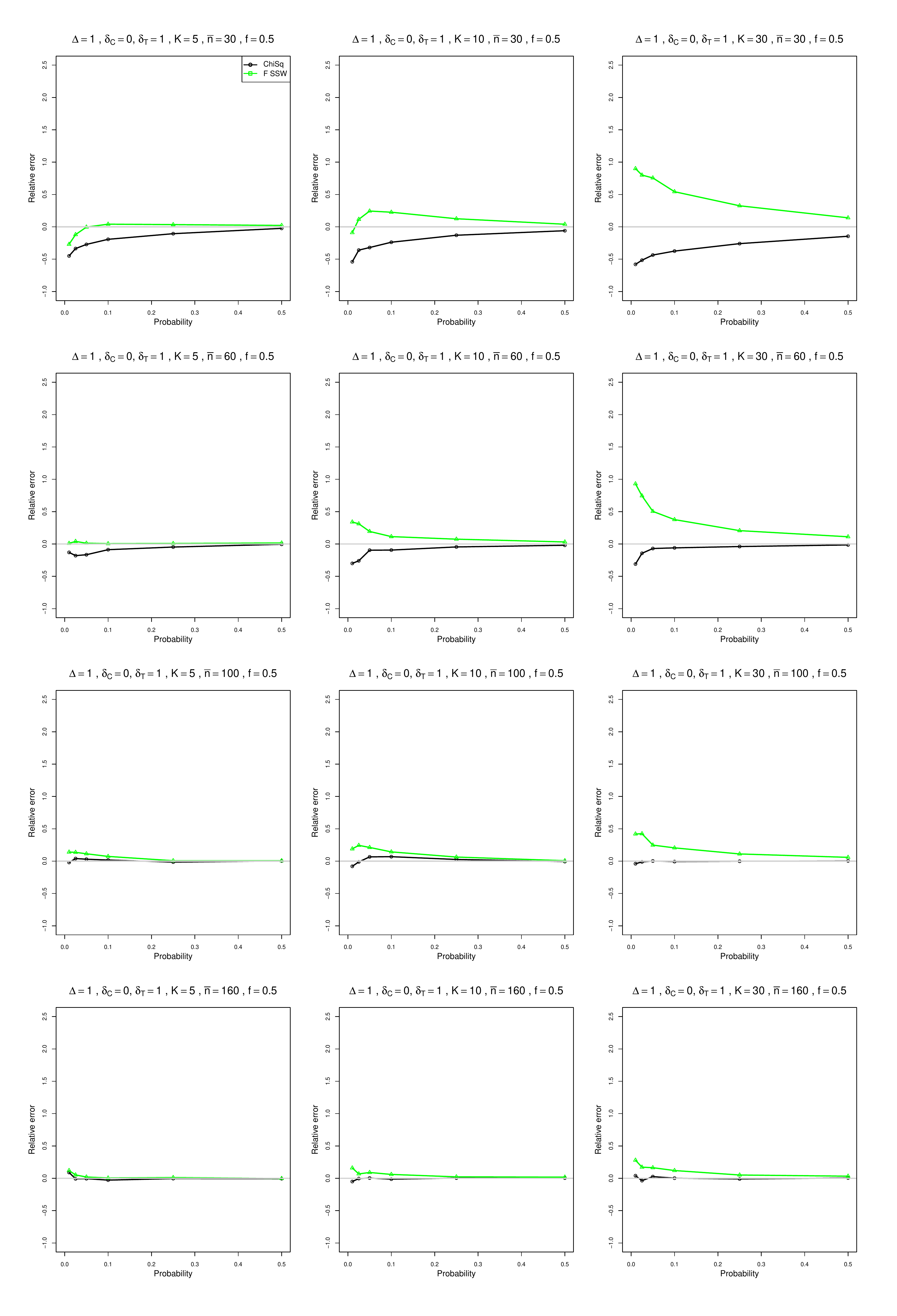}
	\caption{Relative error between the achieved level and the nominal level for two approximations to the null distribution of Q for DSM (Chisq and F SSW) vs upper tail probability, for unequal sample sizes $\bar{n}=30,\;60,\;100$ and $160$, $\delta_{iC} = 0$, $\Delta=1$ and  $f = 0.5$.   }
	\label{Pplot_relative_truncated_deltaC_0deltaT=1_DSM_unequal_sample_sizes.pdf}
\end{figure}

\begin{figure}[ht]
	\centering
	\includegraphics[scale=0.33]{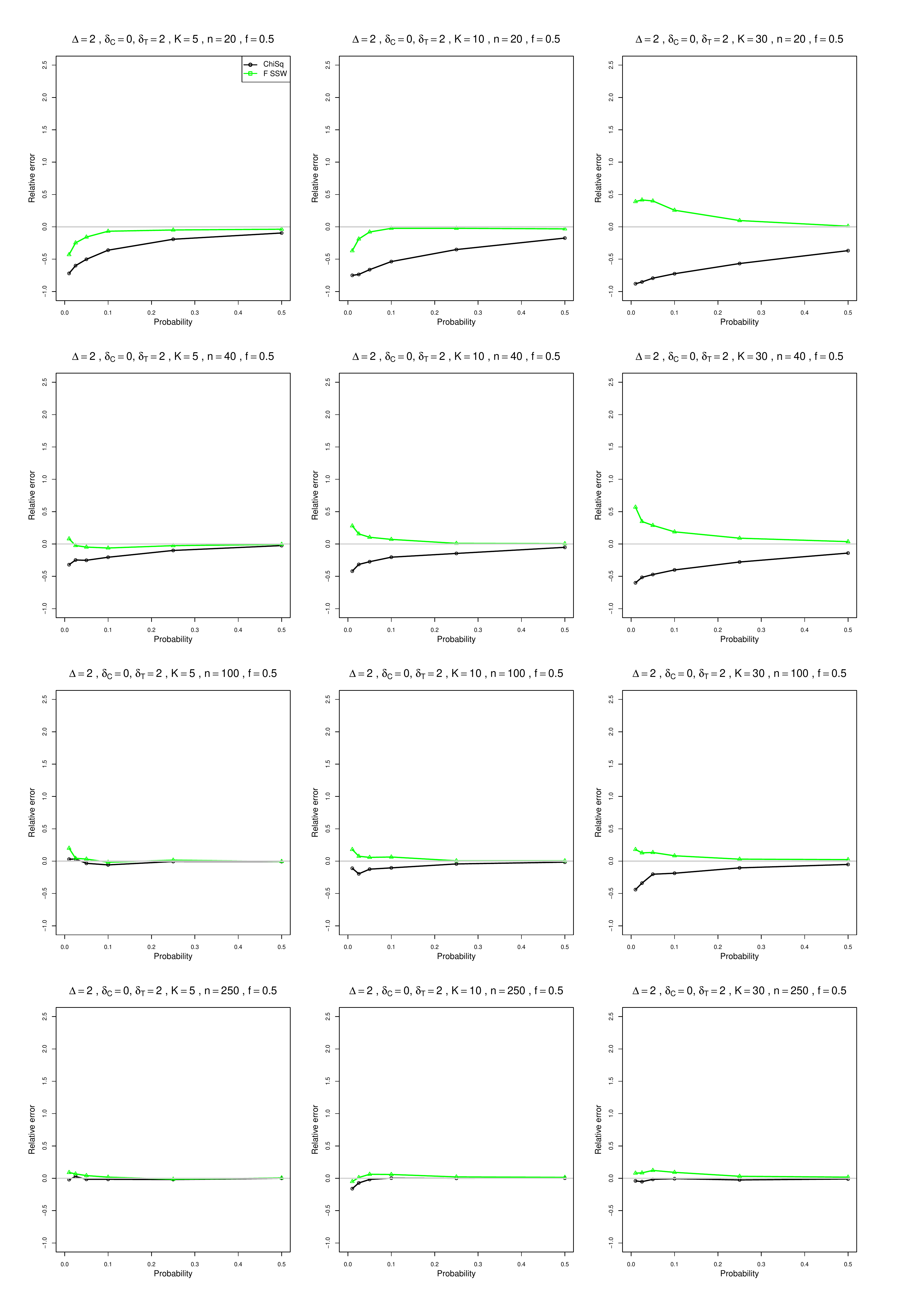}
	\caption{Relative error between the achieved level and the nominal level for two approximations to the null distribution of Q for DSM (Chisq and F SSW) vs upper tail probability, for equal sample sizes $n=20,\;40,\;100$ and $250$, $\delta_{iC} = 0$, $\Delta=2$ and  $f = 0.5$.   }
	\label{Pplot_relative_truncated_deltaC_0deltaT=2_DSM_equal_sample_sizes.pdf}
\end{figure}

\begin{figure}[ht]
	\centering
	\includegraphics[scale=0.33]{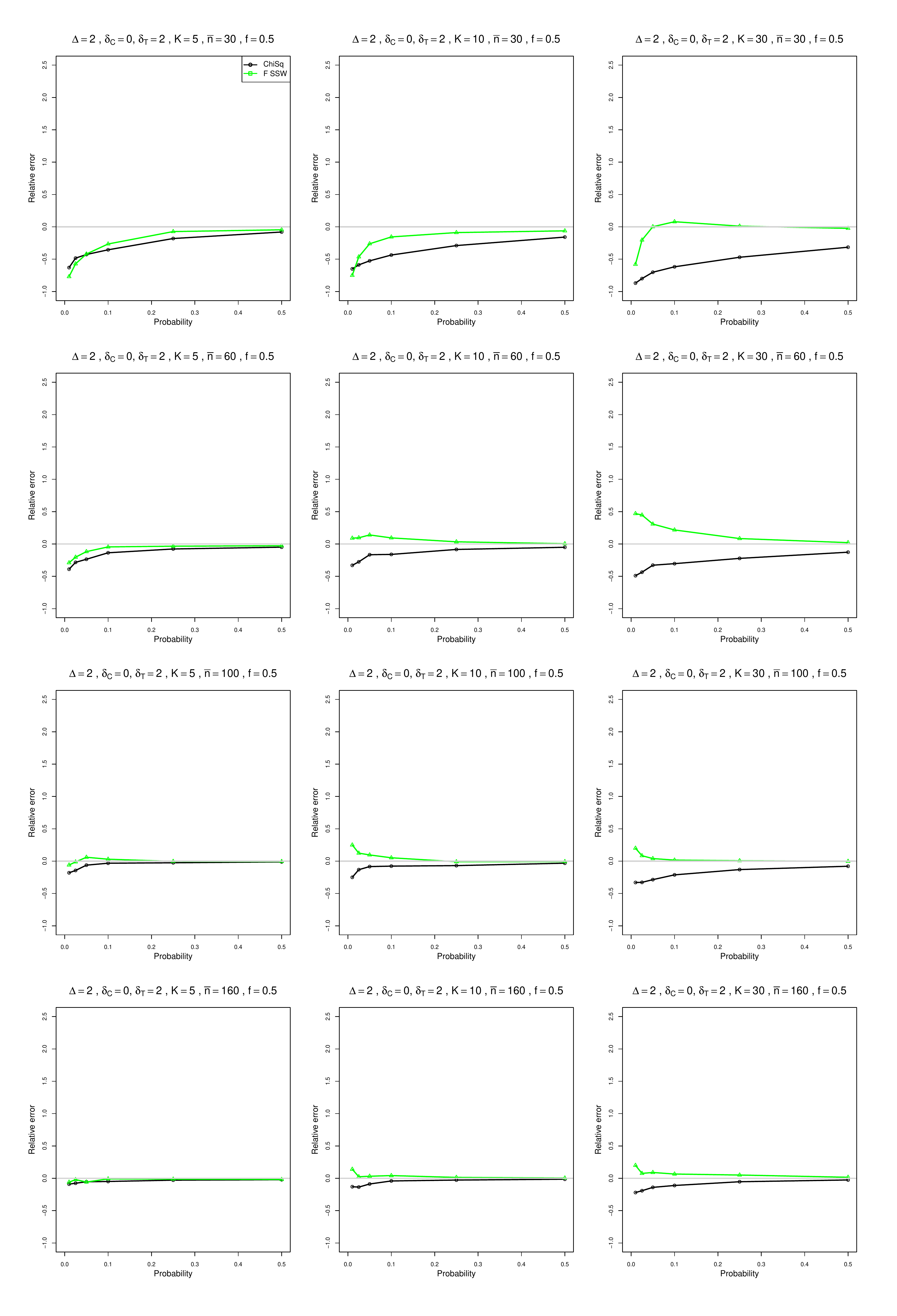}
	\caption{Relative error between the achieved level and the nominal level for two approximations to the null distribution of Q for DSM (Chisq and F SSW) vs upper tail probability, for unequal sample sizes $\bar{n}=30,\;60,\;100$ and $160$, $\delta_{iC} = 0$, $\Delta=2$ and  $f = 0.5$.   }
	\label{Pplot_relative_truncated_deltaC_0deltaT=2_DSM_unequal_sample_sizes.pdf}
\end{figure}


\begin{figure}[ht]
	\centering
	\includegraphics[scale=0.33]{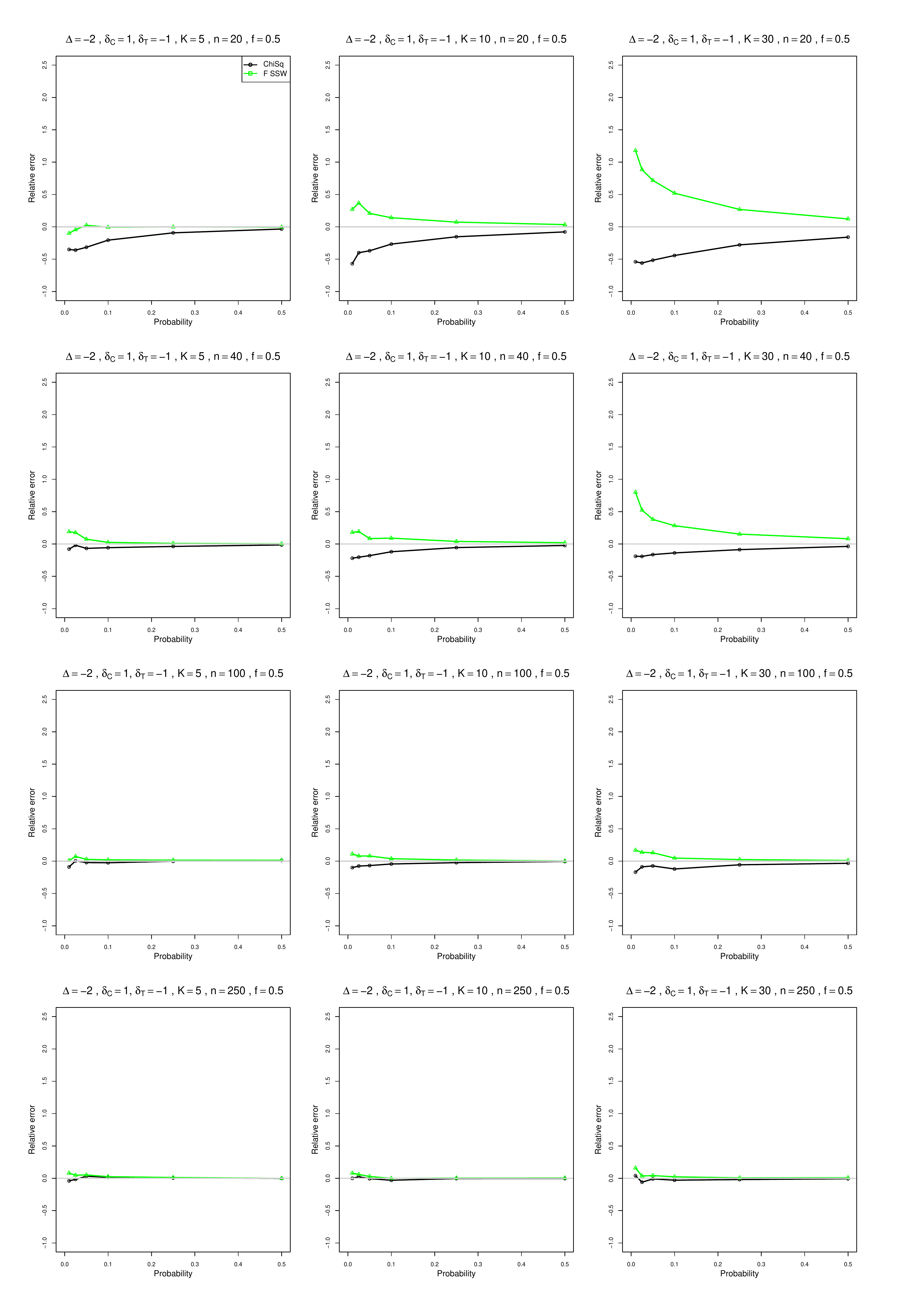}
	\caption{Relative error between the achieved level and the nominal level for two approximations to the null distribution of Q for DSM (Chisq and F SSW) vs upper tail probability, for equal sample sizes $n=20,\;40,\;100$ and $250$, $\delta_{iC} = -1$, $\Delta=-2$ and  $f = 0.5$.   }
	\label{Pplot_relative_truncated_deltaC_1deltaT=-1_DSM_equal_sample_sizes.pdf}
\end{figure}

\begin{figure}[ht]
	\centering
	\includegraphics[scale=0.33]{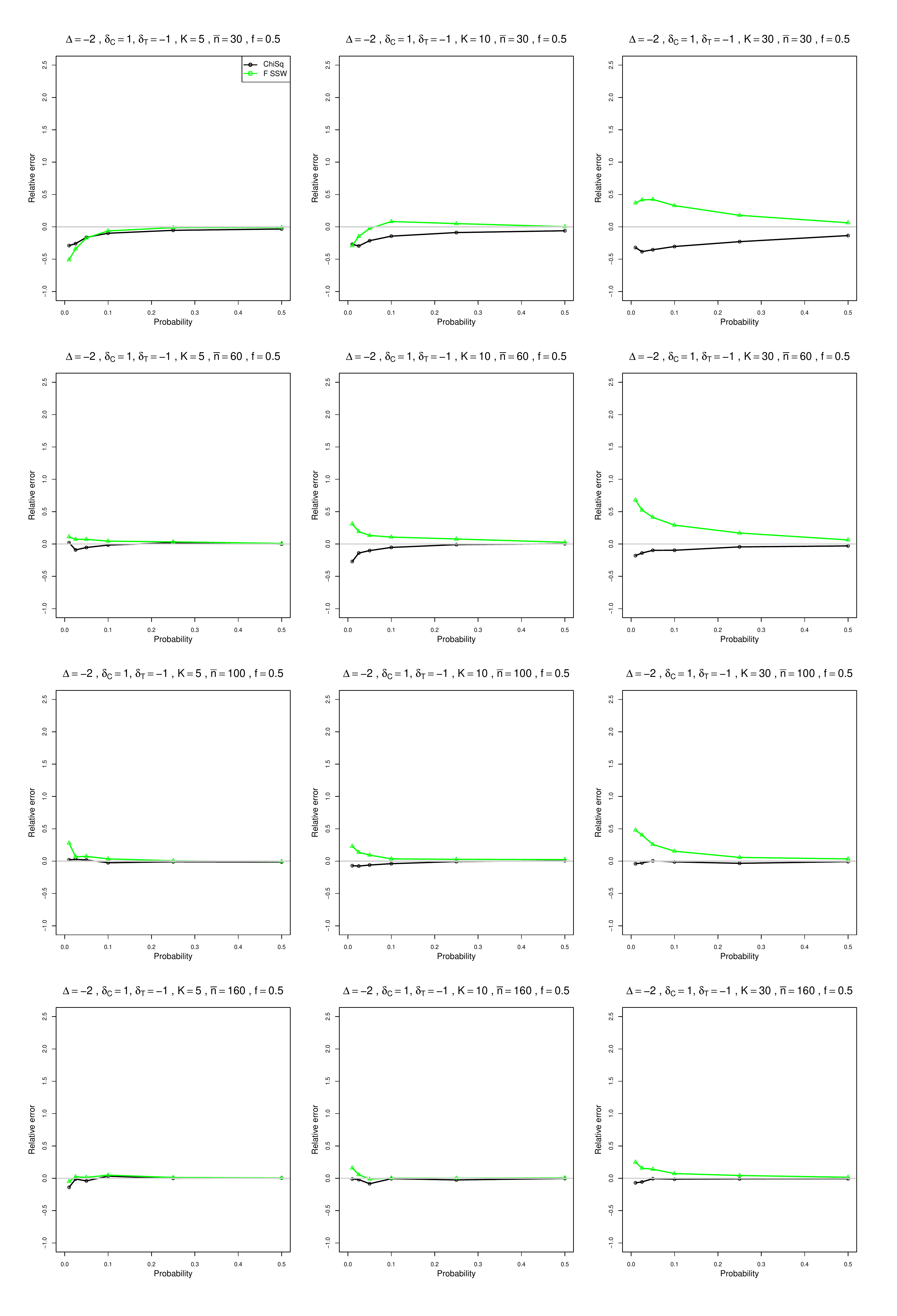}
	\caption{Relative error between the achieved level and the nominal level for two approximations to the null distribution of Q for DSM (Chisq and F SSW) vs upper tail probability, for unequal sample sizes $\bar{n}=30,\;60,\;100$ and $160$, $\delta_{iC} = 1$, $\Delta=-2$ and  $f = 0.5$.   }
	\label{Pplot_relative_truncated_deltaC_1deltaT=-1_DSM_unequal_sample_sizes.pdf}
\end{figure}

\begin{figure}[ht]
	\centering
	\includegraphics[scale=0.33]{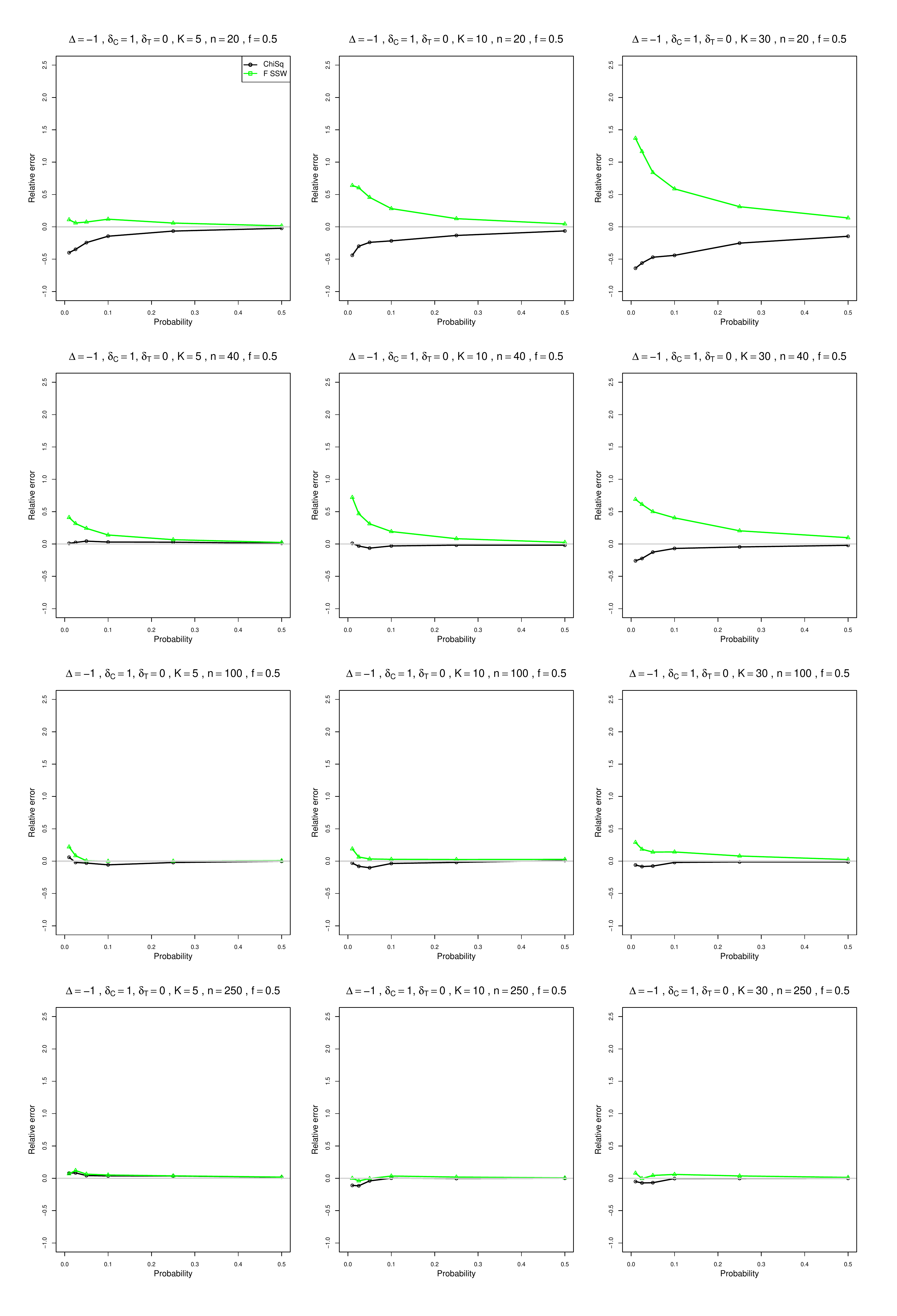}
	\caption{Relative error between the achieved level and the nominal level for two approximations to the null distribution of Q for DSM (Chisq and F SSW) vs upper tail probability, for equal sample sizes $n=20,\;40,\;100$ and $250$, $\delta_{iC} = 1$, $\Delta=-1$ and  $f = 0.5$.   }
	\label{Pplot_relative_truncated_deltaC_1deltaT=0_DSM_equal_sample_sizes.pdf}
\end{figure}

\begin{figure}[ht]
	\centering
	\includegraphics[scale=0.33]{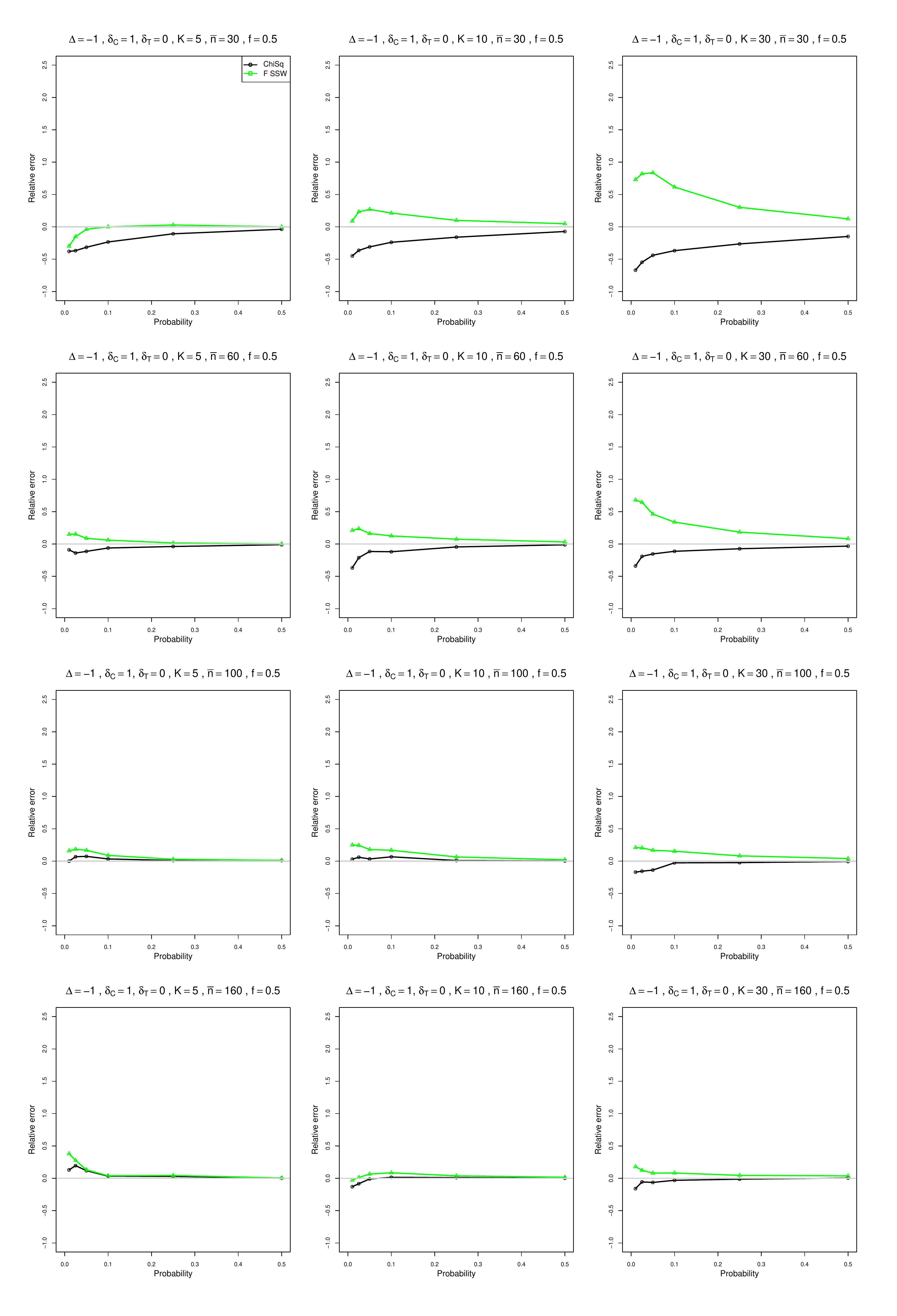}
	\caption{Relative error between the achieved level and the nominal level for two approximations to the null distribution of Q for DSM (Chisq and F SSW) vs upper tail probability, for unequal sample sizes $\bar{n}=30,\;60,\;100$ and $160$, $\delta_{iC} = 1$, $\Delta=-1$ and  $f = 0.5$.   }
	\label{Pplot_relative_truncated_deltaC_1deltaT=0_DSM_unequal_sample_sizes.pdf}
\end{figure}

\begin{figure}[ht]
	\centering
	\includegraphics[scale=0.33]{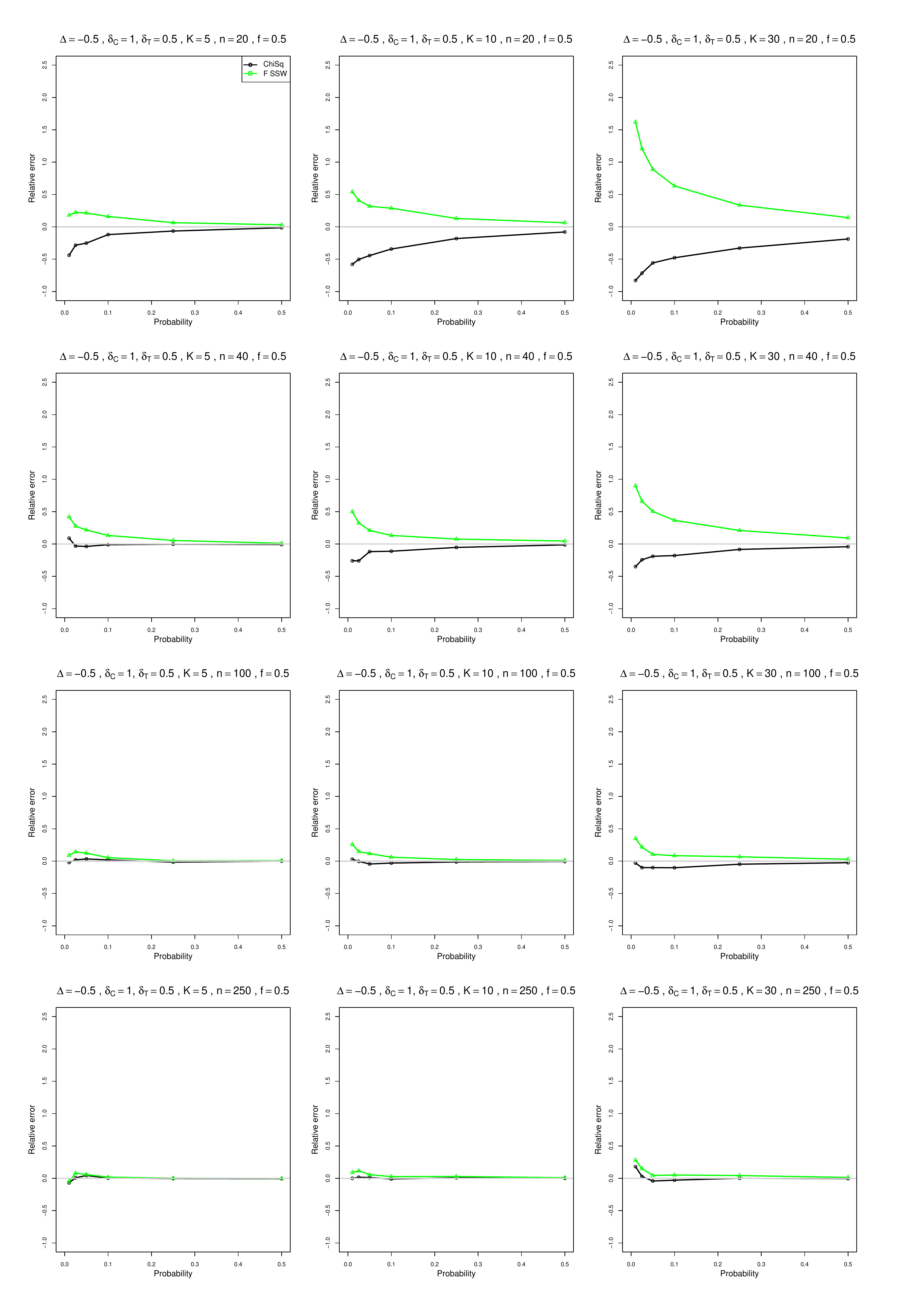}
	\caption{Relative error between the achieved level and the nominal level for two approximations to the null distribution of Q for DSM (Chisq and F SSW) vs upper tail probability, for equal sample sizes $n=20,\;40,\;100$ and $250$, $\delta_{iC} = 1$, $\Delta=-0.5$ and  $f = 0.5$.   }
	\label{Pplot_relative_truncated_deltaC_1deltaT=0.5_DSM_equal_sample_sizes.pdf}
\end{figure}

\begin{figure}[ht]
	\centering
	\includegraphics[scale=0.33]{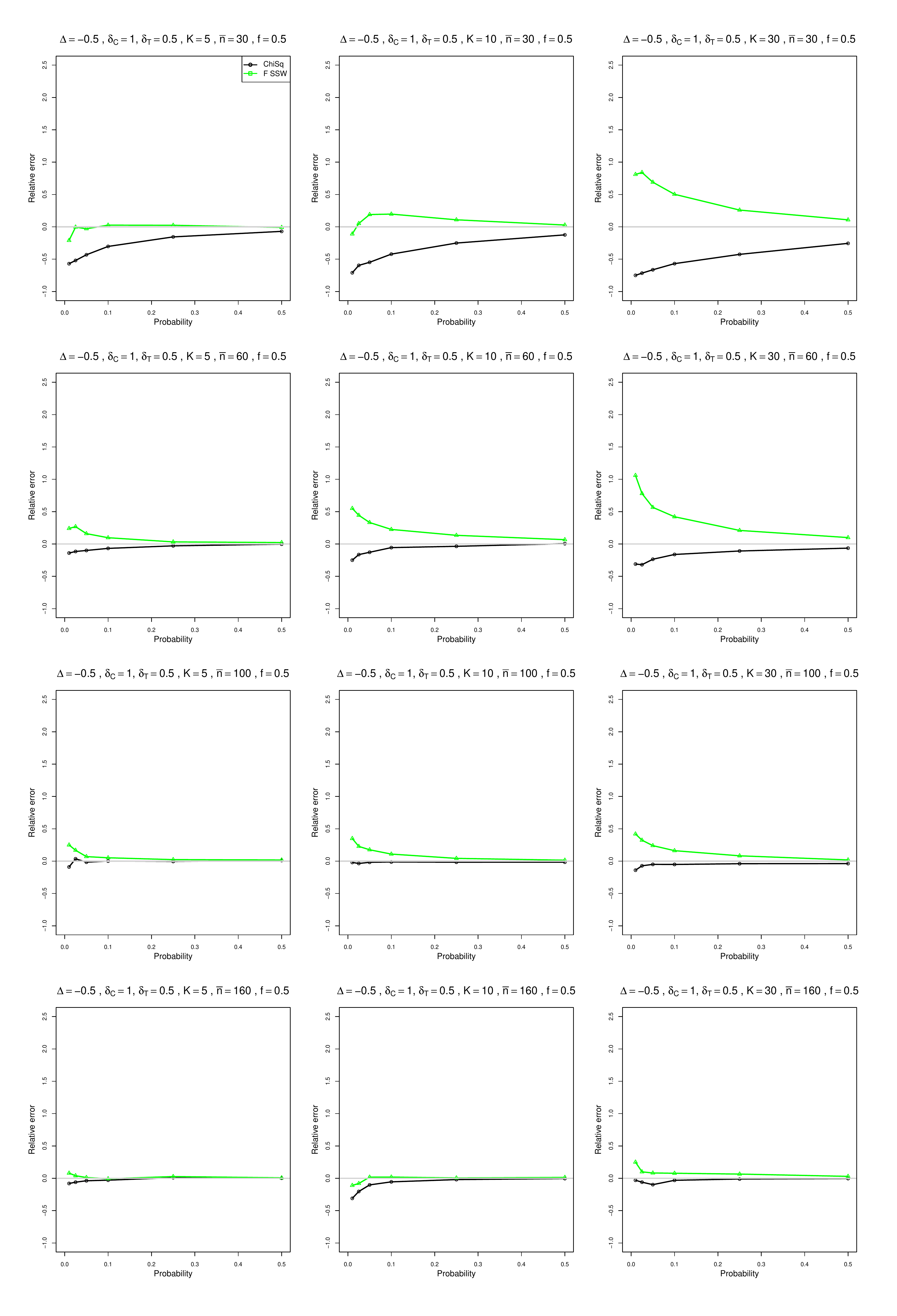}
	\caption{Relative error between the achieved level and the nominal level for two approximations to the null distribution of Q for DSM (Chisq and F SSW) vs upper tail probability, for unequal sample sizes $\bar{n}=30,\;60,\;100$ and $160$, $\delta_{iC} = 1$, $\Delta=-0.5$ and  $f = 0.5$.   }
	\label{Pplot_relative_truncated_deltaC_1deltaT=0.5_DSM_unequal_sample_sizes.pdf}
\end{figure}

\begin{figure}[ht]
	\centering
	\includegraphics[scale=0.33]{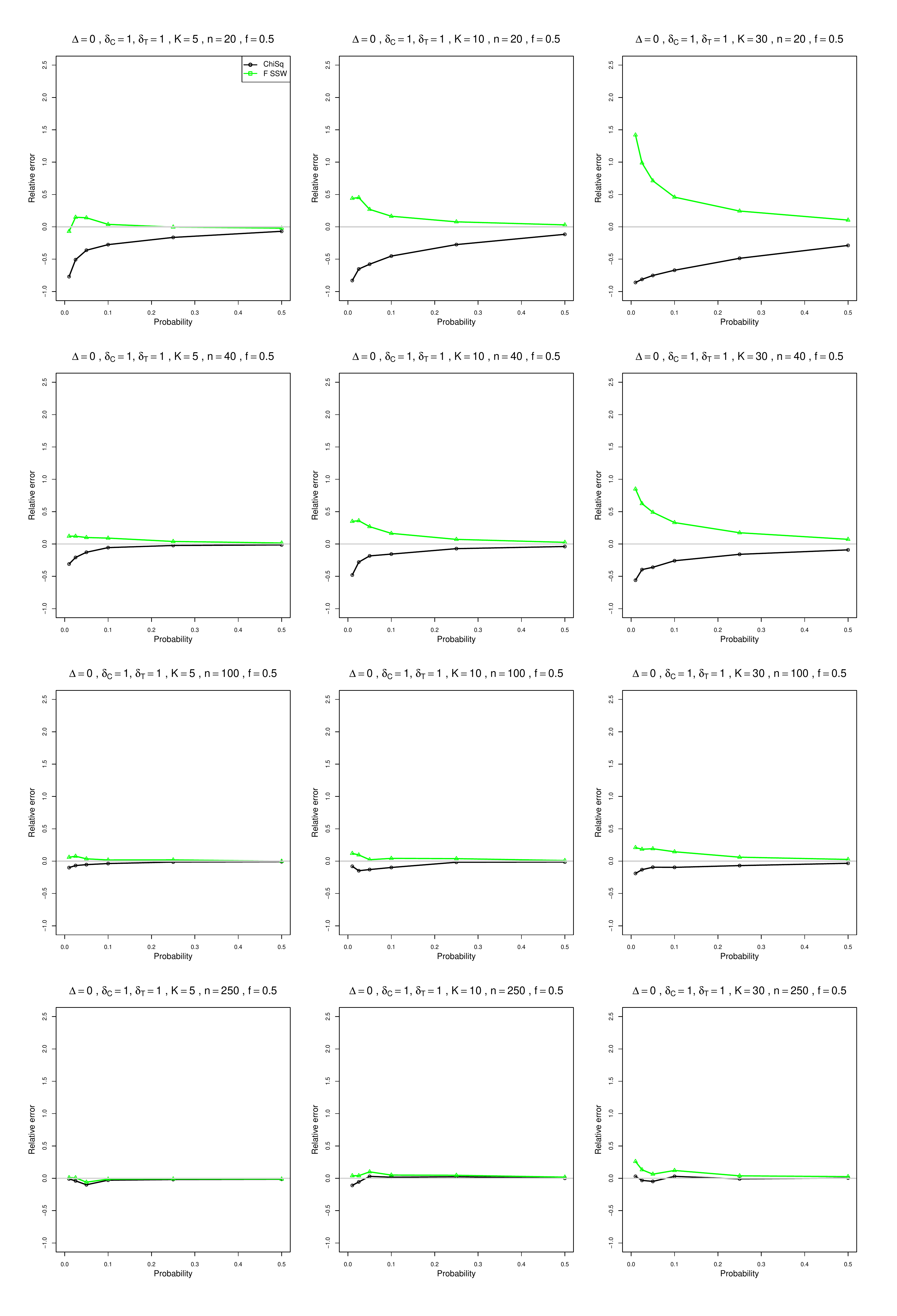}
	\caption{Relative error between the achieved level and the nominal level for two approximations to the null distribution of Q for DSM (Chisq and F SSW) vs upper tail probability, for equal sample sizes $n=20,\;40,\;100$ and $250$, $\delta_{iC} = 1$, $\Delta=0$ and  $f = 0.5$.   }
	\label{Pplot_relative_truncated_deltaC_1deltaT=1_DSM_equal_sample_sizes.pdf}
\end{figure}

\begin{figure}[ht]
	\centering
	\includegraphics[scale=0.33]{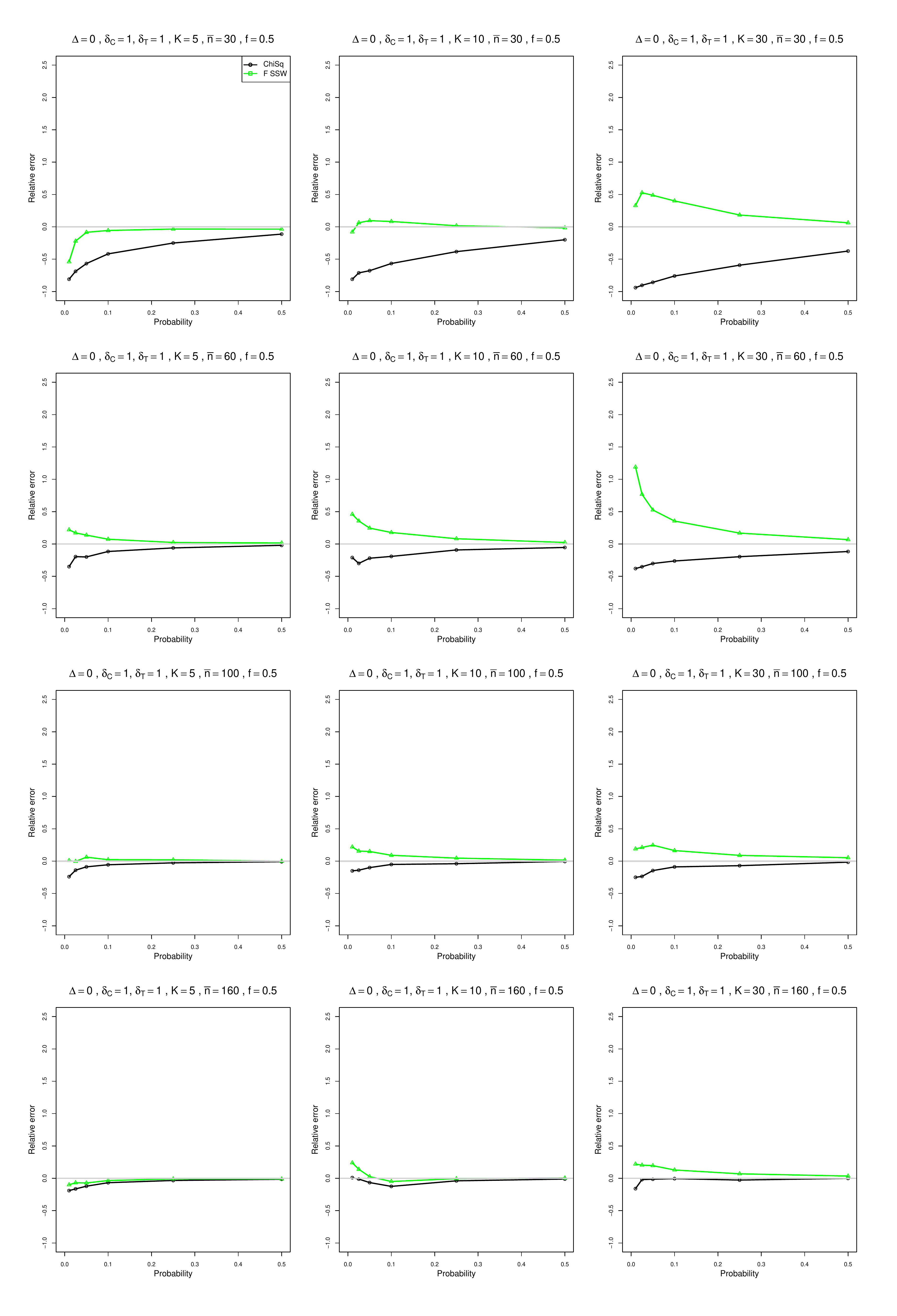}
	\caption{Relative error between the achieved level and the nominal level for two approximations to the null distribution of Q for DSM (Chisq and F SSW) vs upper tail probability, for unequal sample sizes $\bar{n}=30,\;60,\;100$ and $160$, $\delta_{iC} = 1$, $\Delta=0$ and  $f = 0.5$.   }
	\label{Pplot_relative_truncated_deltaC_1deltaT=1_DSM_unequal_sample_sizes.pdf}
\end{figure}

\begin{figure}[ht]
	\centering
	\includegraphics[scale=0.33]{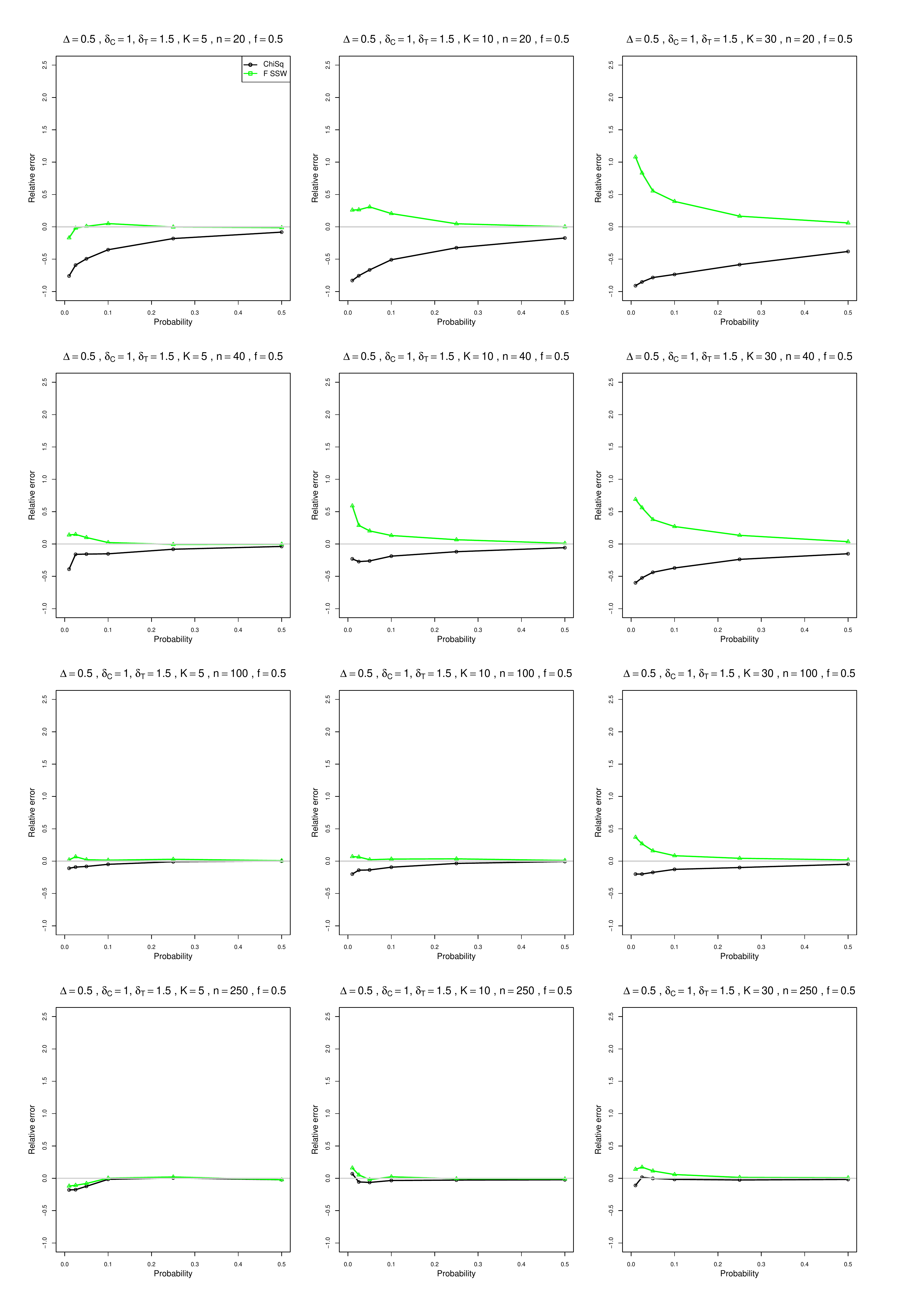}
	\caption{Relative error between the achieved level and the nominal level for two approximations to the null distribution of Q for DSM (Chisq and F SSW) vs upper tail probability, for equal sample sizes $n=20,\;40,\;100$ and $250$, $\delta_{iC} = 1$, $\Delta=0.5$ and  $f = 0.5$.   }
	\label{Pplot_relative_truncated_deltaC_1deltaT=1.5_DSM_equal_sample_sizes.pdf}
\end{figure}

\begin{figure}[ht]
	\centering
	\includegraphics[scale=0.33]{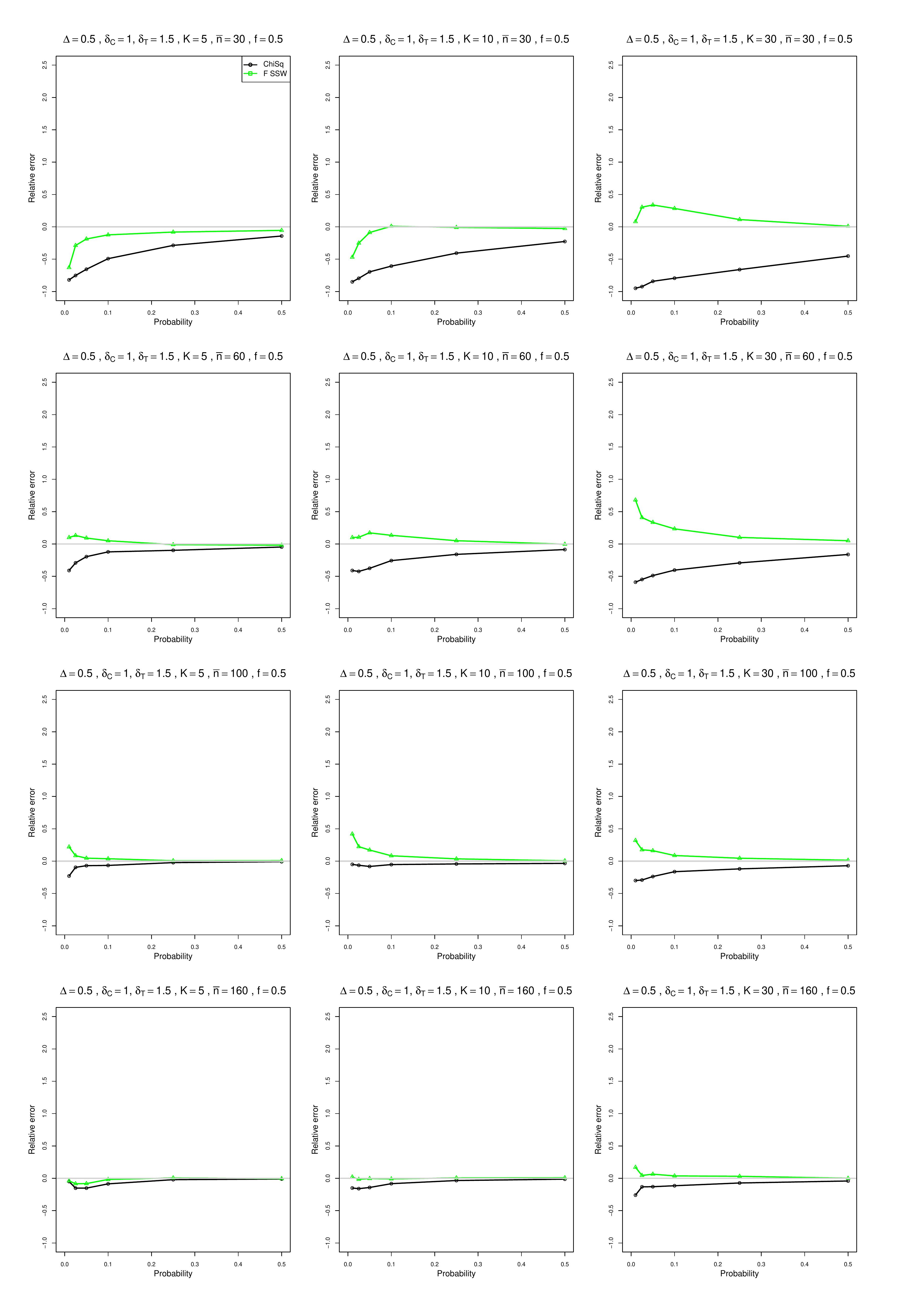}
	\caption{Relative error between the achieved level and the nominal level for two approximations to the null distribution of Q for DSM (Chisq and F SSW) vs upper tail probability, for unequal sample sizes $\bar{n}=30,\;60,\;100$ and $160$, $\delta_{iC} = 1$, $\Delta=0.5$ and  $f = 0.5$.   }
	\label{Pplot_relative_truncated_deltaC_1deltaT=1.5_DSM_unequal_sample_sizes.pdf}
\end{figure}
\clearpage

\begin{figure}[ht]
	\centering
	\includegraphics[scale=0.33]{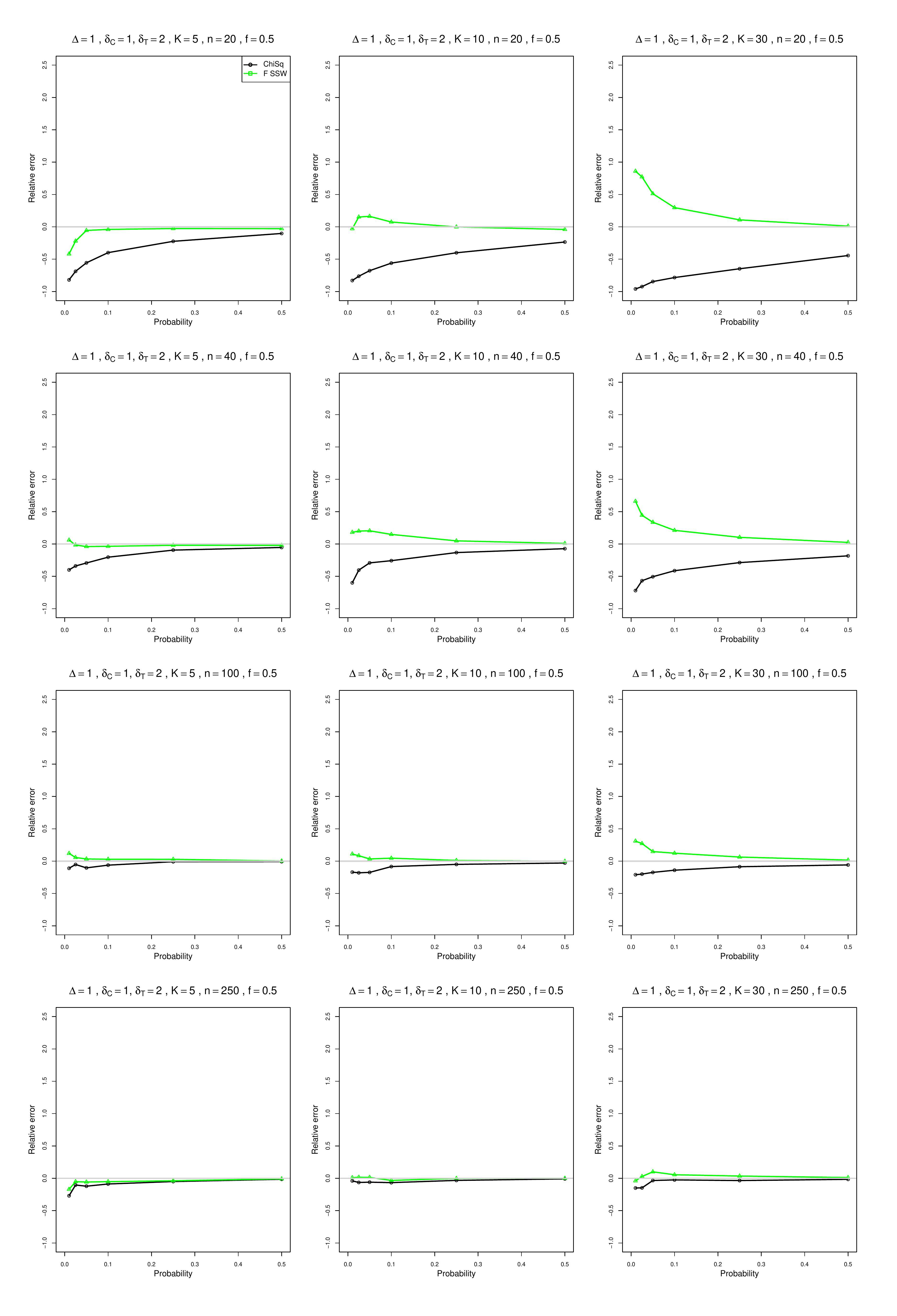}
	\caption{Relative error between the achieved level and the nominal level for two approximations to the null distribution of Q for DSM (Chisq and F SSW) vs upper tail probability, for equal sample sizes $n=20,\;40,\;100$ and $250$, $\delta_{iC} = 1$, $\Delta=1$ and  $f = 0.5$.   }
	\label{Pplot_relative_truncated_deltaC_=1deltaT=2_DSM_equal_sample_sizes.pdf}
\end{figure}

\begin{figure}[ht]
	\centering
	\includegraphics[scale=0.33]{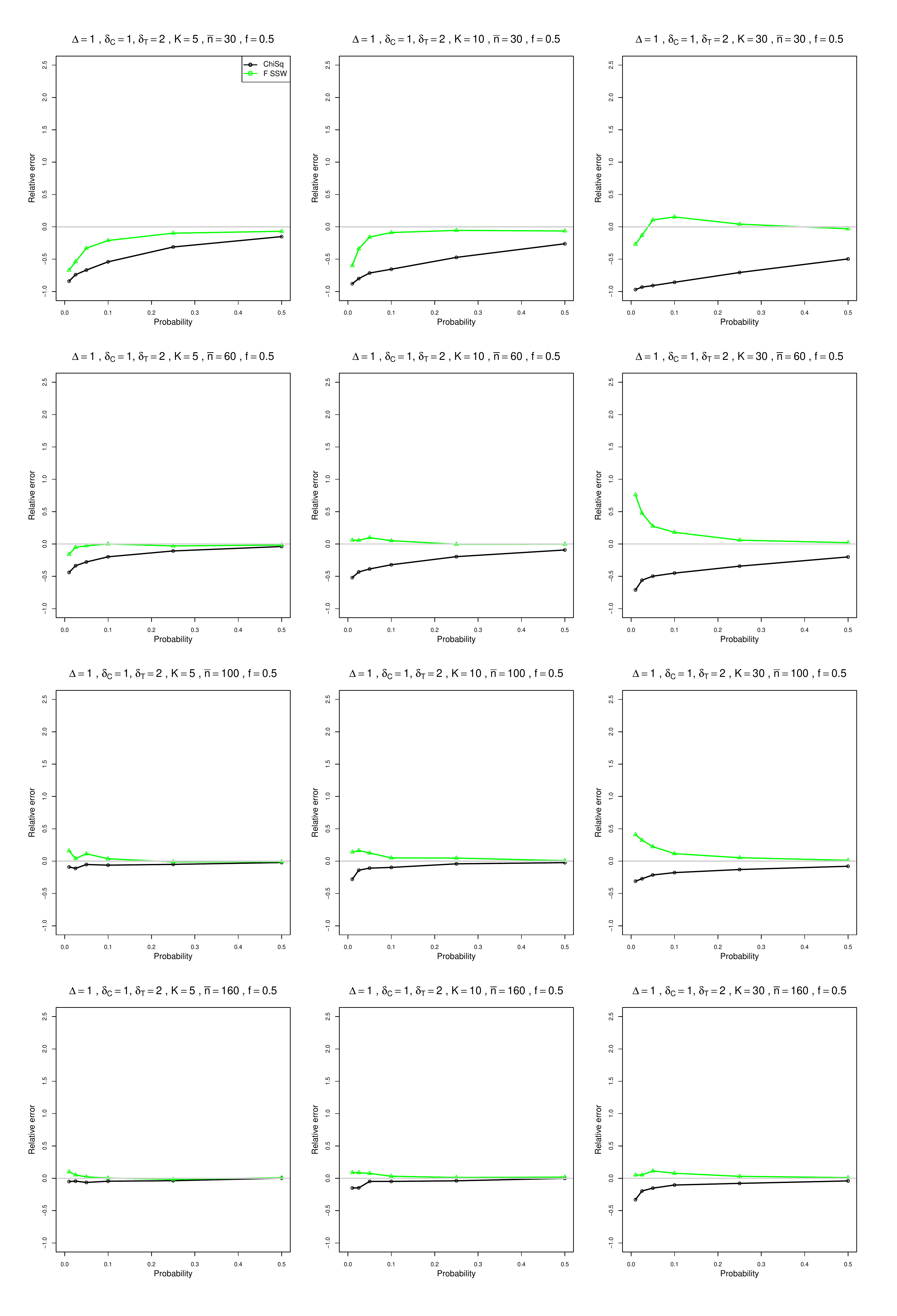}
	\caption{Relative error between the achieved level and the nominal level for two approximations to the null distribution of Q for DSM (Chisq and F SSW) vs upper tail probability, for unequal sample sizes $\bar{n}=30,\;60,\;100$ and $160$, $\delta_{iC} = 1$, $\Delta=1$ and  $f = 0.5$.   }
	\label{Pplot_relative_truncated_deltaC_1deltaT=2_DSM_unequal_sample_sizes.pdf}
\end{figure}

\begin{figure}[ht]
	\centering
	\includegraphics[scale=0.33]{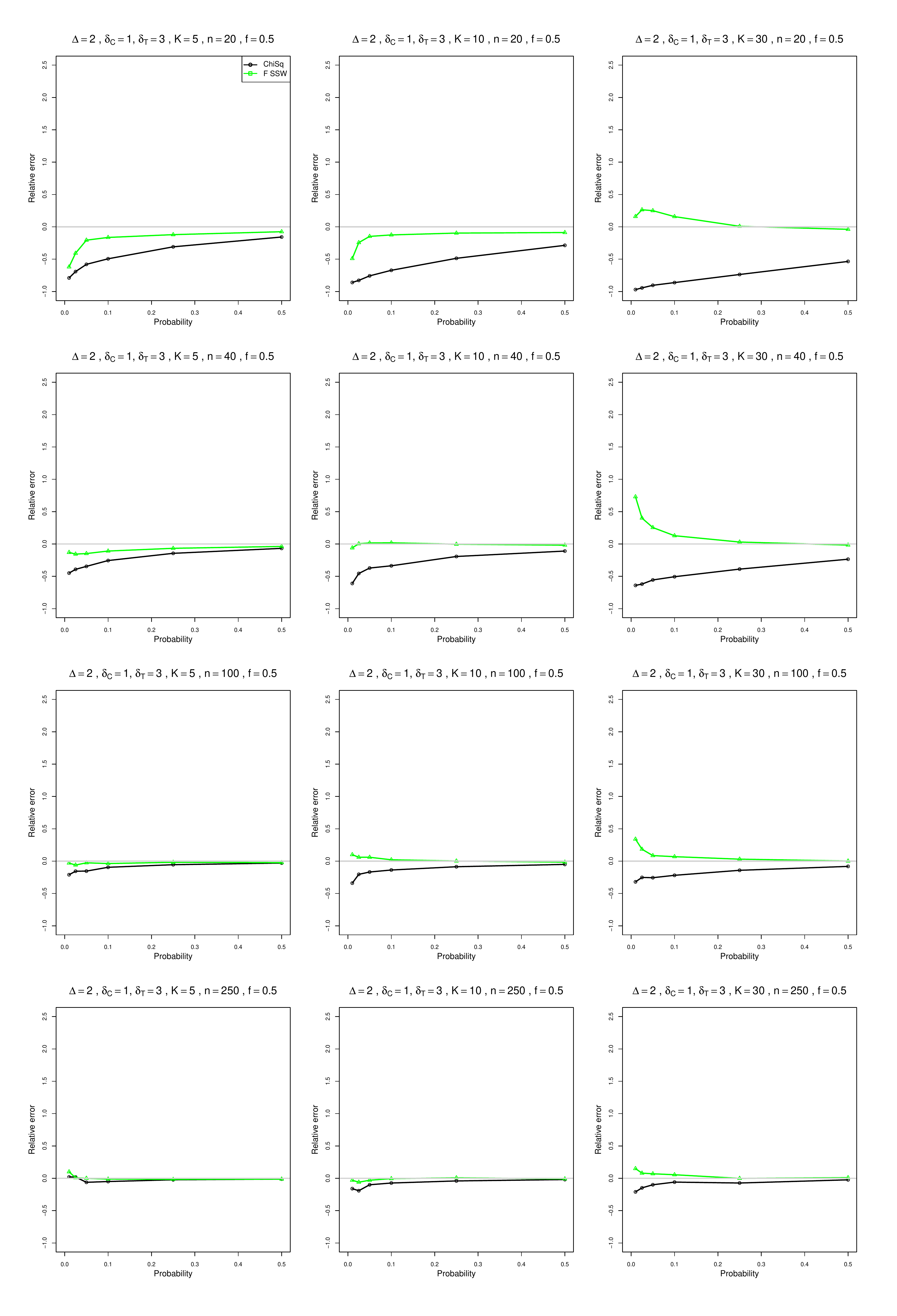}
	\caption{Relative error between the achieved level and the nominal level for two approximations to the null distribution of Q for DSM (Chisq and F SSW) vs upper tail probability, for equal sample sizes $n=20,\;40,\;100$ and $250$, $\delta_{iC} = 1$, $\Delta=2$ and  $f = 0.5$.   }
	\label{Pplot_relative_truncated_deltaC_1deltaT=3_DSM_equal_sample_sizes.pdf}
\end{figure}

\begin{figure}[ht]
	\centering
	\includegraphics[scale=0.33]{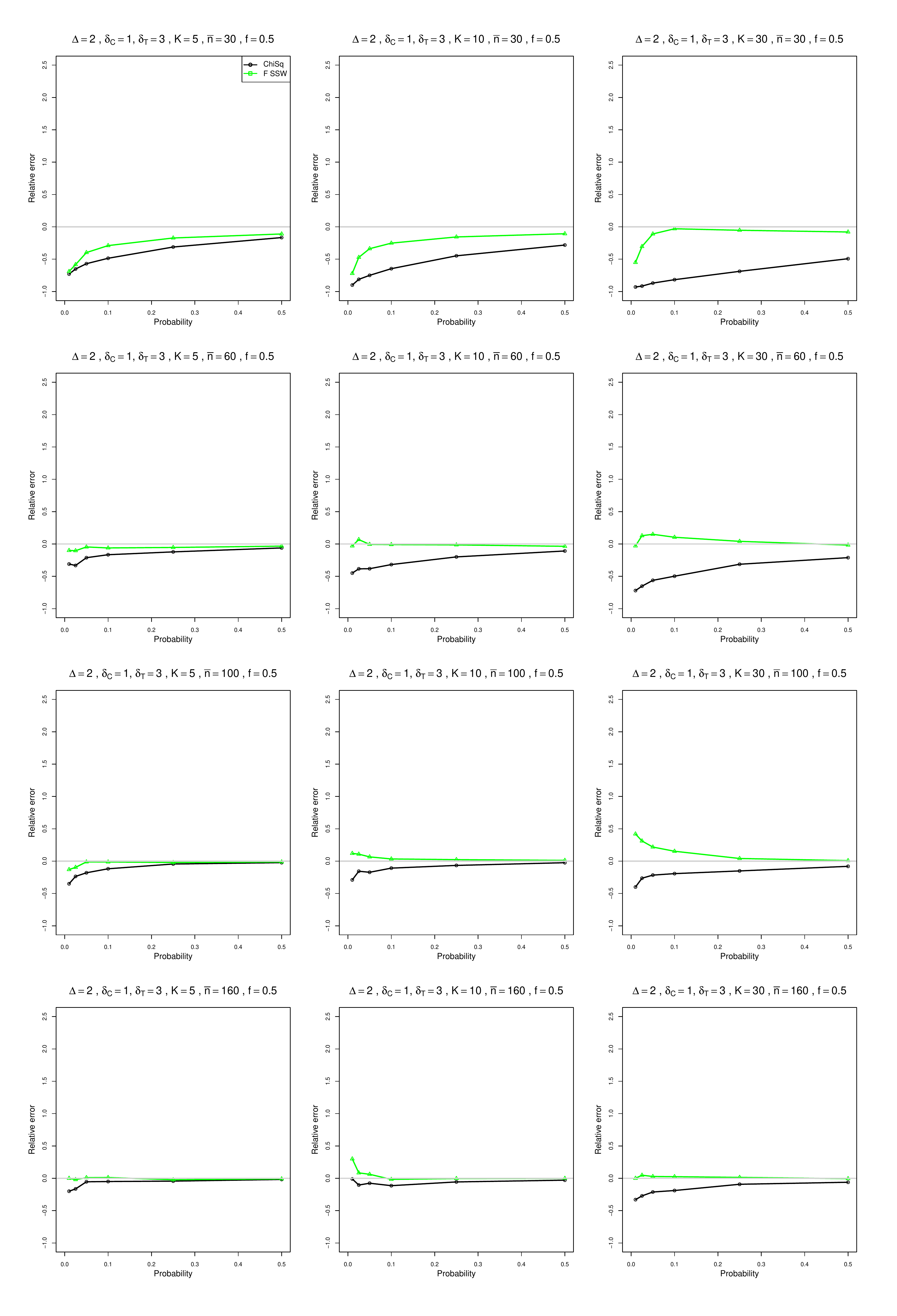}
	\caption{Relative error between the achieved level and the nominal level for two approximations to the null distribution of Q for DSM (Chisq and F SSW) vs upper tail probability, for unequal sample sizes $\bar{n}=30,\;60,\;100$ and $160$, $\delta_{iC} = 1$, $\Delta=2$ and  $f = 0.5$.   }
	\label{Pplot_relative_truncated_deltaC_1deltaT=3_DSM_unequal_sample_sizes.pdf}
\end{figure}


\begin{figure}[ht]
	\centering
	\includegraphics[scale=0.33]{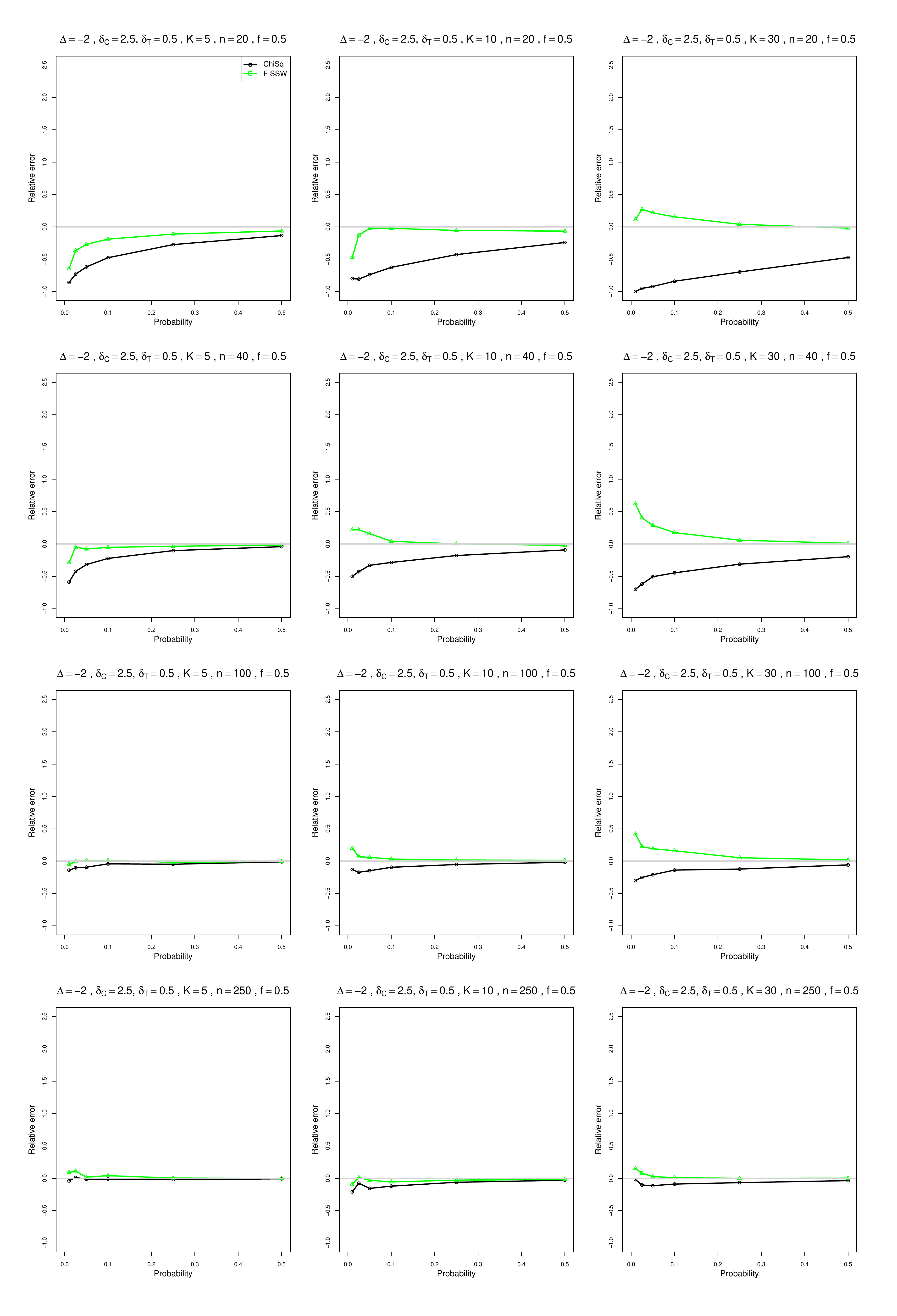}
	\caption{Relative error between the achieved level and the nominal level for two approximations to the null distribution of Q for DSM (Chisq and F SSW) vs upper tail probability, for equal sample sizes $n=20,\;40,\;100$ and $250$, $\delta_{iC} = 2.5$, $\Delta=-2$ and  $f = 0.5$.   }
	\label{Pplot_relative_truncated_deltaC_2.5deltaT=0.5_DSM_equal_sample_sizes.pdf}
\end{figure}

\begin{figure}[ht]
	\centering
	\includegraphics[scale=0.33]{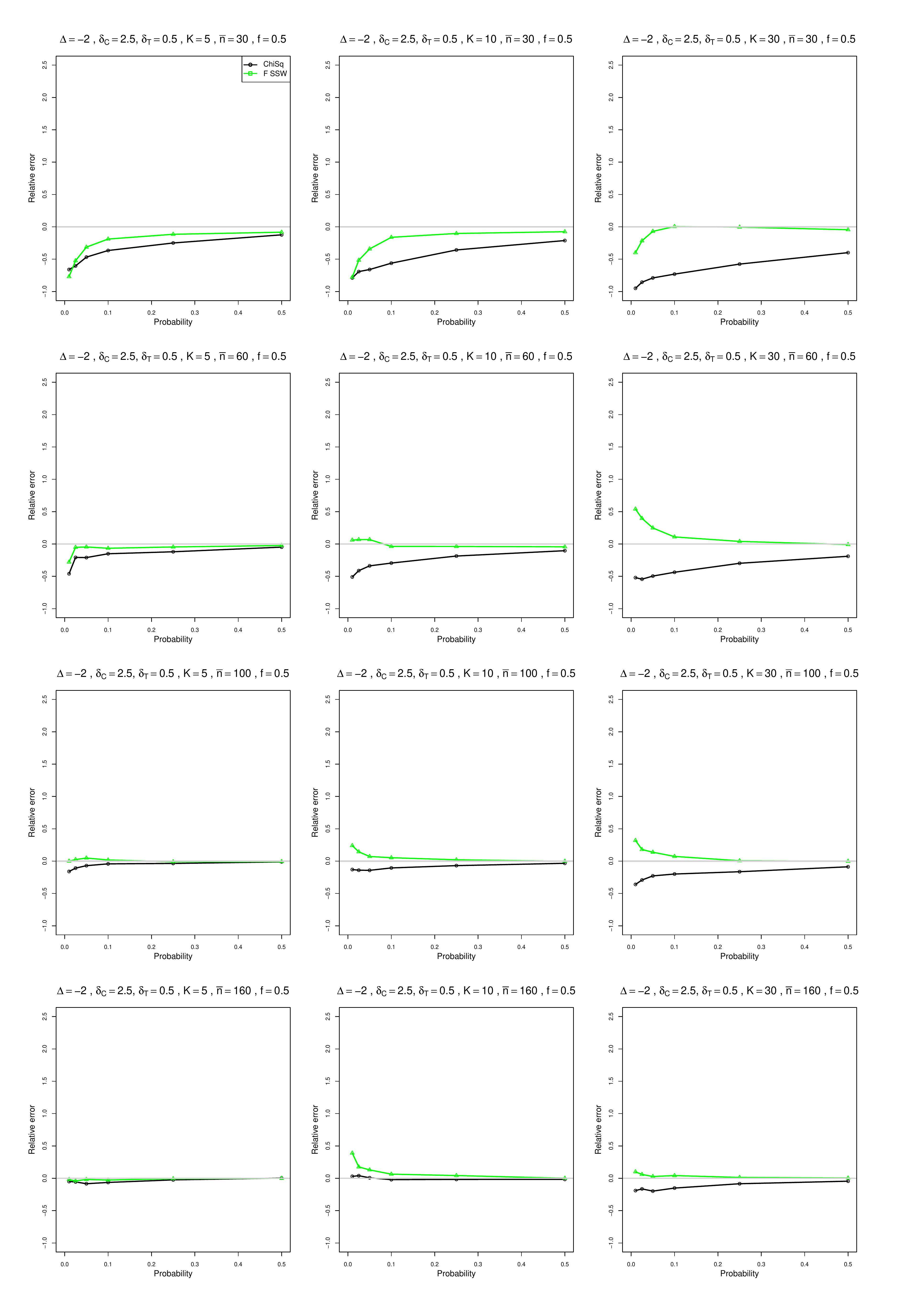}
	\caption{Relative error between the achieved level and the nominal level for two approximations to the null distribution of Q for DSM (Chisq and F SSW) vs upper tail probability, for unequal sample sizes $\bar{n}=30,\;60,\;100$ and $160$, $\delta_{iC} = 2.5$, $\Delta=-2$ and  $f = 0.5$.   }
	\label{Pplot_relative_truncated_deltaC_2.5deltaT=0.5_DSM_unequal_sample_sizes.pdf}
\end{figure}

\begin{figure}[ht]
	\centering
	\includegraphics[scale=0.33]{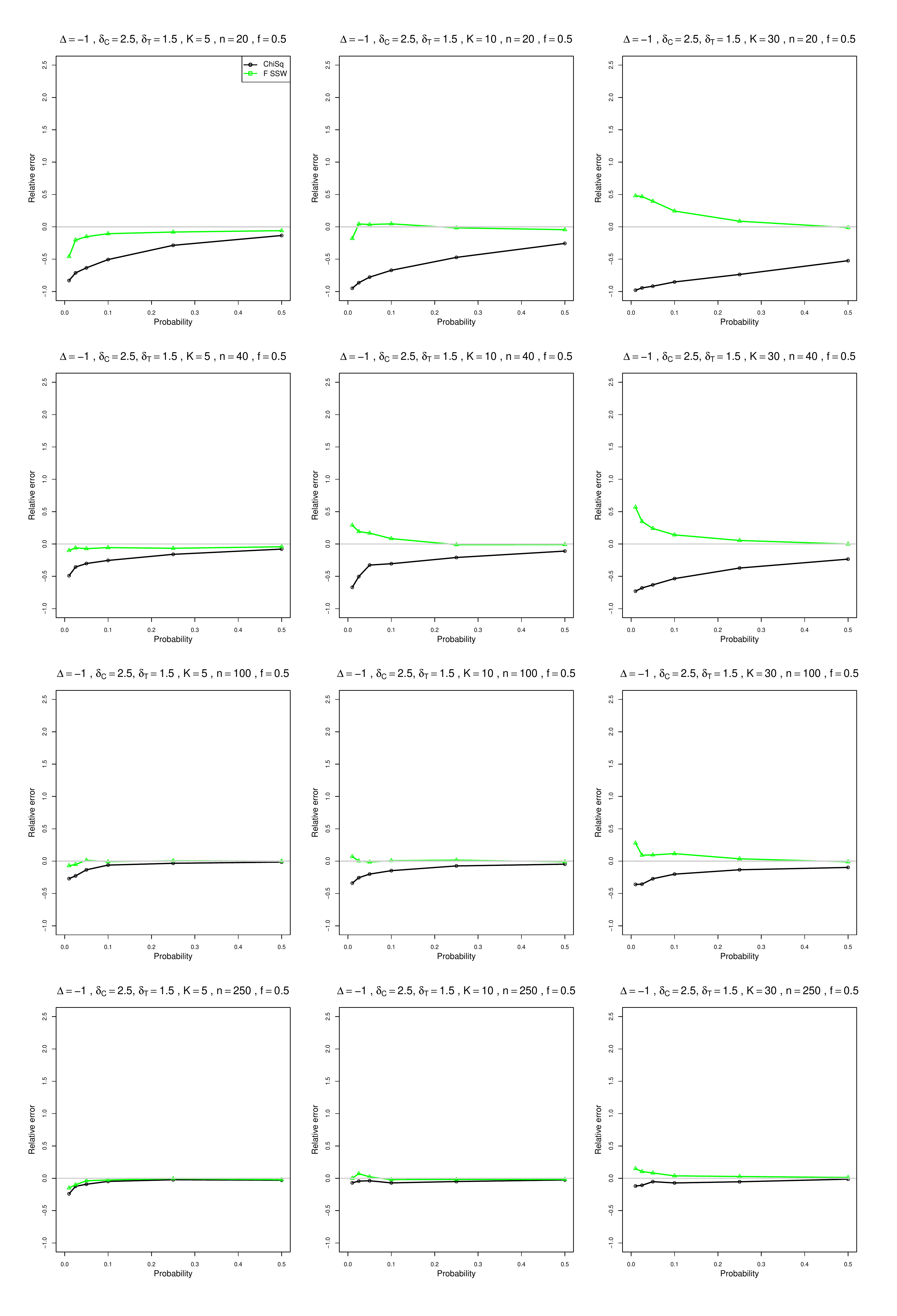}
	\caption{Relative error between the achieved level and the nominal level for two approximations to the null distribution of Q for DSM (Chisq and F SSW) vs upper tail probability, for equal sample sizes $n=20,\;40,\;100$ and $250$, $\delta_{iC} = 2.5$, $\Delta=-1$ and  $f = 0.5$.   }
	\label{Pplot_relative_truncated_deltaC_2.5deltaT=1.5_DSM_equal_sample_sizes.pdf}
\end{figure}

\begin{figure}[ht]
	\centering
	\includegraphics[scale=0.33]{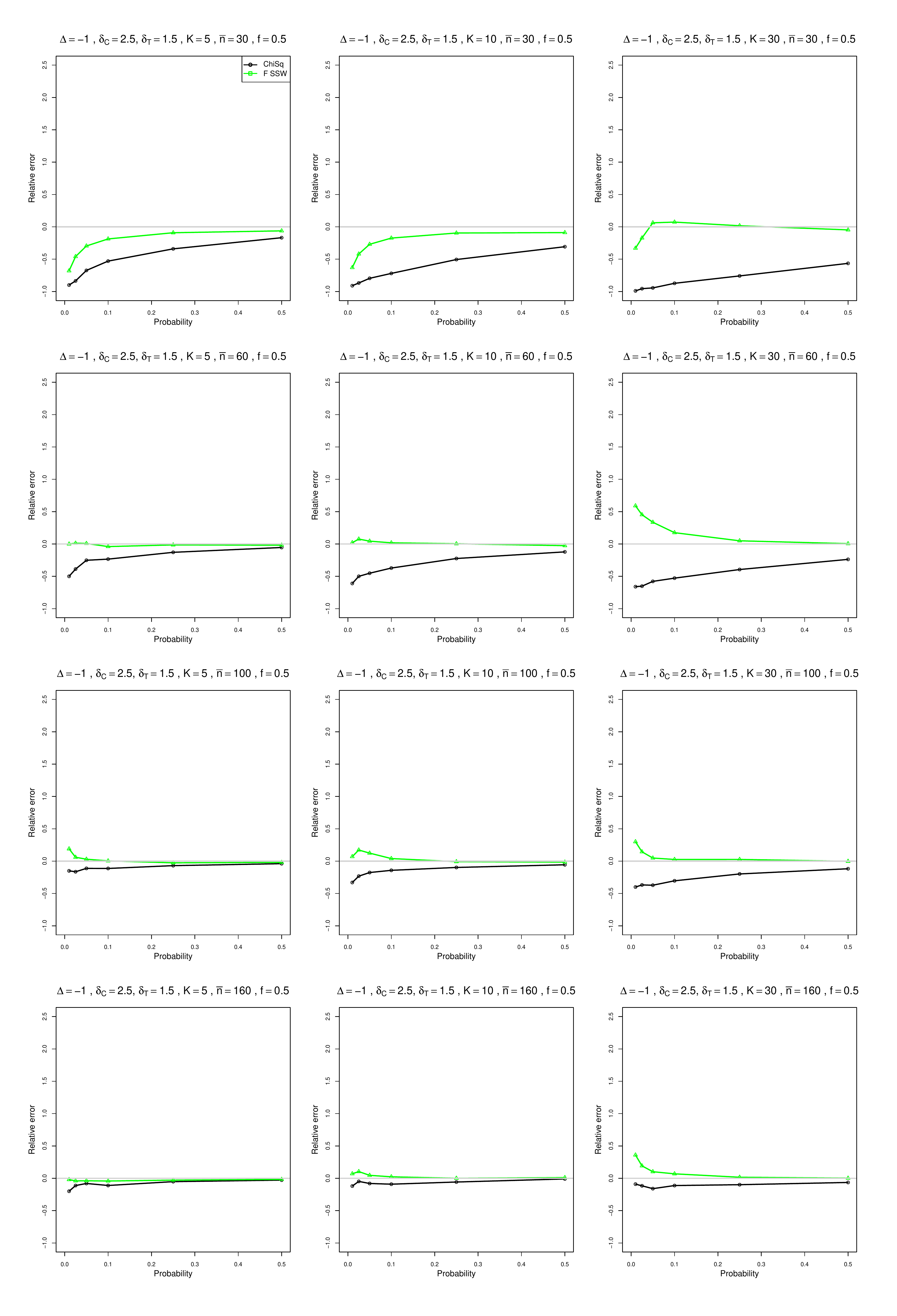}
	\caption{Relative error between the achieved level and the nominal level for two approximations to the null distribution of Q for DSM (Chisq and F SSW) vs upper tail probability, for unequal sample sizes $\bar{n}=30,\;60,\;100$ and $160$, $\delta_{iC} = 2.5$, $\Delta=-1$ and  $f = 0.5$.   }
	\label{Pplot_relative_truncated_deltaC_2.5deltaT=1.5_DSM_unequal_sample_sizes.pdf}
\end{figure}

\begin{figure}[ht]
	\centering
	\includegraphics[scale=0.33]{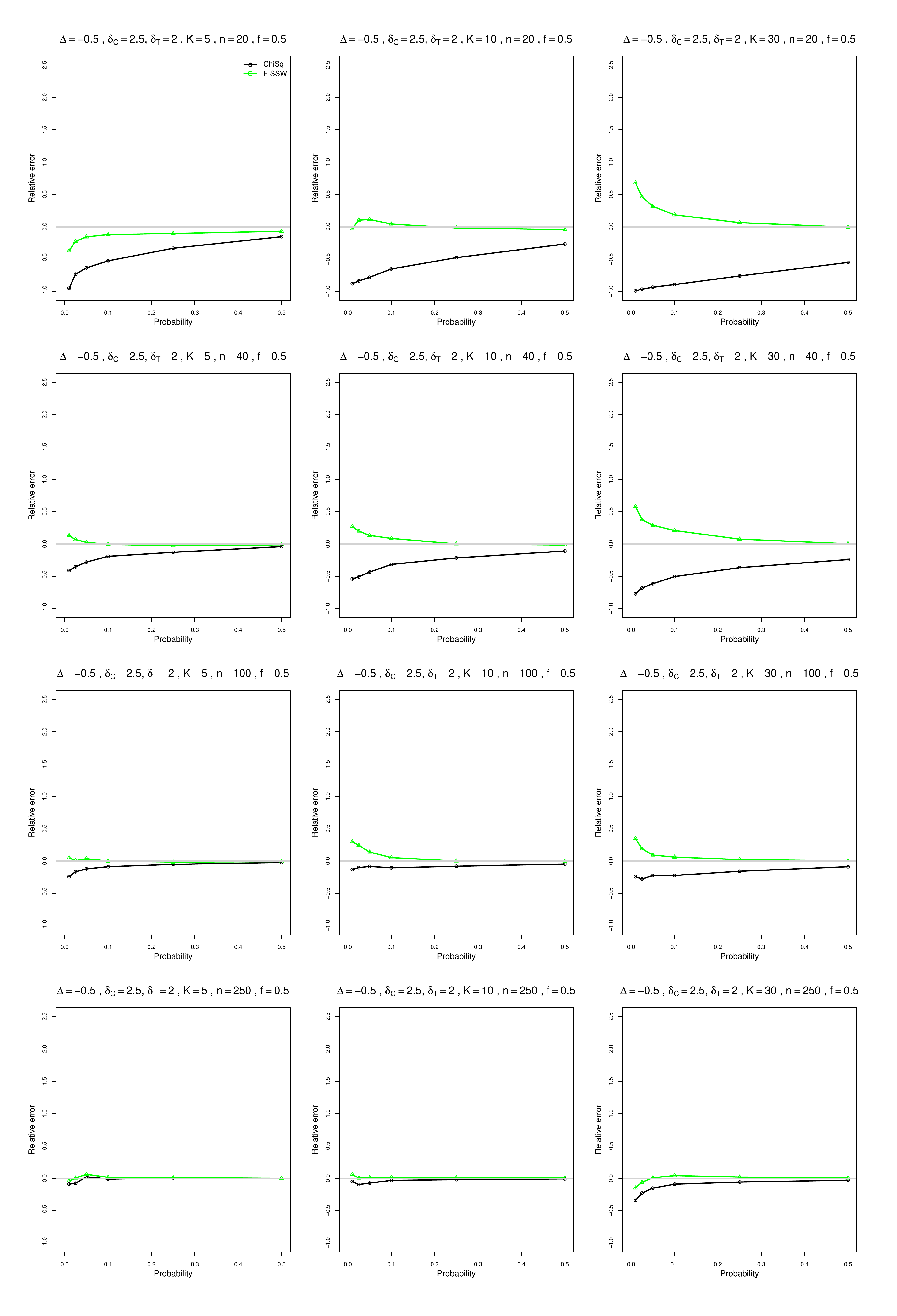}
	\caption{Relative error between the achieved level and the nominal level for two approximations to the null distribution of Q for DSM (Chisq and F SSW) vs upper tail probability, for equal sample sizes $n=20,\;40,\;100$ and $250$, $\delta_{iC} = 2.5$, $\Delta=-0.5$ and  $f = 0.5$.   }
	\label{Pplot_relative_truncated_deltaC_2.5deltaT=2_DSM_equal_sample_sizes.pdf}
\end{figure}

\begin{figure}[ht]
	\centering
	\includegraphics[scale=0.33]{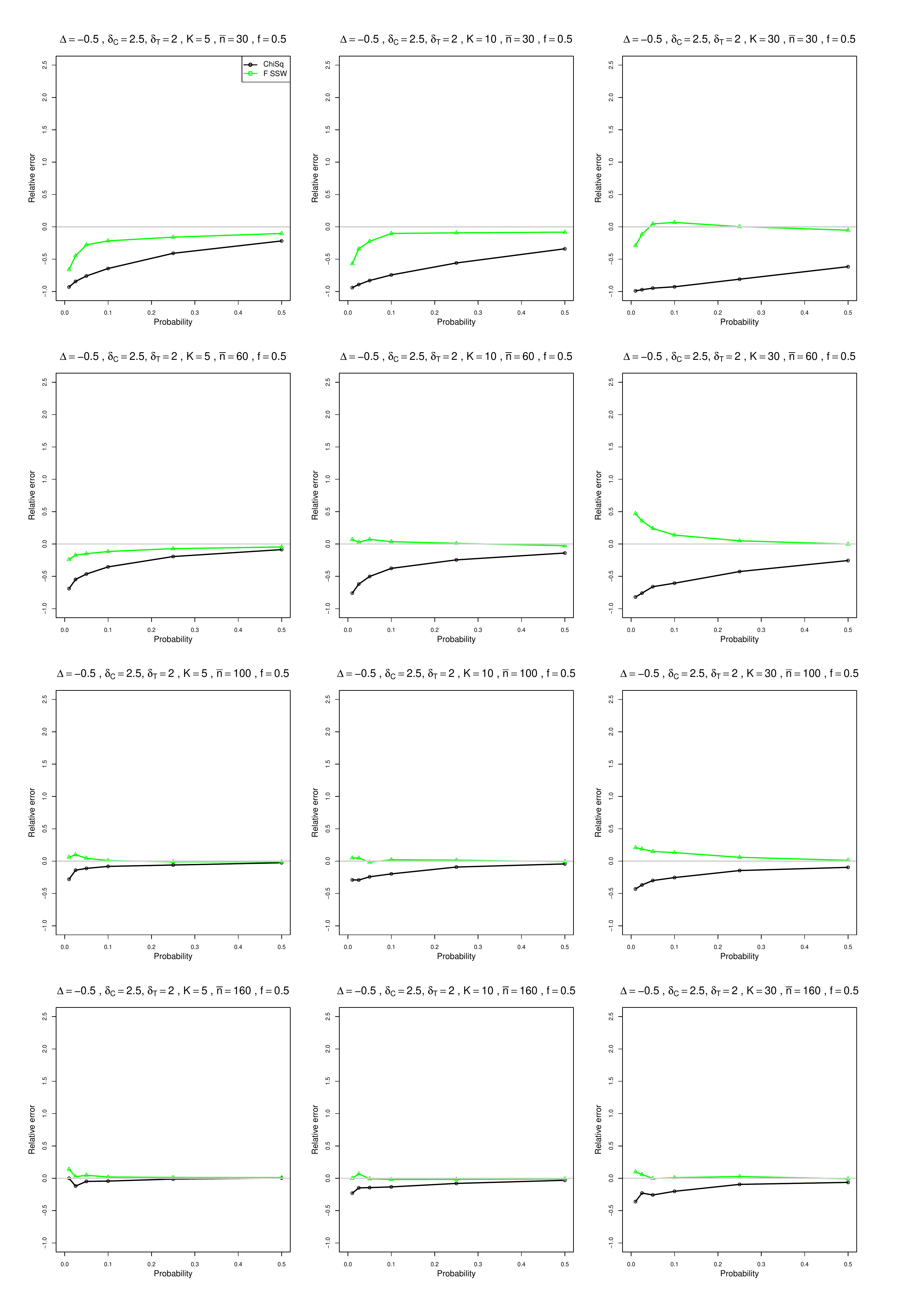}
	\caption{Relative error between the achieved level and the nominal level for two approximations to the null distribution of Q for DSM (Chisq and F SSW) vs upper tail probability, for unequal sample sizes $\bar{n}=30,\;60,\;100$ and $160$, $\delta_{iC} = 2.5$, $\Delta=-0.5$ and  $f = 0.5$.   }
	\label{Pplot_relative_truncated_deltaC_2.5deltaT=2_DSM_unequal_sample_sizes.pdf}
\end{figure}

\begin{figure}[ht]
	\centering
	\includegraphics[scale=0.33]{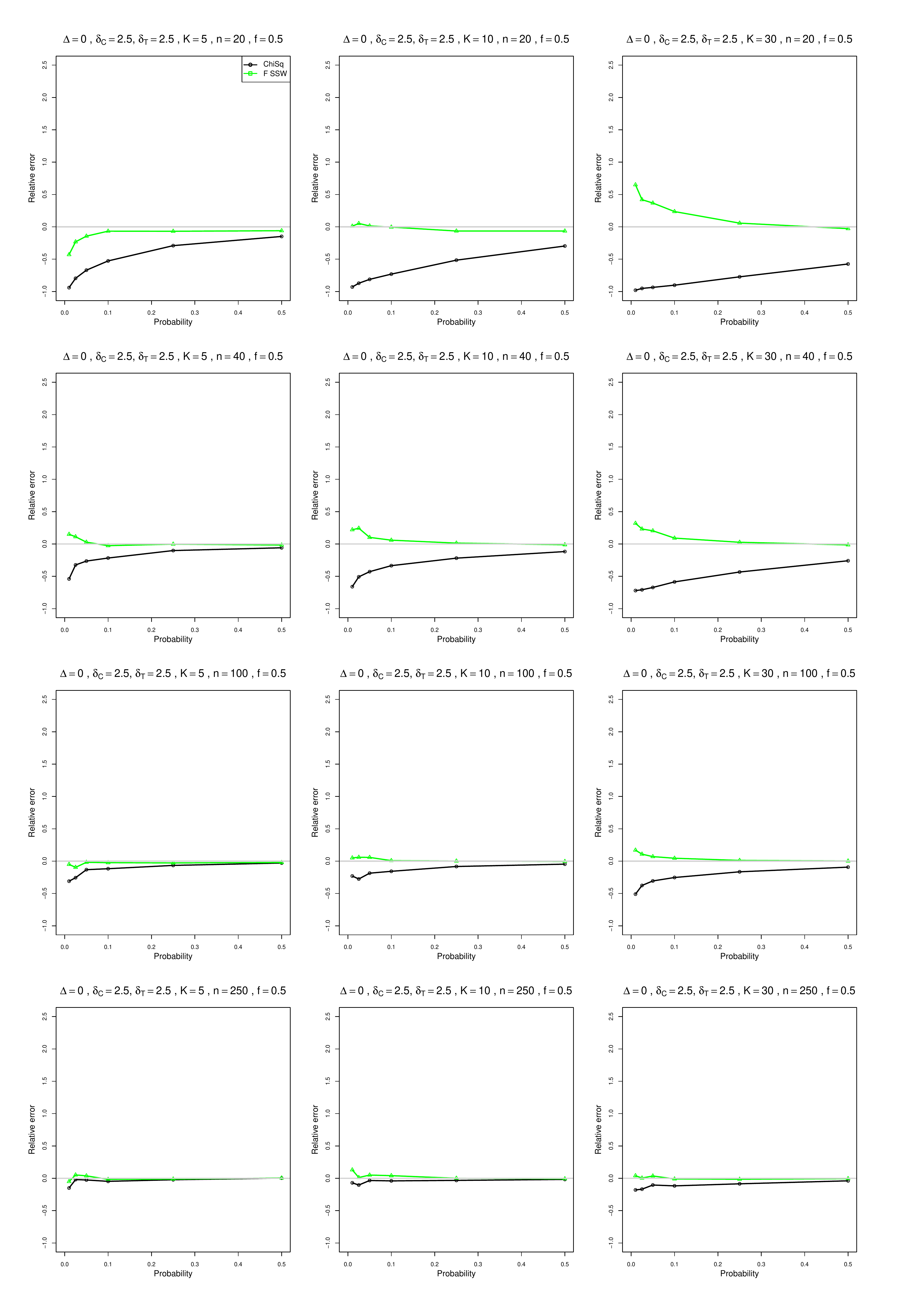}
	\caption{Relative error between the achieved level and the nominal level for two approximations to the null distribution of Q for DSM (Chisq and F SSW) vs upper tail probability, for equal sample sizes $n=20,\;40,\;100$ and $250$, $\delta_{iC} = 2.5$, $\Delta=0$ and  $f = 0.5$.   }
	\label{Pplot_relative_truncated_deltaC_2.5deltaT=2.5_DSM_equal_sample_sizes.pdf}
\end{figure}

\begin{figure}[ht]
	\centering
	\includegraphics[scale=0.33]{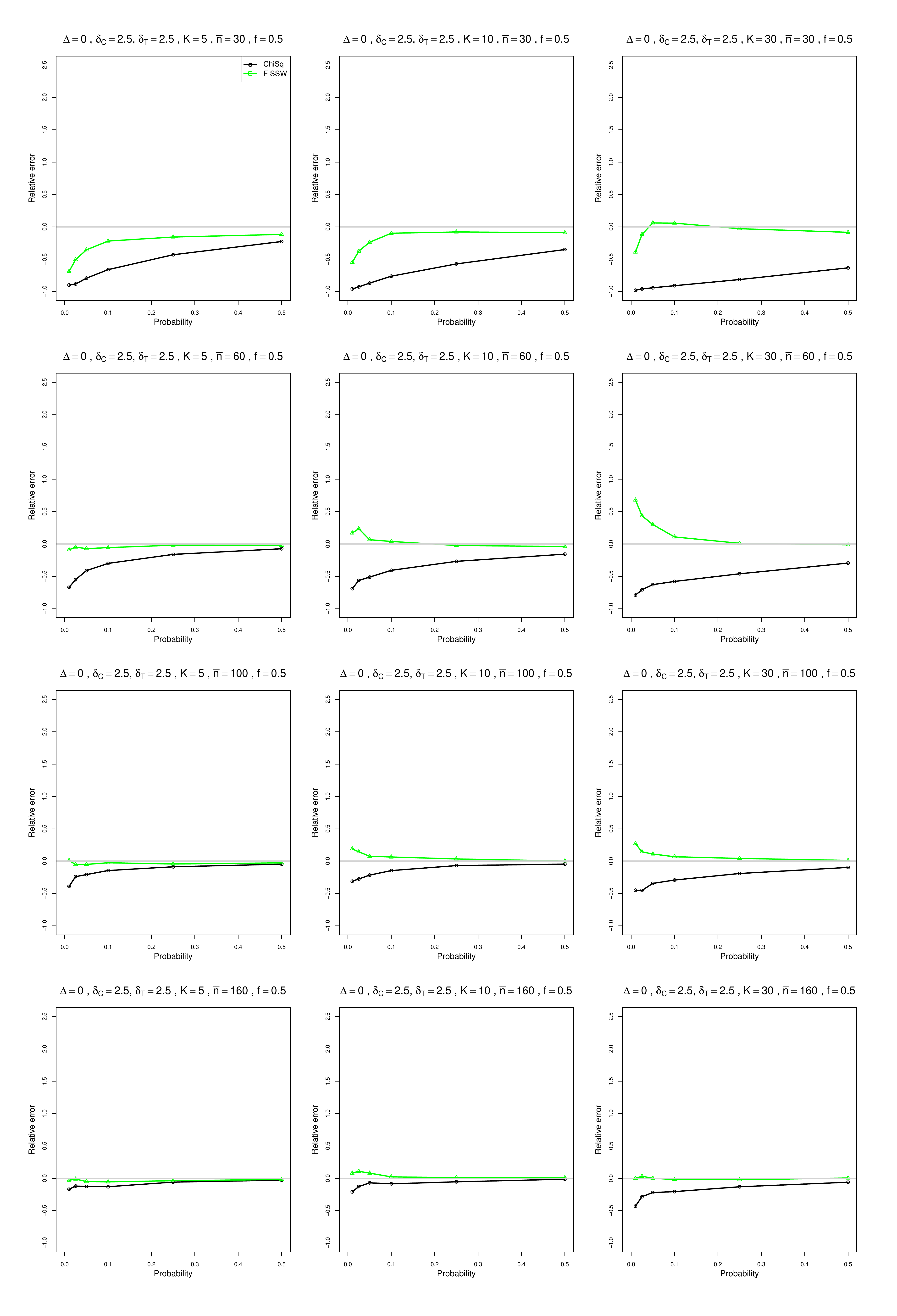}
	\caption{Relative error between the achieved level and the nominal level for two approximations to the null distribution of Q for DSM (Chisq and F SSW) vs upper tail probability, for unequal sample sizes $\bar{n}=30,\;60,\;100$ and $160$, $\delta_{iC} = 2.5$, $\Delta=0$ and  $f = 0.5$.   }
	\label{Pplot_relative_truncated_deltaC_2.5deltaT=2.5_DSM_unequal_sample_sizes.pdf}
\end{figure}

\begin{figure}[ht]
	\centering
	\includegraphics[scale=0.33]{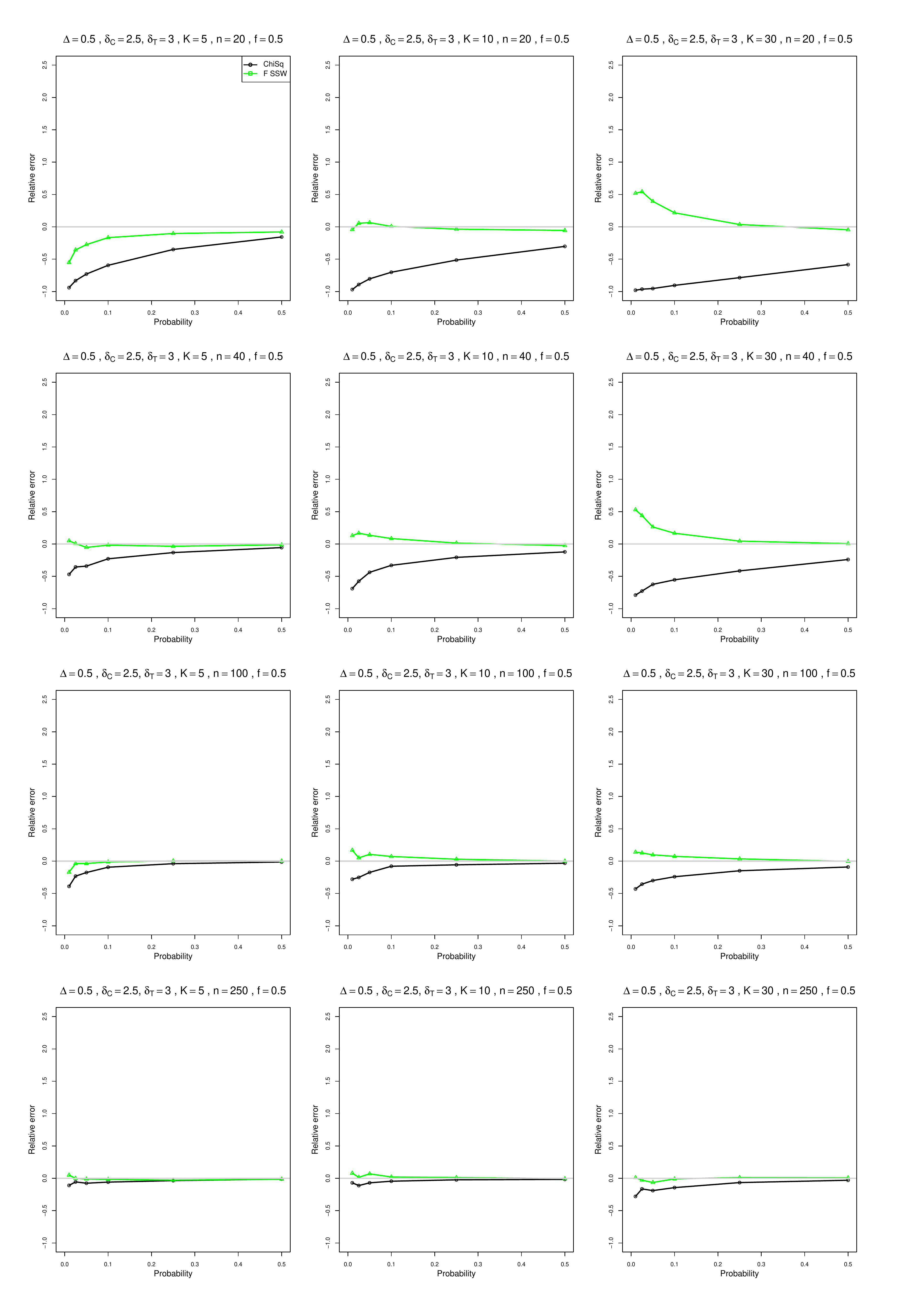}
	\caption{Relative error between the achieved level and the nominal level for two approximations to the null distribution of Q for DSM (Chisq and F SSW) vs upper tail probability, for equal sample sizes $n=20,\;40,\;100$ and $250$, $\delta_{iC} = 2.5$, $\Delta=0.5$ and  $f = 0.5$.   }
	\label{Pplot_relative_truncated_deltaC_2.5deltaT=3_DSM_equal_sample_sizes.pdf}
\end{figure}

\begin{figure}[ht]
	\centering
	\includegraphics[scale=0.33]{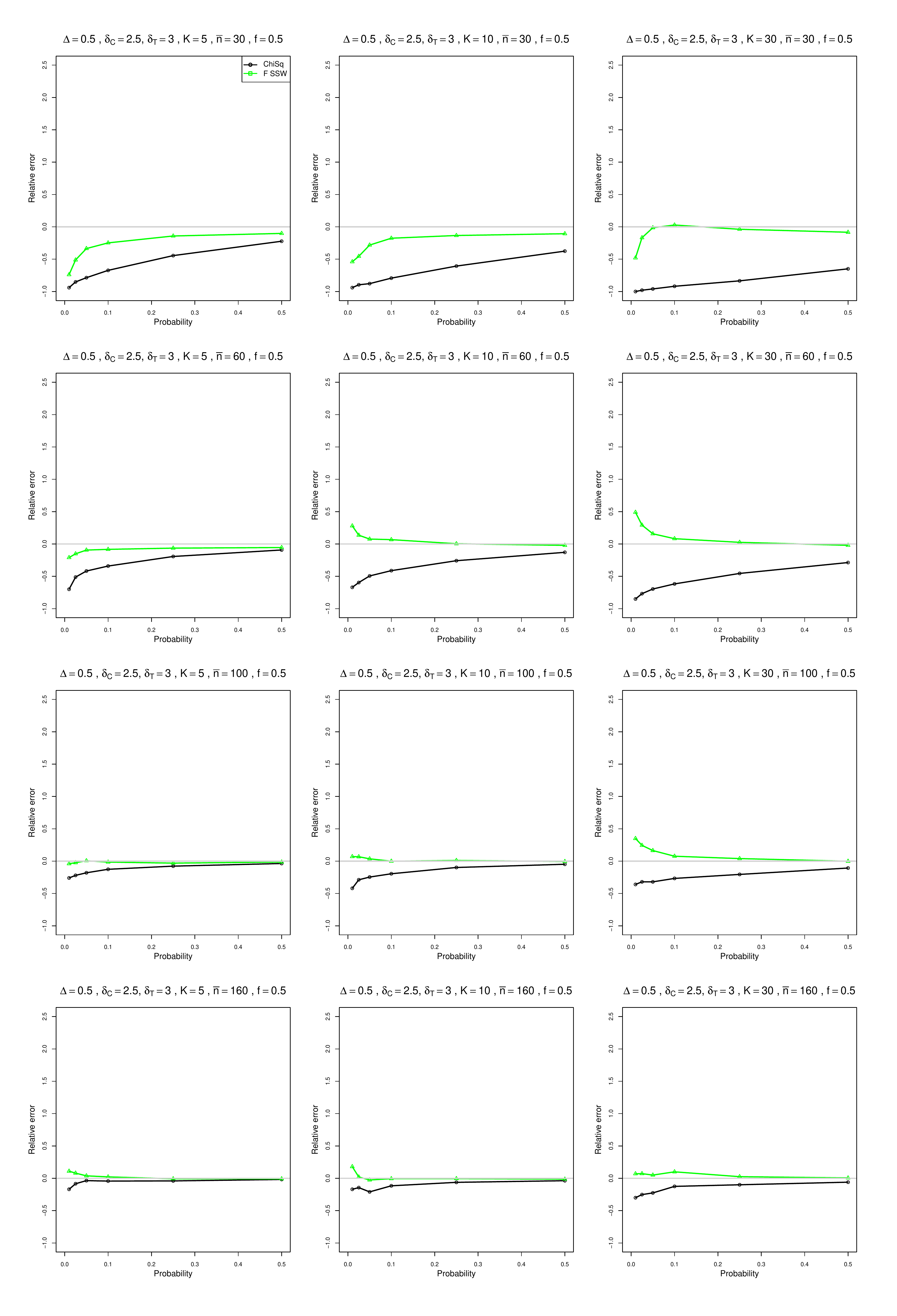}
	\caption{Relative error between the achieved level and the nominal level for two approximations to the null distribution of Q for DSM (Chisq and F SSW) vs upper tail probability, for unequal sample sizes $\bar{n}=30,\;60,\;100$ and $160$, $\delta_{iC} = 2.5$, $\Delta=0.5$ and  $f = 0.5$.   }
	\label{Pplot_relative_truncated_deltaC_2.5deltaT=3_DSM_unequal_sample_sizes.pdf}
\end{figure}

\begin{figure}[ht]
	\centering
	\includegraphics[scale=0.33]{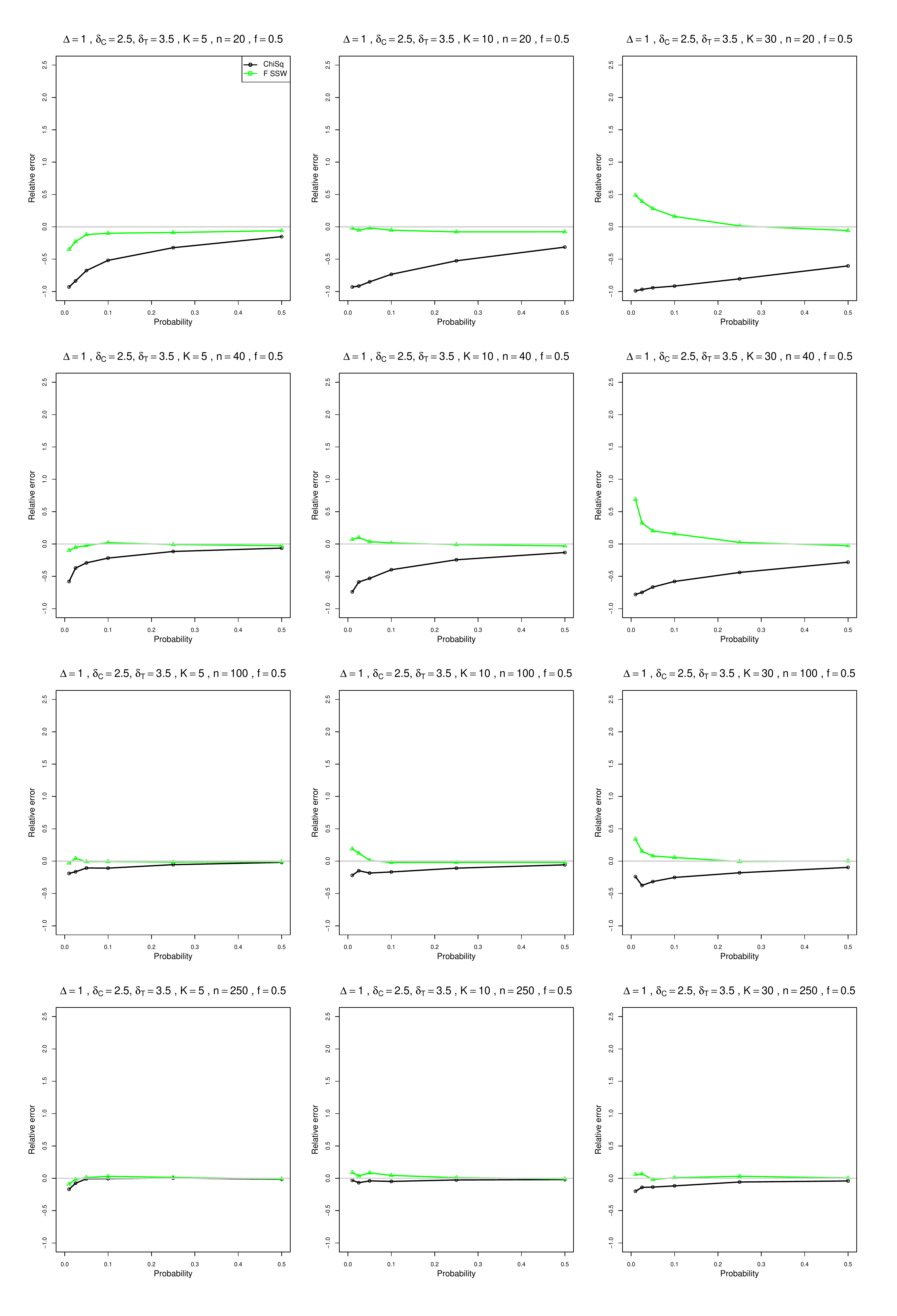}
	\caption{Relative error between the achieved level and the nominal level for two approximations to the null distribution of Q for DSM (Chisq and F SSW) vs upper tail probability, for equal sample sizes $n=20,\;40,\;100$ and $250$, $\delta_{iC} = 2.5$, $\Delta=1$ and  $f = 0.5$.   }
	\label{Pplot_relative_truncated_deltaC_2.5deltaT=3.5_DSM_equal_sample_sizes.pdf}
\end{figure}

\begin{figure}[ht]
	\centering
	\includegraphics[scale=0.33]{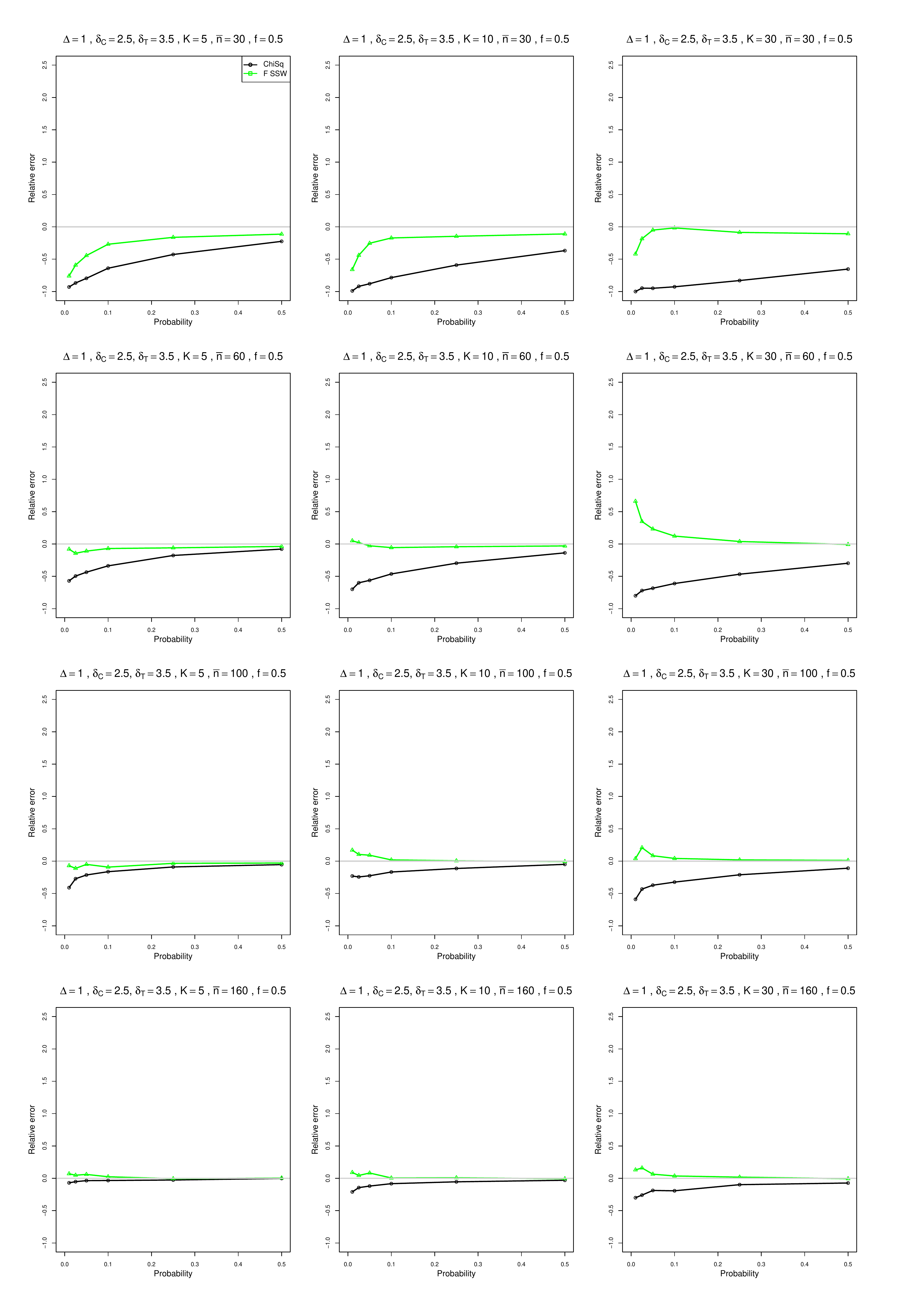}
	\caption{Relative error between the achieved level and the nominal level for two approximations to the null distribution of Q for DSM (Chisq and F SSW) vs upper tail probability, for unequal sample sizes $\bar{n}=30,\;60,\;100$ and $160$, $\delta_{iC} = 2.5$, $\Delta=1$ and  $f = 0.5$.   }
	\label{Pplot_relative_truncated_deltaC_2.5deltaT=3.5_DSM_unequal_sample_sizes.pdf}
\end{figure}

\begin{figure}[ht]
	\centering
	\includegraphics[scale=0.33]{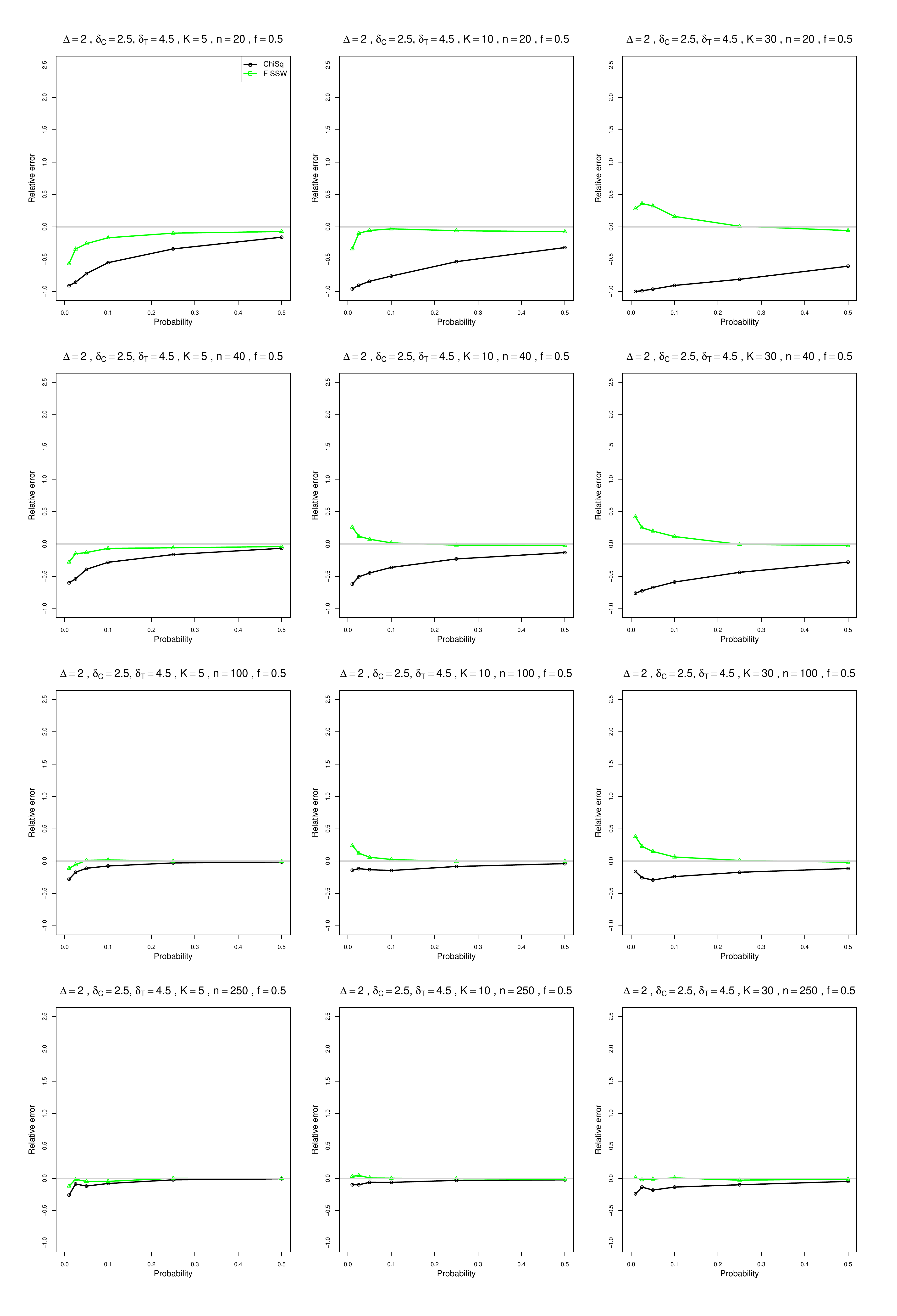}
	\caption{Relative error between the achieved level and the nominal level for two approximations to the null distribution of Q for DSM (Chisq and F SSW) vs upper tail probability, for equal sample sizes $n=20,\;40,\;100$ and $250$, $\delta_{iC} = 2.5$, $\Delta=2$ and  $f = 0.5$.   }
	\label{Pplot_relative_truncated_deltaC_2.5deltaT=2.5_DSM_equal_sample_sizes.pdf}
\end{figure}

\begin{figure}[ht]
	\centering
	\includegraphics[scale=0.33]{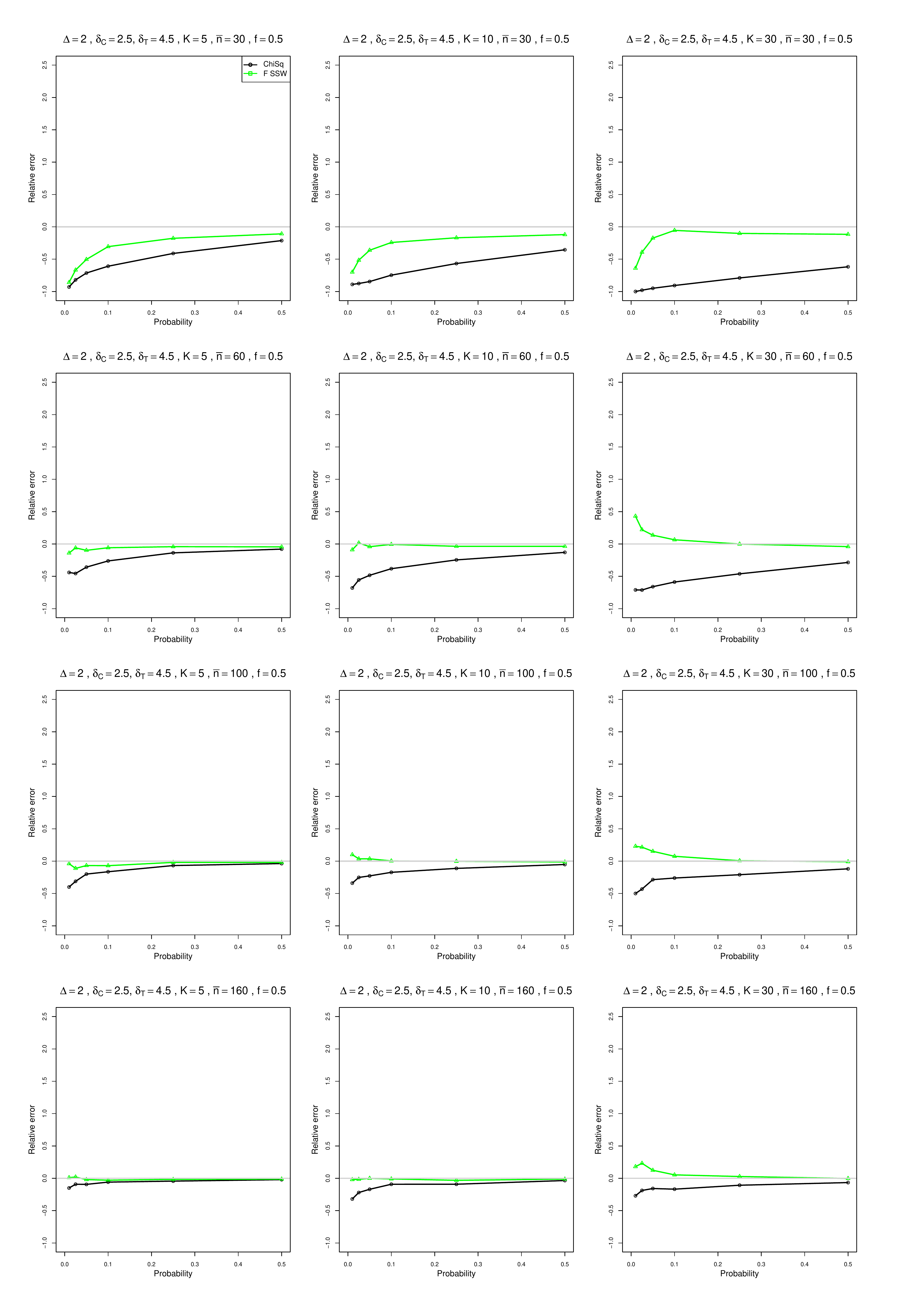}
	\caption{Relative error between the achieved level and the nominal level for two approximations to the null distribution of Q for DSM (Chisq and F SSW) vs upper tail probability, for unequal sample sizes $\bar{n}=30,\;60,\;100$ and $160$, $\delta_{iC} = 2.5$, $\Delta=2$ and  $f = 0.5$.   }
	\label{Pplot_relative_truncated_deltaC_2.5deltaT=4.5_DSM_unequal_sample_sizes.pdf}
\end{figure}


\clearpage

\setcounter{figure}{0}
\setcounter{section}{0}

\section*{Appendix B: Empirical level at $\alpha = .05$, vs $\Delta$, of the test for heterogeneity of DSM ($\tau^2 = 0$ versus $\tau^2 > 0$) based on approximations for the null distribution of $Q$}

Each figure corresponds to a value of the standardized mean  in the Control arm $\delta_{C}$  (= $-2.5$, $-1$, 0, 1, 2.5). \\
The fraction of each study's sample size in the Control arm ($f$) is held constant at 0.5.

For each combination of a value of $n$ (= 20, 40, 100, 250) or  $\bar{n}$ (= 30, 60, 100, 160) and a value of $K$ (= 5, 10, 30), a panel plots the actual level  versus the values $\Delta$ (= $-2$, $-1$, $-0.5$, 0, 0.5, 1, 2) for two approximations to the null distribution of $Q$:
\begin{itemize}
\item ChiSq (Chi-square approximation with $K - 1$ df, inverse-variance weights)
\item F SSW  (Farebrother approximation, effective-sample-size weights)
\end{itemize}

\clearpage
\renewcommand{\thefigure}{B.\arabic{figure}}

\begin{figure}[ht]
	\centering
	\includegraphics[scale=0.33]{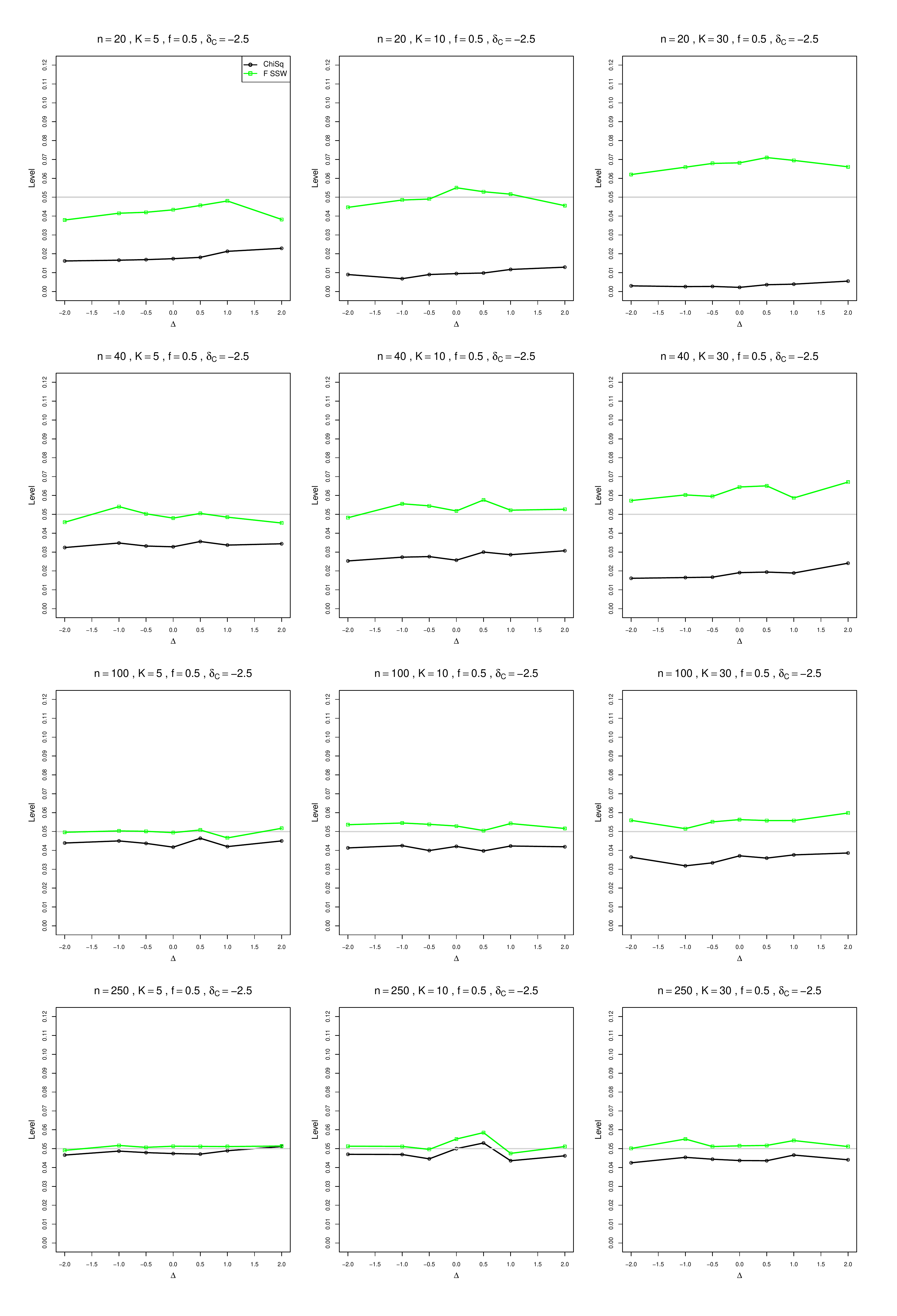}
	\caption{Q for DSM: actual level at $\alpha = .05$ for $\delta_{C}=-2.5$ and $f = .5$, equal sample sizes
		\label{PlotPvalueAtNominal005_deltaC_-2.5_DSM_equal_sample_sizes.pdf}}
\end{figure}

\begin{figure}[ht]
	\centering
	\includegraphics[scale=0.33]{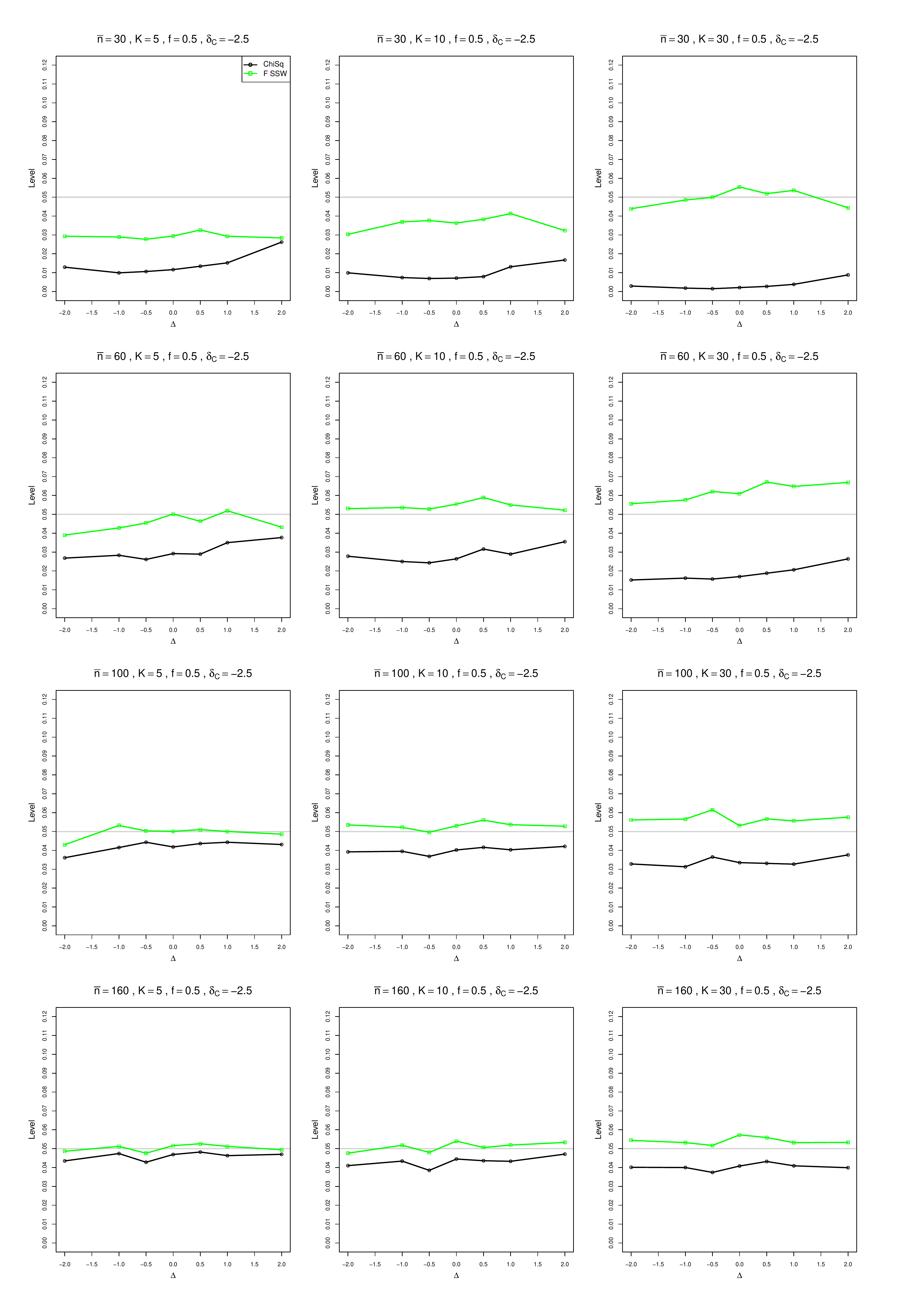}
	\caption{Q for DSM: actual level at $\alpha = .05$ for $\delta_{C}=-2.5$ and $f = .5$, unequal sample sizes
		\label{PlotPvalueAtNominal005_deltaC_-2.5_DSM_unequal_sample_sizes.pdf}}
\end{figure}

\begin{figure}[ht]
	\centering
	\includegraphics[scale=0.33]{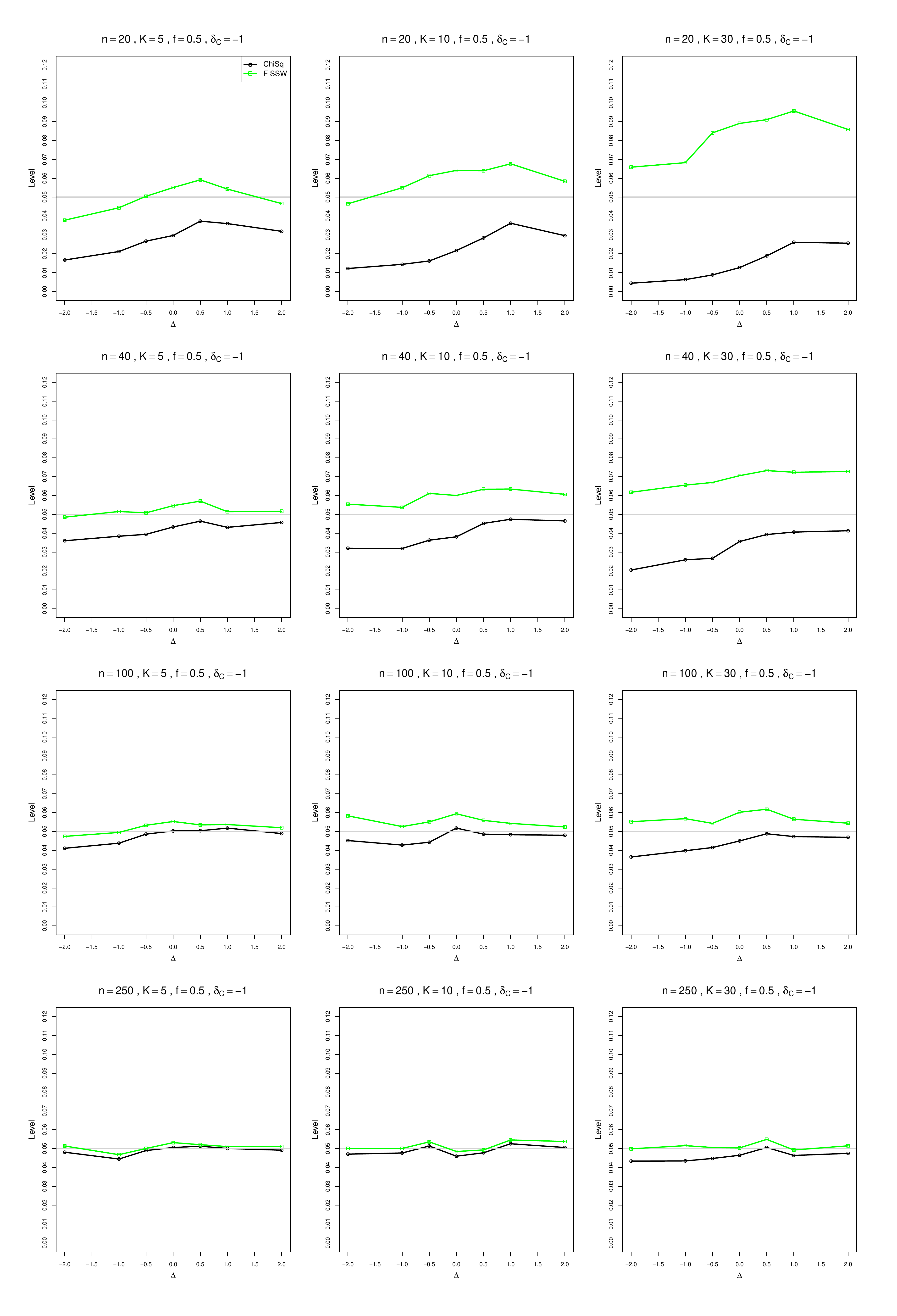}
	\caption{Q for DSM: actual level at $\alpha = .05$ for $\delta_{C}=-1$ and $f = .5$, equal sample sizes
		\label{PlotPvalueAtNominal005_deltaC_-1_DSM_equal_sample_sizes.pdf}}
\end{figure}

\begin{figure}[ht]
	\centering
	\includegraphics[scale=0.33]{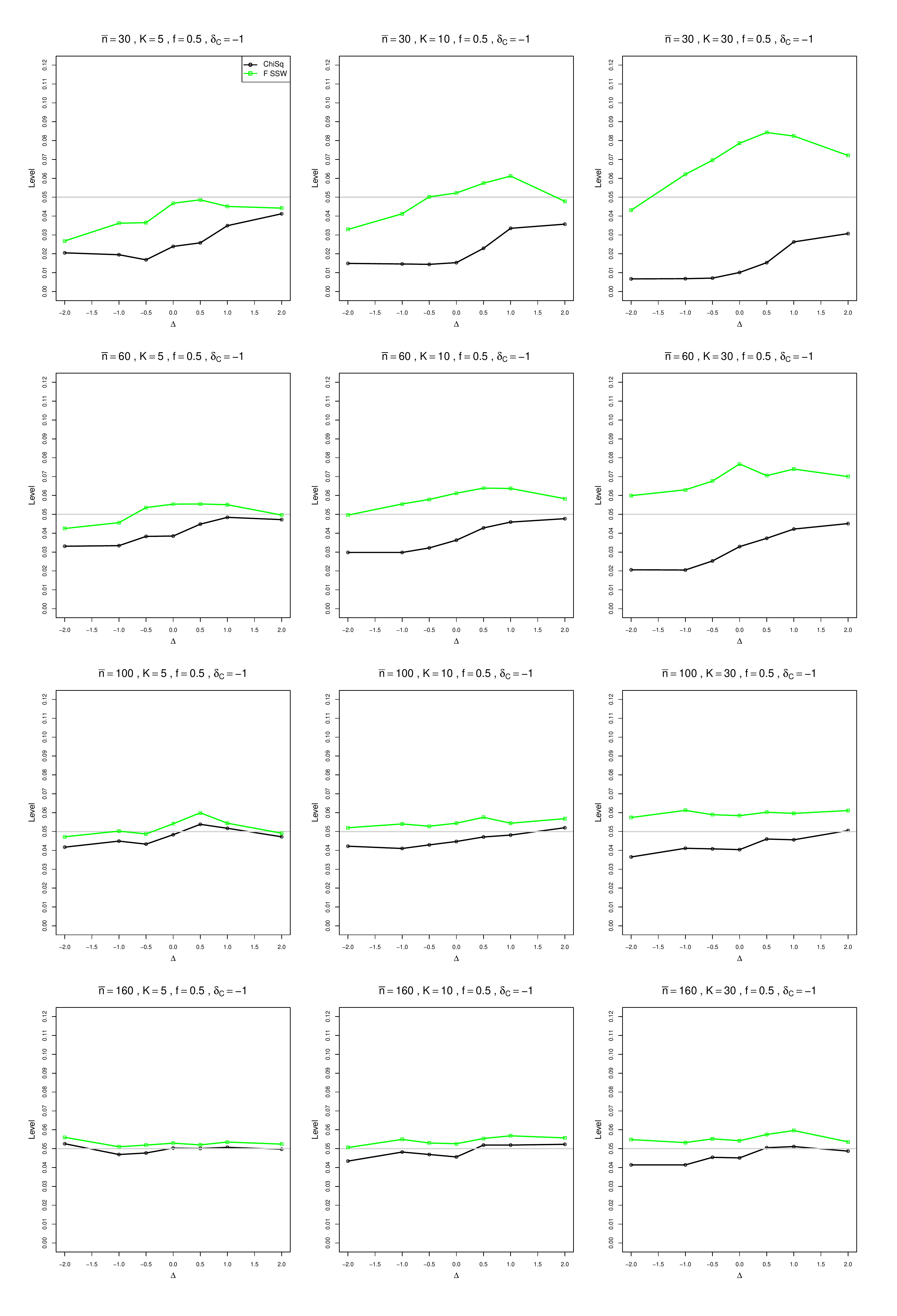}
	\caption{Q for DSM: actual level at $\alpha = .05$ for $\delta_{C}=-1$ and $f = .5$, unequal sample sizes
		\label{PlotPvalueAtNominal005_deltaC_-1_DSM_unequal_sample_sizes.pdf}}
\end{figure}

\begin{figure}[ht]
	\centering
	\includegraphics[scale=0.33]{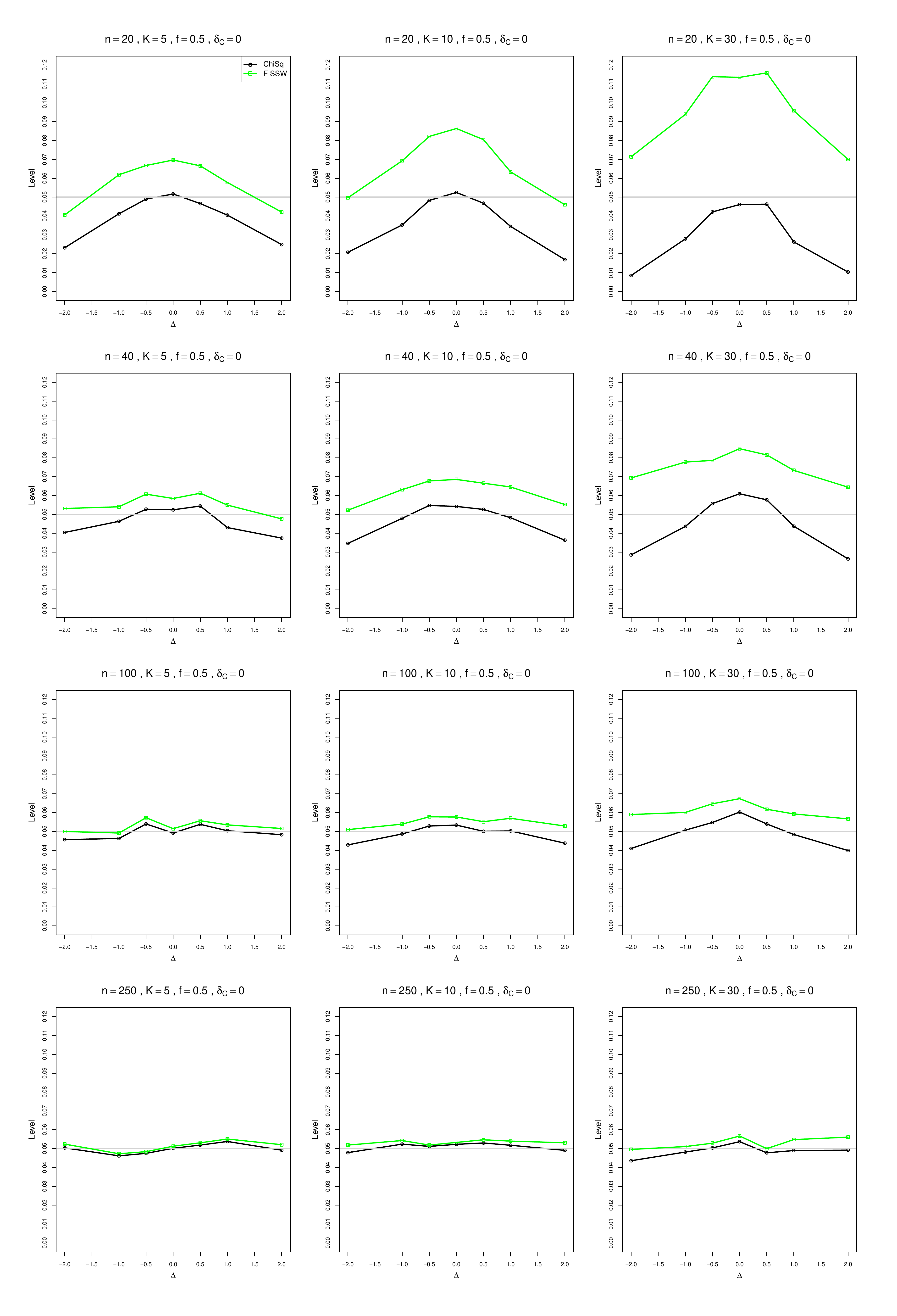}
	\caption{Q for DSM: actual level at $\alpha = .05$ for $\delta_{C}=0$ and $f = .5$, equal sample sizes
		\label{PlotPvalueAtNominal005_deltaC_0_DSM_equal_sample_sizes.pdf}}
\end{figure}

\begin{figure}[ht]
	\centering
	\includegraphics[scale=0.33]{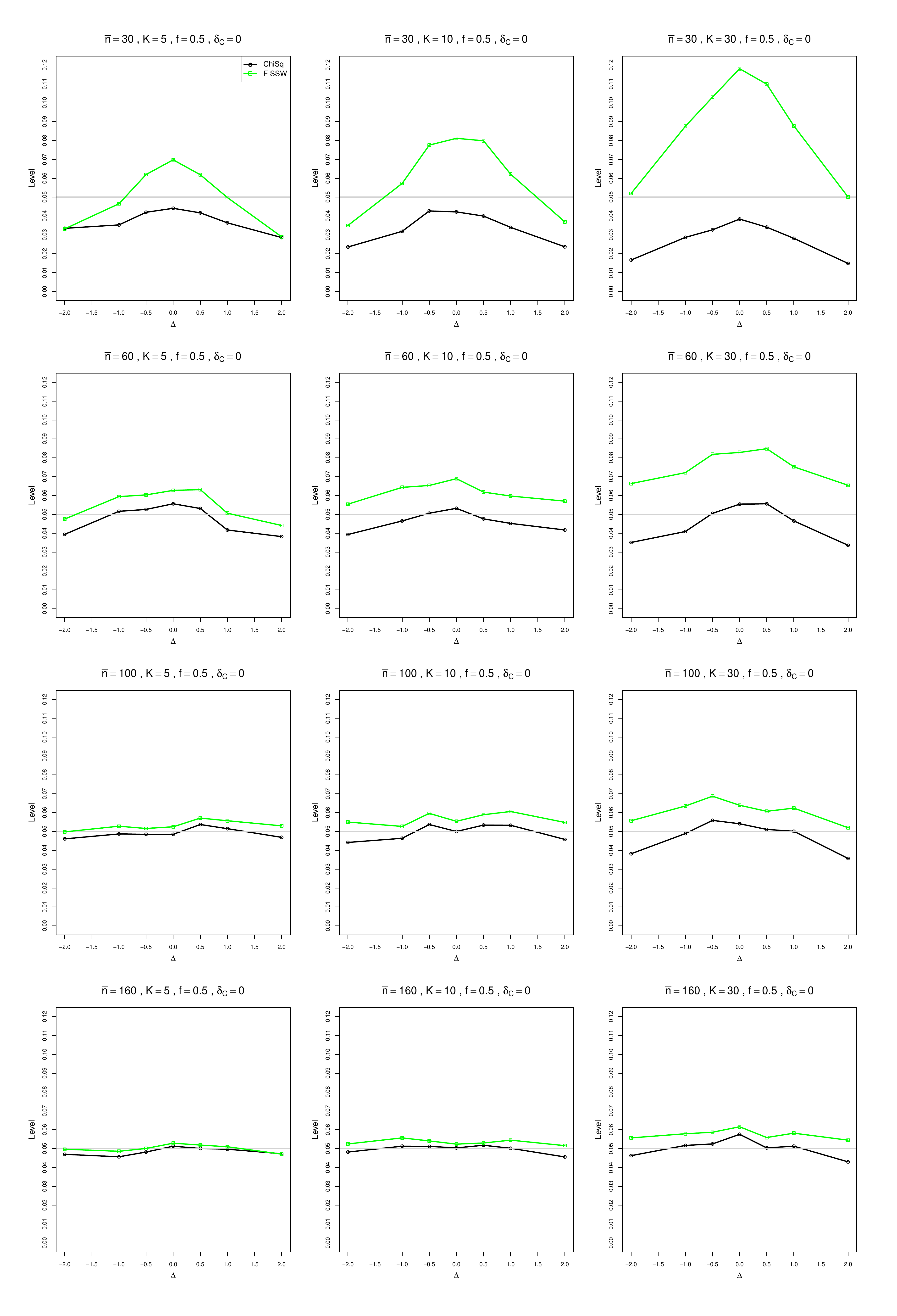}
	\caption{Q for DSM: actual level at $\alpha = .05$ for $\delta_{C}=0$ and $f = .5$, unequal sample sizes
		\label{PlotPvalueAtNominal005_deltaC_0_DSM_unequal_sample_sizes.pdf}}
\end{figure}

\begin{figure}[ht]
	\centering
	\includegraphics[scale=0.33]{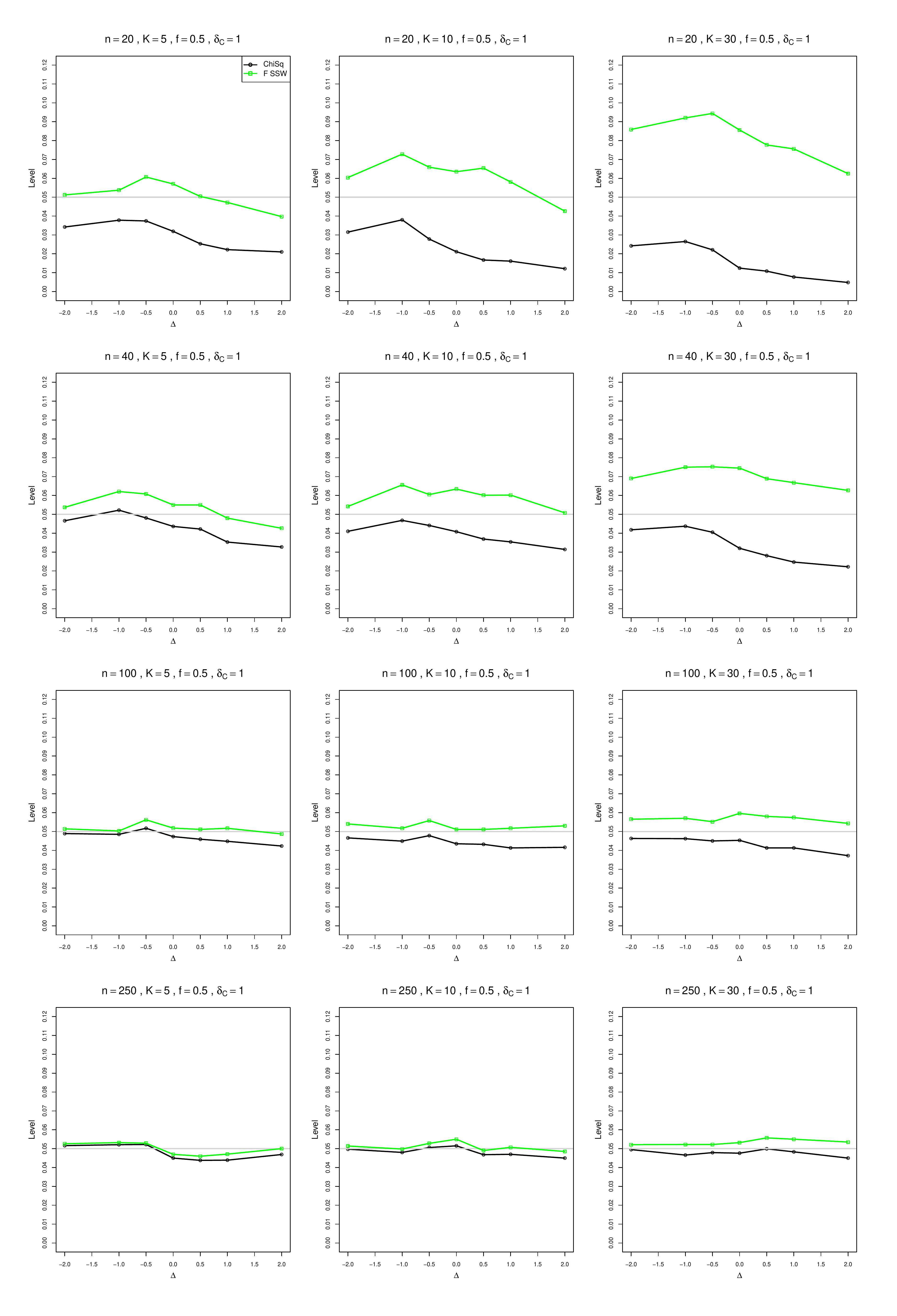}
	\caption{Q for DSM: actual level at $\alpha = .05$ for $\delta_{C}=1$ and $f = .5$, equal sample sizes
		\label{PlotPvalueAtNominal005_deltaC_1_DSM_equal_sample_sizes.pdf}}
\end{figure}

\begin{figure}[ht]
	\centering
	\includegraphics[scale=0.33]{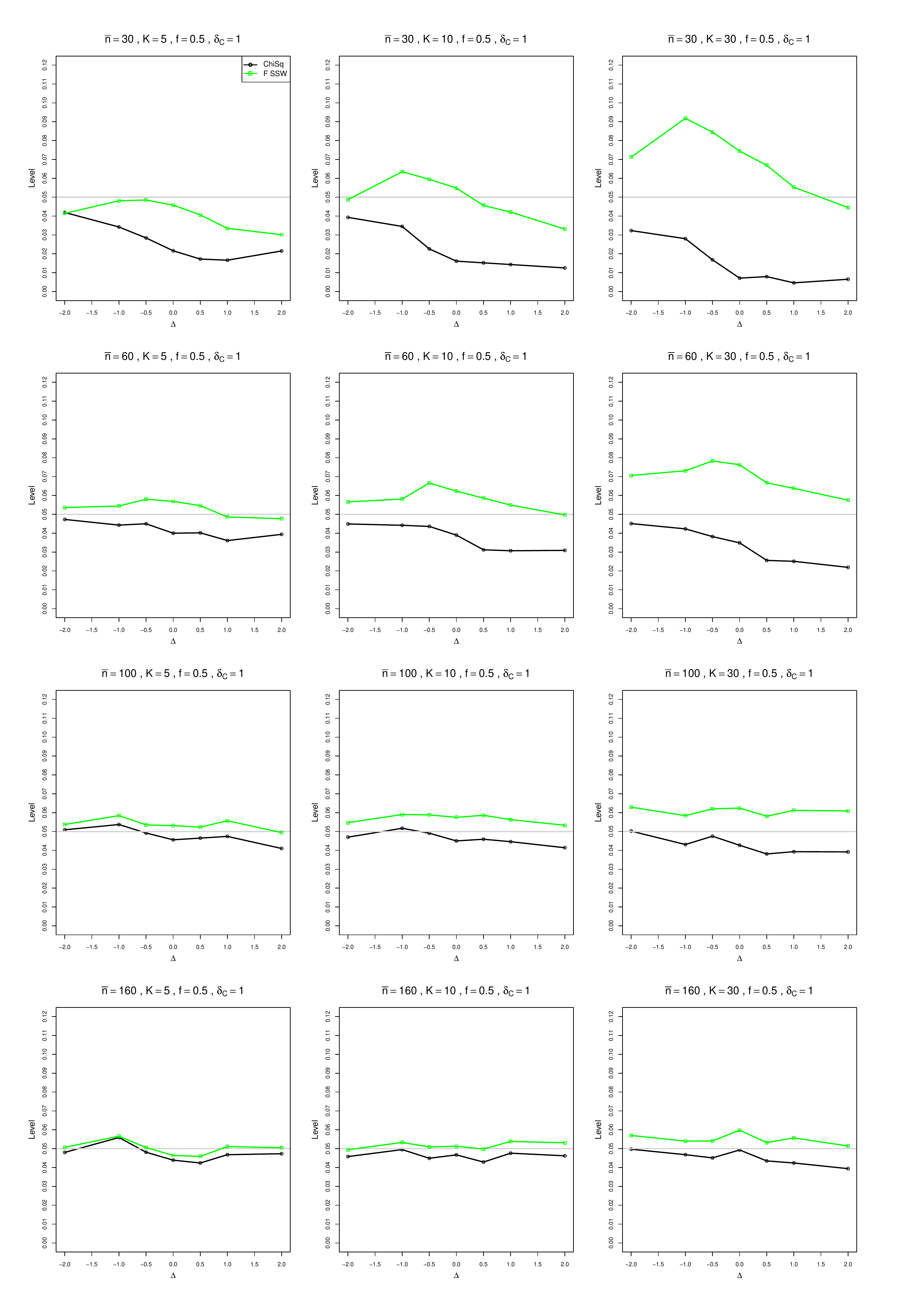}
	\caption{Q for DSM: actual level at $\alpha = .05$ for $\delta_{C}=1$ and $f = .5$, unequal sample sizes
		\label{PlotPvalueAtNominal005_deltaC_1_DSM_unequal_sample_sizes.pdf}}
\end{figure}

\begin{figure}[ht]
	\centering
	\includegraphics[scale=0.33]{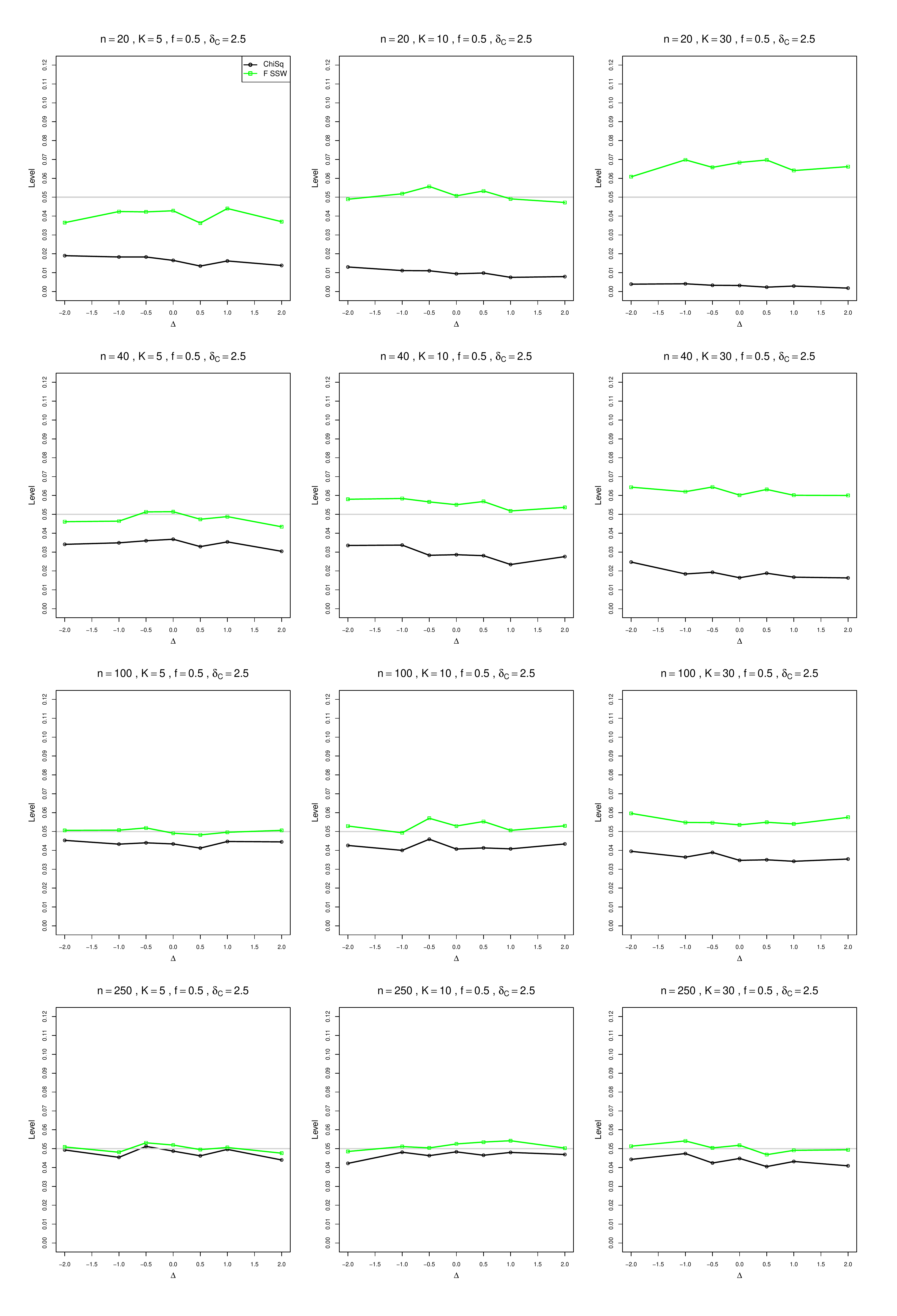}
	\caption{Q for DSM: actual level at $\alpha = .05$ for $\delta_{C}=2.5$ and $f = .5$, equal sample sizes
		\label{PlotPvalueAtNominal005_deltaC_25_DSM_equal_sample_sizes.pdf}}
\end{figure}

\begin{figure}[ht]
	\centering
	\includegraphics[scale=0.33]{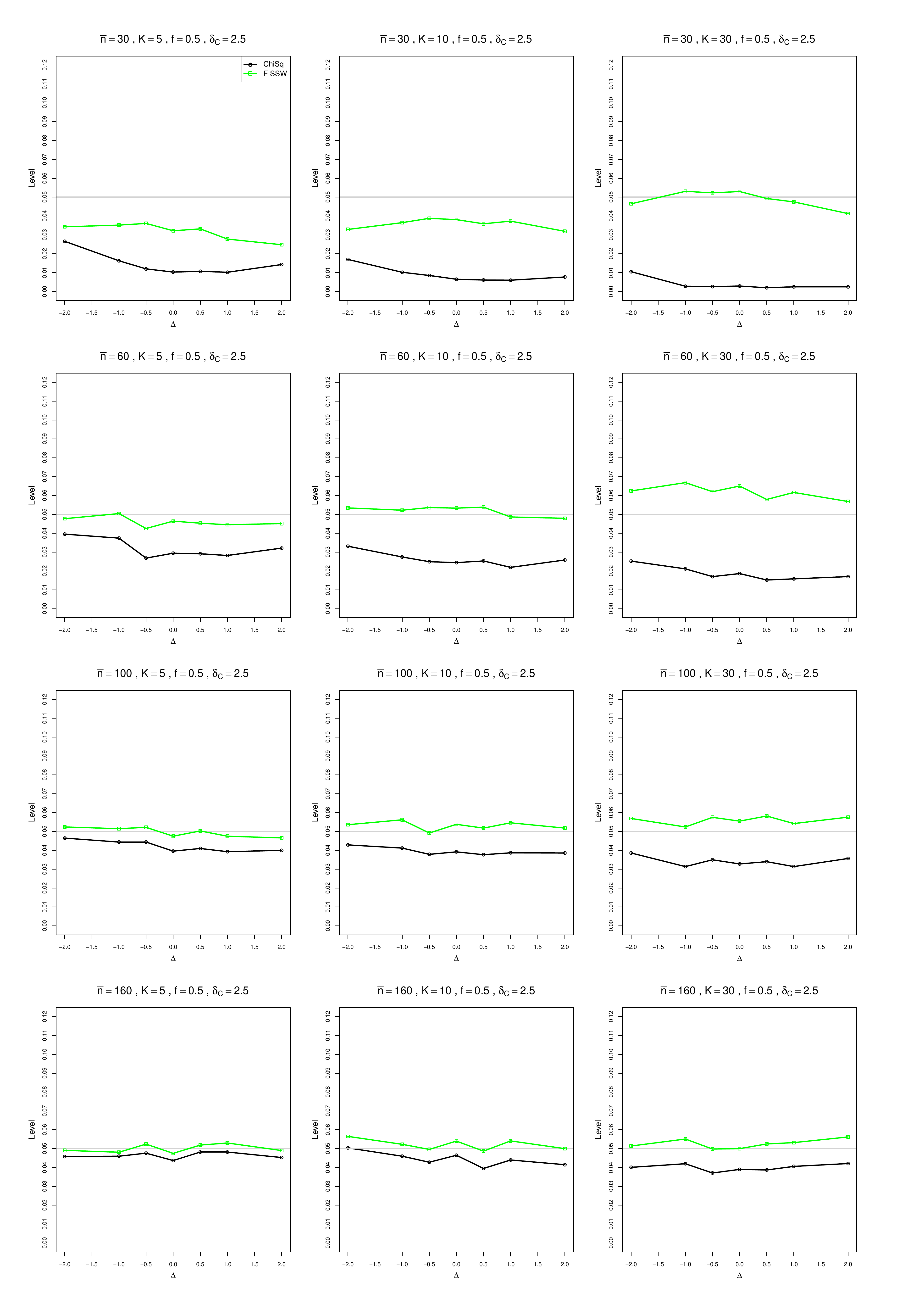}
	\caption{Q for DSM: actual level at $\alpha = .05$ for $\delta_{C}=2.5$ and $f = .5$, unequal sample sizes
		\label{PlotPvalueAtNominal005_deltaC_25_DSM_unequal_sample_sizes.pdf}}
\end{figure}

\clearpage

\setcounter{figure}{0}
\setcounter{section}{0}

\section*{Appendix C: Bias in point estimators of $\tau^2$}

Each figure corresponds to a value of the standardized mean  in the Control arm $\delta_{C}$  (= $-2.5$, $-1$, 0, 1, 2.5)  and a value of the overall DSM $\Delta$ (= $-2$, $-1$, $-0.5$, 0, 0.5, 1, 2) . \\
The fraction of each study's sample size in the Control arm ($f$) is held constant at 0.5.

For each combination of a value of $n$ (= 20, 40, 100, 250) or  $\bar{n}$ (= 30, 60, 100, 160) and a value of $K$ (= 5, 10, 30), a panel plots bias versus $\tau^2$ (= 0, 0.1, 0.5, 1, 1.5).\\
The point estimators of $\tau^2$ are
\begin{itemize}
\item DL (DerSimonian-Laird) method, inverse-variance weights
\item REML method, inverse-variance weights
\item MP (Mandel-Paule) method, inverse-variance weights
\item SSC method, effective-sample-size weights, conditional variance of DSM
\item SMC method, median-unbiased, effective-sample-size weights, conditional variance of DSM
\end{itemize}

\clearpage
\renewcommand{\thefigure}{C.\arabic{figure}}


\begin{figure}[ht]
	\centering
	\includegraphics[scale=0.33]{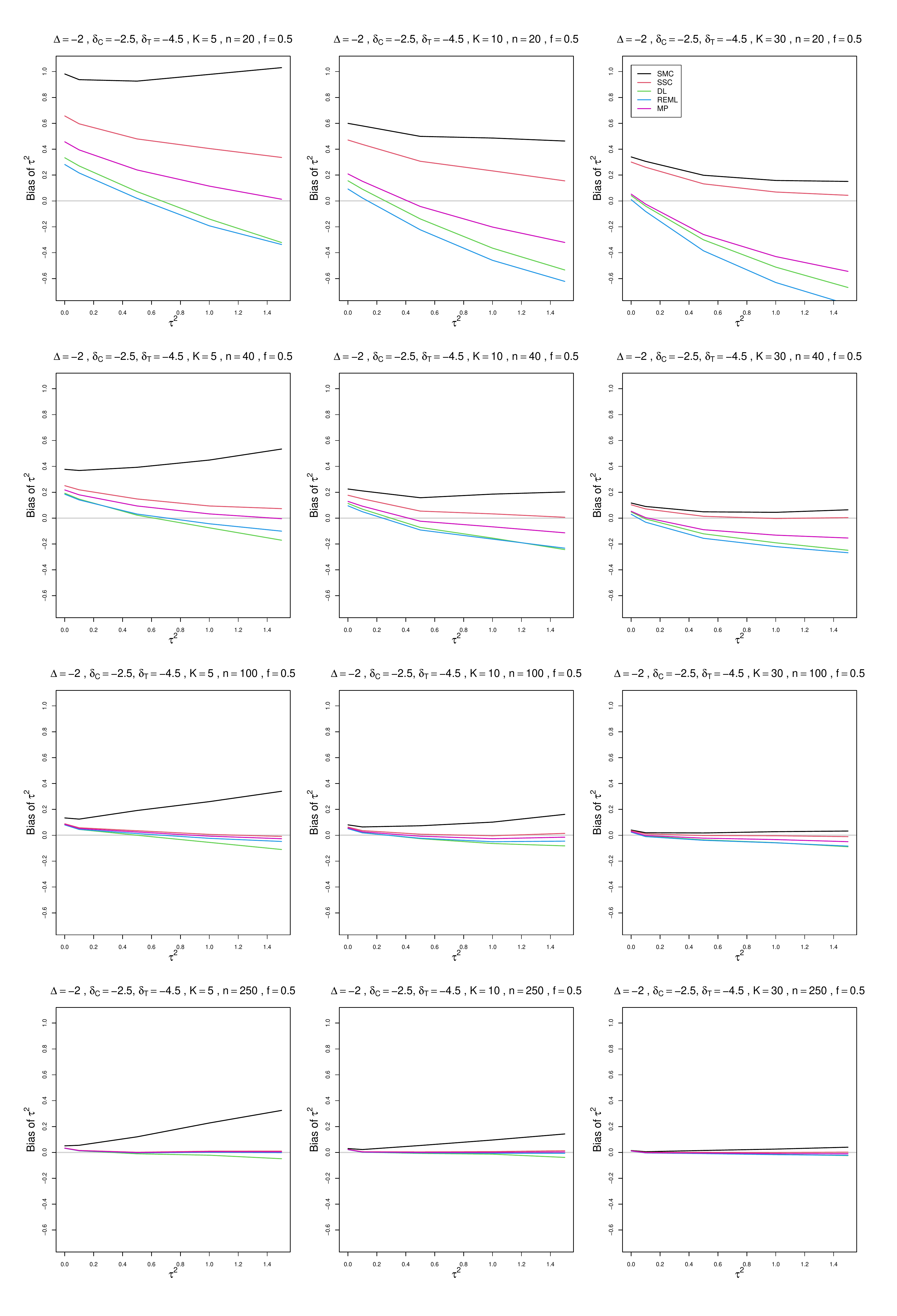}
	\caption{Bias  of estimators of between-study variance of DSM (DL, REML, MP, SMC and SSC ) vs $\tau^2$, for equal sample sizes $n=20,\;40,\;100$ and $250$, $\delta_{iC} = -2.5$, $\Delta=-2$ and  $f = 0.5$.   }
	\label{PlotBiasOfTau2_deltaC_-25deltaT=-4.5_DSM_equal_sample_sizes.pdf}
\end{figure}

\begin{figure}[ht]
	\centering
	\includegraphics[scale=0.33]{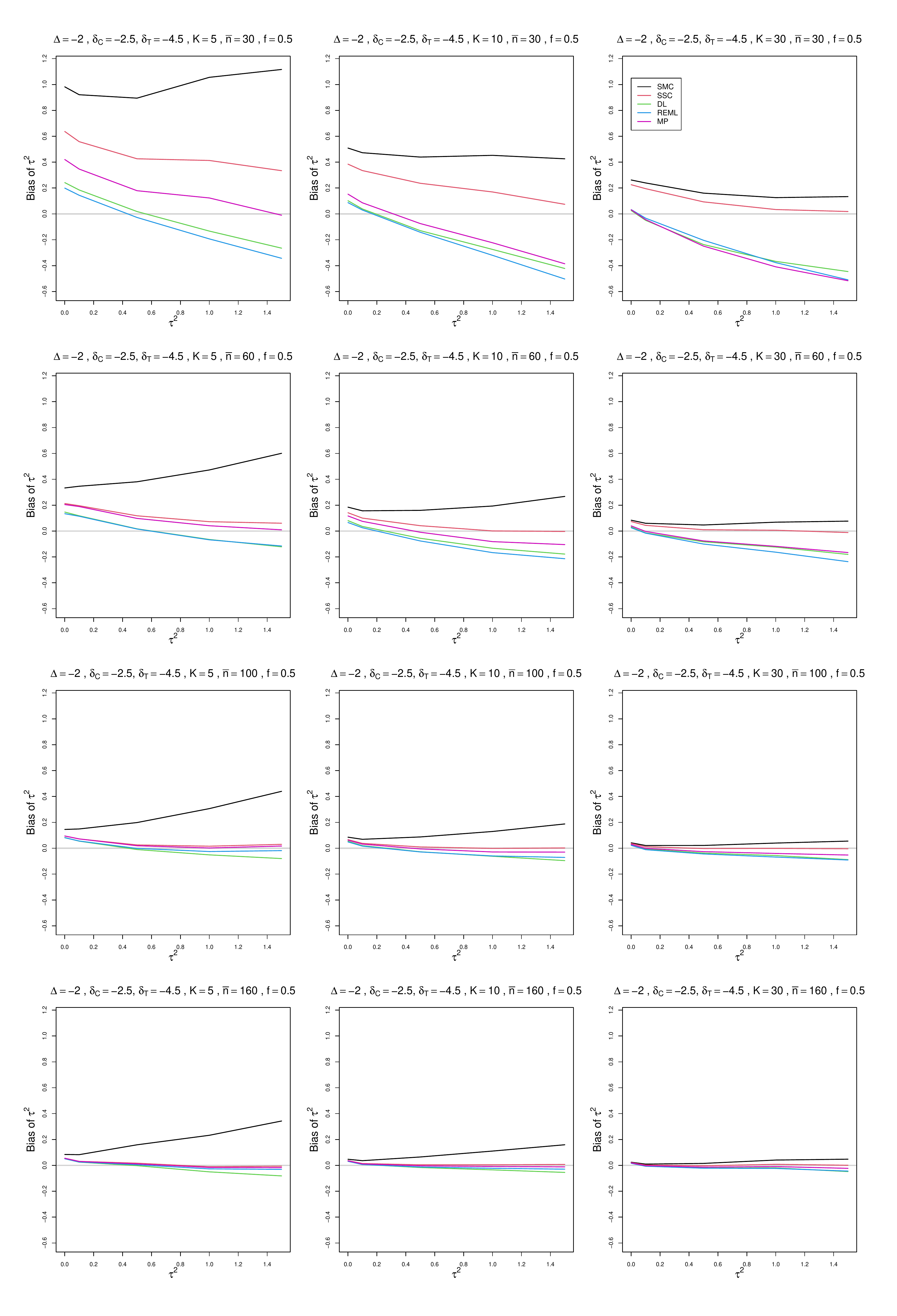}
	\caption{Bias  of estimators of between-study variance of DSM (DL, REML, MP, SMC and SSC ) vs $\tau^2$, for unequal sample sizes $\bar{n}=30,\;60,\;100$ and $160$, $\delta_{iC} = -2.5$, $\Delta=-2$ and  $f = 0.5$.   }
	\label{PlotBiasOfTau2_deltaC_-25deltaT=-4.5_DSM_unequal_sample_sizes.pdf}
\end{figure}

\begin{figure}[ht]
	\centering
	\includegraphics[scale=0.33]{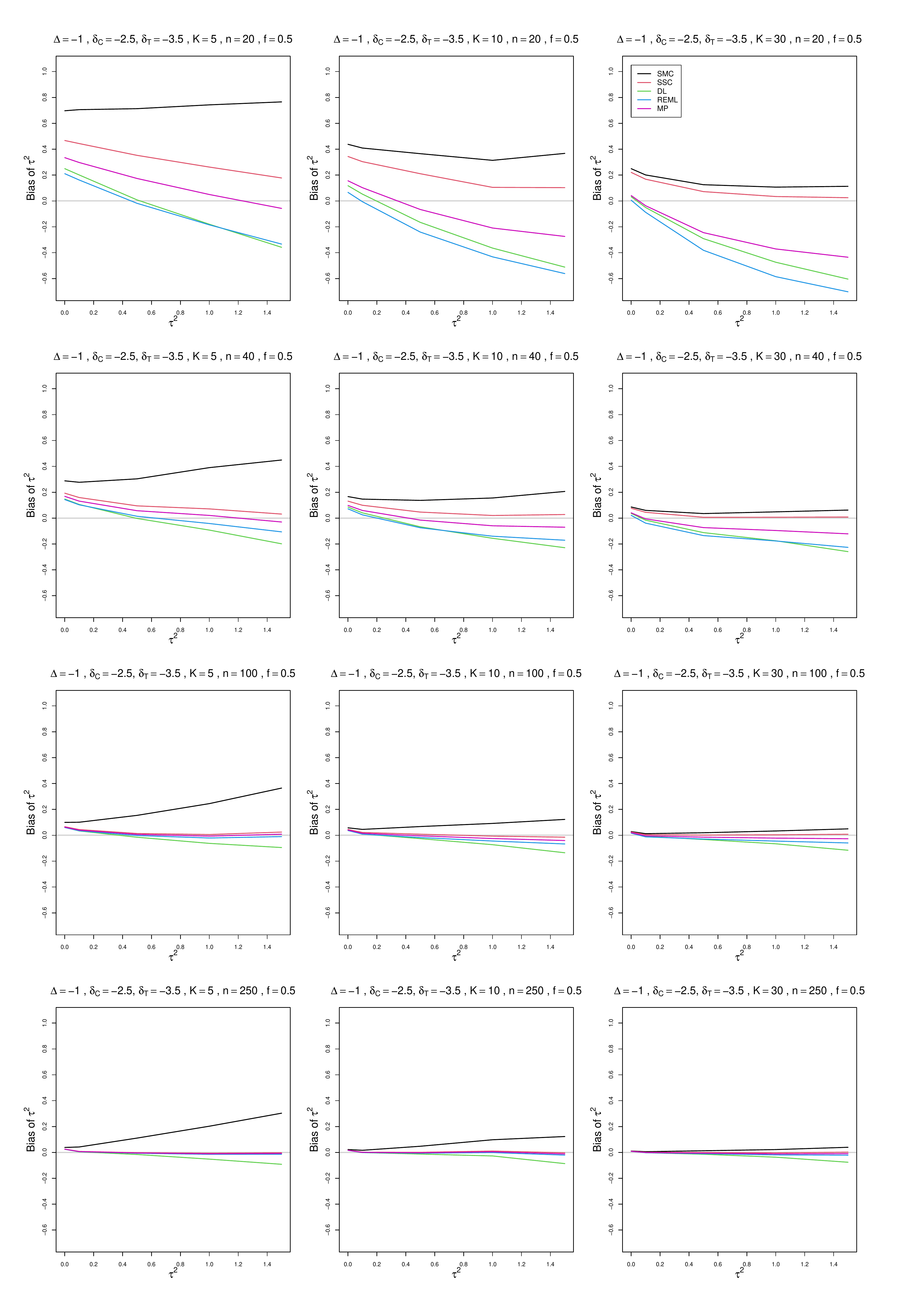}
	\caption{Bias  of estimators of between-study variance of DSM (DL, REML, MP, SMC and SSC ) vs $\tau^2$, for equal sample sizes $n=20,\;40,\;100$ and $250$, $\delta_{iC} = -2.5$, $\Delta=-1$ and  $f = 0.5$.   }
	\label{PlotBiasOfTau2_deltaC_-25deltaT=-3.5_DSM_equal_sample_sizes.pdf}
\end{figure}

\begin{figure}[ht]
	\centering
	\includegraphics[scale=0.33]{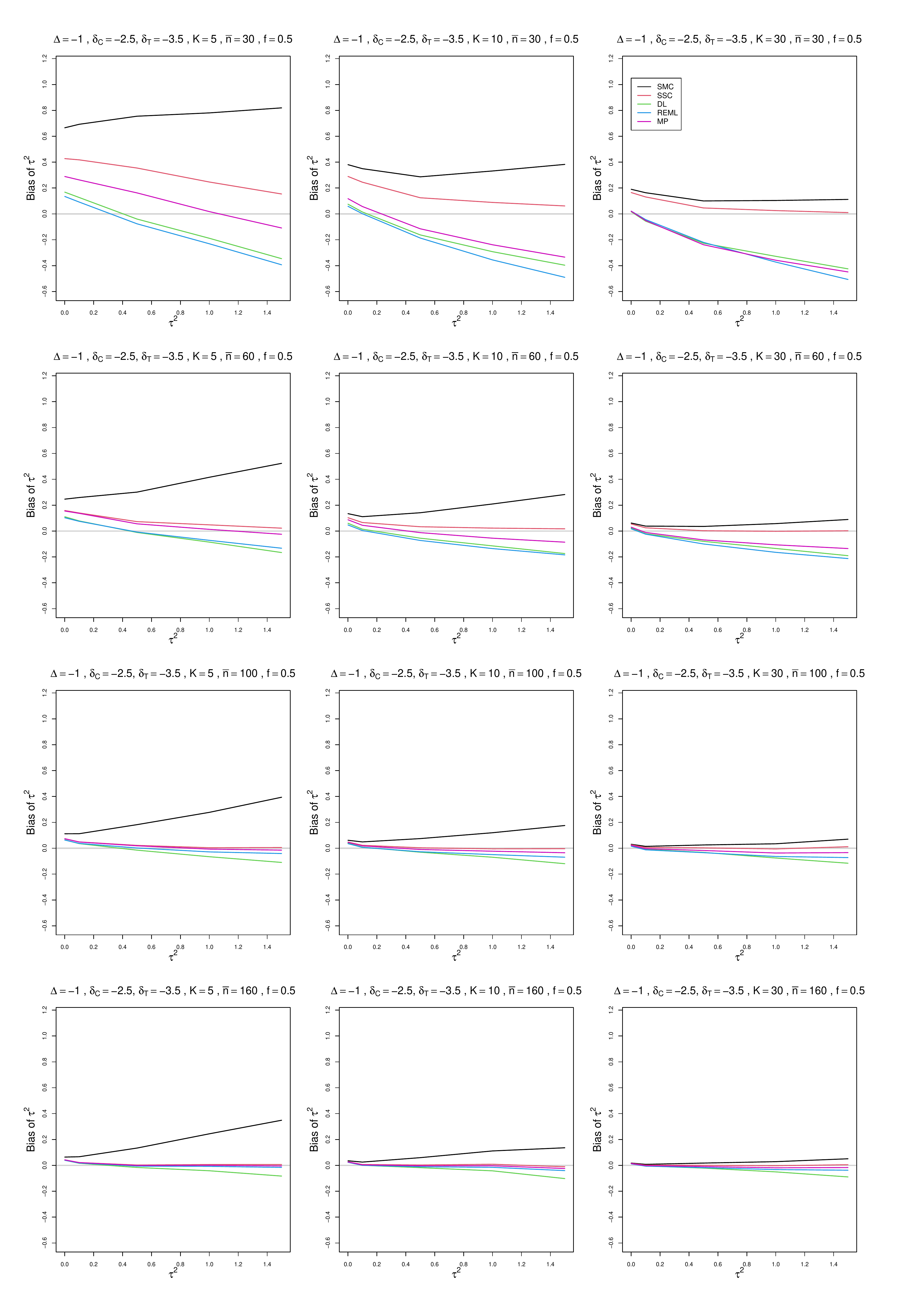}
	\caption{Bias  of estimators of between-study variance of DSM (DL, REML, MP, SMC and SSC ) vs $\tau^2$, for unequal sample sizes $\bar{n}=30,\;60,\;100$ and $160$, $\delta_{iC} = -2.5$, $\Delta=-1$ and  $f = 0.5$.   }
	\label{PlotBiasOfTau2_deltaC_-25deltaT=-3.5_DSM_unequal_sample_sizes.pdf}
\end{figure}

\begin{figure}[ht]
	\centering
	\includegraphics[scale=0.33]{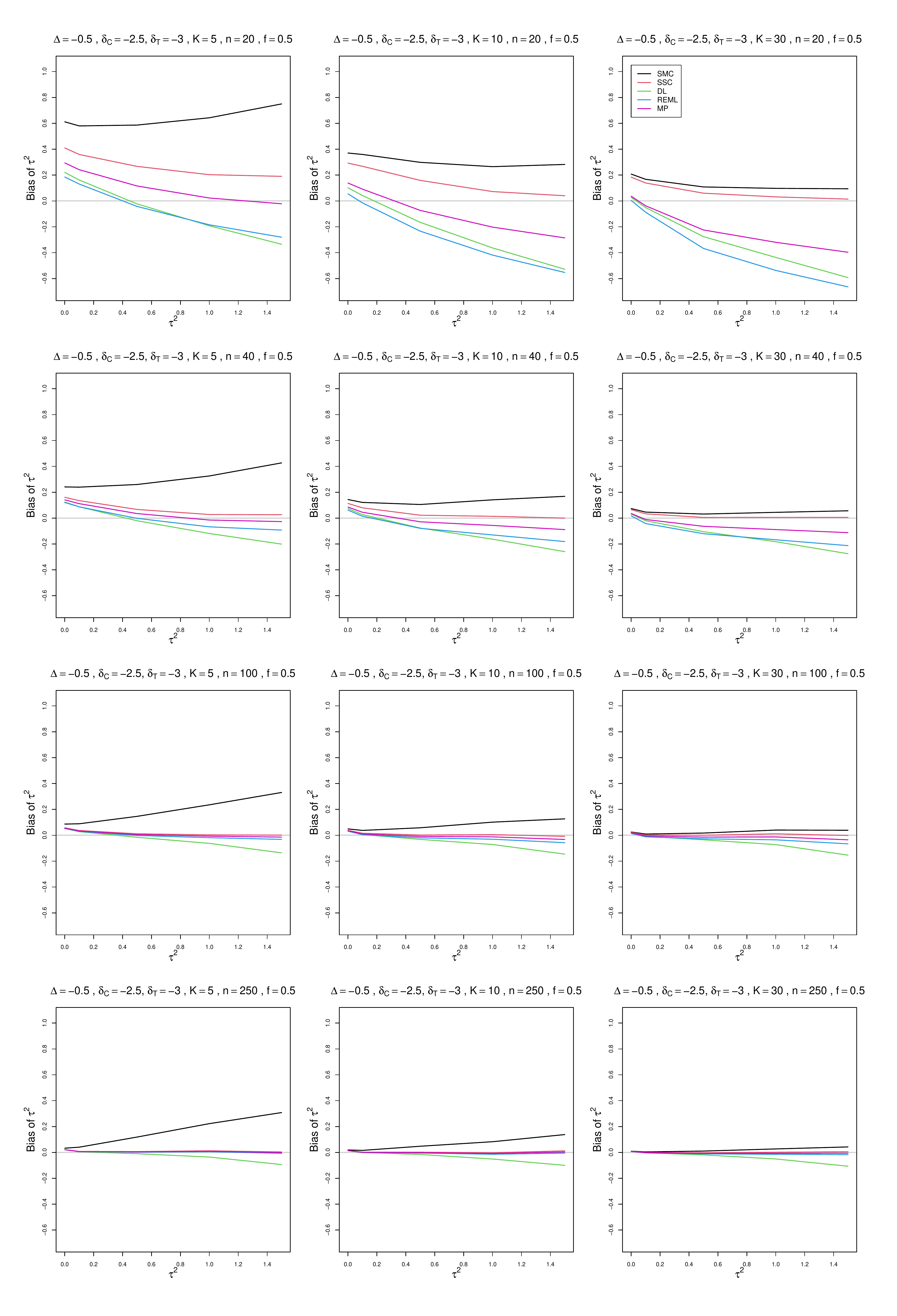}
	\caption{Bias  of estimators of between-study variance of DSM (DL, REML, MP, SMC and SSC ) vs $\tau^2$, for equal sample sizes $n=20,\;40,\;100$ and $250$, $\delta_{iC} = -2.5$, $\Delta=-0.5$ and  $f = 0.5$.   }
	\label{PlotBiasOfTau2_deltaC_-25deltaT=-3_DSM_equal_sample_sizes.pdf}
\end{figure}

\begin{figure}[ht]
	\centering
	\includegraphics[scale=0.33]{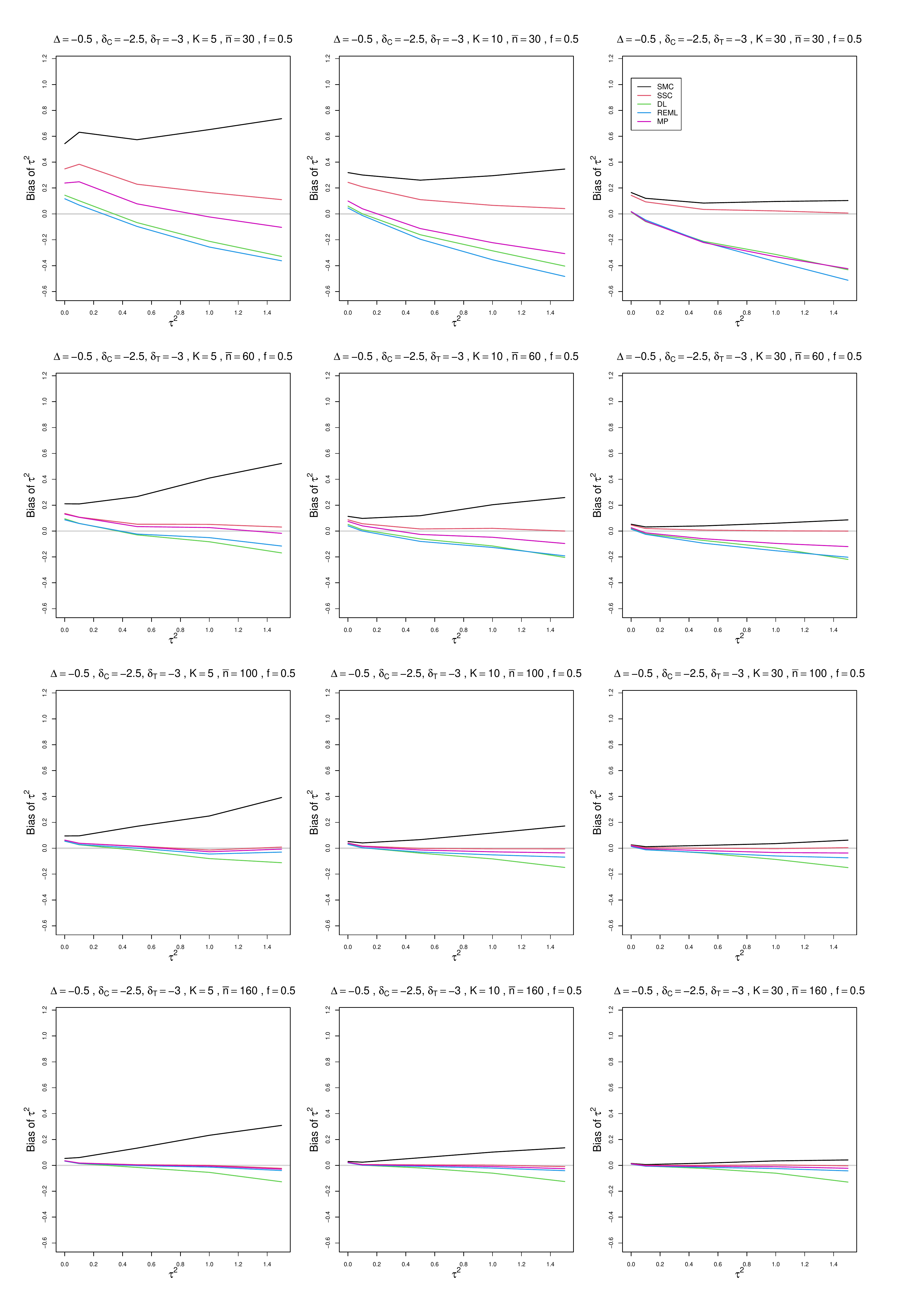}
	\caption{Bias  of estimators of between-study variance of DSM (DL, REML, MP, SMC and SSC ) vs $\tau^2$, for unequal sample sizes $\bar{n}=30,\;60,\;100$ and $160$, $\delta_{iC} = -2.5$, $\Delta=-0.5$ and  $f = 0.5$.   }
	\label{PlotBiasOfTau2_deltaC_-25deltaT=-3_DSM_unequal_sample_sizes.pdf}
\end{figure}

\begin{figure}[ht]
	\centering
	\includegraphics[scale=0.33]{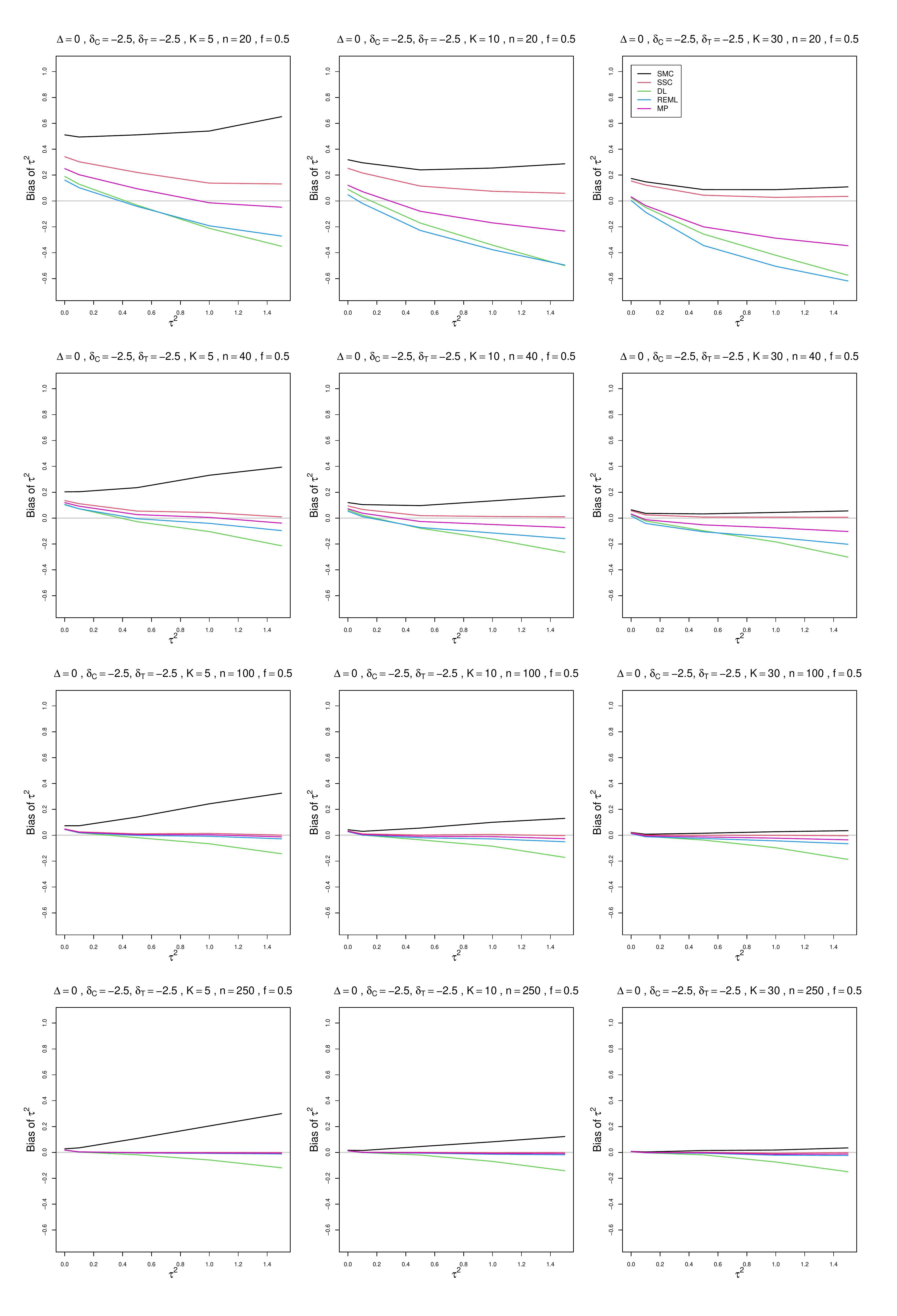}
	\caption{Bias  of estimators of between-study variance of DSM (DL, REML, MP, SMC and SSC ) vs $\tau^2$, for equal sample sizes $n=20,\;40,\;100$ and $250$, $\delta_{iC} = -2.5$, $\Delta=0$ and  $f = 0.5$.   }
	\label{PlotBiasOfTau2_deltaC_-25deltaT=-2.5_DSM_equal_sample_sizes.pdf}
\end{figure}

\begin{figure}[ht]
	\centering
	\includegraphics[scale=0.33]{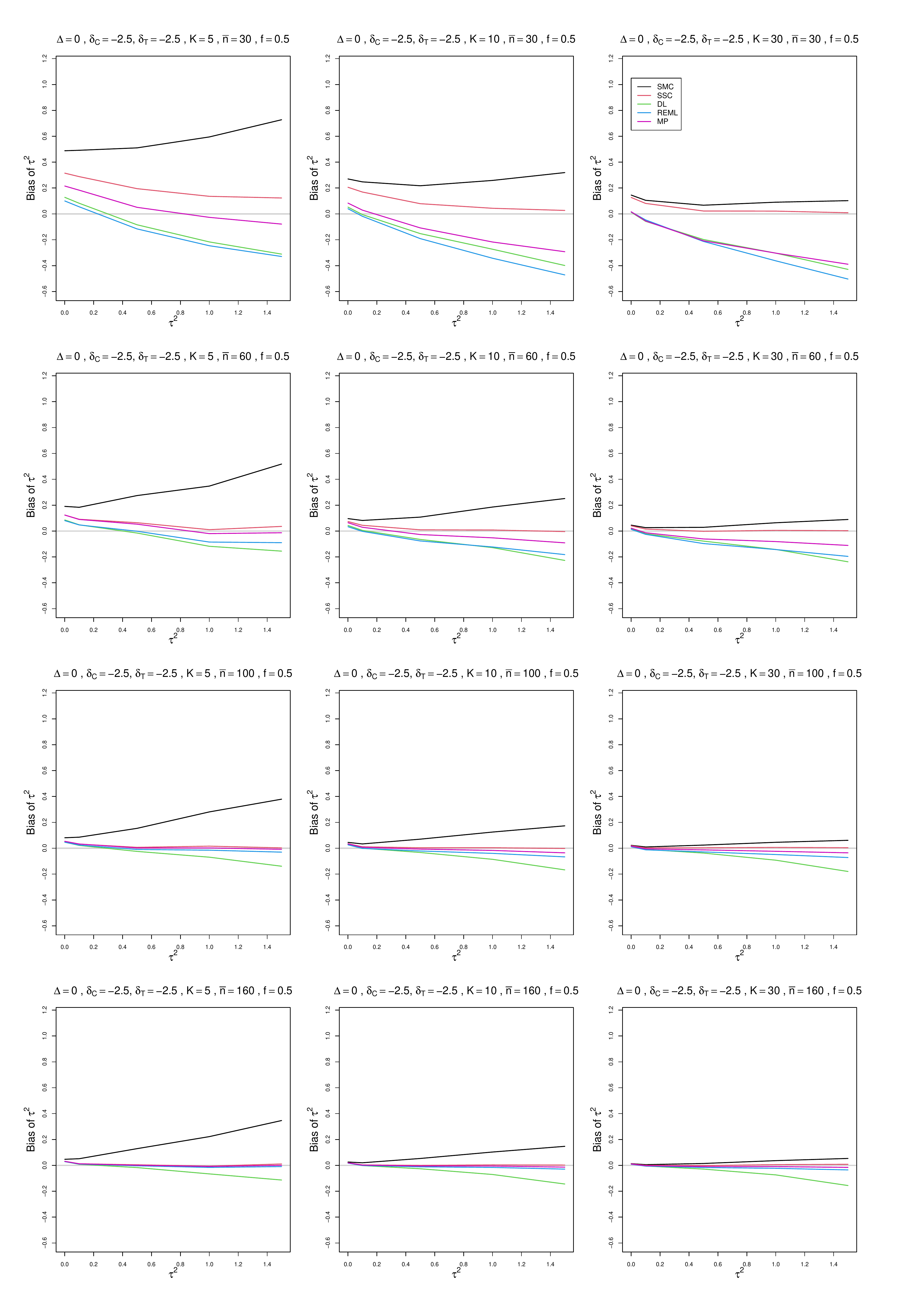}
	\caption{Bias  of estimators of between-study variance of DSM (DL, REML, MP, SMC and SSC ) vs $\tau^2$, for unequal sample sizes $\bar{n}=30,\;60,\;100$ and $160$, $\delta_{iC} = -2.5$, $\Delta=0$ and  $f = 0.5$.   }
	\label{PlotBiasOfTau2_deltaC_-25deltaT=-2.5_DSM_unequal_sample_sizes.pdf}
\end{figure}

\begin{figure}[ht]
	\centering
	\includegraphics[scale=0.33]{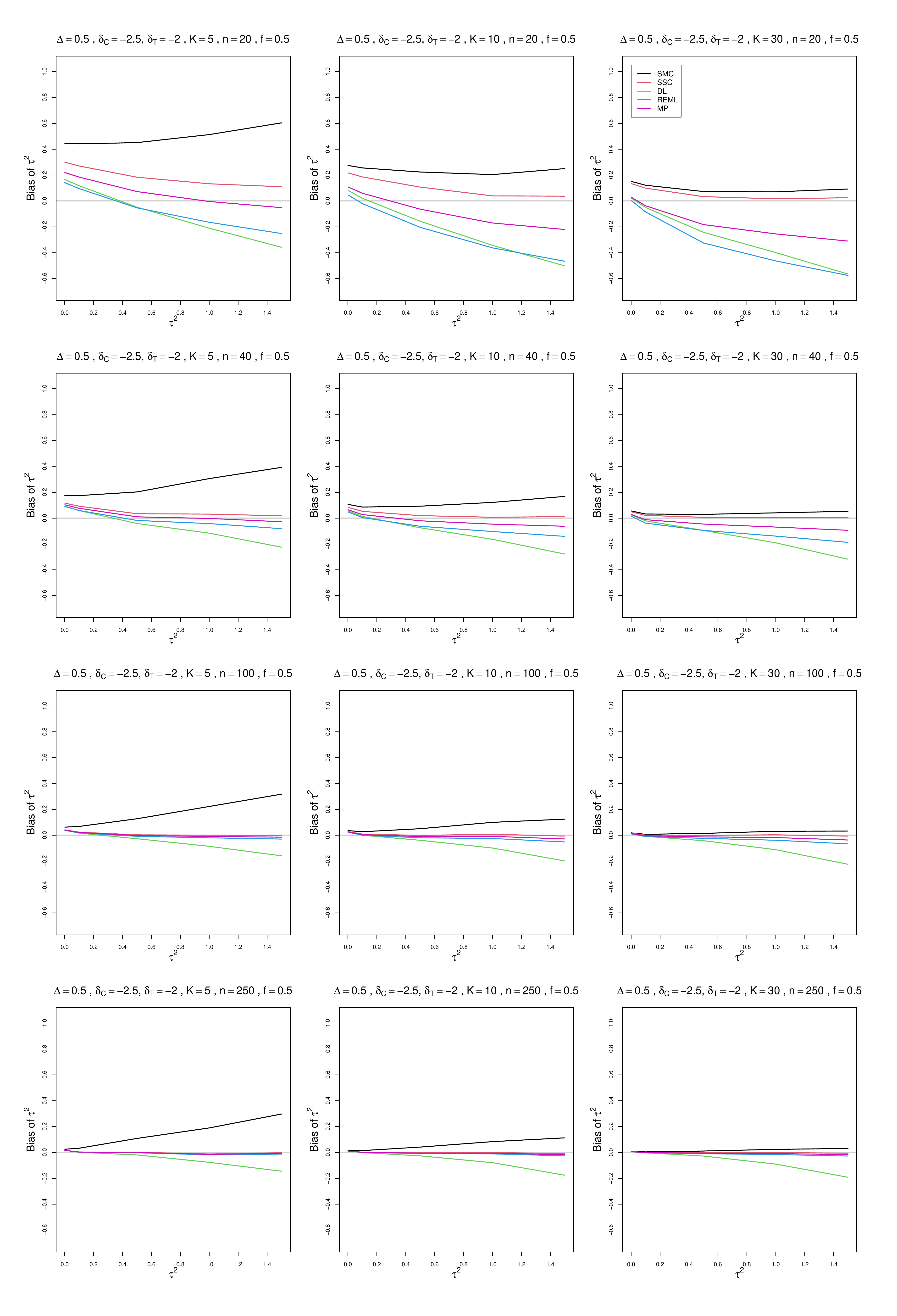}
	\caption{Bias  of estimators of between-study variance of DSM (DL, REML, MP, SMC and SSC ) vs $\tau^2$, for equal sample sizes $n=20,\;40,\;100$ and $250$, $\delta_{iC} = -2.5$, $\Delta=0.5$ and  $f = 0.5$.   }
	\label{PlotBiasOfTau2_deltaC_-25deltaT=-2_DSM_equal_sample_sizes.pdf}
\end{figure}

\begin{figure}[ht]
	\centering
	\includegraphics[scale=0.33]{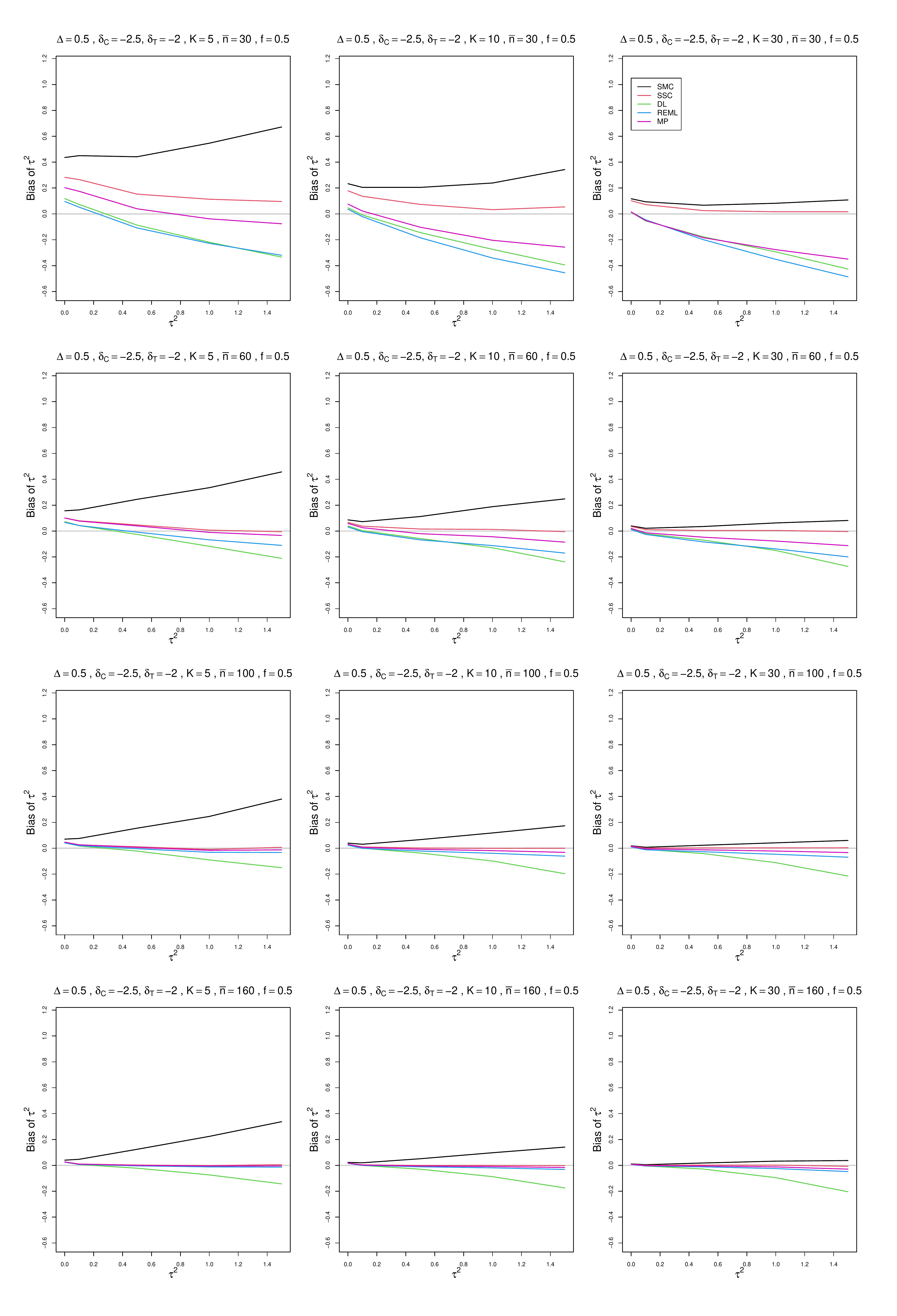}
	\caption{Bias  of estimators of between-study variance of DSM (DL, REML, MP, SMC and SSC ) vs $\tau^2$, for unequal sample sizes $\bar{n}=30,\;60,\;100$ and $160$, $\delta_{iC} = -2.5$, $\Delta=0.5$ and  $f = 0.5$.   }
	\label{PlotBiasOfTau2_deltaC_-25deltaT=-2_DSM_unequal_sample_sizes.pdf}
\end{figure}

\begin{figure}[ht]
	\centering
	\includegraphics[scale=0.33]{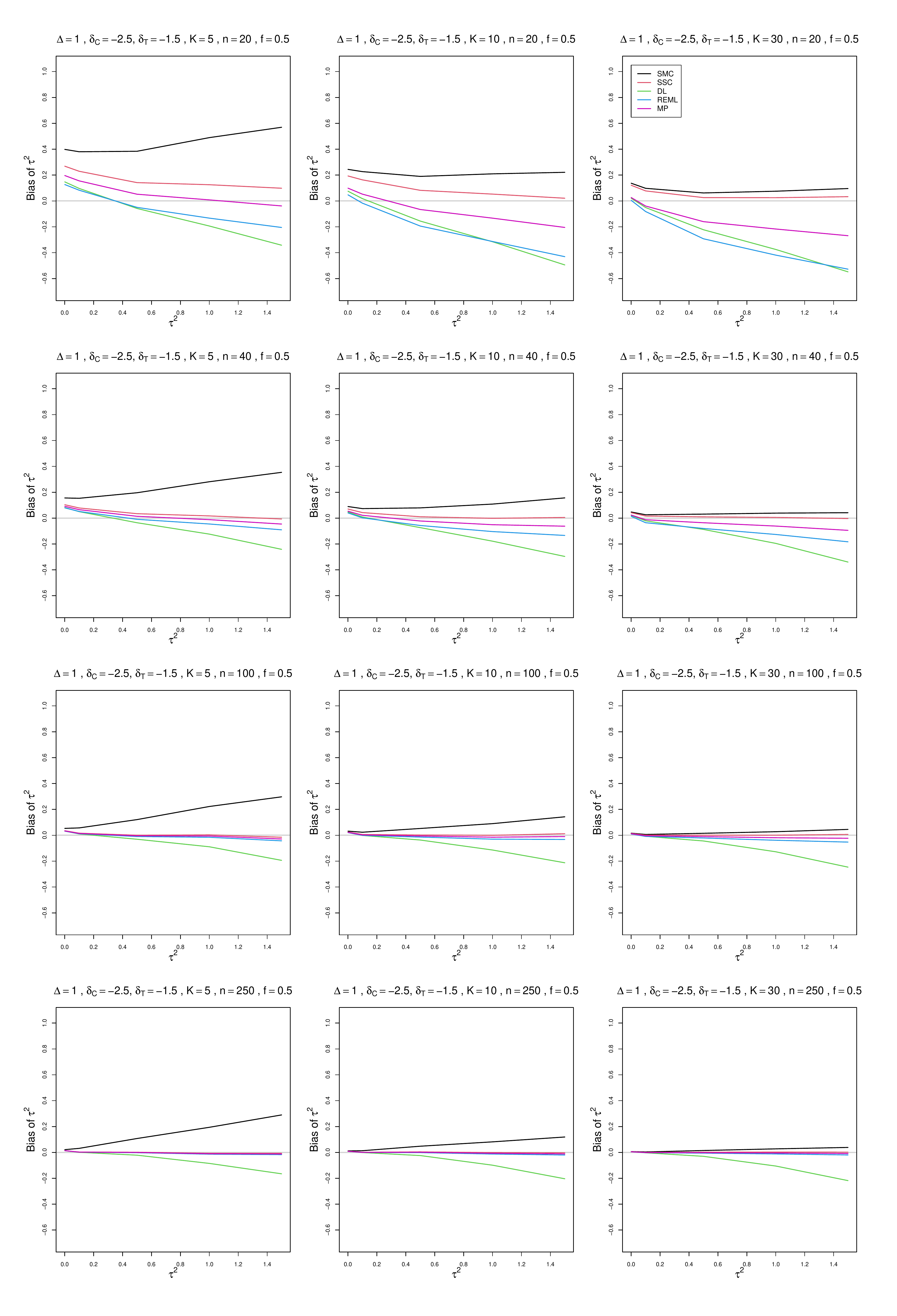}
	\caption{Bias  of estimators of between-study variance of DSM (DL, REML, MP, SMC and SSC ) vs $\tau^2$, for equal sample sizes $n=20,\;40,\;100$ and $250$, $\delta_{iC} = -2.5$, $\Delta=1$ and  $f = 0.5$.   }
	\label{PlotBiasOfTau2_deltaC_-25deltaT=-1.5_DSM_equal_sample_sizes.pdf}
\end{figure}

\begin{figure}[ht]
	\centering
	\includegraphics[scale=0.33]{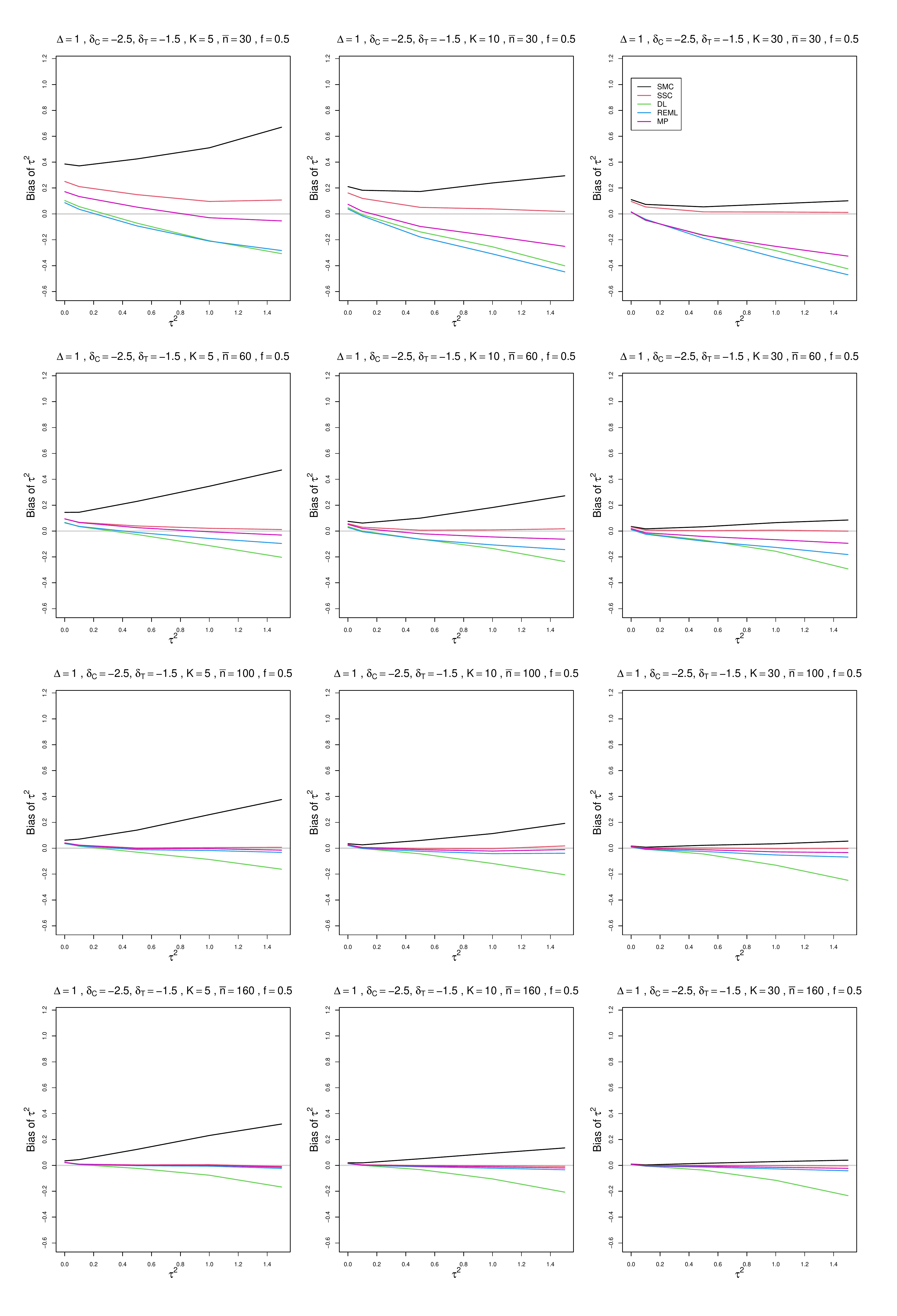}
	\caption{Bias  of estimators of between-study variance of DSM (DL, REML, MP, SMC and SSC ) vs $\tau^2$, for unequal sample sizes $\bar{n}=30,\;60,\;100$ and $160$, $\delta_{iC} = -2.5$, $\Delta=1$ and  $f = 0.5$.   }
	\label{PlotBiasOfTau2_deltaC_-25deltaT=-1.5_DSM_unequal_sample_sizes.pdf}
\end{figure}

\begin{figure}[ht]
	\centering
	\includegraphics[scale=0.33]{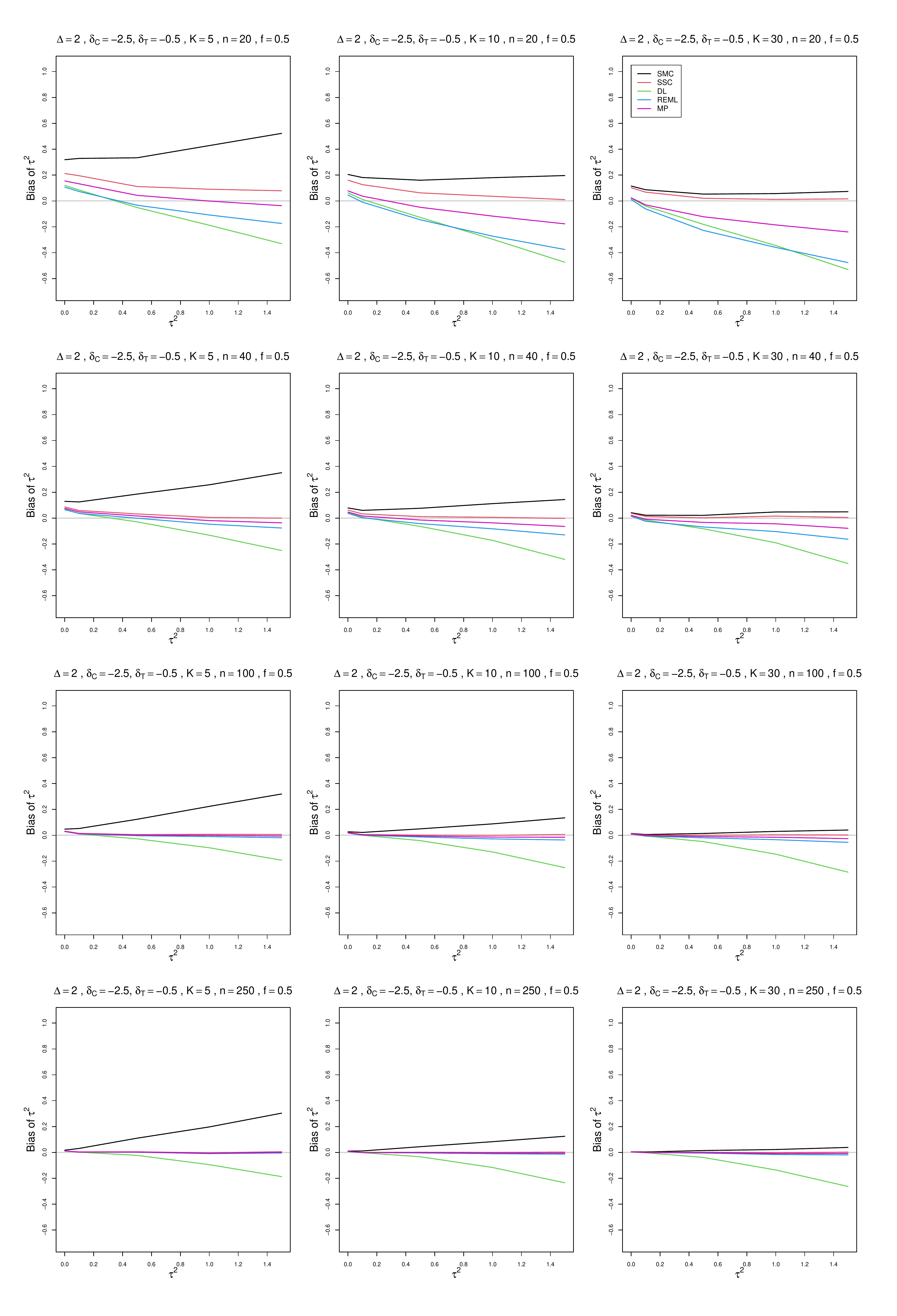}
	\caption{Bias  of estimators of between-study variance of DSM (DL, REML, MP, SMC and SSC ) vs $\tau^2$, for equal sample sizes $n=20,\;40,\;100$ and $250$, $\delta_{iC} = -2.5$, $\Delta=2$ and  $f = 0.5$.   }
	\label{PlotBiasOfTau2_deltaC_-25deltaT=-0.5_DSM_equal_sample_sizes.pdf}
\end{figure}

\begin{figure}[ht]
	\centering
	\includegraphics[scale=0.33]{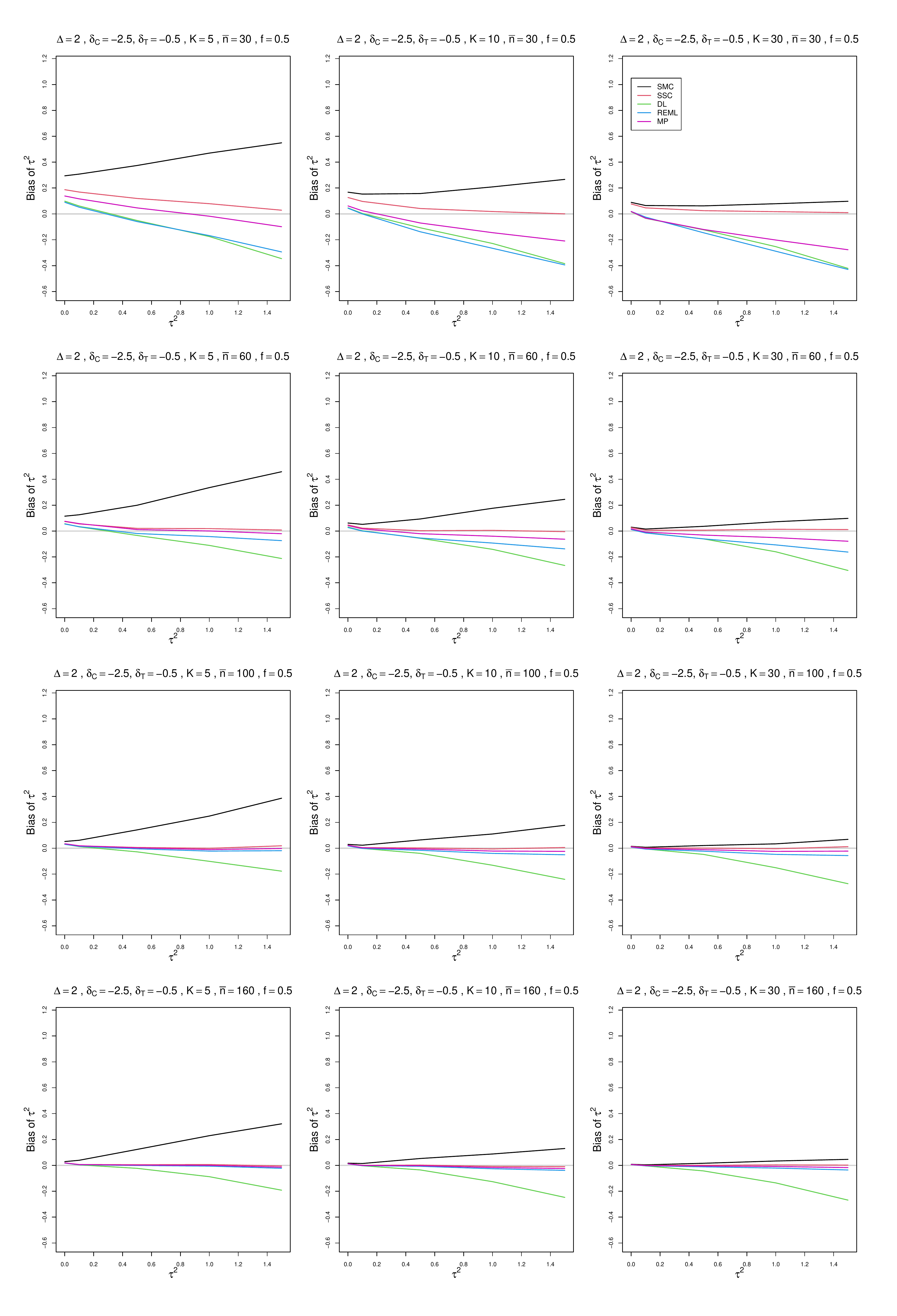}
	\caption{Bias  of estimators of between-study variance of DSM (DL, REML, MP, SMC and SSC ) vs $\tau^2$, for unequal sample sizes $\bar{n}=30,\;60,\;100$ and $160$, $\delta_{iC} = -2.5$, $\Delta=2$ and  $f = 0.5$.   }
	\label{PlotBiasOfTau2_deltaC_-25deltaT=-0.5_DSM_unequal_sample_sizes.pdf}
\end{figure}
\begin{figure}[ht]
	\centering
	\includegraphics[scale=0.33]{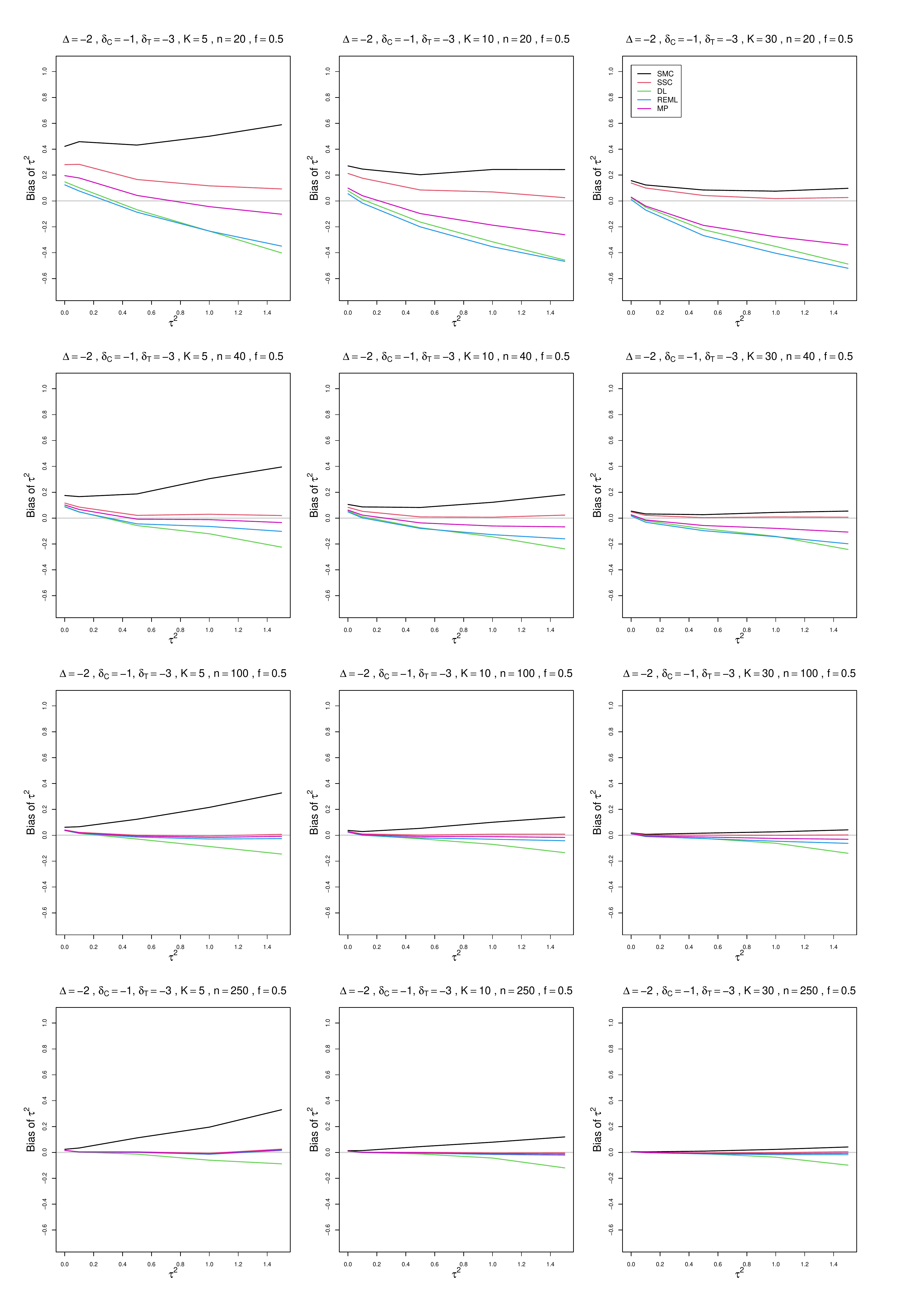}
	\caption{Bias  of estimators of between-study variance of DSM (DL, REML, MP, SMC and SSC ) vs $\tau^2$, for equal sample sizes $n=20,\;40,\;100$ and $250$, $\delta_{iC} = -2.5$, $\Delta=-2$ and  $f = 0.5$.   }
	\label{PlotBiasOfTau2_deltaC_-1deltaT=-3_DSM_equal_sample_sizes.pdf}
\end{figure}

\begin{figure}[ht]
	\centering
	\includegraphics[scale=0.33]{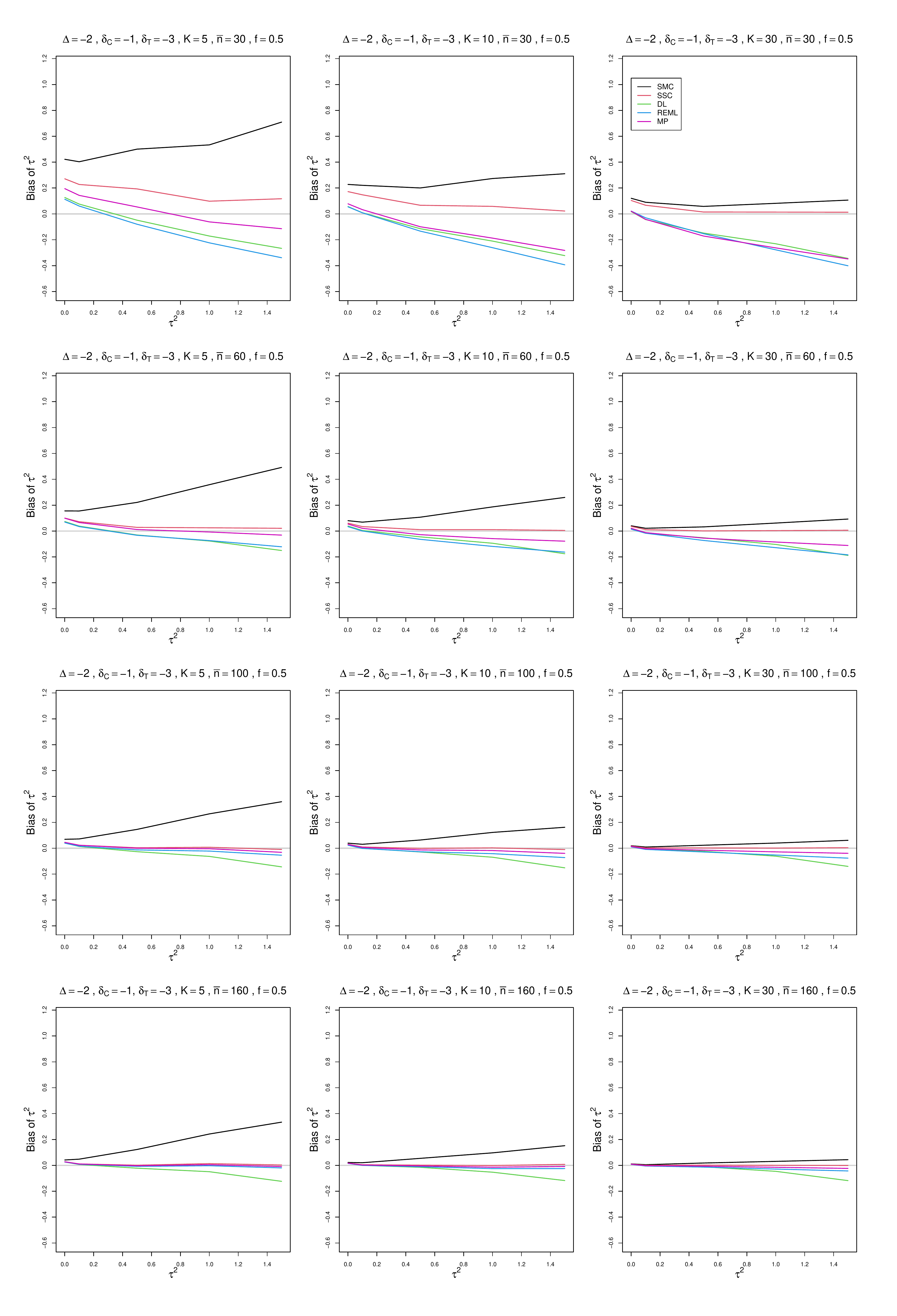}
	\caption{Bias  of estimators of between-study variance of DSM (DL, REML, MP, SMC and SSC ) vs $\tau^2$, for unequal sample sizes $\bar{n}=30,\;60,\;100$ and $160$, $\delta_{iC} = -1$, $\Delta=-2$ and  $f = 0.5$.   }
	\label{PlotBiasOfTau2_deltaC_-1deltaT=-3_DSM_unequal_sample_sizes.pdf}
\end{figure}

\begin{figure}[ht]
	\centering
	\includegraphics[scale=0.33]{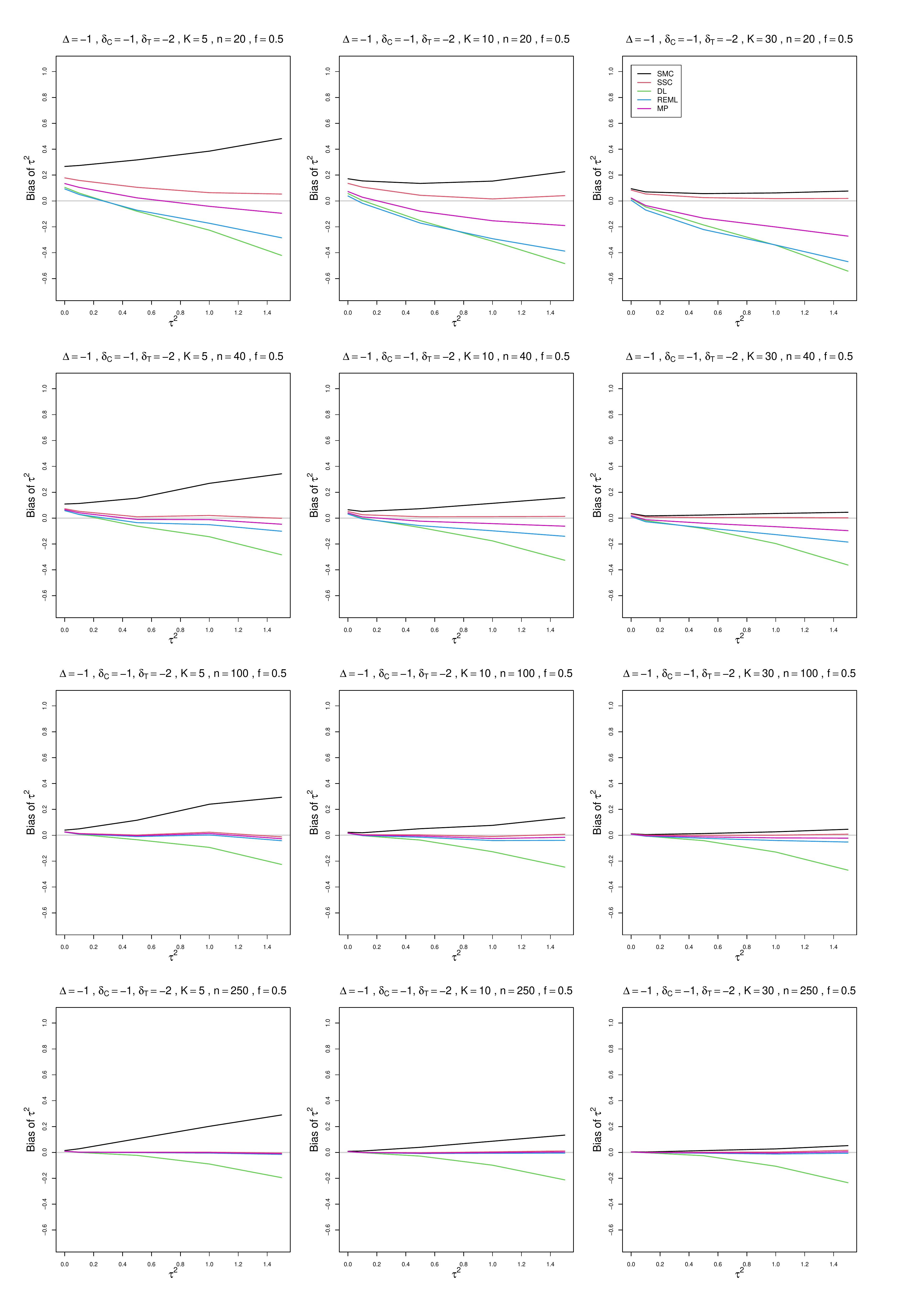}
	\caption{Bias  of estimators of between-study variance of DSM (DL, REML, MP, SMC and SSC ) vs $\tau^2$, for equal sample sizes $n=20,\;40,\;100$ and $250$, $\delta_{iC} = -1$, $\Delta=-1$ and  $f = 0.5$.   }
	\label{PlotBiasOfTau2_deltaC_-1deltaT=-2_DSM_equal_sample_sizes.pdf}
\end{figure}

\begin{figure}[ht]
	\centering
	\includegraphics[scale=0.33]{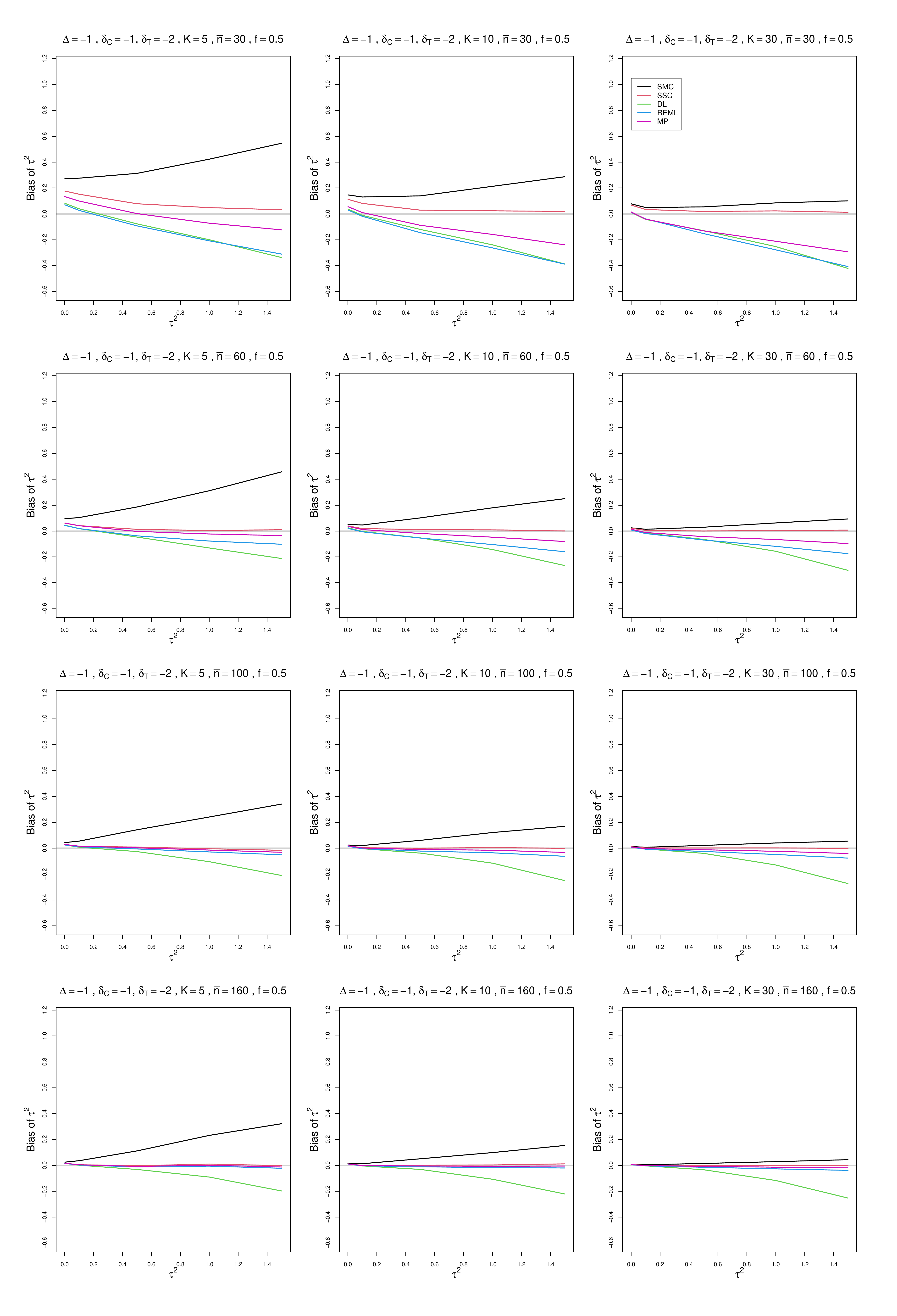}
	\caption{Bias  of estimators of between-study variance of DSM (DL, REML, MP, SMC and SSC ) vs $\tau^2$, for unequal sample sizes $\bar{n}=30,\;60,\;100$ and $160$, $\delta_{iC} = -1$, $\Delta=-1$ and  $f = 0.5$.   }
	\label{PlotBiasOfTau2_deltaC_-1deltaT=-2_DSM_unequal_sample_sizes.pdf}
\end{figure}

\begin{figure}[ht]
	\centering
	\includegraphics[scale=0.33]{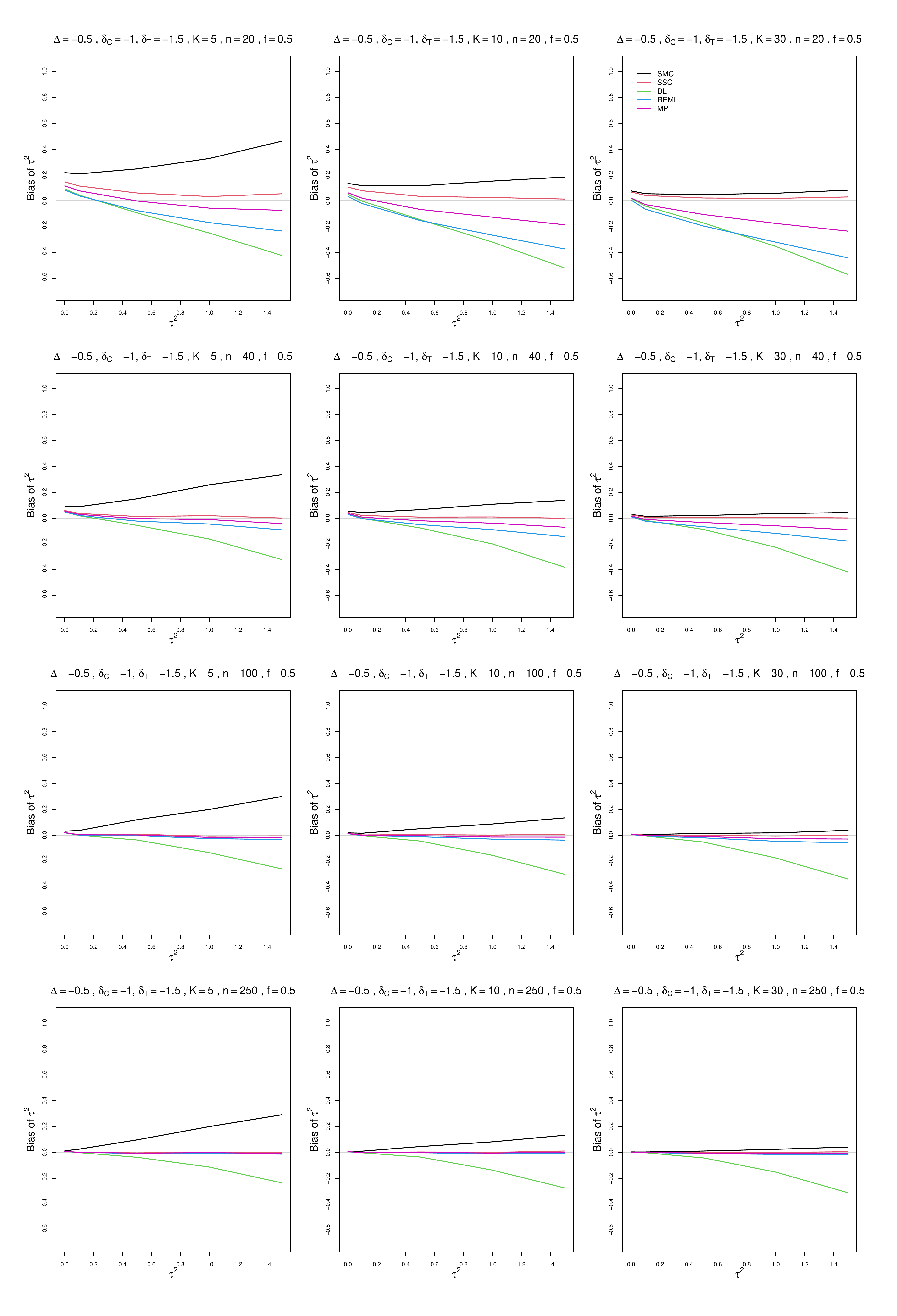}
	\caption{Bias  of estimators of between-study variance of DSM (DL, REML, MP, SMC and SSC ) vs $\tau^2$, for equal sample sizes $n=20,\;40,\;100$ and $250$, $\delta_{iC} = -1$, $\Delta=-0.5$ and  $f = 0.5$.   }
	\label{PlotBiasOfTau2_deltaC_-1deltaT=-1.5_DSM_equal_sample_sizes.pdf}
\end{figure}

\begin{figure}[ht]
	\centering
	\includegraphics[scale=0.33]{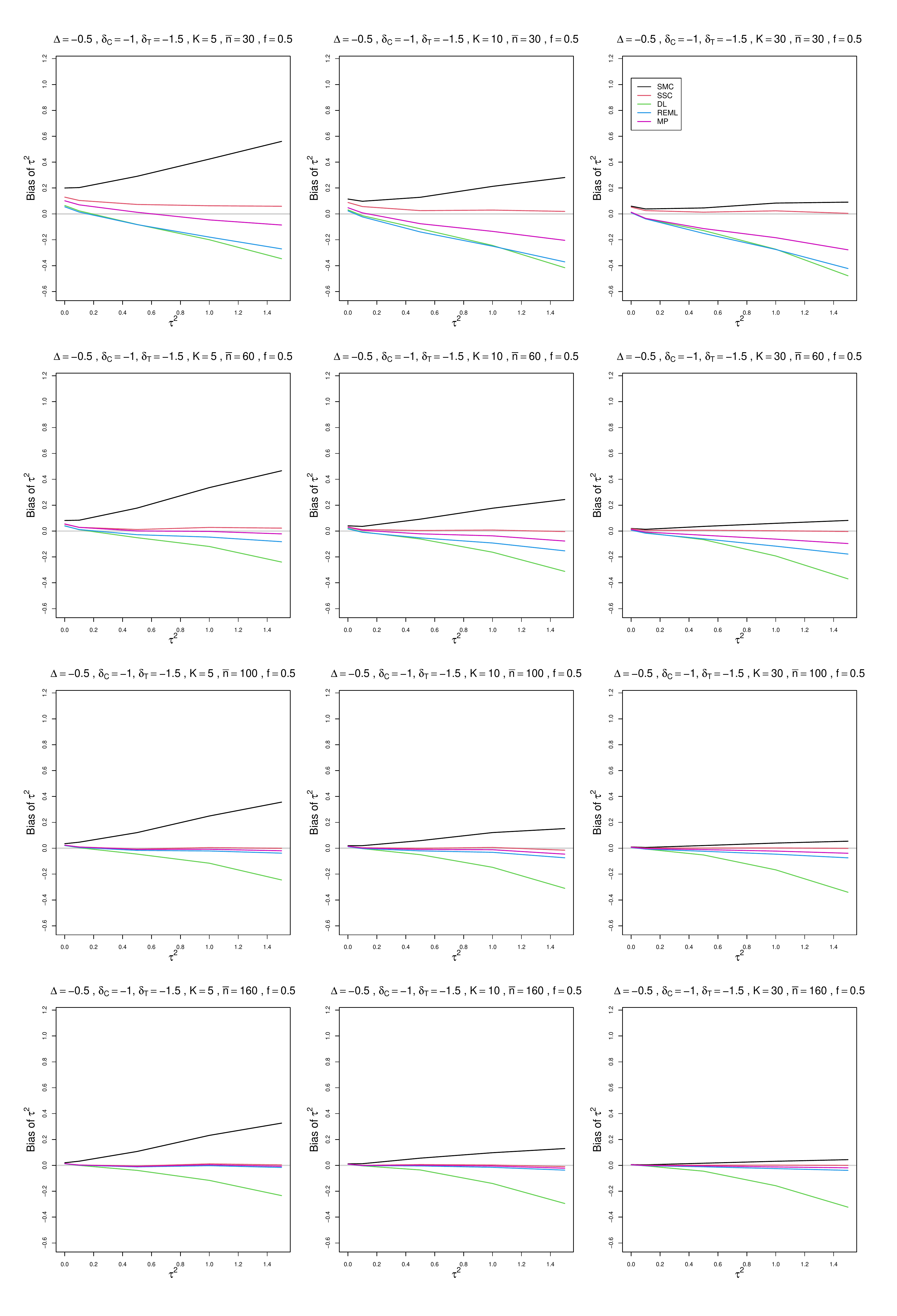}
	\caption{Bias  of estimators of between-study variance of DSM (DL, REML, MP, SMC and SSC ) vs $\tau^2$, for unequal sample sizes $\bar{n}=30,\;60,\;100$ and $160$, $\delta_{iC} = -1$, $\Delta=-0.5$ and  $f = 0.5$.   }
	\label{PlotBiasOfTau2_deltaC_-1deltaT=-1,5_DSM_unequal_sample_sizes.pdf}
\end{figure}

\begin{figure}[ht]
	\centering
	\includegraphics[scale=0.33]{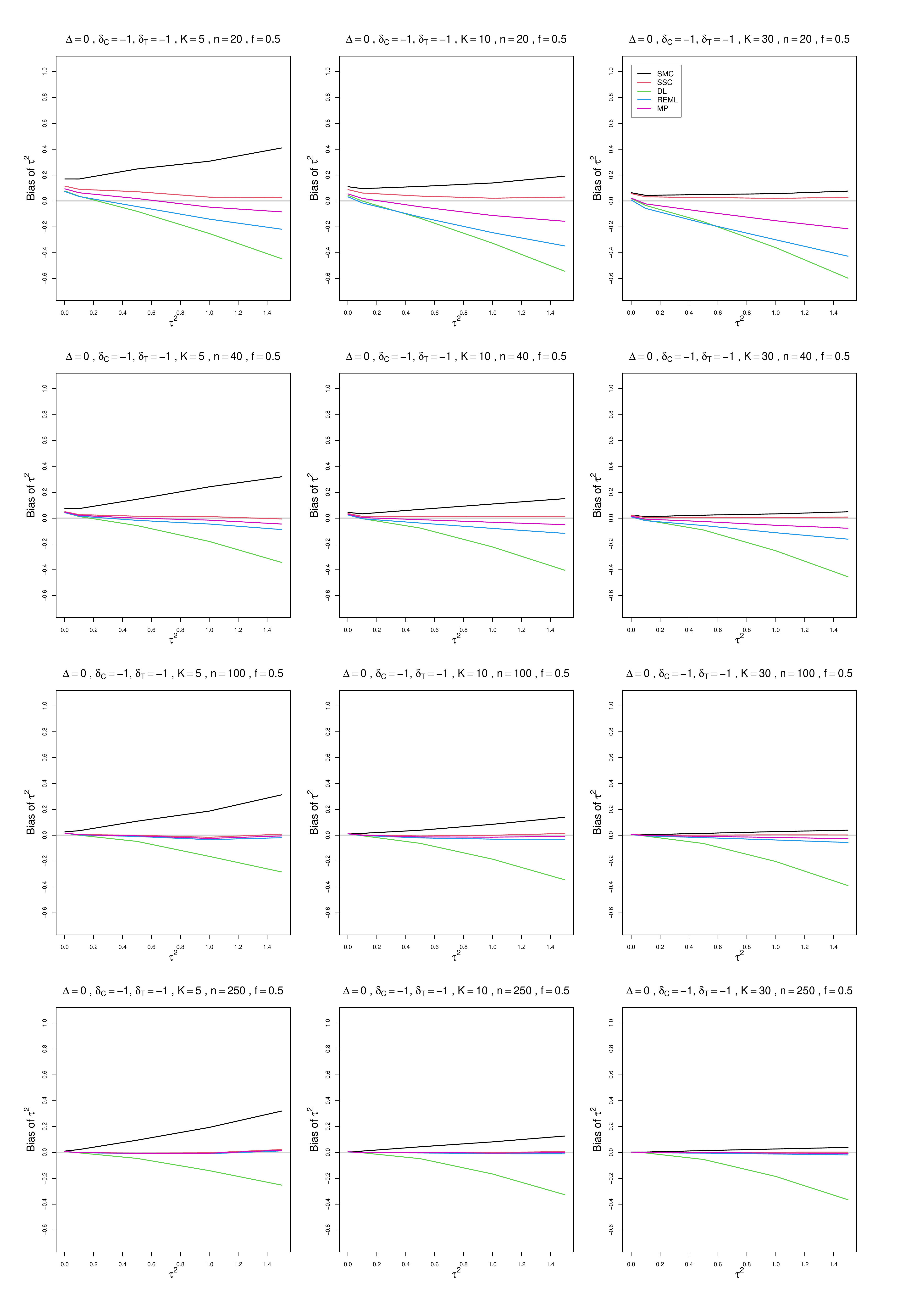}
	\caption{Bias  of estimators of between-study variance of DSM (DL, REML, MP, SMC and SSC ) vs $\tau^2$, for equal sample sizes $n=20,\;40,\;100$ and $250$, $\delta_{iC} = -1$, $\Delta=0$ and  $f = 0.5$.   }
	\label{PlotBiasOfTau2_deltaC_-1deltaT=-1_DSM_equal_sample_sizes.pdf}
\end{figure}

\begin{figure}[ht]
	\centering
	\includegraphics[scale=0.33]{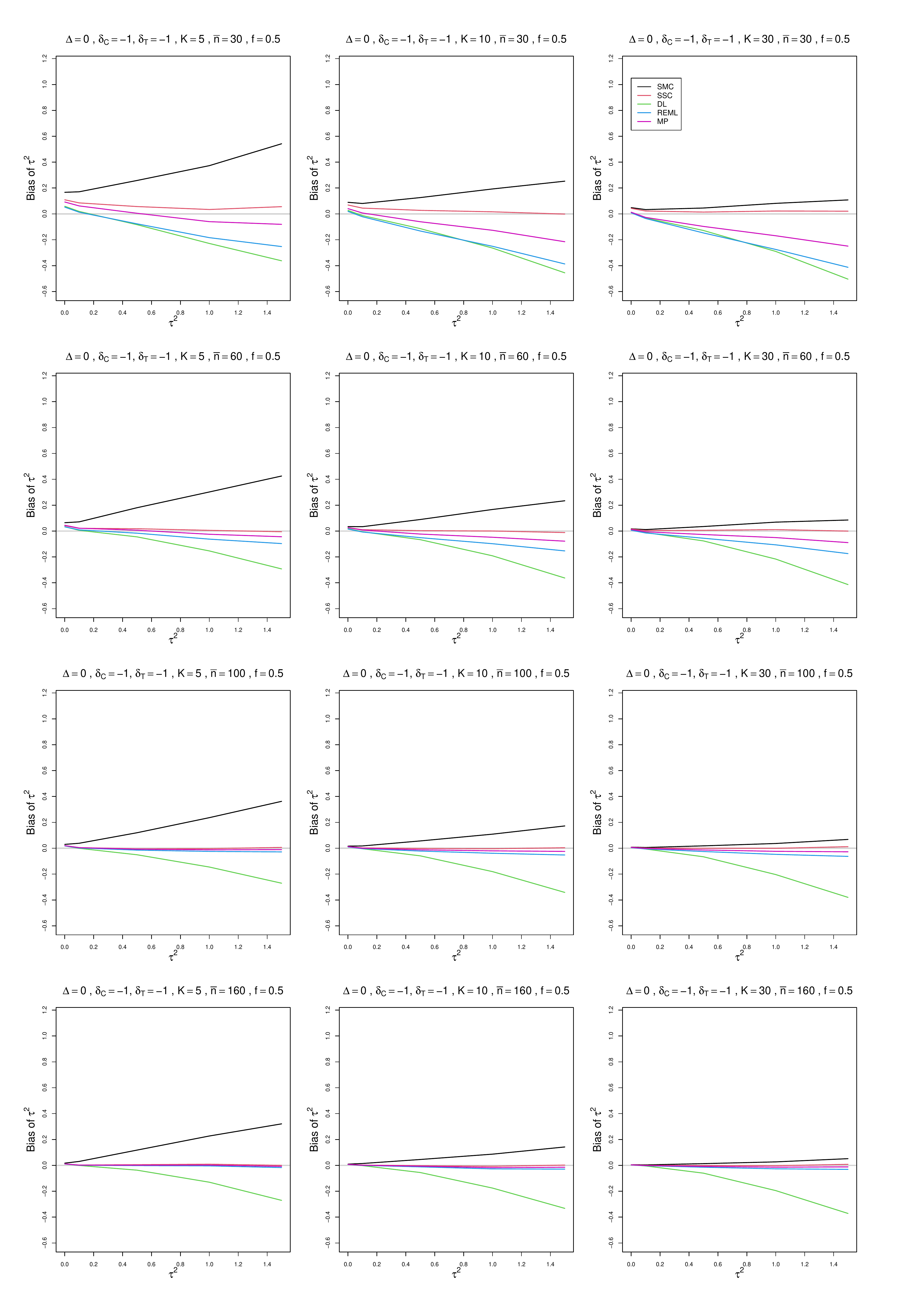}
	\caption{Bias  of estimators of between-study variance of DSM (DL, REML, MP, SMC and SSC ) vs $\tau^2$, for unequal sample sizes $\bar{n}=30,\;60,\;100$ and $160$, $\delta_{iC} = -1$, $\Delta=0$ and  $f = 0.5$.   }
	\label{PlotBiasOfTau2_deltaC_-1deltaT=-1_DSM_unequal_sample_sizes.pdf}
\end{figure}

\begin{figure}[ht]
	\centering
	\includegraphics[scale=0.33]{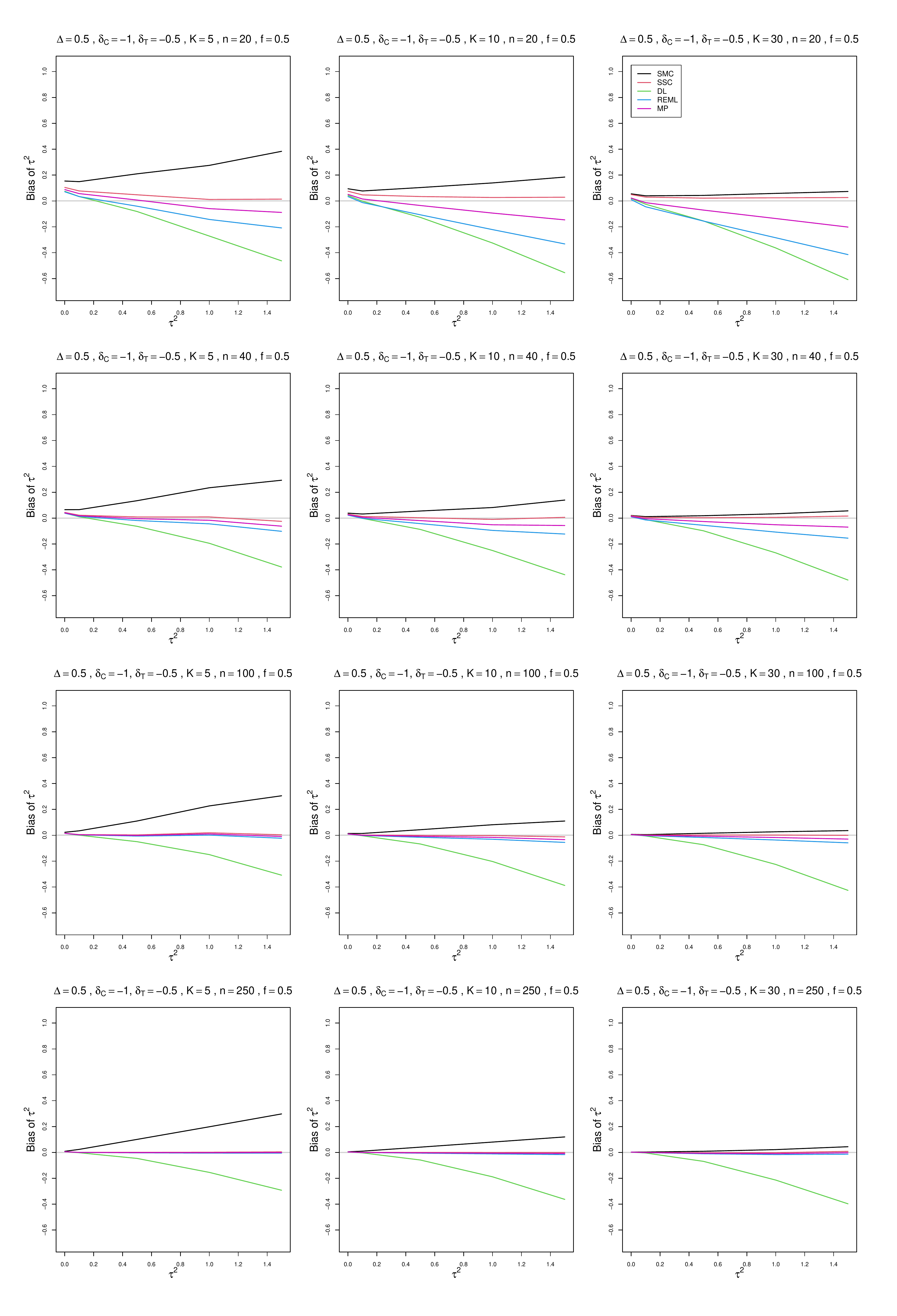}
	\caption{Bias  of estimators of between-study variance of DSM (DL, REML, MP, SMC and SSC ) vs $\tau^2$, for equal sample sizes $n=20,\;40,\;100$ and $250$, $\delta_{iC} = -1$, $\Delta=0.5$ and  $f = 0.5$.   }
	\label{PlotBiasOfTau2_deltaC_-1deltaT=-0.5_DSM_equal_sample_sizes.pdf}
\end{figure}

\begin{figure}[ht]
	\centering
	\includegraphics[scale=0.33]{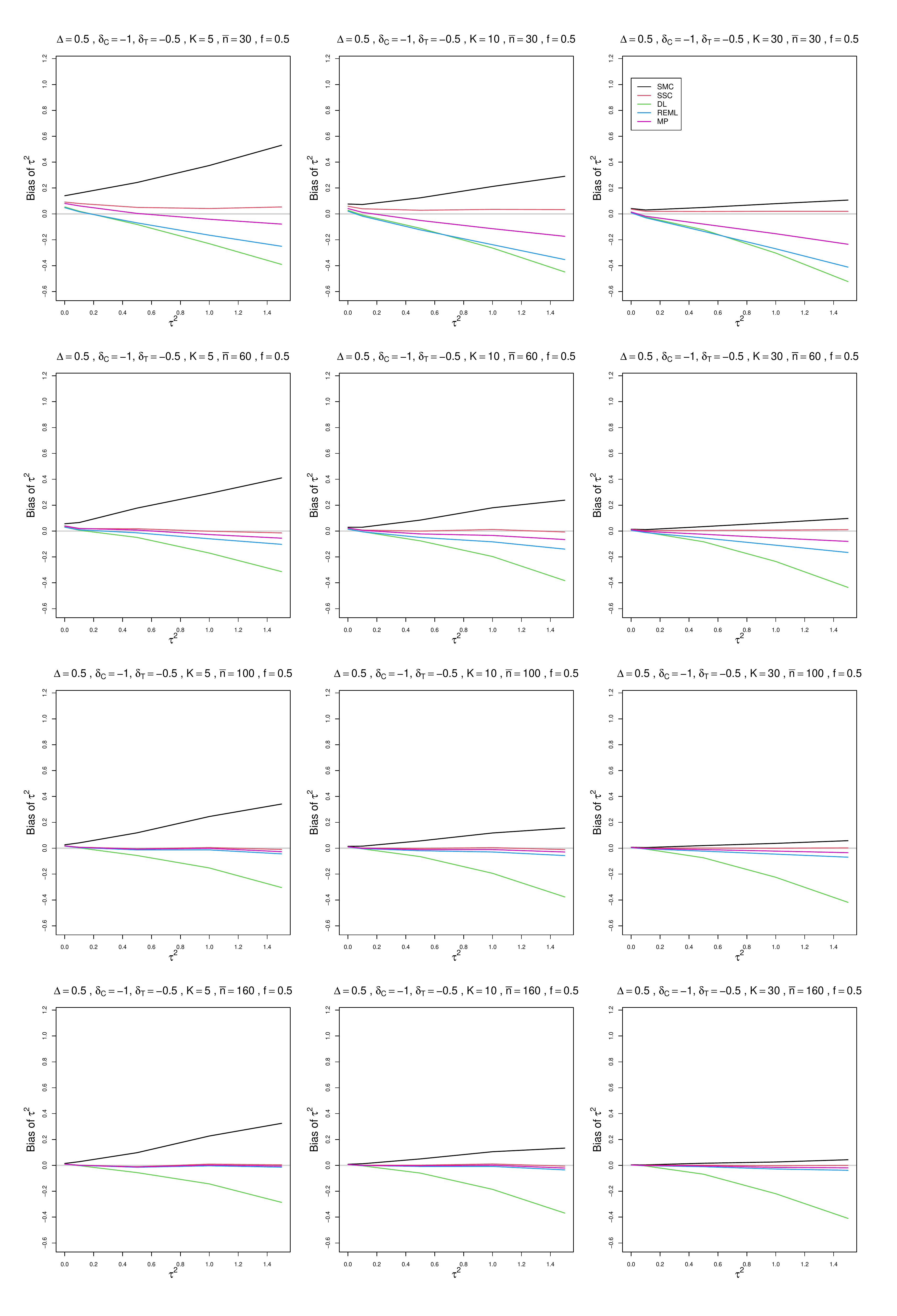}
	\caption{Bias  of estimators of between-study variance of DSM (DL, REML, MP, SMC and SSC ) vs $\tau^2$, for unequal sample sizes $\bar{n}=30,\;60,\;100$ and $160$, $\delta_{iC} = -1$, $\Delta=0.5$ and  $f = 0.5$.   }
	\label{PlotBiasOfTau2_deltaC_-1deltaT=-0,5_DSM_unequal_sample_sizes.pdf}
\end{figure}

\begin{figure}[ht]
	\centering
	\includegraphics[scale=0.33]{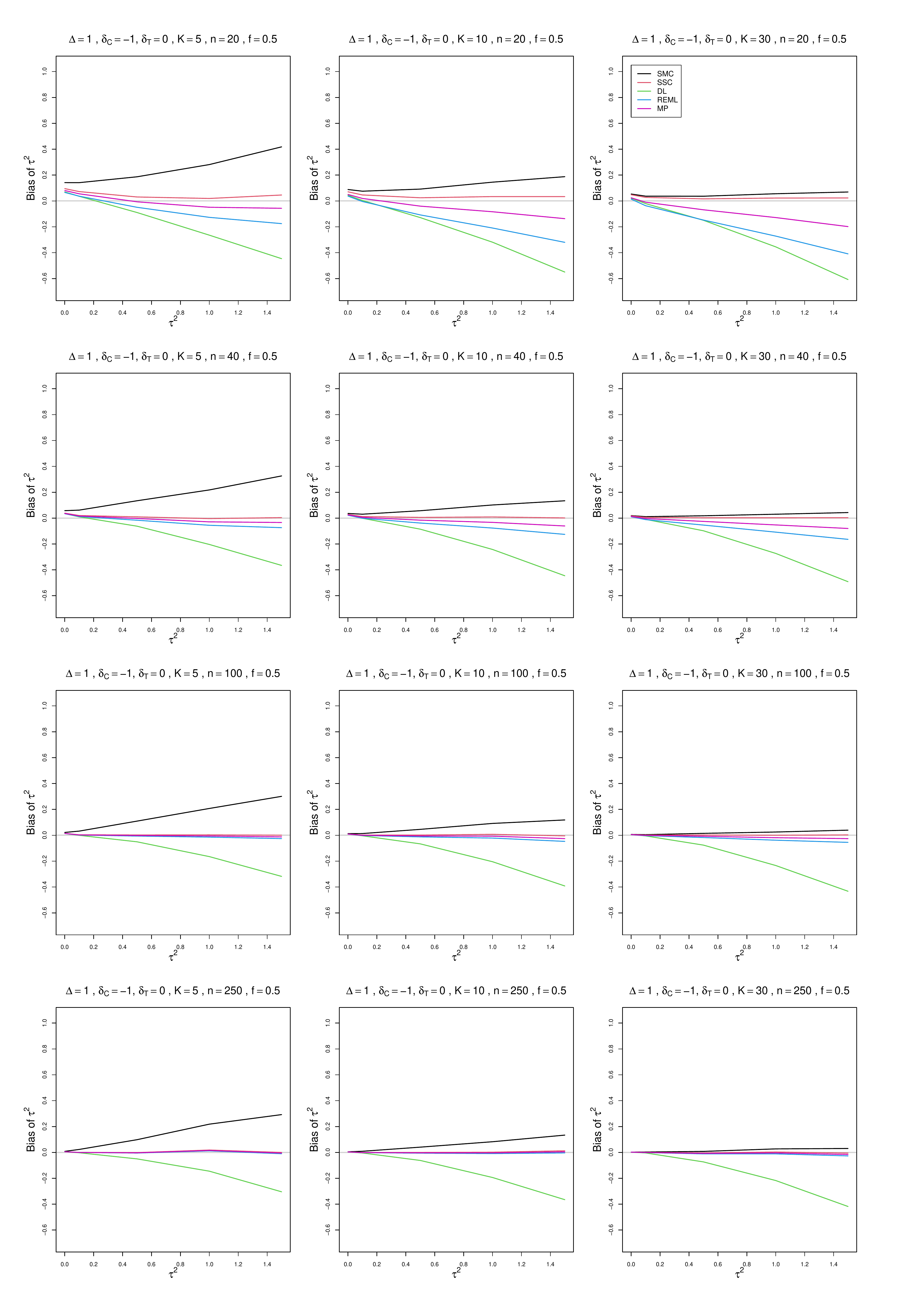}
	\caption{Bias  of estimators of between-study variance of DSM (DL, REML, MP, SMC and SSC ) vs $\tau^2$, for equal sample sizes $n=20,\;40,\;100$ and $250$, $\delta_{iC} = -1$, $\Delta=1$ and  $f = 0.5$.   }
	\label{PlotBiasOfTau2_deltaC_--1deltaT=0_DSM_equal_sample_sizes.pdf}
\end{figure}

\begin{figure}[ht]
	\centering
	\includegraphics[scale=0.33]{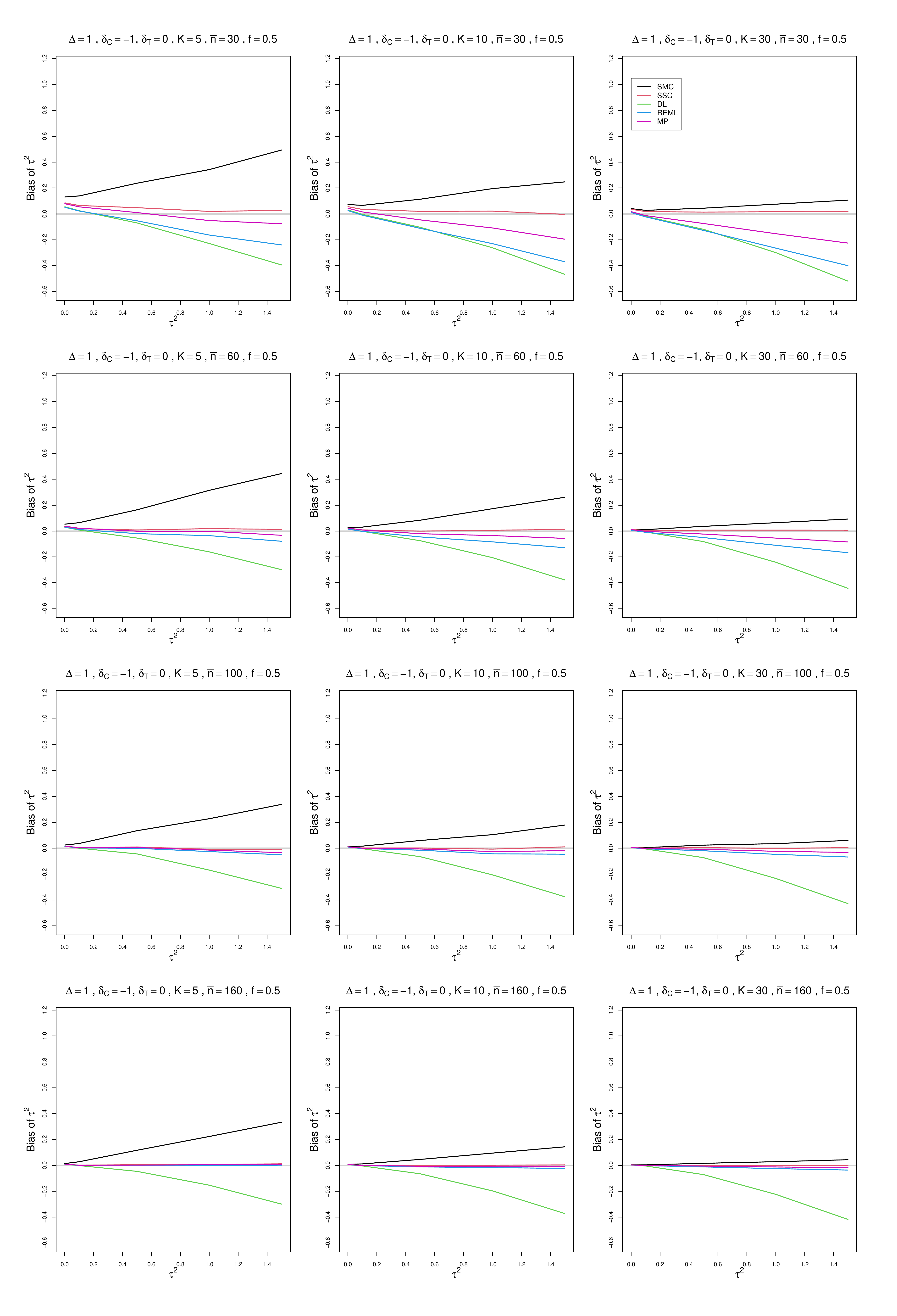}
	\caption{Bias  of estimators of between-study variance of DSM (DL, REML, MP, SMC and SSC ) vs $\tau^2$, for unequal sample sizes $\bar{n}=30,\;60,\;100$ and $160$, $\delta_{iC} = -1$, $\Delta=1$ and  $f = 0.5$.   }
	\label{PlotBiasOfTau2_deltaC_-1deltaT=0_DSM_unequal_sample_sizes.pdf}
\end{figure}

\begin{figure}[ht]
	\centering
	\includegraphics[scale=0.33]{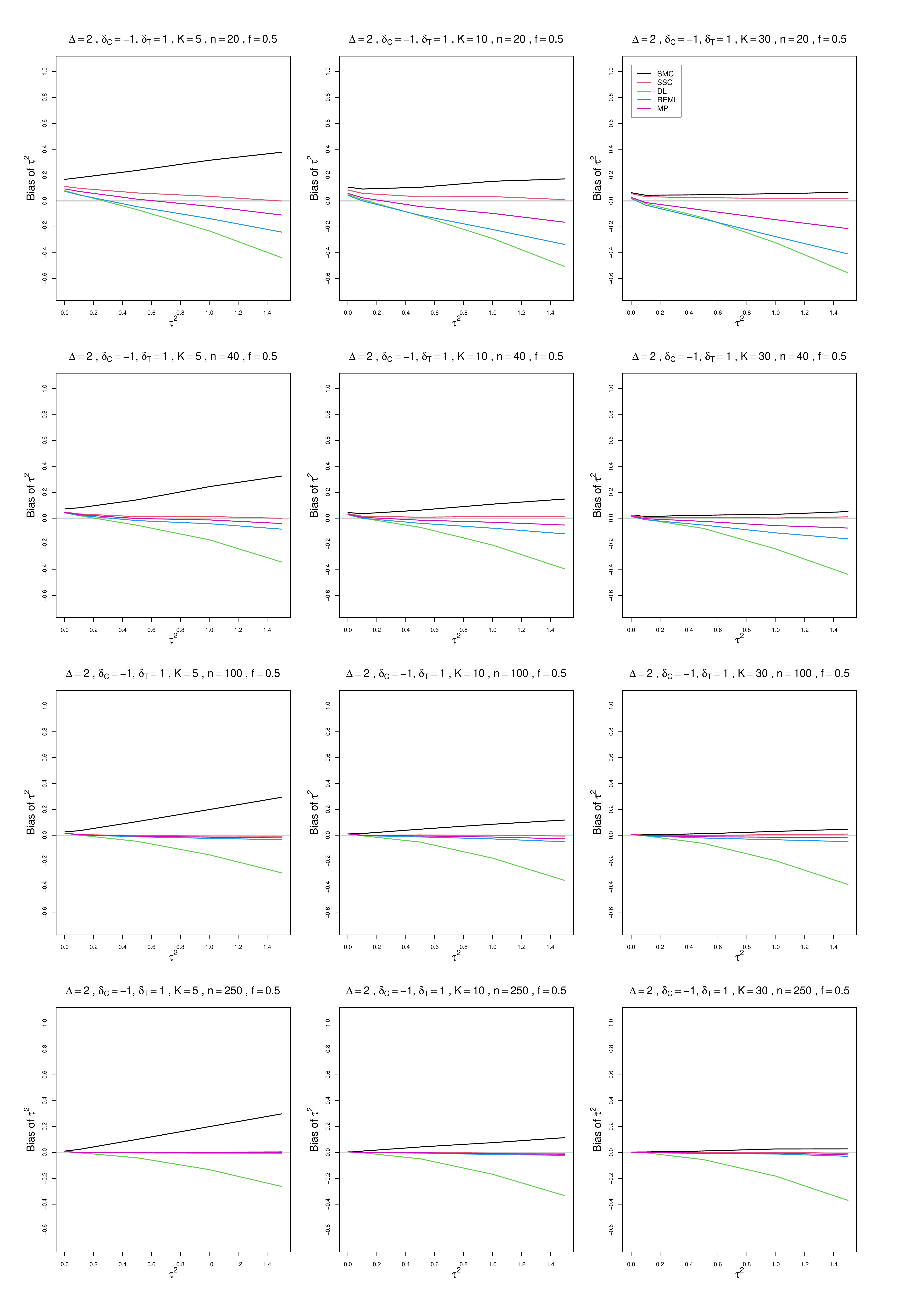}
	\caption{Bias  of estimators of between-study variance of DSM (DL, REML, MP, SMC and SSC ) vs $\tau^2$, for equal sample sizes $n=20,\;40,\;100$ and $250$, $\delta_{iC} = -1$, $\Delta=2$ and  $f = 0.5$.   }
	\label{PlotBiasOfTau2_deltaC_-1deltaT=1_DSM_equal_sample_sizes.pdf}
\end{figure}

\begin{figure}[ht]
	\centering
	\includegraphics[scale=0.33]{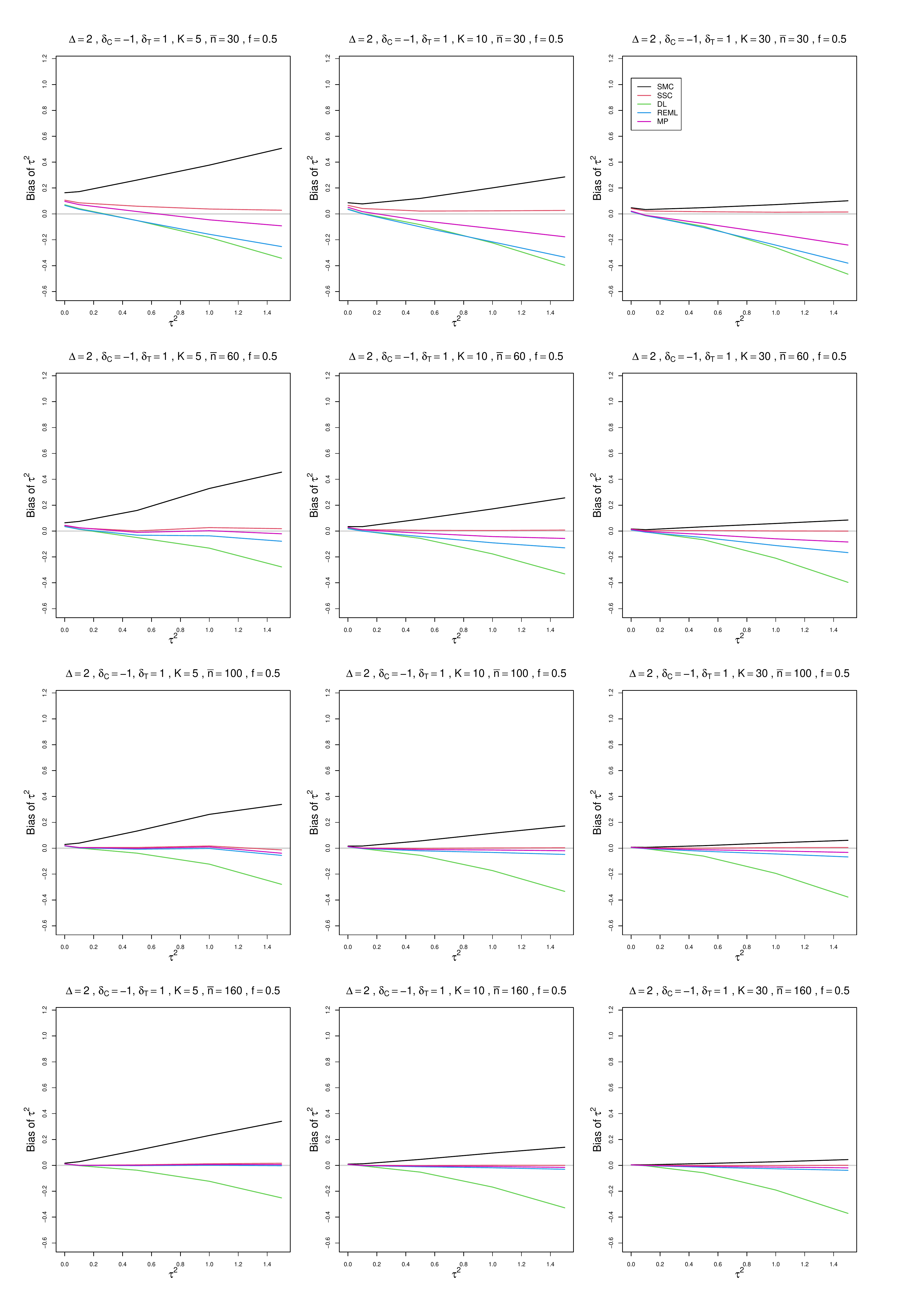}
	\caption{Bias  of estimators of between-study variance of DSM (DL, REML, MP, SMC and SSC ) vs $\tau^2$, for unequal sample sizes $\bar{n}=30,\;60,\;100$ and $160$, $\delta_{iC} = -1$, $\Delta=2$ and  $f = 0.5$.   }
	\label{PlotBiasOfTau2_deltaC_-1deltaT=1_DSM_unequal_sample_sizes.pdf}
\end{figure}
\clearpage

\begin{figure}[ht]
	\centering
	\includegraphics[scale=0.33]{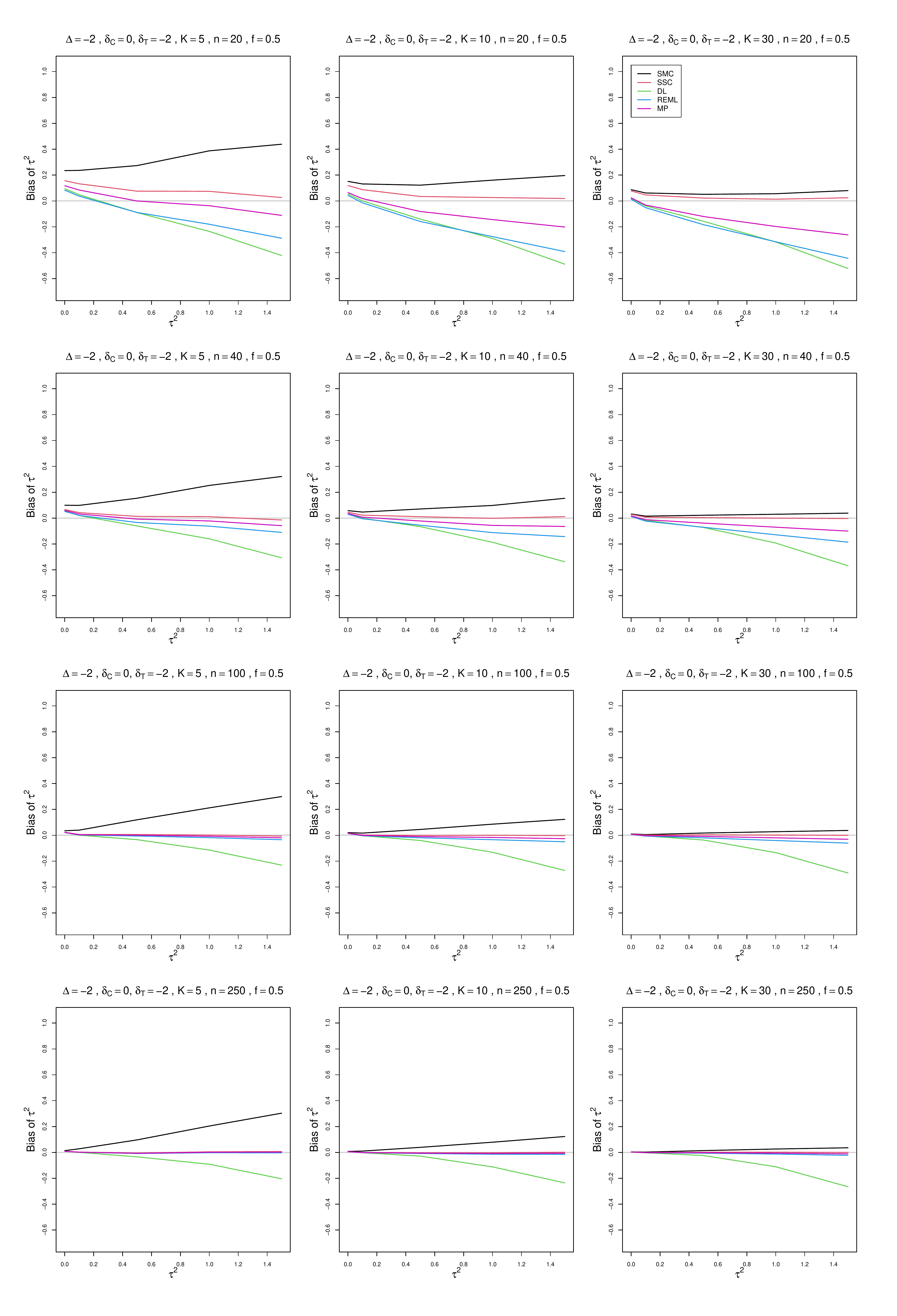}
	\caption{Bias  of estimators of between-study variance of DSM (DL, REML, MP, SMC and SSC ) vs $\tau^2$, for equal sample sizes $n=20,\;40,\;100$ and $250$, $\delta_{iC} = 0$, $\Delta=-2$ and  $f = 0.5$.   }
	\label{PlotBiasOfTau2_deltaC_0deltaT=-2_DSM_equal_sample_sizes.pdf}
\end{figure}

\begin{figure}[ht]
	\centering
	\includegraphics[scale=0.33]{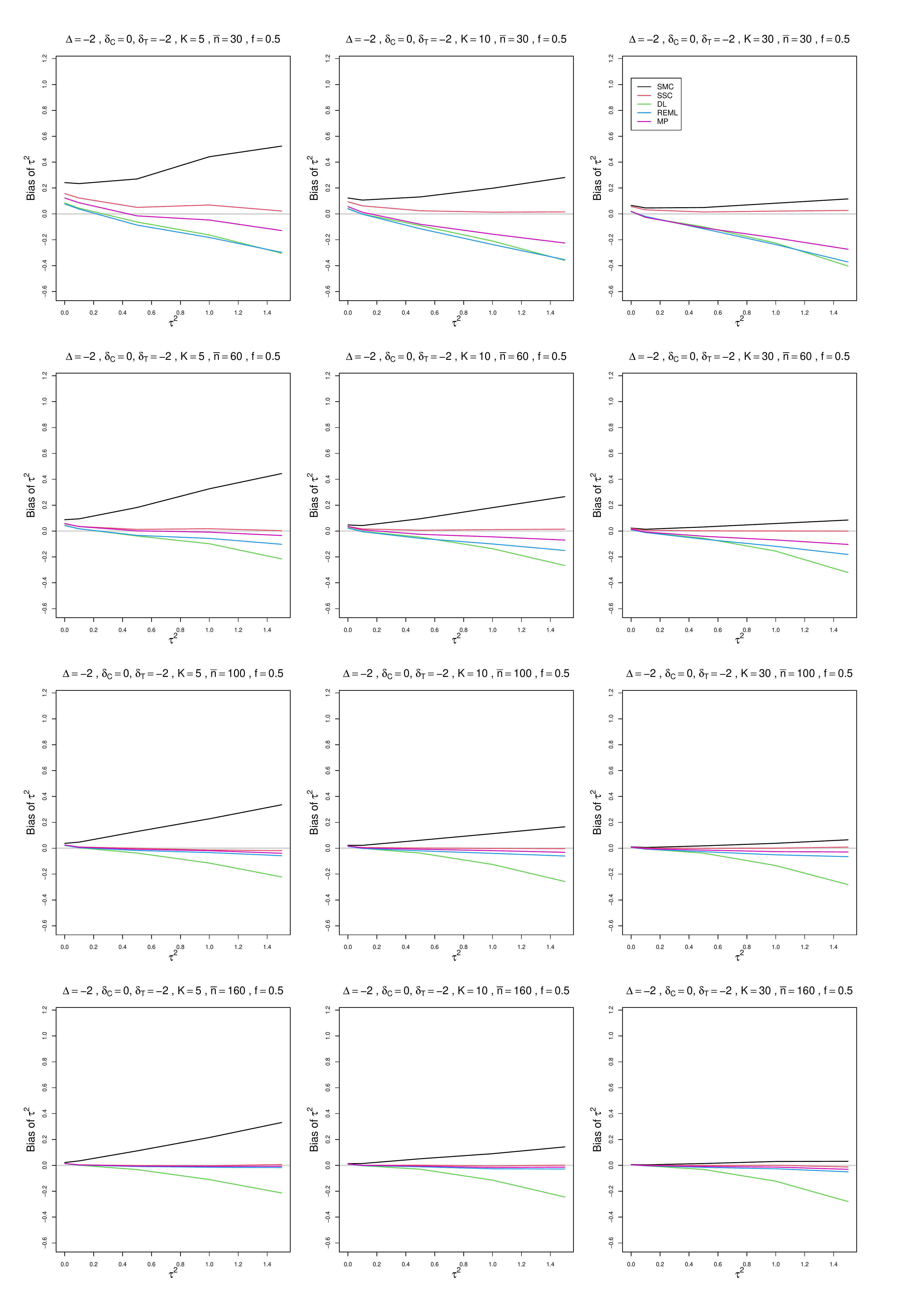}
	\caption{Bias  of estimators of between-study variance of DSM (DL, REML, MP, SMC and SSC ) vs $\tau^2$, for unequal sample sizes $\bar{n}=30,\;60,\;100$ and $160$, $\delta_{iC} = 0$, $\Delta=-2$ and  $f = 0.5$.   }
	\label{PlotBiasOfTau2_deltaC_-1deltaT=-3_DSM_unequal_sample_sizes.pdf}
\end{figure}

\begin{figure}[ht]
	\centering
	\includegraphics[scale=0.33]{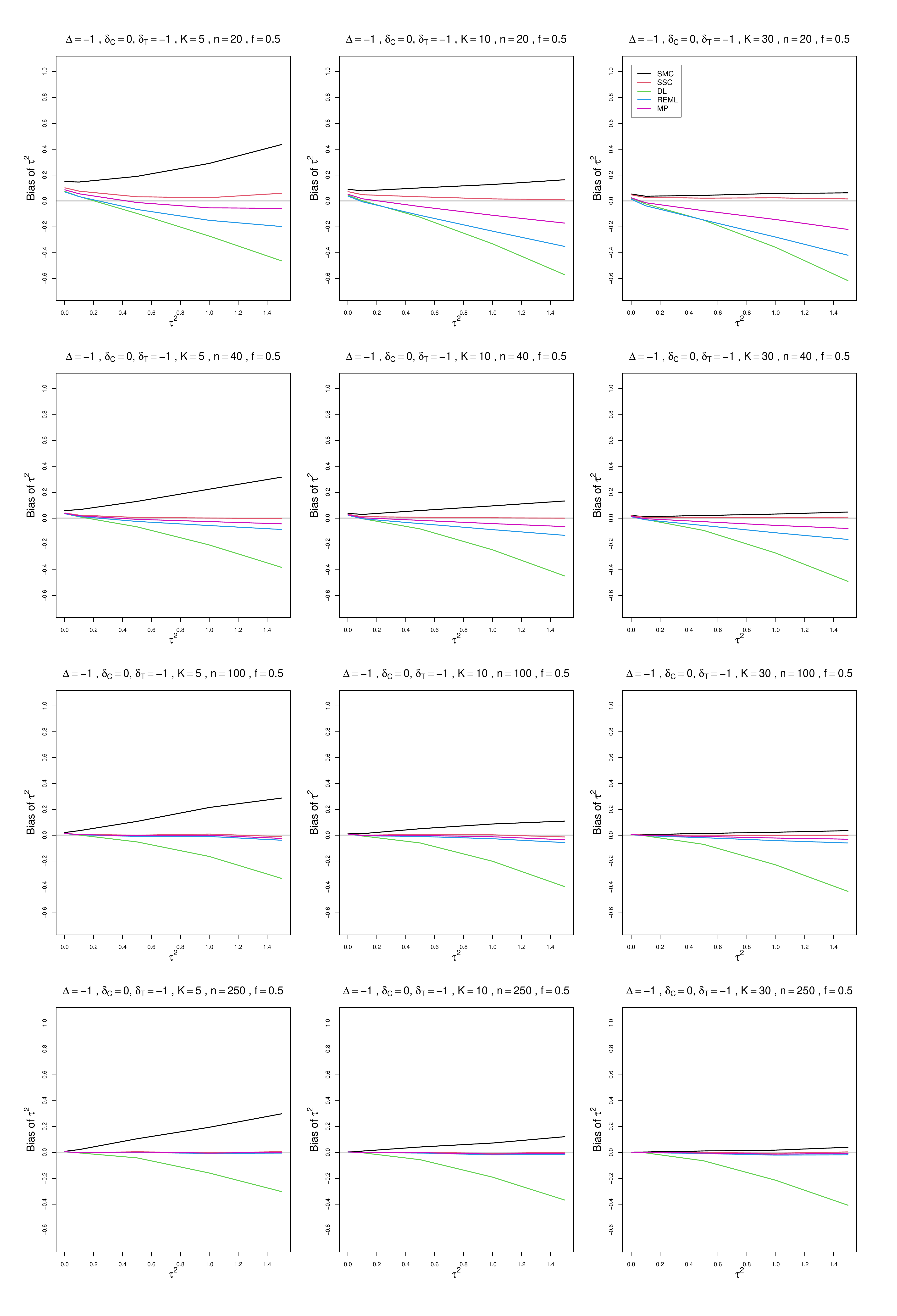}
	\caption{Bias  of estimators of between-study variance of DSM (DL, REML, MP, SMC and SSC ) vs $\tau^2$, for equal sample sizes $n=20,\;40,\;100$ and $250$, $\delta_{iC} = 0$, $\Delta=-1$ and  $f = 0.5$.   }
	\label{PlotBiasOfTau2_deltaC_-0deltaT=-1_DSM_equal_sample_sizes.pdf}
\end{figure}

\begin{figure}[ht]
	\centering
	\includegraphics[scale=0.33]{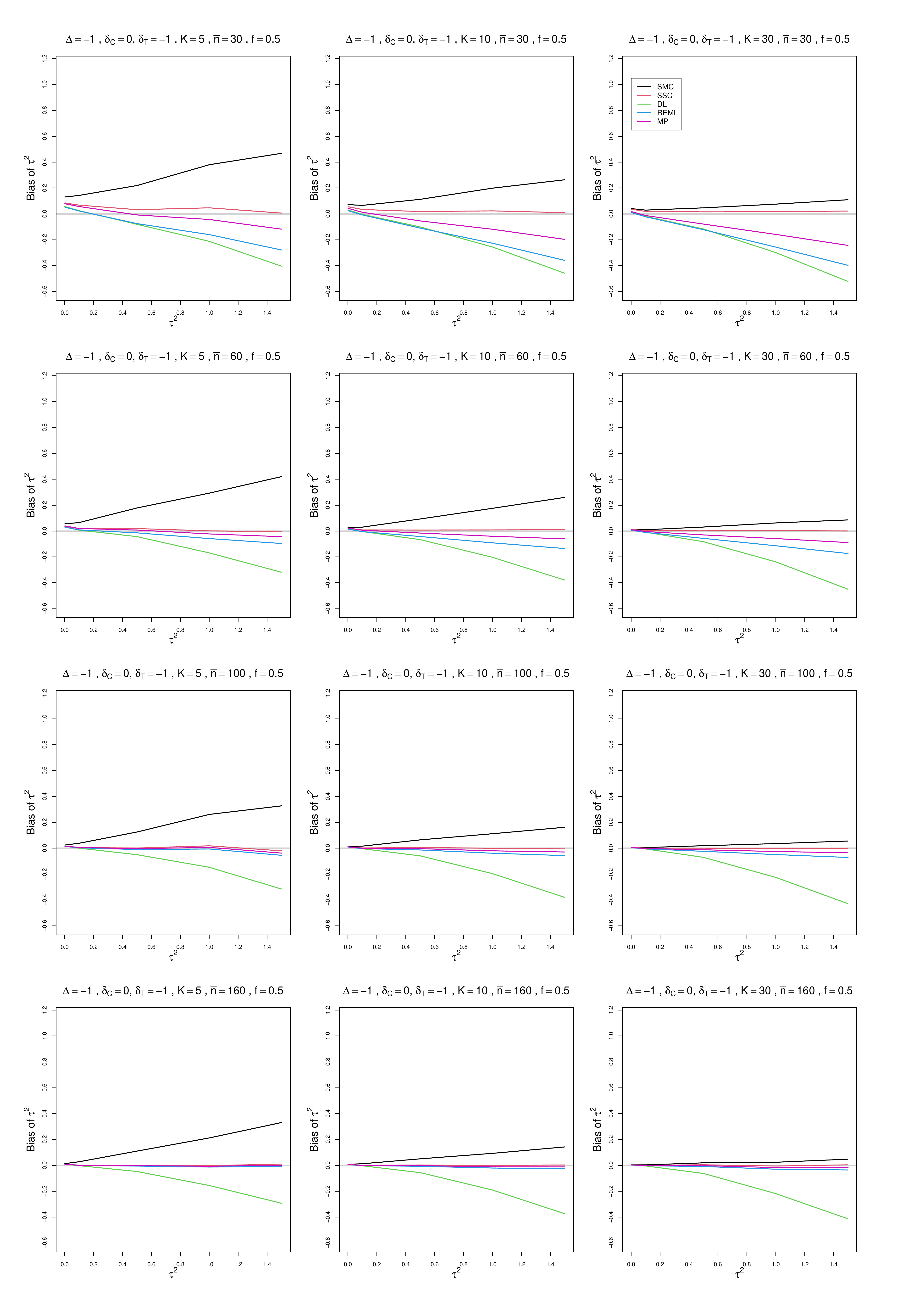}
	\caption{Bias  of estimators of between-study variance of DSM (DL, REML, MP, SMC and SSC ) vs $\tau^2$, for unequal sample sizes $\bar{n}=30,\;60,\;100$ and $160$, $\delta_{iC} = 0$, $\Delta=-1$ and  $f = 0.5$.   }
	\label{PlotBiasOfTau2_deltaC_0deltaT=-1_DSM_unequal_sample_sizes.pdf}
\end{figure}

\begin{figure}[ht]
	\centering
	\includegraphics[scale=0.33]{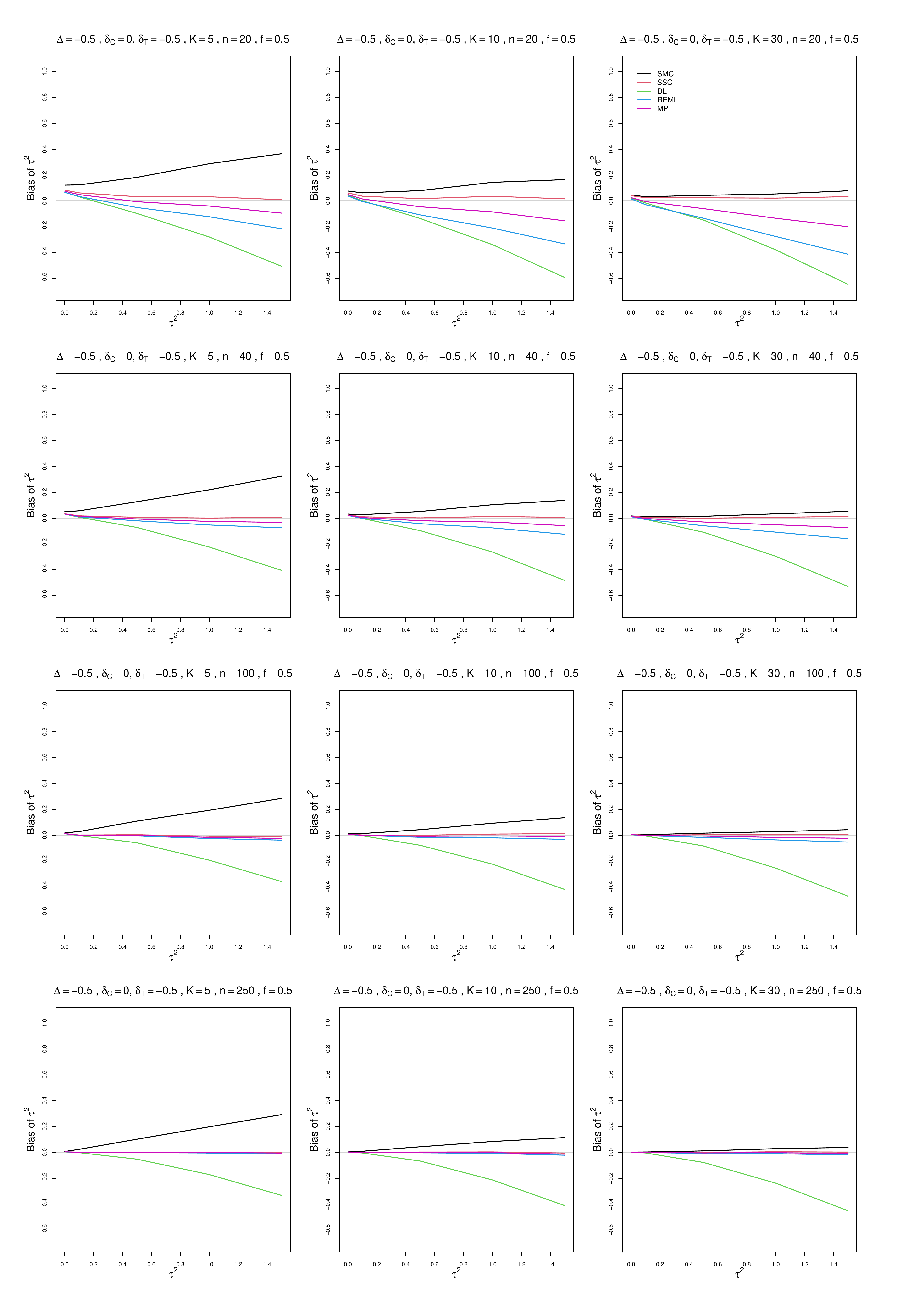}
	\caption{Bias  of estimators of between-study variance of DSM (DL, REML, MP, SMC and SSC ) vs $\tau^2$, for equal sample sizes $n=20,\;40,\;100$ and $250$, $\delta_{iC} = 0$, $\Delta=-0.5$ and  $f = 0.5$.   }
	\label{PlotBiasOfTau2_deltaC_-0deltaT=-0.5_DSM_equal_sample_sizes.pdf}
\end{figure}

\begin{figure}[ht]
	\centering
	\includegraphics[scale=0.33]{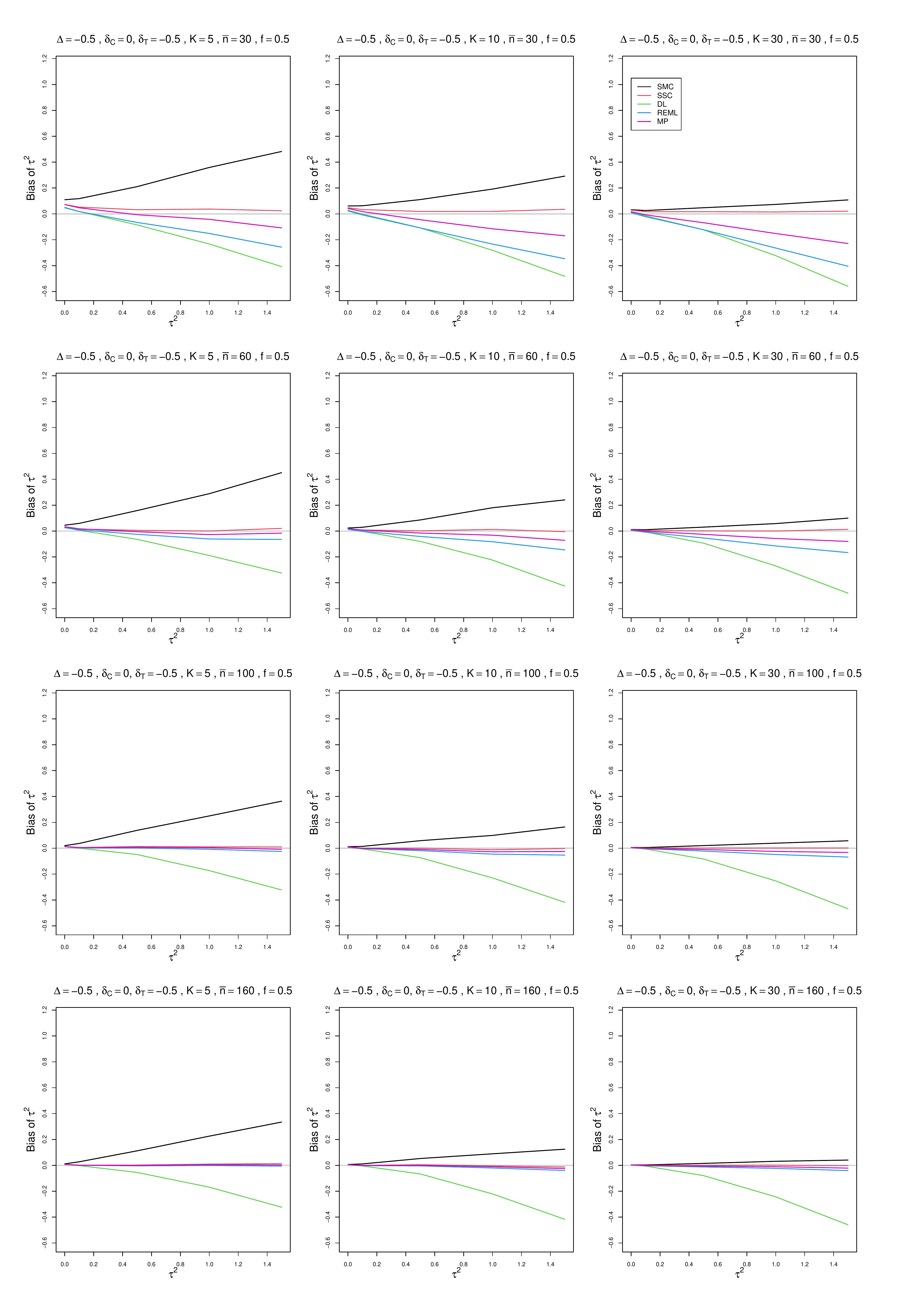}
	\caption{Bias  of estimators of between-study variance of DSM (DL, REML, MP, SMC and SSC ) vs $\tau^2$, for unequal sample sizes $\bar{n}=30,\;60,\;100$ and $160$, $\delta_{iC} = 0$, $\Delta=-0.5$ and  $f = 0.5$.   }
	\label{PlotBiasOfTau2_deltaC_0deltaT=-0.5_DSM_unequal_sample_sizes.pdf}
\end{figure}

\begin{figure}[ht]
	\centering
	\includegraphics[scale=0.33]{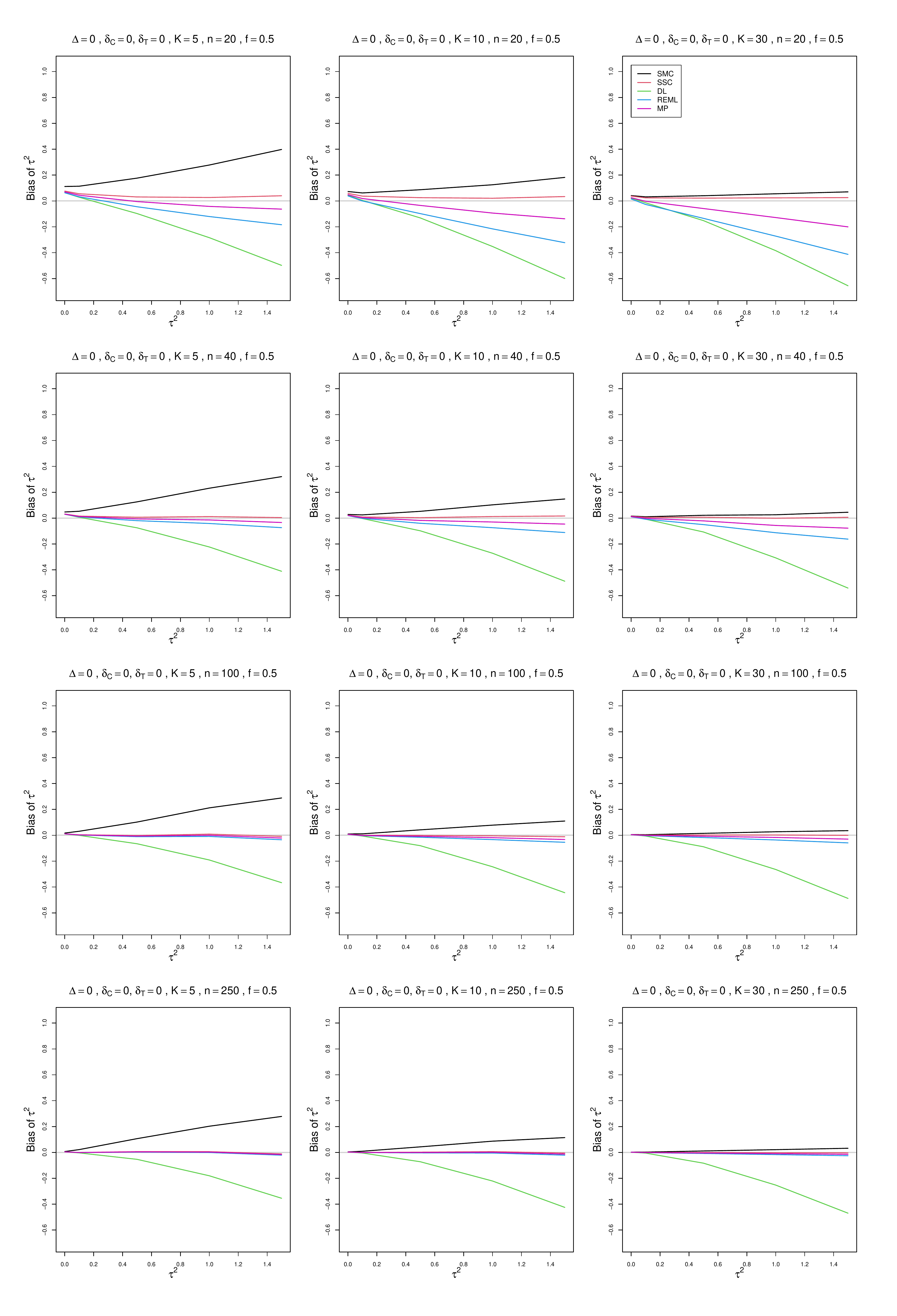}
	\caption{Bias  of estimators of between-study variance of DSM (DL, REML, MP, SMC and SSC ) vs $\tau^2$, for equal sample sizes $n=20,\;40,\;100$ and $250$, $\delta_{iC} = 0$, $\Delta=0$ and  $f = 0.5$.   }
	\label{PlotBiasOfTau2_deltaC_0deltaT=0_DSM_equal_sample_sizes.pdf}
\end{figure}

\begin{figure}[ht]
	\centering
	\includegraphics[scale=0.33]{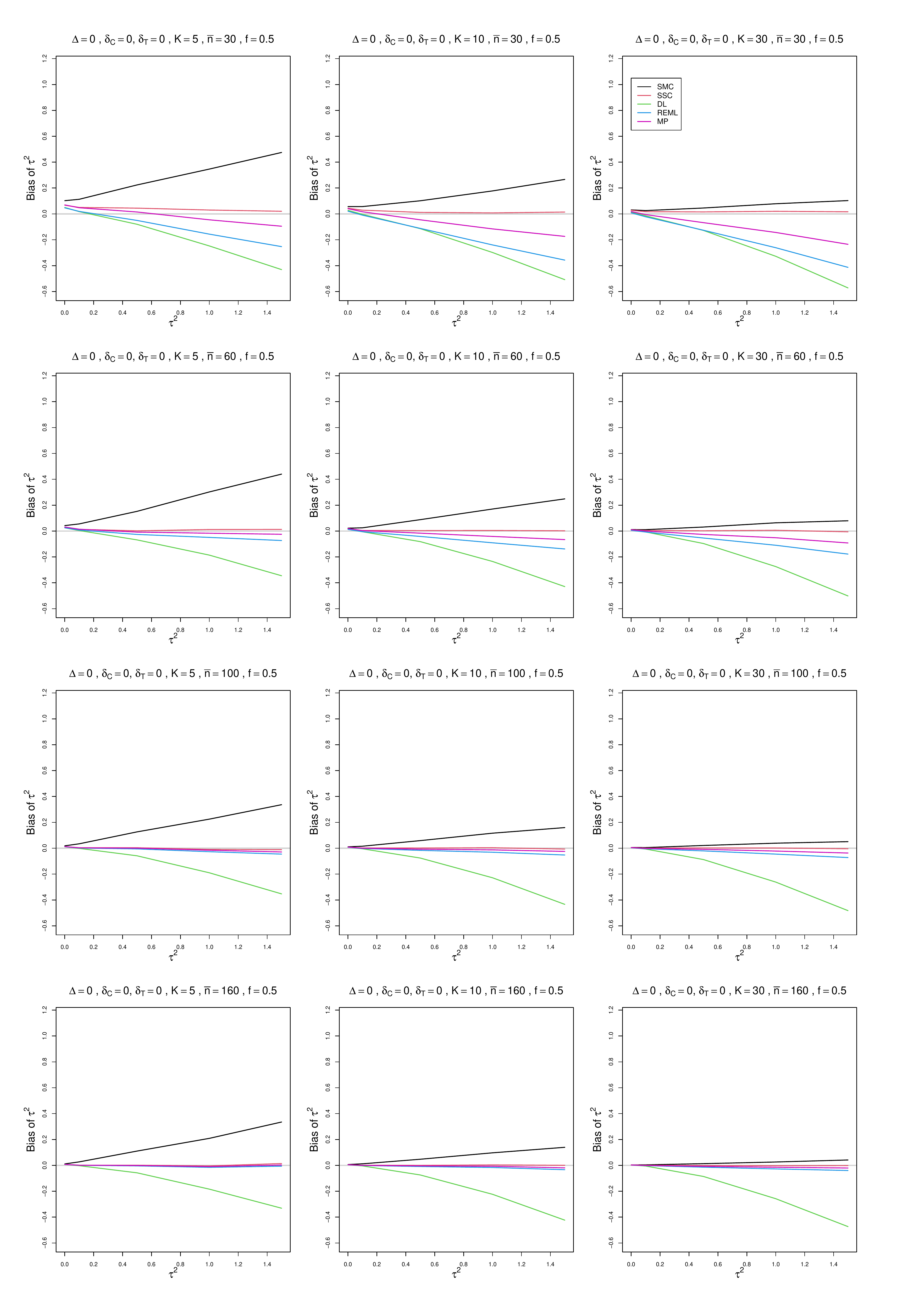}
	\caption{Bias  of estimators of between-study variance of DSM (DL, REML, MP, SMC and SSC ) vs $\tau^2$, for unequal sample sizes $\bar{n}=30,\;60,\;100$ and $160$, $\delta_{iC} = 0$, $\Delta=0$ and  $f = 0.5$.   }
	\label{PlotBiasOfTau2_deltaC_0deltaT=0_DSM_unequal_sample_sizes.pdf}
\end{figure}

\begin{figure}[ht]
	\centering
	\includegraphics[scale=0.33]{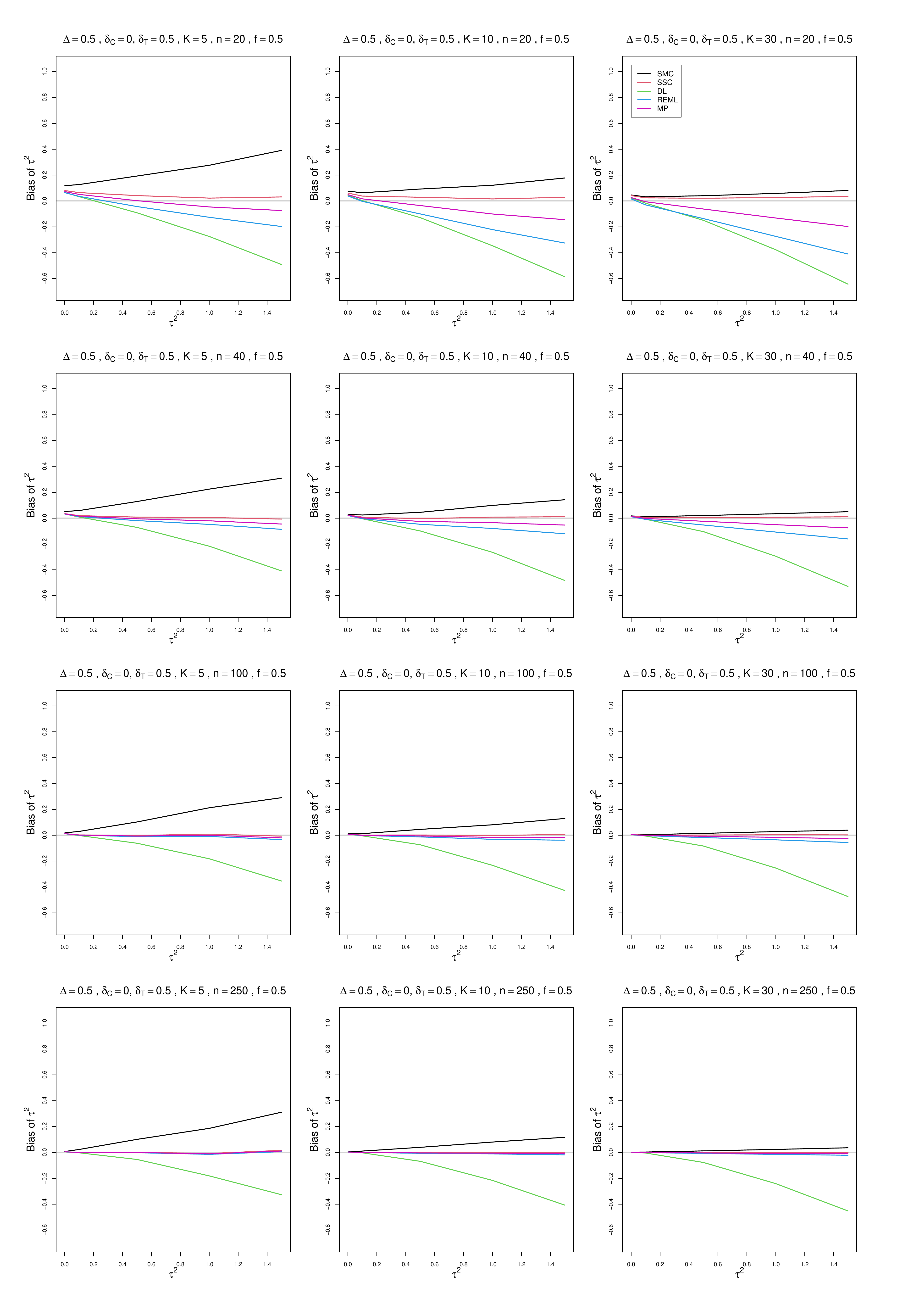}
	\caption{Bias  of estimators of between-study variance of DSM (DL, REML, MP, SMC and SSC ) vs $\tau^2$, for equal sample sizes $n=20,\;40,\;100$ and $250$, $\delta_{iC} = 0$, $\Delta=0.5$ and  $f = 0.5$.   }
	\label{PlotBiasOfTau2_deltaC_0deltaT=0.5_DSM_equal_sample_sizes.pdf}
\end{figure}

\begin{figure}[ht]
	\centering
	\includegraphics[scale=0.33]{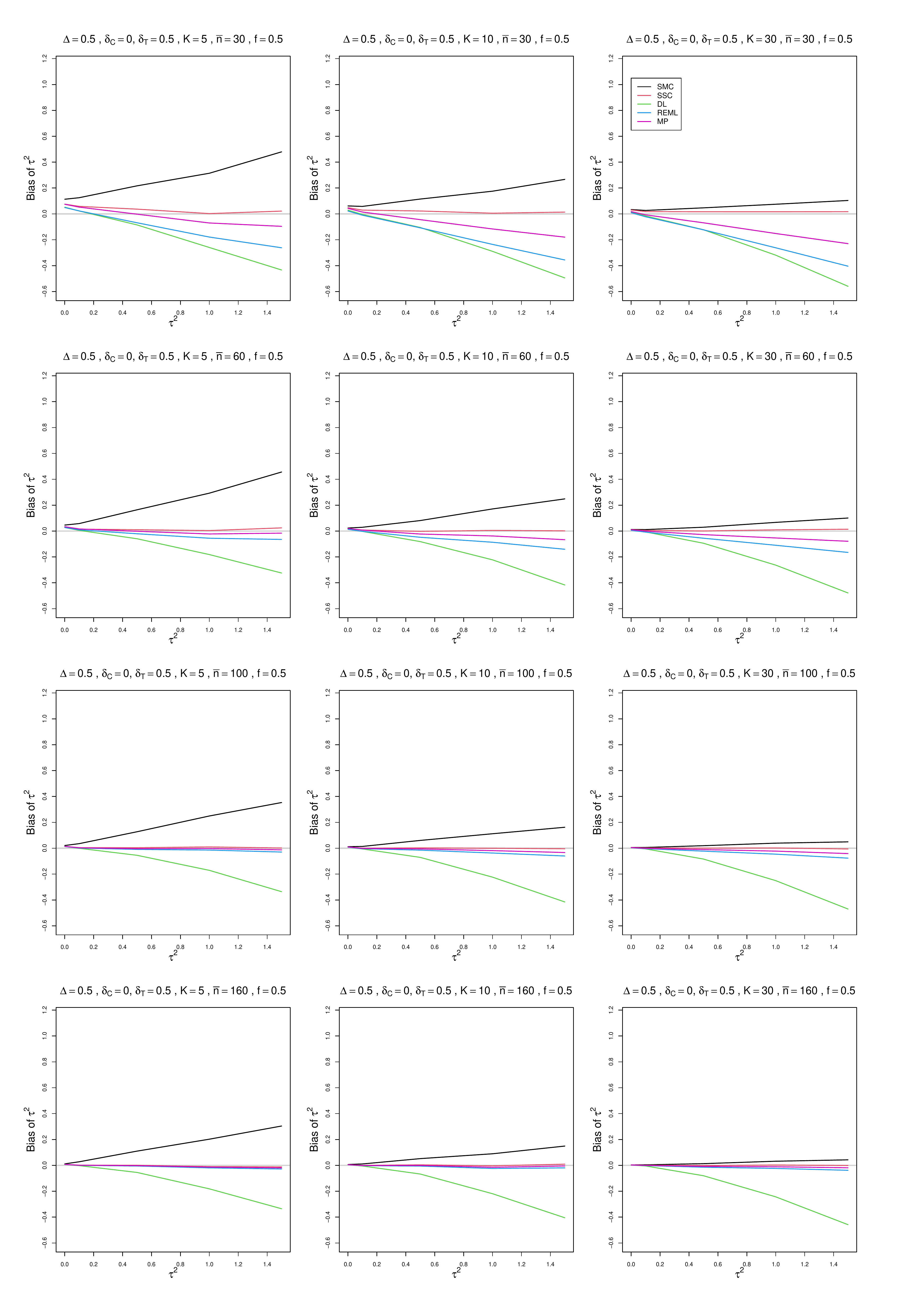}
	\caption{Bias  of estimators of between-study variance of DSM (DL, REML, MP, SMC and SSC ) vs $\tau^2$, for unequal sample sizes $\bar{n}=30,\;60,\;100$ and $160$, $\delta_{iC} = 0$, $\Delta=0.5$ and  $f = 0.5$.   }
	\label{PlotBiasOfTau2_deltaC_0deltaT=-0.5_DSM_unequal_sample_sizes.pdf}
\end{figure}

\begin{figure}[ht]
	\centering
	\includegraphics[scale=0.33]{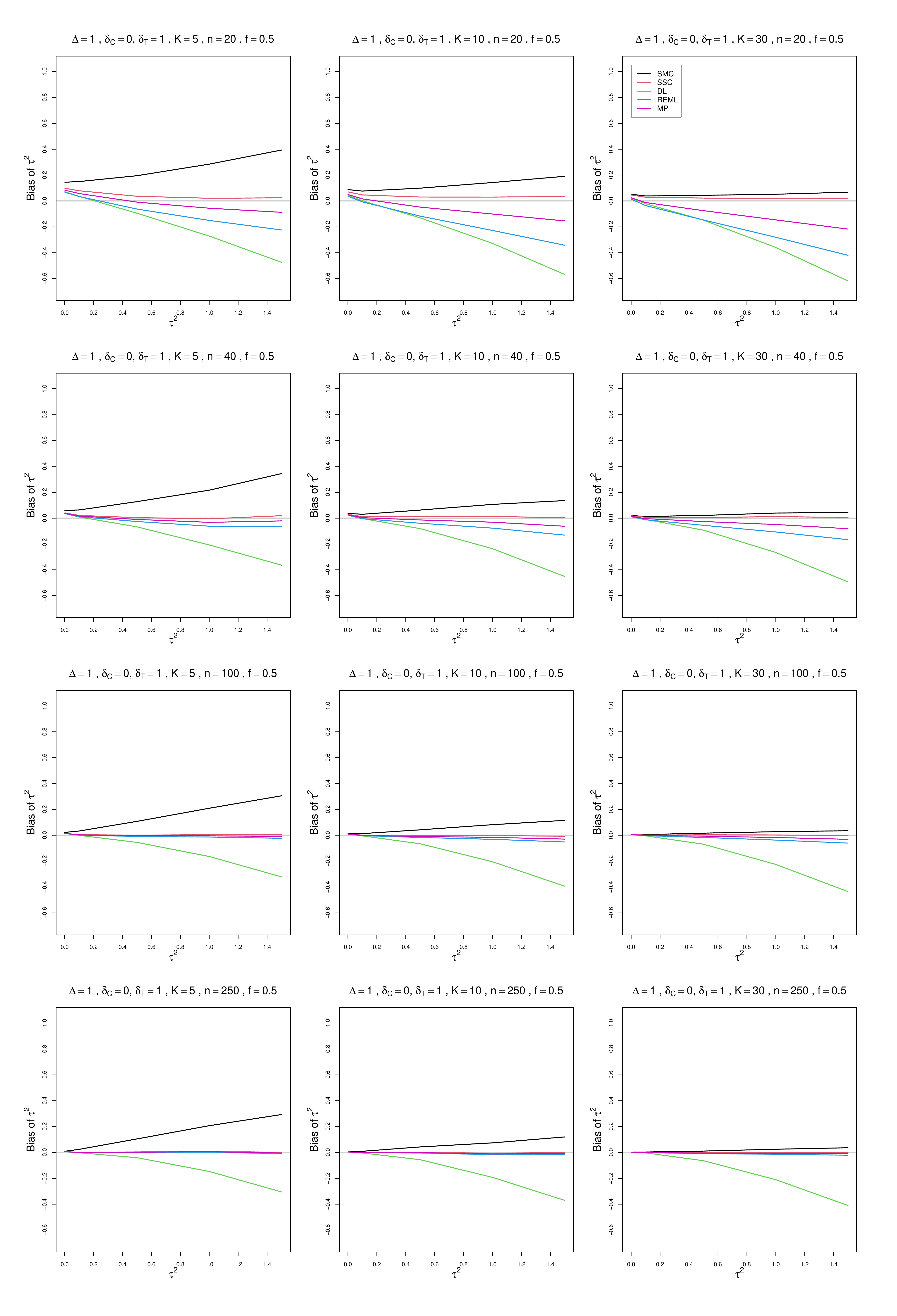}
	\caption{Bias  of estimators of between-study variance of DSM (DL, REML, MP, SMC and SSC ) vs $\tau^2$, for equal sample sizes $n=20,\;40,\;100$ and $250$, $\delta_{iC} = 0$, $\Delta=1$ and  $f = 0.5$.   }
	\label{PlotBiasOfTau2_deltaC_=0deltaT=1_DSM_equal_sample_sizes.pdf}
\end{figure}

\begin{figure}[ht]
	\centering
	\includegraphics[scale=0.33]{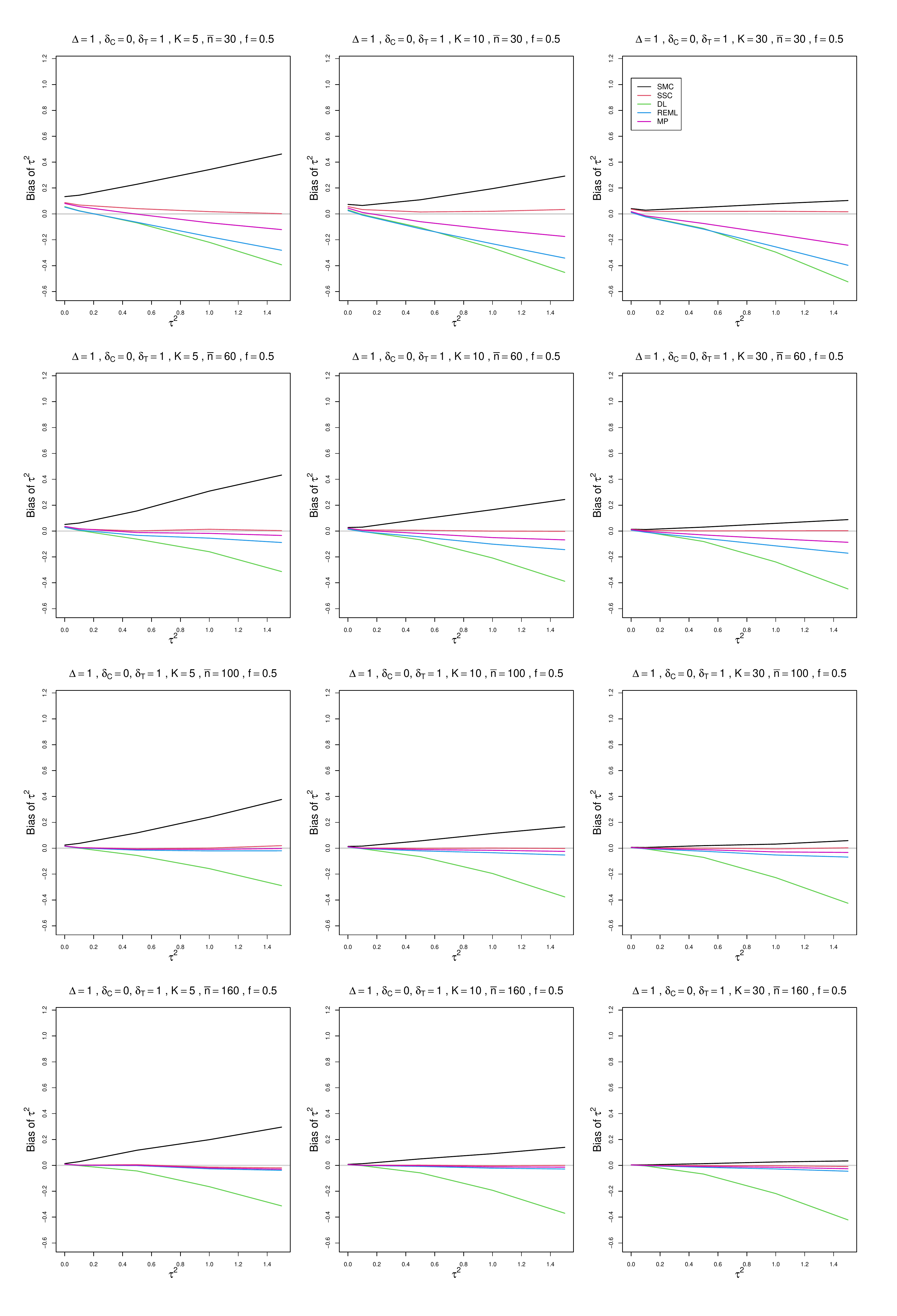}
	\caption{Bias  of estimators of between-study variance of DSM (DL, REML, MP, SMC and SSC ) vs $\tau^2$, for unequal sample sizes $\bar{n}=30,\;60,\;100$ and $160$, $\delta_{iC} = 0$, $\Delta=1$ and  $f = 0.5$.   }
	\label{PlotBiasOfTau2_deltaC_0deltaT=1_DSM_unequal_sample_sizes.pdf}
\end{figure}

\begin{figure}[ht]
	\centering
	\includegraphics[scale=0.33]{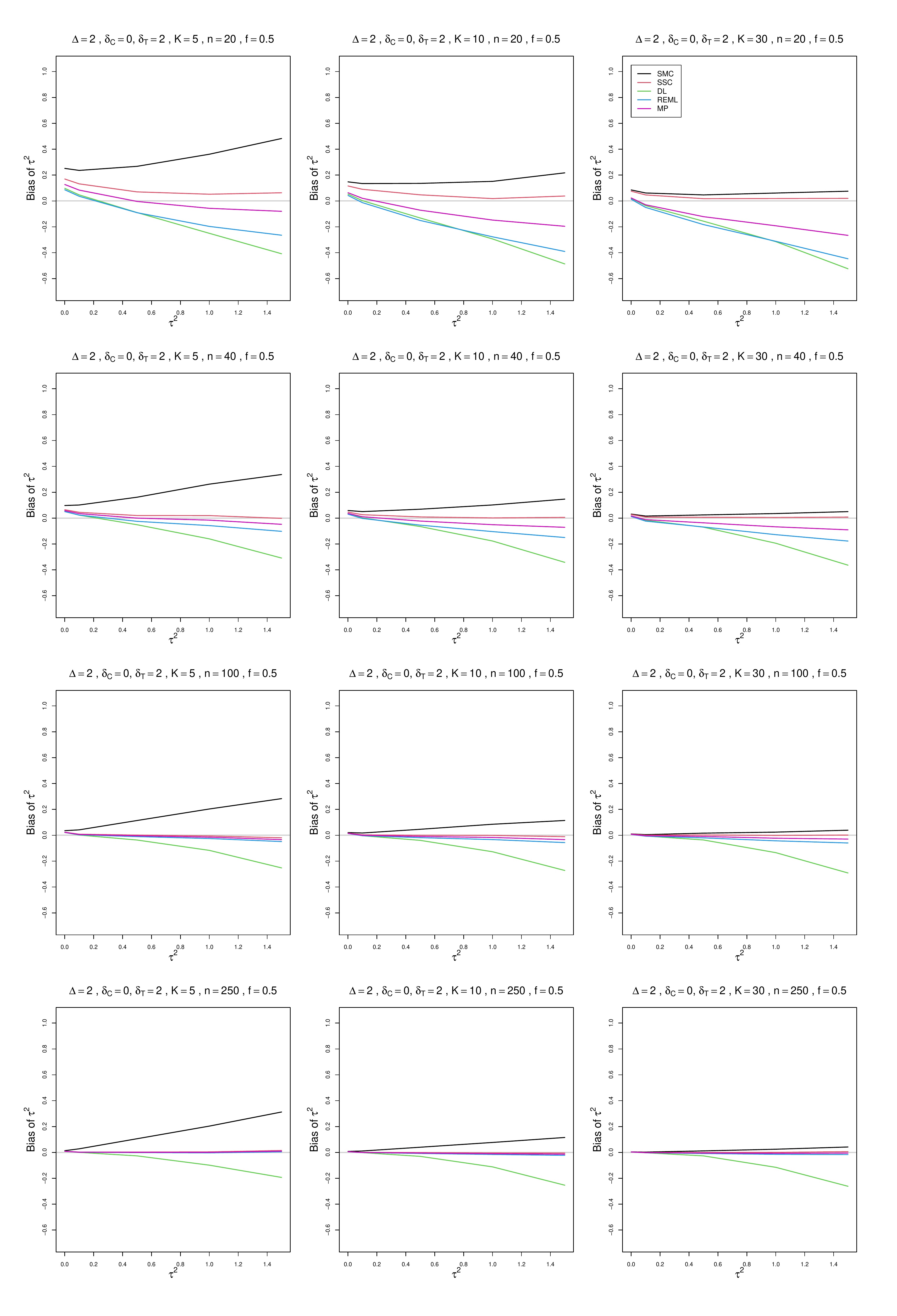}
	\caption{Bias  of estimators of between-study variance of DSM (DL, REML, MP, SMC and SSC ) vs $\tau^2$, for equal sample sizes $n=20,\;40,\;100$ and $250$, $\delta_{iC} = 0$, $\Delta=2$ and  $f = 0.5$.   }
	\label{PlotBiasOfTau2_deltaC_0deltaT=2_DSM_equal_sample_sizes.pdf}
\end{figure}

\begin{figure}[ht]
	\centering
	\includegraphics[scale=0.33]{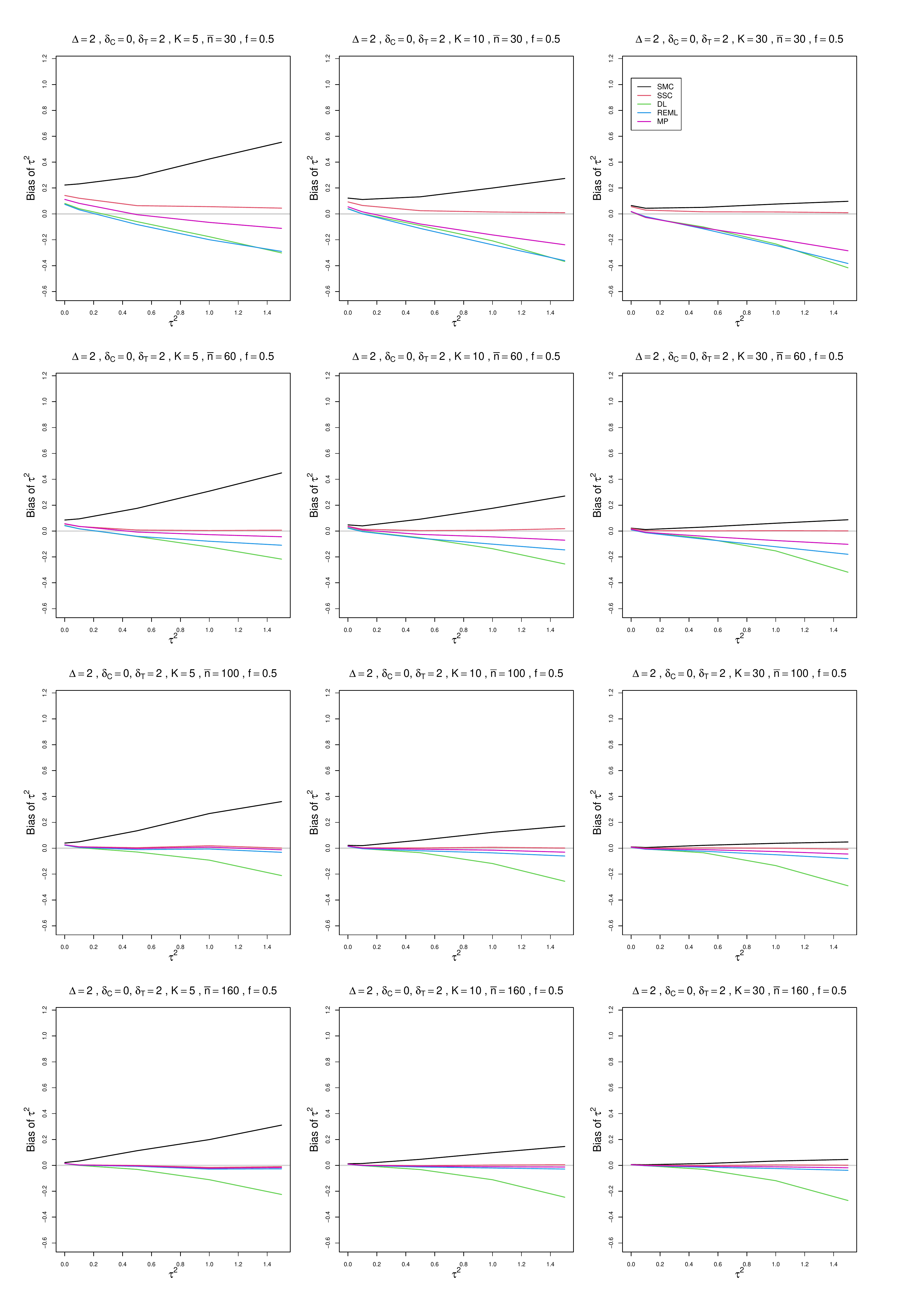}
	\caption{Bias  of estimators of between-study variance of DSM (DL, REML, MP, SMC and SSC ) vs $\tau^2$, for unequal sample sizes $\bar{n}=30,\;60,\;100$ and $160$, $\delta_{iC} = 0$, $\Delta=2$ and  $f = 0.5$.   }
	\label{PlotBiasOfTau2_deltaC_0deltaT=2_DSM_unequal_sample_sizes.pdf}
\end{figure}
\clearpage

\begin{figure}[ht]
	\centering
	\includegraphics[scale=0.33]{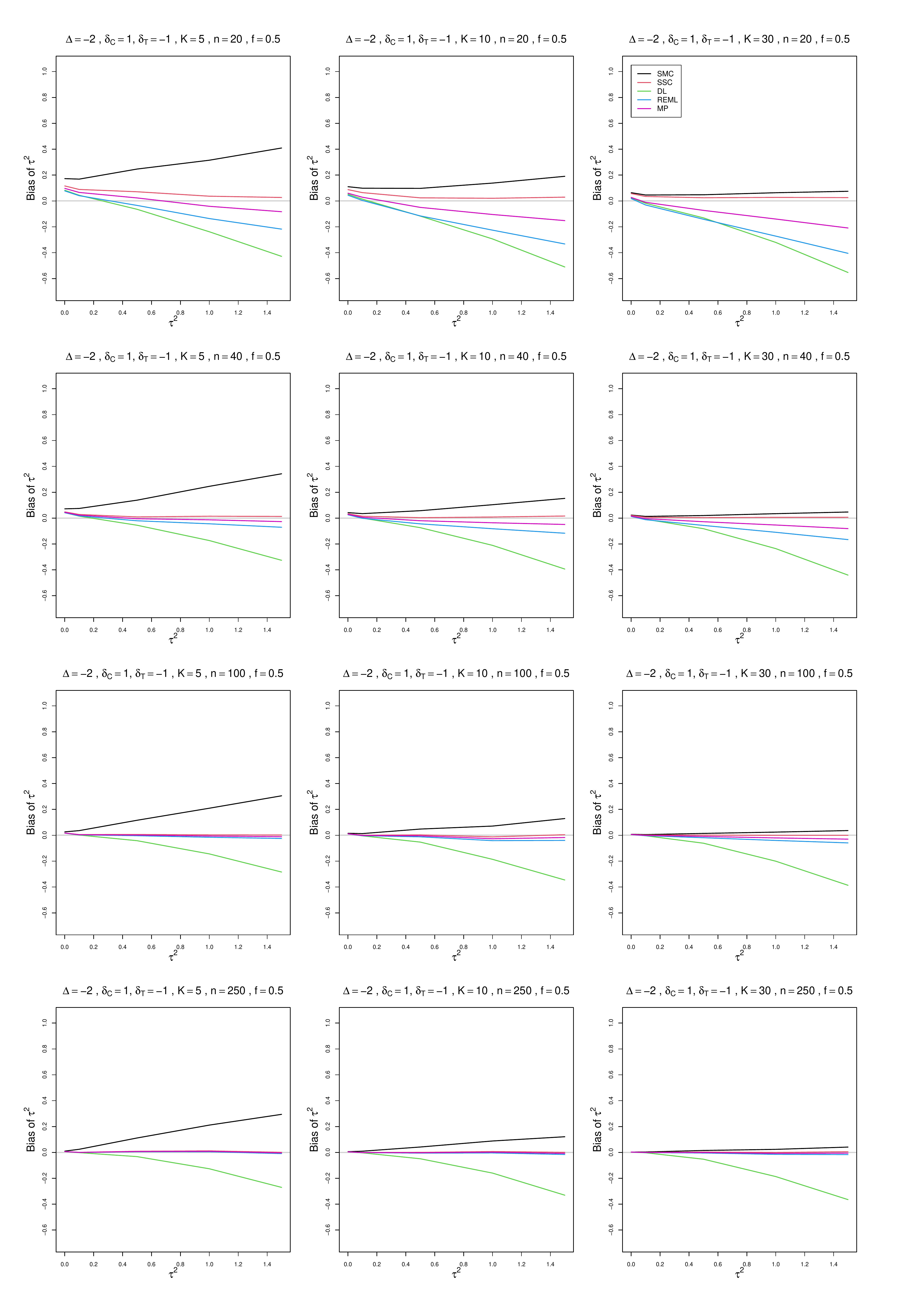}
	\caption{Bias  of estimators of between-study variance of DSM (DL, REML, MP, SMC and SSC ) vs $\tau^2$, for equal sample sizes $n=20,\;40,\;100$ and $250$, $\delta_{iC} = -1$, $\Delta=-2$ and  $f = 0.5$.   }
	\label{PlotBiasOfTau2_deltaC_1deltaT=-1_DSM_equal_sample_sizes.pdf}
\end{figure}

\begin{figure}[ht]
	\centering
	\includegraphics[scale=0.33]{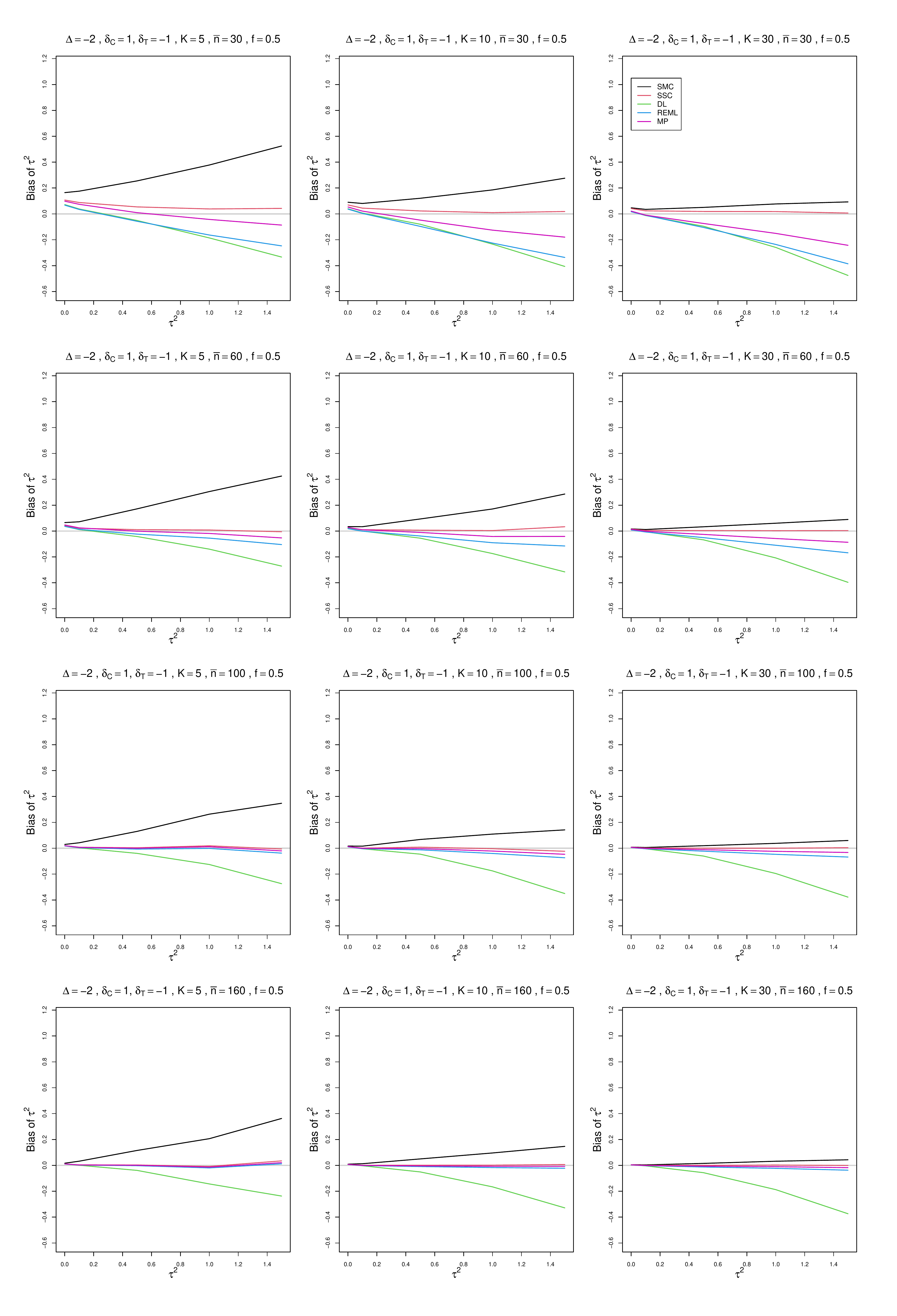}
	\caption{Bias  of estimators of between-study variance of DSM (DL, REML, MP, SMC and SSC ) vs $\tau^2$, for unequal sample sizes $\bar{n}=30,\;60,\;100$ and $160$, $\delta_{iC} = 1$, $\Delta=-2$ and  $f = 0.5$.   }
	\label{PlotBiasOfTau2_deltaC_1deltaT=-1_DSM_unequal_sample_sizes.pdf}
\end{figure}

\begin{figure}[ht]
	\centering
	\includegraphics[scale=0.33]{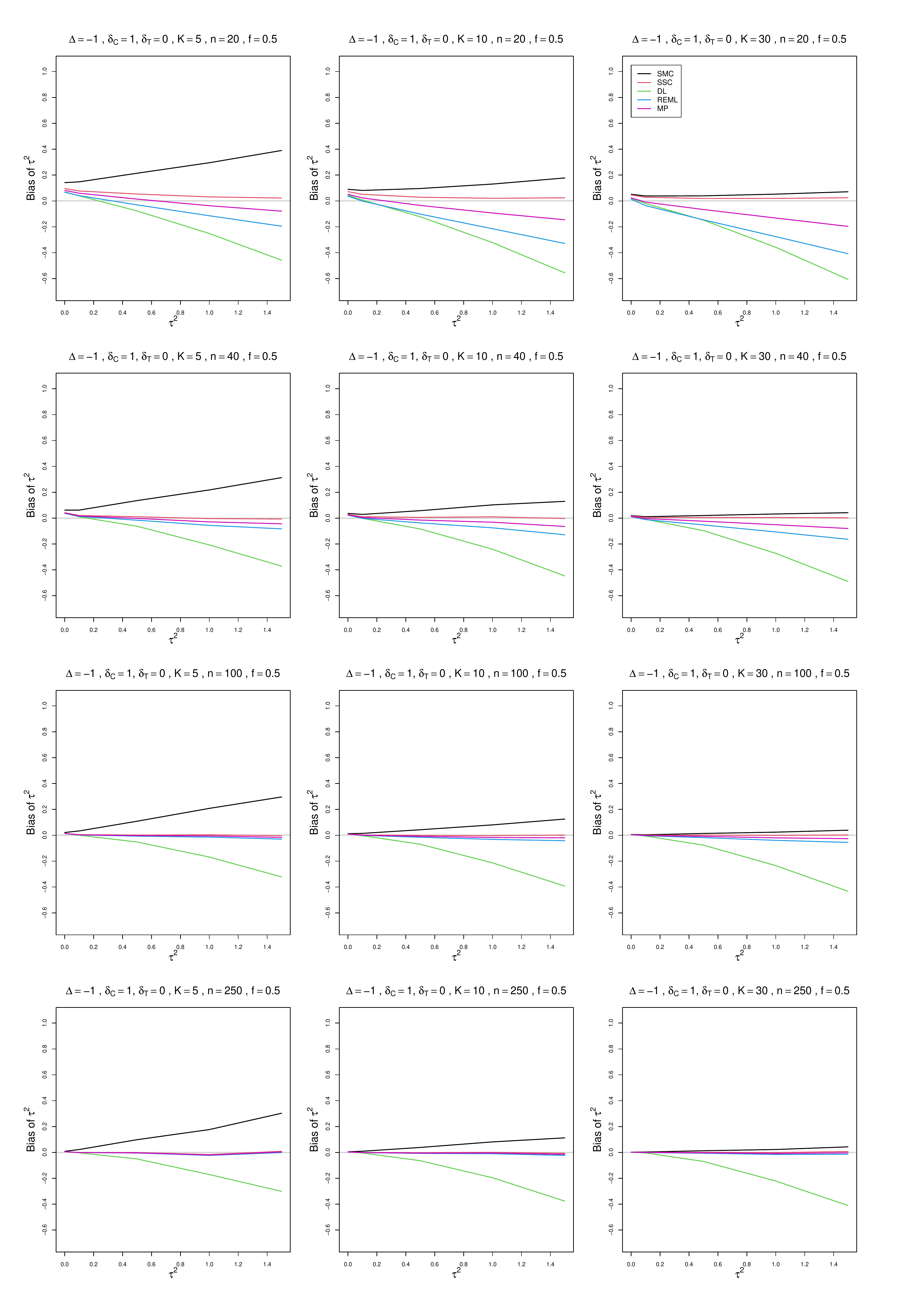}
	\caption{Bias  of estimators of between-study variance of DSM (DL, REML, MP, SMC and SSC ) vs $\tau^2$, for equal sample sizes $n=20,\;40,\;100$ and $250$, $\delta_{iC} = 1$, $\Delta=-1$ and  $f = 0.5$.   }
	\label{PlotBiasOfTau2_deltaC_1deltaT=0_DSM_equal_sample_sizes.pdf}
\end{figure}

\begin{figure}[ht]
	\centering
	\includegraphics[scale=0.33]{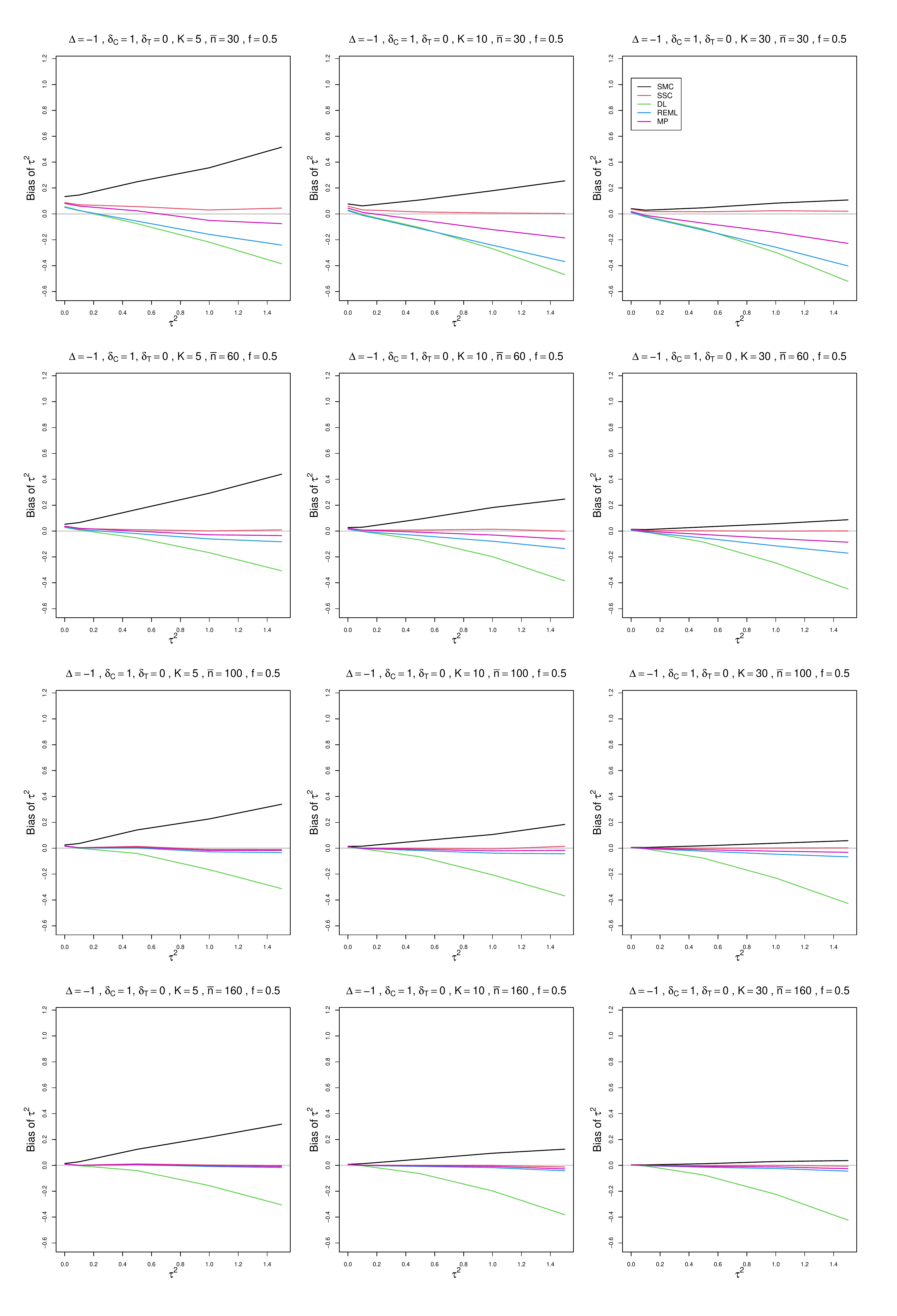}
	\caption{Bias  of estimators of between-study variance of DSM (DL, REML, MP, SMC and SSC ) vs $\tau^2$, for unequal sample sizes $\bar{n}=30,\;60,\;100$ and $160$, $\delta_{iC} = 1$, $\Delta=-1$ and  $f = 0.5$.   }
	\label{PlotBiasOfTau2_deltaC_1deltaT=0_DSM_unequal_sample_sizes.pdf}
\end{figure}

\begin{figure}[ht]
	\centering
	\includegraphics[scale=0.33]{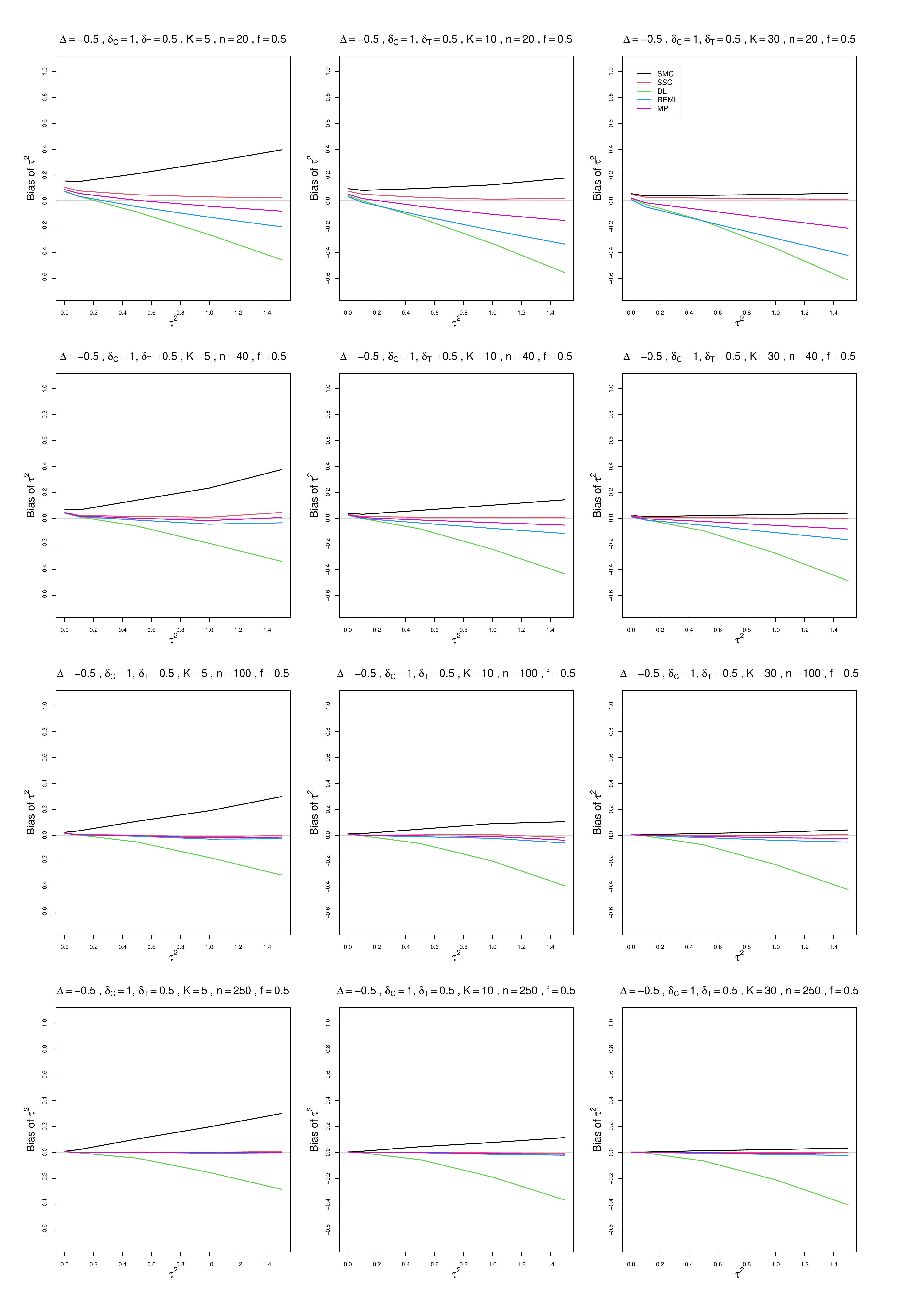}
	\caption{Bias  of estimators of between-study variance of DSM (DL, REML, MP, SMC and SSC ) vs $\tau^2$, for equal sample sizes $n=20,\;40,\;100$ and $250$, $\delta_{iC} = 1$, $\Delta=-0.5$ and  $f = 0.5$.   }
	\label{PlotBiasOfTau2_deltaC_1deltaT=0.5_DSM_equal_sample_sizes.pdf}
\end{figure}

\begin{figure}[ht]
	\centering
	\includegraphics[scale=0.33]{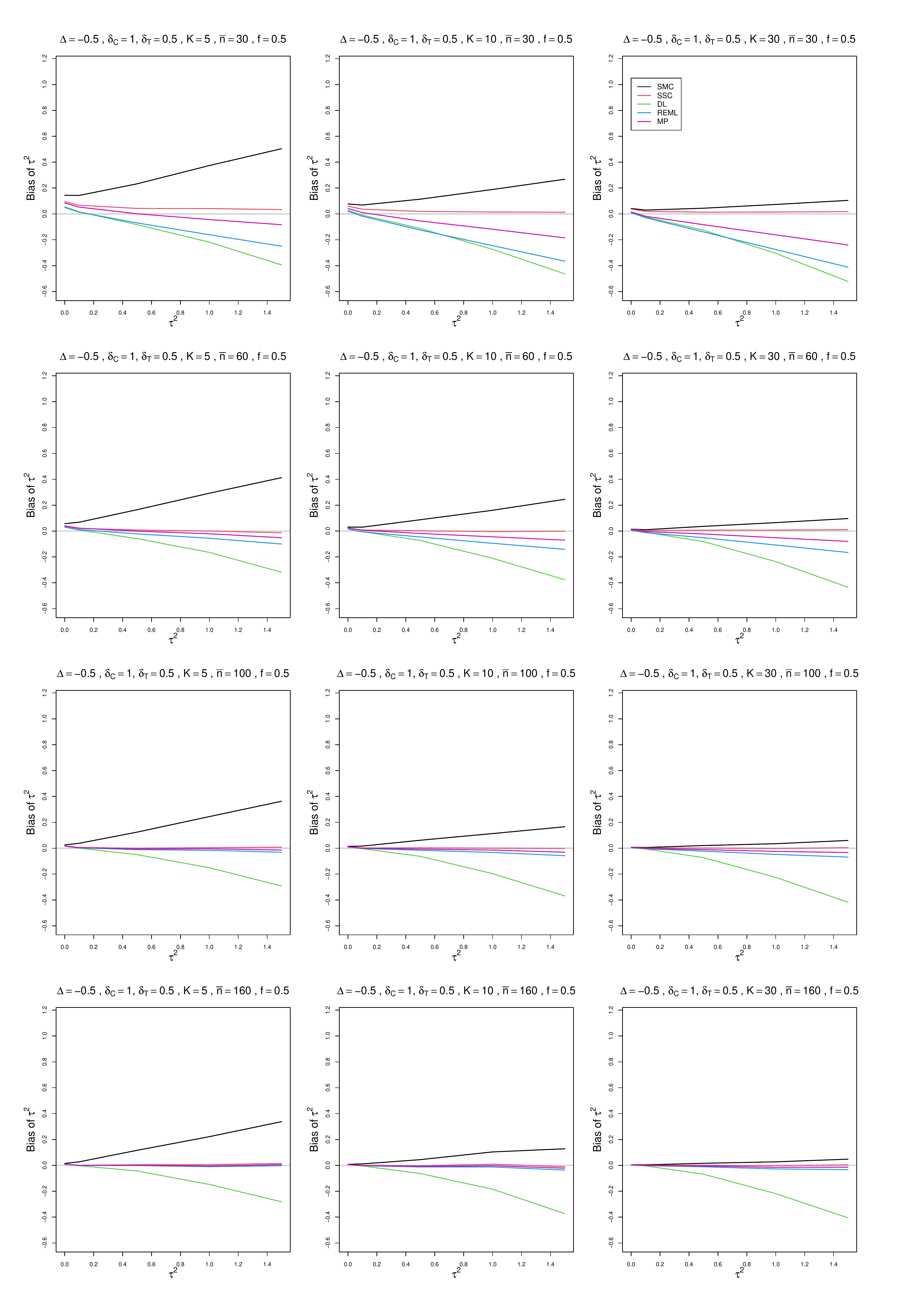}
	\caption{Bias  of estimators of between-study variance of DSM (DL, REML, MP, SMC and SSC ) vs $\tau^2$, for unequal sample sizes $\bar{n}=30,\;60,\;100$ and $160$, $\delta_{iC} = 1$, $\Delta=-0.5$ and  $f = 0.5$.   }
	\label{PlotBiasOfTau2_deltaC_1deltaT=0.5_DSM_unequal_sample_sizes.pdf}
\end{figure}

\begin{figure}[ht]
	\centering
	\includegraphics[scale=0.33]{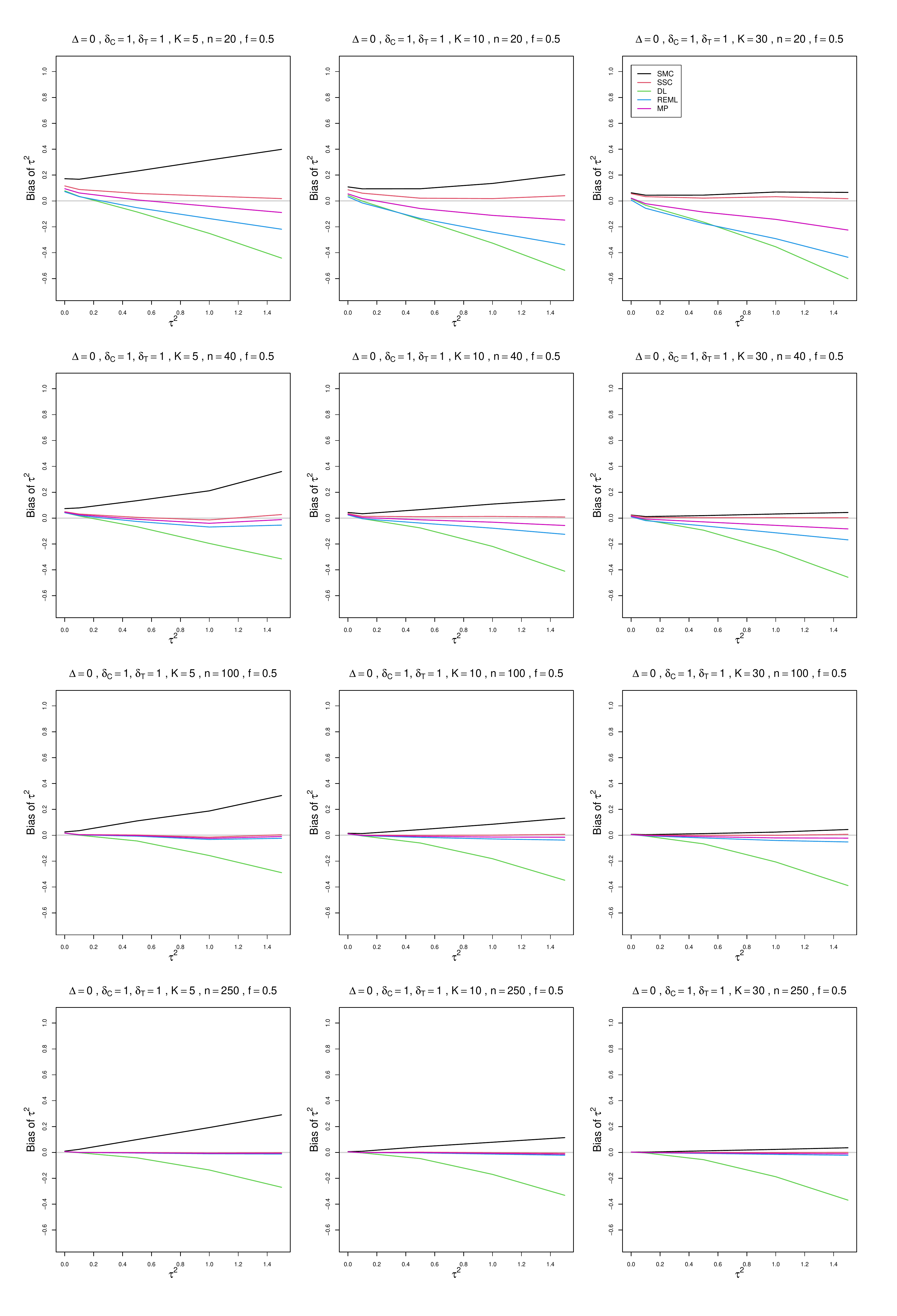}
	\caption{Bias  of estimators of between-study variance of DSM (DL, REML, MP, SMC and SSC ) vs $\tau^2$, for equal sample sizes $n=20,\;40,\;100$ and $250$, $\delta_{iC} = 1$, $\Delta=0$ and  $f = 0.5$.   }
	\label{PlotBiasOfTau2_deltaC_1deltaT=1_DSM_equal_sample_sizes.pdf}
\end{figure}

\begin{figure}[ht]
	\centering
	\includegraphics[scale=0.33]{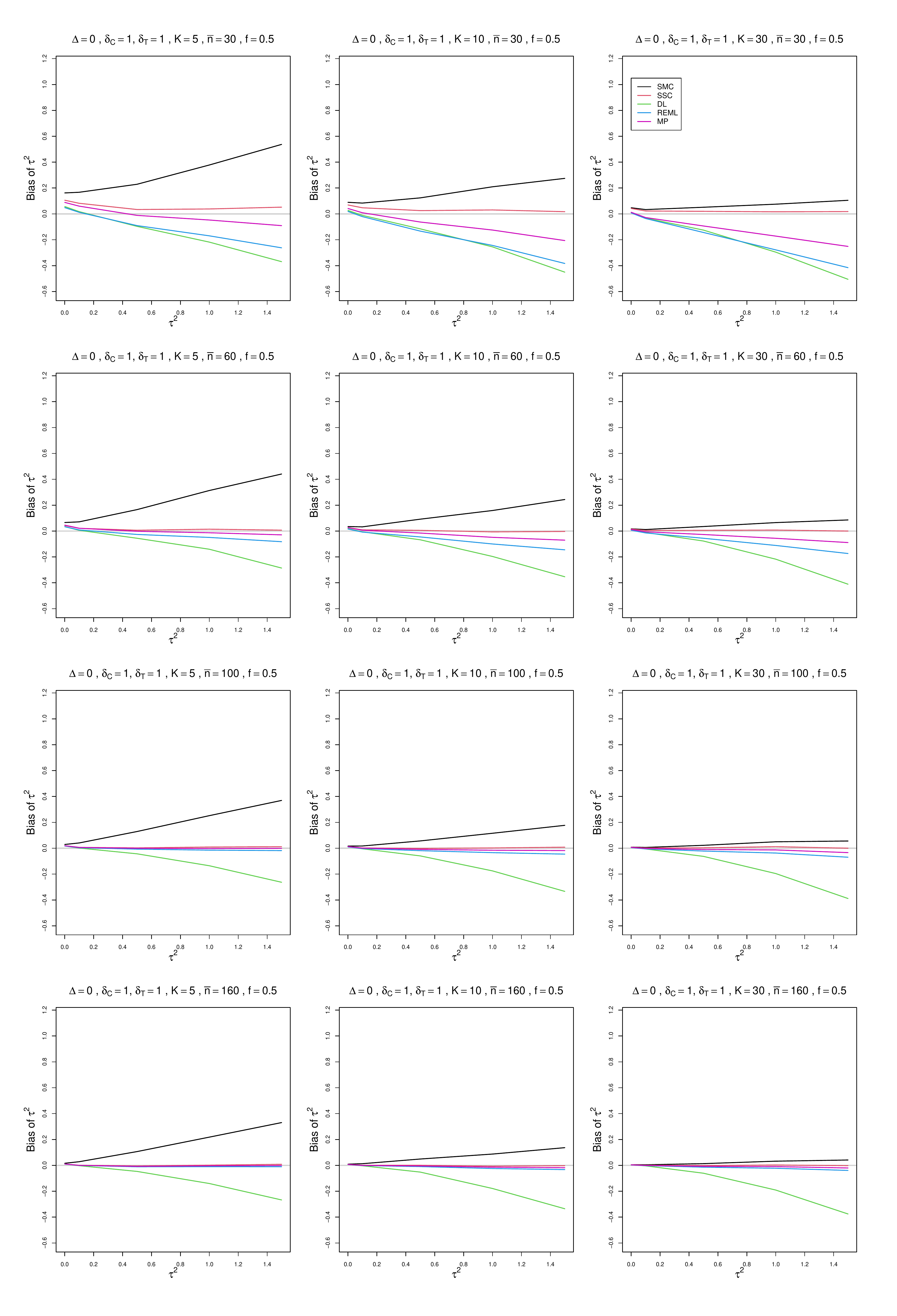}
	\caption{Bias  of estimators of between-study variance of DSM (DL, REML, MP, SMC and SSC ) vs $\tau^2$, for unequal sample sizes $\bar{n}=30,\;60,\;100$ and $160$, $\delta_{iC} = 1$, $\Delta=0$ and  $f = 0.5$.   }
	\label{PlotBiasOfTau2_deltaC_1deltaT=1_DSM_unequal_sample_sizes.pdf}
\end{figure}

\begin{figure}[ht]
	\centering
	\includegraphics[scale=0.33]{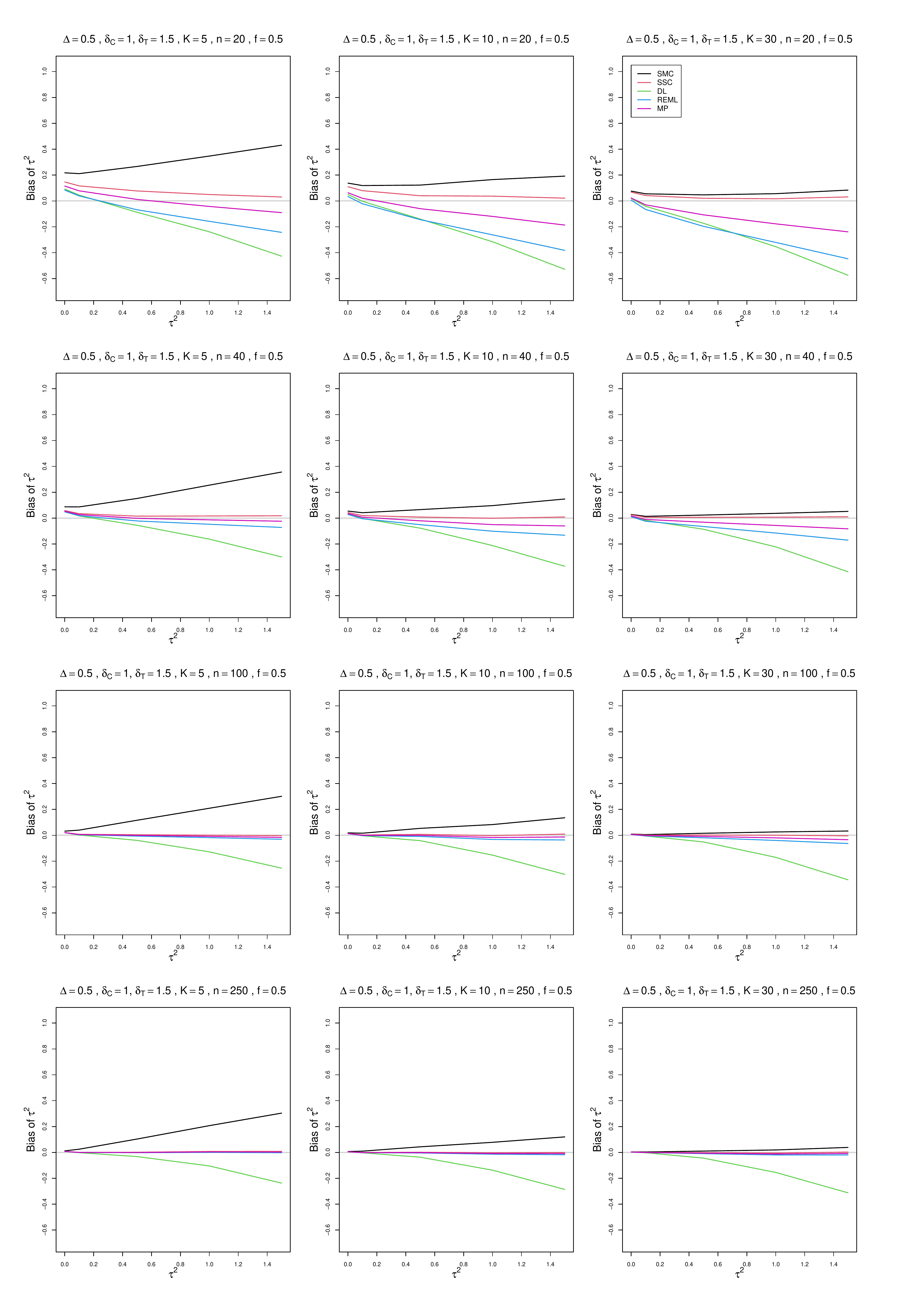}
	\caption{Bias  of estimators of between-study variance of DSM (DL, REML, MP, SMC and SSC ) vs $\tau^2$, for equal sample sizes $n=20,\;40,\;100$ and $250$, $\delta_{iC} = 1$, $\Delta=0.5$ and  $f = 0.5$.   }
	\label{PlotBiasOfTau2_deltaC_1deltaT=1.5_DSM_equal_sample_sizes.pdf}
\end{figure}

\begin{figure}[ht]
	\centering
	\includegraphics[scale=0.33]{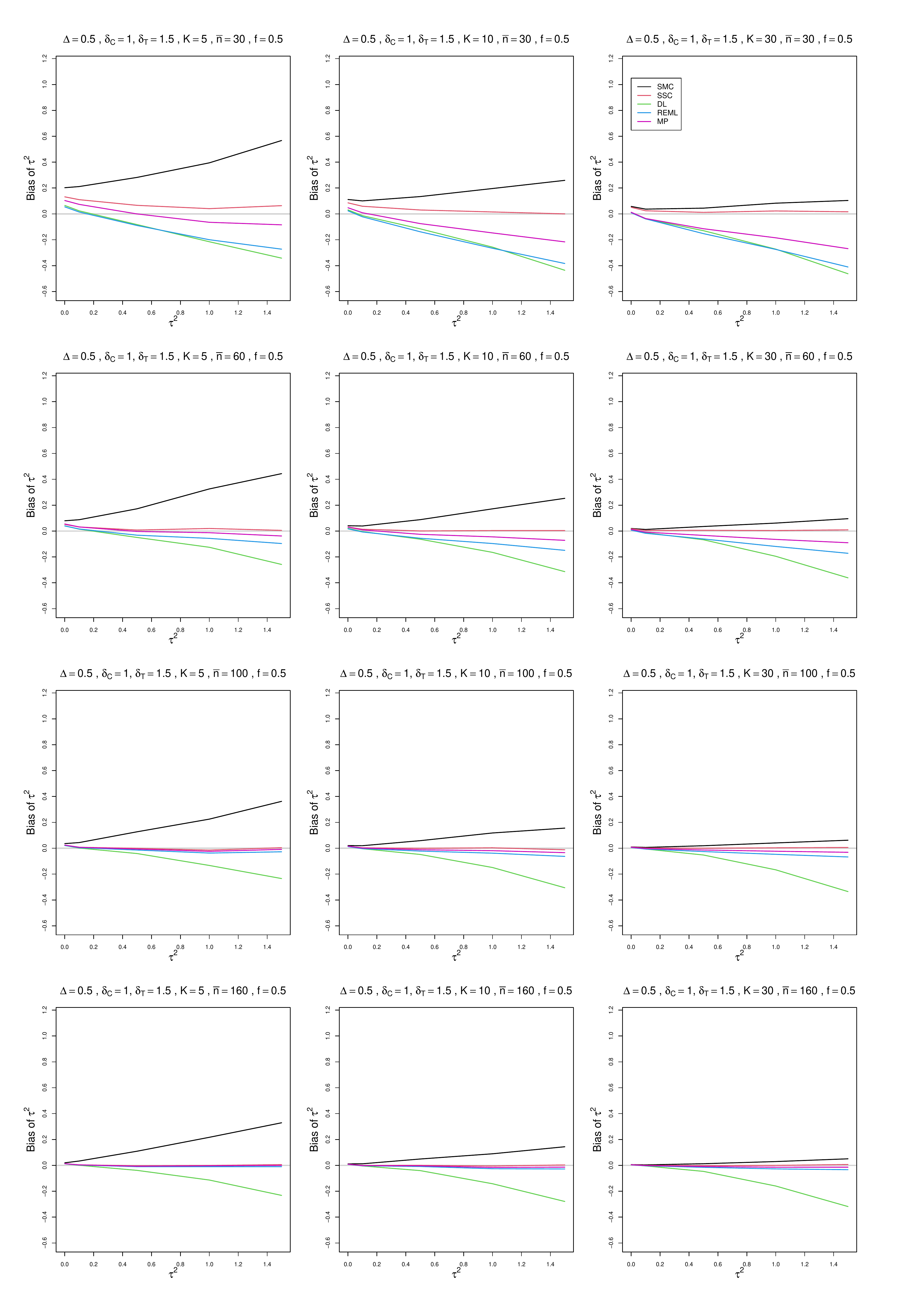}
	\caption{Bias  of estimators of between-study variance of DSM (DL, REML, MP, SMC and SSC ) vs $\tau^2$, for unequal sample sizes $\bar{n}=30,\;60,\;100$ and $160$, $\delta_{iC} = 1$, $\Delta=0.5$ and  $f = 0.5$.   }
	\label{PlotBiasOfTau2_deltaC_1deltaT=1.5_DSM_unequal_sample_sizes.pdf}
\end{figure}

\begin{figure}[ht]
	\centering
	\includegraphics[scale=0.33]{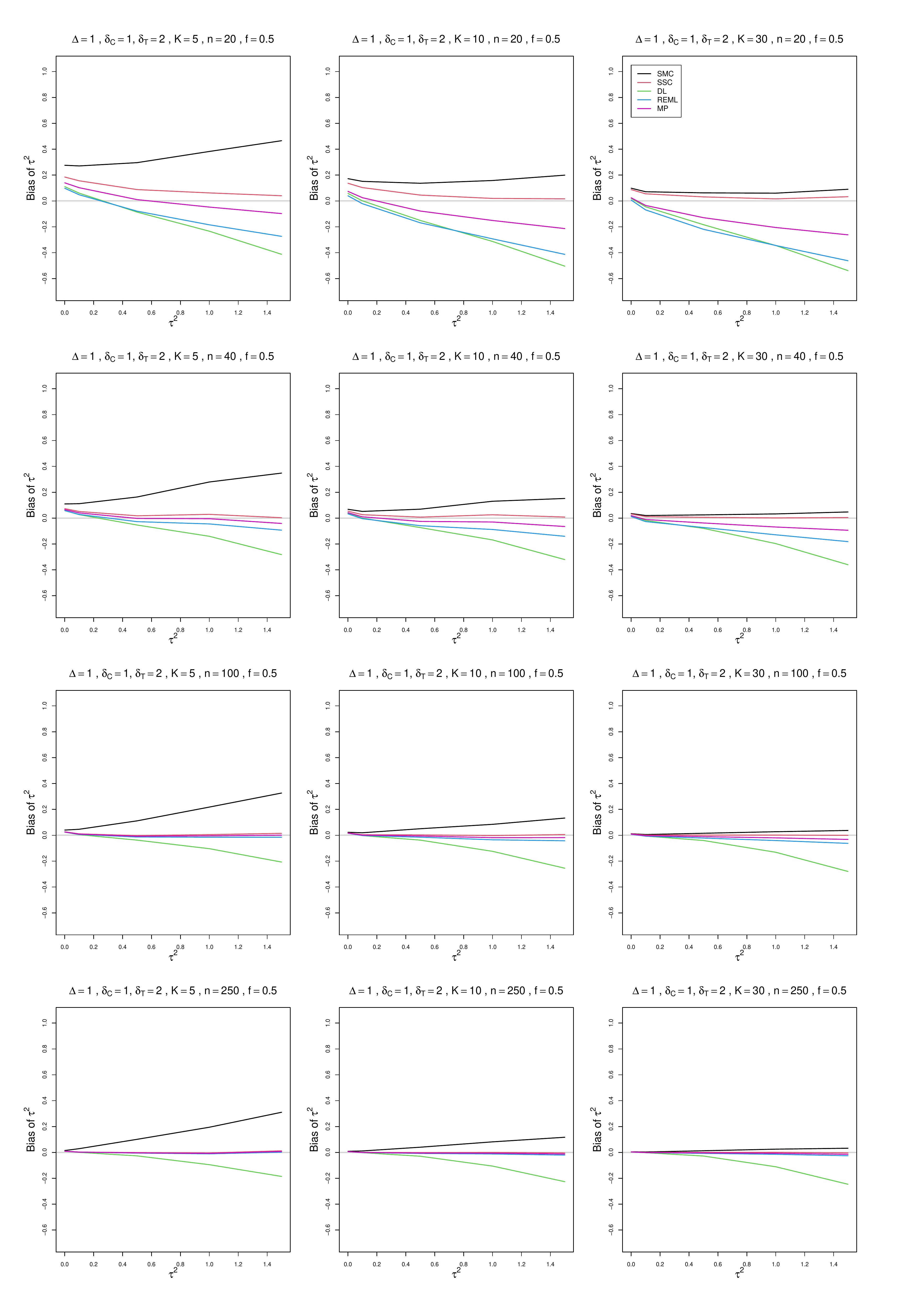}
	\caption{Bias  of estimators of between-study variance of DSM (DL, REML, MP, SMC and SSC ) vs $\tau^2$, for equal sample sizes $n=20,\;40,\;100$ and $250$, $\delta_{iC} = 1$, $\Delta=1$ and  $f = 0.5$.   }
	\label{PlotBiasOfTau2_deltaC_=1deltaT=2_DSM_equal_sample_sizes.pdf}
\end{figure}

\begin{figure}[ht]
	\centering
	\includegraphics[scale=0.33]{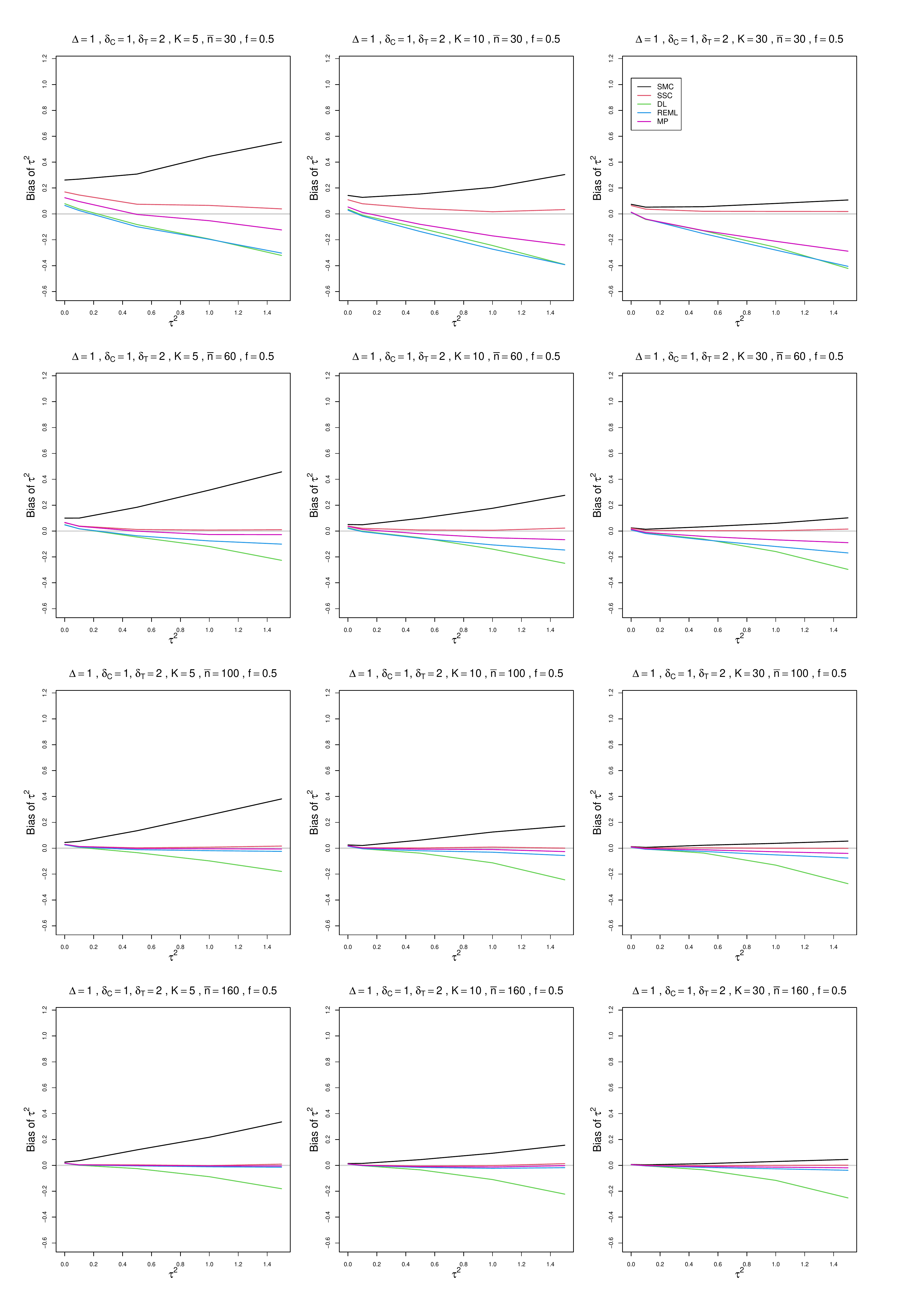}
	\caption{Bias  of estimators of between-study variance of DSM (DL, REML, MP, SMC and SSC ) vs $\tau^2$, for unequal sample sizes $\bar{n}=30,\;60,\;100$ and $160$, $\delta_{iC} = 1$, $\Delta=1$ and  $f = 0.5$.   }
	\label{PlotBiasOfTau2_deltaC_1deltaT=2_DSM_unequal_sample_sizes.pdf}
\end{figure}

\begin{figure}[ht]
	\centering
	\includegraphics[scale=0.33]{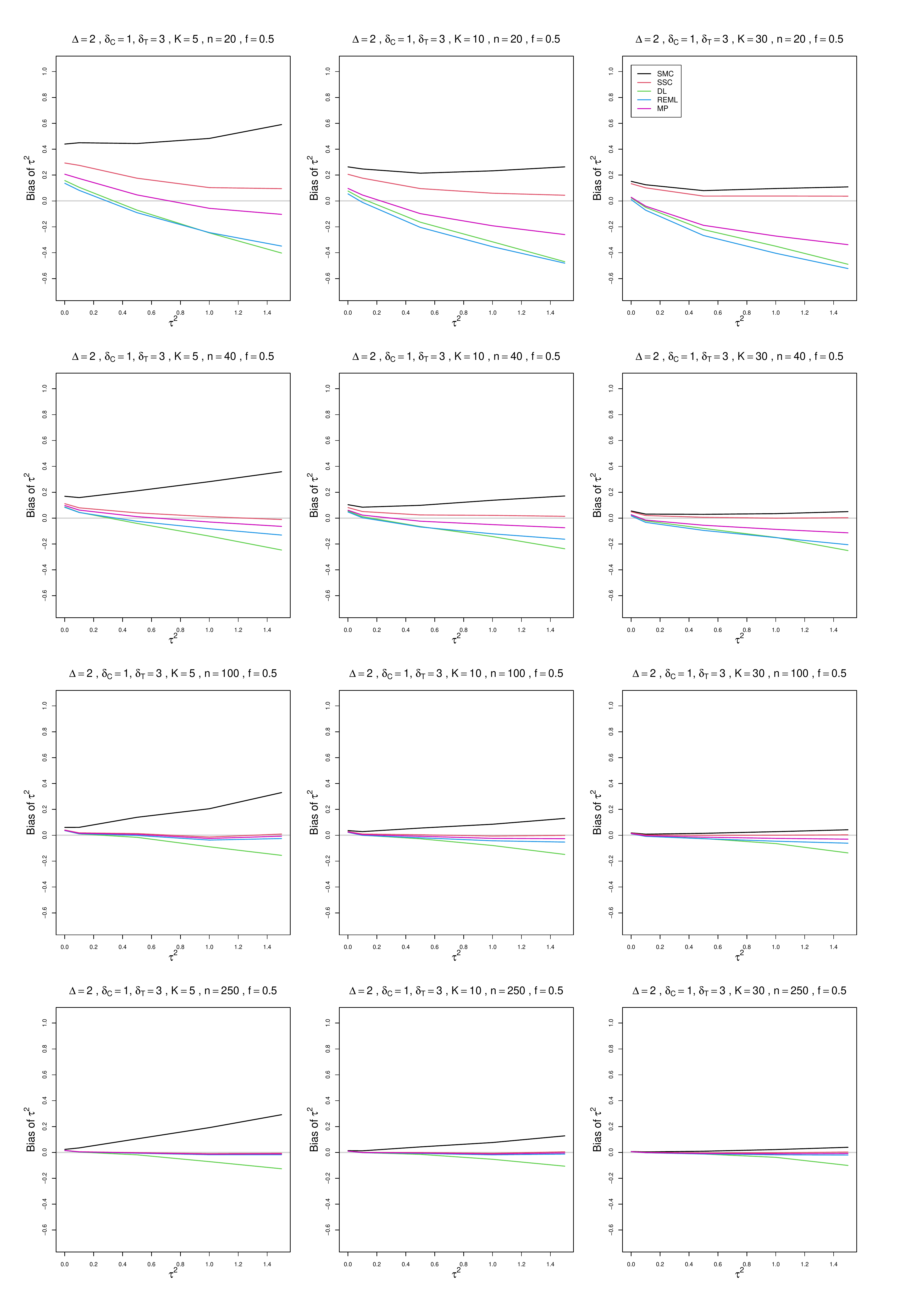}
	\caption{Bias  of estimators of between-study variance of DSM (DL, REML, MP, SMC and SSC ) vs $\tau^2$, for equal sample sizes $n=20,\;40,\;100$ and $250$, $\delta_{iC} = 1$, $\Delta=2$ and  $f = 0.5$.   }
	\label{PlotBiasOfTau2_deltaC_1deltaT=3_DSM_equal_sample_sizes.pdf}
\end{figure}

\begin{figure}[ht]
	\centering
	\includegraphics[scale=0.33]{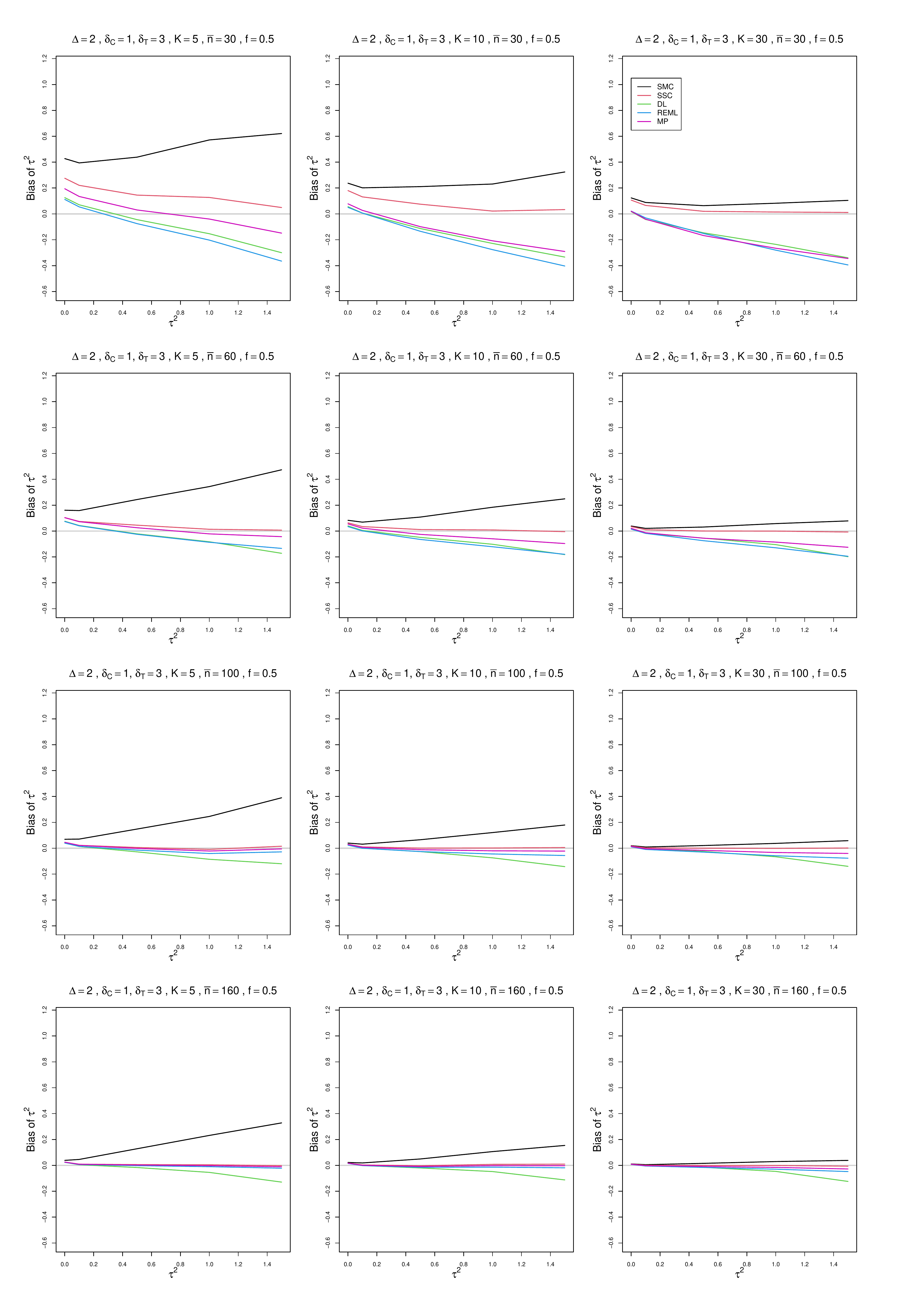}
	\caption{Bias  of estimators of between-study variance of DSM (DL, REML, MP, SMC and SSC ) vs $\tau^2$, for unequal sample sizes $\bar{n}=30,\;60,\;100$ and $160$, $\delta_{iC} = 1$, $\Delta=2$ and  $f = 0.5$.   }
	\label{PlotBiasOfTau2_deltaC_1deltaT=3_DSM_unequal_sample_sizes.pdf}
\end{figure}

\clearpage

\begin{figure}[ht]
	\centering
	\includegraphics[scale=0.33]{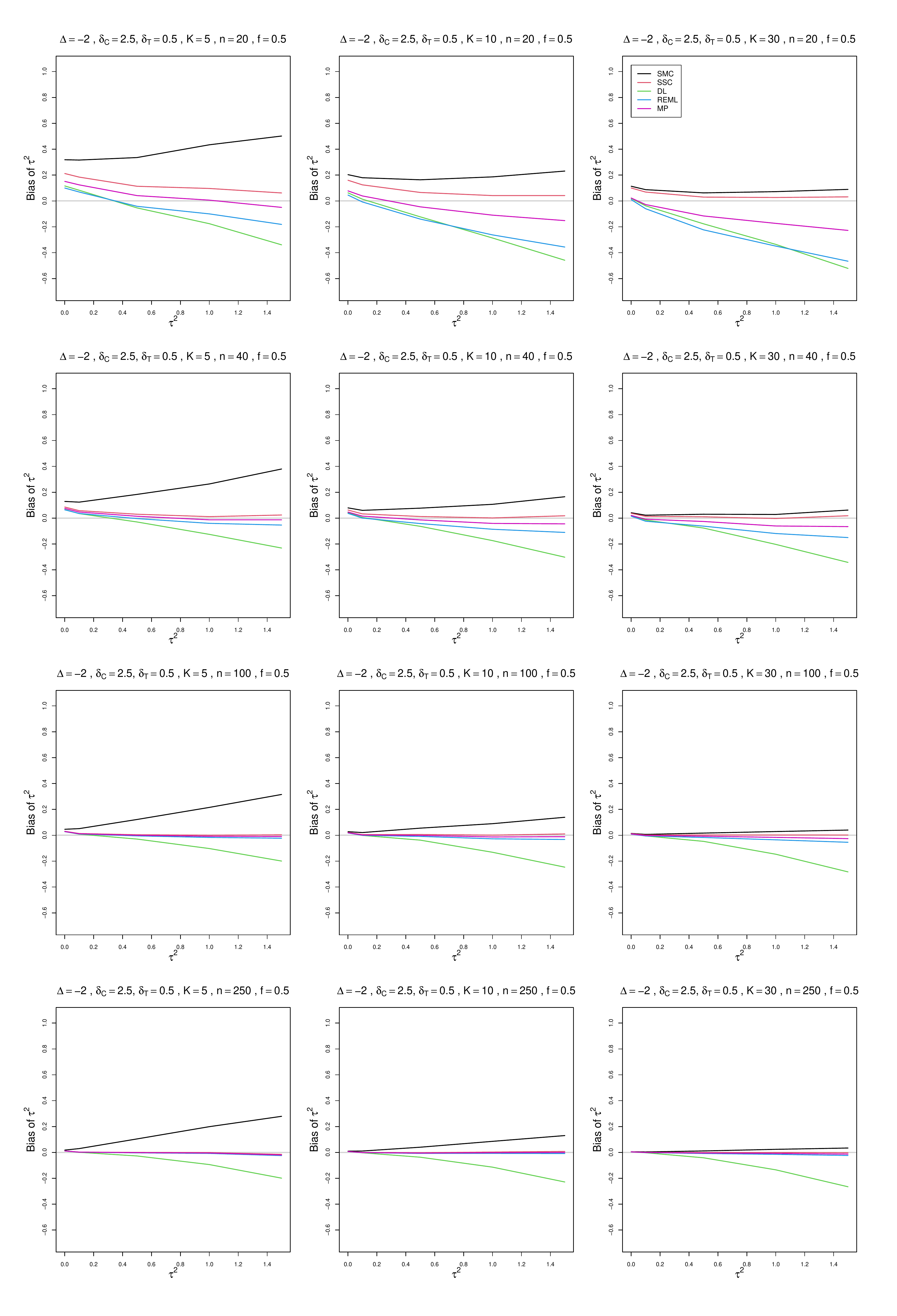}
	\caption{Bias  of estimators of between-study variance of DSM (DL, REML, MP, SMC and SSC ) vs $\tau^2$, for equal sample sizes $n=20,\;40,\;100$ and $250$, $\delta_{iC} = 2.5$, $\Delta=-2$ and  $f = 0.5$.   }
	\label{PlotBiasOfTau2_deltaC_2.5deltaT=0.5_DSM_equal_sample_sizes.pdf}
\end{figure}

\begin{figure}[ht]
	\centering
	\includegraphics[scale=0.33]{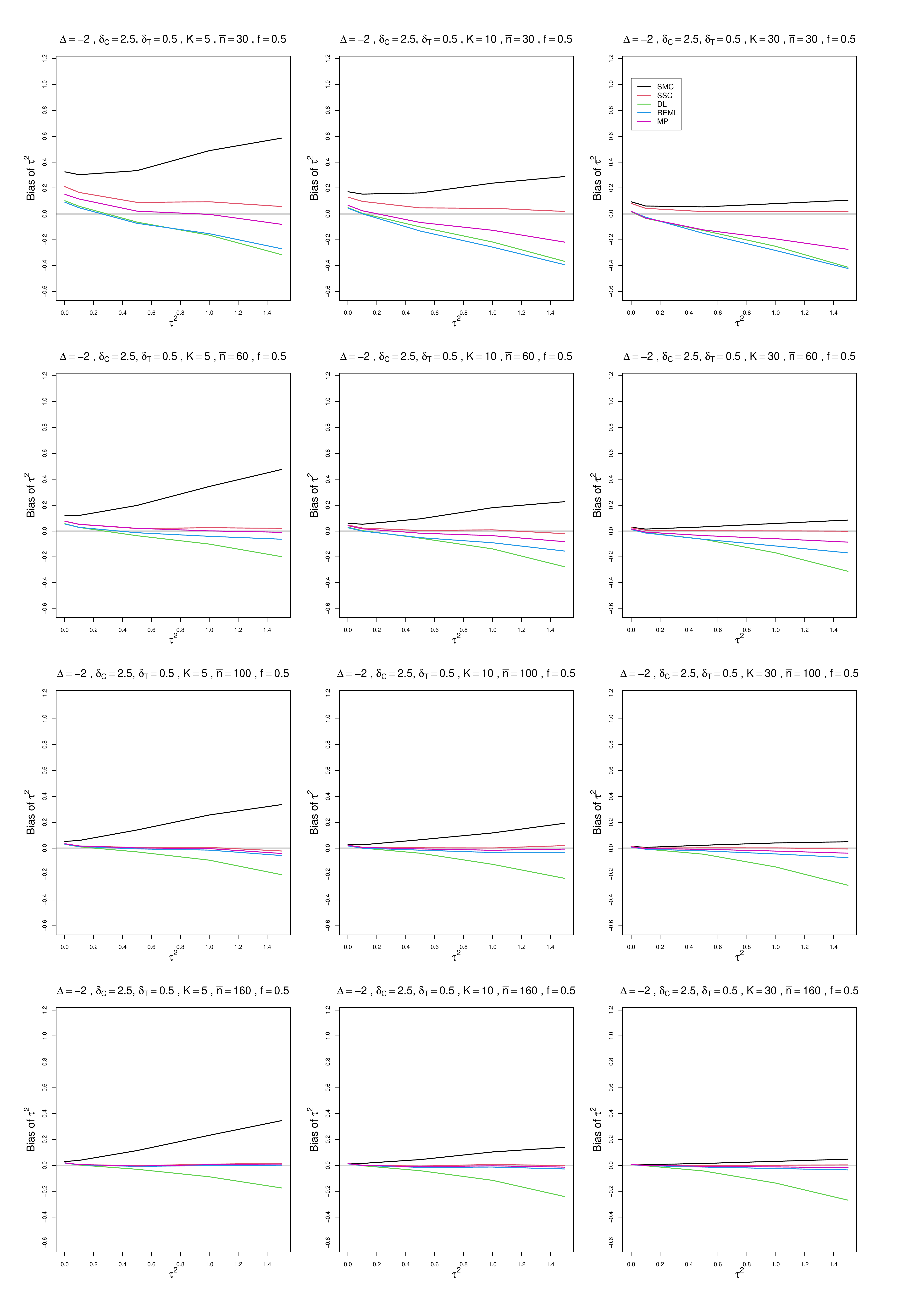}
	\caption{Bias  of estimators of between-study variance of DSM (DL, REML, MP, SMC and SSC ) vs $\tau^2$, for unequal sample sizes $\bar{n}=30,\;60,\;100$ and $160$, $\delta_{iC} = 2.5$, $\Delta=-2$ and  $f = 0.5$.   }
	\label{PlotBiasOfTau2_deltaC_2.5deltaT=0.5_DSM_unequal_sample_sizes.pdf}
\end{figure}

\begin{figure}[ht]
	\centering
	\includegraphics[scale=0.33]{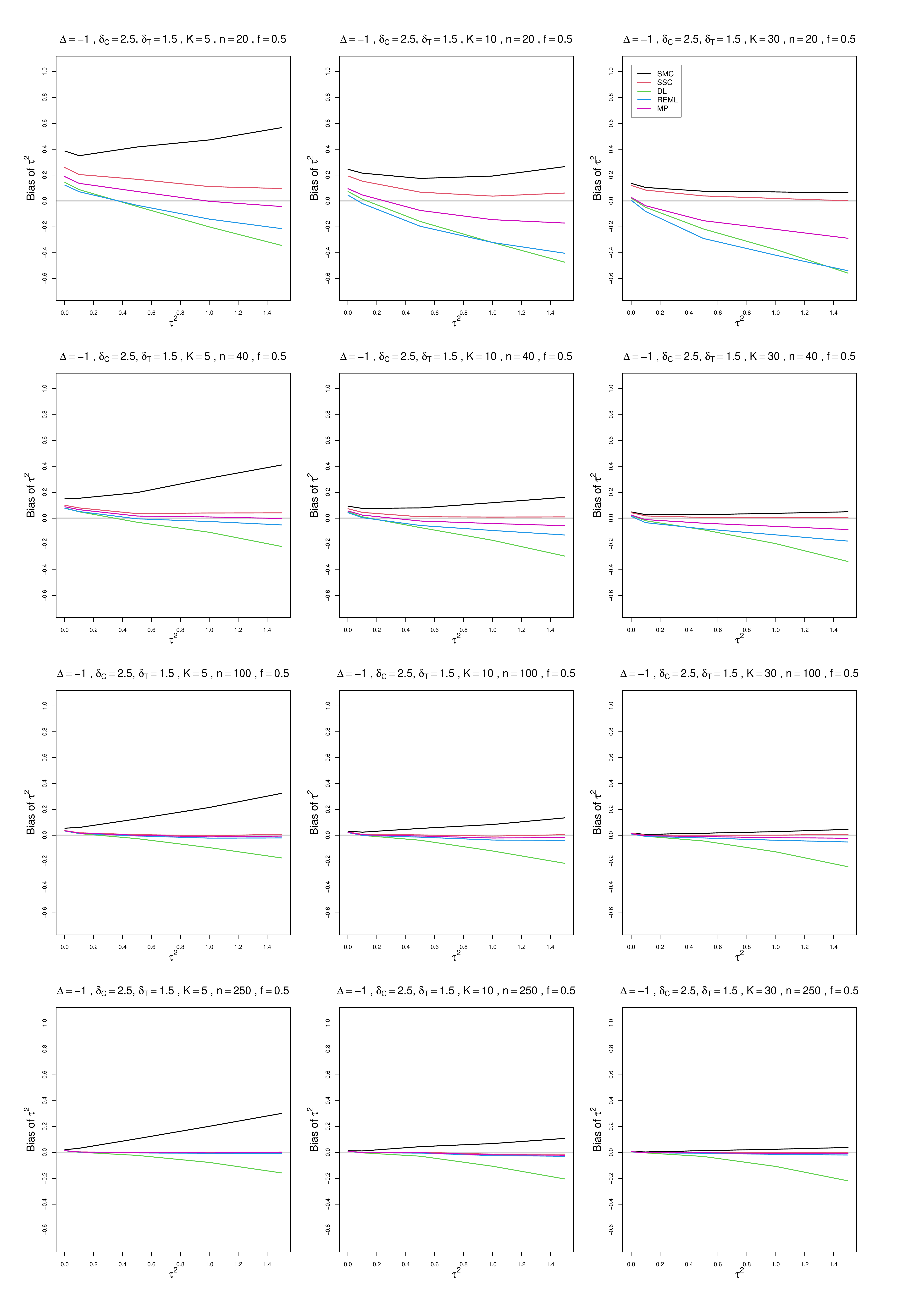}
	\caption{Bias  of estimators of between-study variance of DSM (DL, REML, MP, SMC and SSC ) vs $\tau^2$, for equal sample sizes $n=20,\;40,\;100$ and $250$, $\delta_{iC} = 2.5$, $\Delta=-1$ and  $f = 0.5$.   }
	\label{PlotBiasOfTau2_deltaC_2.5deltaT=1.5_DSM_equal_sample_sizes.pdf}
\end{figure}

\begin{figure}[ht]
	\centering
	\includegraphics[scale=0.33]{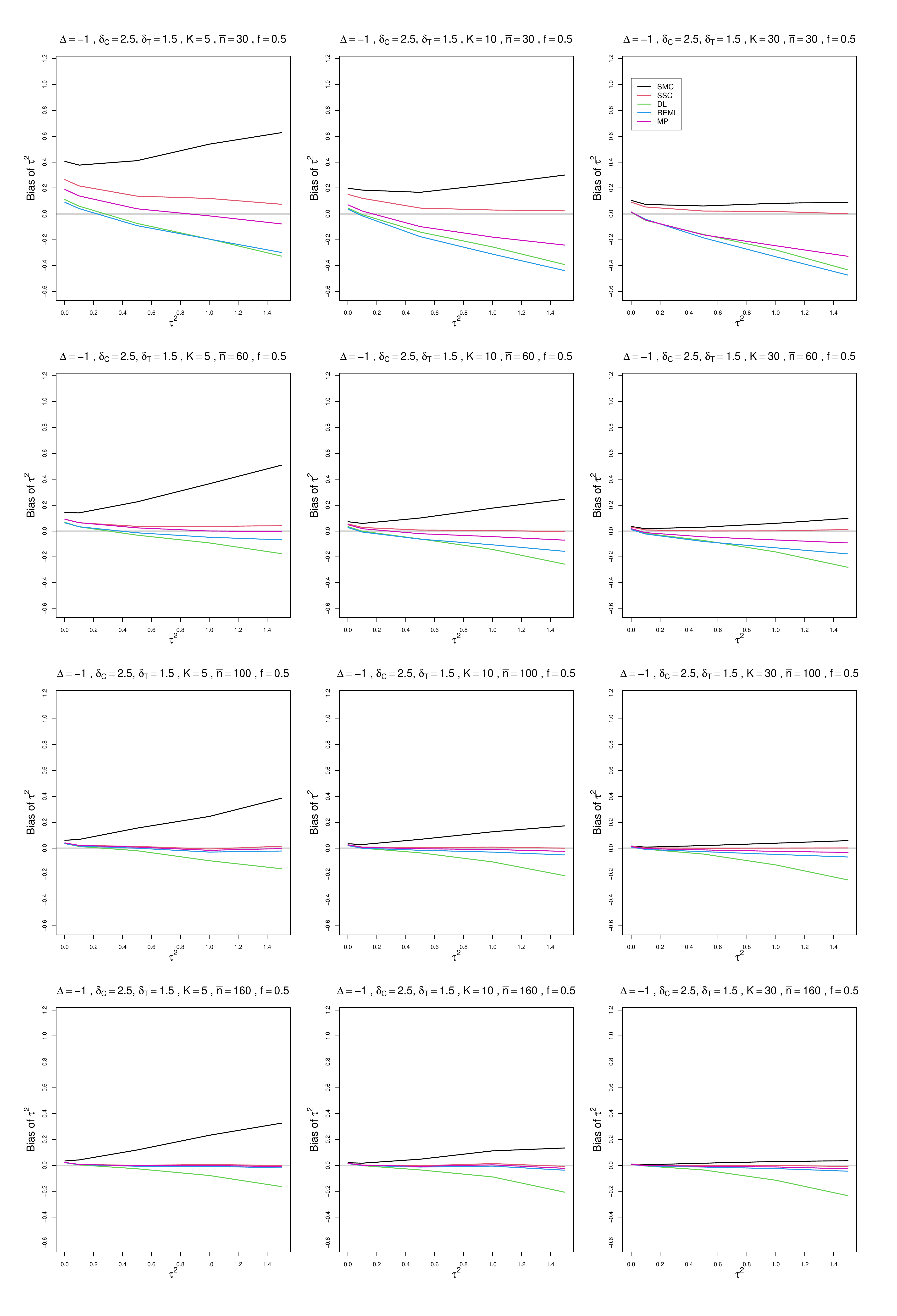}
	\caption{Bias  of estimators of between-study variance of DSM (DL, REML, MP, SMC and SSC ) vs $\tau^2$, for unequal sample sizes $\bar{n}=30,\;60,\;100$ and $160$, $\delta_{iC} = 2.5$, $\Delta=-1$ and  $f = 0.5$.   }
	\label{PlotBiasOfTau2_deltaC_2.5deltaT=1.5_DSM_unequal_sample_sizes.pdf}
\end{figure}

\begin{figure}[ht]
	\centering
	\includegraphics[scale=0.33]{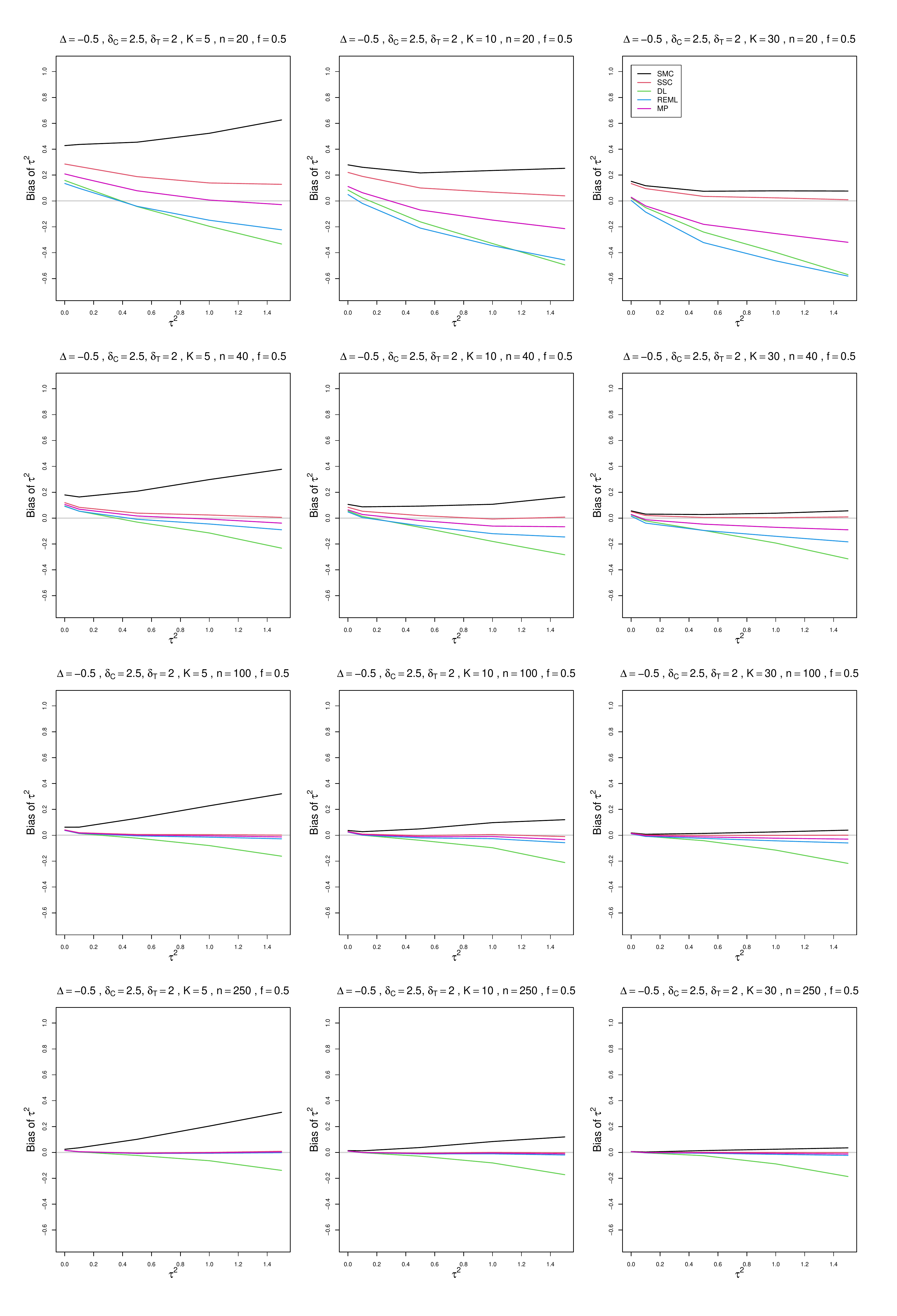}
	\caption{Bias  of estimators of between-study variance of DSM (DL, REML, MP, SMC and SSC ) vs $\tau^2$, for equal sample sizes $n=20,\;40,\;100$ and $250$, $\delta_{iC} = 2.5$, $\Delta=-0.5$ and  $f = 0.5$.   }
	\label{PlotBiasOfTau2_deltaC_2.5deltaT=2_DSM_equal_sample_sizes.pdf}
\end{figure}

\begin{figure}[ht]
	\centering
	\includegraphics[scale=0.33]{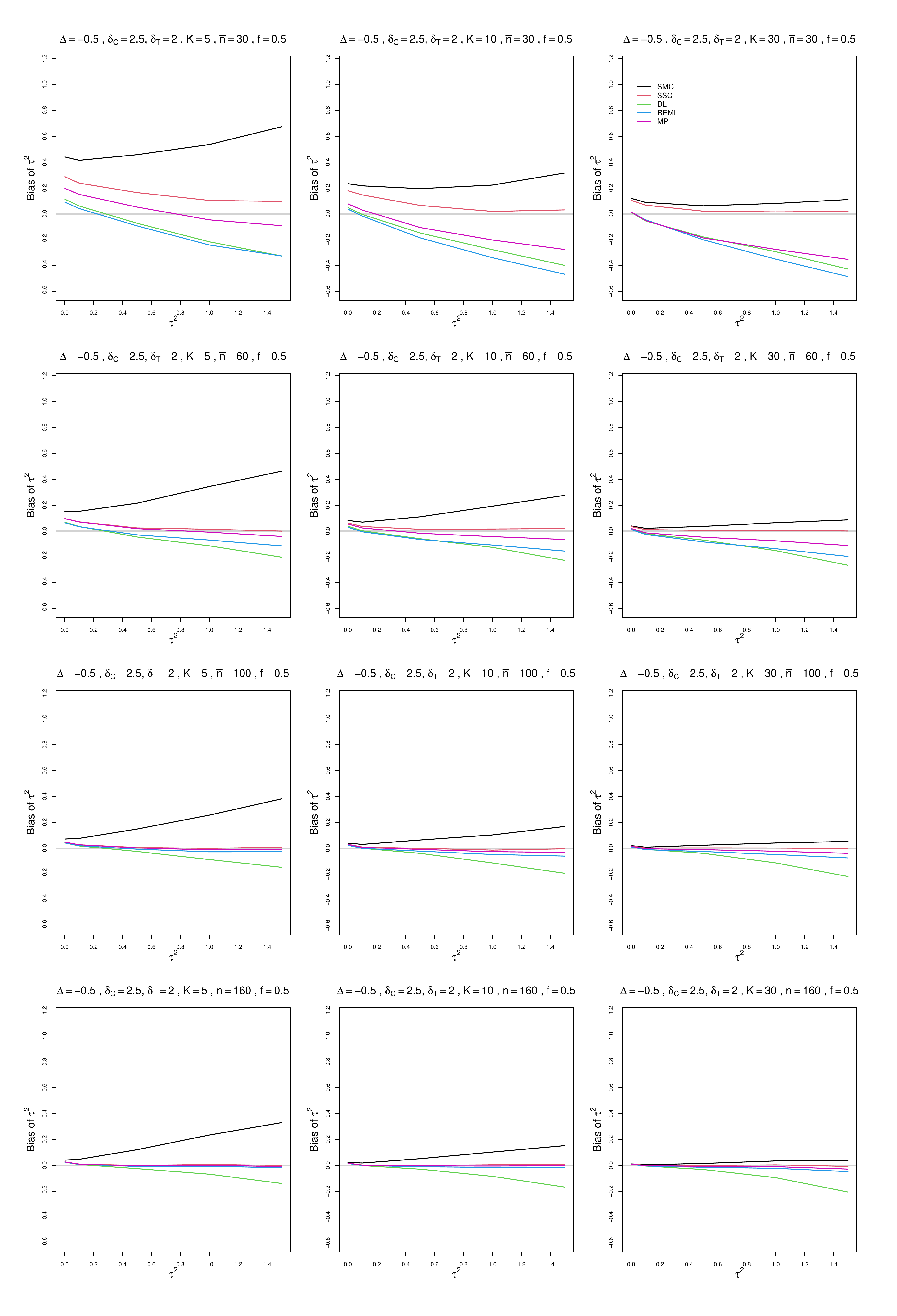}
	\caption{Bias  of estimators of between-study variance of DSM (DL, REML, MP, SMC and SSC ) vs $\tau^2$, for unequal sample sizes $\bar{n}=30,\;60,\;100$ and $160$, $\delta_{iC} = 2.5$, $\Delta=-0.5$ and  $f = 0.5$.   }
	\label{PlotBiasOfTau2_deltaC_2.5deltaT=2_DSM_unequal_sample_sizes.pdf}
\end{figure}

\begin{figure}[ht]
	\centering
	\includegraphics[scale=0.33]{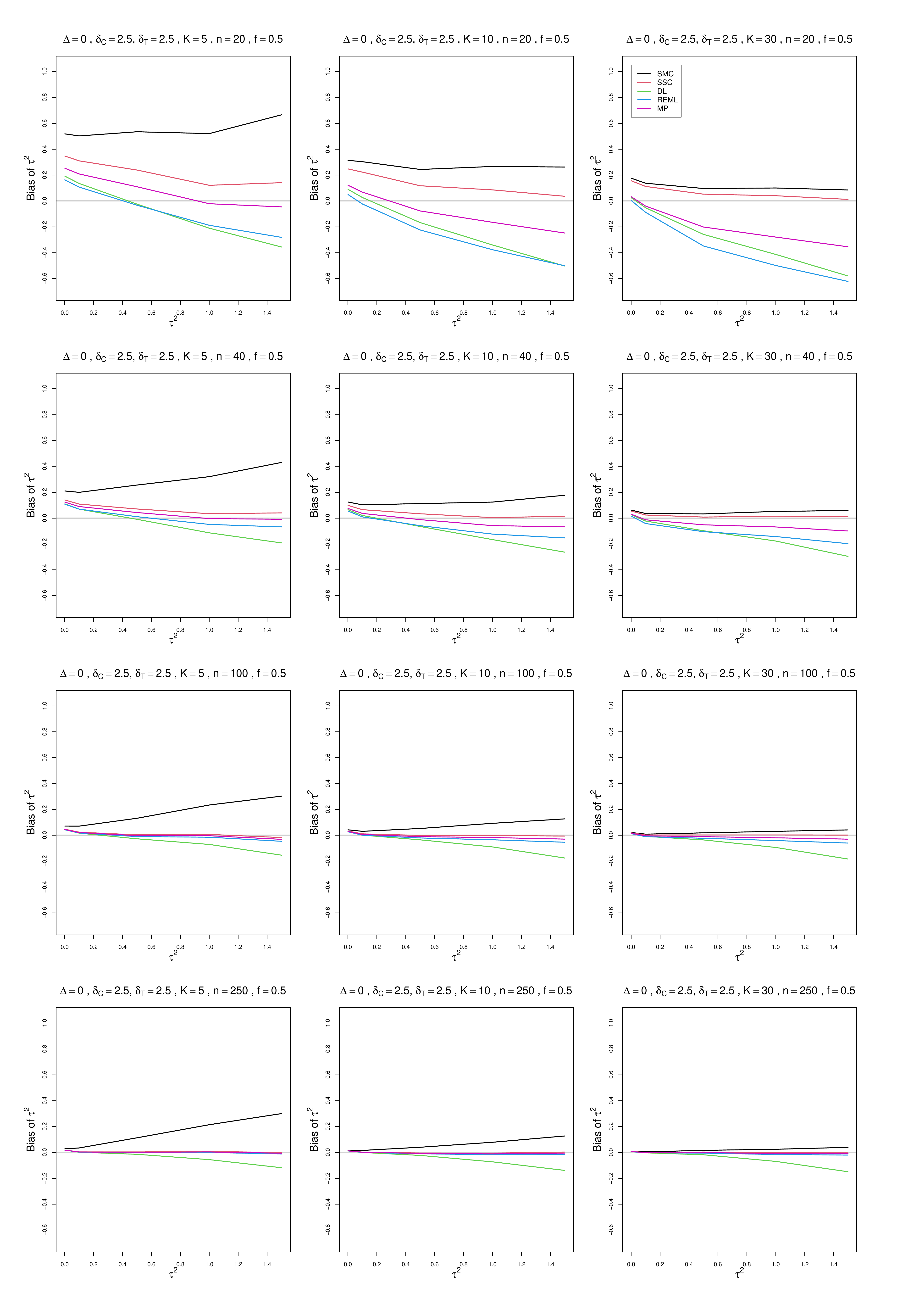}
	\caption{Bias  of estimators of between-study variance of DSM (DL, REML, MP, SMC and SSC ) vs $\tau^2$, for equal sample sizes $n=20,\;40,\;100$ and $250$, $\delta_{iC} = 2.5$, $\Delta=0$ and  $f = 0.5$.   }
	\label{PlotBiasOfTau2_deltaC_2.5deltaT=2.5_DSM_equal_sample_sizes.pdf}
\end{figure}

\begin{figure}[ht]
	\centering
	\includegraphics[scale=0.33]{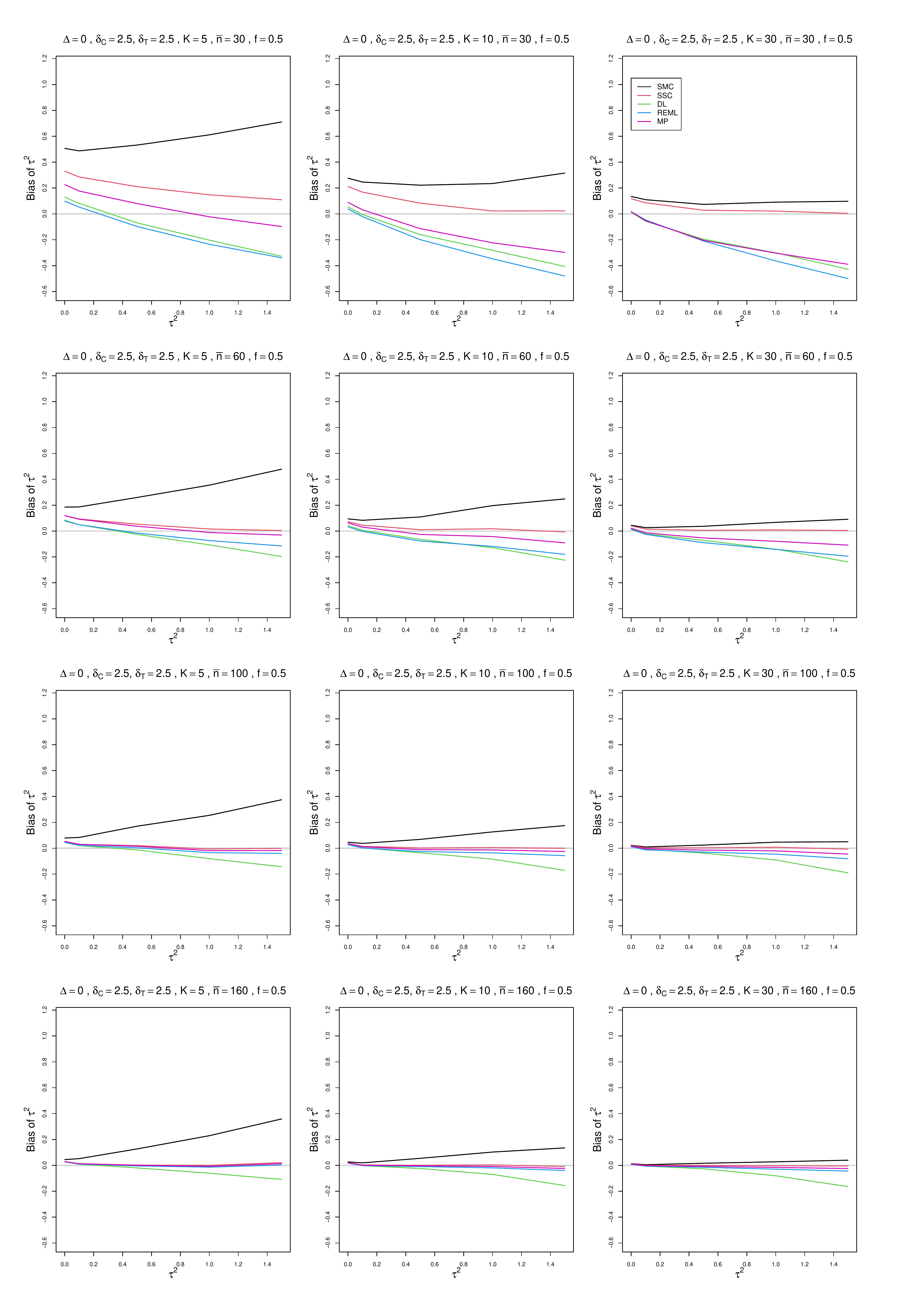}
	\caption{Bias  of estimators of between-study variance of DSM (DL, REML, MP, SMC and SSC ) vs $\tau^2$, for unequal sample sizes $\bar{n}=30,\;60,\;100$ and $160$, $\delta_{iC} = 2.5$, $\Delta=0$ and  $f = 0.5$.   }
	\label{PlotBiasOfTau2_deltaC_2.5deltaT=2.5_DSM_unequal_sample_sizes.pdf}
\end{figure}

\begin{figure}[ht]
	\centering
	\includegraphics[scale=0.33]{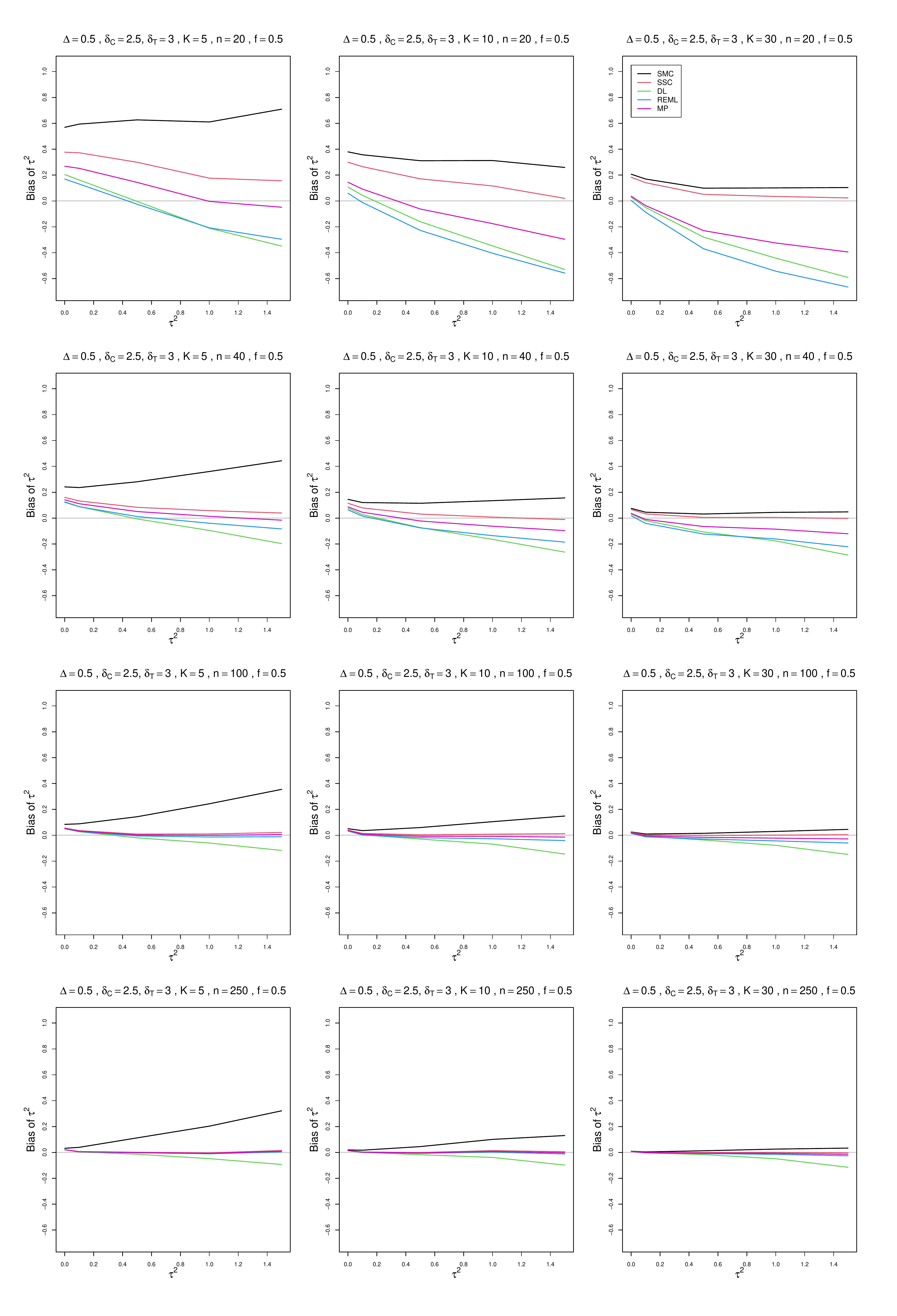}
	\caption{Bias  of estimators of between-study variance of DSM (DL, REML, MP, SMC and SSC ) vs $\tau^2$, for equal sample sizes $n=20,\;40,\;100$ and $250$, $\delta_{iC} = 2.5$, $\Delta=0.5$ and  $f = 0.5$.   }
	\label{PlotBiasOfTau2_deltaC_2.5deltaT=3_DSM_equal_sample_sizes.pdf}
\end{figure}

\begin{figure}[ht]
	\centering
	\includegraphics[scale=0.33]{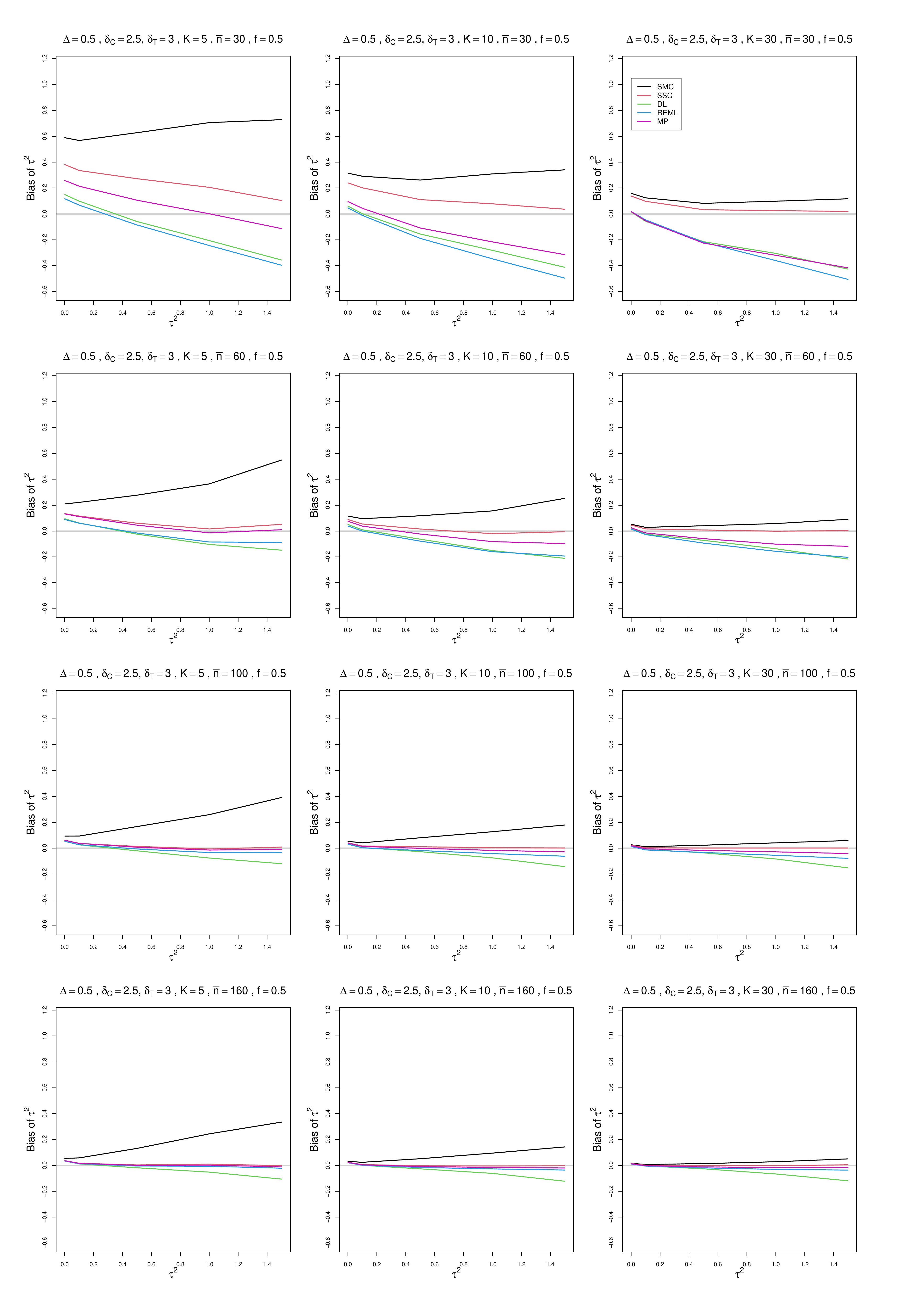}
	\caption{Bias  of estimators of between-study variance of DSM (DL, REML, MP, SMC and SSC ) vs $\tau^2$, for unequal sample sizes $\bar{n}=30,\;60,\;100$ and $160$, $\delta_{iC} = 2.5$, $\Delta=0.5$ and  $f = 0.5$.   }
	\label{PlotBiasOfTau2_deltaC_2.5deltaT=3_DSM_unequal_sample_sizes.pdf}
\end{figure}

\begin{figure}[ht]
	\centering
	\includegraphics[scale=0.33]{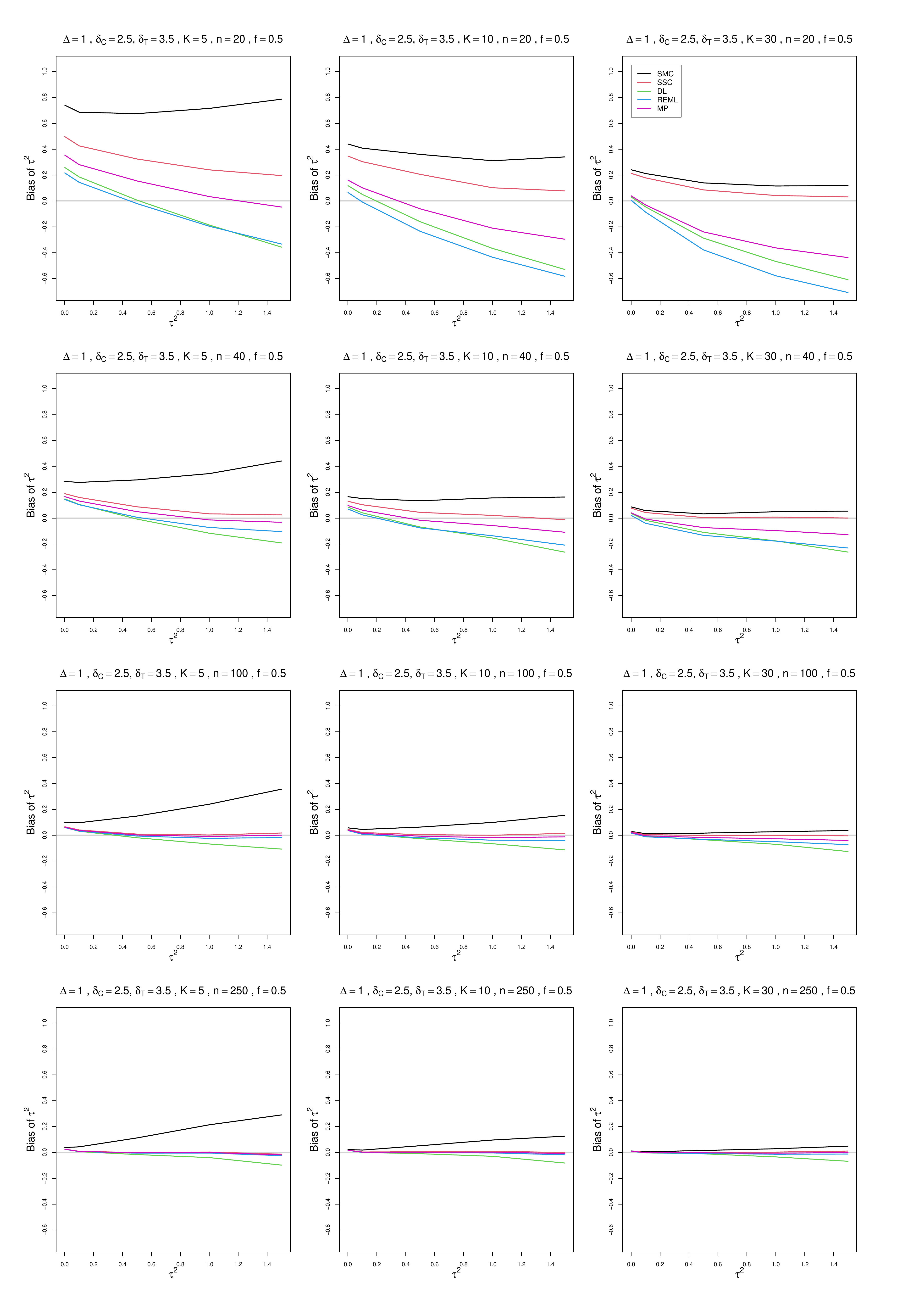}
	\caption{Bias  of estimators of between-study variance of DSM (DL, REML, MP, SMC and SSC ) vs $\tau^2$, for equal sample sizes $n=20,\;40,\;100$ and $250$, $\delta_{iC} = 2.5$, $\Delta=1$ and  $f = 0.5$.   }
	\label{PlotBiasOfTau2_deltaC_2.5deltaT=3.5_DSM_equal_sample_sizes.pdf}
\end{figure}

\begin{figure}[ht]
	\centering
	\includegraphics[scale=0.33]{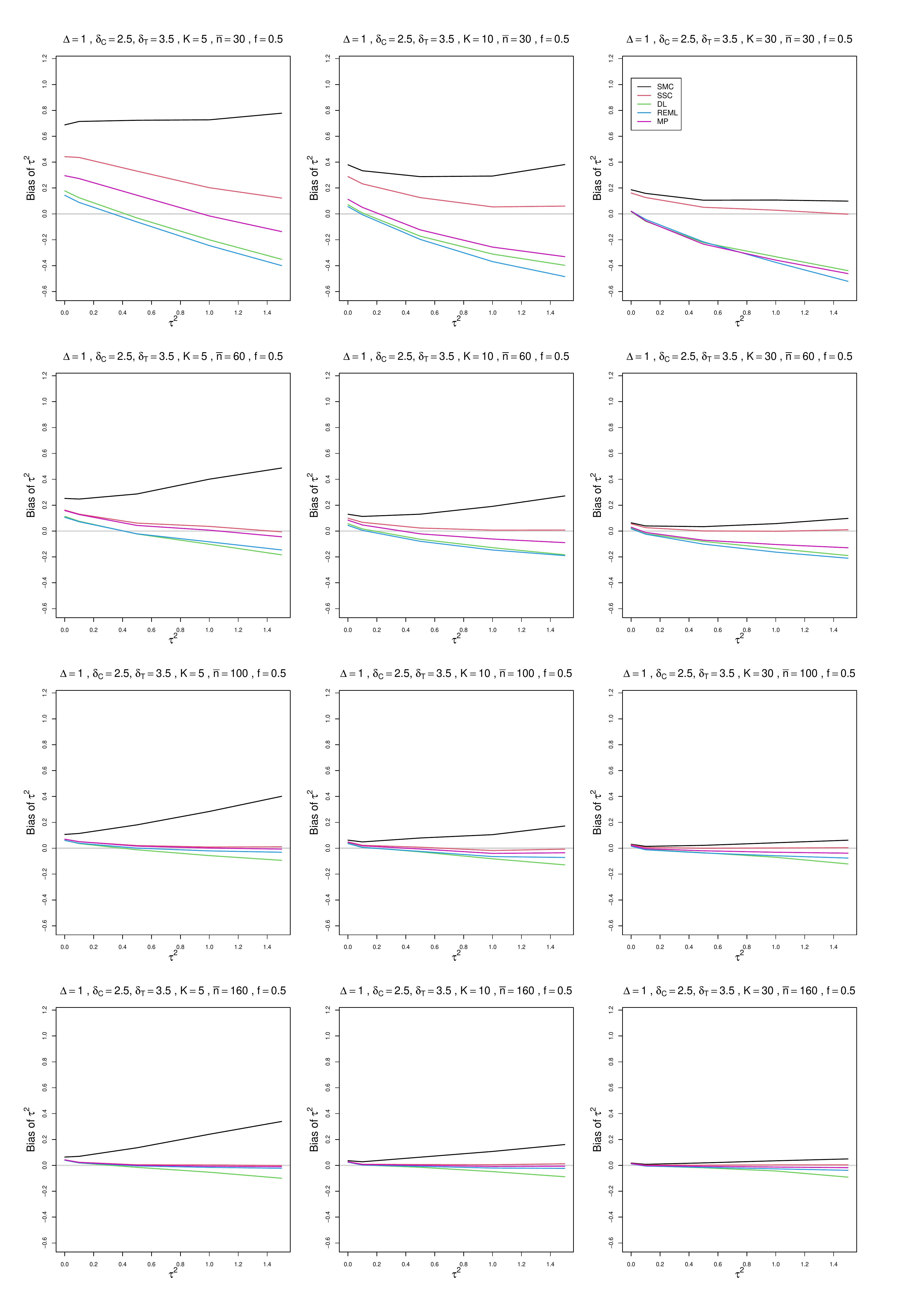}
	\caption{Bias  of estimators of between-study variance of DSM (DL, REML, MP, SMC and SSC ) vs $\tau^2$, for unequal sample sizes $\bar{n}=30,\;60,\;100$ and $160$, $\delta_{iC} = 2.5$, $\Delta=1$ and  $f = 0.5$.   }
	\label{PlotBiasOfTau2_deltaC_2.5deltaT=3.5_DSM_unequal_sample_sizes.pdf}
\end{figure}

\begin{figure}[ht]
	\centering
	\includegraphics[scale=0.33]{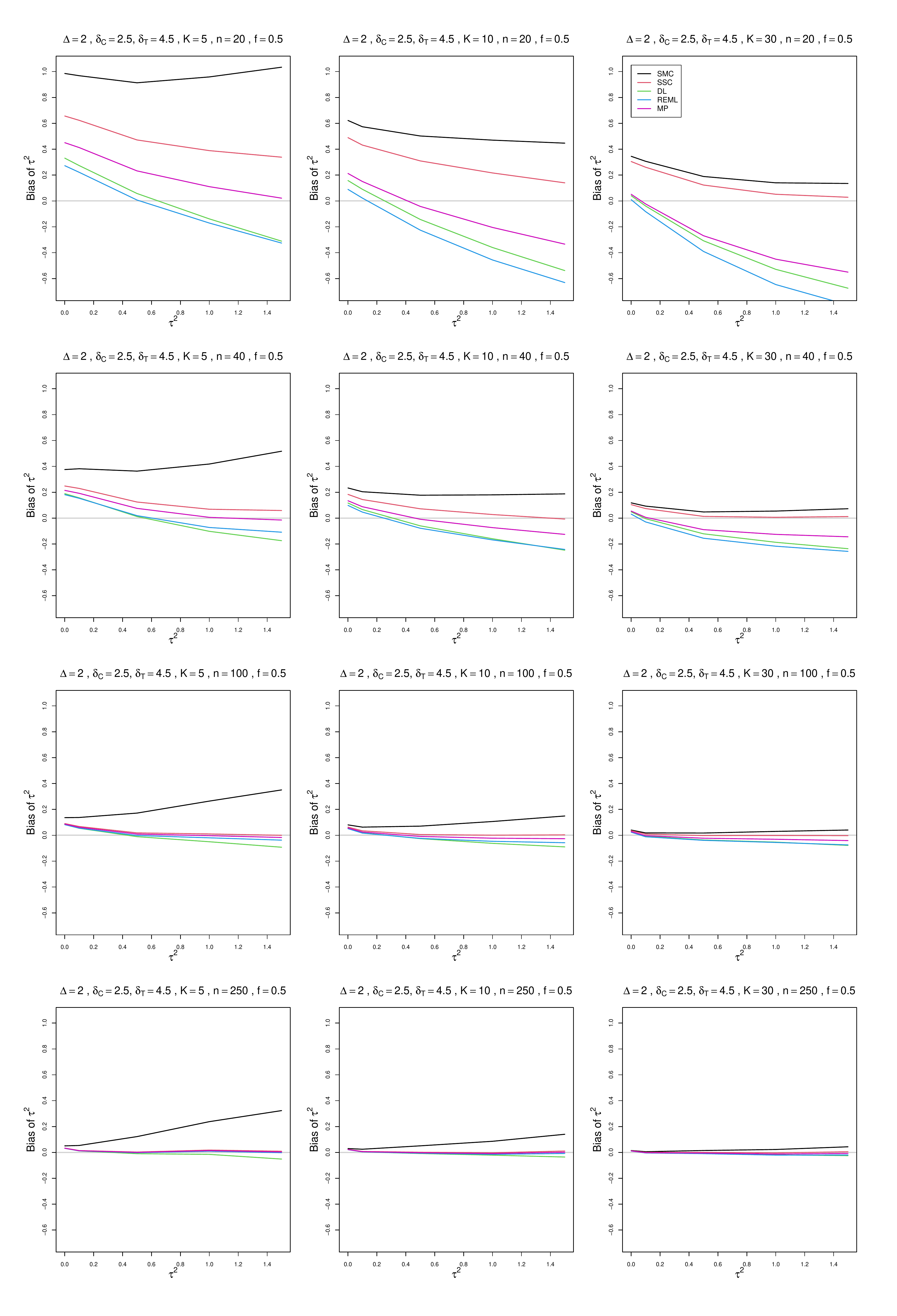}
	\caption{Bias  of estimators of between-study variance of DSM (DL, REML, MP, SMC and SSC ) vs $\tau^2$, for equal sample sizes $n=20,\;40,\;100$ and $250$, $\delta_{iC} = 2.5$, $\Delta=2$ and  $f = 0.5$.   }
	\label{PlotBiasOfTau2_deltaC_2.5deltaT=2.5_DSM_equal_sample_sizes.pdf}
\end{figure}

\begin{figure}[ht]
	\centering
	\includegraphics[scale=0.33]{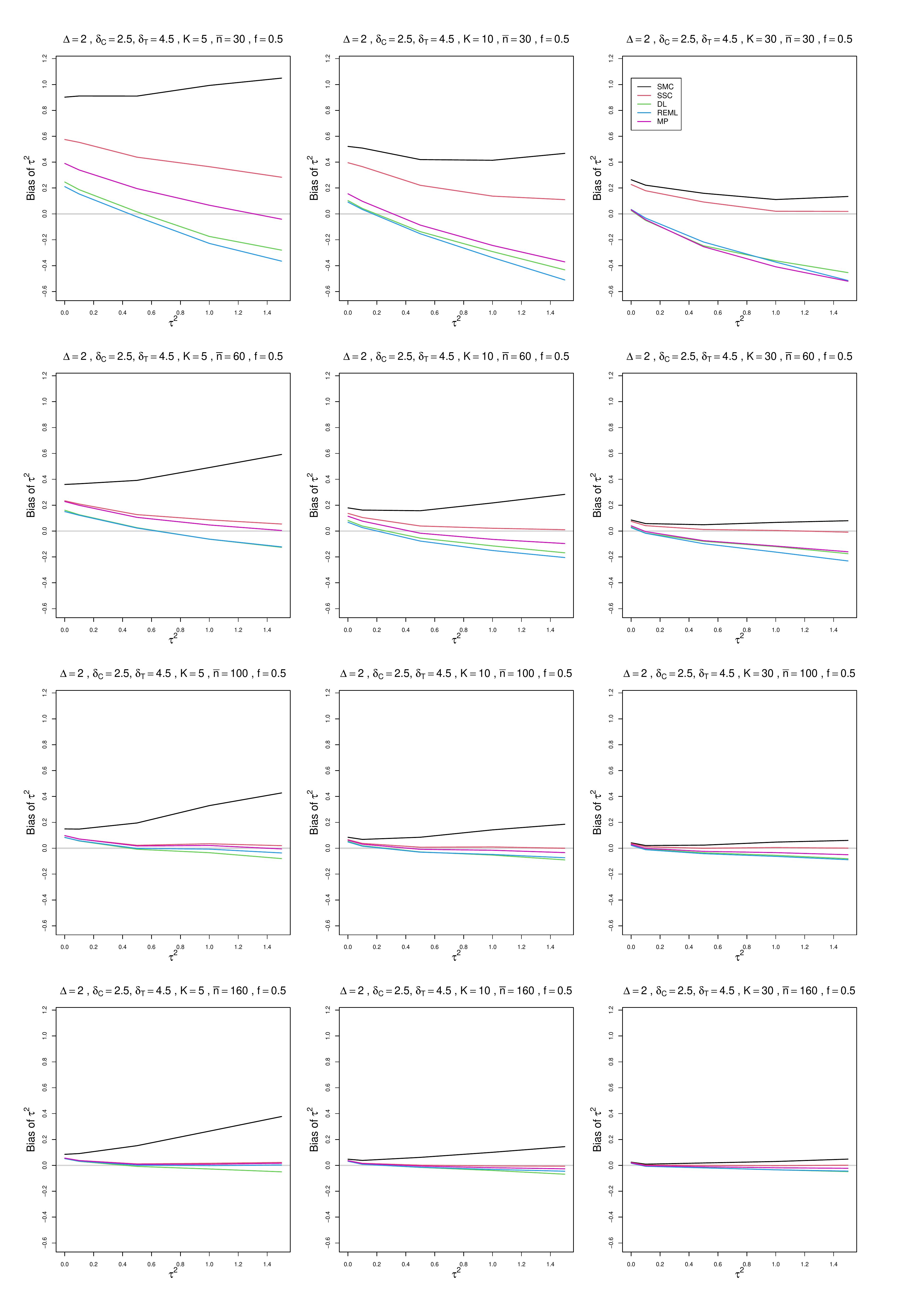}
	\caption{Bias  of estimators of between-study variance of DSM (DL, REML, MP, SMC and SSC ) vs $\tau^2$, for unequal sample sizes $\bar{n}=30,\;60,\;100$ and $160$, $\delta_{iC} = 2.5$, $\Delta=2$ and  $f = 0.5$.   }
	\label{PlotBiasOfTau2_deltaC_2.5deltaT=4.5_DSM_unequal_sample_sizes.pdf}
\end{figure}


\clearpage

\section*{Appendix D: Coverage of 95\% confidence intervals for $\tau^2$}

Each figure corresponds to a value of the standardized mean  in the Control arm $\delta_{C}$  (= $-2.5$, $-1$, 0, 1, 2.5)  and a value of the overall DSM $\Delta$ (= $-2$, $-1$, $-0.5$, 0, 0.5, 1, 2) . \\
The fraction of each study's sample size in the Control arm ($f$) is held constant at 0.5.

For each combination of a value of $n$ (= 20, 40, 100, 250) or  $\bar{n}$ (= 30, 60, 100, 160) and a value of $K$ (= 5, 10, 30), a panel plots  coverage of 95\% confidence intervals versus $\tau^2$ (= 0, 0.1, 0.5, 1, 1.5).\\
The confidence intervals for $\tau^2$ are
\begin{itemize}
\item PL (Profile likelihood), inverse-variance weights
\item QP (Q-profile), inverse-variance weights
\item FPC (based on Farebrother approximation), effective-sample-size weights, conditional variance of DSM
\end{itemize}

\clearpage
\setcounter{figure}{0}
\renewcommand{\thefigure}{D.\arabic{figure}}


\begin{figure}[ht]
	\centering
	\includegraphics[scale=0.33]{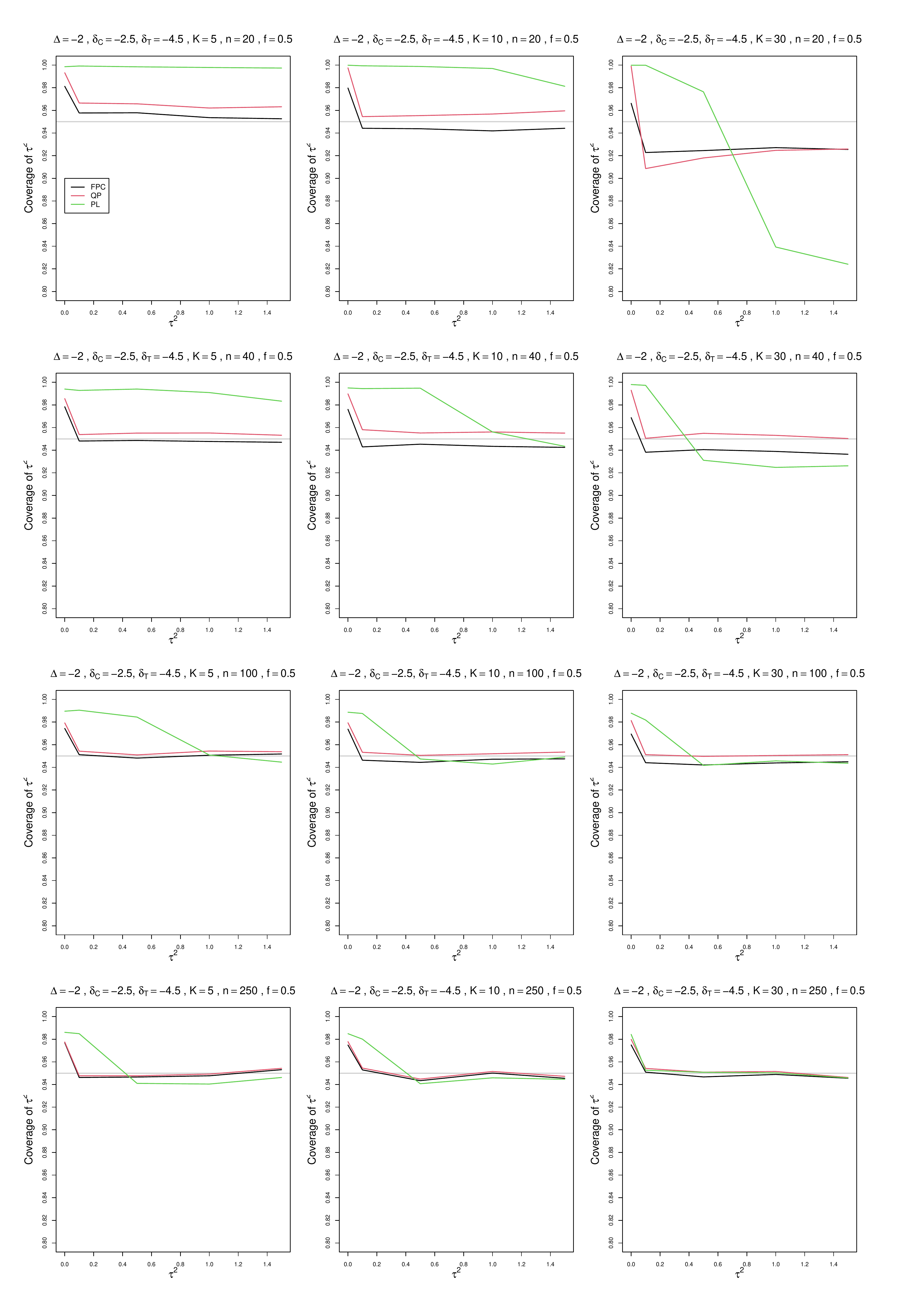}
	\caption{Coverage of PL, QP, and  FPC 95\% confidence intervals for between-study variance of DSM vs $\tau^2$, for equal sample sizes $n = 20,\;40,\;100$ and $250$, $\delta_{iC} = -2.5$, $\Delta=-2$ and  $f = 0.5$.   }
	\label{PlotCoverageOfTau2_deltaC_-25deltaT=-4.5_DSM_equal_sample_sizes.pdf}
\end{figure}

\begin{figure}[ht]
	\centering
	\includegraphics[scale=0.33]{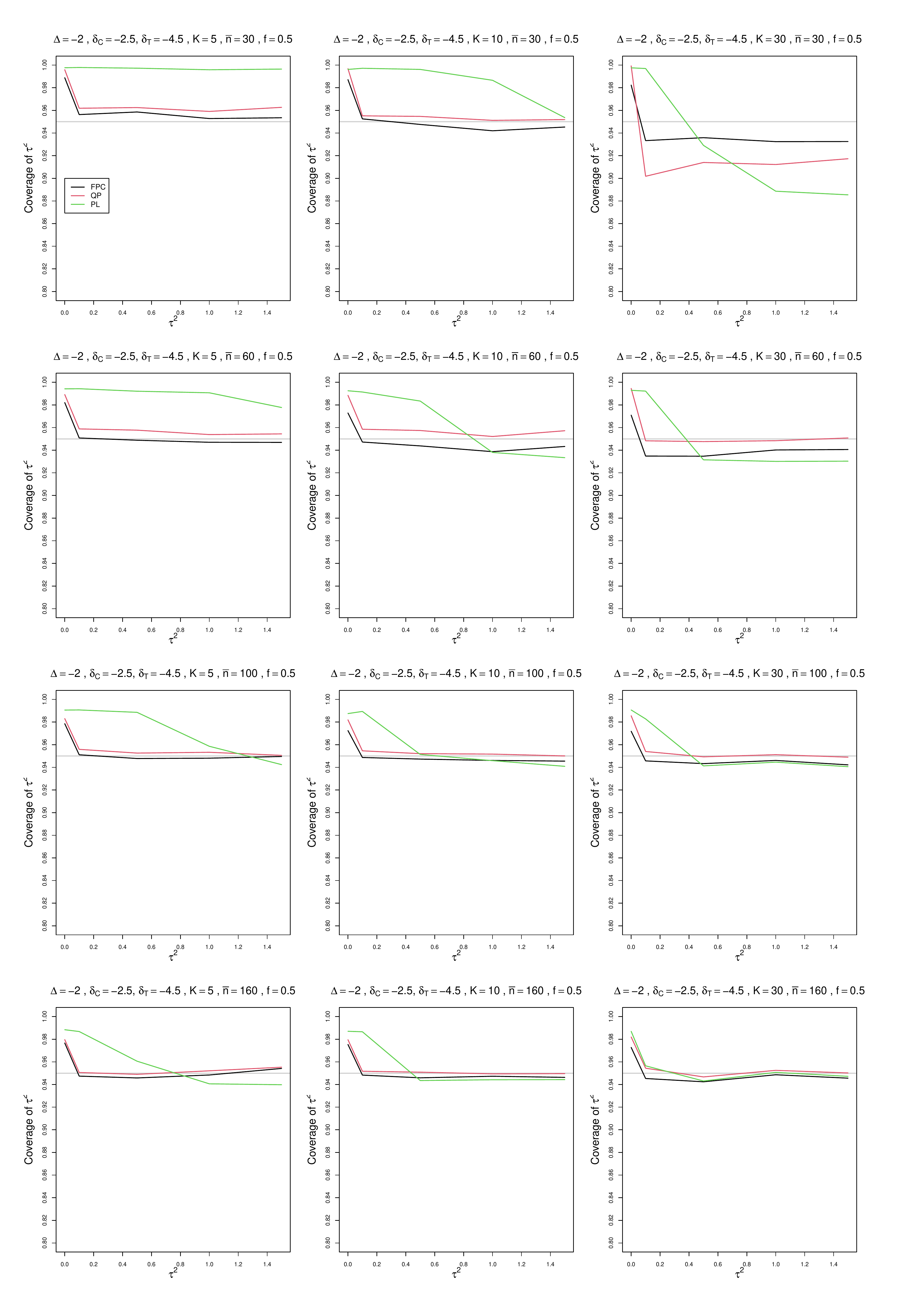}
	\caption{Coverage of PL, QP, and  FPC 95\% confidence intervals for between-study variance of DSM vs $\tau^2$, for unequal sample sizes $\bar{n}=30,\;60,\;100$ and $160$, $\delta_{iC} = -2.5$, $\Delta=-2$ and  $f = 0.5$.   }
	\label{PlotCoverageOfTau2_deltaC_-25deltaT=-4.5_DSM_unequal_sample_sizes.pdf}
\end{figure}

\begin{figure}[ht]
	\centering
	\includegraphics[scale=0.33]{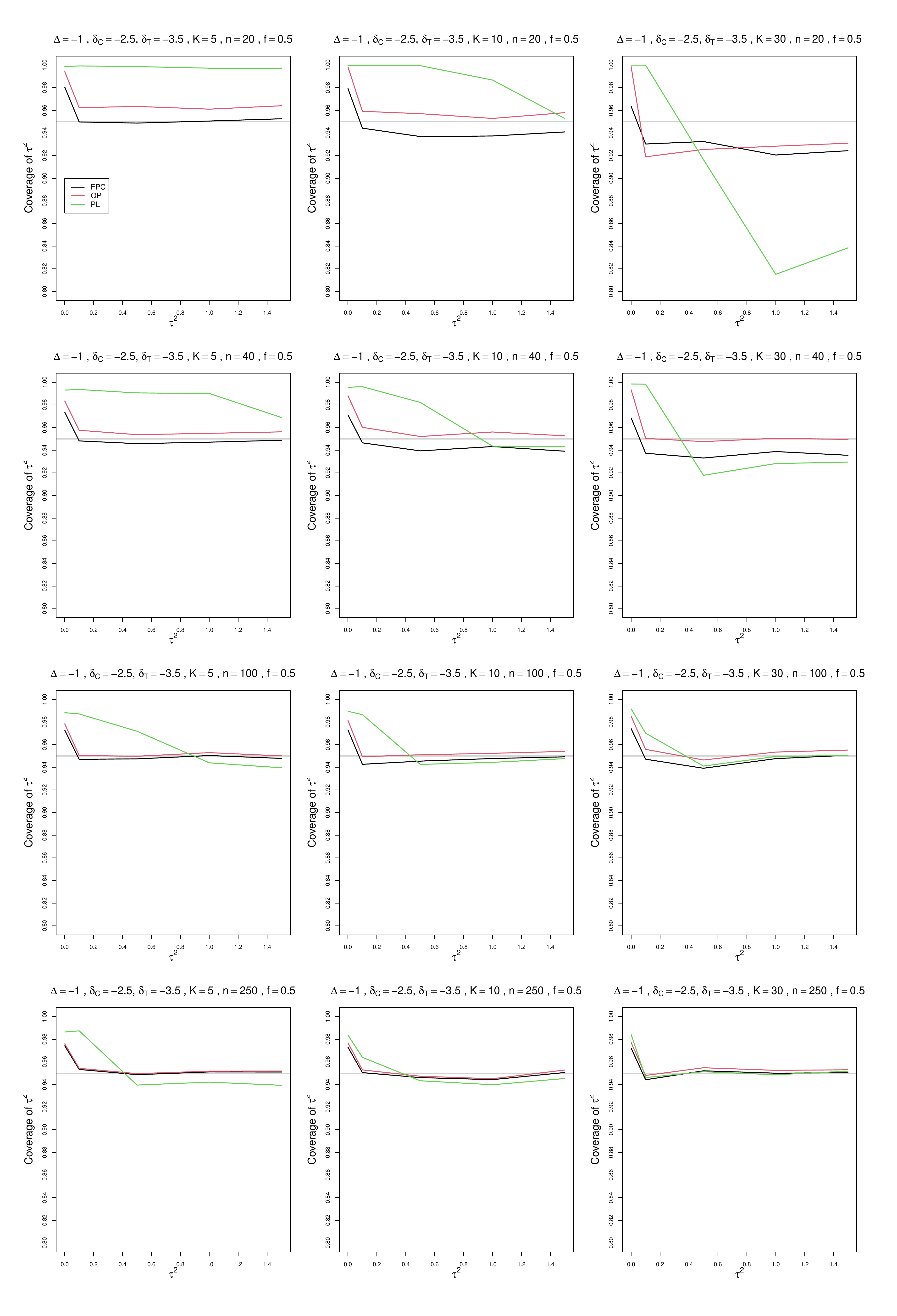}
	\caption{Coverage of PL, QP, and  FPC 95\% confidence intervals for between-study variance of DSM  vs $\tau^2$, for equal sample sizes $n=20,\;40,\;100$ and $250$, $\delta_{iC} = -2.5$, $\Delta=-1$ and  $f = 0.5$.   }
	\label{PlotCoverageOfTau2_deltaC_-25deltaT=-3.5_DSM_equal_sample_sizes.pdf}
\end{figure}

\begin{figure}[ht]
	\centering
	\includegraphics[scale=0.33]{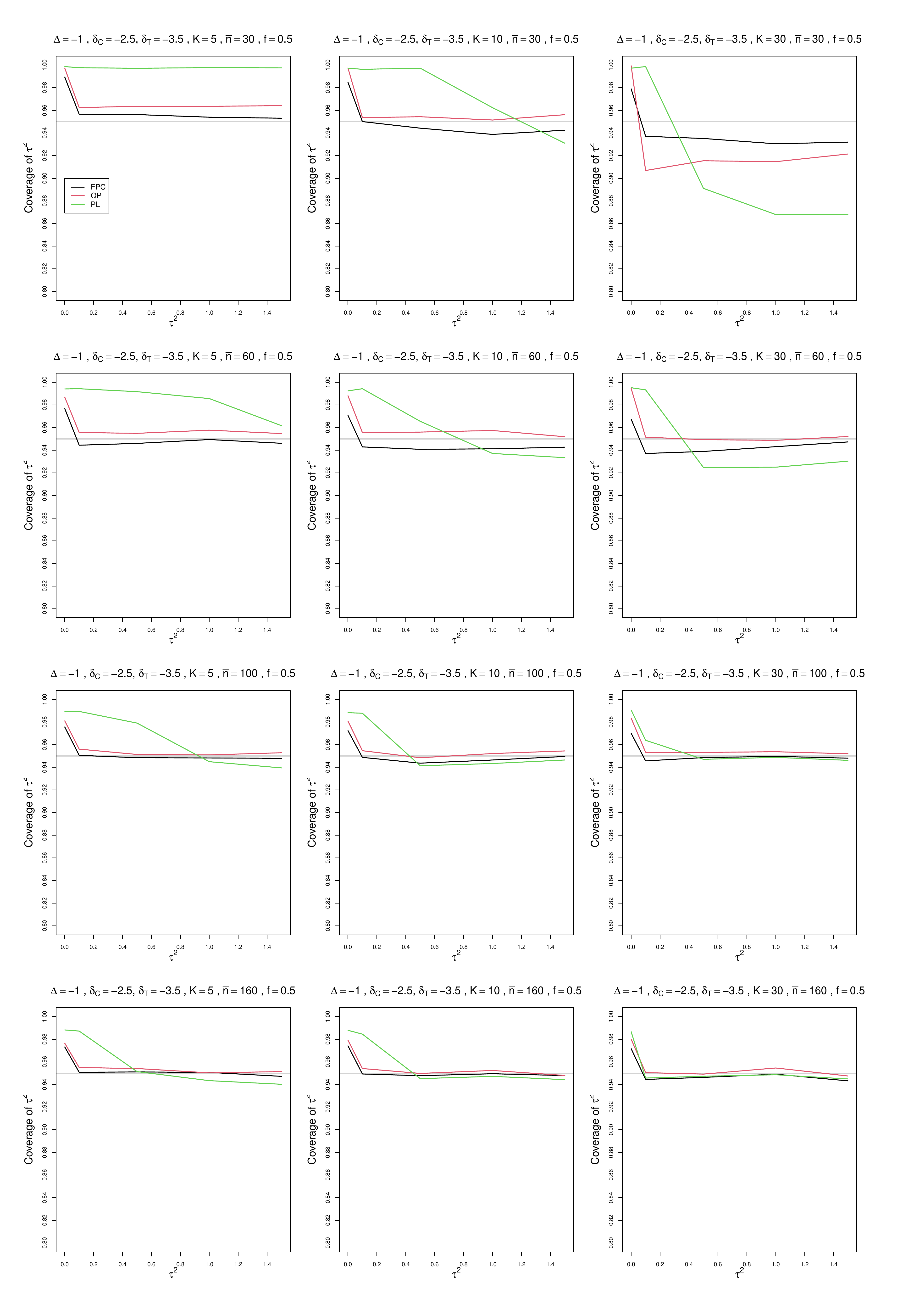}
	\caption{Coverage of PL, QP, and  FPC 95\% confidence intervals for between-study variance of DSM  vs $\tau^2$, for unequal sample sizes $\bar{n}=30,\;60,\;100$ and $160$, $\delta_{iC} = -2.5$, $\Delta=-1$ and  $f = 0.5$.   }
	\label{PlotCoverageOfTau2_deltaC_-25deltaT=-3.5_DSM_unequal_sample_sizes.pdf}
\end{figure}

\begin{figure}[ht]
	\centering
	\includegraphics[scale=0.33]{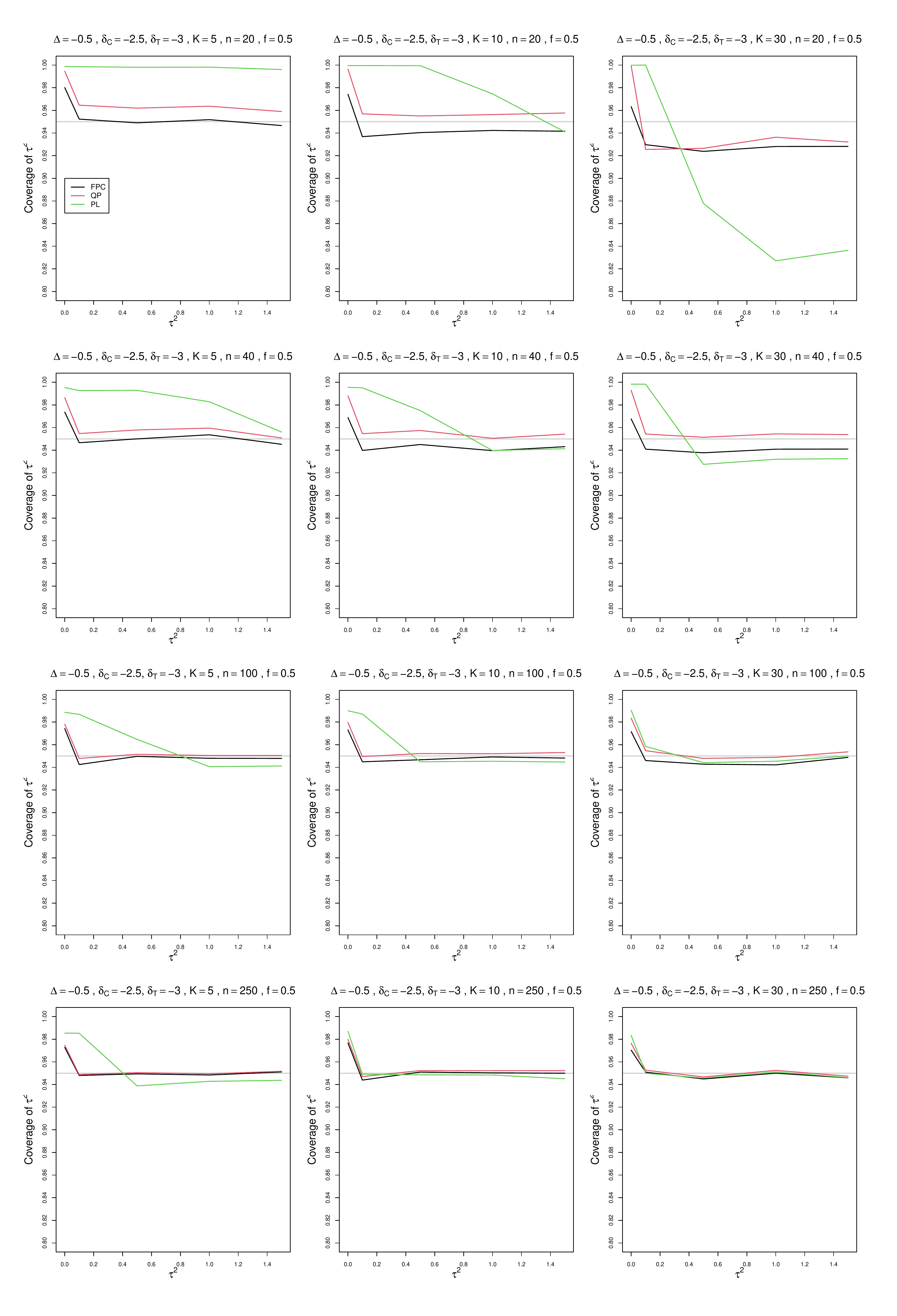}
	\caption{Coverage of PL, QP, and  FPC 95\% confidence intervals for between-study variance of DSM  vs $\tau^2$, for equal sample sizes $n=20,\;40,\;100$ and $250$, $\delta_{iC} = -2.5$, $\Delta=-0.5$ and  $f = 0.5$.   }
	\label{PlotCoverageOfTau2_deltaC_-25deltaT=-3_DSM_equal_sample_sizes.pdf}
\end{figure}

\begin{figure}[ht]
	\centering
	\includegraphics[scale=0.33]{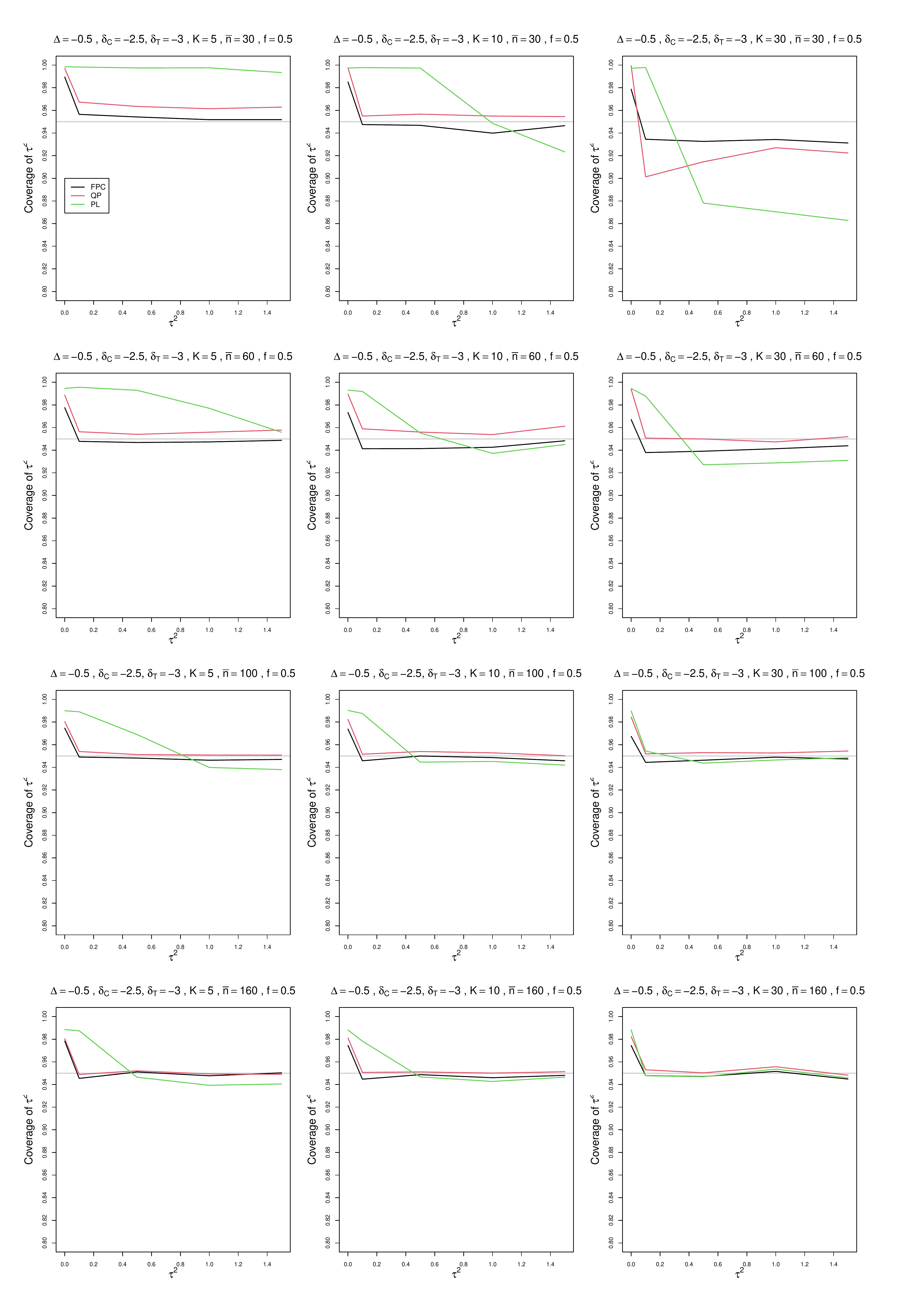}
	\caption{Coverage of PL, QP, and  FPC 95\% confidence intervals for between-study variance of DSM  vs $\tau^2$, for unequal sample sizes $\bar{n}=30,\;60,\;100$ and $160$, $\delta_{iC} = -2.5$, $\Delta=-0.5$ and  $f = 0.5$.   }
	\label{PlotCoverageOfTau2_deltaC_-25deltaT=-3_DSM_unequal_sample_sizes.pdf}
\end{figure}

\begin{figure}[ht]
	\centering
	\includegraphics[scale=0.33]{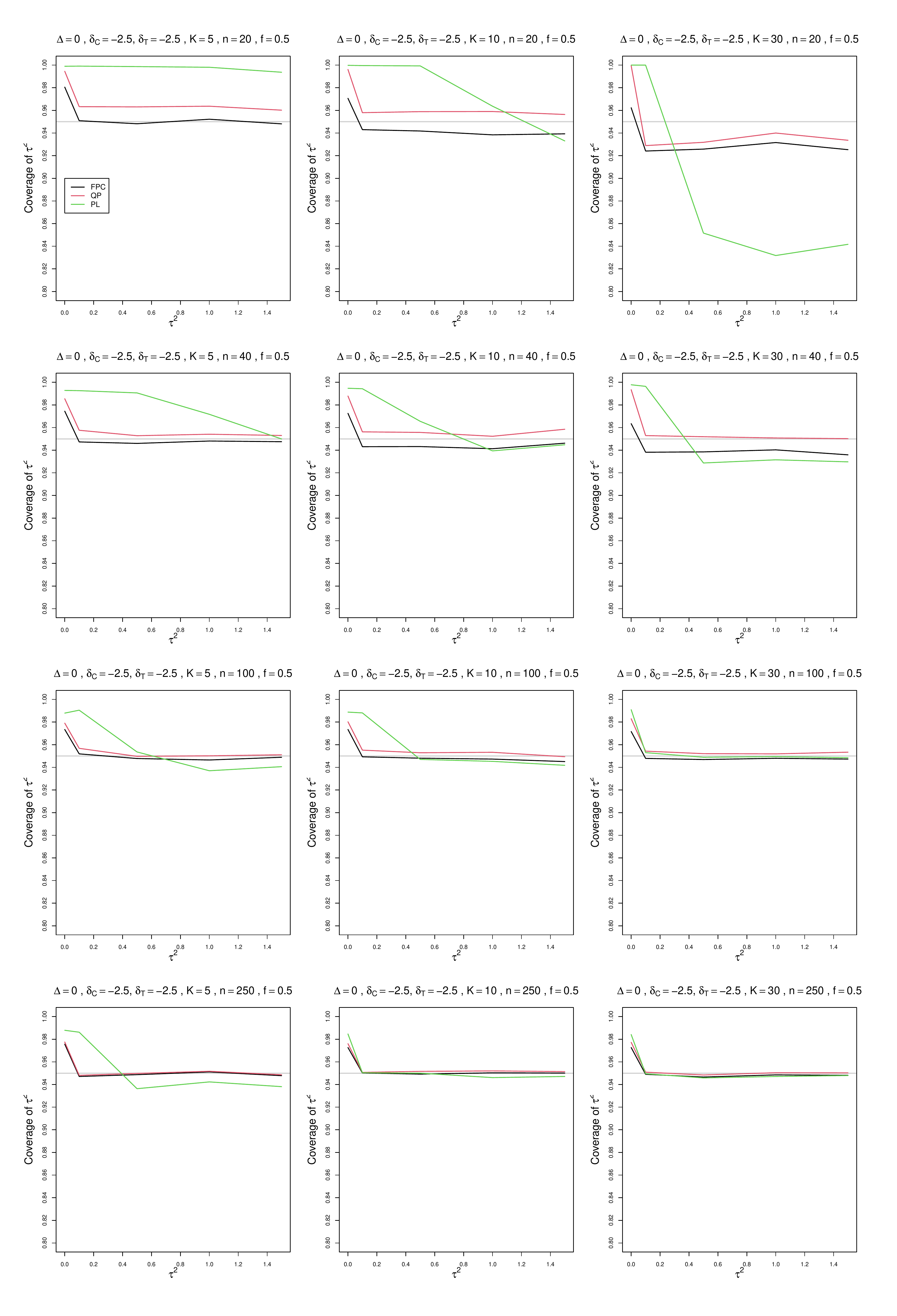}
	\caption{Coverage of PL, QP, and  FPC 95\% confidence intervals for between-study variance of DSM (DL, REML, MP, SMC and SSC ) vs $\tau^2$, for equal sample sizes $n=20,\;40,\;100$ and $250$, $\delta_{iC} = -2.5$, $\Delta=0$ and  $f = 0.5$.   }
	\label{PlotCoverageOfTau2_deltaC_-25deltaT=-2.5_DSM_equal_sample_sizes.pdf}
\end{figure}

\begin{figure}[ht]
	\centering
	\includegraphics[scale=0.33]{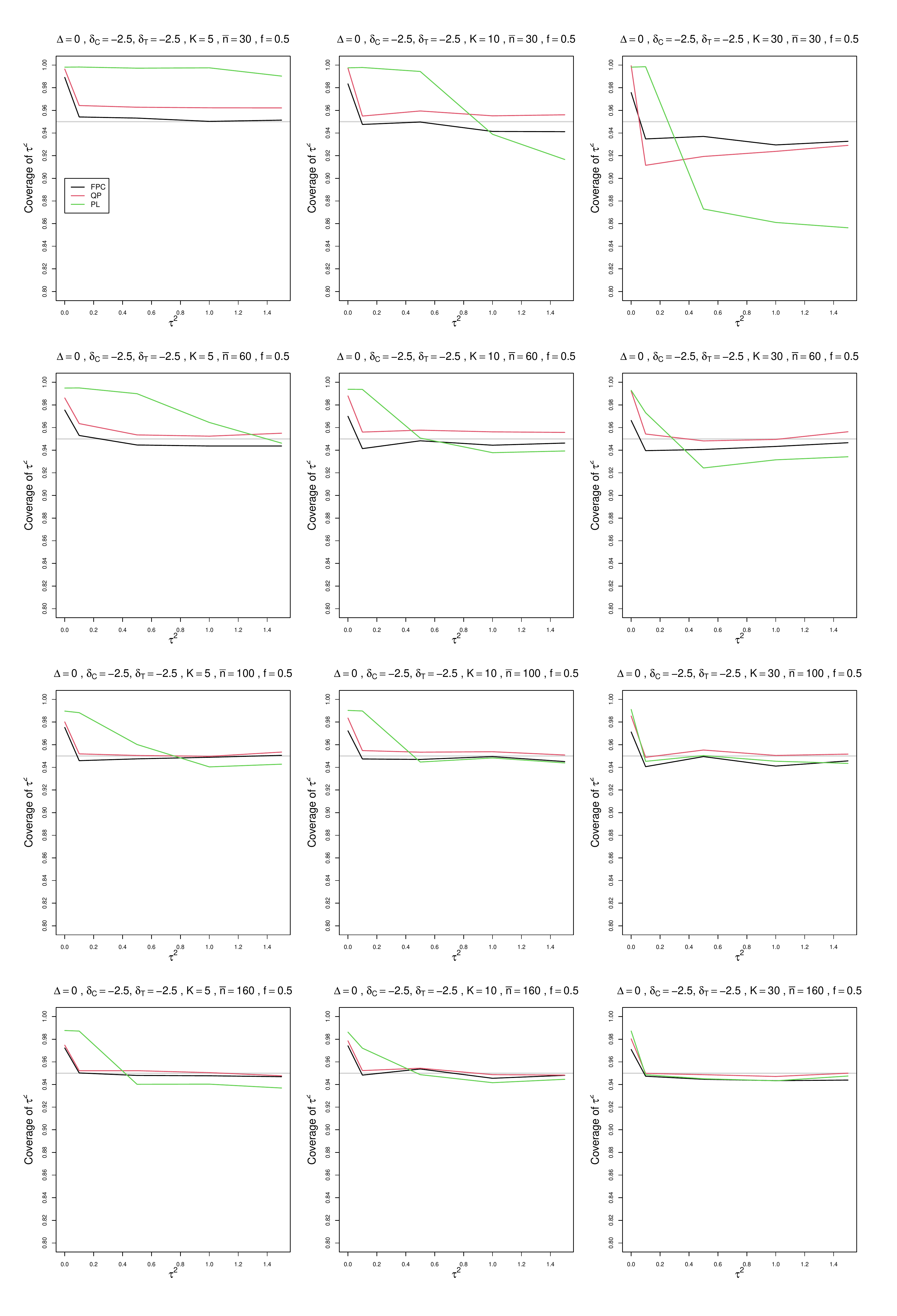}
	\caption{Coverage of PL, QP, and  FPC 95\% confidence intervals for between-study variance of DSM   vs $\tau^2$, for unequal sample sizes $\bar{n}=30,\;60,\;100$ and $160$, $\delta_{iC} = -2.5$, $\Delta=0$ and  $f = 0.5$.   }
	\label{PlotCoverageOfTau2_deltaC_-25deltaT=-2.5_DSM_unequal_sample_sizes.pdf}
\end{figure}

\begin{figure}[ht]
	\centering
	\includegraphics[scale=0.33]{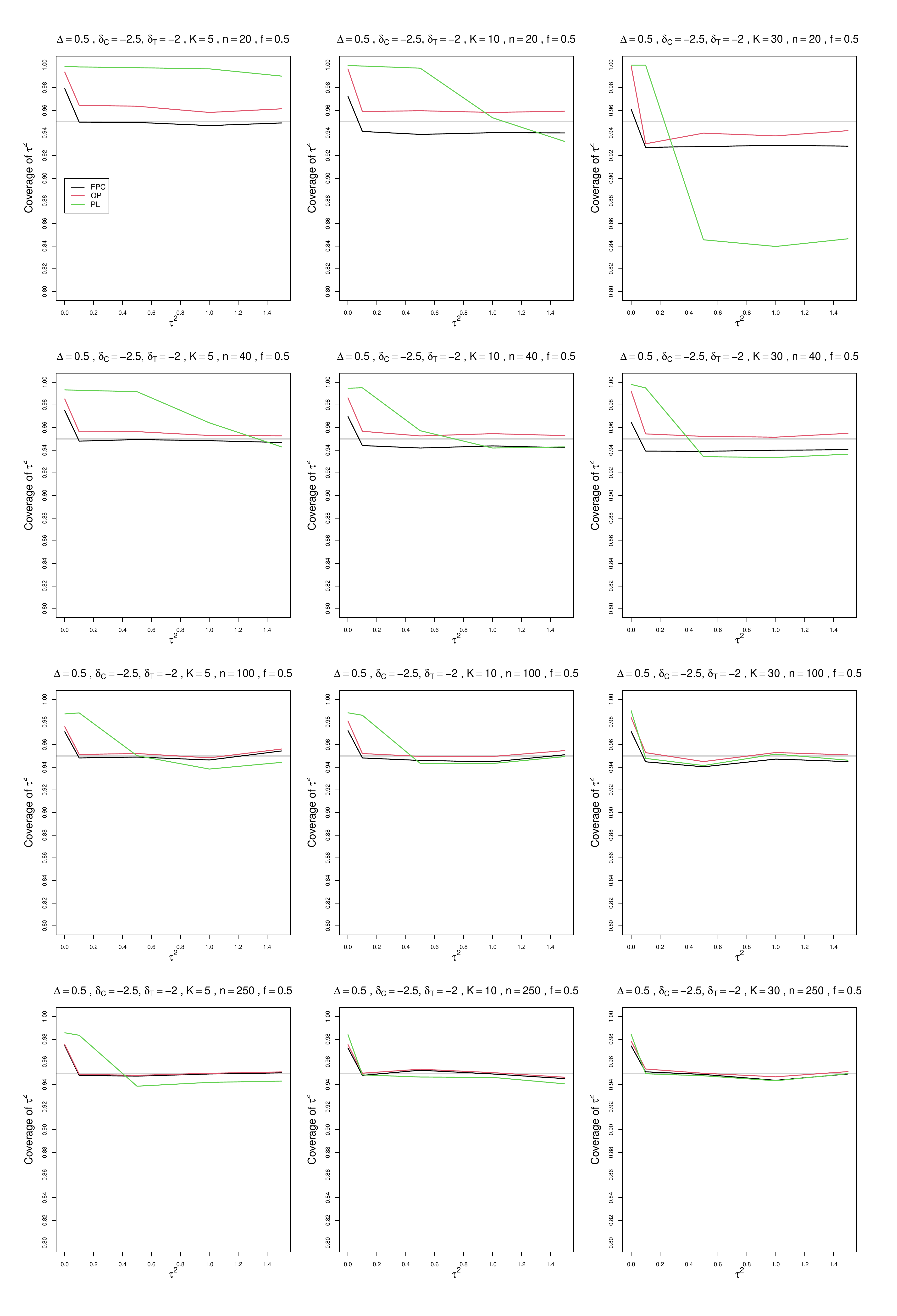}
	\caption{Coverage of PL, QP, and  FPC 95\% confidence intervals for between-study variance of DSM   vs $\tau^2$, for equal sample sizes $n=20,\;40,\;100$ and $250$, $\delta_{iC} = -2.5$, $\Delta=0.5$ and  $f = 0.5$.   }
	\label{PlotCoverageOfTau2_deltaC_-25deltaT=-2_DSM_equal_sample_sizes.pdf}
\end{figure}

\begin{figure}[ht]
	\centering
	\includegraphics[scale=0.33]{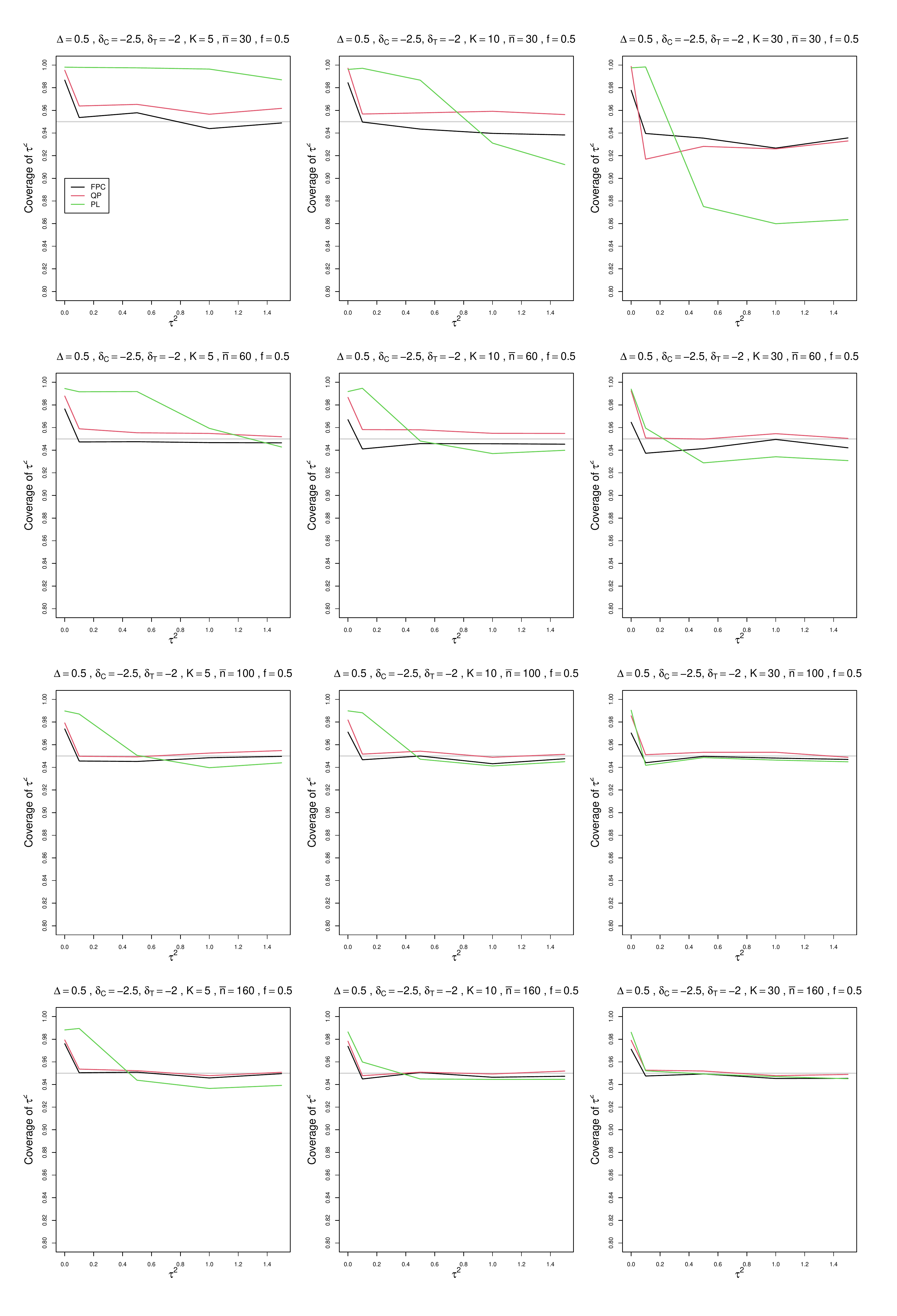}
	\caption{Coverage of PL, QP, and  FPC 95\% confidence intervals for between-study variance of DSM   vs $\tau^2$, for unequal sample sizes $\bar{n}=30,\;60,\;100$ and $160$, $\delta_{iC} = -2.5$, $\Delta=0.5$ and  $f = 0.5$.   }
	\label{PlotCoverageOfTau2_deltaC_-25deltaT=-2_DSM_unequal_sample_sizes.pdf}
\end{figure}

\begin{figure}[ht]
	\centering
	\includegraphics[scale=0.33]{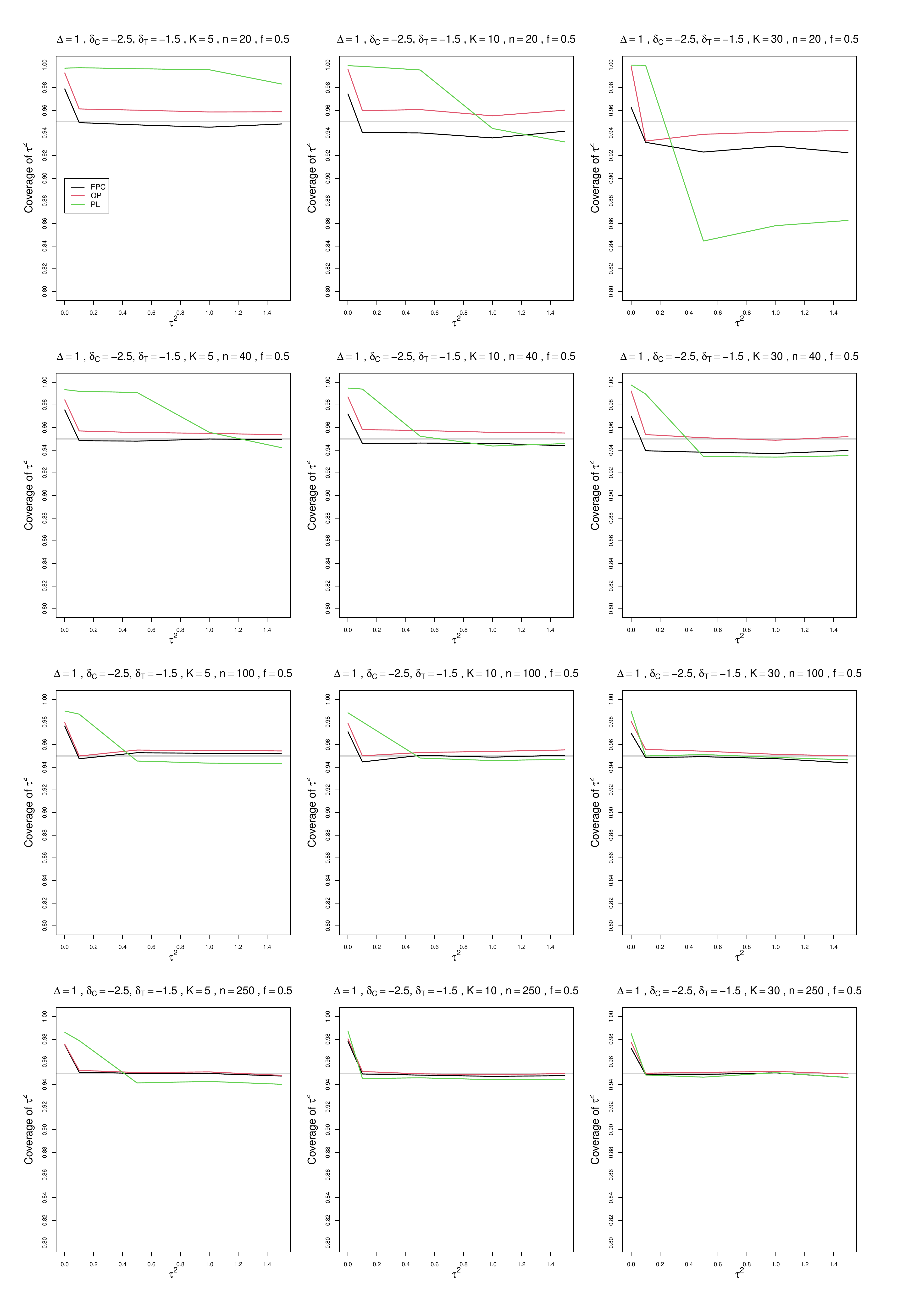}
	\caption{Coverage of PL, QP, and  FPC 95\% confidence intervals for between-study variance of DSM   vs $\tau^2$, for equal sample sizes $n=20,\;40,\;100$ and $250$, $\delta_{iC} = -2.5$, $\Delta=1$ and  $f = 0.5$.   }
	\label{PlotCoverageOfTau2_deltaC_-25deltaT=-1.5_DSM_equal_sample_sizes.pdf}
\end{figure}

\begin{figure}[ht]
	\centering
	\includegraphics[scale=0.33]{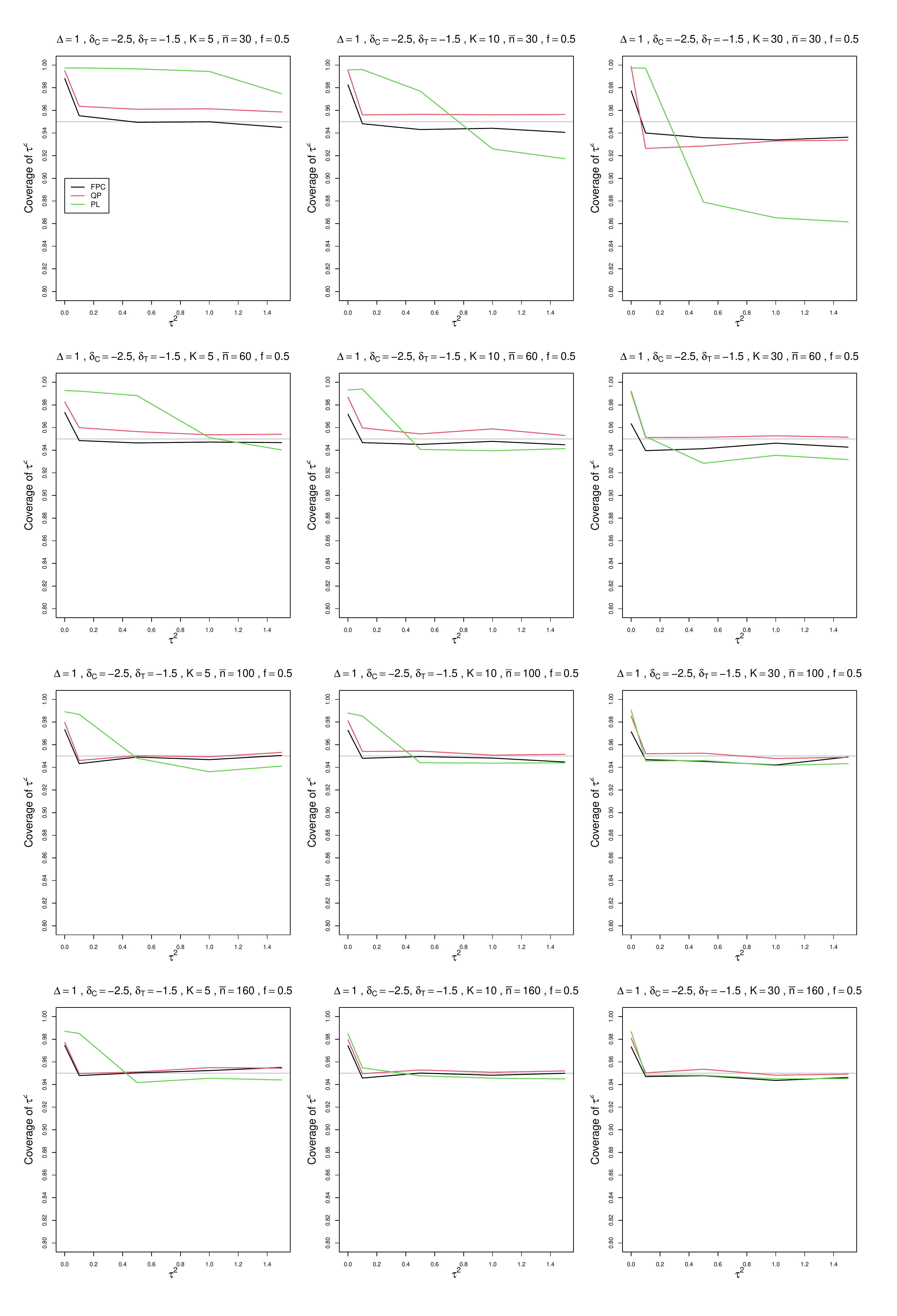}
	\caption{Coverage of PL, QP, and  FPC 95\% confidence intervals for between-study variance of DSM   vs $\tau^2$, for unequal sample sizes $\bar{n}=30,\;60,\;100$ and $160$, $\delta_{iC} = -2.5$, $\Delta=1$ and  $f = 0.5$.   }
	\label{PlotCoverageOfTau2_deltaC_-25deltaT=-1.5_DSM_unequal_sample_sizes.pdf}
\end{figure}

\begin{figure}[ht]
	\centering
	\includegraphics[scale=0.33]{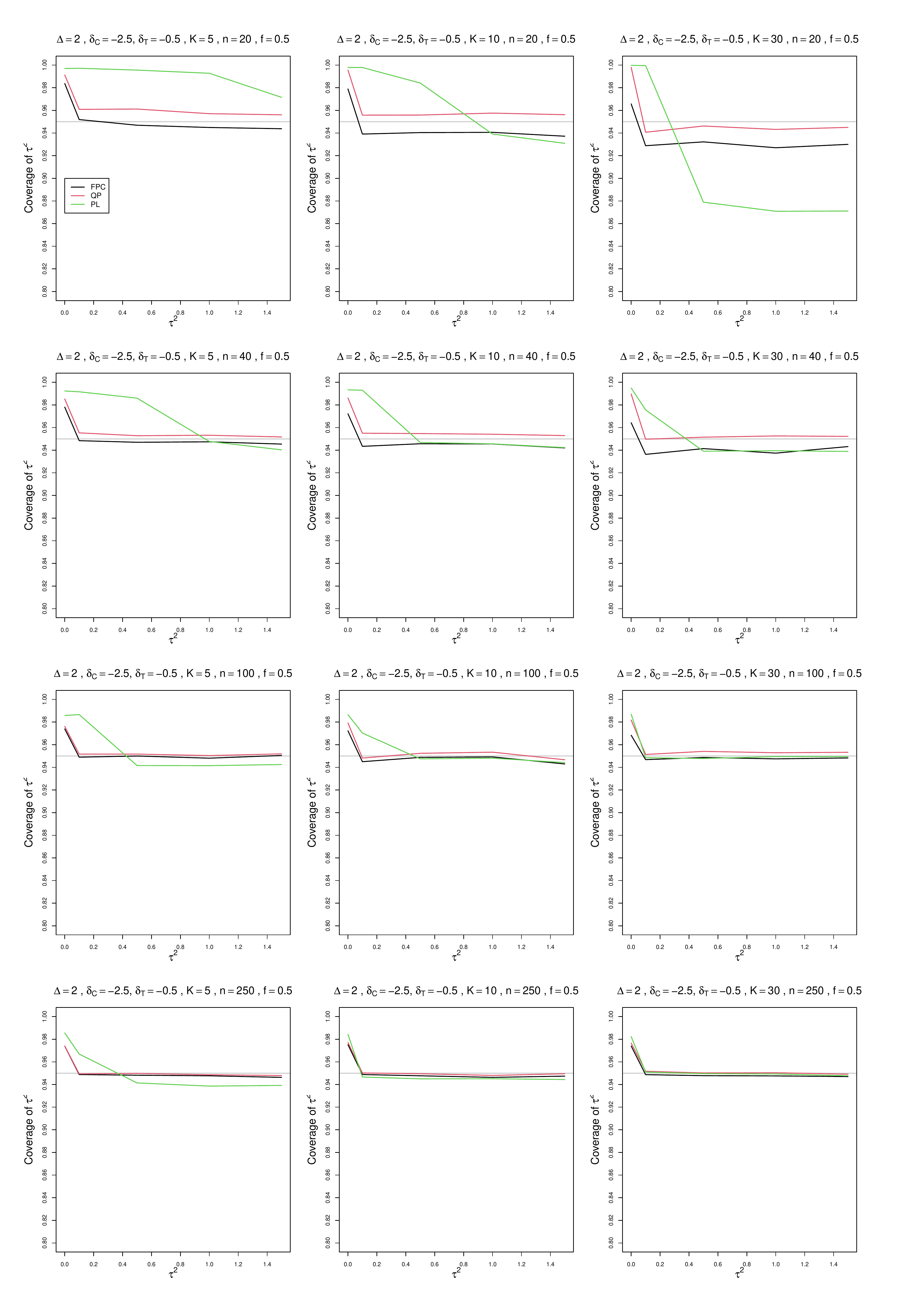}
	\caption{Coverage of PL, QP, and  FPC 95\% confidence intervals for between-study variance of DSM   vs $\tau^2$, for equal sample sizes $n=20,\;40,\;100$ and $250$, $\delta_{iC} = -2.5$, $\Delta=2$ and  $f = 0.5$.   }
	\label{PlotCoverageOfTau2_deltaC_-25deltaT=-0.5_DSM_equal_sample_sizes.pdf}
\end{figure}

\begin{figure}[ht]
	\centering
	\includegraphics[scale=0.33]{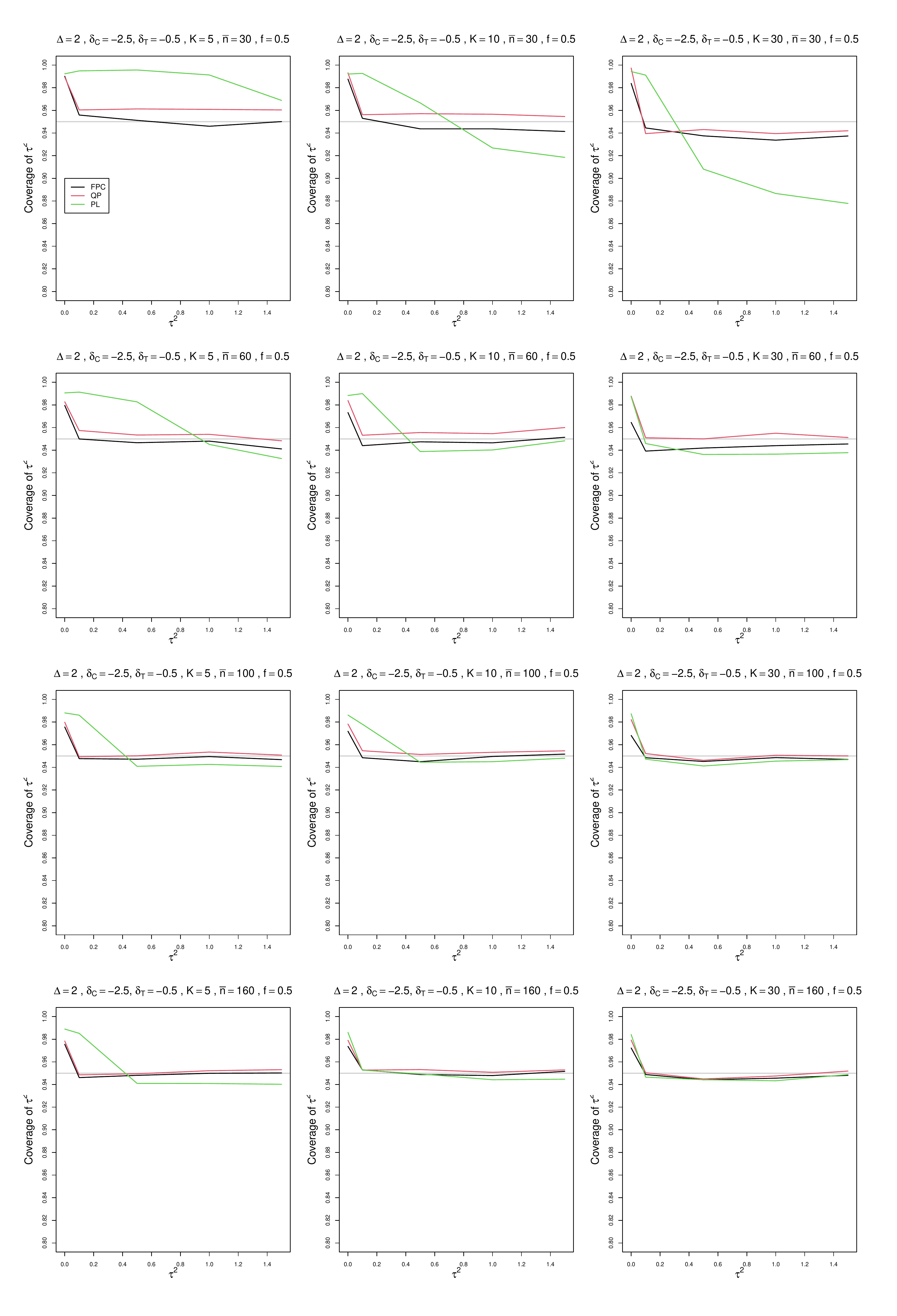}
	\caption{Coverage of PL, QP, and  FPC 95\% confidence intervals for between-study variance of DSM   vs $\tau^2$, for unequal sample sizes $\bar{n}=30,\;60,\;100$ and $160$, $\delta_{iC} = -2.5$, $\Delta=2$ and  $f = 0.5$.   }
	\label{PlotCoverageOfTau2_deltaC_-25deltaT=-0.5_DSM_unequal_sample_sizes.pdf}
\end{figure}

\clearpage
\begin{figure}[ht]
	\centering
	\includegraphics[scale=0.33]{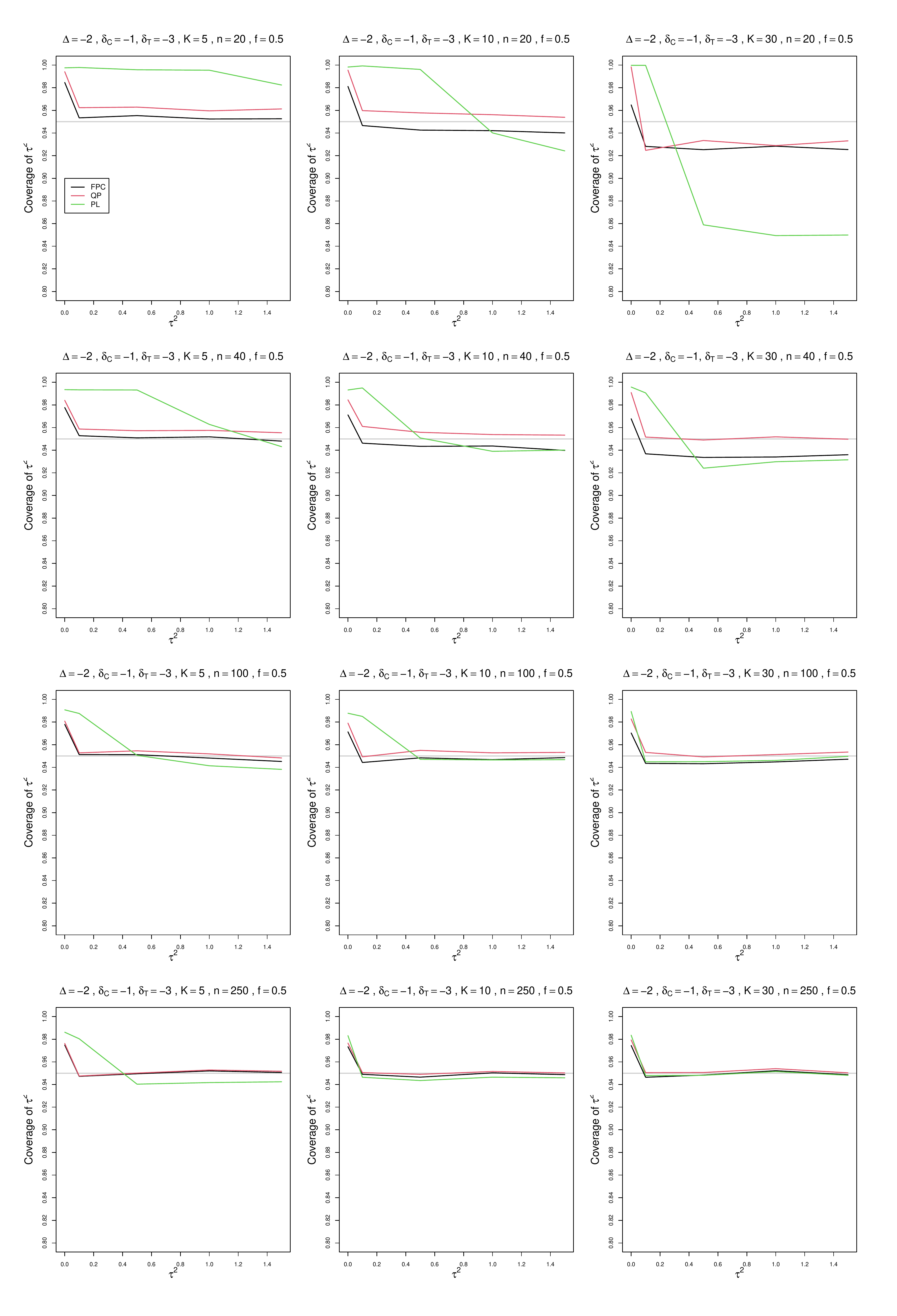}
	\caption{Coverage of PL, QP, and  FPC 95\% confidence intervals for between-study variance of DSM   vs $\tau^2$, for equal sample sizes $n=20,\;40,\;100$ and $250$, $\delta_{iC} = -2.5$, $\Delta=-2$ and  $f = 0.5$.   }
	\label{PlotCoverageOfTau2_deltaC_-1deltaT=-3_DSM_equal_sample_sizes.pdf}
\end{figure}

\begin{figure}[ht]
	\centering
	\includegraphics[scale=0.33]{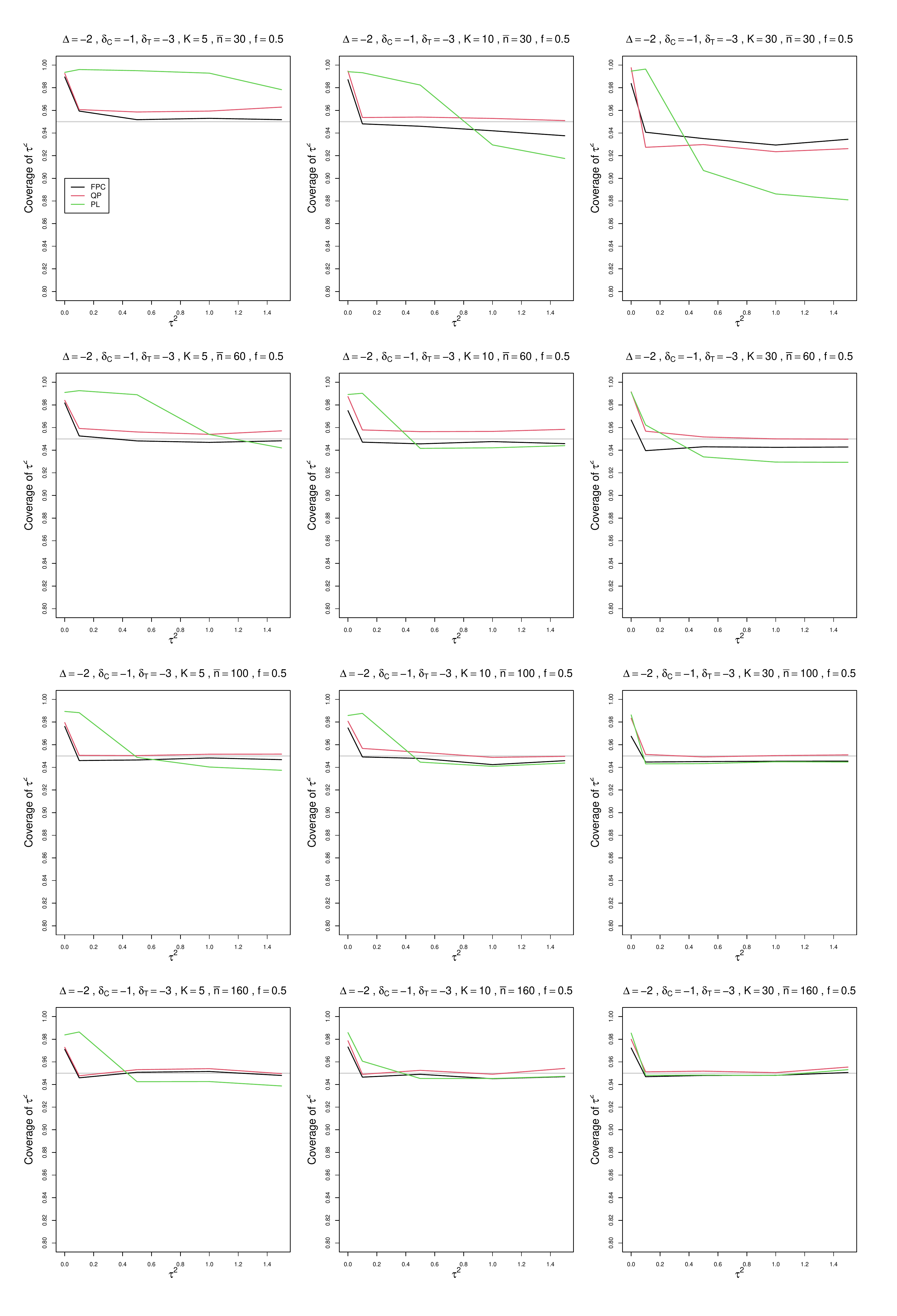}
	\caption{Coverage of PL, QP, and  FPC 95\% confidence intervals for between-study variance of DSM   vs $\tau^2$, for unequal sample sizes $\bar{n}=30,\;60,\;100$ and $160$, $\delta_{iC} = -1$, $\Delta=-2$ and  $f = 0.5$.   }
	\label{PlotCoverageOfTau2_deltaC_-1deltaT=-3_DSM_unequal_sample_sizes.pdf}
\end{figure}

\begin{figure}[ht]
	\centering
	\includegraphics[scale=0.33]{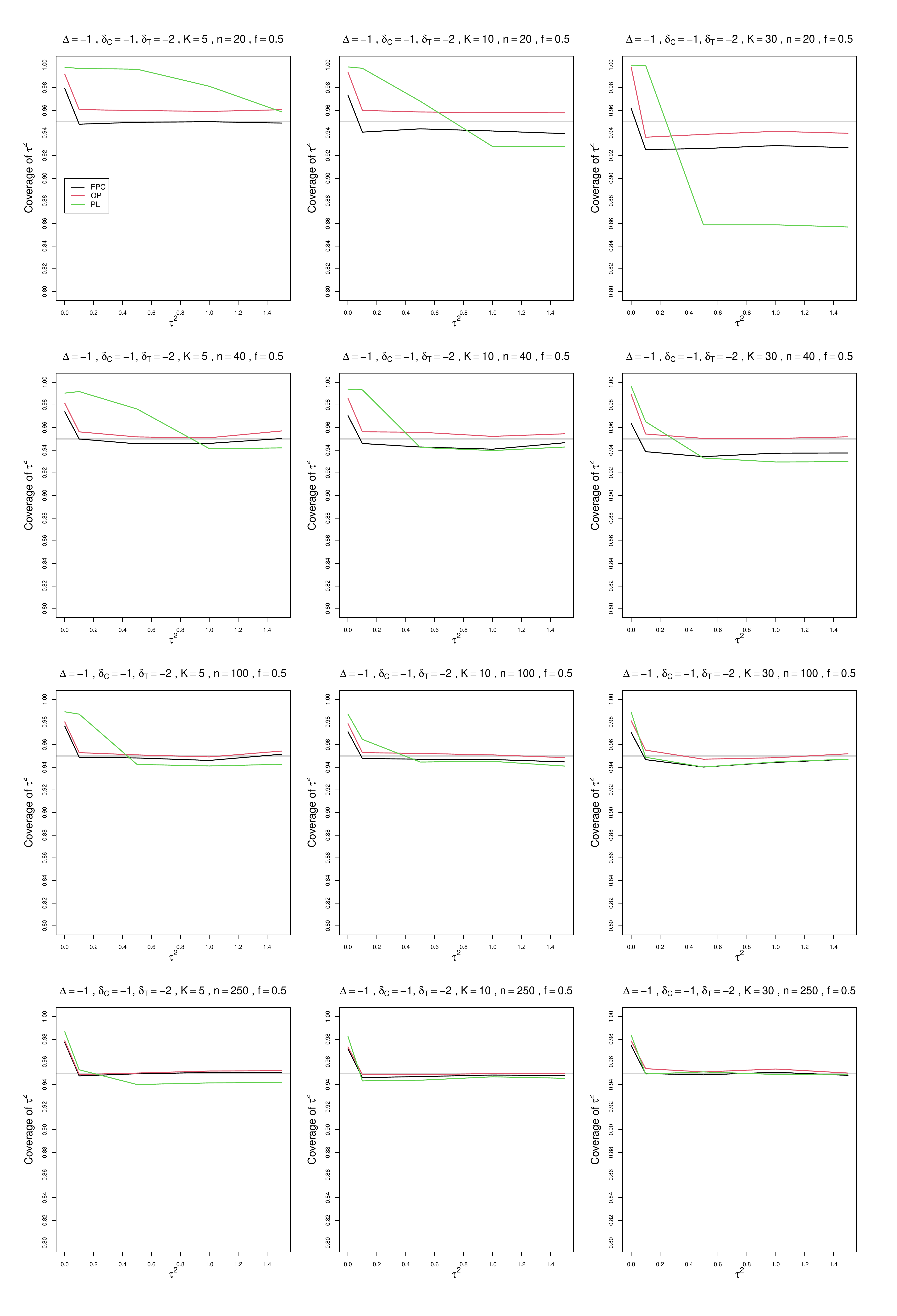}
	\caption{Coverage of PL, QP, and  FPC 95\% confidence intervals for between-study variance of DSM   vs $\tau^2$, for equal sample sizes $n=20,\;40,\;100$ and $250$, $\delta_{iC} = -1$, $\Delta=-1$ and  $f = 0.5$.   }
	\label{PlotCoverageOfTau2_deltaC_-1deltaT=-2_DSM_equal_sample_sizes.pdf}
\end{figure}

\begin{figure}[ht]
	\centering
	\includegraphics[scale=0.33]{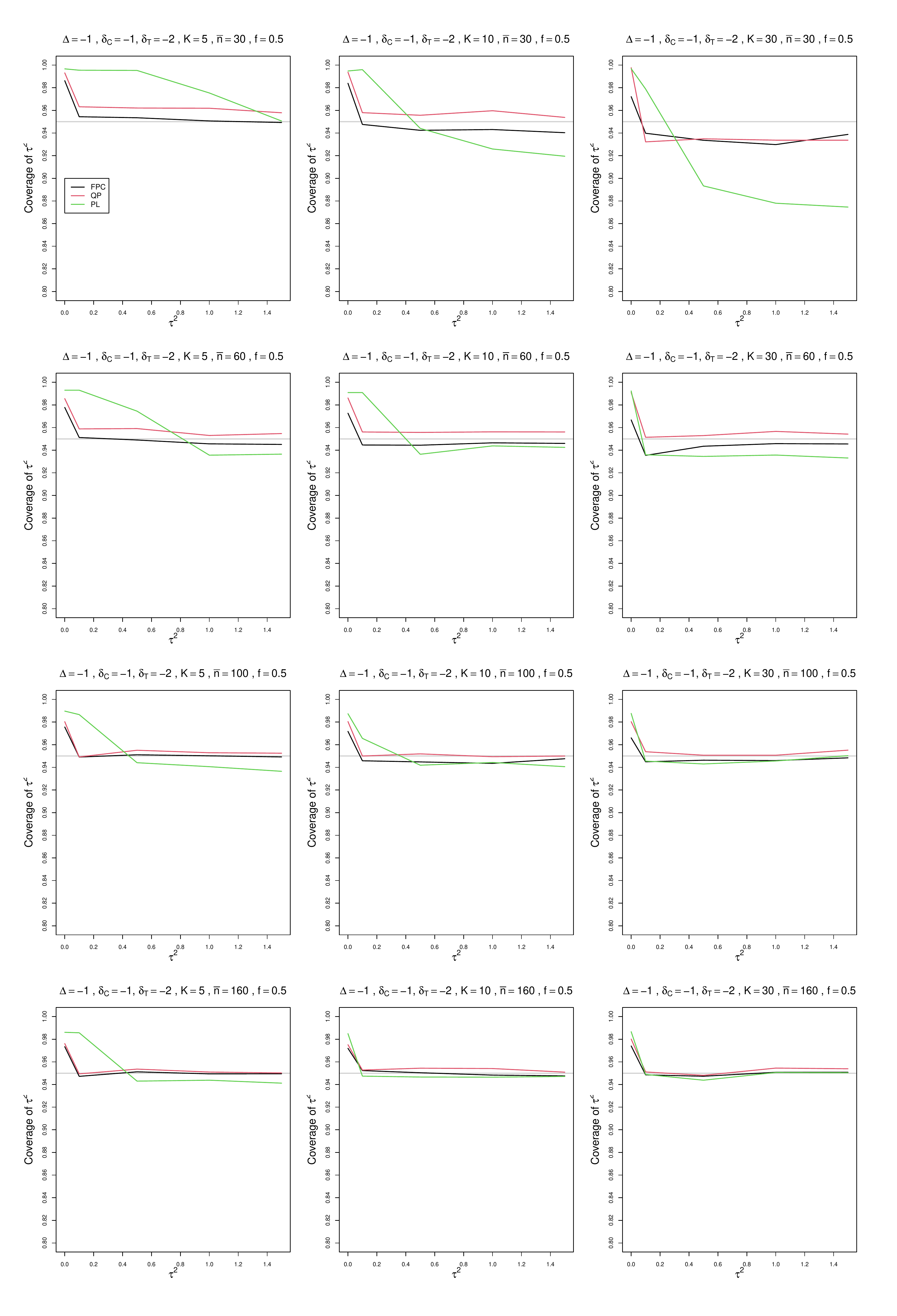}
	\caption{Coverage of PL, QP, and  FPC 95\% confidence intervals for between-study variance of DSM   vs $\tau^2$, for unequal sample sizes $\bar{n}=30,\;60,\;100$ and $160$, $\delta_{iC} = -1$, $\Delta=-1$ and  $f = 0.5$.   }
	\label{PlotCoverageOfTau2_deltaC_-1deltaT=-2_DSM_unequal_sample_sizes.pdf}
\end{figure}

\begin{figure}[ht]
	\centering
	\includegraphics[scale=0.33]{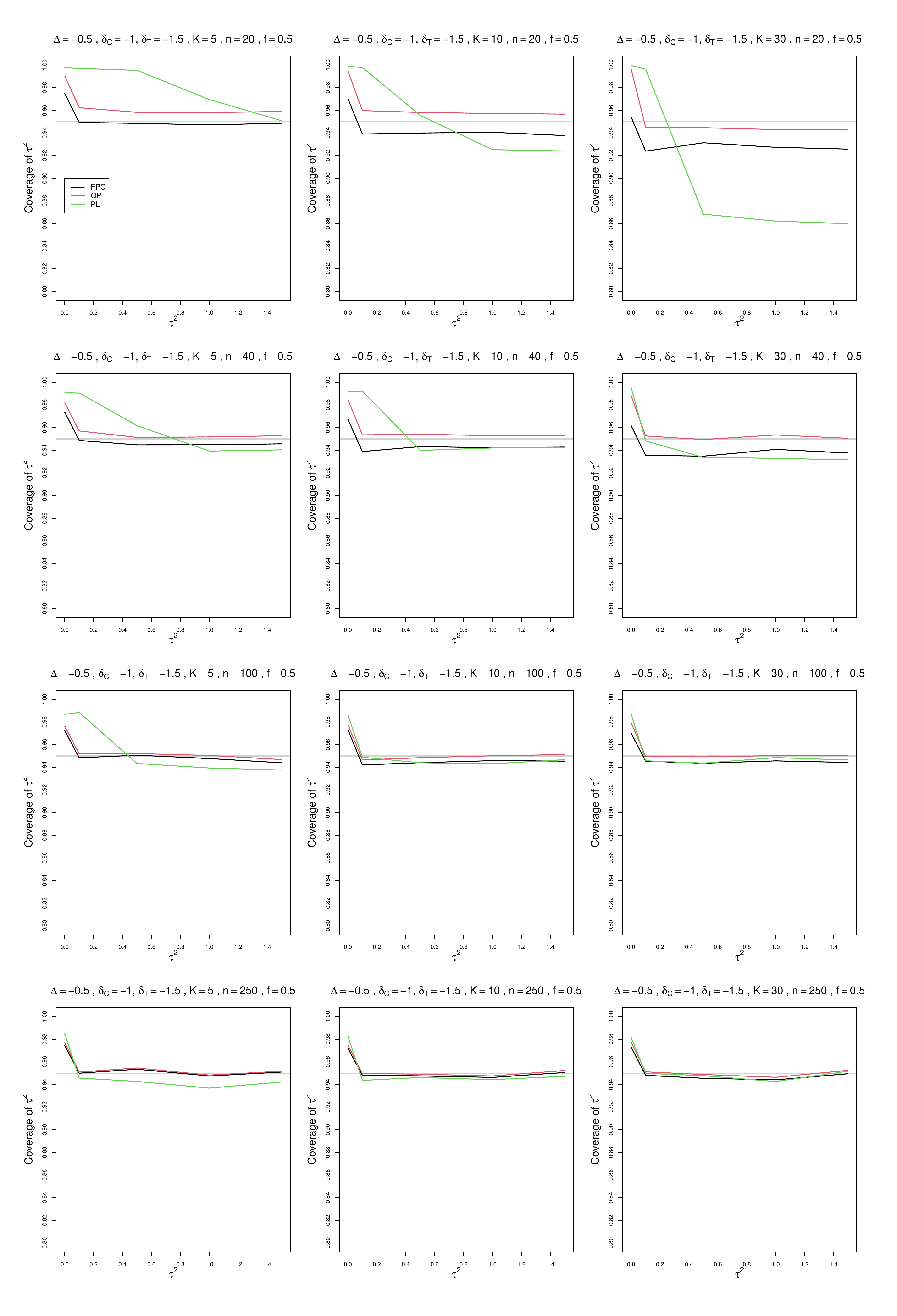}
	\caption{Coverage of PL, QP, and  FPC 95\% confidence intervals for between-study variance of DSM   vs $\tau^2$, for equal sample sizes $n=20,\;40,\;100$ and $250$, $\delta_{iC} = -1$, $\Delta=-0.5$ and  $f = 0.5$.   }
	\label{PlotCoverageOfTau2_deltaC_-1deltaT=-1.5_DSM_equal_sample_sizes.pdf}
\end{figure}

\begin{figure}[ht]
	\centering
	\includegraphics[scale=0.33]{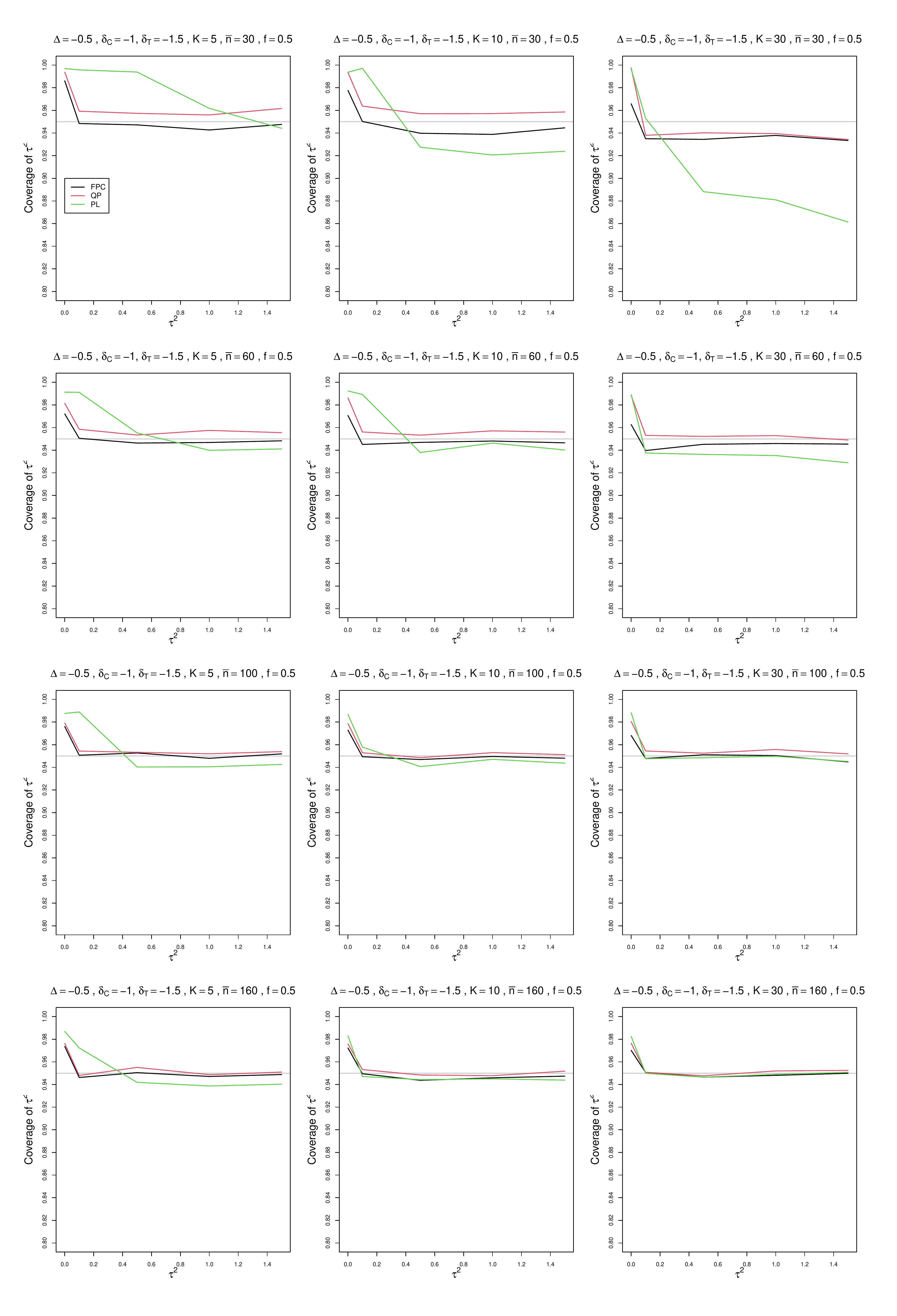}
	\caption{Coverage of PL, QP, and  FPC 95\% confidence intervals for between-study variance of DSM   vs $\tau^2$, for unequal sample sizes $\bar{n}=30,\;60,\;100$ and $160$, $\delta_{iC} = -1$, $\Delta=-0.5$ and  $f = 0.5$.   }
	\label{PlotCoverageOfTau2_deltaC_-1deltaT=-1,5_DSM_unequal_sample_sizes.pdf}
\end{figure}

\begin{figure}[ht]
	\centering
	\includegraphics[scale=0.33]{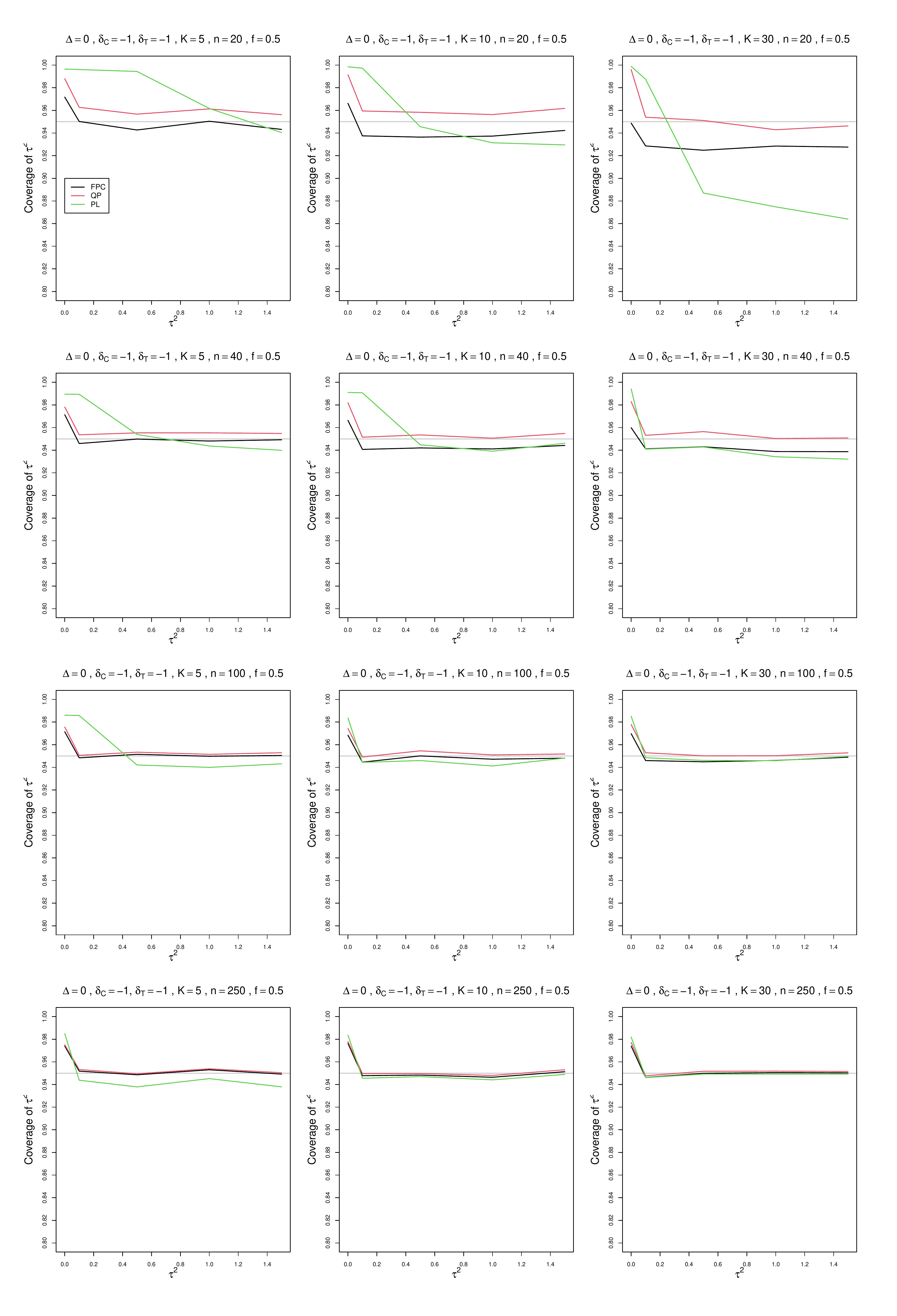}
	\caption{Coverage of PL, QP, and  FPC 95\% confidence intervals for between-study variance of DSM   vs $\tau^2$, for equal sample sizes $n=20,\;40,\;100$ and $250$, $\delta_{iC} = -1$, $\Delta=0$ and  $f = 0.5$.   }
	\label{PlotCoverageOfTau2_deltaC_-1deltaT=-1_DSM_equal_sample_sizes.pdf}
\end{figure}

\begin{figure}[ht]
	\centering
	\includegraphics[scale=0.33]{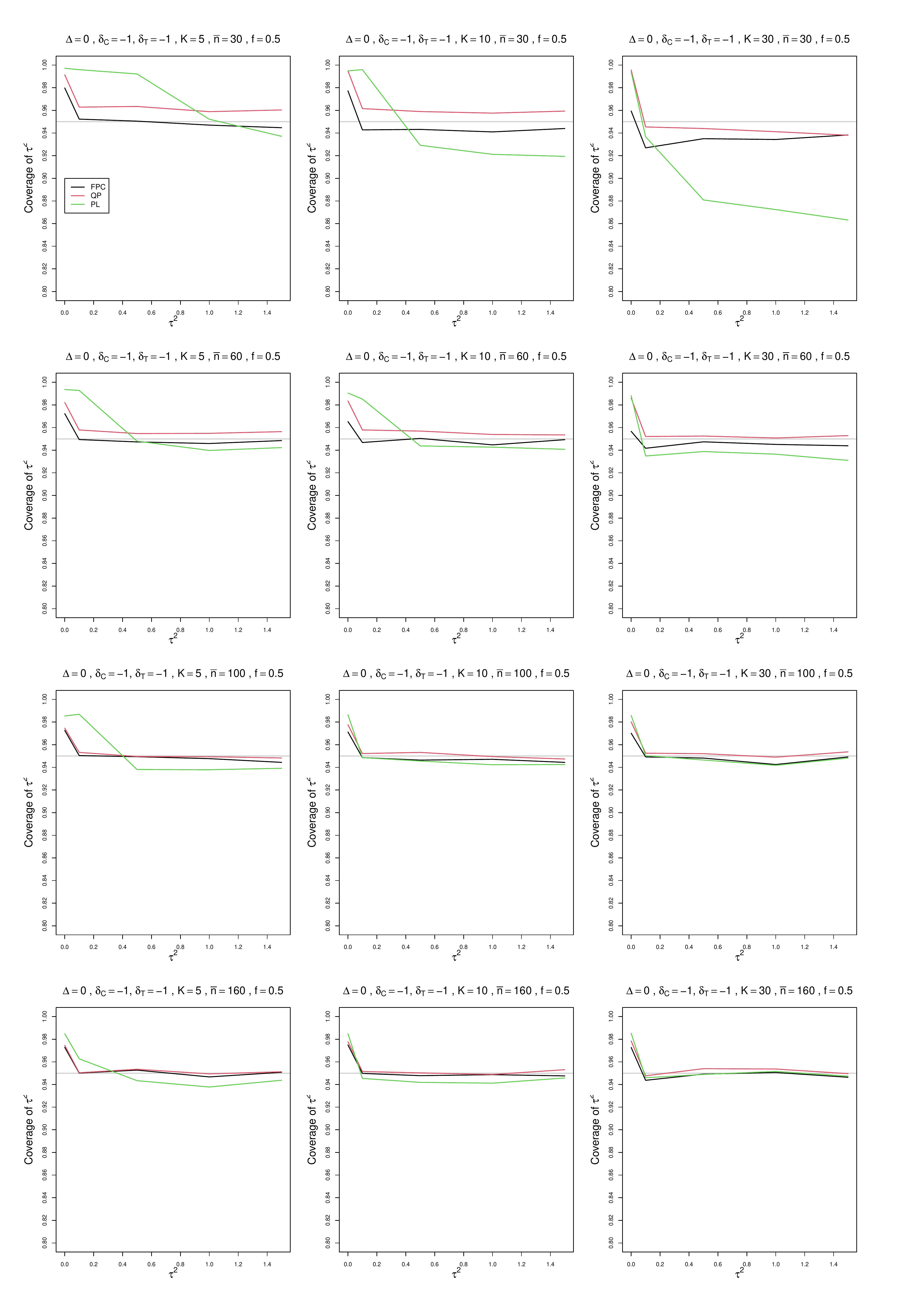}
	\caption{Coverage of PL, QP, and  FPC 95\% confidence intervals for between-study variance of DSM   vs $\tau^2$, for unequal sample sizes $\bar{n}=30,\;60,\;100$ and $160$, $\delta_{iC} = -1$, $\Delta=0$ and  $f = 0.5$.   }
	\label{PlotCoverageOfTau2_deltaC_-1deltaT=-1_DSM_unequal_sample_sizes.pdf}
\end{figure}

\begin{figure}[ht]
	\centering
	\includegraphics[scale=0.33]{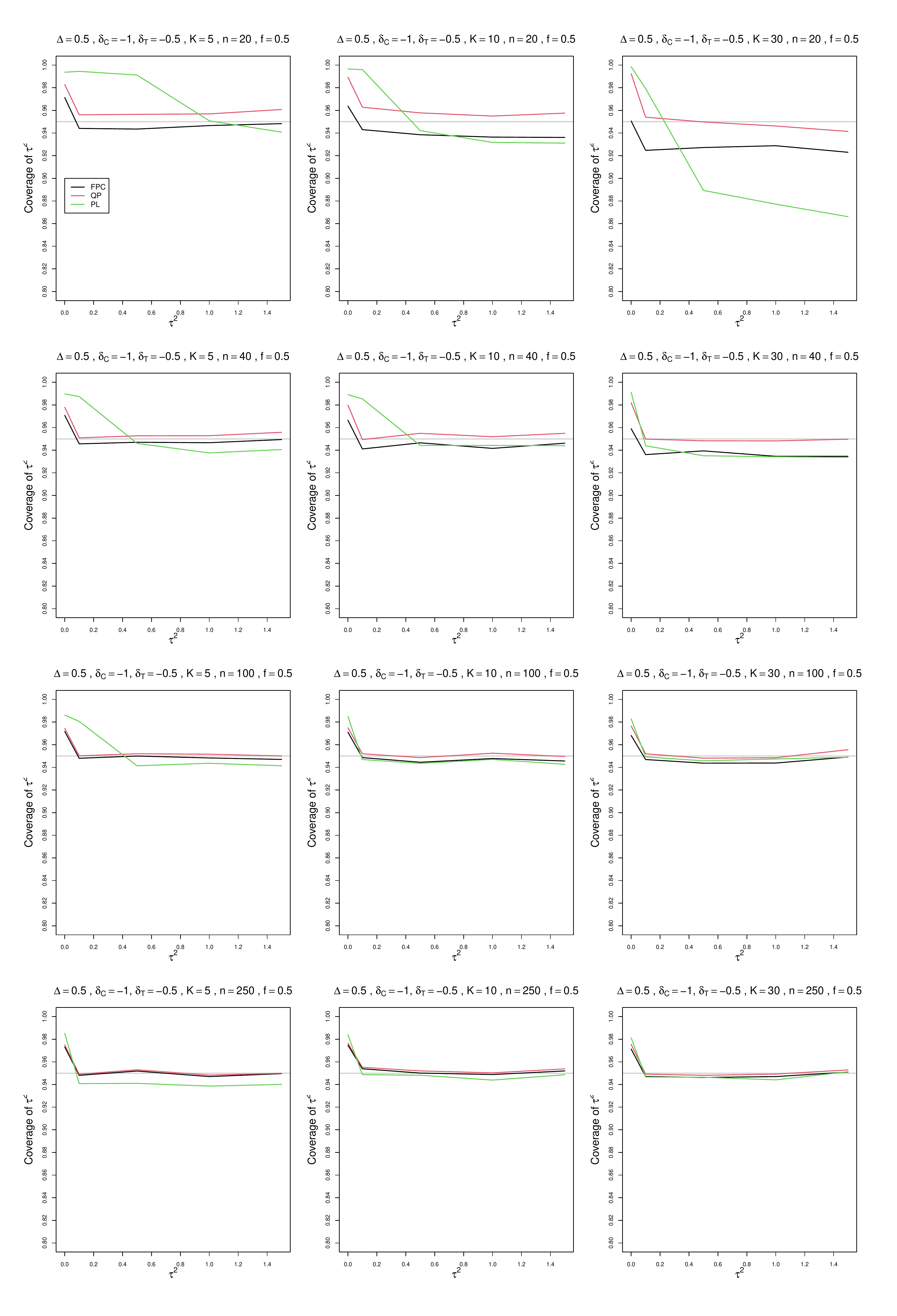}
	\caption{Coverage of PL, QP, and  FPC 95\% confidence intervals for between-study variance of DSM   vs $\tau^2$, for equal sample sizes $n=20,\;40,\;100$ and $250$, $\delta_{iC} = -1$, $\Delta=0.5$ and  $f = 0.5$.   }
	\label{PlotCoverageOfTau2_deltaC_-1deltaT=-0.5_DSM_equal_sample_sizes.pdf}
\end{figure}

\begin{figure}[ht]
	\centering
	\includegraphics[scale=0.33]{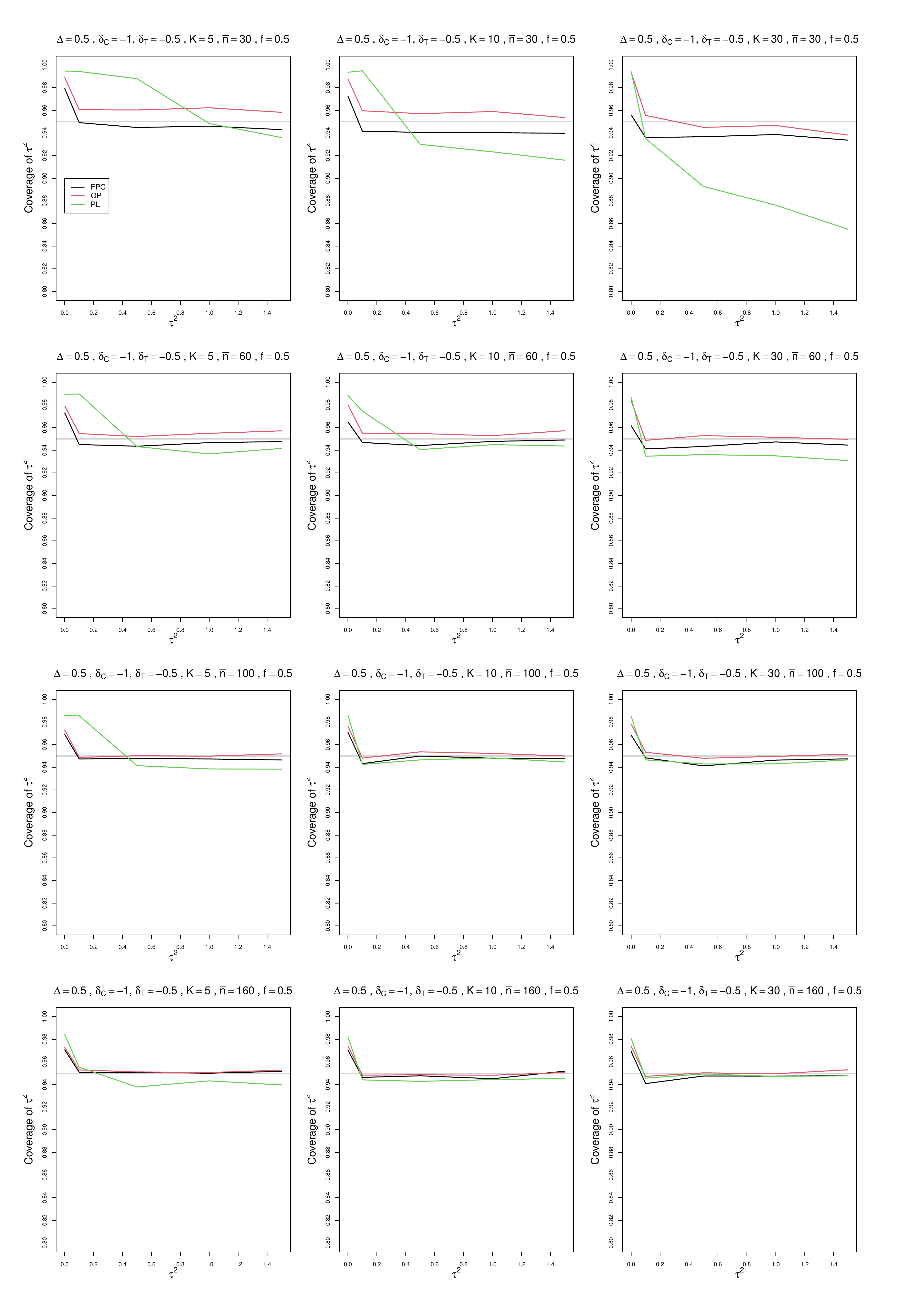}
	\caption{Coverage of PL, QP, and  FPC 95\% confidence intervals for between-study variance of DSM   vs $\tau^2$, for unequal sample sizes $\bar{n}=30,\;60,\;100$ and $160$, $\delta_{iC} = -1$, $\Delta=0.5$ and  $f = 0.5$.   }
	\label{PlotCoverageOfTau2_deltaC_-1deltaT=-0,5_DSM_unequal_sample_sizes.pdf}
\end{figure}

\begin{figure}[ht]
	\centering
	\includegraphics[scale=0.33]{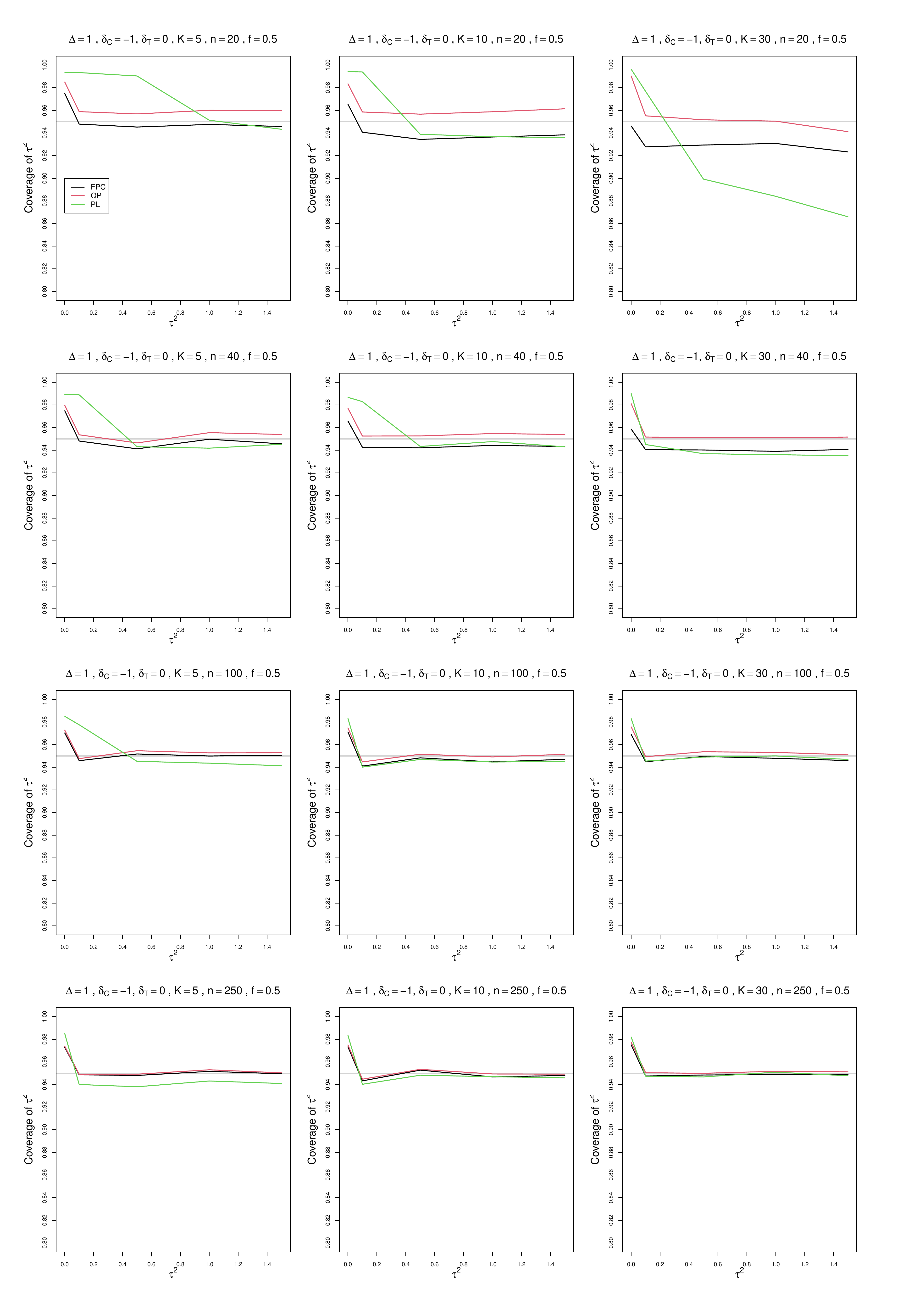}
	\caption{Coverage of PL, QP, and  FPC 95\% confidence intervals for between-study variance of DSM   vs $\tau^2$, for equal sample sizes $n=20,\;40,\;100$ and $250$, $\delta_{iC} = -1$, $\Delta=1$ and  $f = 0.5$.   }
	\label{PlotCoverageOfTau2_deltaC_--1deltaT=0_DSM_equal_sample_sizes.pdf}
\end{figure}

\begin{figure}[ht]
	\centering
	\includegraphics[scale=0.33]{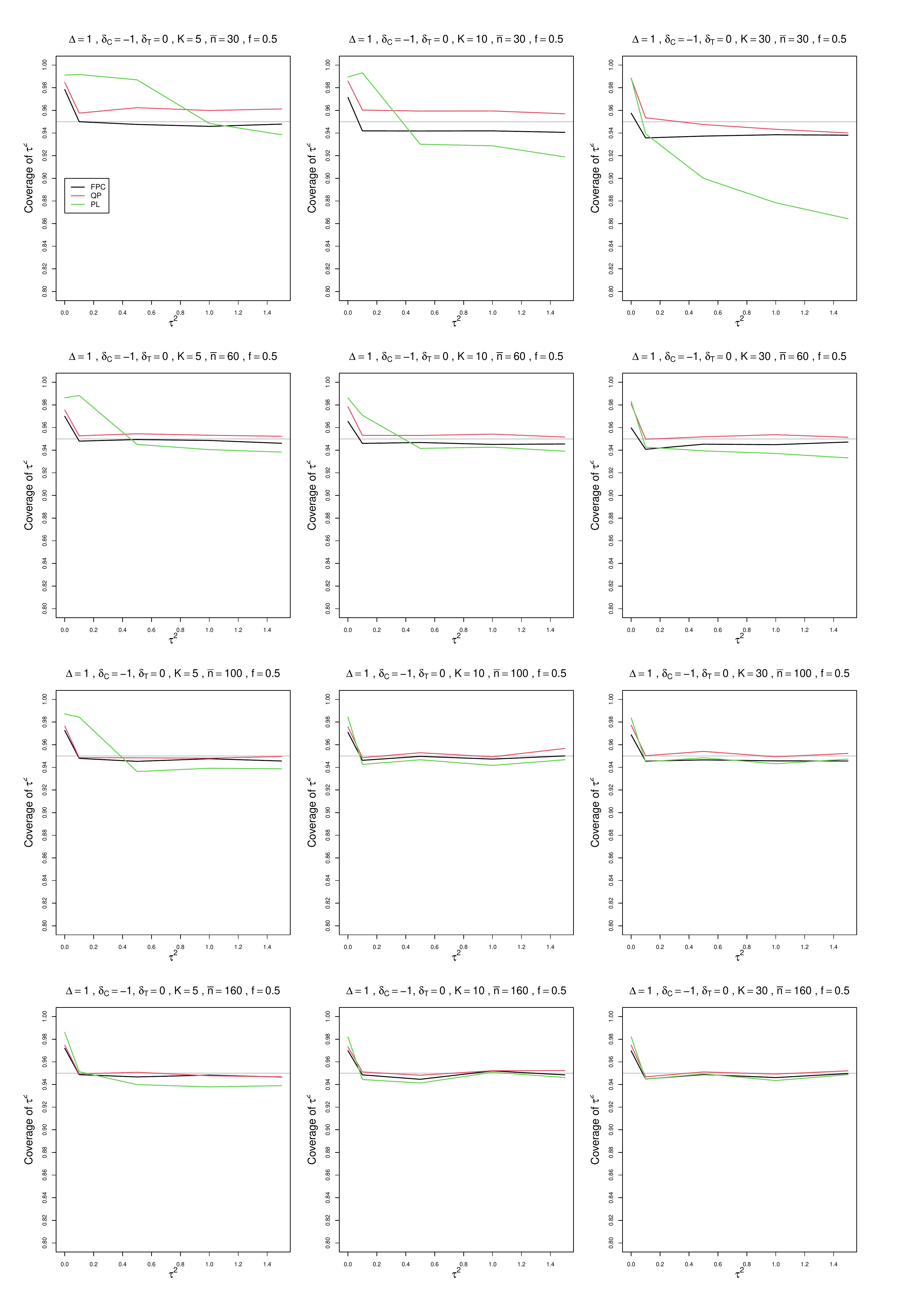}
	\caption{Coverage of PL, QP, and  FPC 95\% confidence intervals for between-study variance of DSM   vs $\tau^2$, for unequal sample sizes $\bar{n}=30,\;60,\;100$ and $160$, $\delta_{iC} = -1$, $\Delta=1$ and  $f = 0.5$.   }
	\label{PlotCoverageOfTau2_deltaC_-1deltaT=0_DSM_unequal_sample_sizes.pdf}
\end{figure}

\begin{figure}[ht]
	\centering
	\includegraphics[scale=0.33]{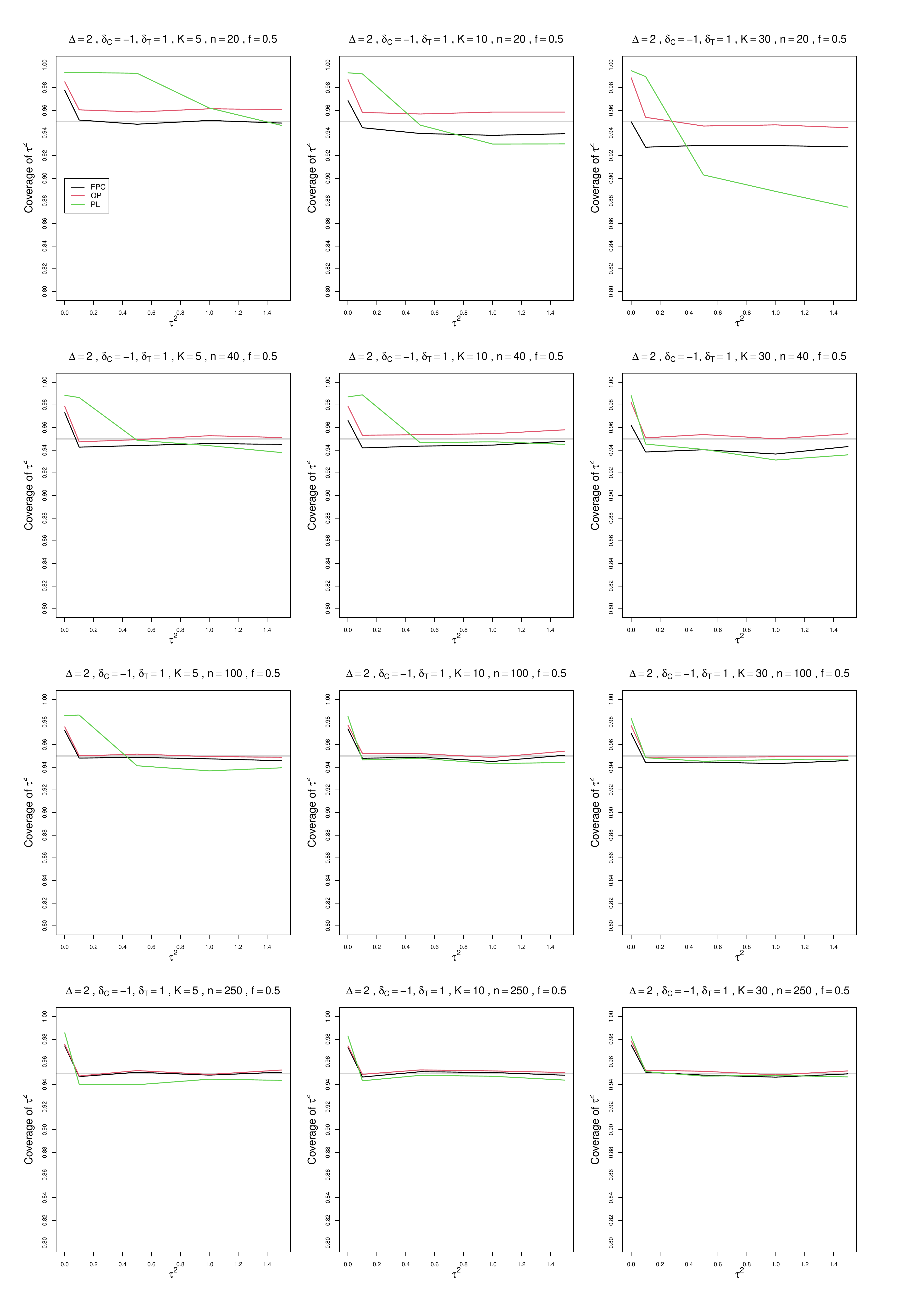}
	\caption{Coverage of PL, QP, and  FPC 95\% confidence intervals for between-study variance of DSM   vs $\tau^2$, for equal sample sizes $n=20,\;40,\;100$ and $250$, $\delta_{iC} = -1$, $\Delta=2$ and  $f = 0.5$.   }
	\label{PlotCoverageOfTau2_deltaC_-1deltaT=1_DSM_equal_sample_sizes.pdf}
\end{figure}

\begin{figure}[ht]
	\centering
	\includegraphics[scale=0.33]{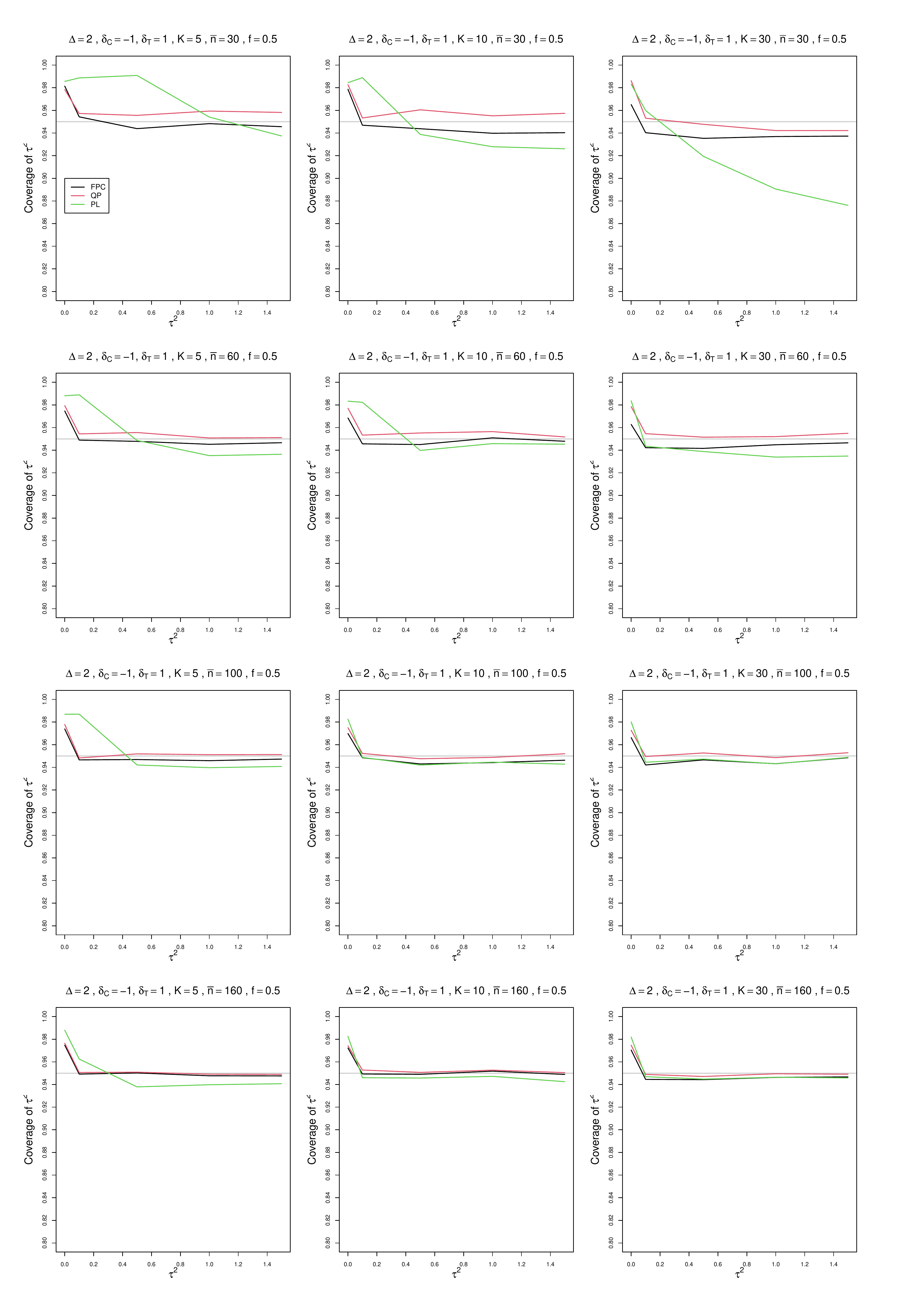}
	\caption{Coverage of PL, QP, and  FPC 95\% confidence intervals for between-study variance of DSM   vs $\tau^2$, for unequal sample sizes $\bar{n}=30,\;60,\;100$ and $160$, $\delta_{iC} = -1$, $\Delta=2$ and  $f = 0.5$.   }
	\label{PlotCoverageOfTau2_deltaC_-1deltaT=1_DSM_unequal_sample_sizes.pdf}
\end{figure}


\begin{figure}[ht]
	\centering
	\includegraphics[scale=0.33]{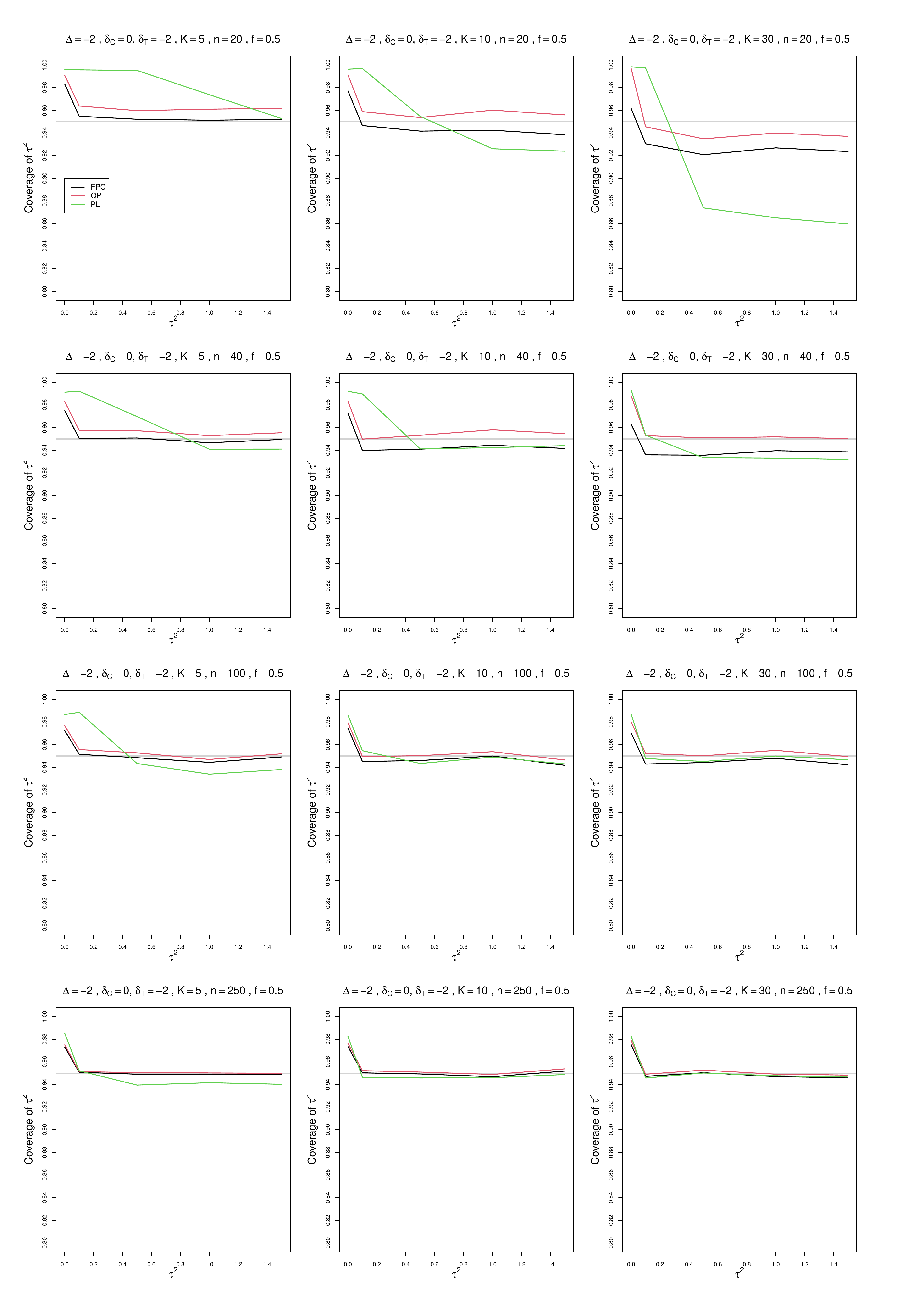}
	\caption{Coverage of PL, QP, and  FPC 95\% confidence intervals for between-study variance of DSM   vs $\tau^2$, for equal sample sizes $n=20,\;40,\;100$ and $250$, $\delta_{iC} = 0$, $\Delta=-2$ and  $f = 0.5$.   }
	\label{PlotCoverageOfTau2_deltaC_0deltaT=-2_DSM_equal_sample_sizes.pdf}
\end{figure}

\begin{figure}[ht]
	\centering
	\includegraphics[scale=0.33]{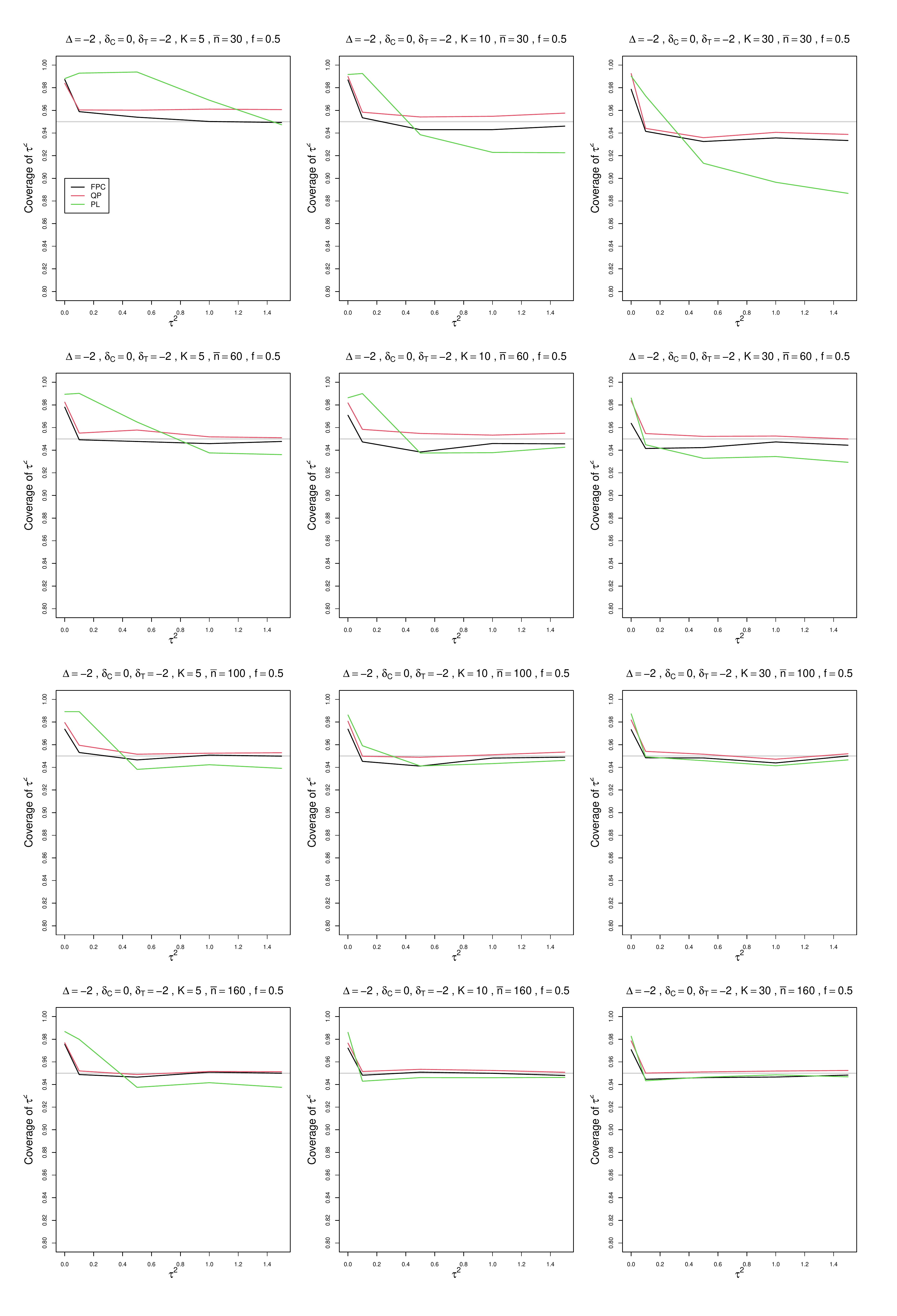}
	\caption{Coverage of PL, QP, and  FPC 95\% confidence intervals for between-study variance of DSM   vs $\tau^2$, for unequal sample sizes $\bar{n}=30,\;60,\;100$ and $160$, $\delta_{iC} = 0$, $\Delta=-2$ and  $f = 0.5$.   }
	\label{PlotCoverageOfTau2_deltaC_-1deltaT=-3_DSM_unequal_sample_sizes.pdf}
\end{figure}

\begin{figure}[ht]
	\centering
	\includegraphics[scale=0.33]{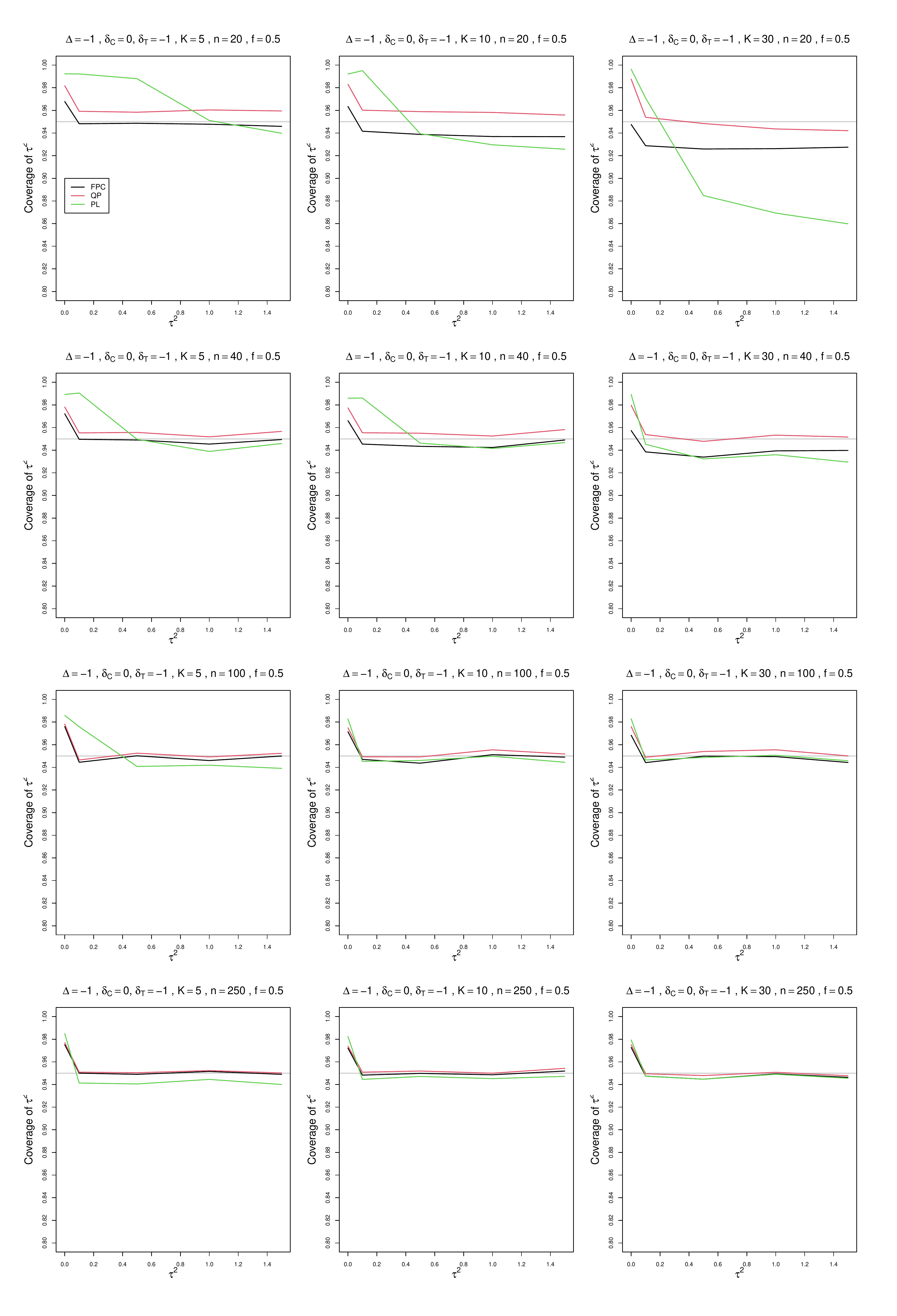}
	\caption{Coverage of PL, QP, and  FPC 95\% confidence intervals for between-study variance of DSM   vs $\tau^2$, for equal sample sizes $n=20,\;40,\;100$ and $250$, $\delta_{iC} = 0$, $\Delta=-1$ and  $f = 0.5$.   }
	\label{PlotCoverageOfTau2_deltaC_-0deltaT=-1_DSM_equal_sample_sizes.pdf}
\end{figure}

\begin{figure}[ht]
	\centering
	\includegraphics[scale=0.33]{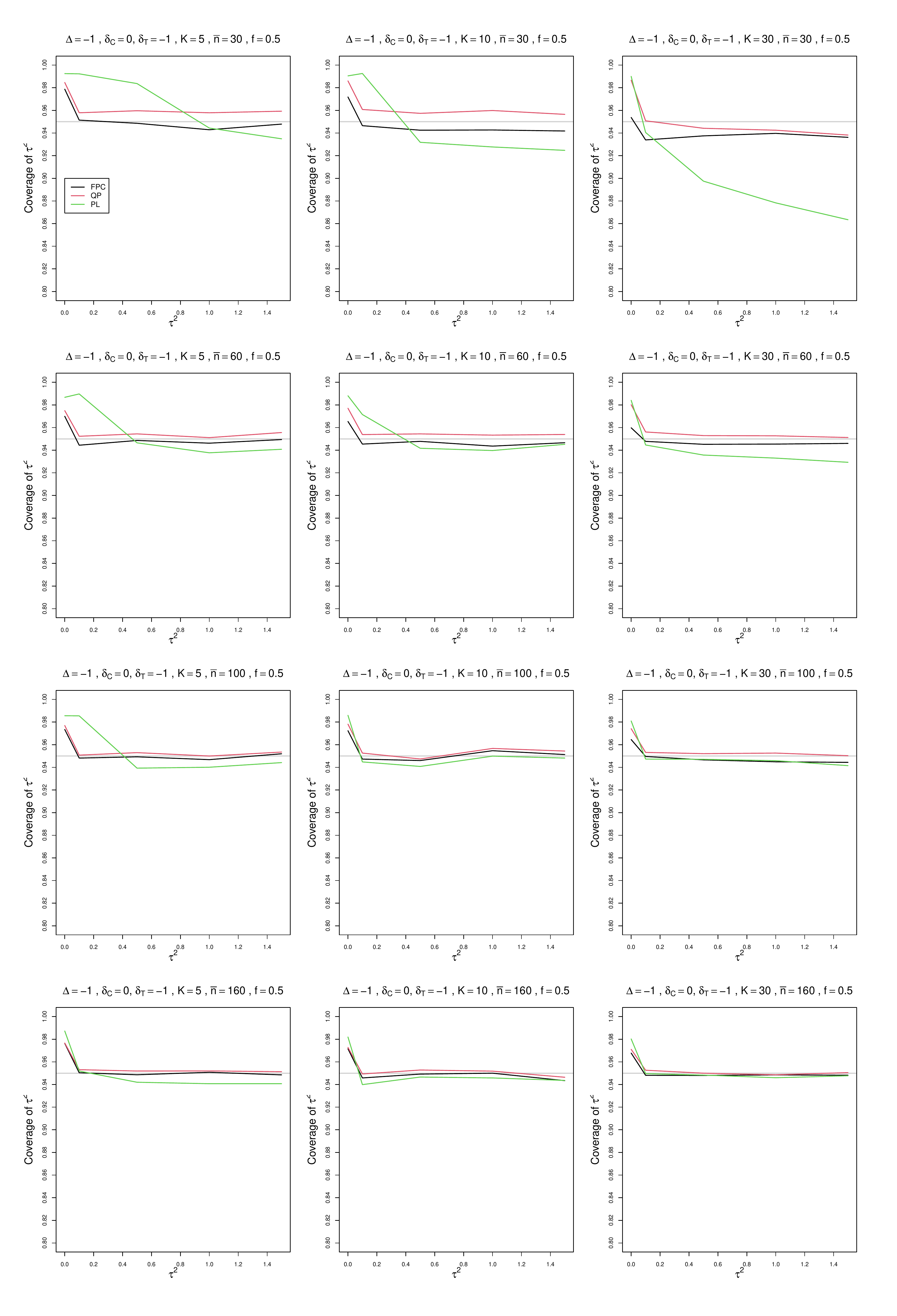}
	\caption{Coverage of PL, QP, and  FPC 95\% confidence intervals for between-study variance of DSM   vs $\tau^2$, for unequal sample sizes $\bar{n}=30,\;60,\;100$ and $160$, $\delta_{iC} = 0$, $\Delta=-1$ and  $f = 0.5$.   }
	\label{PlotCoverageOfTau2_deltaC_0deltaT=-1_DSM_unequal_sample_sizes.pdf}
\end{figure}

\begin{figure}[ht]
	\centering
	\includegraphics[scale=0.33]{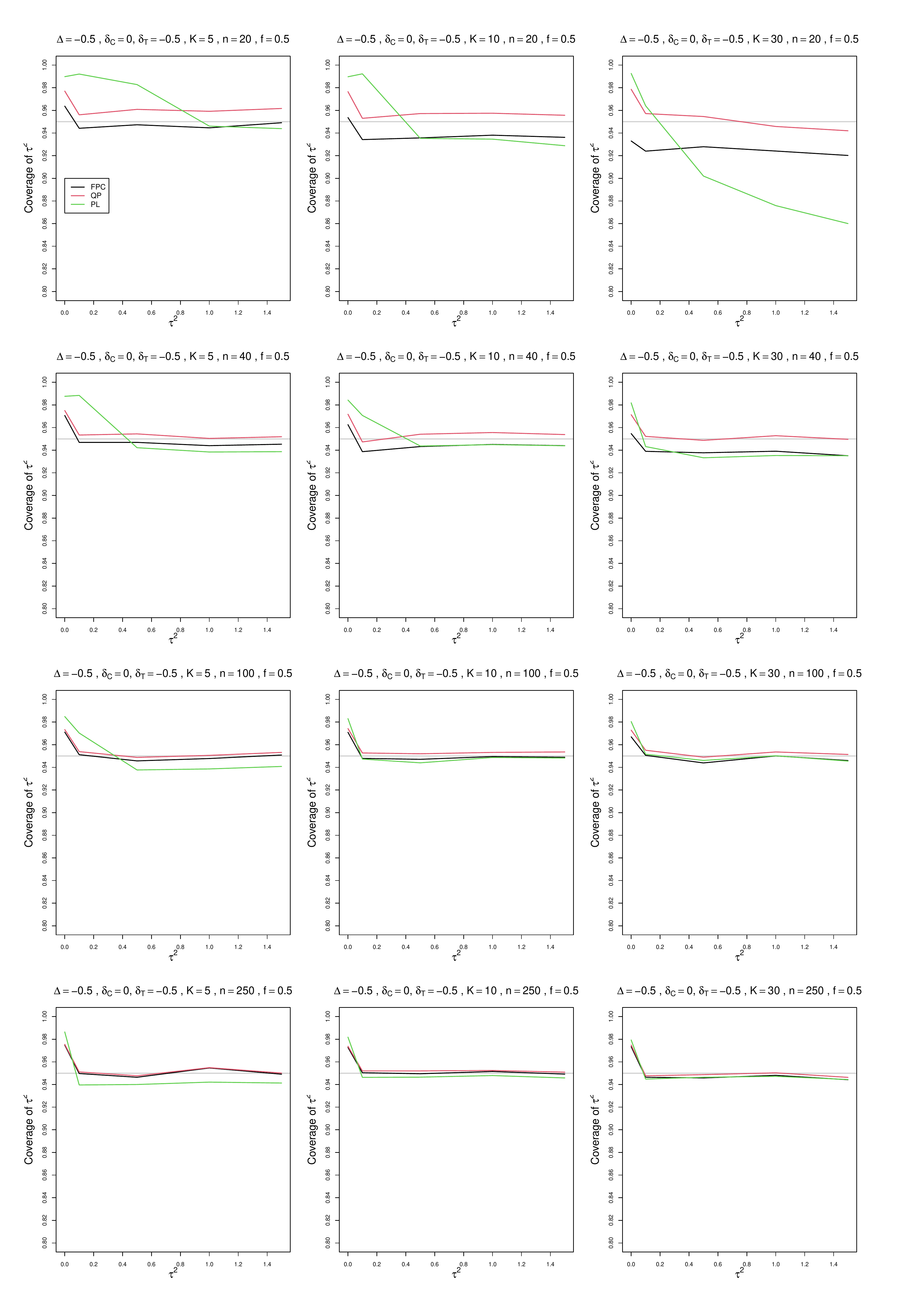}
	\caption{Coverage of PL, QP, and  FPC 95\% confidence intervals for between-study variance of DSM   vs $\tau^2$, for equal sample sizes $n=20,\;40,\;100$ and $250$, $\delta_{iC} = 0$, $\Delta=-0.5$ and  $f = 0.5$.   }
	\label{PlotCoverageOfTau2_deltaC_-0deltaT=-0.5_DSM_equal_sample_sizes.pdf}
\end{figure}

\begin{figure}[ht]
	\centering
	\includegraphics[scale=0.33]{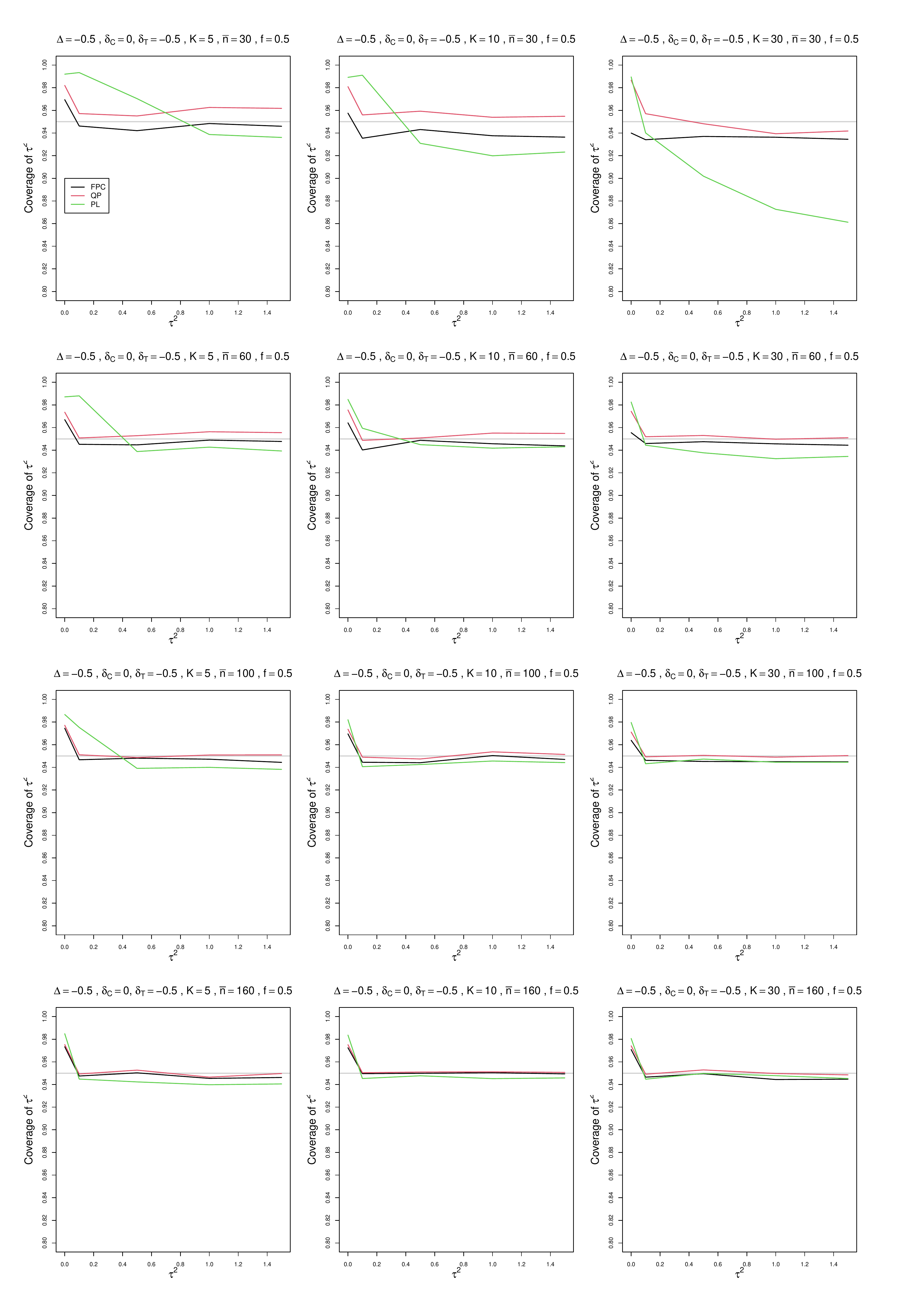}
	\caption{Coverage of PL, QP, and  FPC 95\% confidence intervals for between-study variance of DSM   vs $\tau^2$, for unequal sample sizes $\bar{n}=30,\;60,\;100$ and $160$, $\delta_{iC} = 0$, $\Delta=-0.5$ and  $f = 0.5$.   }
	\label{PlotCoverageOfTau2_deltaC_0deltaT=-0.5_DSM_unequal_sample_sizes.pdf}
\end{figure}

\begin{figure}[ht]
	\centering
	\includegraphics[scale=0.33]{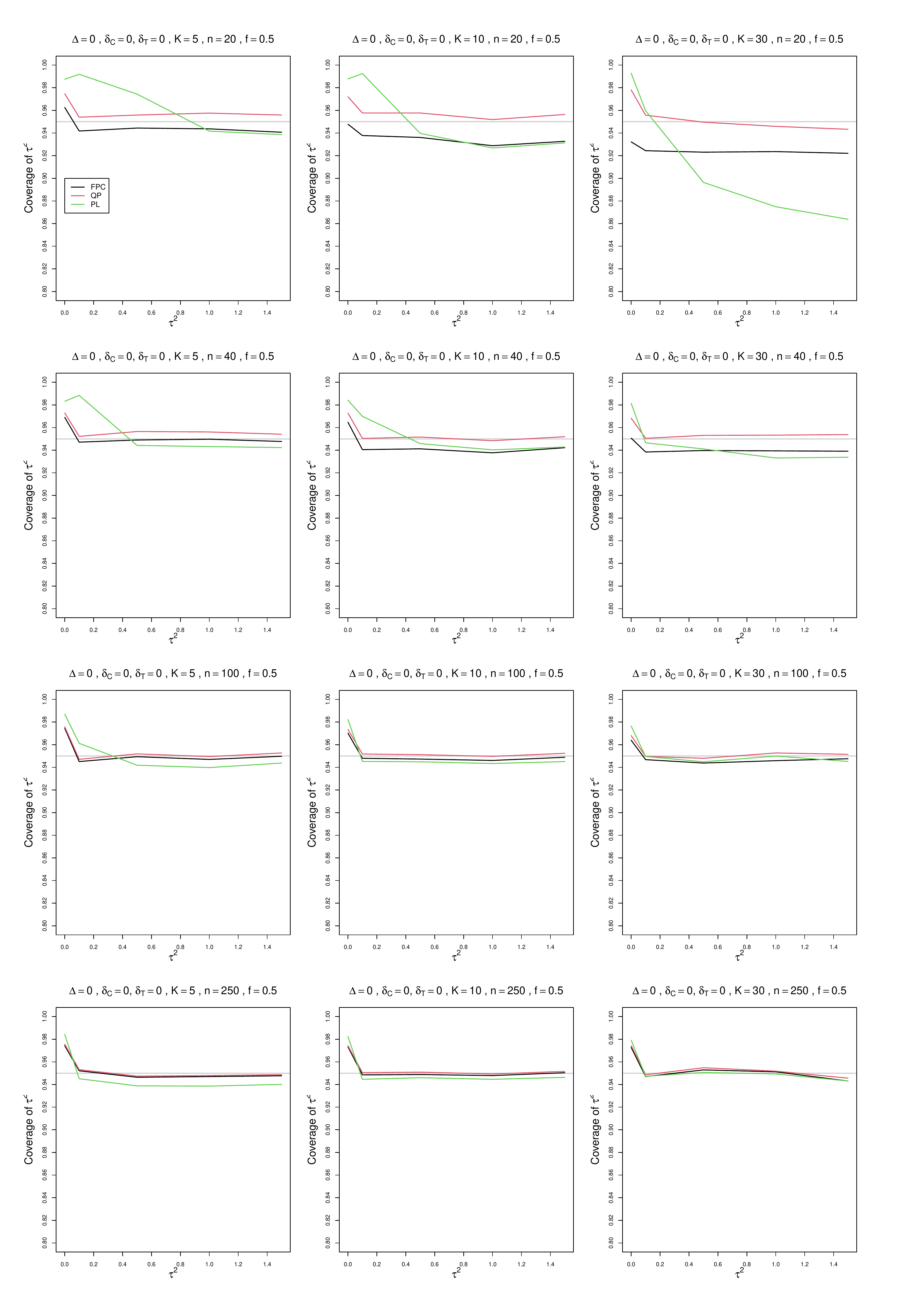}
	\caption{Coverage of PL, QP, and  FPC 95\% confidence intervals for between-study variance of DSM   vs $\tau^2$, for equal sample sizes $n=20,\;40,\;100$ and $250$, $\delta_{iC} = 0$, $\Delta=0$ and  $f = 0.5$.   }
	\label{PlotCoverageOfTau2_deltaC_0deltaT=0_DSM_equal_sample_sizes.pdf}
\end{figure}

\begin{figure}[ht]
	\centering
	\includegraphics[scale=0.33]{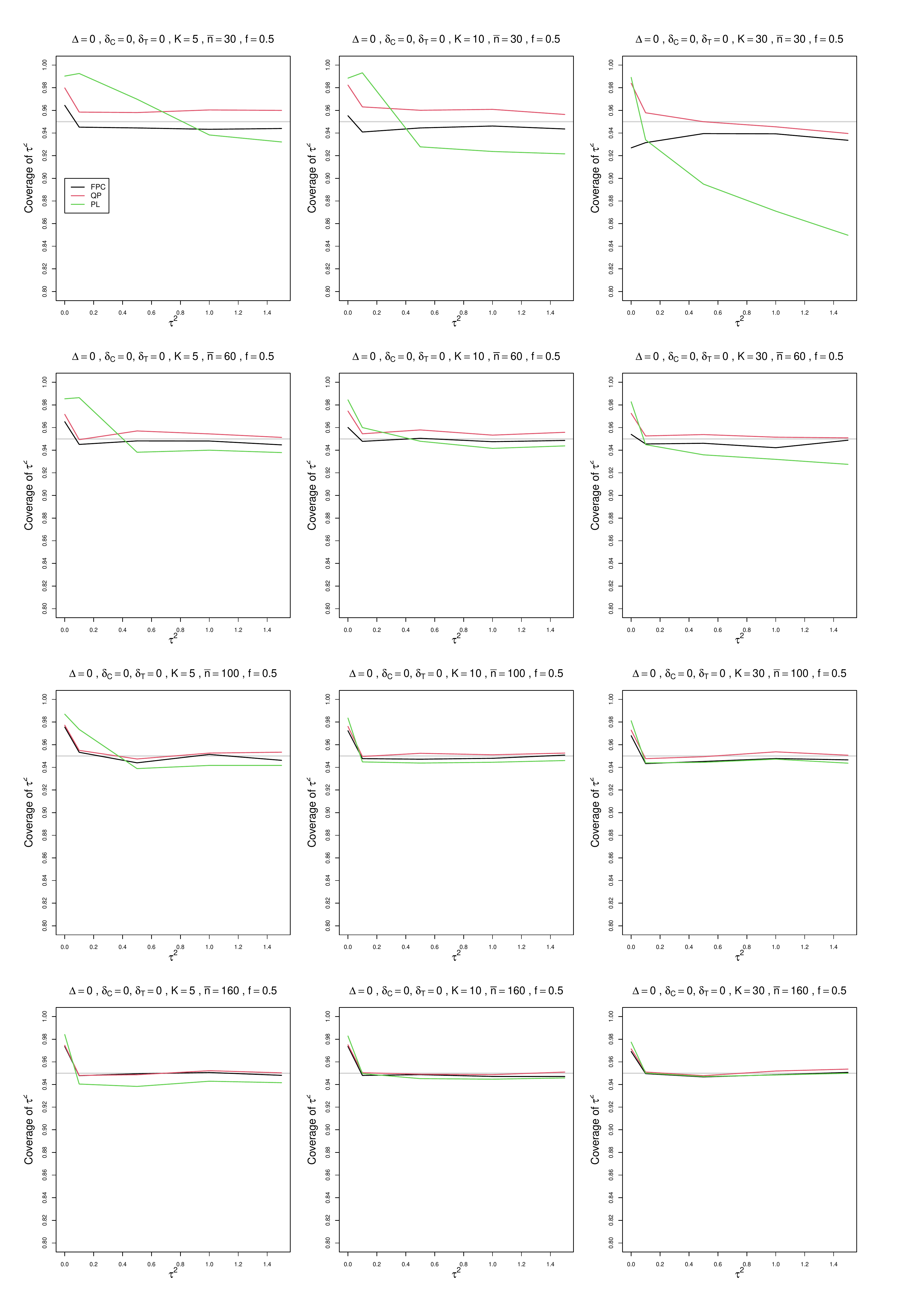}
	\caption{Coverage of PL, QP, and  FPC 95\% confidence intervals for between-study variance of DSM   vs $\tau^2$, for unequal sample sizes $\bar{n}=30,\;60,\;100$ and $160$, $\delta_{iC} = 0$, $\Delta=0$ and  $f = 0.5$.   }
	\label{PlotCoverageOfTau2_deltaC_0deltaT=0_DSM_unequal_sample_sizes.pdf}
\end{figure}

\begin{figure}[ht]
	\centering
	\includegraphics[scale=0.33]{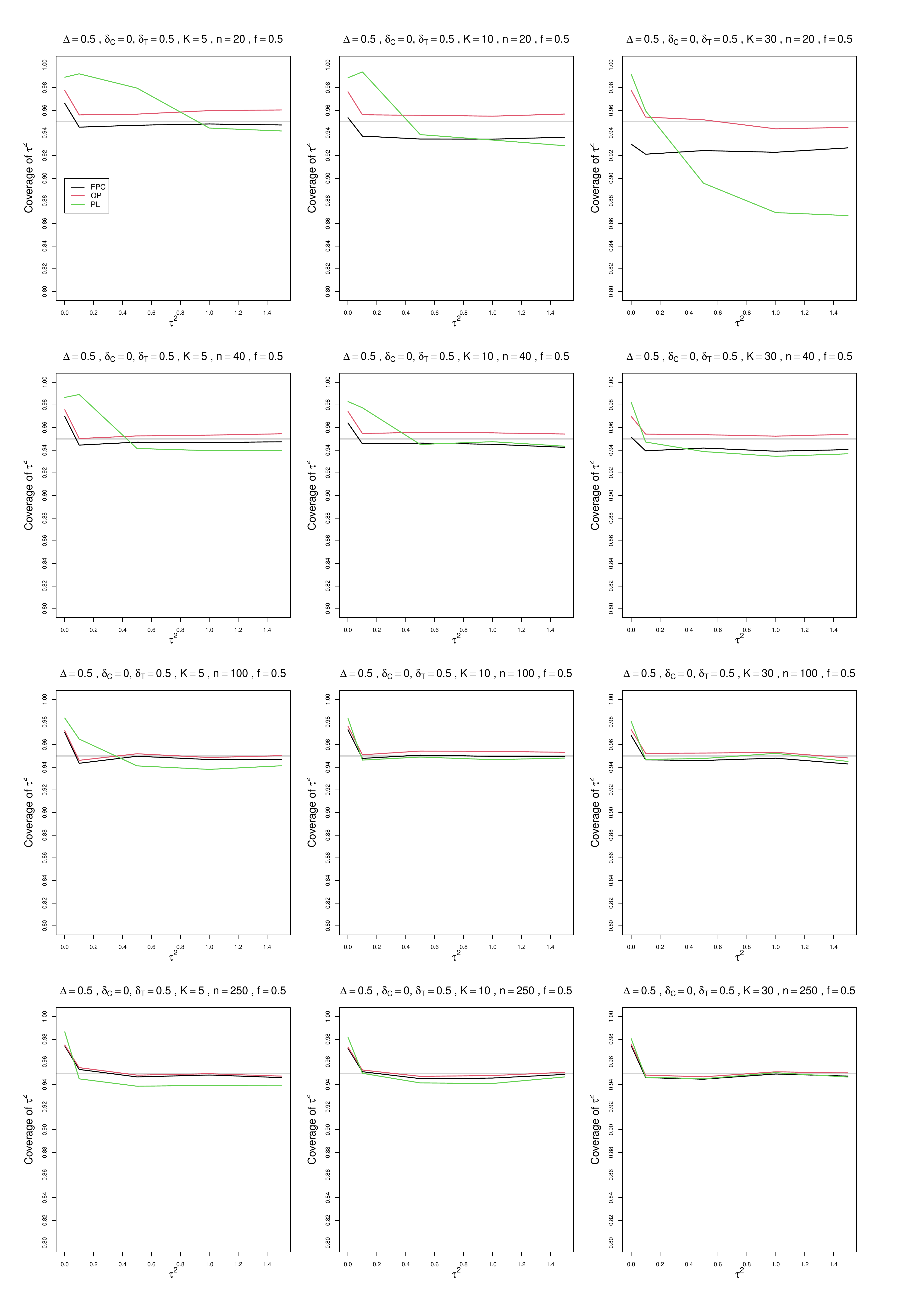}
	\caption{Coverage of PL, QP, and  FPC 95\% confidence intervals for between-study variance of DSM   vs $\tau^2$, for equal sample sizes $n=20,\;40,\;100$ and $250$, $\delta_{iC} = 0$, $\Delta=0.5$ and  $f = 0.5$.   }
	\label{PlotCoverageOfTau2_deltaC_0deltaT=0.5_DSM_equal_sample_sizes.pdf}
\end{figure}

\begin{figure}[ht]
	\centering
	\includegraphics[scale=0.33]{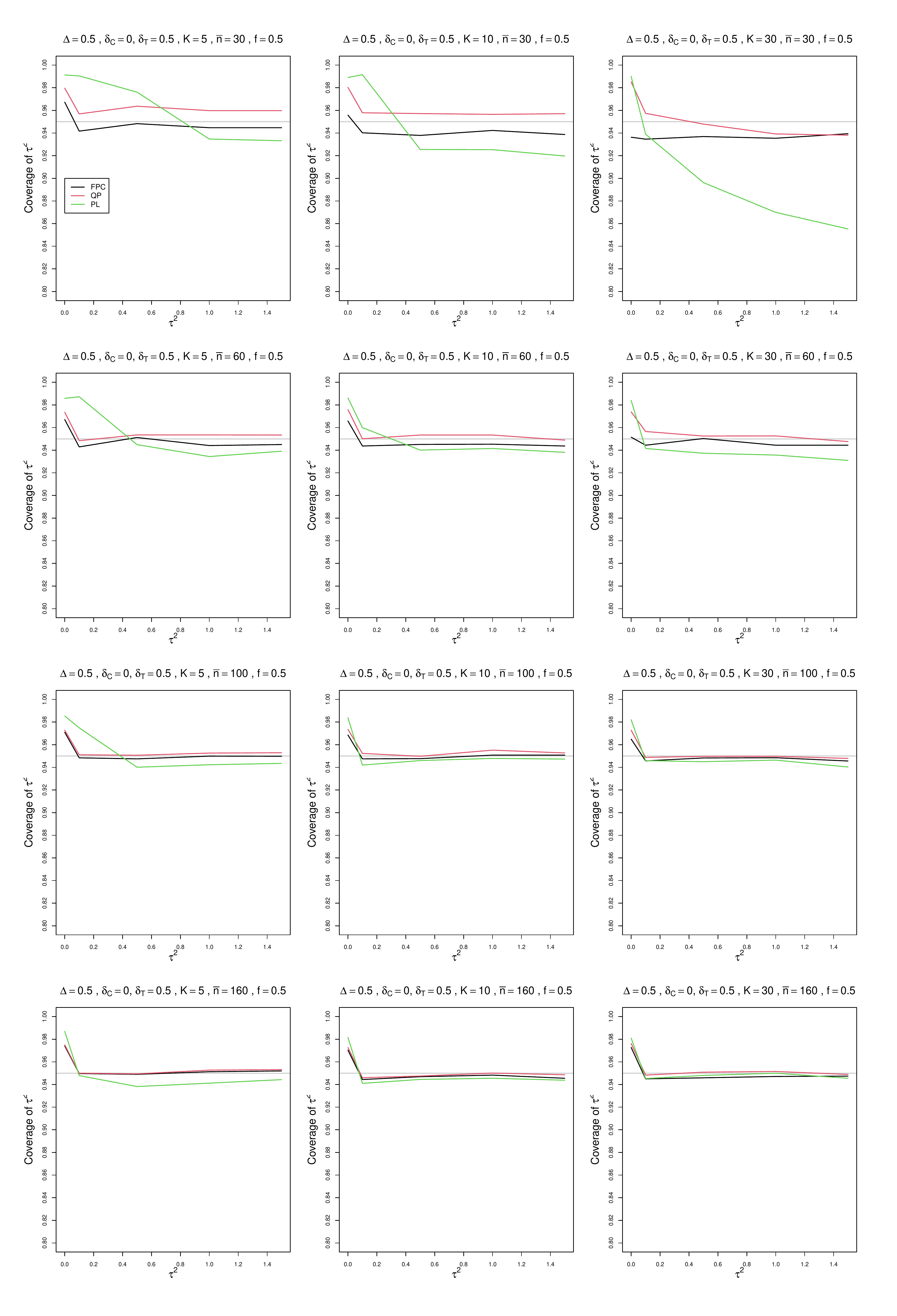}
	\caption{Coverage of PL, QP, and  FPC 95\% confidence intervals for between-study variance of DSM   vs $\tau^2$, for unequal sample sizes $\bar{n}=30,\;60,\;100$ and $160$, $\delta_{iC} = 0$, $\Delta=0.5$ and  $f = 0.5$.   }
	\label{PlotCoverageOfTau2_deltaC_0deltaT=-0.5_DSM_unequal_sample_sizes.pdf}
\end{figure}

\begin{figure}[ht]
	\centering
	\includegraphics[scale=0.33]{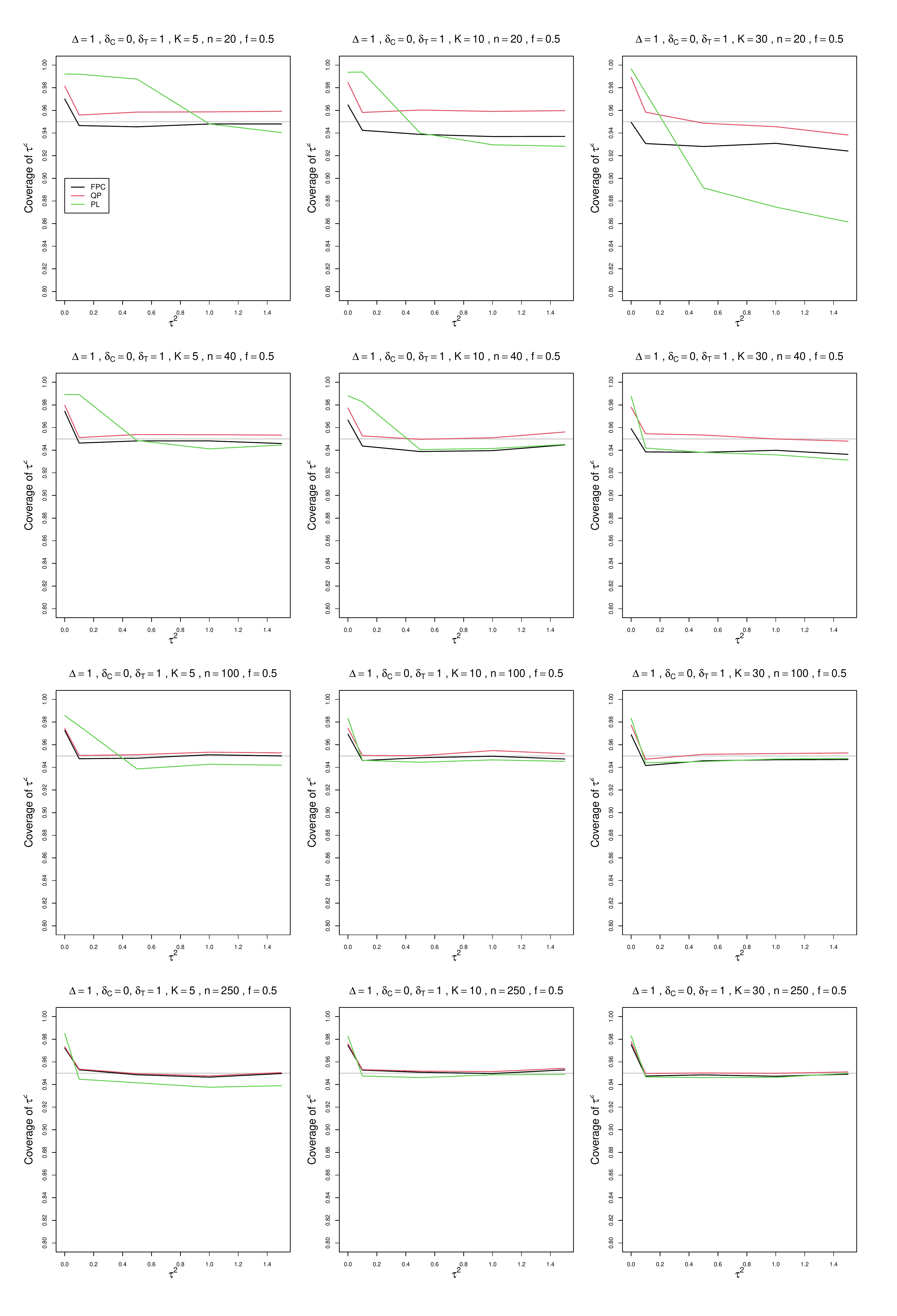}
	\caption{Coverage of PL, QP, and  FPC 95\% confidence intervals for between-study variance of DSM   vs $\tau^2$, for equal sample sizes $n=20,\;40,\;100$ and $250$, $\delta_{iC} = 0$, $\Delta=1$ and  $f = 0.5$.   }
	\label{PlotCoverageOfTau2_deltaC_=0deltaT=1_DSM_equal_sample_sizes.pdf}
\end{figure}

\begin{figure}[ht]
	\centering
	\includegraphics[scale=0.33]{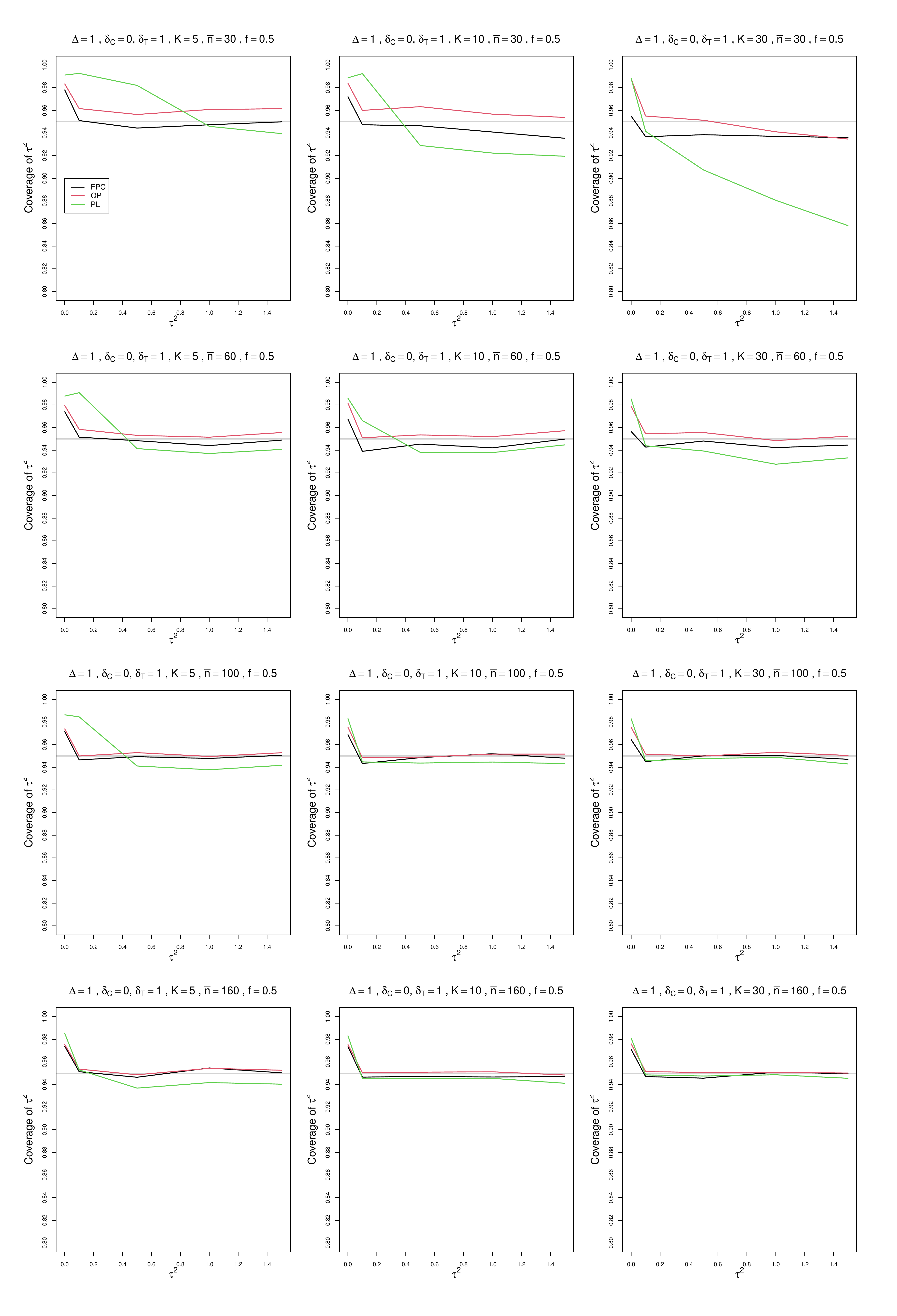}
	\caption{Coverage of PL, QP, and  FPC 95\% confidence intervals for between-study variance of DSM   vs $\tau^2$, for unequal sample sizes $\bar{n}=30,\;60,\;100$ and $160$, $\delta_{iC} = 0$, $\Delta=1$ and  $f = 0.5$.   }
	\label{PlotCoverageOfTau2_deltaC_0deltaT=1_DSM_unequal_sample_sizes.pdf}
\end{figure}

\begin{figure}[ht]
	\centering
	\includegraphics[scale=0.33]{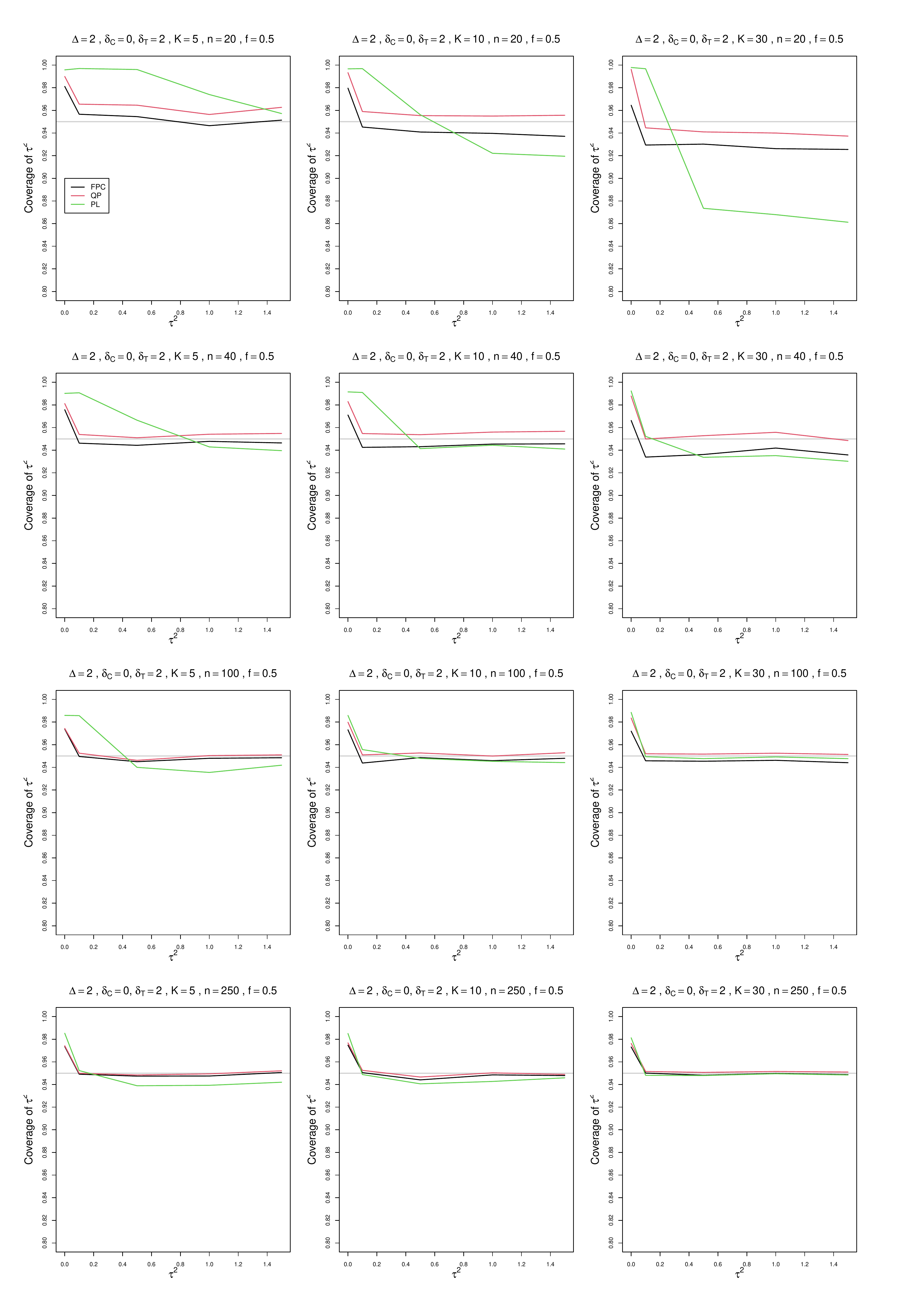}
	\caption{Coverage of PL, QP, and  FPC 95\% confidence intervals for between-study variance of DSM   vs $\tau^2$, for equal sample sizes $n=20,\;40,\;100$ and $250$, $\delta_{iC} = 0$, $\Delta=2$ and  $f = 0.5$.   }
	\label{PlotCoverageOfTau2_deltaC_0deltaT=2_DSM_equal_sample_sizes.pdf}
\end{figure}

\begin{figure}[ht]
	\centering
	\includegraphics[scale=0.33]{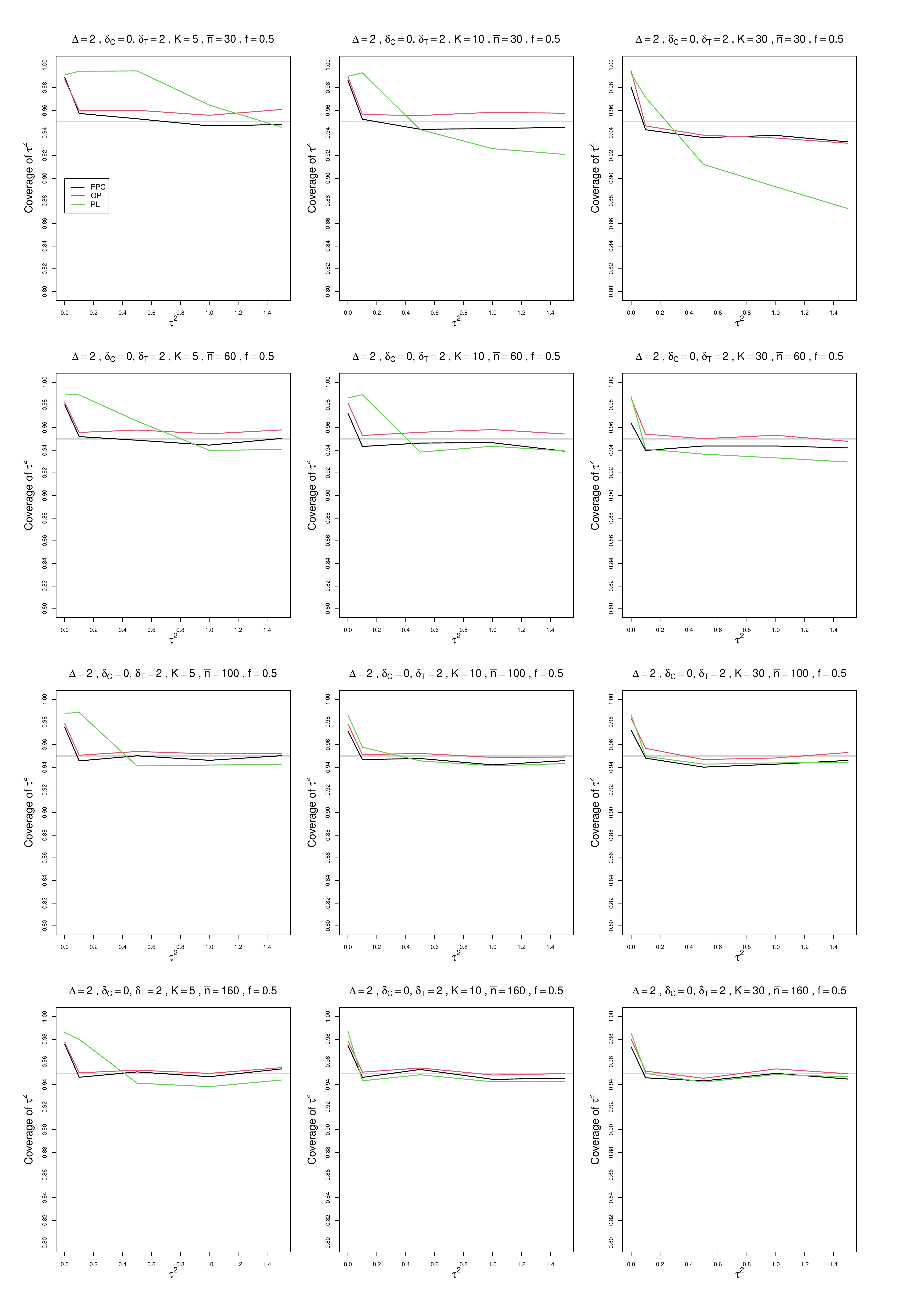}
	\caption{Coverage of PL, QP, and  FPC 95\% confidence intervals for between-study variance of DSM   vs $\tau^2$, for unequal sample sizes $\bar{n}=30,\;60,\;100$ and $160$, $\delta_{iC} = 0$, $\Delta=2$ and  $f = 0.5$.   }
	\label{PlotCoverageOfTau2_deltaC_0deltaT=2_DSM_unequal_sample_sizes.pdf}
\end{figure}


\begin{figure}[ht]
	\centering
	\includegraphics[scale=0.33]{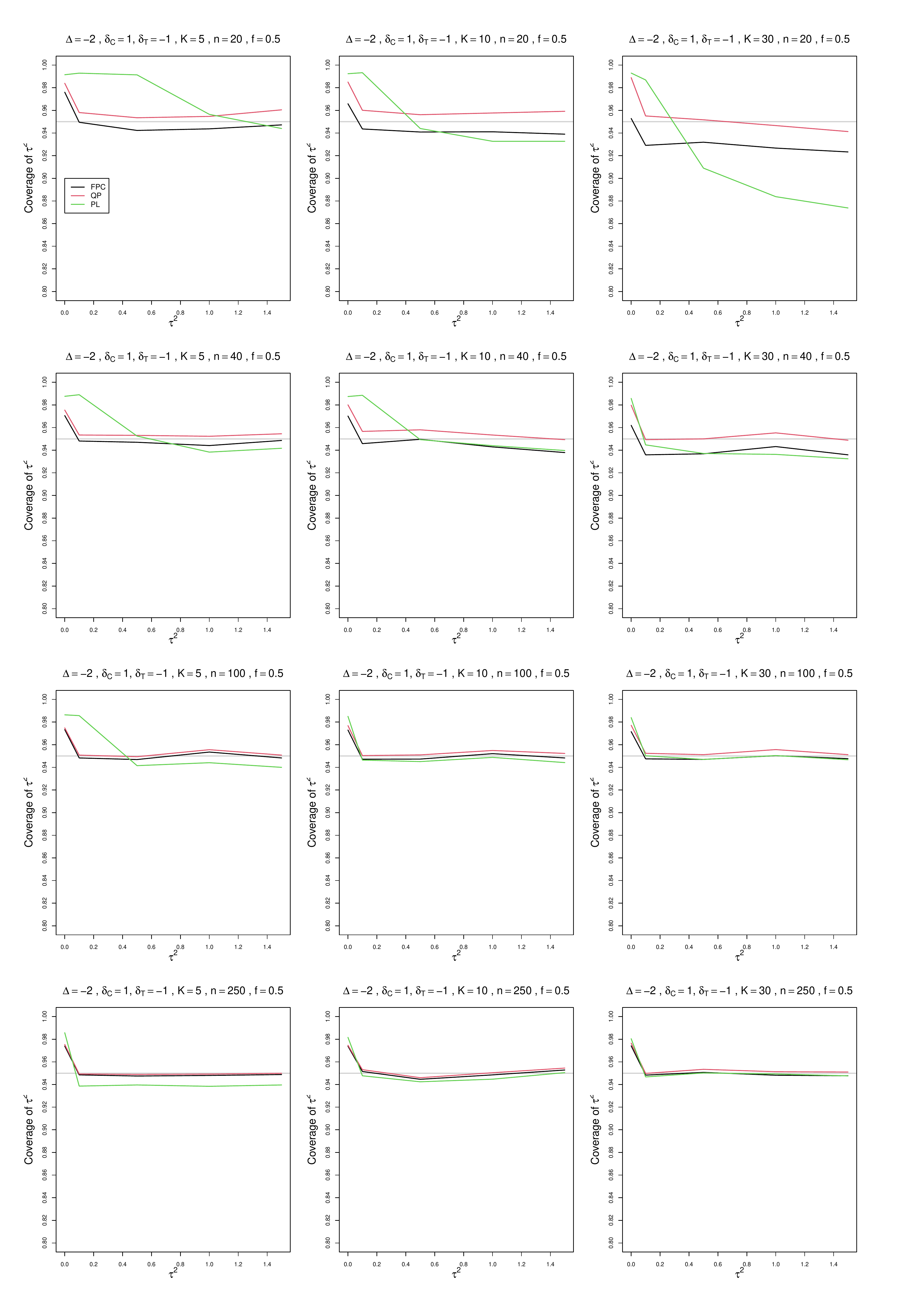}
	\caption{Coverage of PL, QP, and  FPC 95\% confidence intervals for between-study variance of DSM   vs $\tau^2$, for equal sample sizes $n=20,\;40,\;100$ and $250$, $\delta_{iC} = -1$, $\Delta=-2$ and  $f = 0.5$.   }
	\label{PlotCoverageOfTau2_deltaC_1deltaT=-1_DSM_equal_sample_sizes.pdf}
\end{figure}

\begin{figure}[ht]
	\centering
	\includegraphics[scale=0.33]{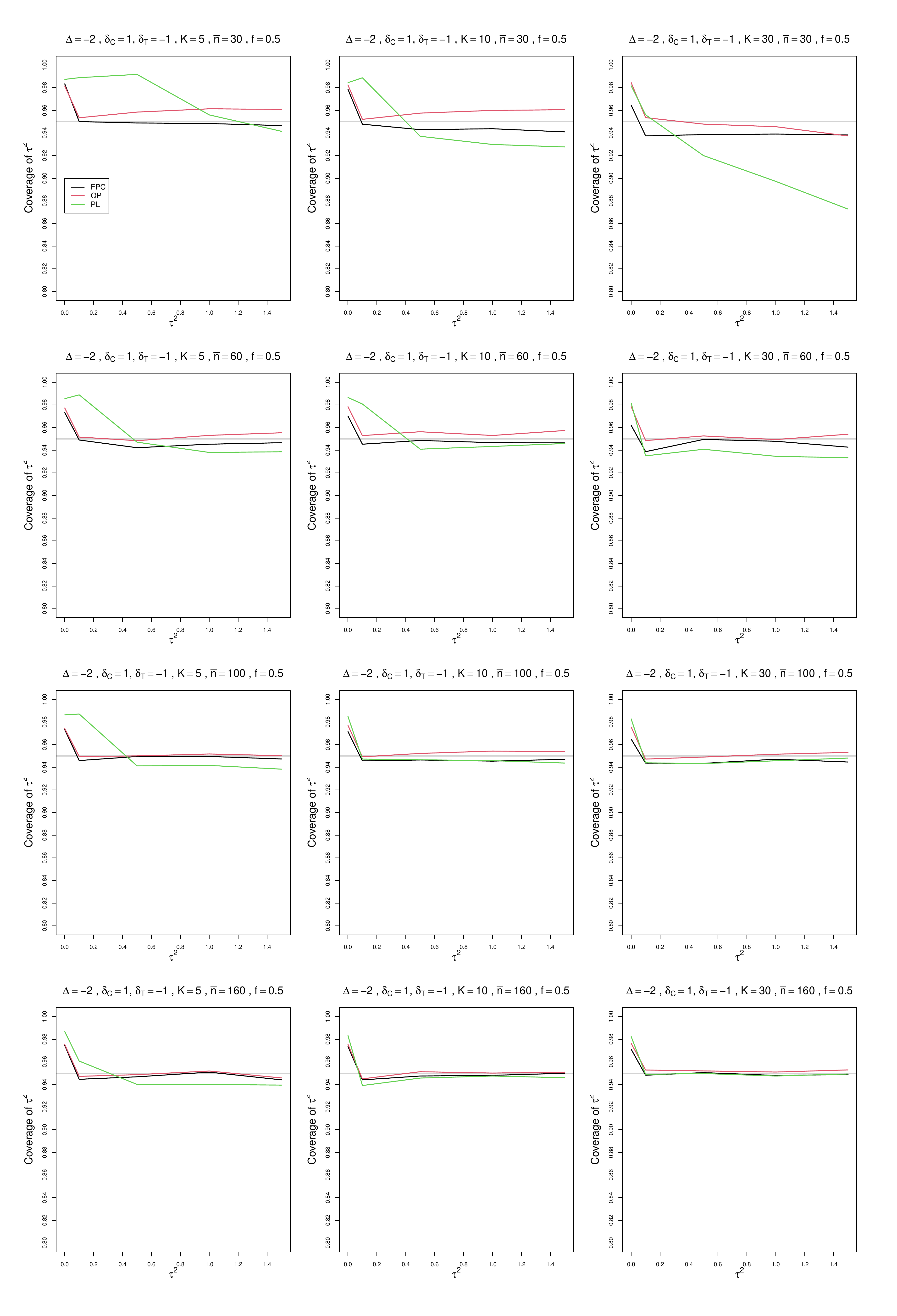}
	\caption{Coverage of PL, QP, and  FPC 95\% confidence intervals for between-study variance of DSM   vs $\tau^2$, for unequal sample sizes $\bar{n}=30,\;60,\;100$ and $160$, $\delta_{iC} = 1$, $\Delta=-2$ and  $f = 0.5$.   }
	\label{PlotCoverageOfTau2_deltaC_1deltaT=-1_DSM_unequal_sample_sizes.pdf}
\end{figure}

\begin{figure}[ht]
	\centering
	\includegraphics[scale=0.33]{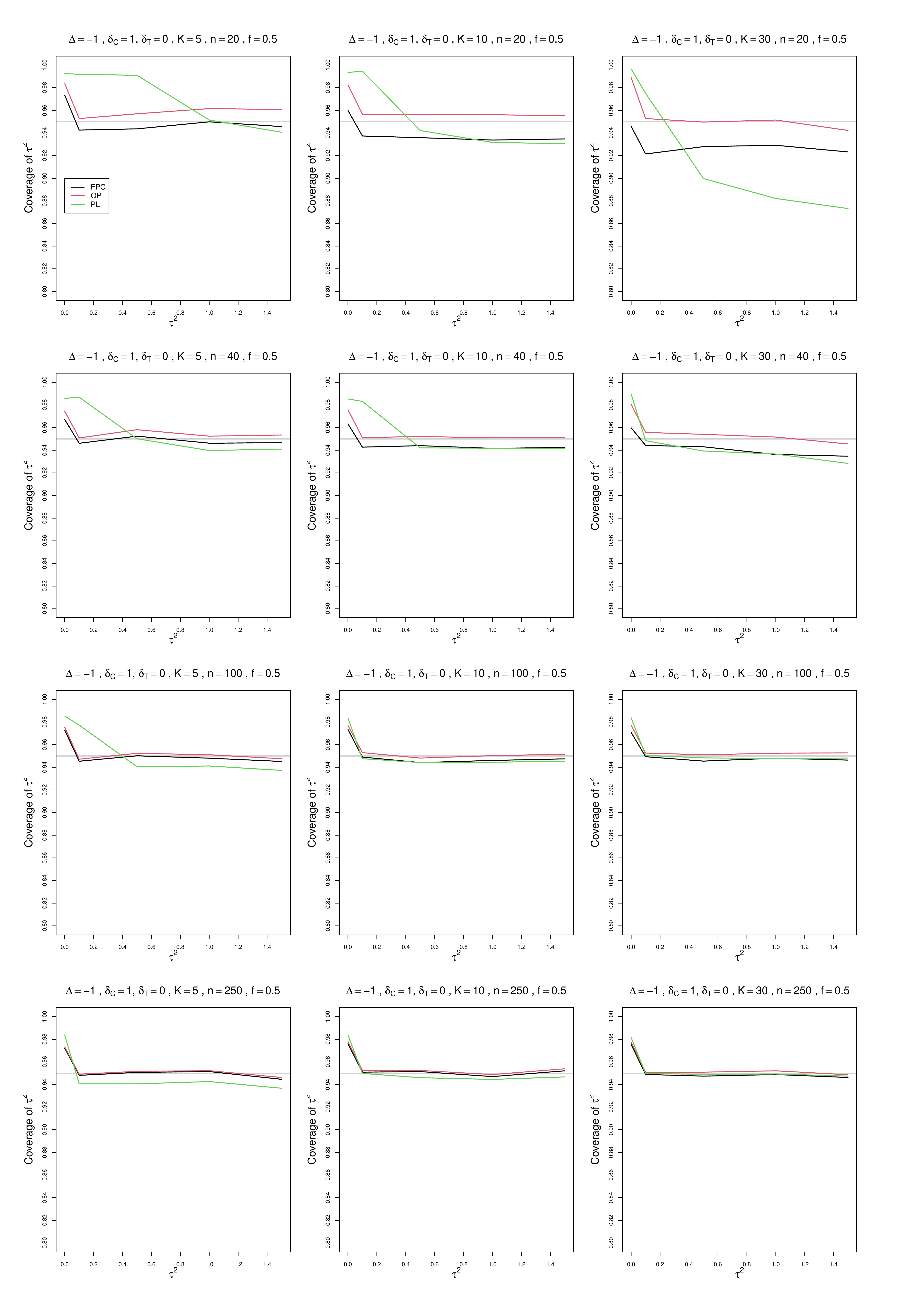}
	\caption{Coverage of PL, QP, and  FPC 95\% confidence intervals for between-study variance of DSM   vs $\tau^2$, for equal sample sizes $n=20,\;40,\;100$ and $250$, $\delta_{iC} = 1$, $\Delta=-1$ and  $f = 0.5$.   }
	\label{PlotCoverageOfTau2_deltaC_1deltaT=0_DSM_equal_sample_sizes.pdf}
\end{figure}

\begin{figure}[ht]
	\centering
	\includegraphics[scale=0.33]{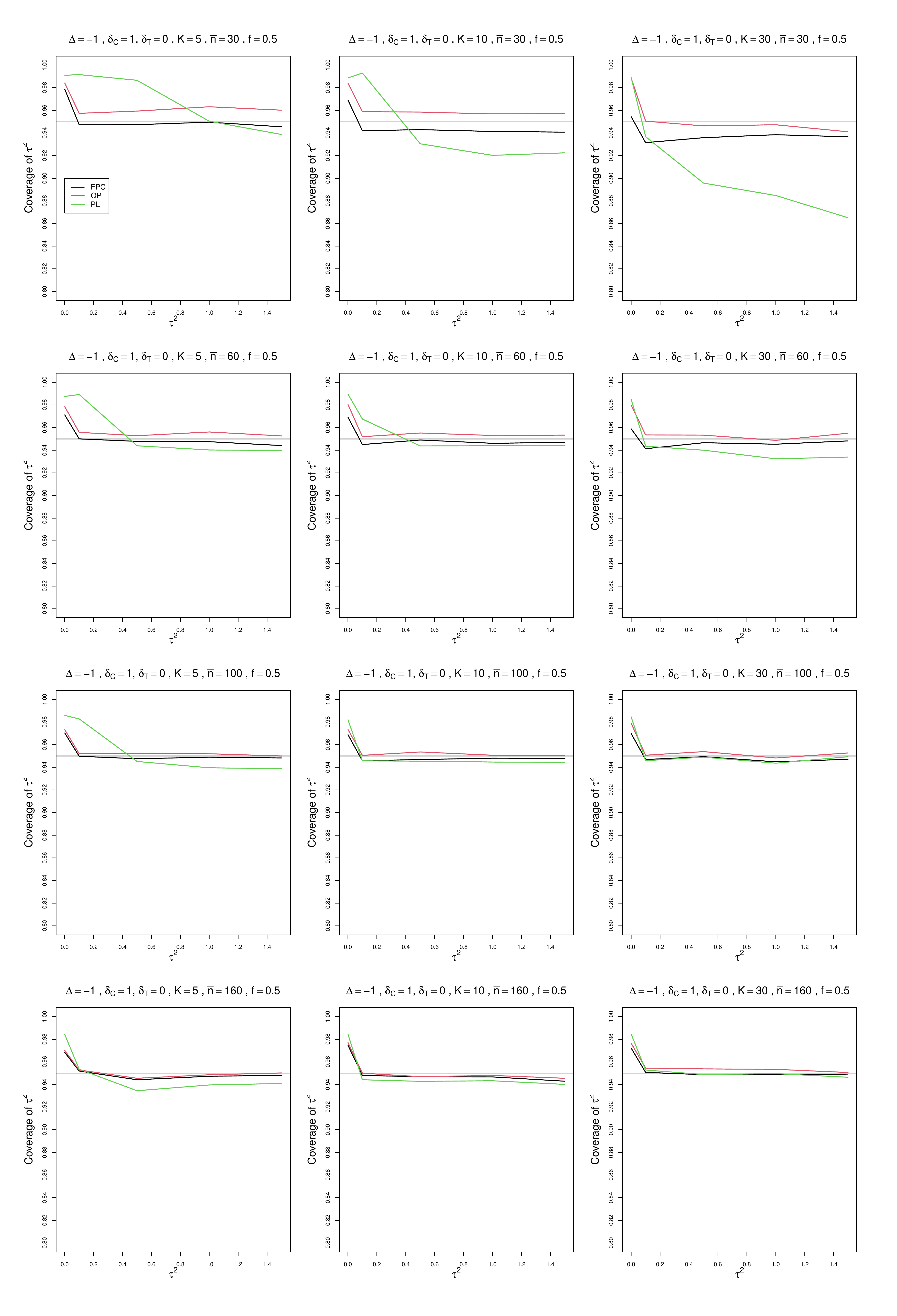}
	\caption{Coverage of PL, QP, and  FPC 95\% confidence intervals for between-study variance of DSM   vs $\tau^2$, for unequal sample sizes $\bar{n}=30,\;60,\;100$ and $160$, $\delta_{iC} = 1$, $\Delta=-1$ and  $f = 0.5$.   }
	\label{PlotCoverageOfTau2_deltaC_1deltaT=0_DSM_unequal_sample_sizes.pdf}
\end{figure}

\begin{figure}[ht]
	\centering
	\includegraphics[scale=0.33]{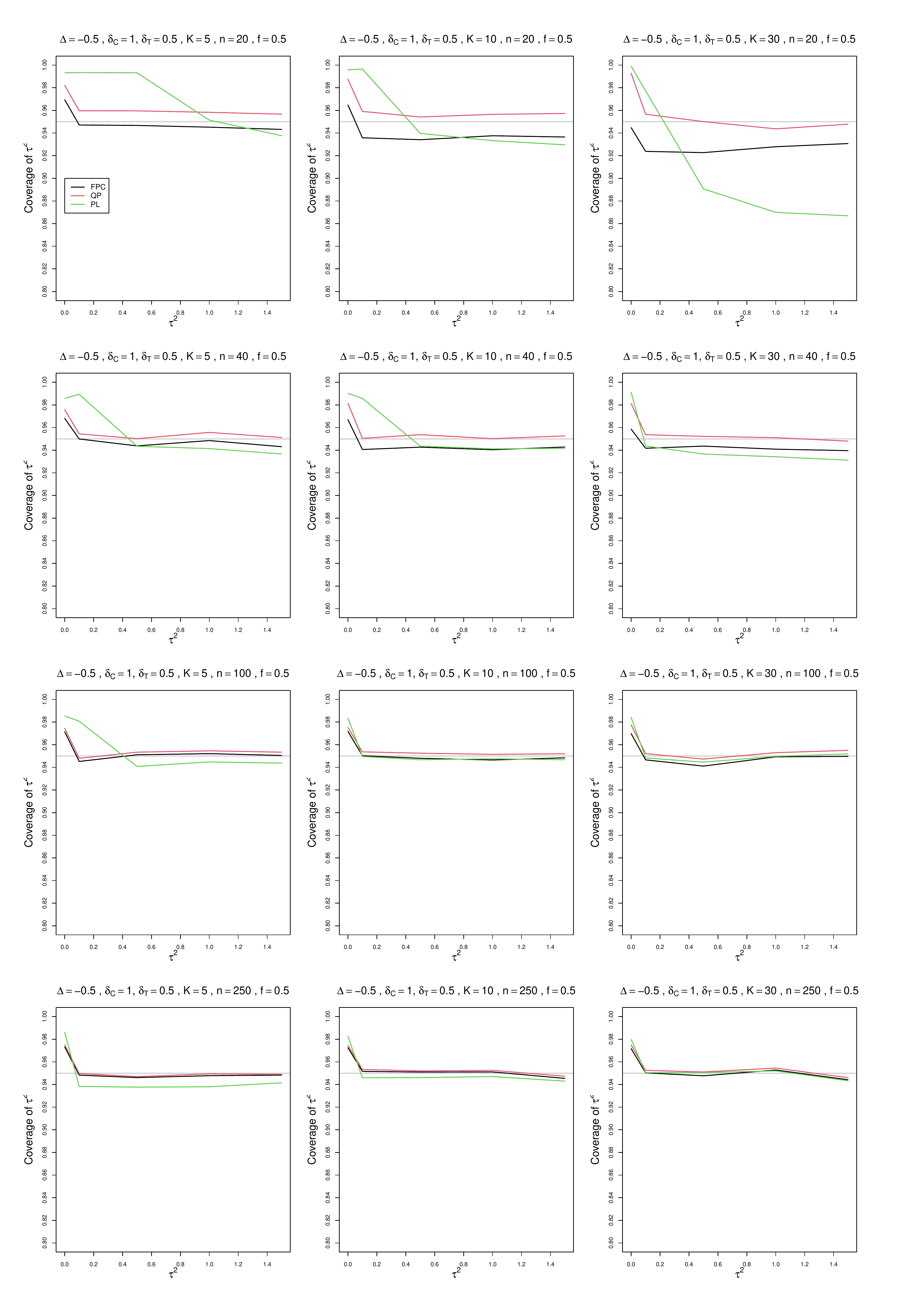}
	\caption{Coverage of PL, QP, and  FPC 95\% confidence intervals for between-study variance of DSM   vs $\tau^2$, for equal sample sizes $n=20,\;40,\;100$ and $250$, $\delta_{iC} = 1$, $\Delta=-0.5$ and  $f = 0.5$.   }
	\label{PlotCoverageOfTau2_deltaC_1deltaT=0.5_DSM_equal_sample_sizes.pdf}
\end{figure}

\begin{figure}[ht]
	\centering
	\includegraphics[scale=0.33]{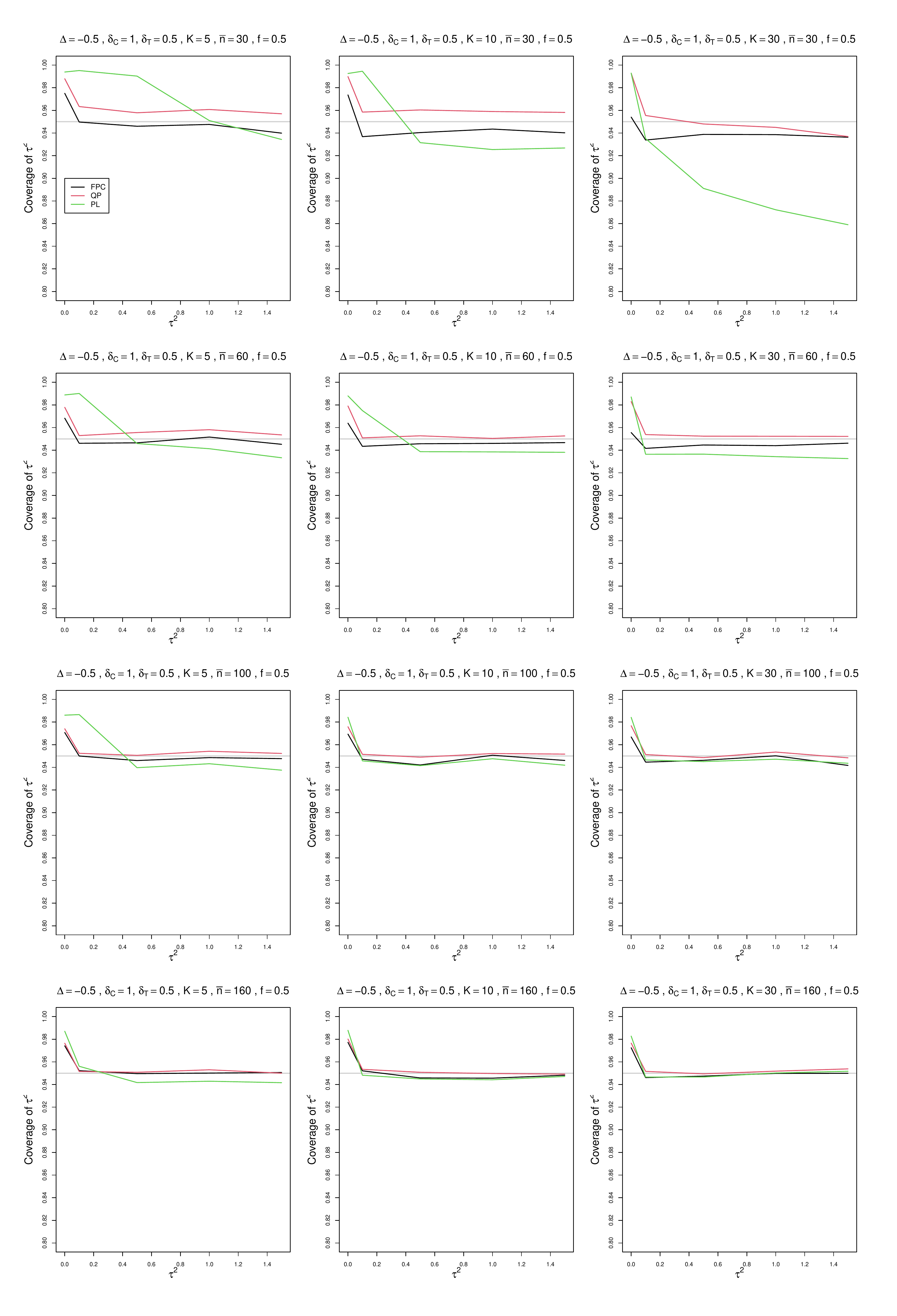}
	\caption{Coverage of PL, QP, and  FPC 95\% confidence intervals for between-study variance of DSM   vs $\tau^2$, for unequal sample sizes $\bar{n}=30,\;60,\;100$ and $160$, $\delta_{iC} = 1$, $\Delta=-0.5$ and  $f = 0.5$.   }
	\label{PlotCoverageOfTau2_deltaC_1deltaT=0.5_DSM_unequal_sample_sizes.pdf}
\end{figure}

\begin{figure}[ht]
	\centering
	\includegraphics[scale=0.33]{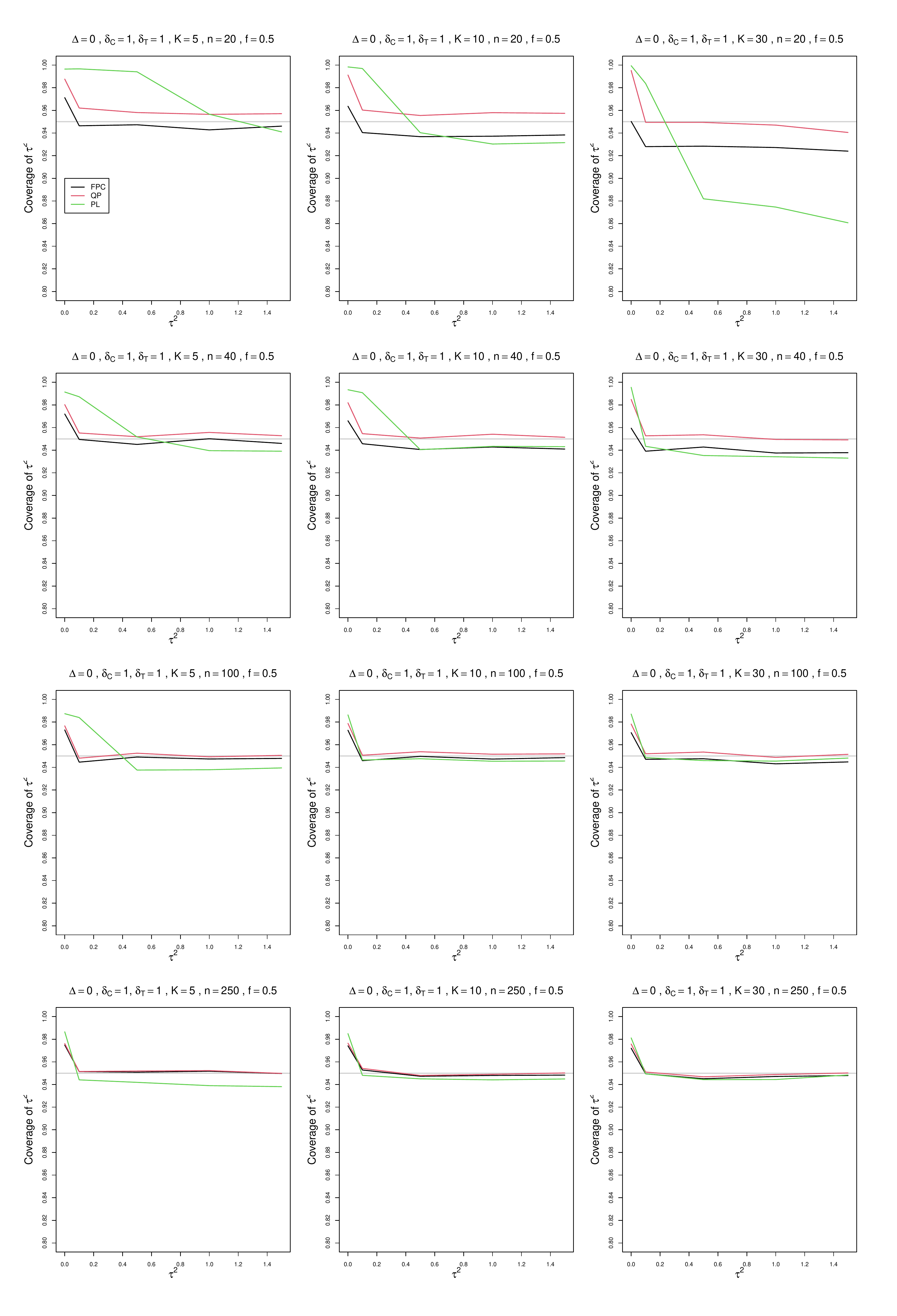}
	\caption{Coverage of PL, QP, and  FPC 95\% confidence intervals for between-study variance of DSM   vs $\tau^2$, for equal sample sizes $n=20,\;40,\;100$ and $250$, $\delta_{iC} = 1$, $\Delta=0$ and  $f = 0.5$.   }
	\label{PlotCoverageOfTau2_deltaC_1deltaT=1_DSM_equal_sample_sizes.pdf}
\end{figure}

\begin{figure}[ht]
	\centering
	\includegraphics[scale=0.33]{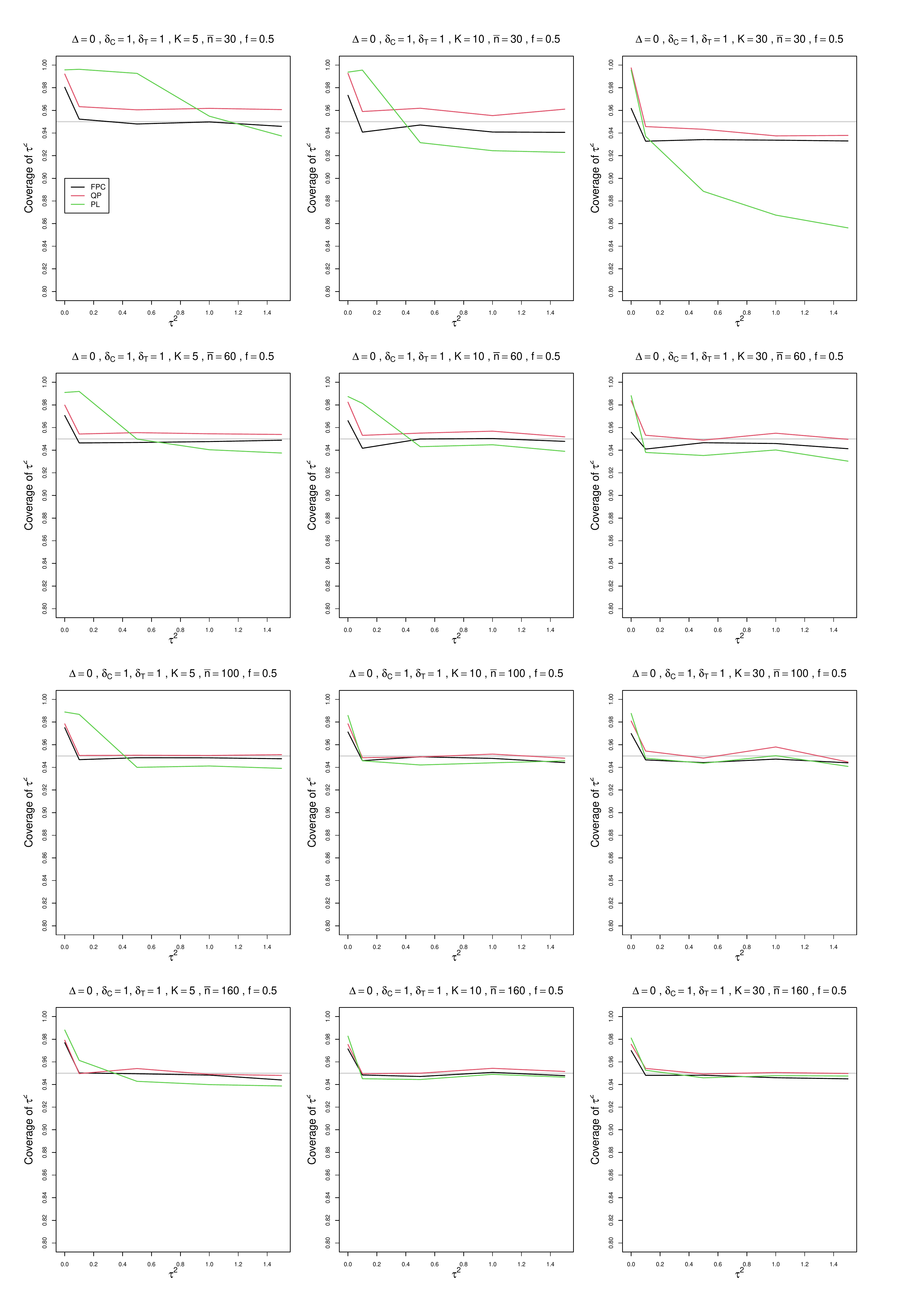}
	\caption{Coverage of PL, QP, and  FPC 95\% confidence intervals for between-study variance of DSM   vs $\tau^2$, for unequal sample sizes $\bar{n}=30,\;60,\;100$ and $160$, $\delta_{iC} = 1$, $\Delta=0$ and  $f = 0.5$.   }
	\label{PlotCoverageOfTau2_deltaC_1deltaT=1_DSM_unequal_sample_sizes.pdf}
\end{figure}

\begin{figure}[ht]
	\centering
	\includegraphics[scale=0.33]{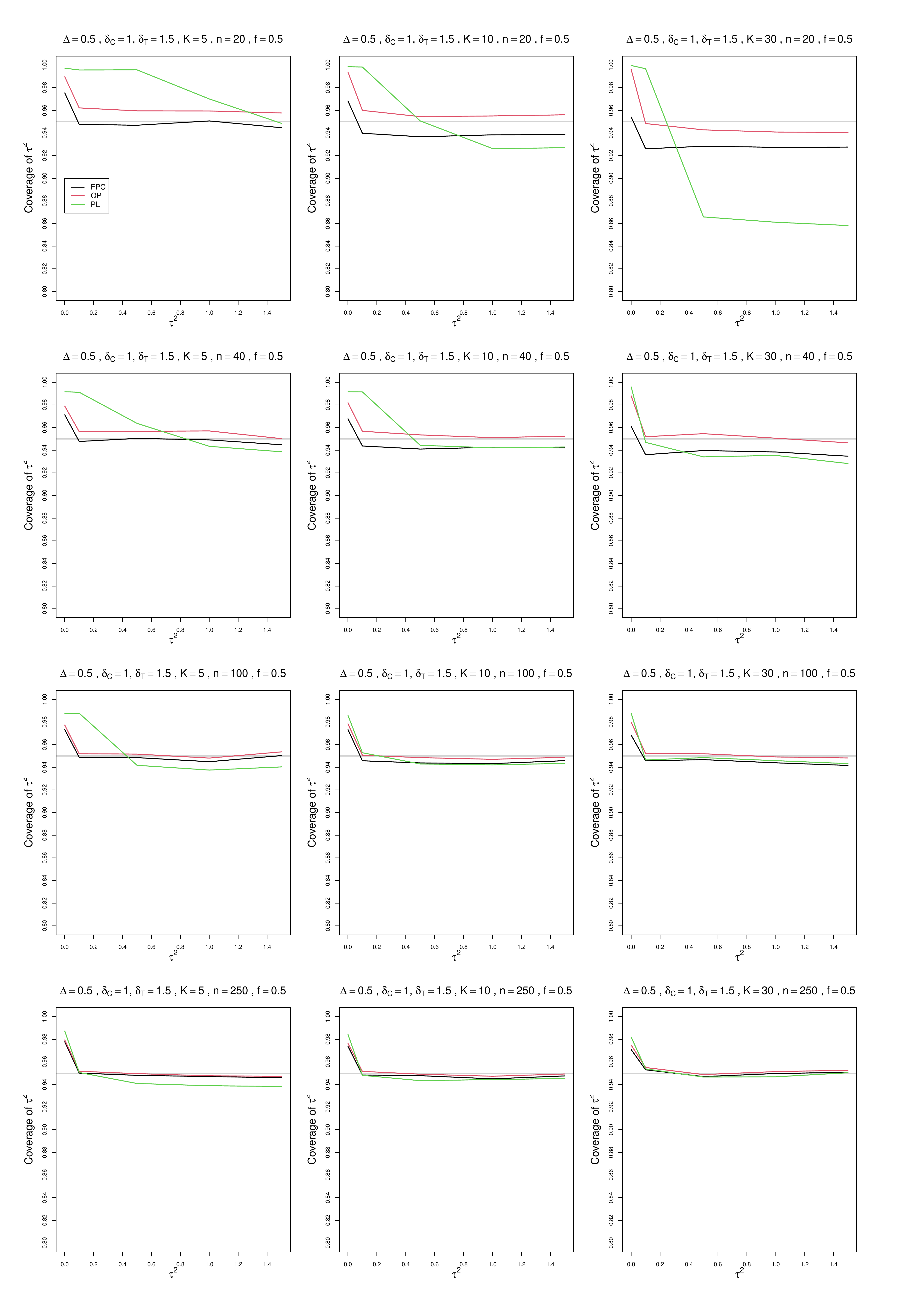}
	\caption{Coverage of PL, QP, and  FPC 95\% confidence intervals for between-study variance of DSM   vs $\tau^2$, for equal sample sizes $n=20,\;40,\;100$ and $250$, $\delta_{iC} = 1$, $\Delta=0.5$ and  $f = 0.5$.   }
	\label{PlotCoverageOfTau2_deltaC_1deltaT=1.5_DSM_equal_sample_sizes.pdf}
\end{figure}

\begin{figure}[ht]
	\centering
	\includegraphics[scale=0.33]{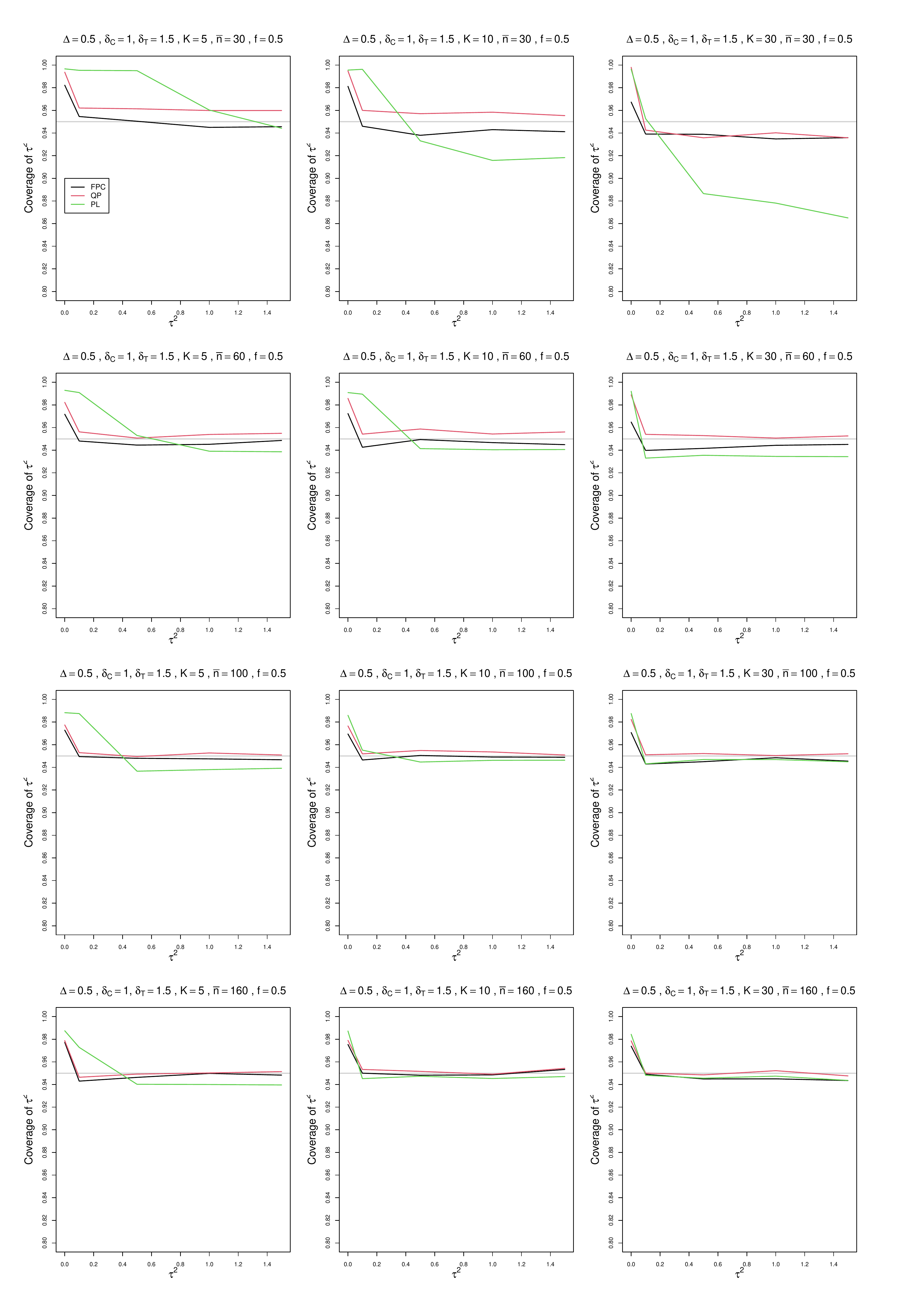}
	\caption{Coverage of PL, QP, and  FPC 95\% confidence intervals for between-study variance of DSM   vs $\tau^2$, for unequal sample sizes $\bar{n}=30,\;60,\;100$ and $160$, $\delta_{iC} = 1$, $\Delta=0.5$ and  $f = 0.5$.   }
	\label{PlotCoverageOfTau2_deltaC_1deltaT=1.5_DSM_unequal_sample_sizes.pdf}
\end{figure}

\begin{figure}[ht]
	\centering
	\includegraphics[scale=0.33]{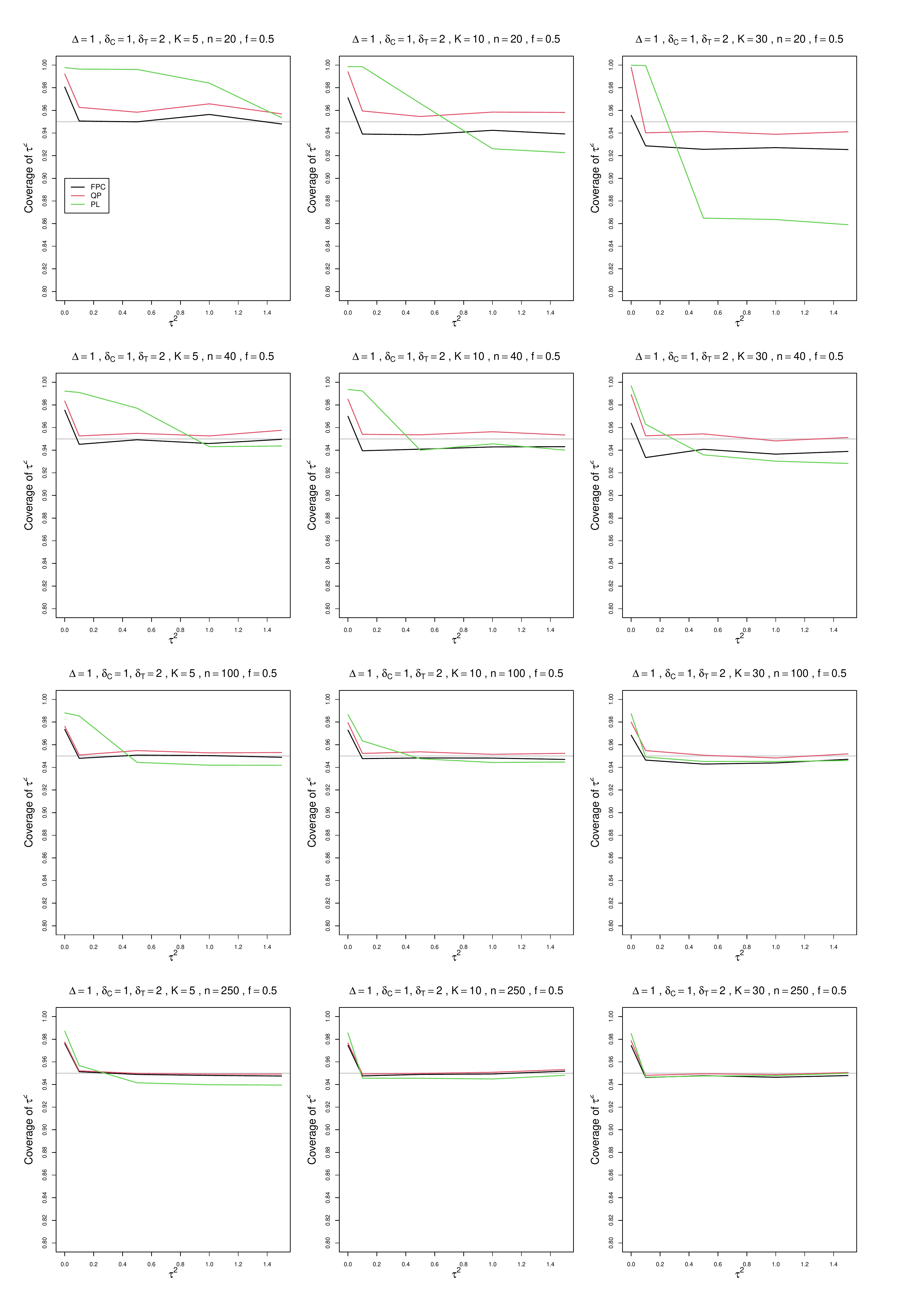}
	\caption{Coverage of PL, QP, and  FPC 95\% confidence intervals for between-study variance of DSM   vs $\tau^2$, for equal sample sizes $n=20,\;40,\;100$ and $250$, $\delta_{iC} = 1$, $\Delta=1$ and  $f = 0.5$.   }
	\label{PlotCoverageOfTau2_deltaC_=1deltaT=2_DSM_equal_sample_sizes.pdf}
\end{figure}

\begin{figure}[ht]
	\centering
	\includegraphics[scale=0.33]{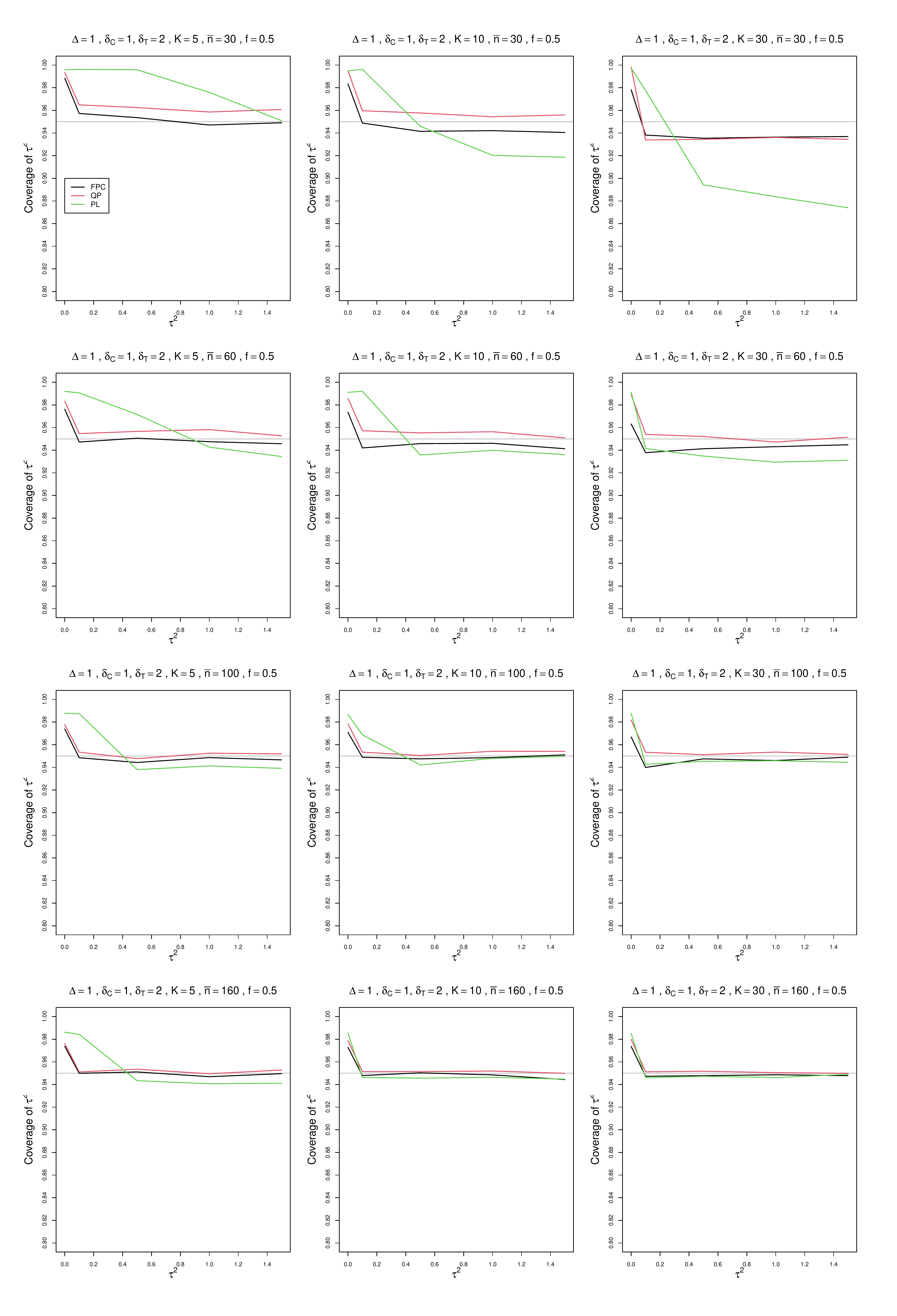}
	\caption{Coverage of PL, QP, and  FPC 95\% confidence intervals for between-study variance of DSM   vs $\tau^2$, for unequal sample sizes $\bar{n}=30,\;60,\;100$ and $160$, $\delta_{iC} = 1$, $\Delta=1$ and  $f = 0.5$.   }
	\label{PlotCoverageOfTau2_deltaC_1deltaT=2_DSM_unequal_sample_sizes.pdf}
\end{figure}

\begin{figure}[ht]
	\centering
	\includegraphics[scale=0.33]{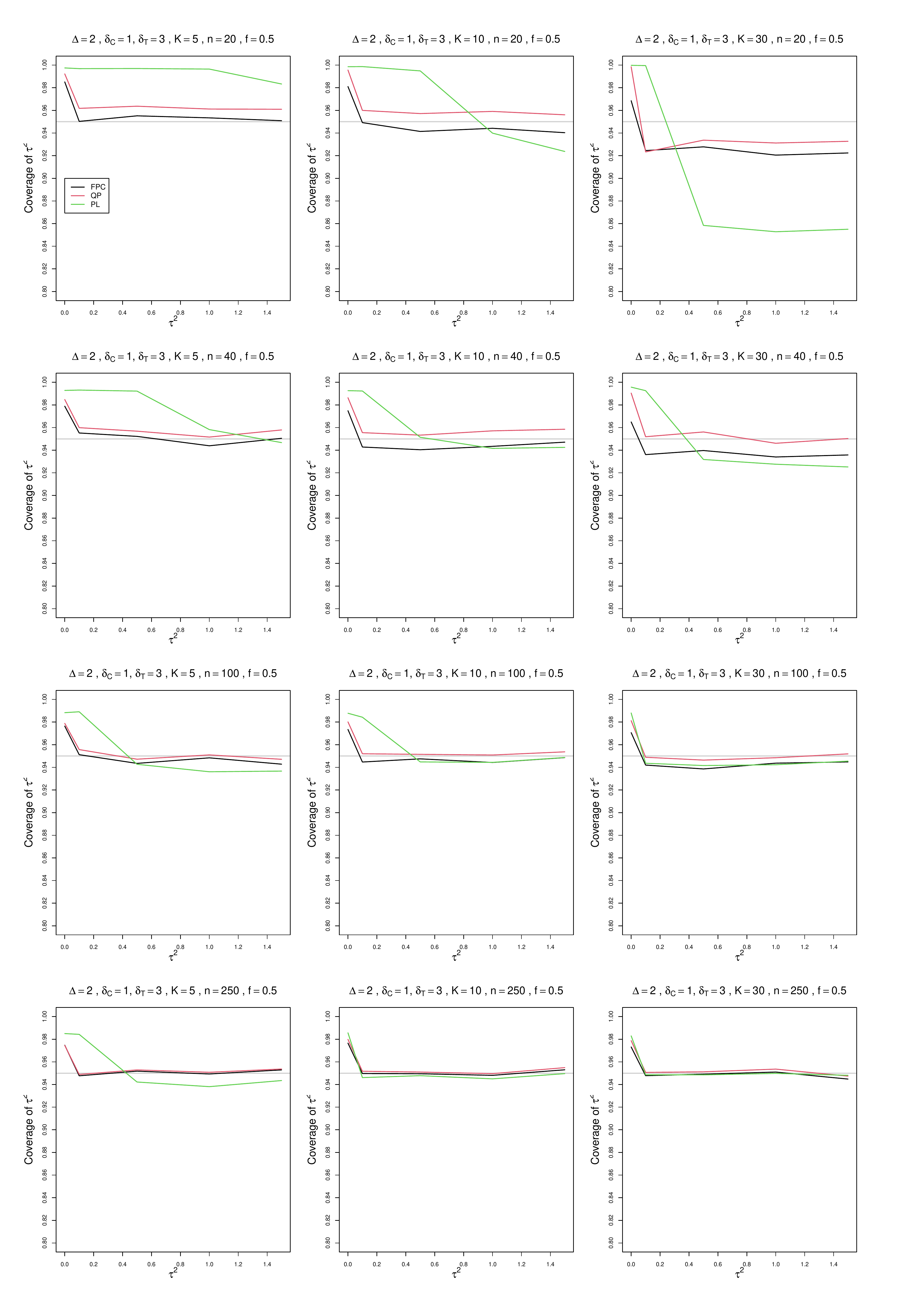}
	\caption{Coverage of PL, QP, and  FPC 95\% confidence intervals for between-study variance of DSM   vs $\tau^2$, for equal sample sizes $n=20,\;40,\;100$ and $250$, $\delta_{iC} = 1$, $\Delta=2$ and  $f = 0.5$.   }
	\label{PlotCoverageOfTau2_deltaC_1deltaT=3_DSM_equal_sample_sizes.pdf}
\end{figure}

\begin{figure}[ht]
	\centering
	\includegraphics[scale=0.33]{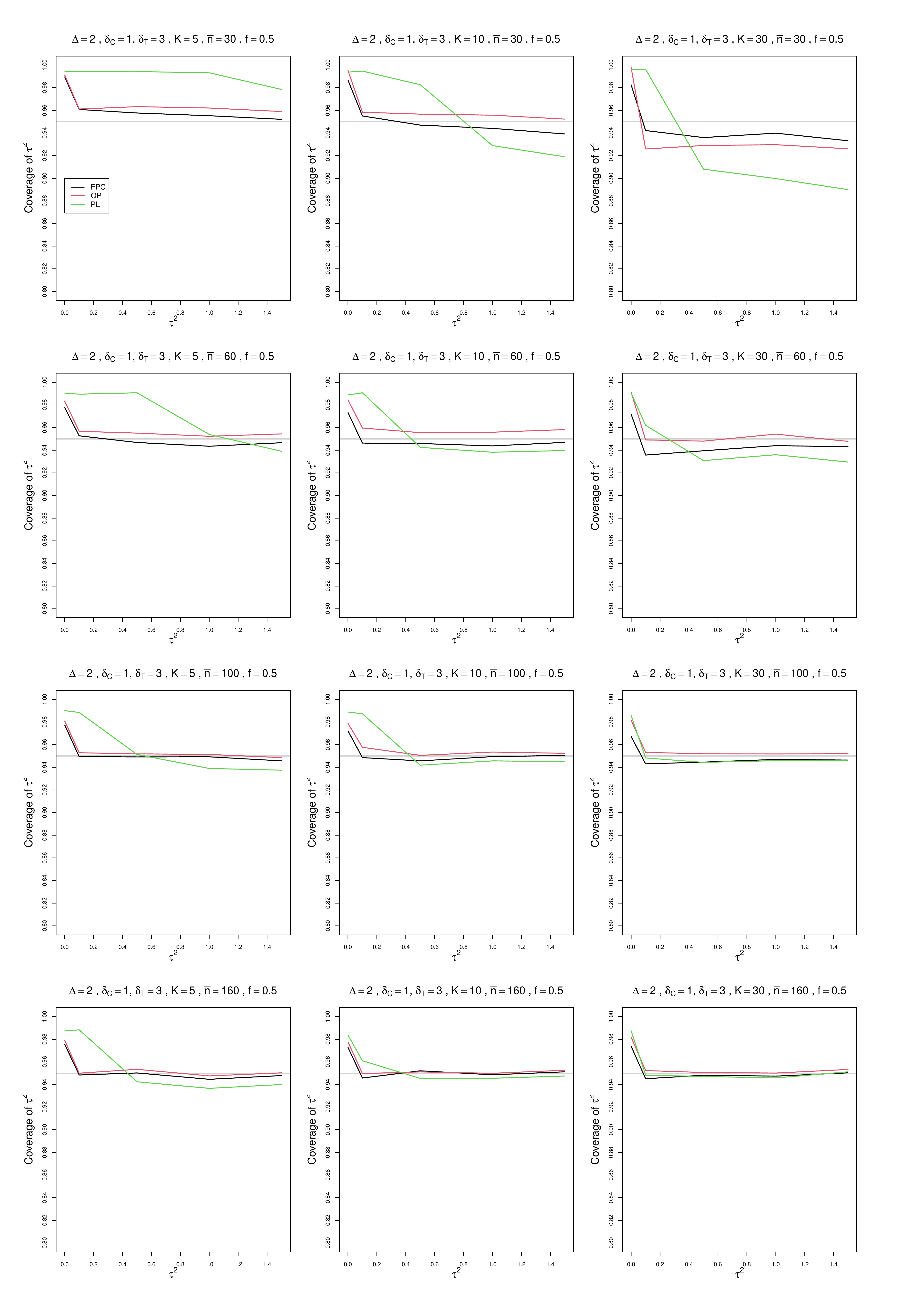}
	\caption{Coverage of PL, QP, and  FPC 95\% confidence intervals for between-study variance of DSM   vs $\tau^2$, for unequal sample sizes $\bar{n}=30,\;60,\;100$ and $160$, $\delta_{iC} = 1$, $\Delta=2$ and  $f = 0.5$.   }
	\label{PlotCoverageOfTau2_deltaC_1deltaT=3_DSM_unequal_sample_sizes.pdf}
\end{figure}

\clearpage

\begin{figure}[ht]
	\centering
	\includegraphics[scale=0.33]{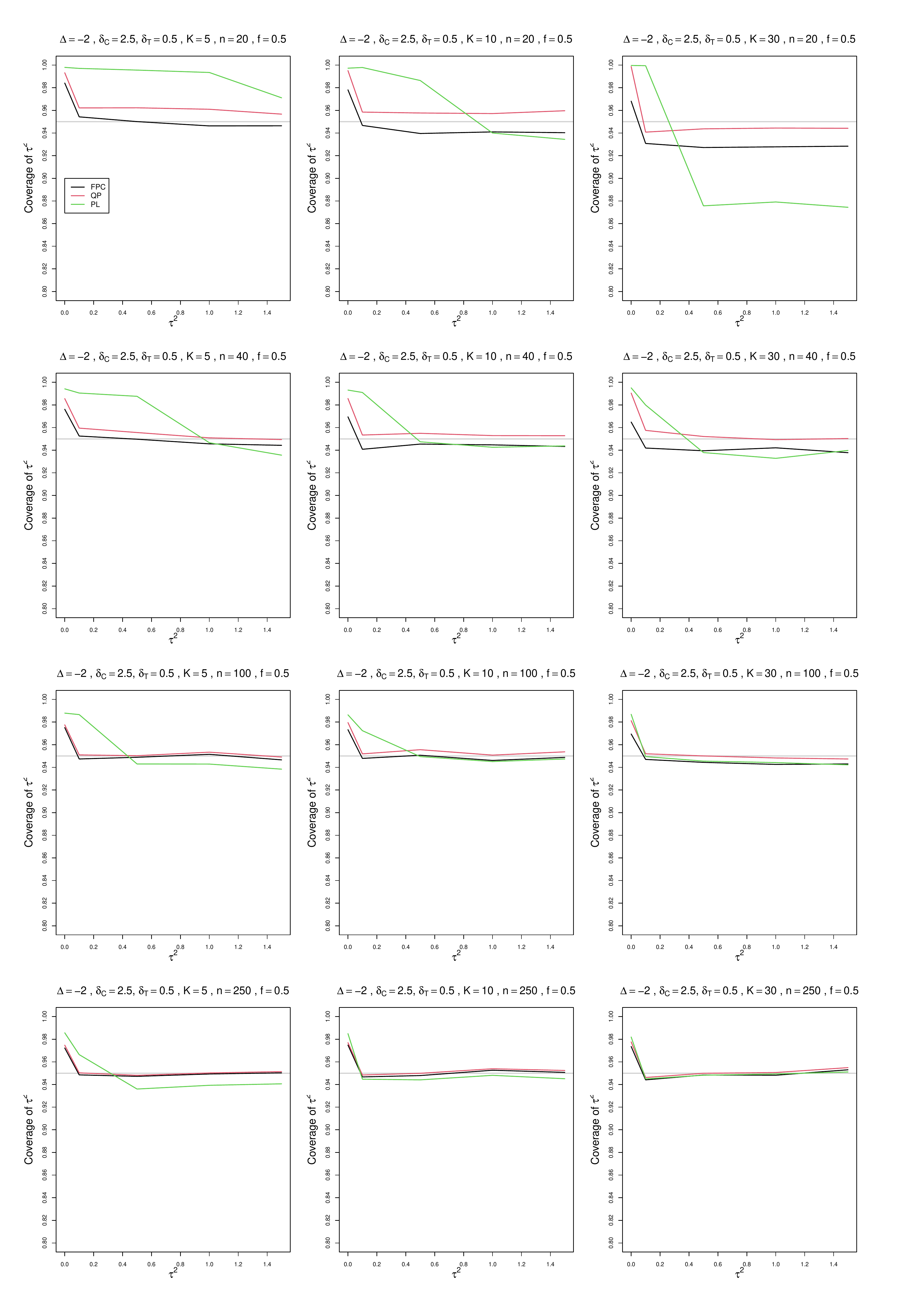}
	\caption{Coverage of PL, QP, and  FPC 95\% confidence intervals for between-study variance of DSM   vs $\tau^2$, for equal sample sizes $n=20,\;40,\;100$ and $250$, $\delta_{iC} = 2.5$, $\Delta=-2$ and  $f = 0.5$.   }
	\label{PlotCoverageOfTau2_deltaC_2.5deltaT=0.5_DSM_equal_sample_sizes.pdf}
\end{figure}

\begin{figure}[ht]
	\centering
	\includegraphics[scale=0.33]{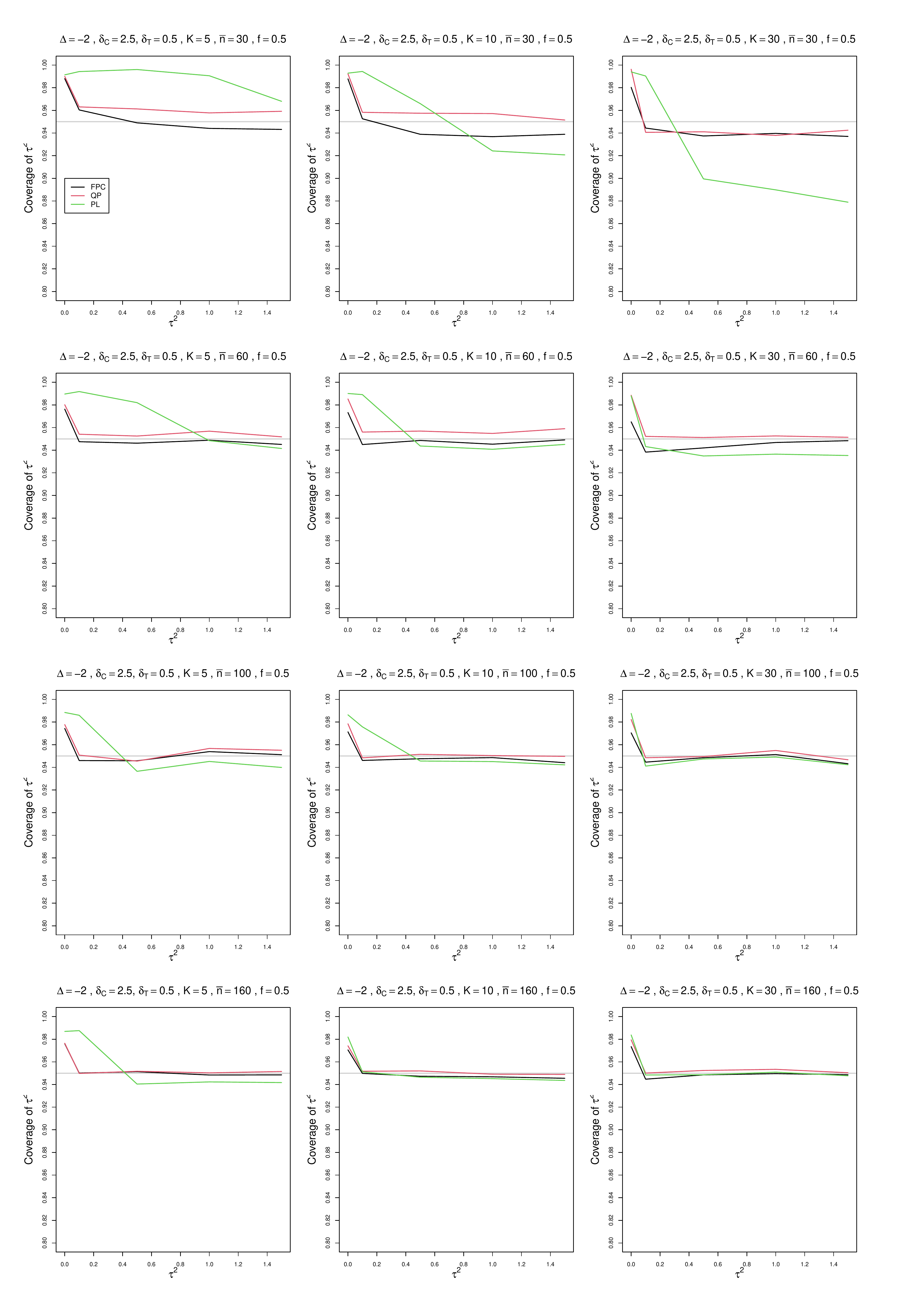}
	\caption{Coverage of PL, QP, and  FPC 95\% confidence intervals for between-study variance of DSM   vs $\tau^2$, for unequal sample sizes $\bar{n}=30,\;60,\;100$ and $160$, $\delta_{iC} = 2.5$, $\Delta=-2$ and  $f = 0.5$.   }
	\label{PlotCoverageOfTau2_deltaC_2.5deltaT=0.5_DSM_unequal_sample_sizes.pdf}
\end{figure}

\begin{figure}[ht]
	\centering
	\includegraphics[scale=0.33]{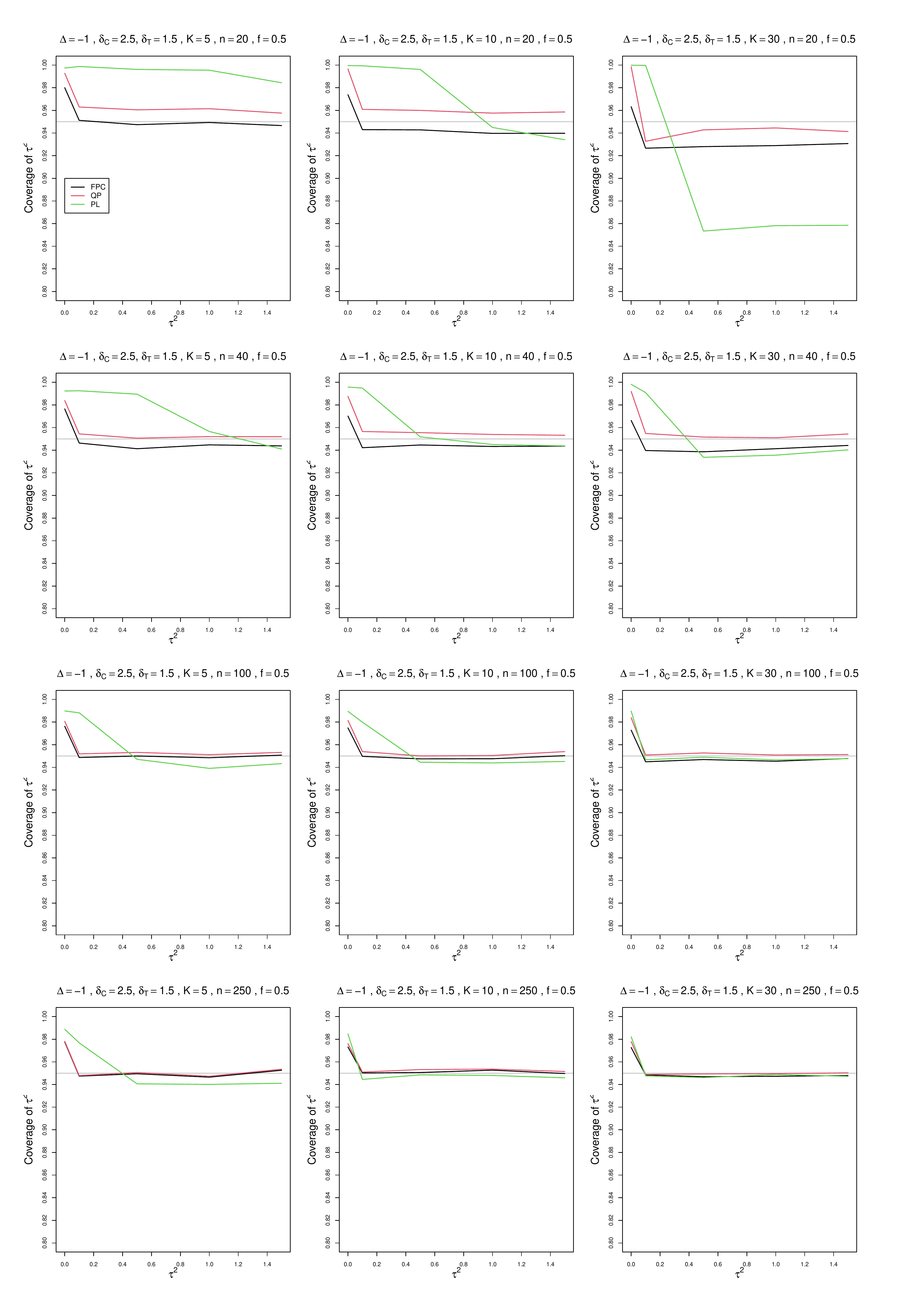}
	\caption{Coverage of PL, QP, and  FPC 95\% confidence intervals for between-study variance of DSM   vs $\tau^2$, for equal sample sizes $n=20,\;40,\;100$ and $250$, $\delta_{iC} = 2.5$, $\Delta=-1$ and  $f = 0.5$.   }
	\label{PlotCoverageOfTau2_deltaC_2.5deltaT=1.5_DSM_equal_sample_sizes.pdf}
\end{figure}

\begin{figure}[ht]
	\centering
	\includegraphics[scale=0.33]{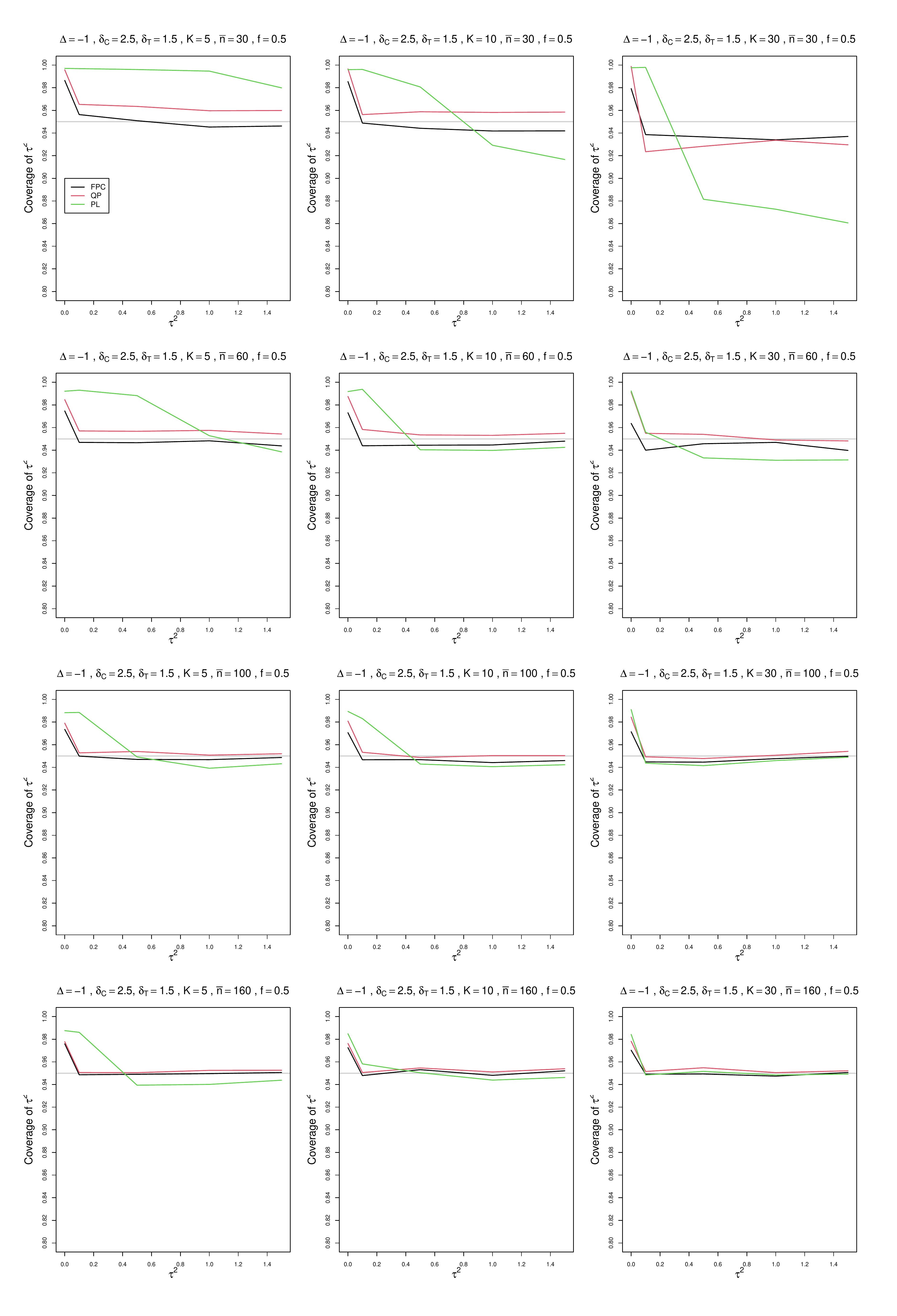}
	\caption{Coverage of PL, QP, and  FPC 95\% confidence intervals for between-study variance of DSM   vs $\tau^2$, for unequal sample sizes $\bar{n}=30,\;60,\;100$ and $160$, $\delta_{iC} = 2.5$, $\Delta=-1$ and  $f = 0.5$.   }
	\label{PlotCoverageOfTau2_deltaC_2.5deltaT=1.5_DSM_unequal_sample_sizes.pdf}
\end{figure}

\begin{figure}[ht]
	\centering
	\includegraphics[scale=0.33]{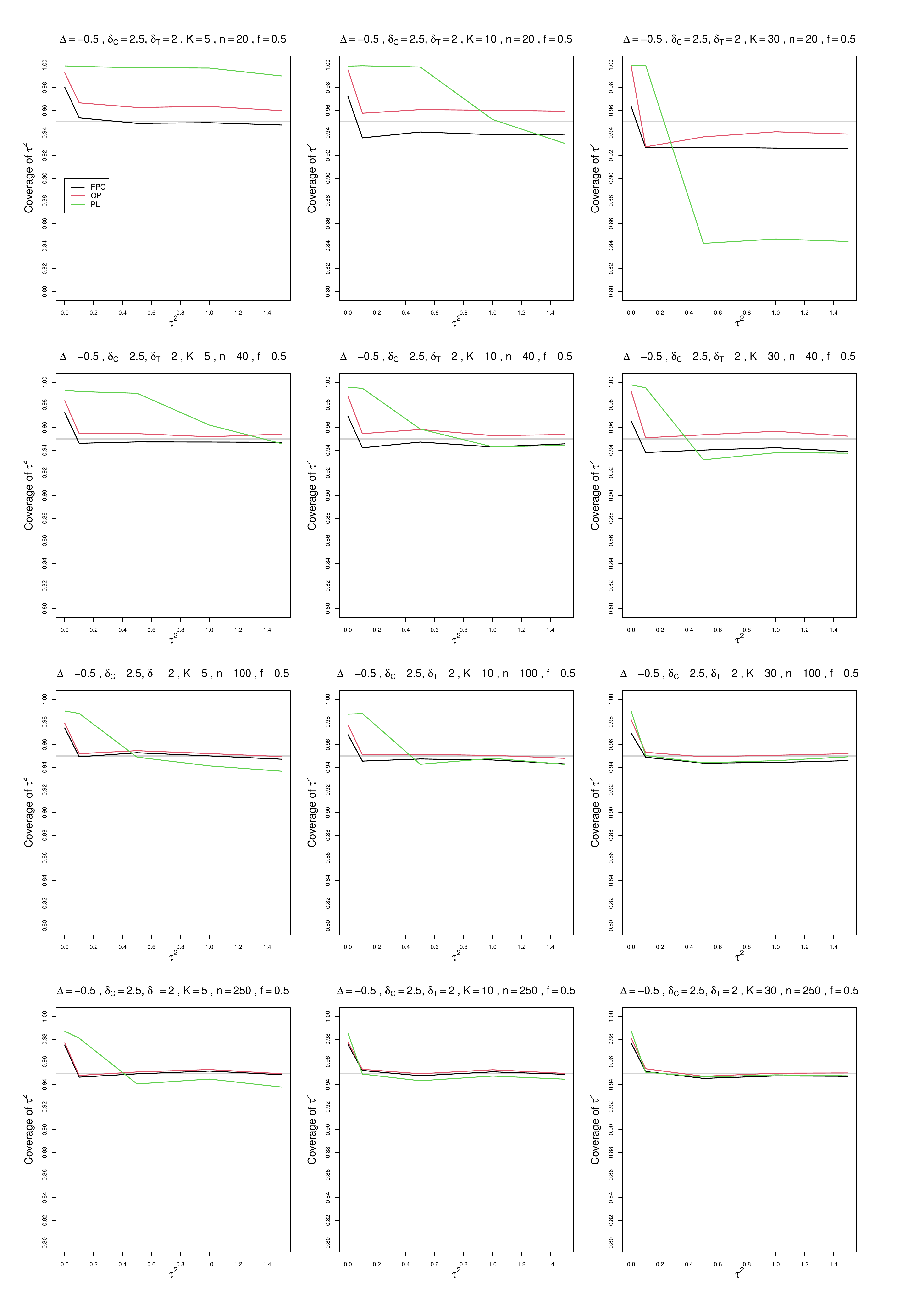}
	\caption{Coverage of PL, QP, and  FPC 95\% confidence intervals for between-study variance of DSM   vs $\tau^2$, for equal sample sizes $n=20,\;40,\;100$ and $250$, $\delta_{iC} = 2.5$, $\Delta=-0.5$ and  $f = 0.5$.   }
	\label{PlotCoverageOfTau2_deltaC_2.5deltaT=2_DSM_equal_sample_sizes.pdf}
\end{figure}

\begin{figure}[ht]
	\centering
	\includegraphics[scale=0.33]{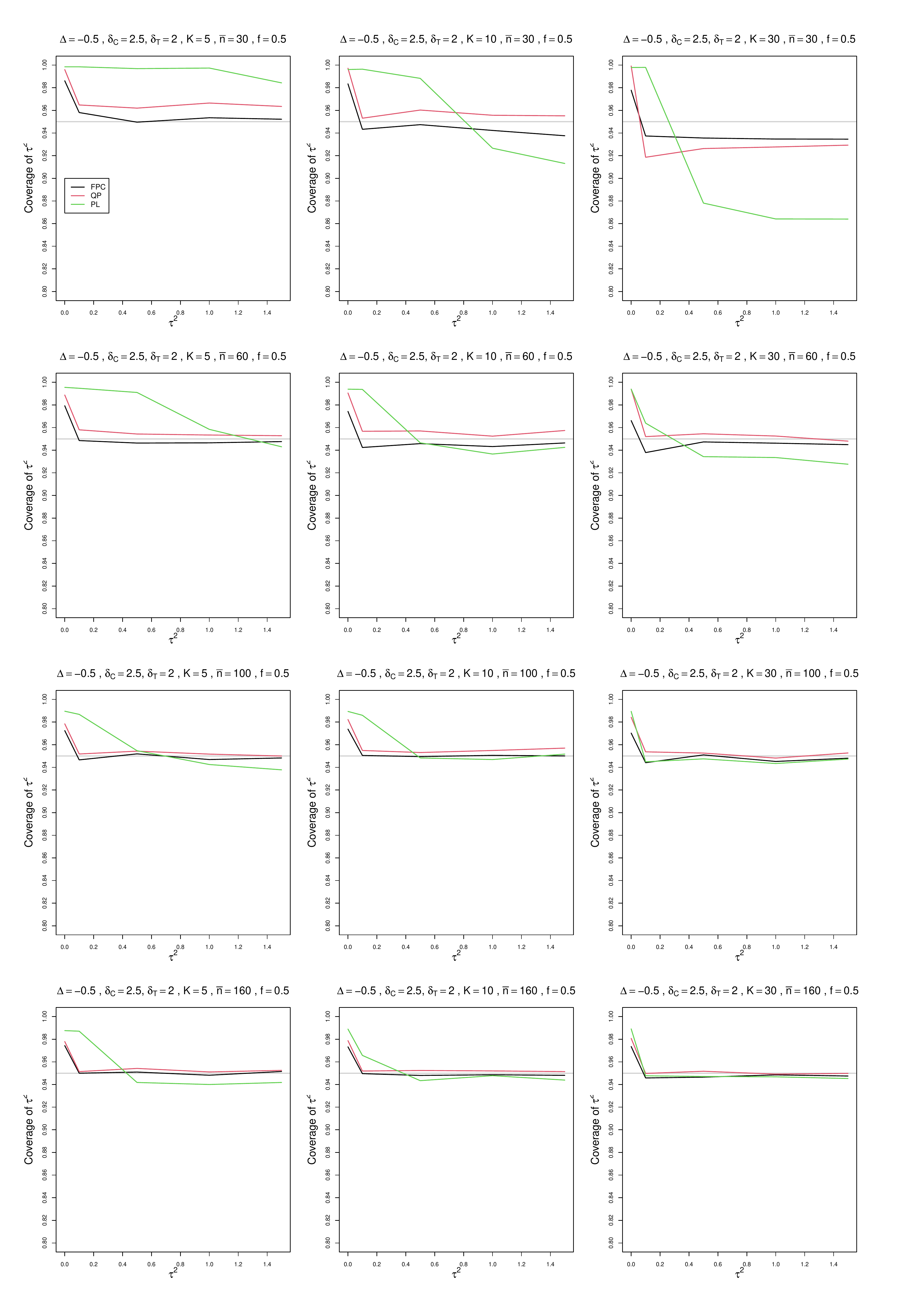}
	\caption{Coverage of PL, QP, and  FPC 95\% confidence intervals for between-study variance of DSM   vs $\tau^2$, for unequal sample sizes $\bar{n}=30,\;60,\;100$ and $160$, $\delta_{iC} = 2.5$, $\Delta=-0.5$ and  $f = 0.5$.   }
	\label{PlotCoverageOfTau2_deltaC_2.5deltaT=2_DSM_unequal_sample_sizes.pdf}
\end{figure}

\begin{figure}[ht]
	\centering
	\includegraphics[scale=0.33]{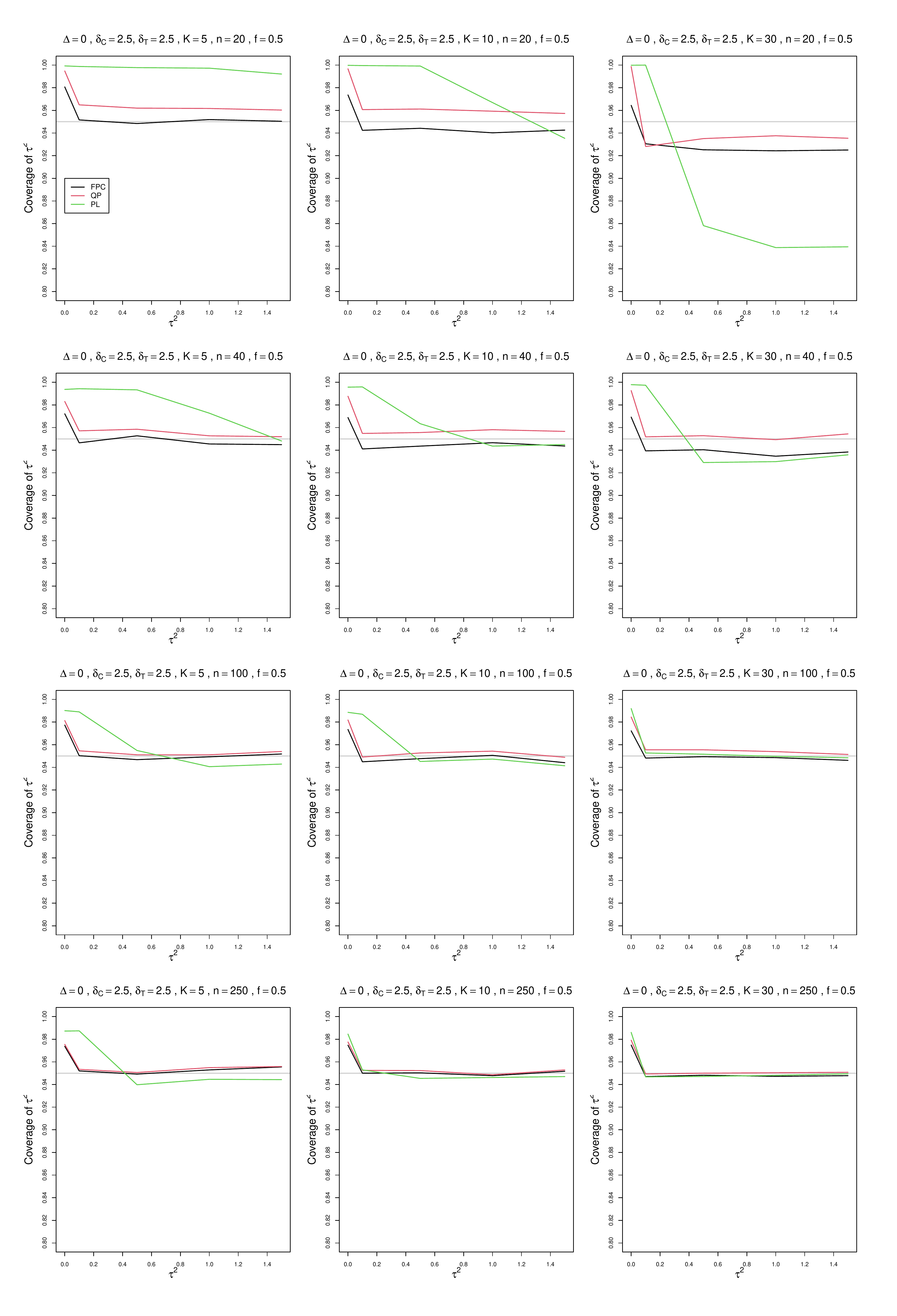}
	\caption{Coverage of PL, QP, and  FPC 95\% confidence intervals for between-study variance of DSM   vs $\tau^2$, for equal sample sizes $n=20,\;40,\;100$ and $250$, $\delta_{iC} = 2.5$, $\Delta=0$ and  $f = 0.5$.   }
	\label{PlotCoverageOfTau2_deltaC_2.5deltaT=2.5_DSM_equal_sample_sizes.pdf}
\end{figure}

\begin{figure}[ht]
	\centering
	\includegraphics[scale=0.33]{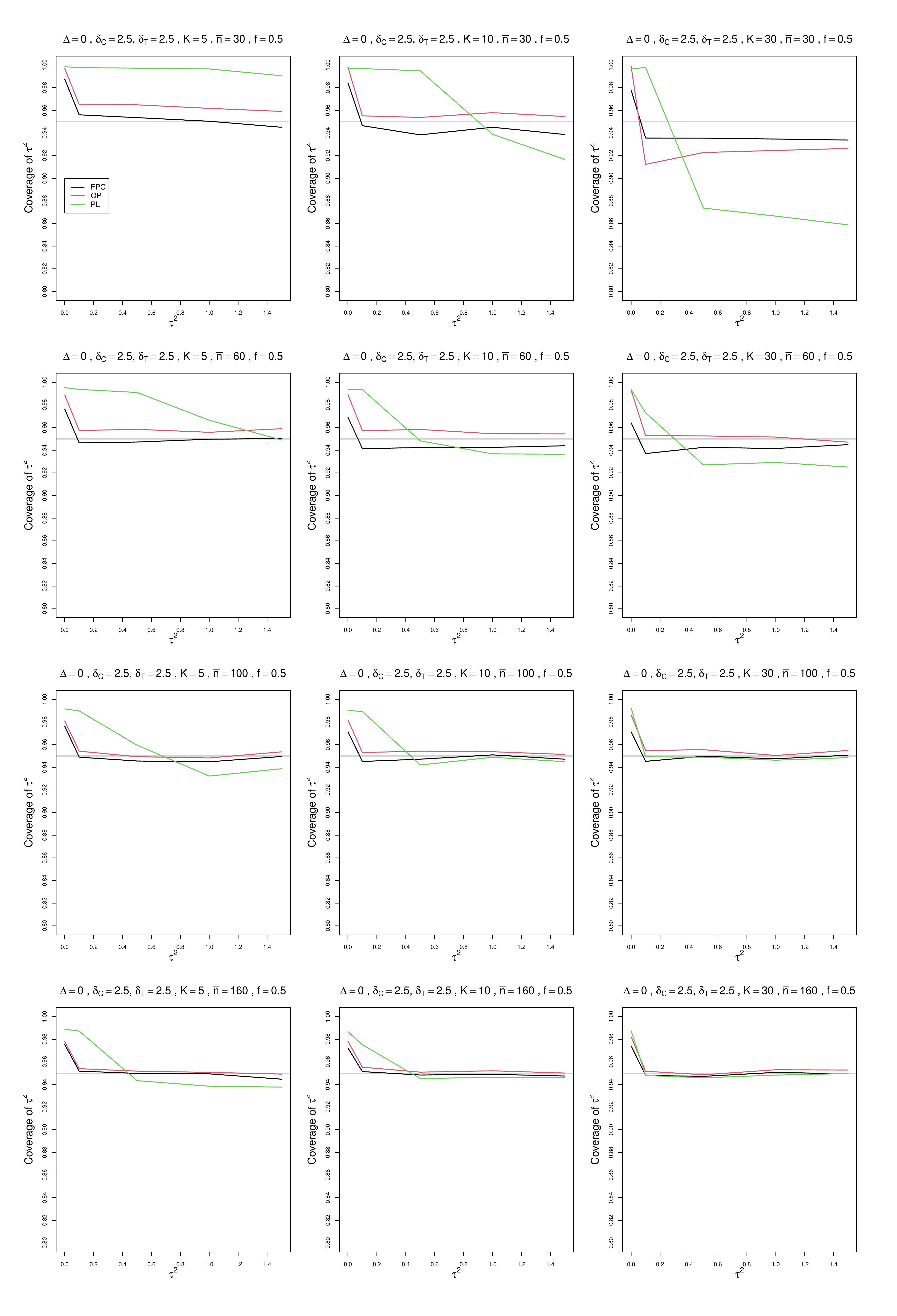}
	\caption{Coverage of PL, QP, and  FPC 95\% confidence intervals for between-study variance of DSM   vs $\tau^2$, for unequal sample sizes $\bar{n}=30,\;60,\;100$ and $160$, $\delta_{iC} = 2.5$, $\Delta=0$ and  $f = 0.5$.   }
	\label{PlotCoverageOfTau2_deltaC_2.5deltaT=2.5_DSM_unequal_sample_sizes.pdf}
\end{figure}

\begin{figure}[ht]
	\centering
	\includegraphics[scale=0.33]{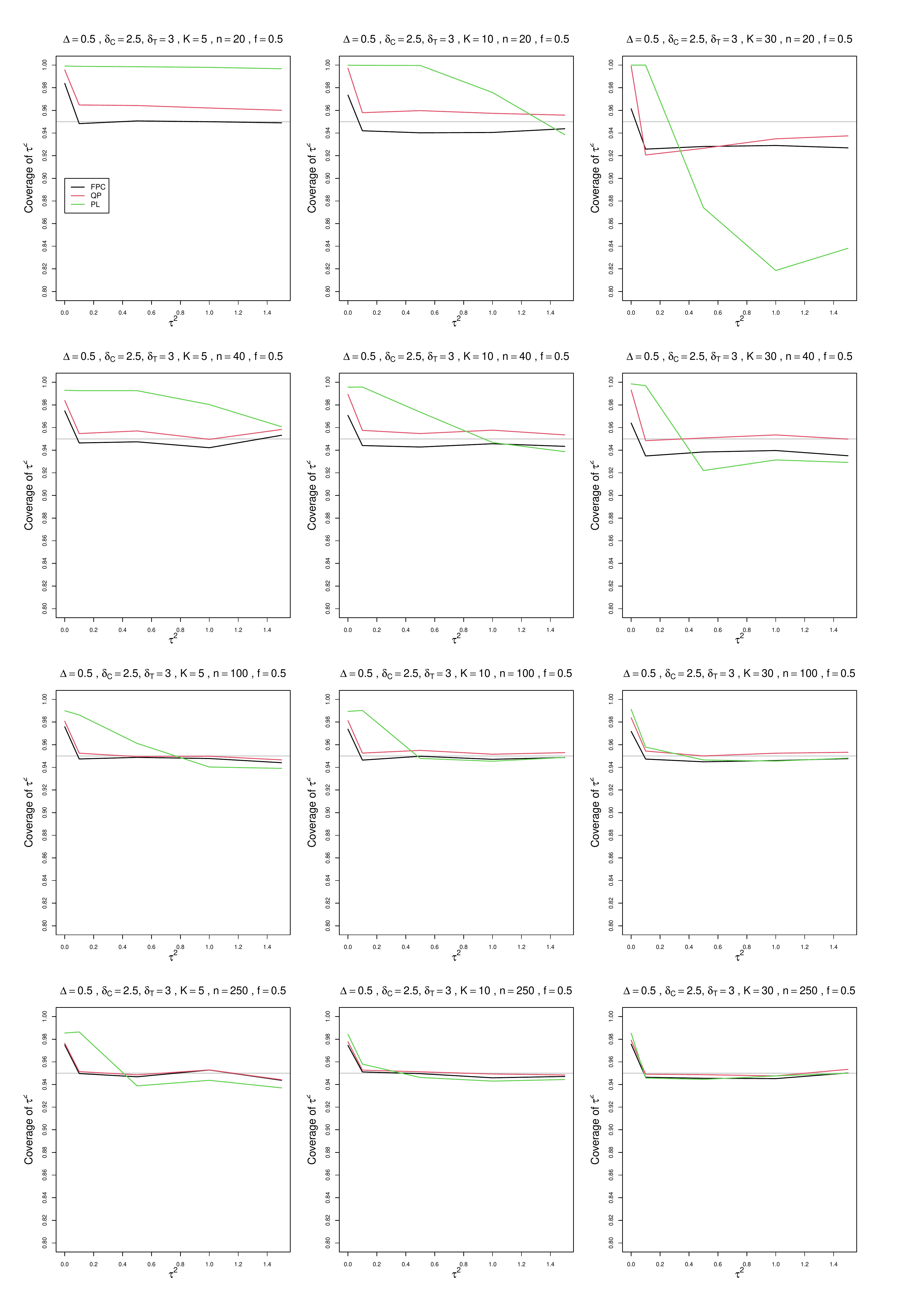}
	\caption{Coverage of PL, QP, and  FPC 95\% confidence intervals for between-study variance of DSM   vs $\tau^2$, for equal sample sizes $n=20,\;40,\;100$ and $250$, $\delta_{iC} = 2.5$, $\Delta=0.5$ and  $f = 0.5$.   }
	\label{PlotCoverageOfTau2_deltaC_2.5deltaT=3_DSM_equal_sample_sizes.pdf}
\end{figure}

\begin{figure}[ht]
	\centering
	\includegraphics[scale=0.33]{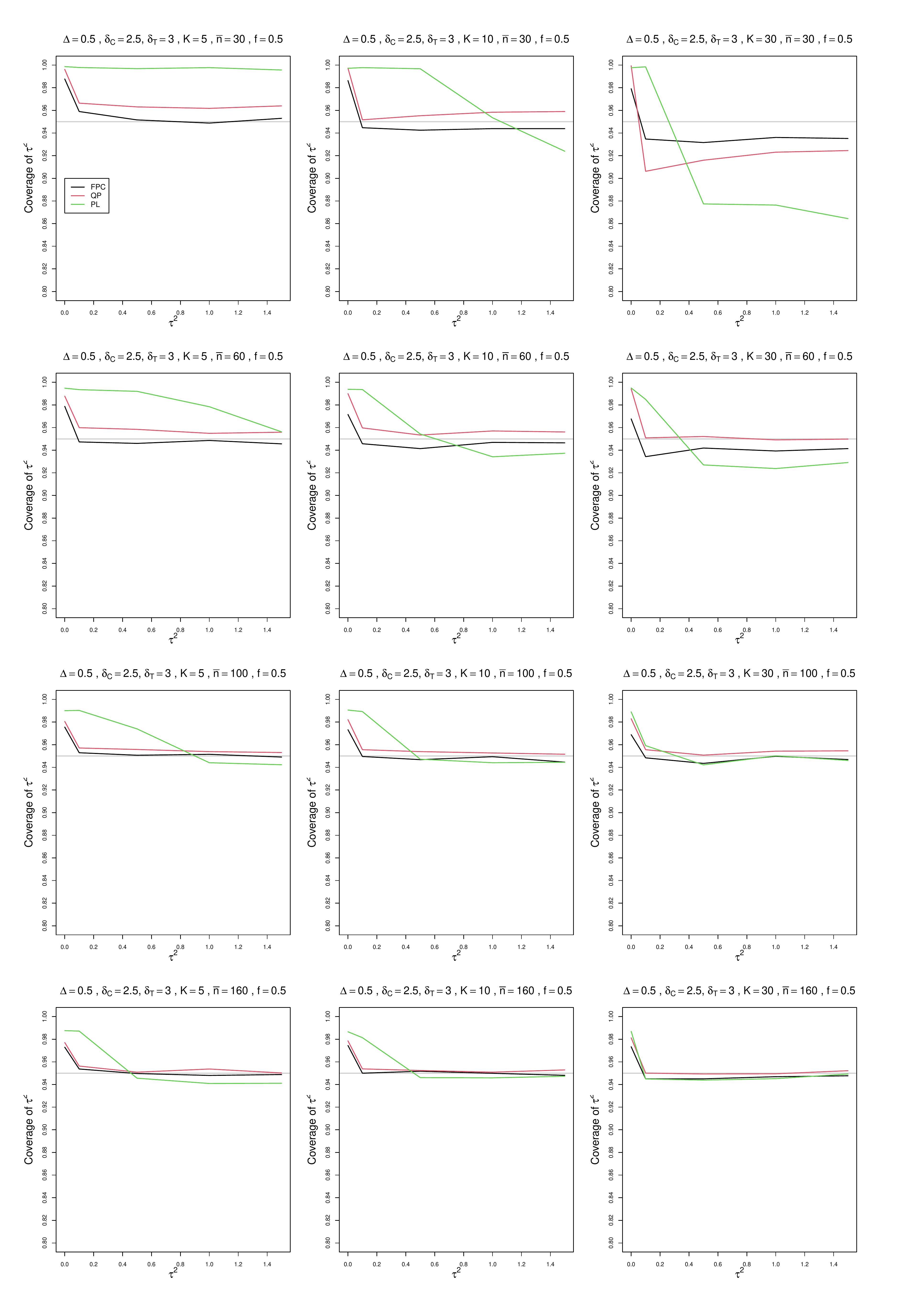}
	\caption{Coverage of PL, QP, and  FPC 95\% confidence intervals for between-study variance of DSM   vs $\tau^2$, for unequal sample sizes $\bar{n}=30,\;60,\;100$ and $160$, $\delta_{iC} = 2.5$, $\Delta=0.5$ and  $f = 0.5$.   }
	\label{PlotCoverageOfTau2_deltaC_2.5deltaT=3_DSM_unequal_sample_sizes.pdf}
\end{figure}

\begin{figure}[ht]
	\centering
	\includegraphics[scale=0.33]{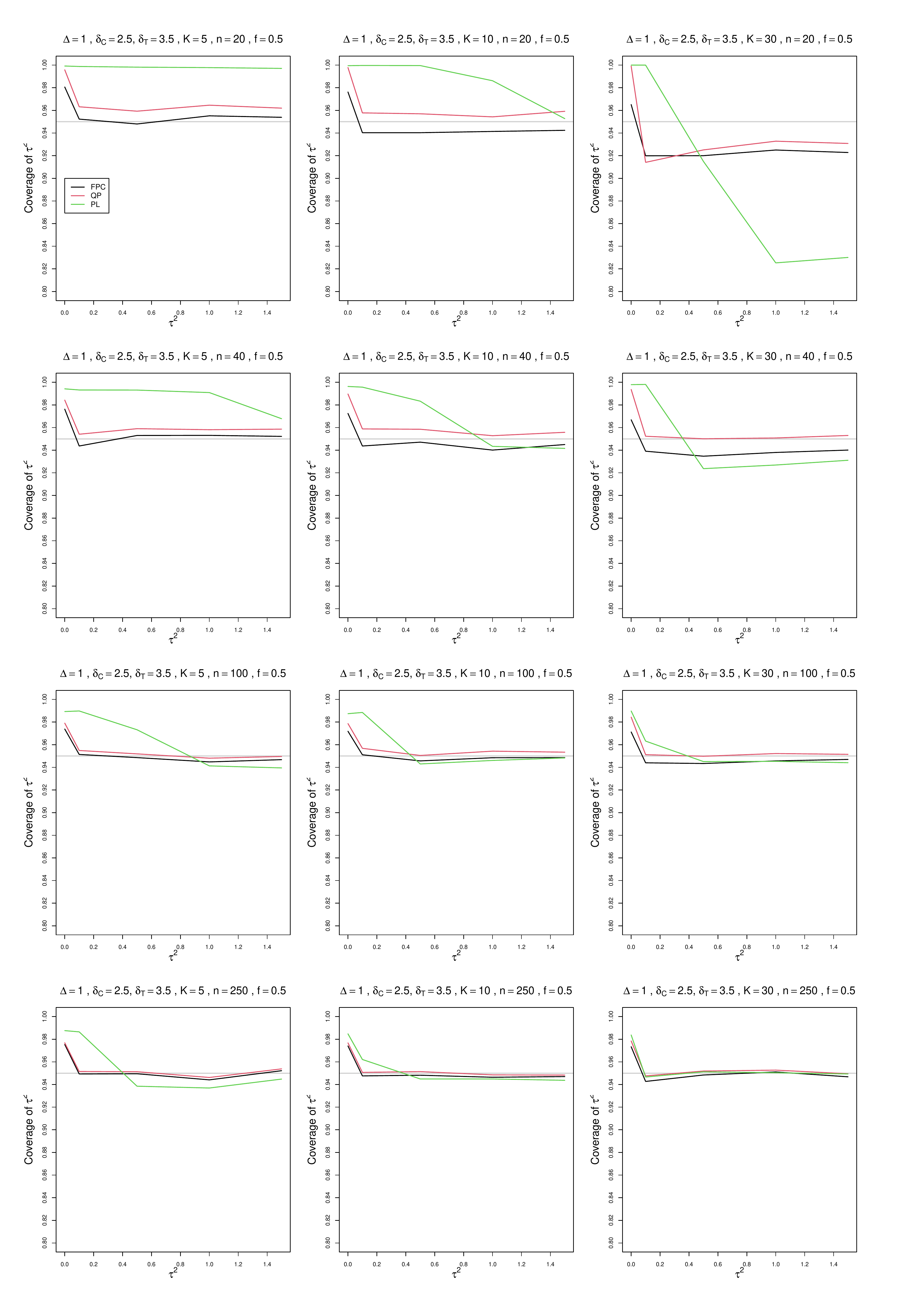}
	\caption{Coverage of PL, QP, and  FPC 95\% confidence intervals for between-study variance of DSM   vs $\tau^2$, for equal sample sizes $n=20,\;40,\;100$ and $250$, $\delta_{iC} = 2.5$, $\Delta=1$ and  $f = 0.5$.   }
	\label{PlotCoverageOfTau2_deltaC_2.5deltaT=3.5_DSM_equal_sample_sizes.pdf}
\end{figure}

\begin{figure}[ht]
	\centering
	\includegraphics[scale=0.33]{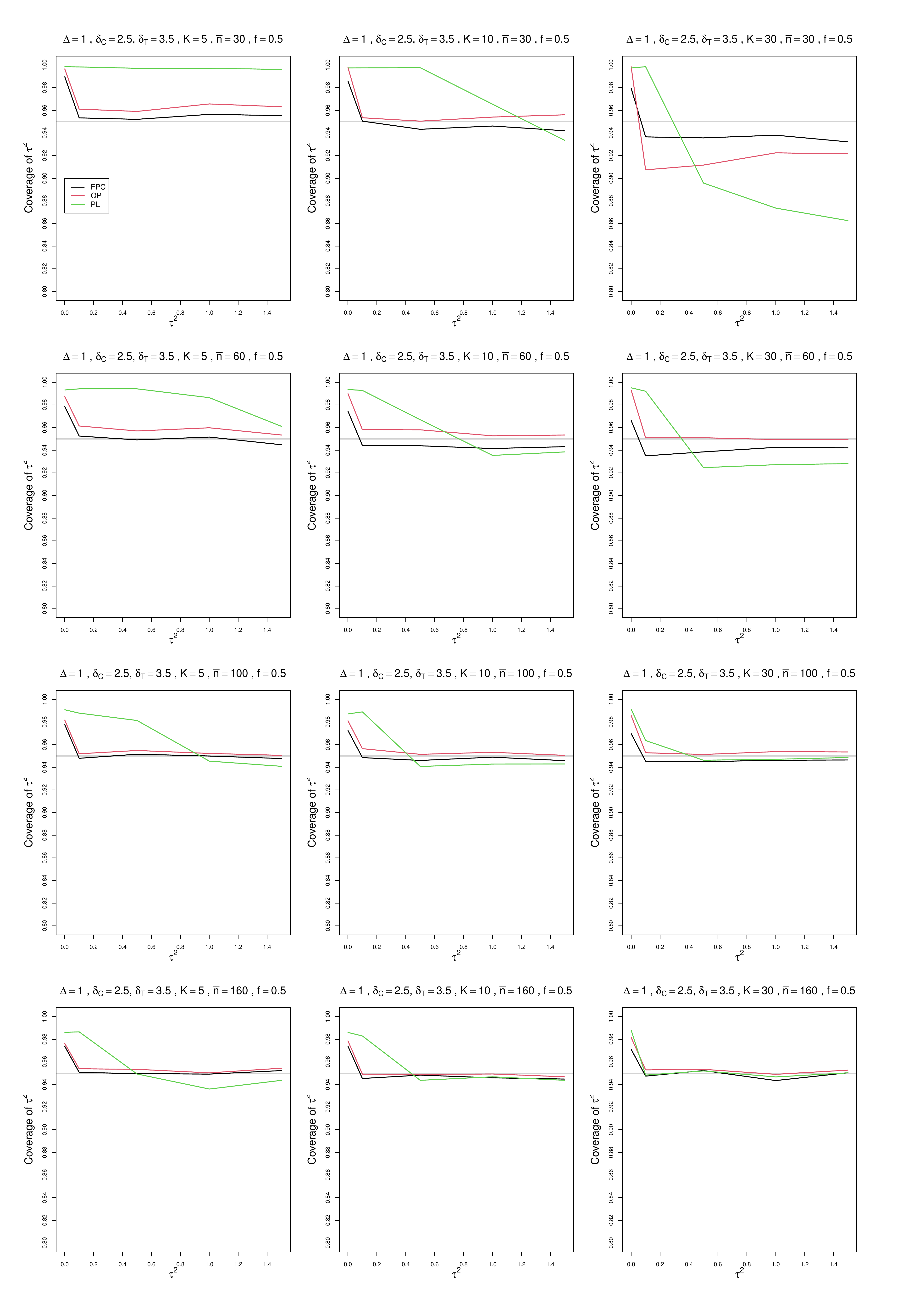}
	\caption{Coverage of PL, QP, and  FPC 95\% confidence intervals for between-study variance of DSM   vs $\tau^2$, for unequal sample sizes $\bar{n}=30,\;60,\;100$ and $160$, $\delta_{iC} = 2.5$, $\Delta=1$ and  $f = 0.5$.   }
	\label{PlotCoverageOfTau2_deltaC_2.5deltaT=3.5_DSM_unequal_sample_sizes.pdf}
\end{figure}

\begin{figure}[ht]
	\centering
	\includegraphics[scale=0.33]{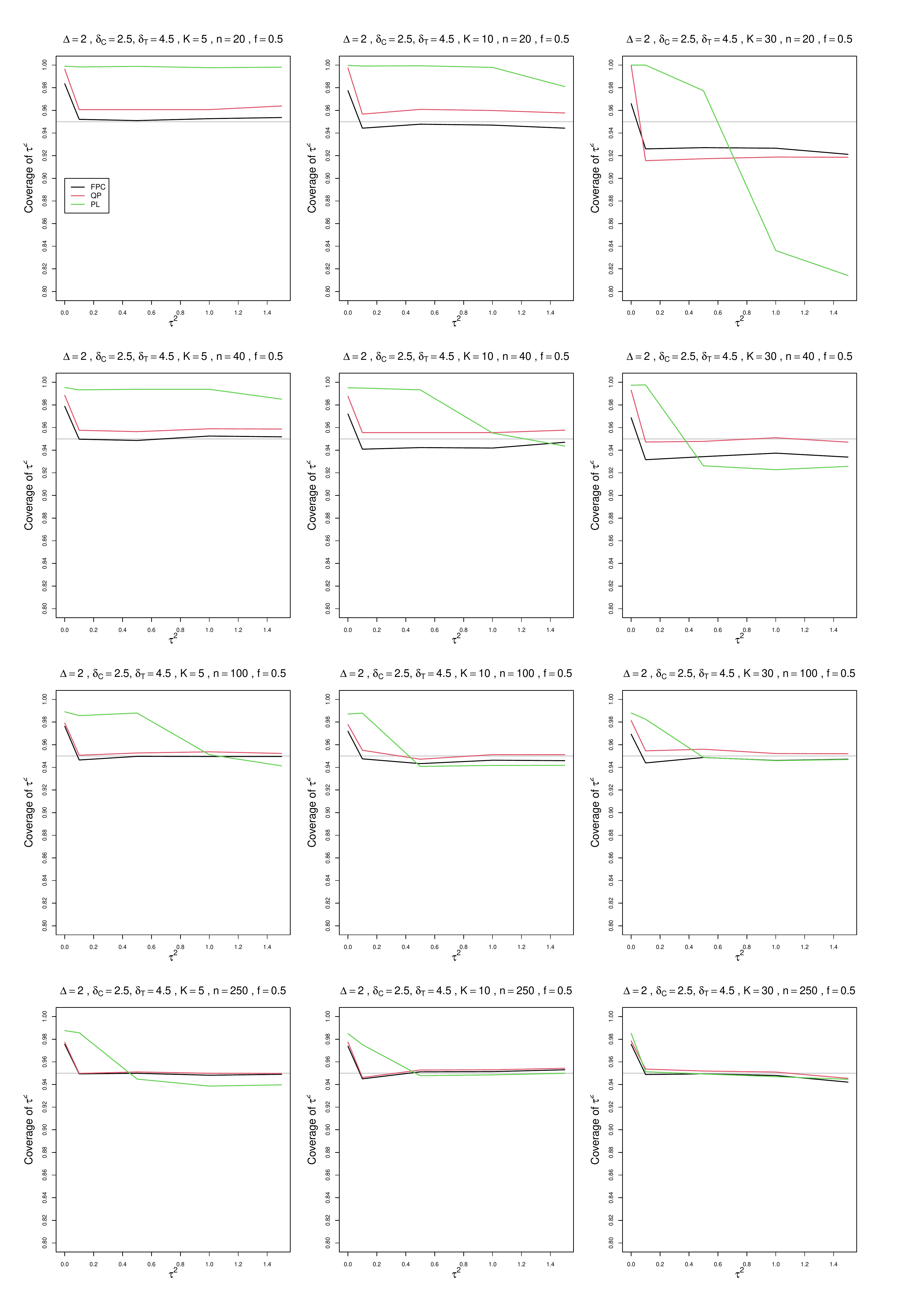}
	\caption{Coverage of PL, QP, and  FPC 95\% confidence intervals for between-study variance of DSM   vs $\tau^2$, for equal sample sizes $n=20,\;40,\;100$ and $250$, $\delta_{iC} = 2.5$, $\Delta=2$ and  $f = 0.5$.   }
	\label{PlotCoverageOfTau2_deltaC_2.5deltaT=2.5_DSM_equal_sample_sizes.pdf}
\end{figure}

\begin{figure}[ht]
	\centering
	\includegraphics[scale=0.33]{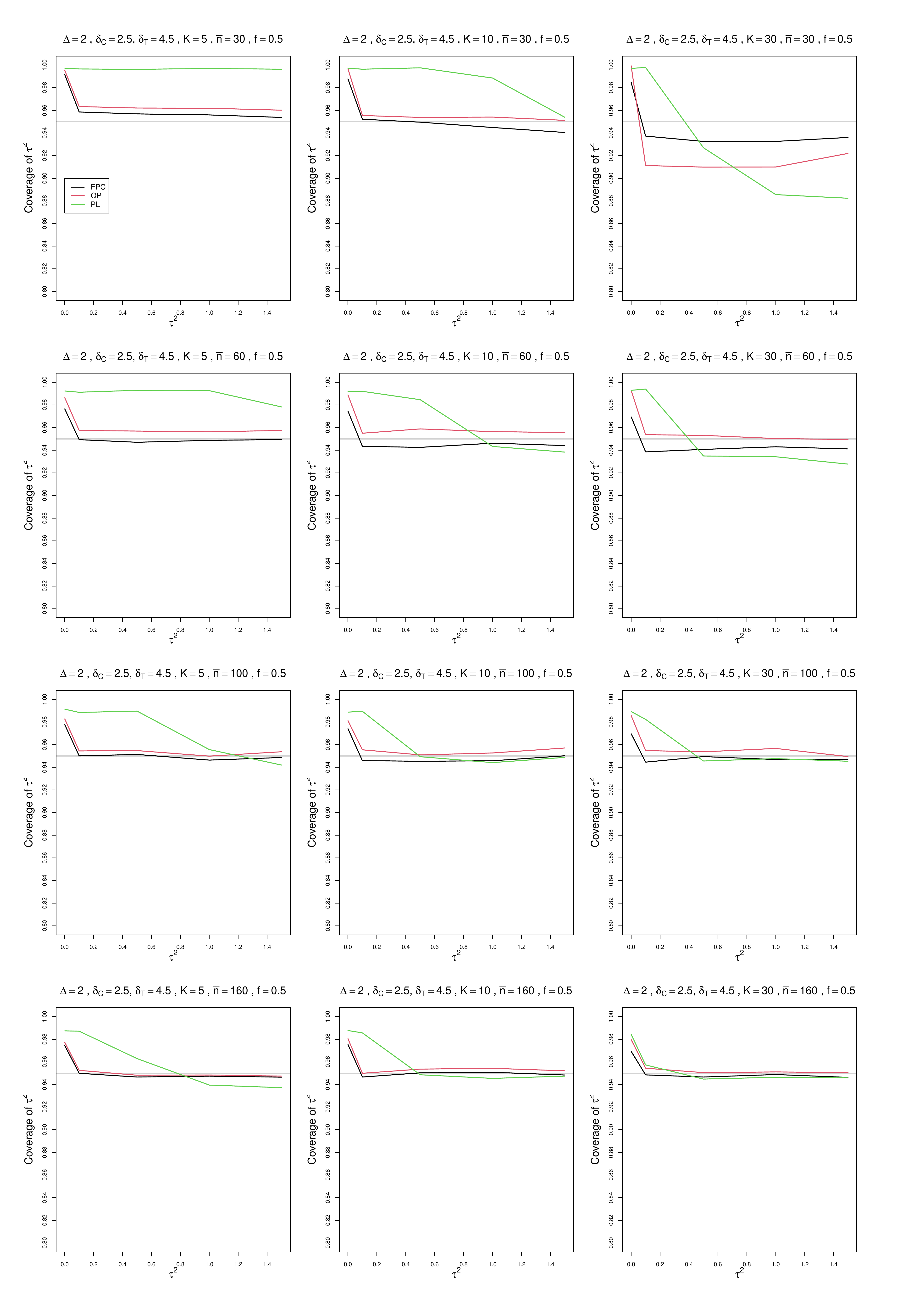}
	\caption{Coverage of PL, QP, and  FPC 95\% confidence intervals for between-study variance of DSM   vs $\tau^2$, for unequal sample sizes $\bar{n}=30,\;60,\;100$ and $160$, $\delta_{iC} = 2.5$, $\Delta=2$ and  $f = 0.5$.   }
	\label{PlotCoverageOfTau2_deltaC_2.5deltaT=4.5_DSM_unequal_sample_sizes.pdf}
\end{figure}


\clearpage

\section*{Appendix E: Bias in point estimation of $\Delta$}


Each figure corresponds to a value of the standardized mean  in the Control arm $\delta_{C}$  (= $-2.5$, $-1$, 0, 1, 2.5)  and a value of the overall DSM $\Delta$ (= $-2$, $-1$, $-0.5$, 0, 0.5, 1, 2) . \\
The fraction of each study's sample size in the Control arm ($f$) is held constant at 0.5.

For each combination of a value of $n$ (= 20, 40, 100, 250) or  $\bar{n}$ (= 30, 60, 100, 160) and a value of $K$ (= 5, 10, 30), a panel plots bias versus $\tau^2$ (= 0, 0.1, 0.5, 1, 1.5).\\
The point estimators of $\Delta$ are
\begin{itemize}
\item DL (DerSimonian-Laird) method, inverse-variance weights
\item REML method, inverse-variance weights
\item MP (Mandel-Paule) method, inverse-variance weights
\item SSW method, effective-sample-size weights
\end{itemize}

\setcounter{figure}{0}
\setcounter{section}{0}

\clearpage
\renewcommand{\thefigure}{E.\arabic{figure}}


\begin{figure}[ht]
	\centering
	\includegraphics[scale=0.33]{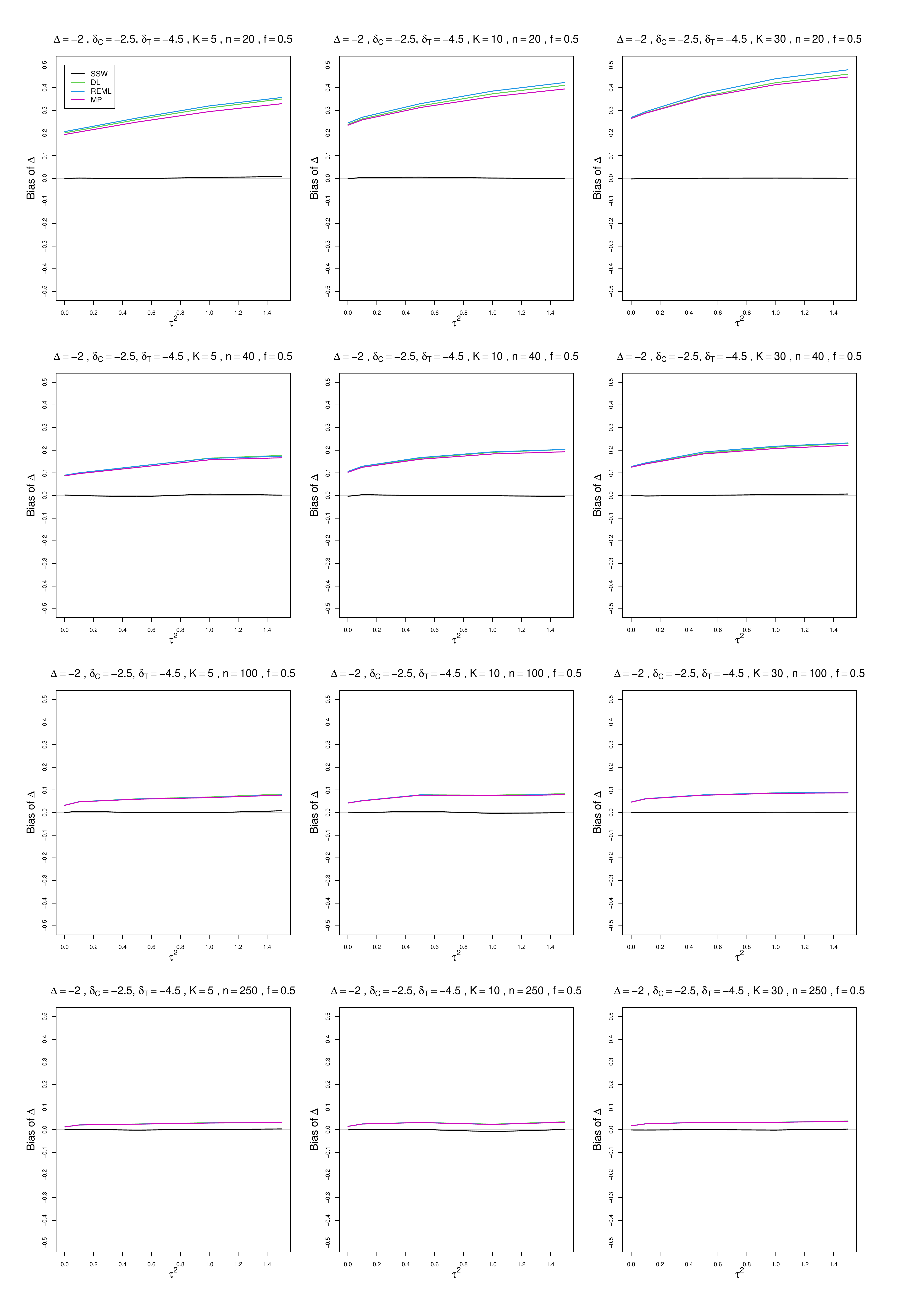}
	\caption{Bias of  estimators of  overall effect measure $\Delta$  (DL, REML, MP, MP and SSW estimators) vs $\tau^2$, for equal sample sizes $n=20,\;40,\;100$ and $250$, $\delta_{iC} = -2.5$, $\Delta=-2$ and  $f = 0.5$.   }
	\label{PlotBiasOfDelta_deltaC_-25deltaT=-4.5_DSM_equal_sample_sizes.pdf}
\end{figure}

\begin{figure}[ht]
	\centering
	\includegraphics[scale=0.33]{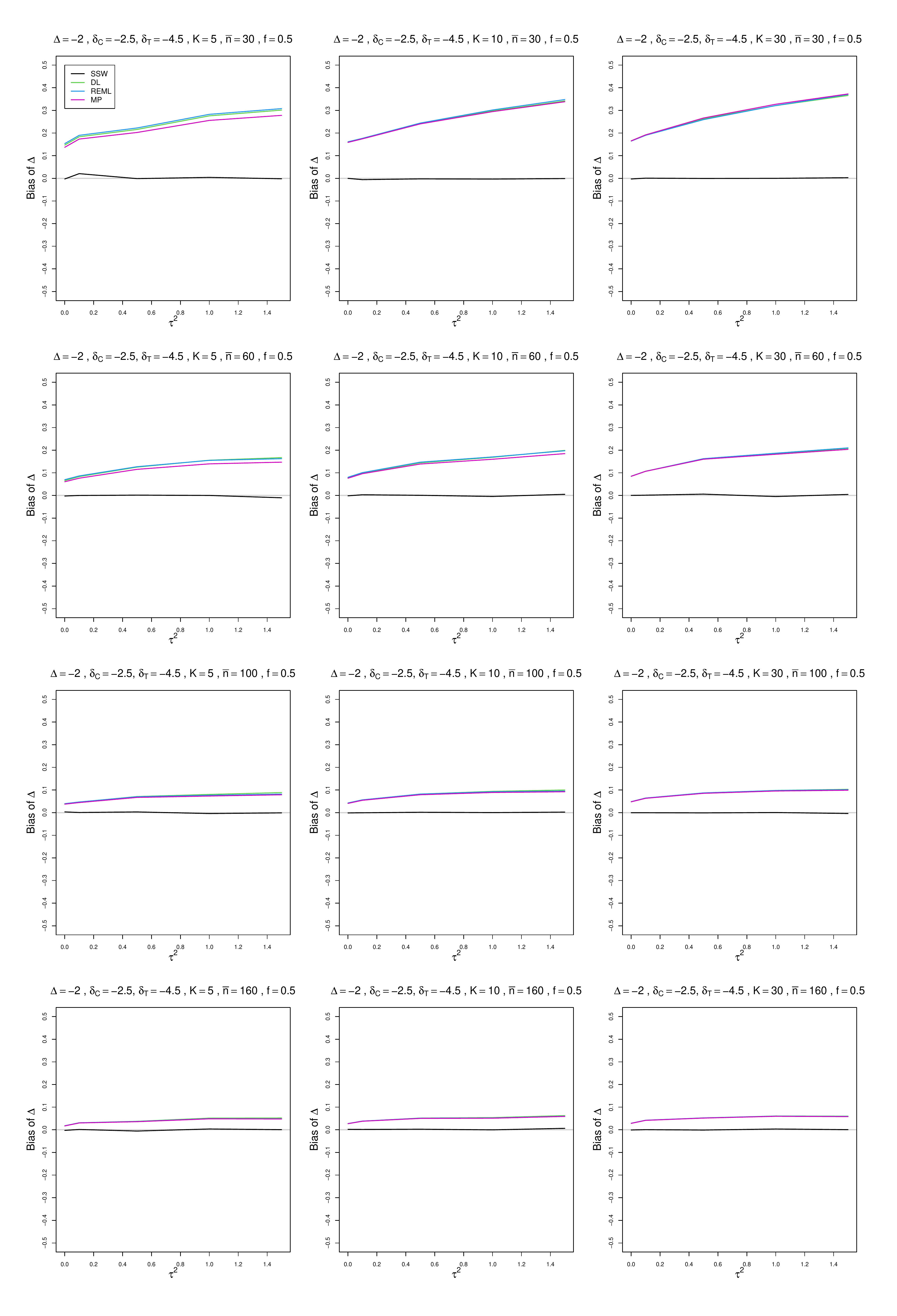}
	\caption{Bias  of estimators of overall effect measure $\Delta$ (DL, REML, MP  and SSW ) vs $\tau^2$, for unequal sample sizes $\bar{n}=30,\;60,\;100$ and $160$, $\delta_{iC} = -2.5$, $\Delta=-2$ and  $f = 0.5$.   }
	\label{PlotBiasOfDelta_deltaC_-25deltaT=-4.5_DSM_unequal_sample_sizes.pdf}
\end{figure}

\begin{figure}[ht]
	\centering
	\includegraphics[scale=0.33]{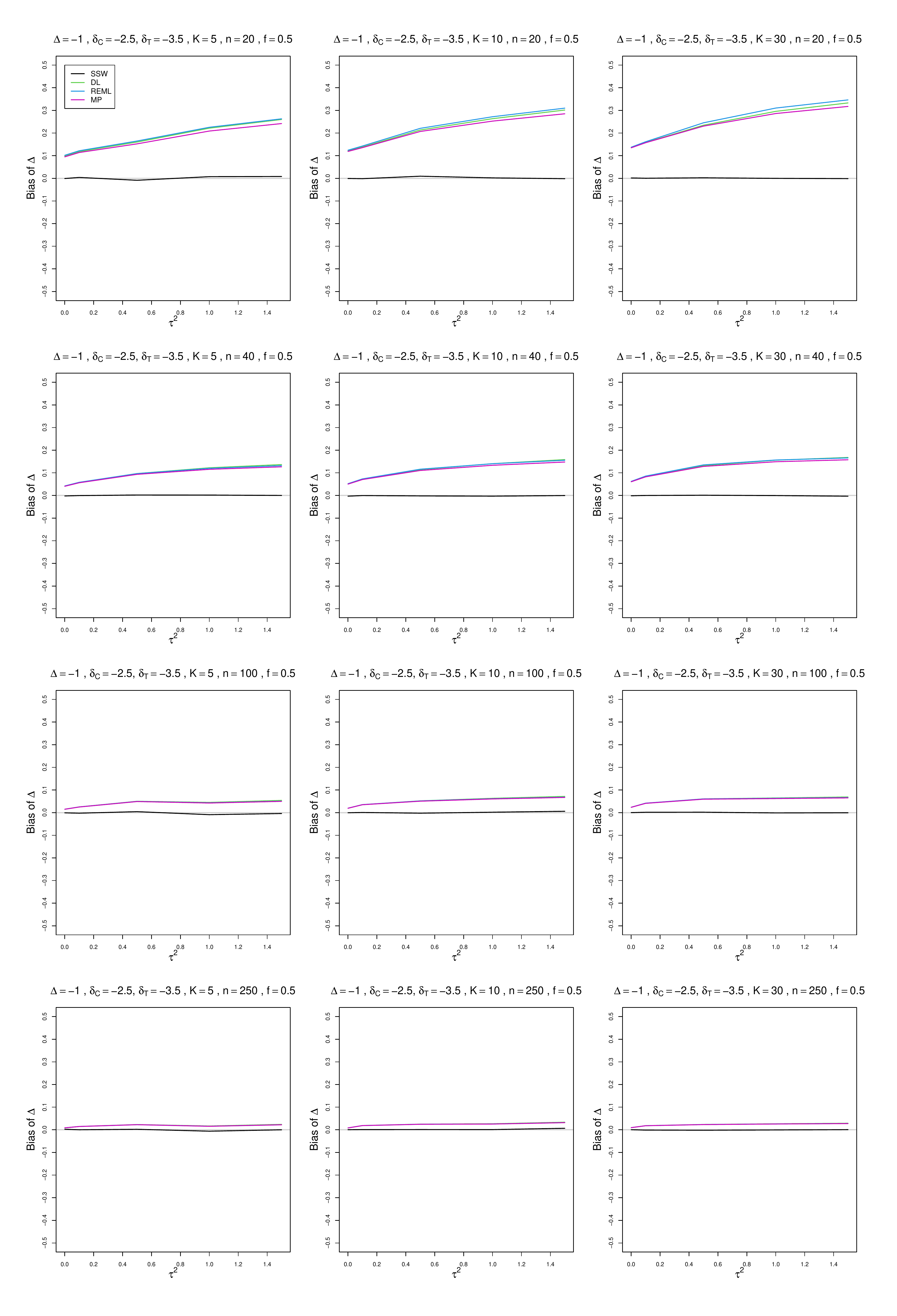}
	\caption{Bias  of estimators of overall effect measure $\Delta$ (DL, REML, MP and SSW ) vs $\tau^2$, for equal sample sizes $n=20,\;40,\;100$ and $250$, $\delta_{iC} = -2.5$, $\Delta=-1$ and  $f = 0.5$.   }
	\label{PlotBiasOfDelta_deltaC_-25deltaT=-3.5_DSM_equal_sample_sizes.pdf}
\end{figure}

\begin{figure}[ht]
	\centering
	\includegraphics[scale=0.33]{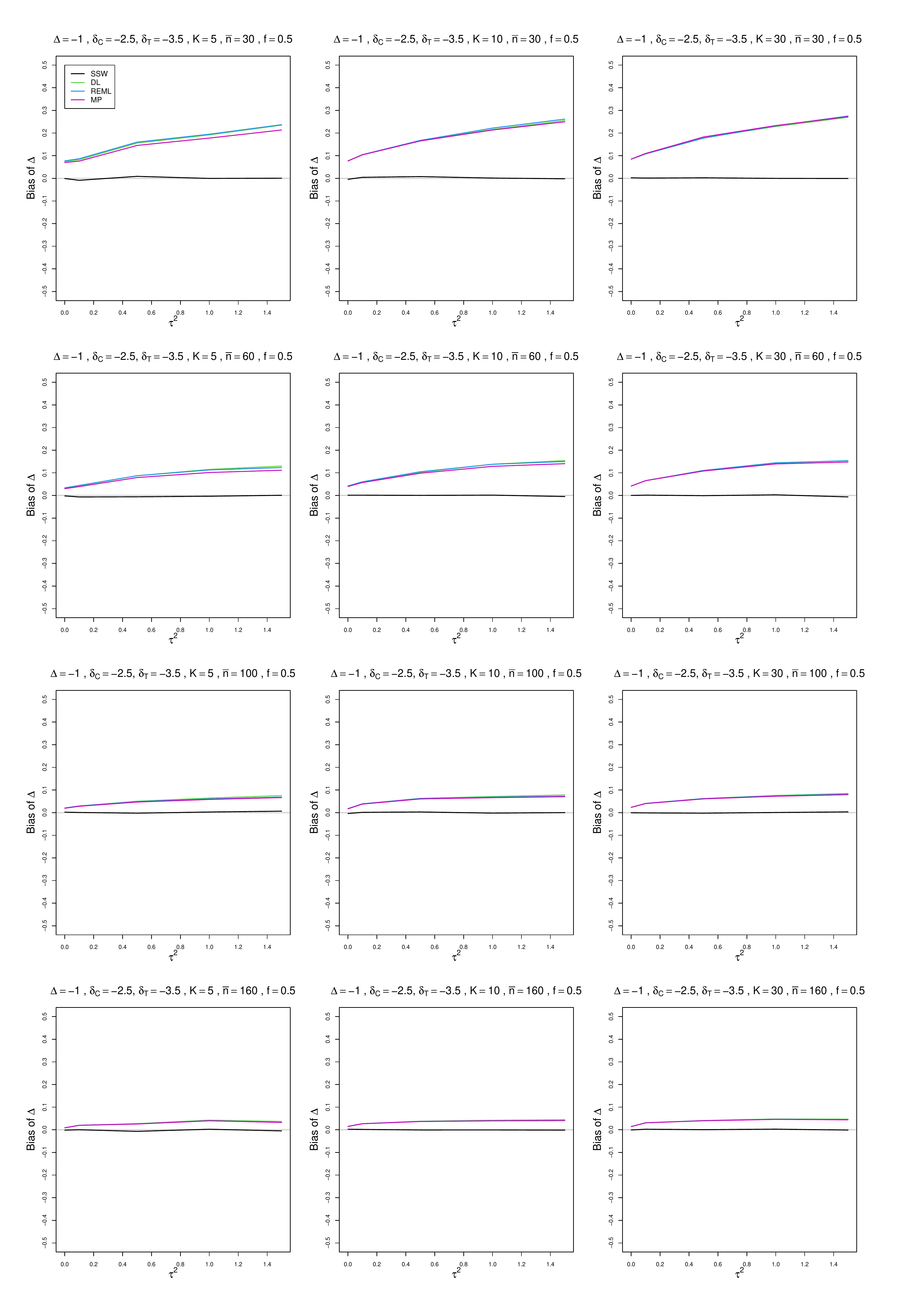}
	\caption{Bias  of estimators of overall effect measure $\Delta$ (DL, REML, MP, and SSW ) vs $\tau^2$, for unequal sample sizes $\bar{n}=30,\;60,\;100$ and $160$, $\delta_{iC} = -2.5$, $\Delta=-1$ and  $f = 0.5$.   }
	\label{PlotBiasOfDelta_deltaC_-25deltaT=-3.5_DSM_unequal_sample_sizes.pdf}
\end{figure}

\begin{figure}[ht]
	\centering
	\includegraphics[scale=0.33]{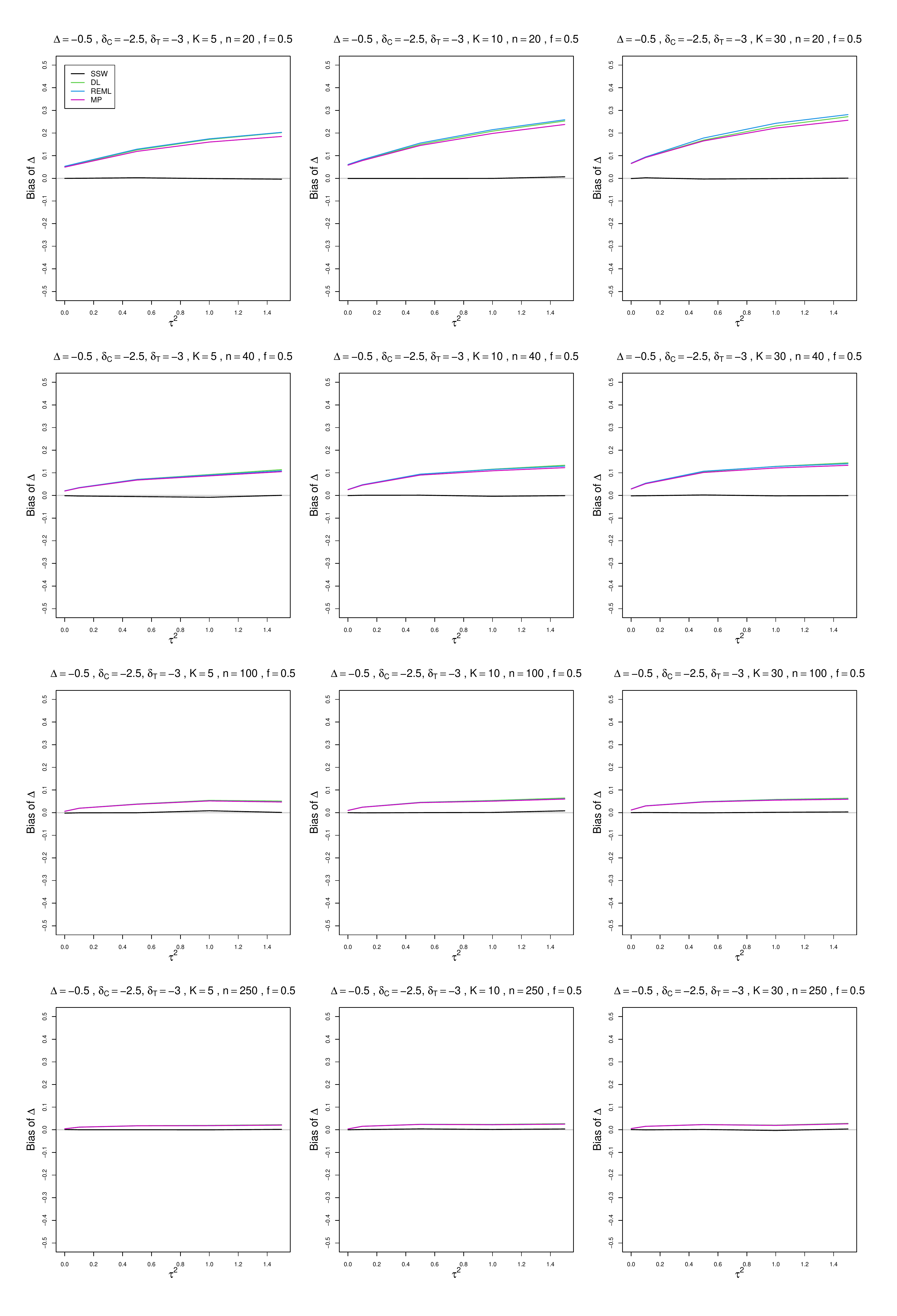}
	\caption{Bias  of estimators of overall effect measure $\Delta$ (DL, REML, MP and SSW ) vs $\tau^2$, for equal sample sizes $n=20,\;40,\;100$ and $250$, $\delta_{iC} = -2.5$, $\Delta=-0.5$ and  $f = 0.5$.   }
	\label{PlotBiasOfDelta_deltaC_-25deltaT=-3_DSM_equal_sample_sizes.pdf}
\end{figure}

\begin{figure}[ht]
	\centering
	\includegraphics[scale=0.33]{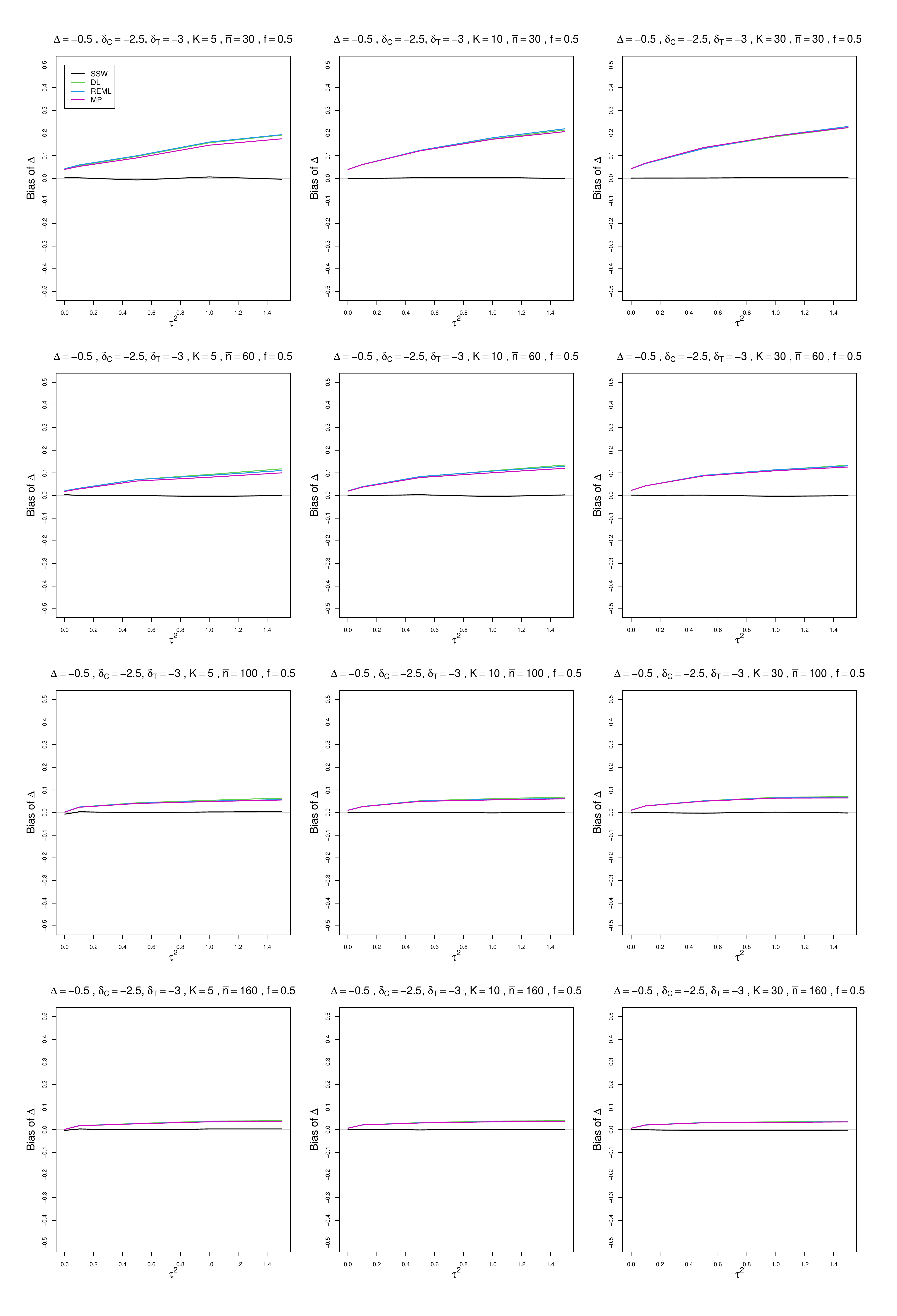}
	\caption{Bias  of estimators of overall effect measure $\Delta$ (DL, REML, MP and SSW ) vs $\tau^2$, for unequal sample sizes $\bar{n}=30,\;60,\;100$ and $160$, $\delta_{iC} = -2.5$, $\Delta=-0.5$ and  $f = 0.5$.   }
	\label{PlotBiasOfDelta_deltaC_-25deltaT=-3_DSM_unequal_sample_sizes.pdf}
\end{figure}

\begin{figure}[ht]
	\centering
	\includegraphics[scale=0.33]{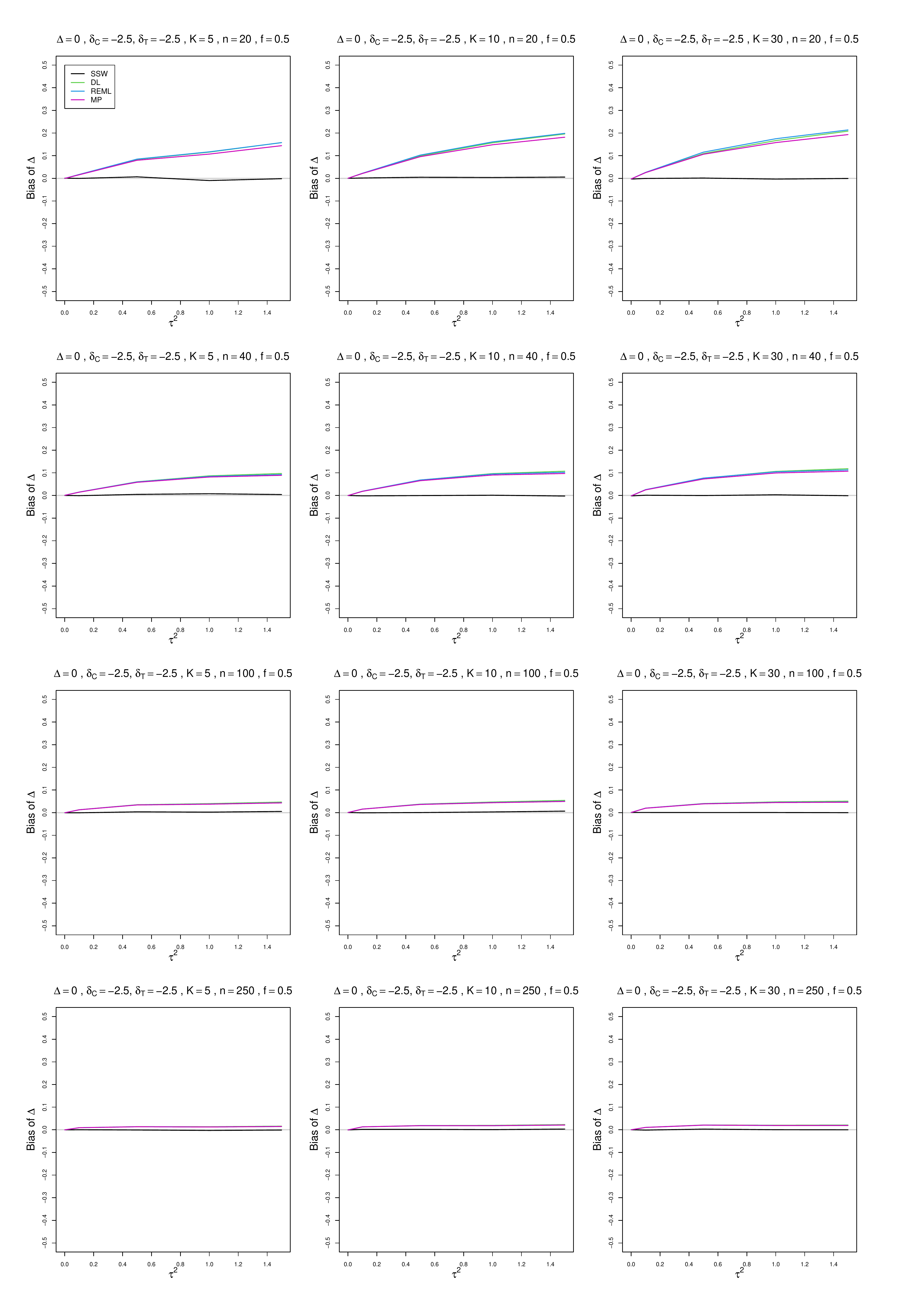}
	\caption{Bias  of estimators of overall effect measure $\Delta$ (DL, REML, MP and SSW ) vs $\tau^2$, for equal sample sizes $n=20,\;40,\;100$ and $250$, $\delta_{iC} = -2.5$, $\Delta=0$ and  $f = 0.5$.   }
	\label{PlotBiasOfDelta_deltaC_-25deltaT=-2.5_DSM_equal_sample_sizes.pdf}
\end{figure}

\begin{figure}[ht]
	\centering
	\includegraphics[scale=0.33]{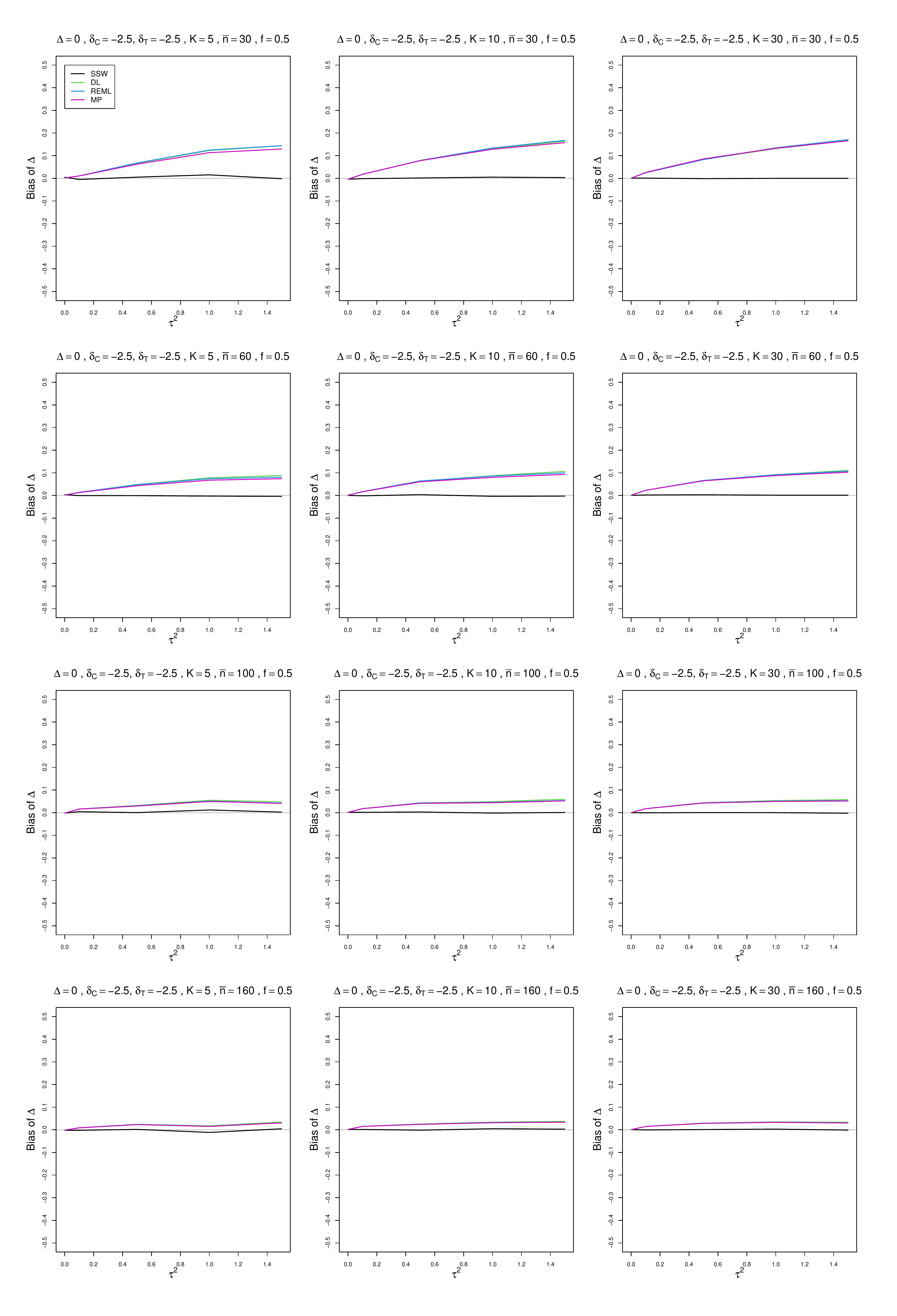}
	\caption{Bias  of estimators of overall effect measure $\Delta$ (DL, REML, MP and SSW ) vs $\tau^2$, for unequal sample sizes $\bar{n}=30,\;60,\;100$ and $160$, $\delta_{iC} = -2.5$, $\Delta=0$ and  $f = 0.5$.   }
	\label{PlotBiasOfDelta_deltaC_-25deltaT=-2.5_DSM_unequal_sample_sizes.pdf}
\end{figure}

\begin{figure}[ht]
	\centering
	\includegraphics[scale=0.33]{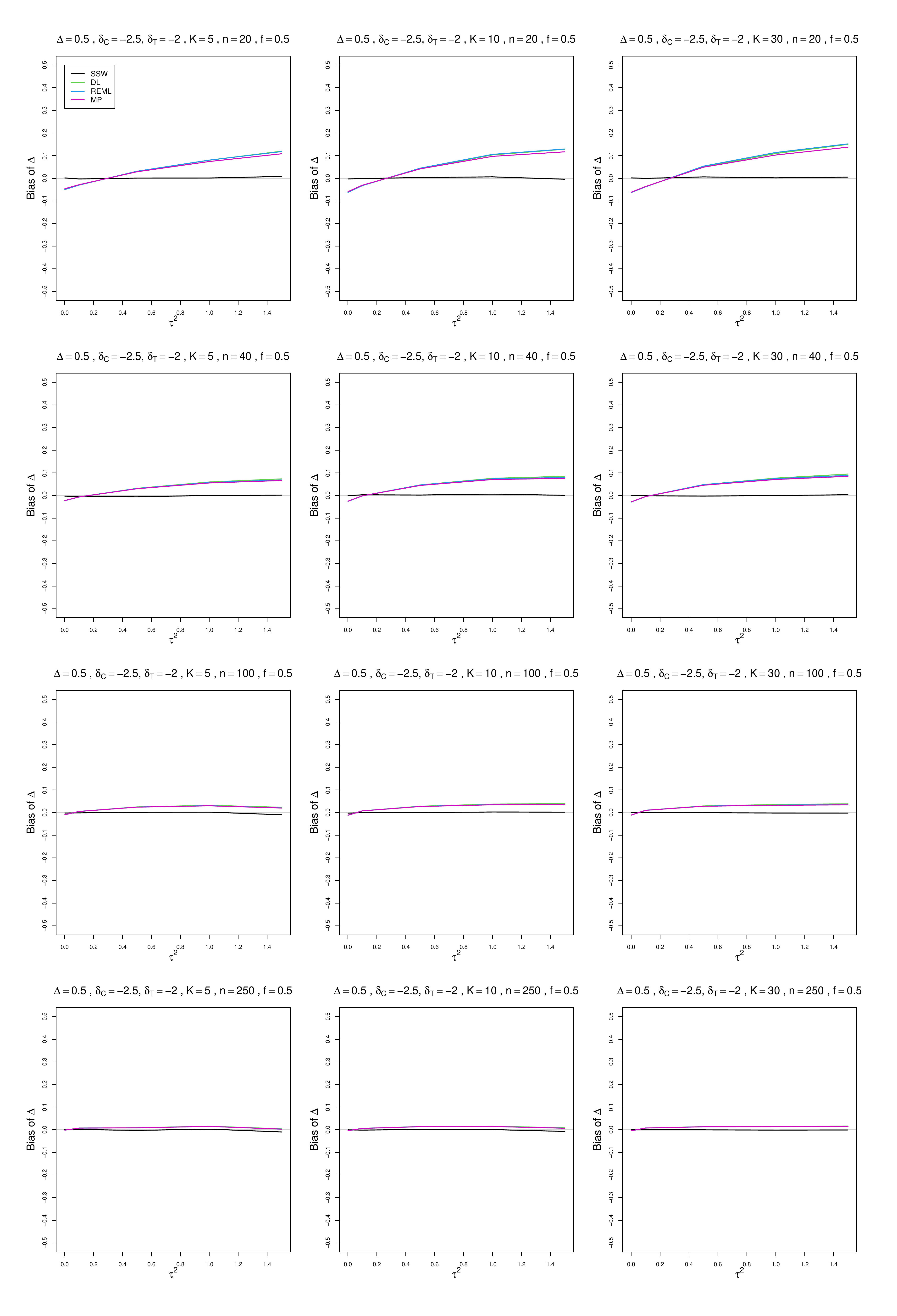}
	\caption{Bias  of estimators of overall effect measure $\Delta$ (DL, REML, MP and SSW) vs $\tau^2$, for equal sample sizes $n=20,\;40,\;100$ and $250$, $\delta_{iC} = -2.5$, $\Delta=0.5$ and  $f = 0.5$.   }
	\label{PlotBiasOfDelta_deltaC_-25deltaT=-2_DSM_equal_sample_sizes.pdf}
\end{figure}

\begin{figure}[ht]
	\centering
	\includegraphics[scale=0.33]{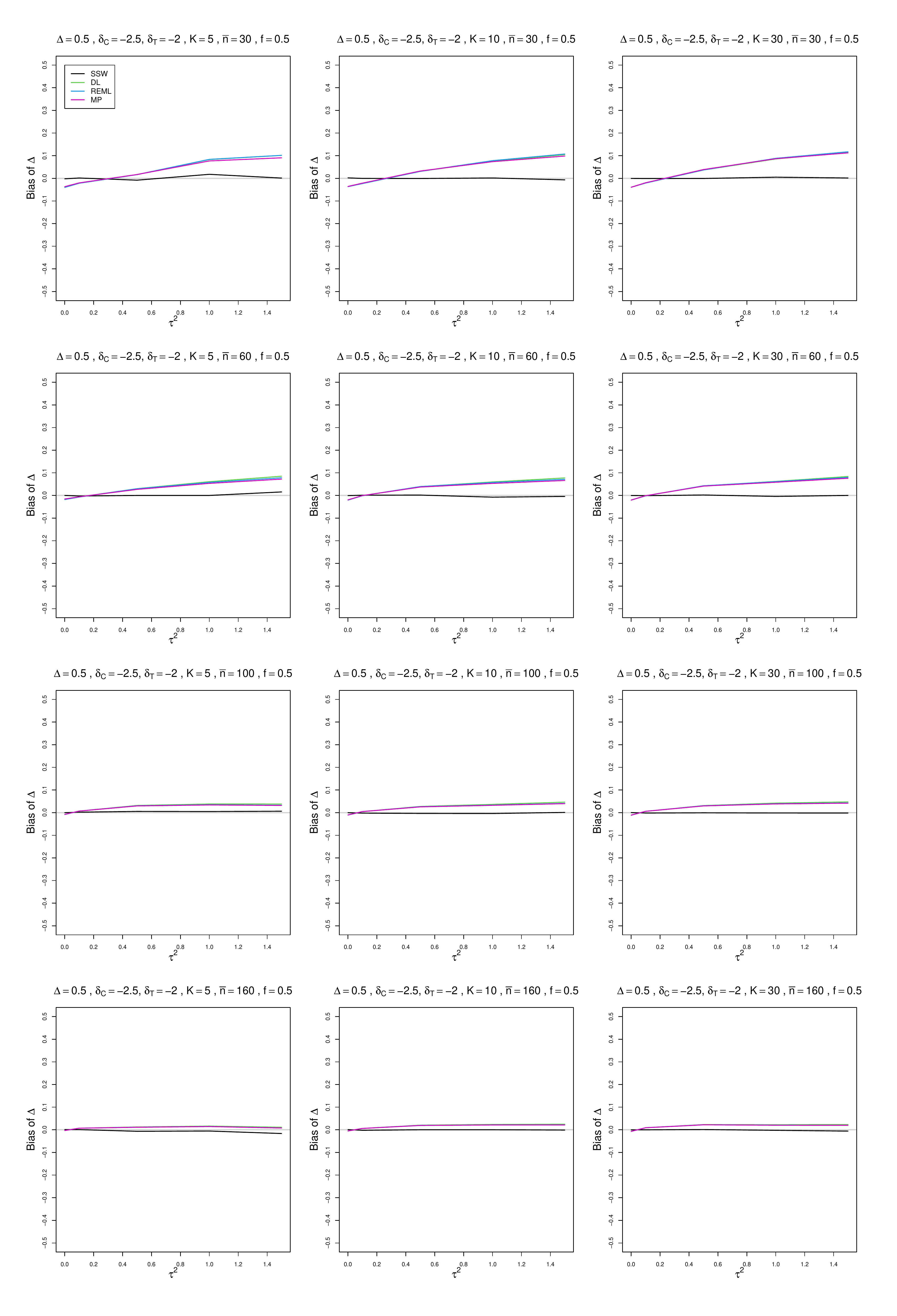}
	\caption{Bias  of estimators of overall effect measure $\Delta$ (DL, REML, MP and SSW) vs $\tau^2$, for unequal sample sizes $\bar{n}=30,\;60,\;100$ and $160$, $\delta_{iC} = -2.5$, $\Delta=0.5$ and  $f = 0.5$.   }
	\label{PlotBiasOfDelta_deltaC_-25deltaT=-2_DSM_unequal_sample_sizes.pdf}
\end{figure}

\begin{figure}[ht]
	\centering
	\includegraphics[scale=0.33]{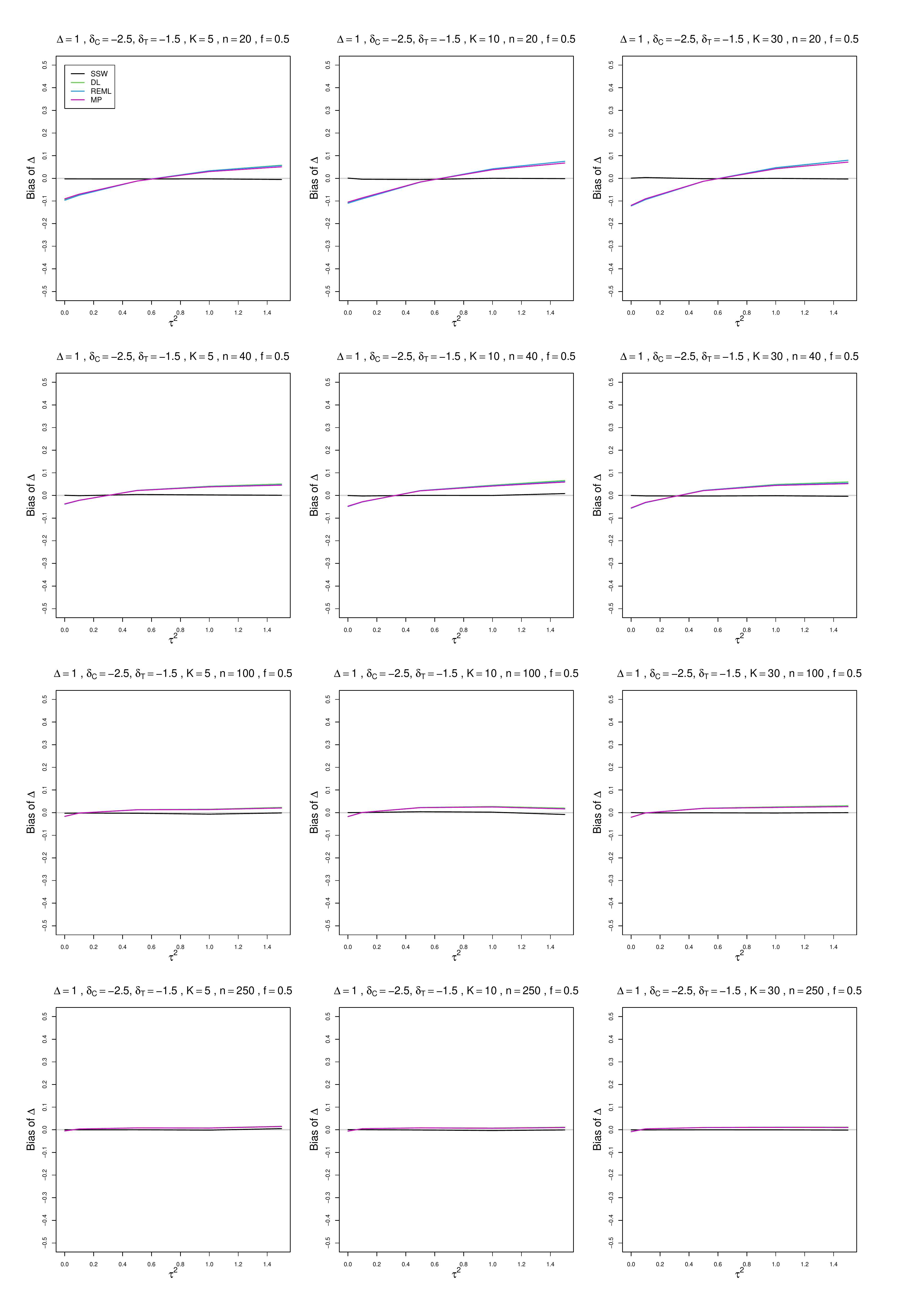}
	\caption{Bias  of estimators of overall effect measure $\Delta$ (DL, REML, MP and SSW) vs $\tau^2$, for equal sample sizes $n=20,\;40,\;100$ and $250$, $\delta_{iC} = -2.5$, $\Delta=1$ and  $f = 0.5$.   }
	\label{PlotBiasOfDelta_deltaC_-25deltaT=-1.5_DSM_equal_sample_sizes.pdf}
\end{figure}

\begin{figure}[ht]
	\centering
	\includegraphics[scale=0.33]{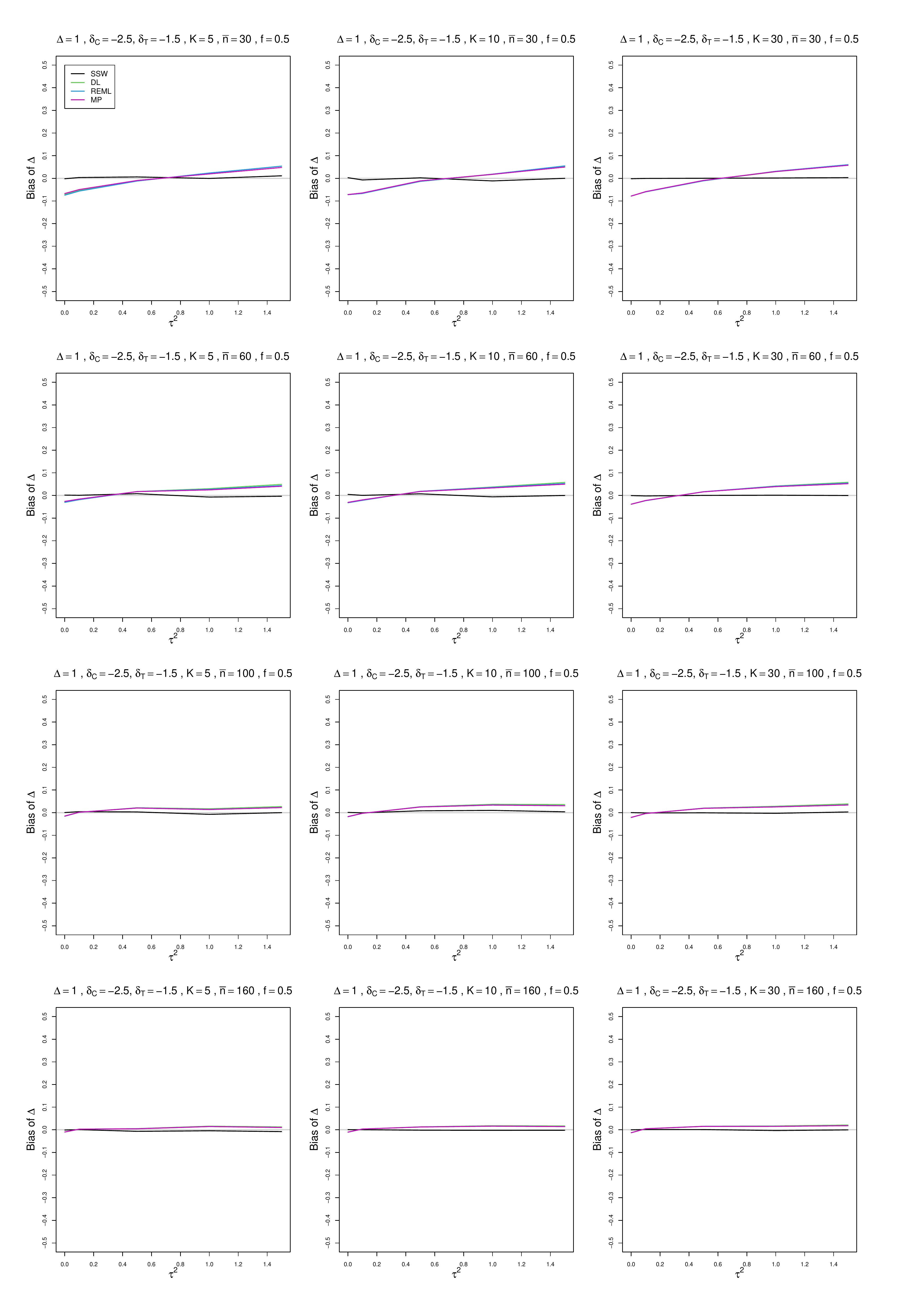}
	\caption{Bias  of estimators of overall effect measure $\Delta$ (DL, REML, MP and SSW) vs $\tau^2$, for unequal sample sizes $\bar{n}=30,\;60,\;100$ and $160$, $\delta_{iC} = -2.5$, $\Delta=1$ and  $f = 0.5$.   }
	\label{PlotBiasOfDelta_deltaC_-25deltaT=-1.5_DSM_unequal_sample_sizes.pdf}
\end{figure}

\begin{figure}[ht]
	\centering
	\includegraphics[scale=0.33]{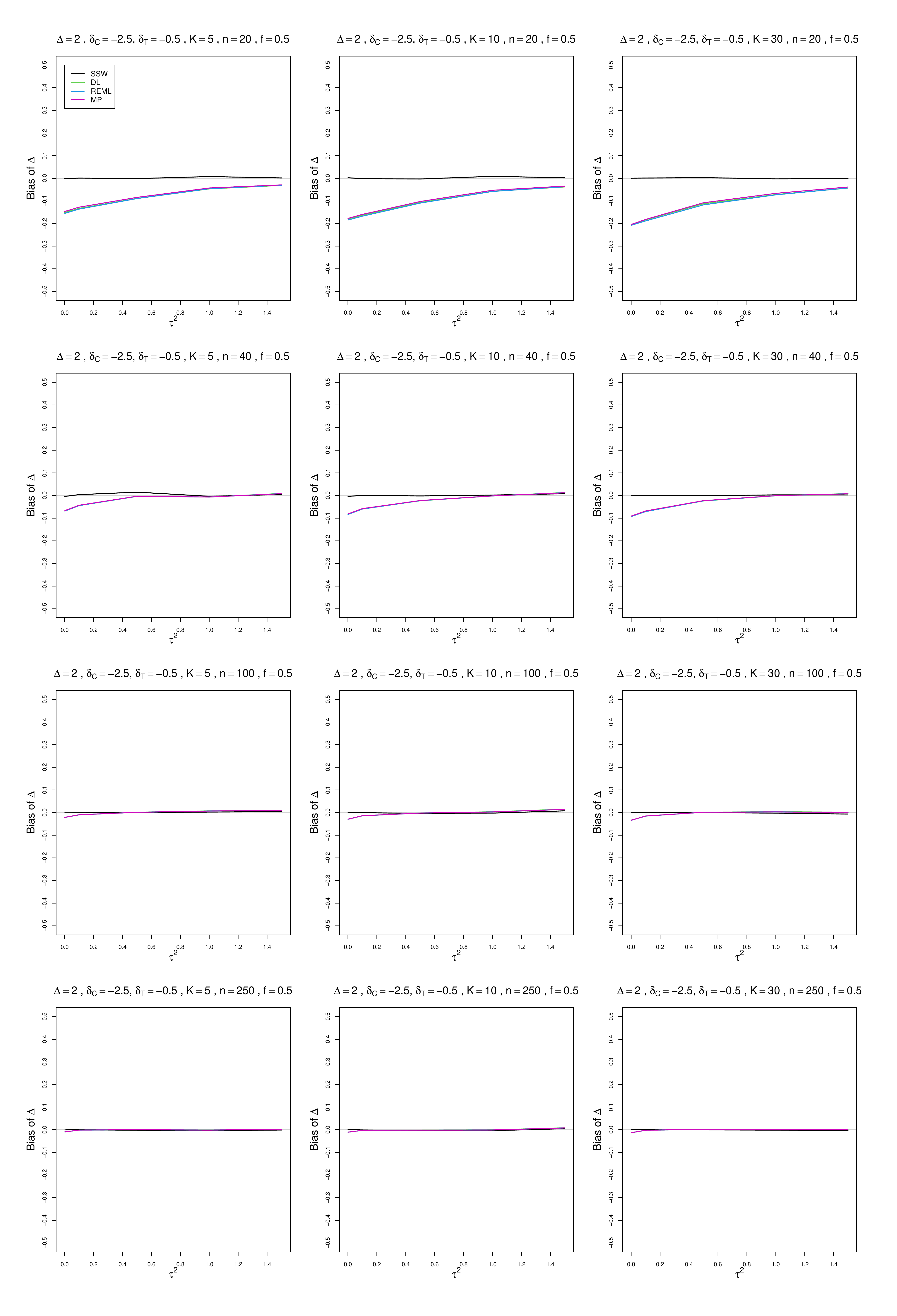}
	\caption{Bias  of estimators of overall effect measure $\Delta$ (DL, REML, MP and SSW) vs $\tau^2$, for equal sample sizes $n=20,\;40,\;100$ and $250$, $\delta_{iC} = -2.5$, $\Delta=2$ and  $f = 0.5$.   }
	\label{PlotBiasOfDelta_deltaC_-25deltaT=-0.5_DSM_equal_sample_sizes.pdf}
\end{figure}

\begin{figure}[ht]
	\centering
	\includegraphics[scale=0.33]{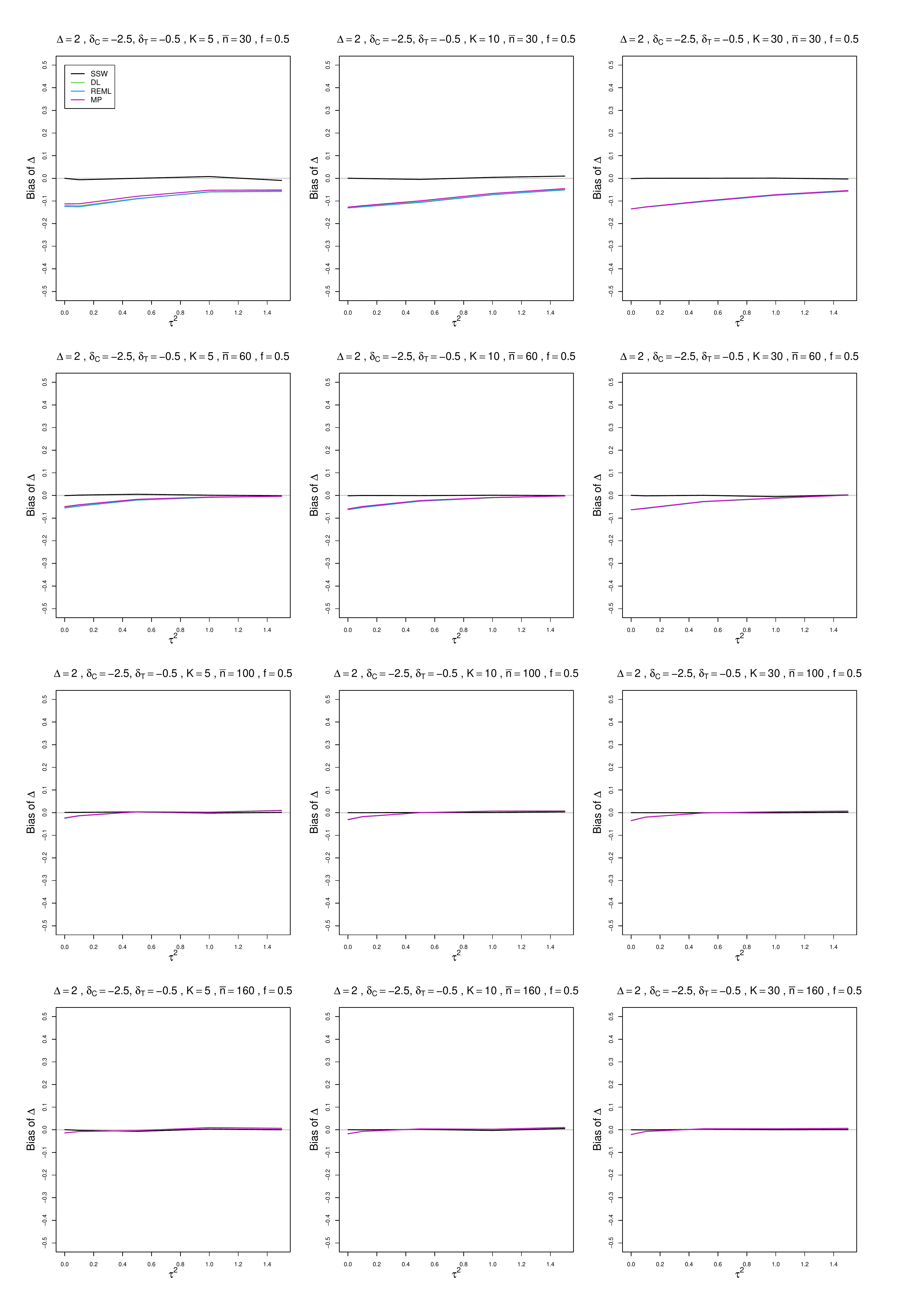}
	\caption{Bias  of estimators of overall effect measure $\Delta$ (DL, REML, MP and SSW) vs $\tau^2$, for unequal sample sizes $\bar{n}=30,\;60,\;100$ and $160$, $\delta_{iC} = -2.5$, $\Delta=2$ and  $f = 0.5$.   }
	\label{PlotBiasOfDelta_deltaC_-25deltaT=-0.5_DSM_unequal_sample_sizes.pdf}
\end{figure}
\begin{figure}[ht]
	\centering
	\includegraphics[scale=0.33]{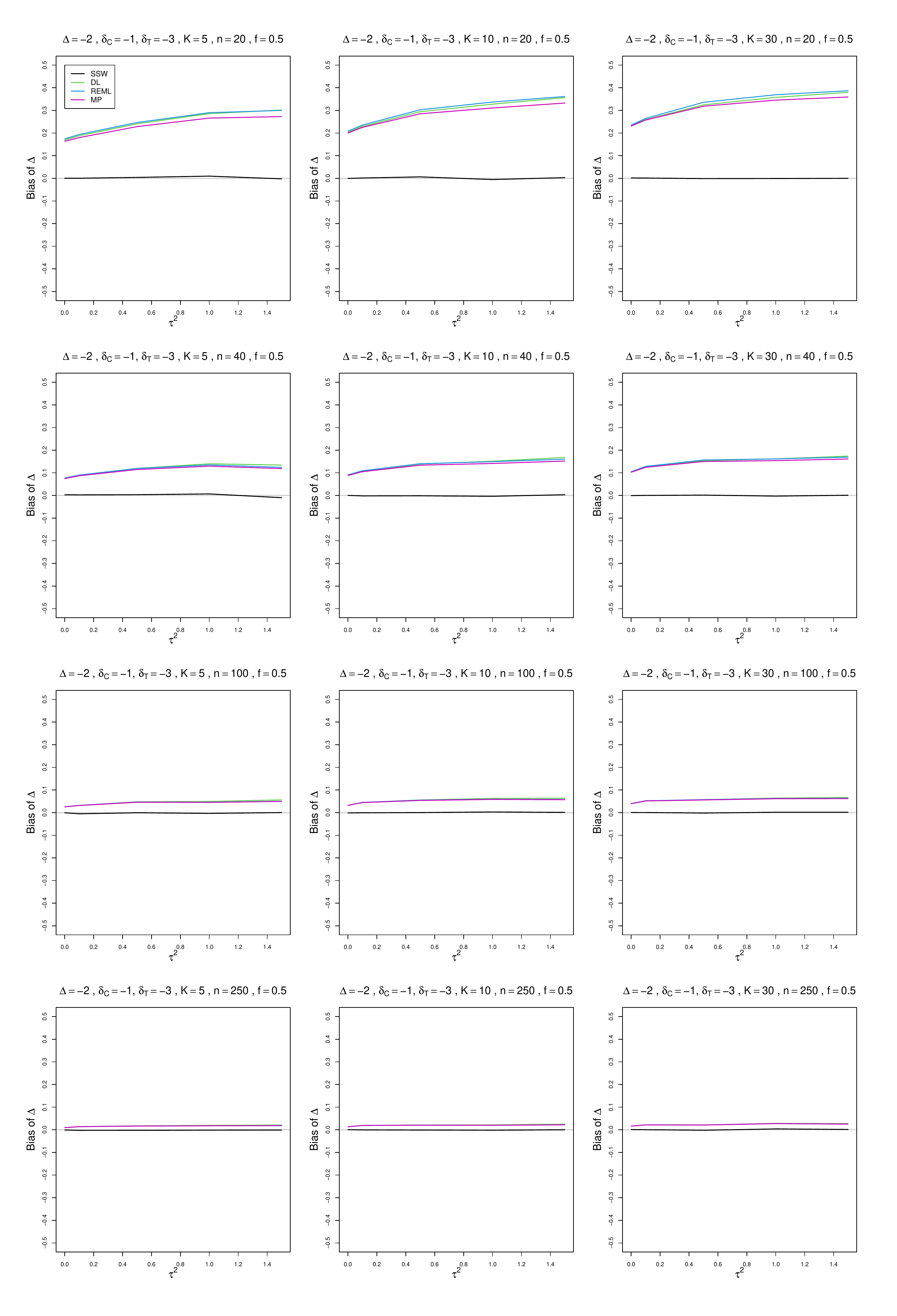}
	\caption{Bias  of estimators of overall effect measure $\Delta$ (DL, REML, MP and SSW) vs $\tau^2$, for equal sample sizes $n=20,\;40,\;100$ and $250$, $\delta_{iC} = -2.5$, $\Delta=-2$ and  $f = 0.5$.   }
	\label{PlotBiasOfDelta_deltaC_-1deltaT=-3_DSM_equal_sample_sizes.pdf}
\end{figure}

\begin{figure}[ht]
	\centering
	\includegraphics[scale=0.33]{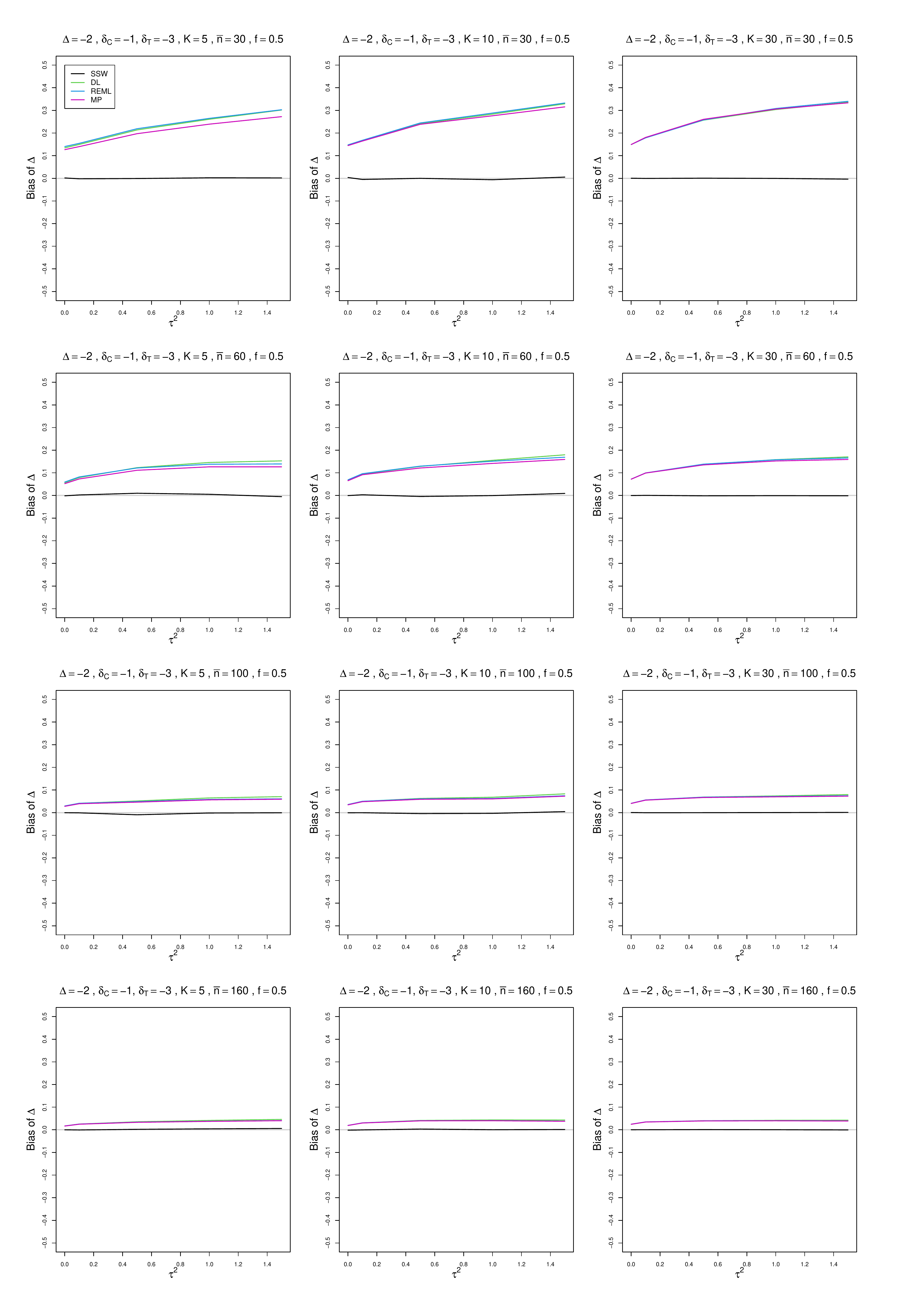}
	\caption{Bias  of estimators of overall effect measure $\Delta$ (DL, REML, MP and SSW) vs $\tau^2$, for unequal sample sizes $\bar{n}=30,\;60,\;100$ and $160$, $\delta_{iC} = -1$, $\Delta=-2$ and  $f = 0.5$.   }
	\label{PlotBiasOfDelta_deltaC_-1deltaT=-3_DSM_unequal_sample_sizes.pdf}
\end{figure}

\begin{figure}[ht]
	\centering
	\includegraphics[scale=0.33]{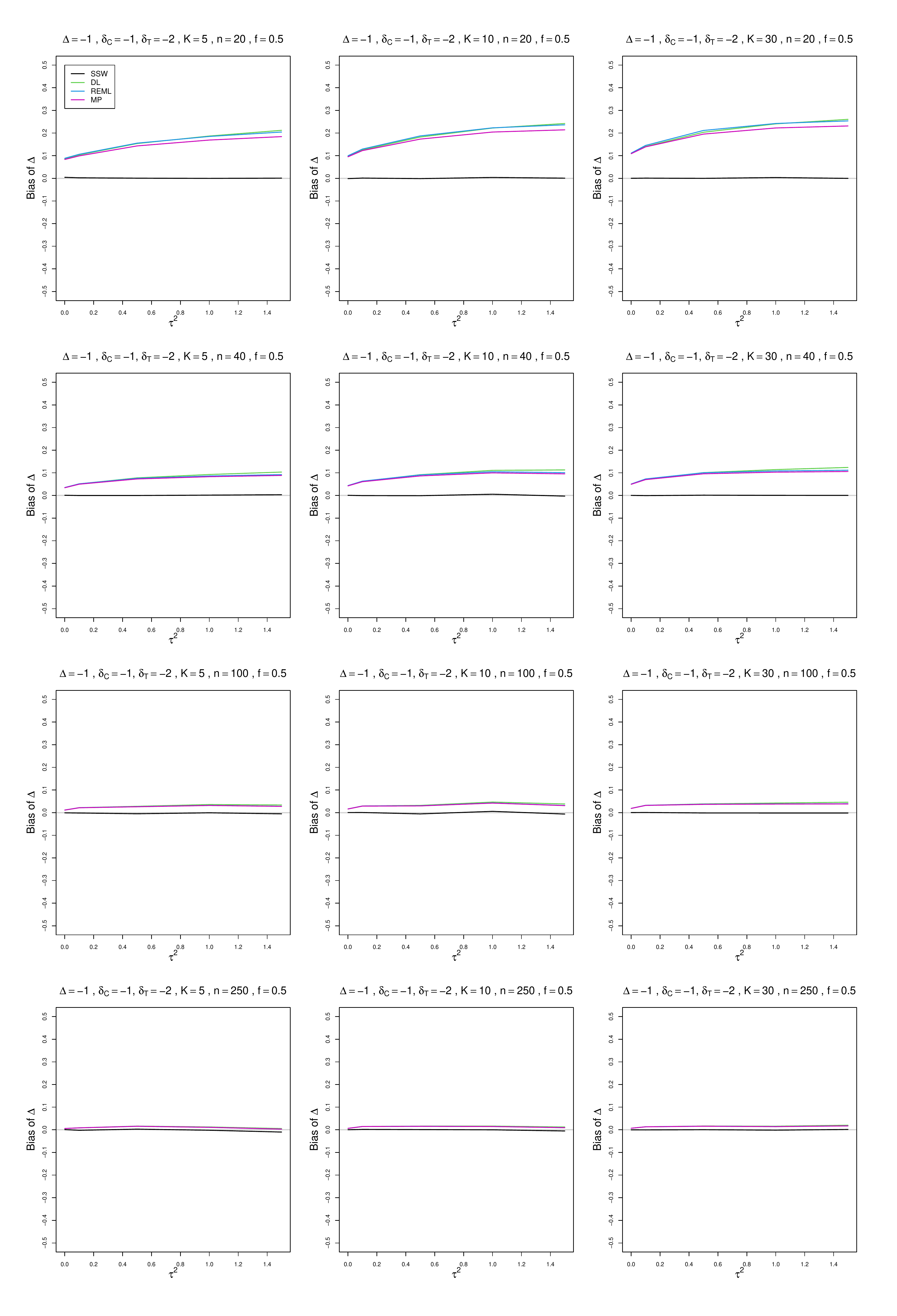}
	\caption{Bias  of estimators of overall effect measure $\Delta$ (DL, REML, MP and SSW) vs $\tau^2$, for equal sample sizes $n=20,\;40,\;100$ and $250$, $\delta_{iC} = -1$, $\Delta=-1$ and  $f = 0.5$.   }
	\label{PlotBiasOfDelta_deltaC_-1deltaT=-2_DSM_equal_sample_sizes.pdf}
\end{figure}

\begin{figure}[ht]
	\centering
	\includegraphics[scale=0.33]{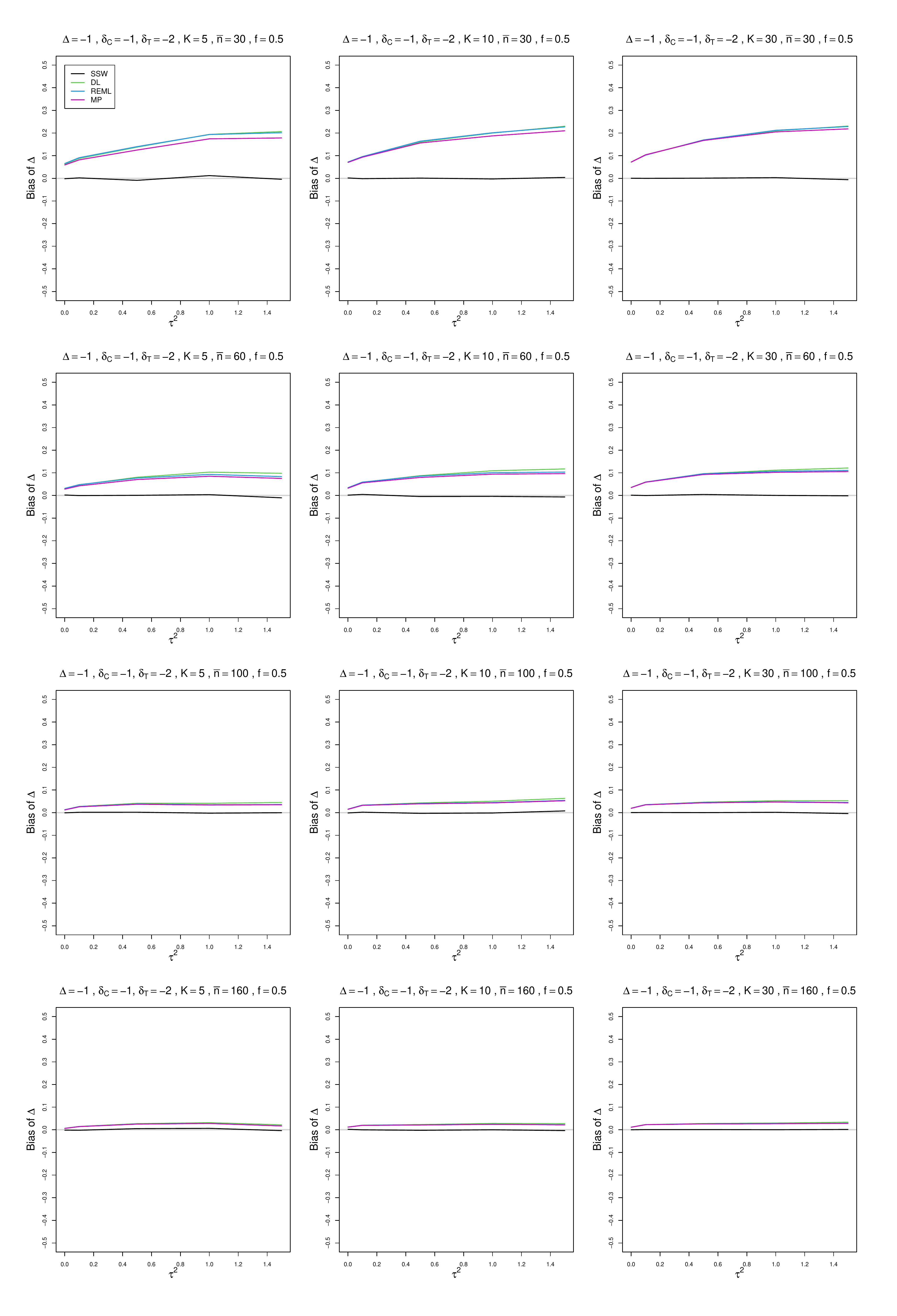}
	\caption{Bias  of estimators of overall effect measure $\Delta$ (DL, REML, MP and SSW ) vs $\tau^2$, for unequal sample sizes $\bar{n}=30,\;60,\;100$ and $160$, $\delta_{iC} = -1$, $\Delta=-1$ and  $f = 0.5$.   }
	\label{PlotBiasOfDelta_deltaC_-1deltaT=-2_DSM_unequal_sample_sizes.pdf}
\end{figure}

\begin{figure}[ht]
	\centering
	\includegraphics[scale=0.33]{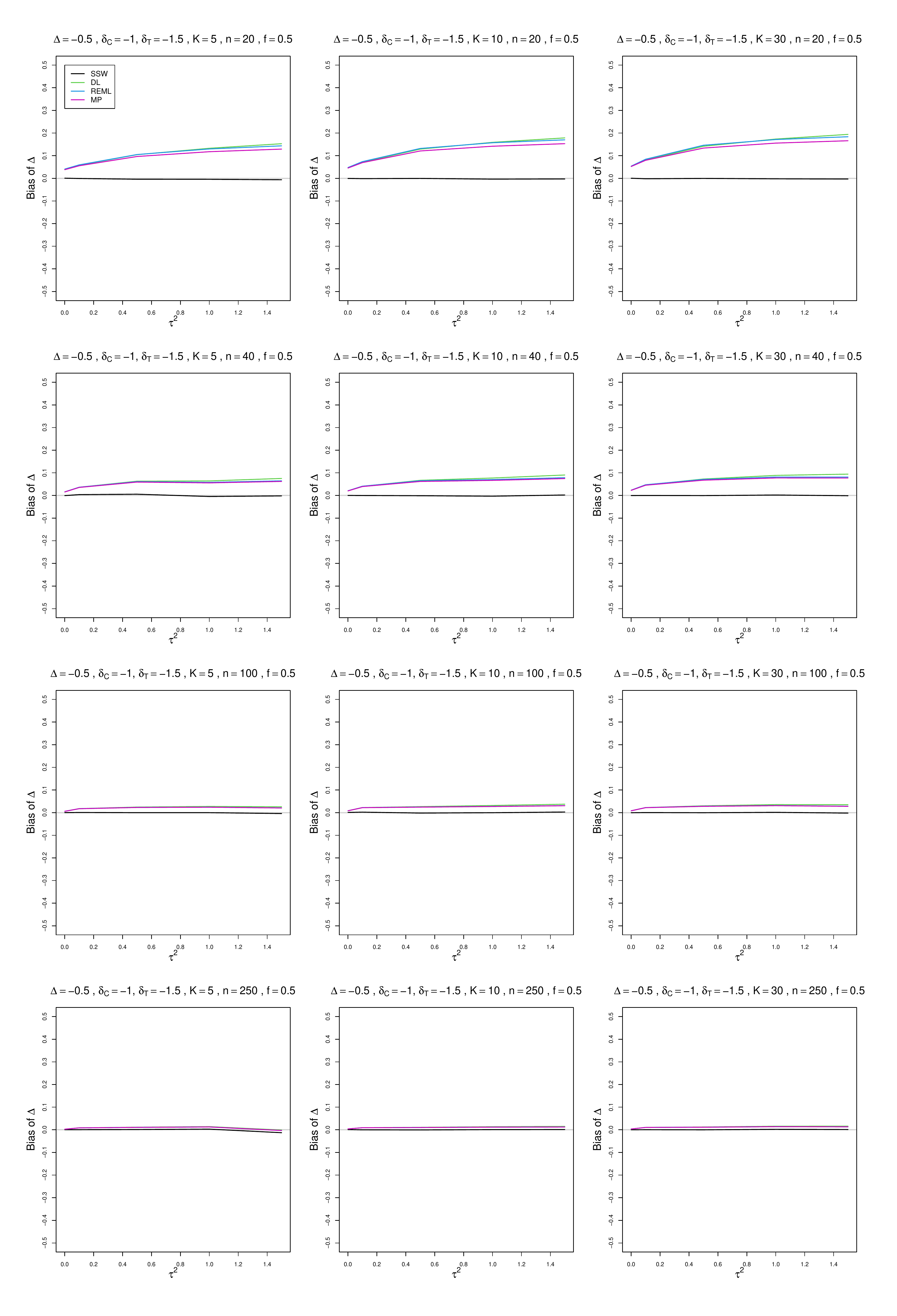}
	\caption{Bias  of estimators of overall effect measure $\Delta$ (DL, REML, MP and SSW) vs $\tau^2$, for equal sample sizes $n=20,\;40,\;100$ and $250$, $\delta_{iC} = -1$, $\Delta=-0.5$ and  $f = 0.5$.   }
	\label{PlotBiasOfDelta_deltaC_-1deltaT=-1.5_DSM_equal_sample_sizes.pdf}
\end{figure}

\begin{figure}[ht]
	\centering
	\includegraphics[scale=0.33]{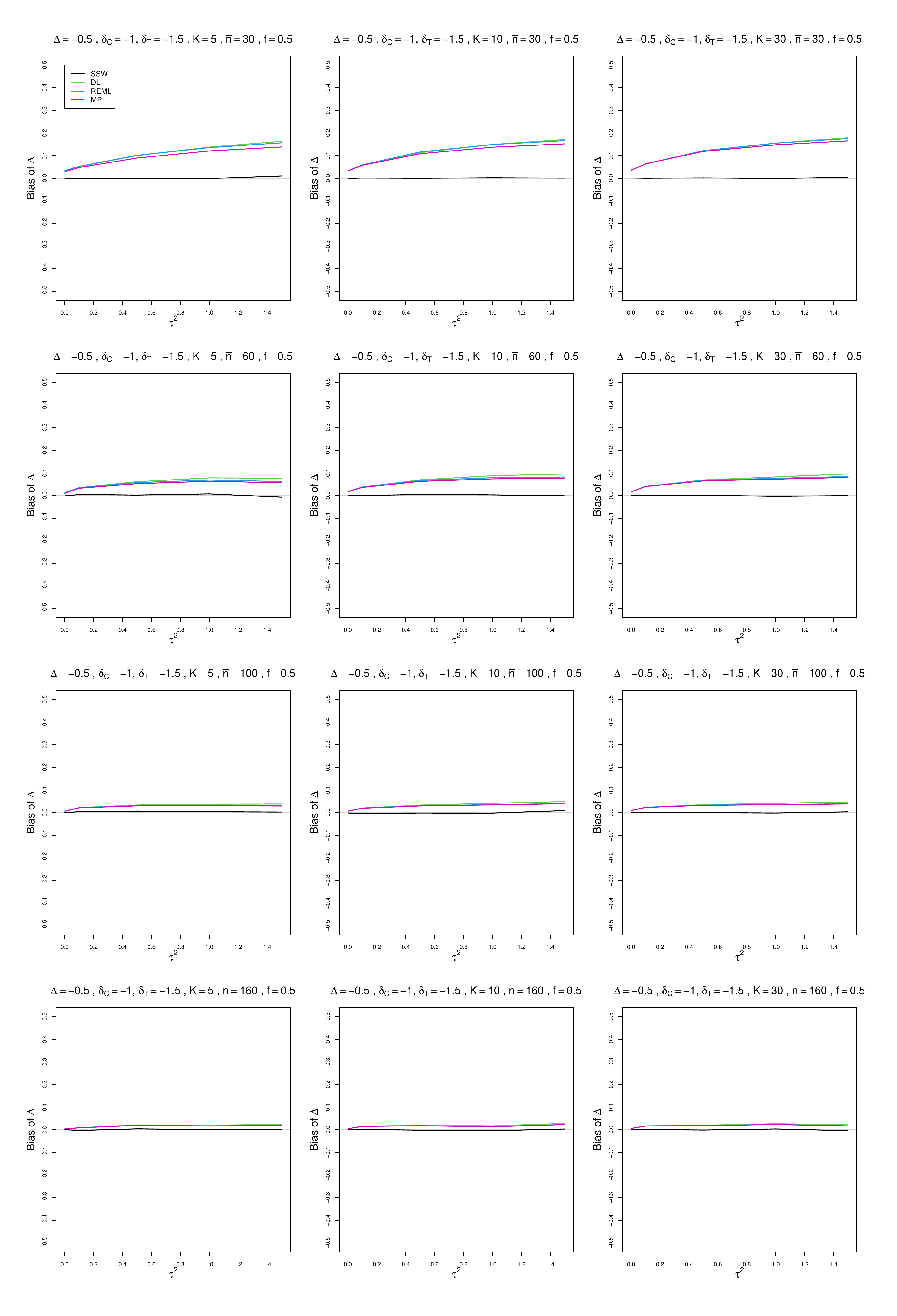}
	\caption{Bias  of estimators of overall effect measure $\Delta$ (DL, REML, MP and SSW) vs $\tau^2$, for unequal sample sizes $\bar{n}=30,\;60,\;100$ and $160$, $\delta_{iC} = -1$, $\Delta=-0.5$ and  $f = 0.5$.   }
	\label{PlotBiasOfDelta_deltaC_-1deltaT=-1,5_DSM_unequal_sample_sizes.pdf}
\end{figure}

\begin{figure}[ht]
	\centering
	\includegraphics[scale=0.33]{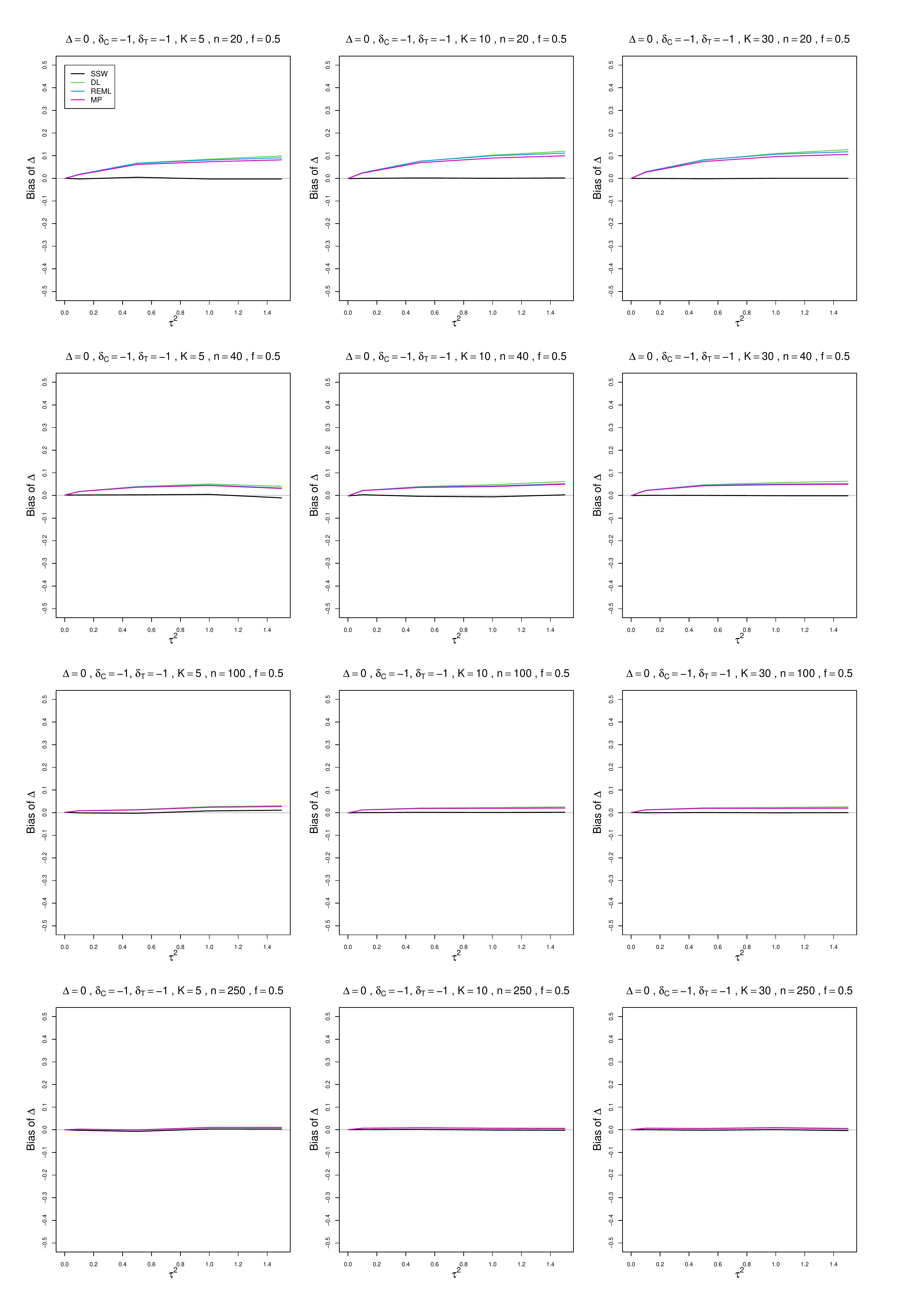}
	\caption{Bias  of estimators of overall effect measure $\Delta$ (DL, REML, MP and SSW) vs $\tau^2$, for equal sample sizes $n=20,\;40,\;100$ and $250$, $\delta_{iC} = -1$, $\Delta=0$ and  $f = 0.5$.   }
	\label{PlotBiasOfDelta_deltaC_-1deltaT=-1_DSM_equal_sample_sizes.pdf}
\end{figure}

\begin{figure}[ht]
	\centering
	\includegraphics[scale=0.33]{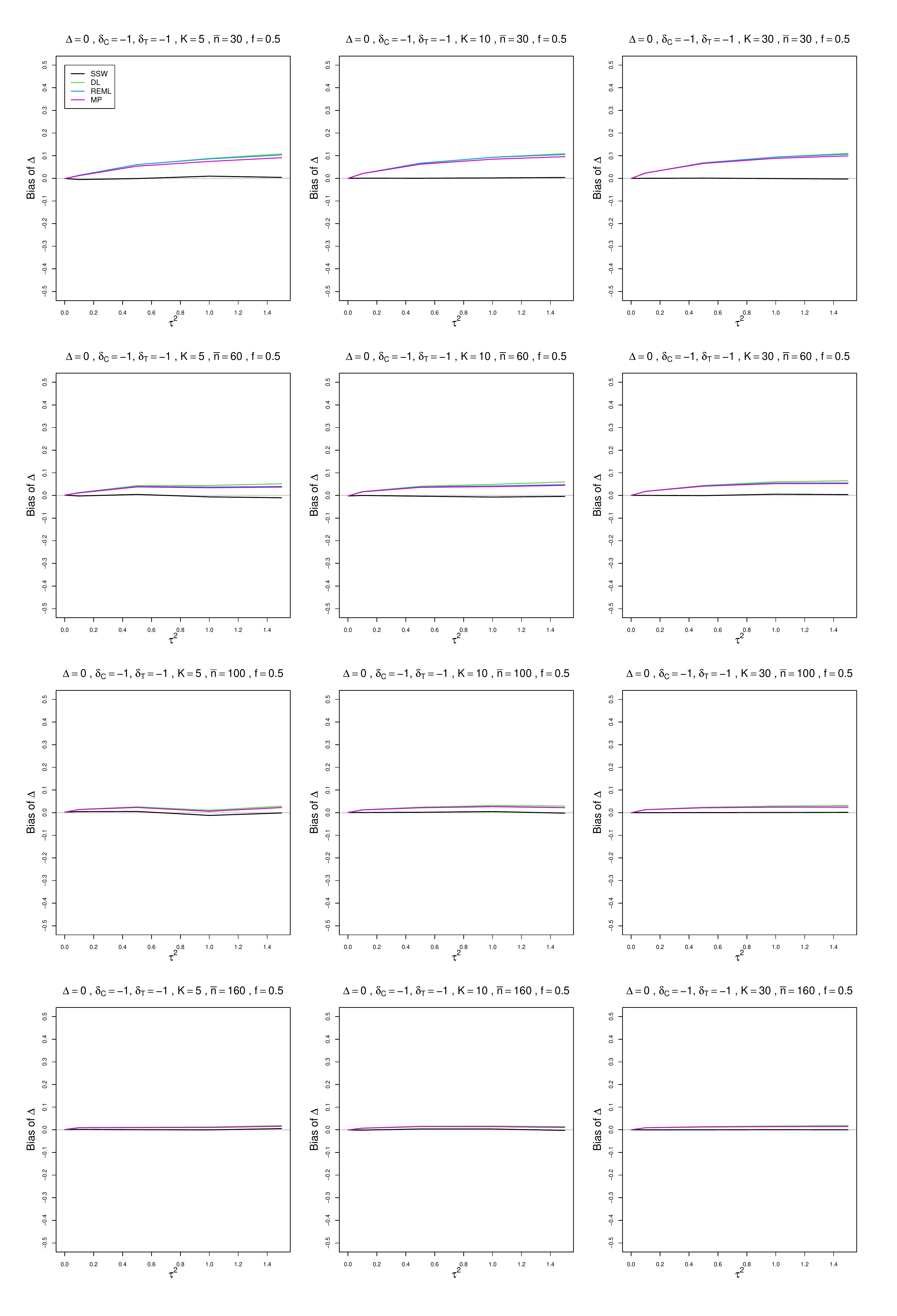}
	\caption{Bias  of estimators of overall effect measure $\Delta$ (DL, REML, MP and SSW ) vs $\tau^2$, for unequal sample sizes $\bar{n}=30,\;60,\;100$ and $160$, $\delta_{iC} = -1$, $\Delta=0$ and  $f = 0.5$.   }
	\label{PlotBiasOfDelta_deltaC_-1deltaT=-1_DSM_unequal_sample_sizes.pdf}
\end{figure}

\begin{figure}[ht]
	\centering
	\includegraphics[scale=0.33]{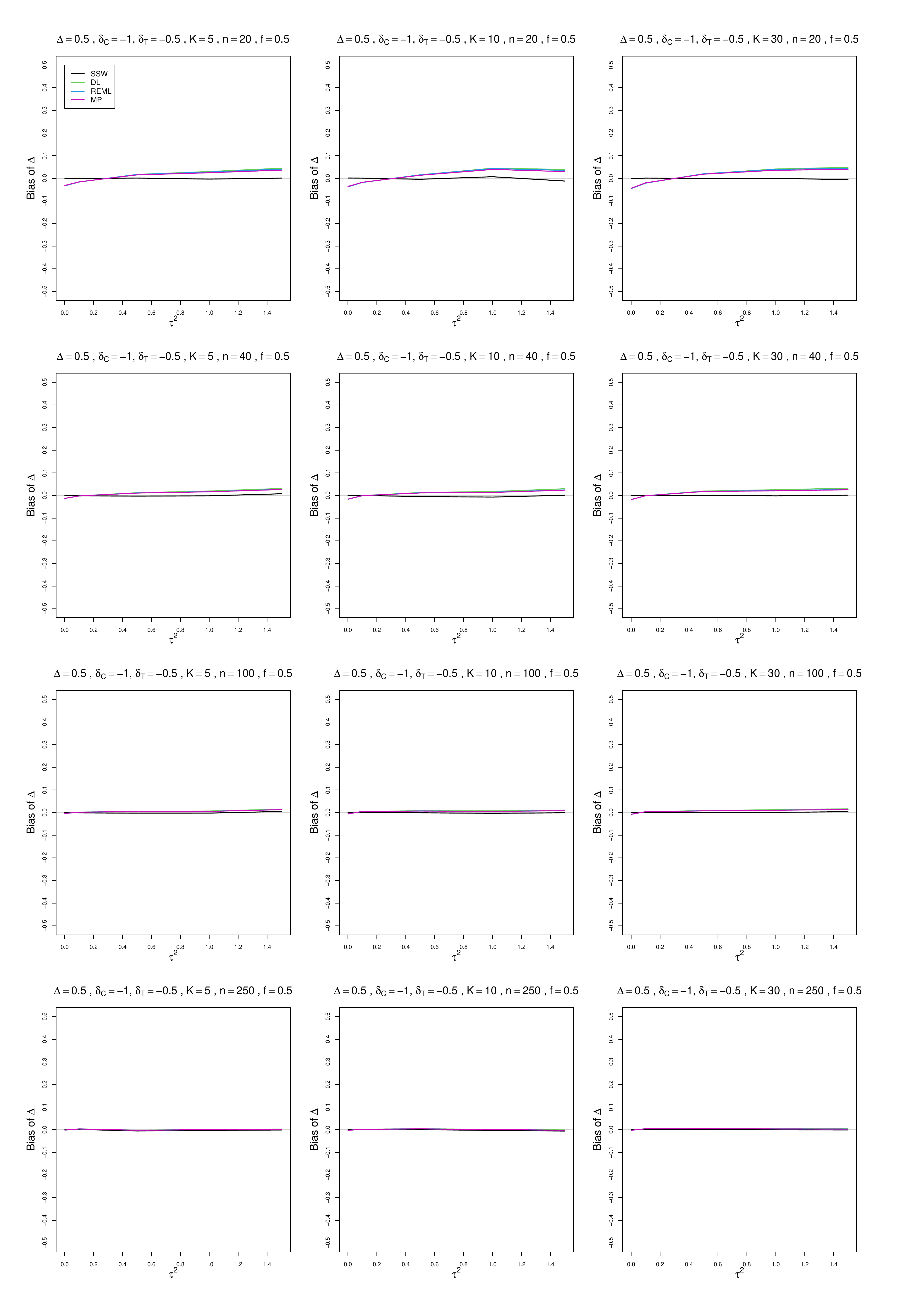}
	\caption{Bias  of estimators of overall effect measure $\Delta$ (DL, REML, MP and SSW) vs $\tau^2$, for equal sample sizes $n=20,\;40,\;100$ and $250$, $\delta_{iC} = -1$, $\Delta=0.5$ and  $f = 0.5$.   }
	\label{PlotBiasOfDelta_deltaC_-1deltaT=-0.5_DSM_equal_sample_sizes.pdf}
\end{figure}

\begin{figure}[ht]
	\centering
	\includegraphics[scale=0.33]{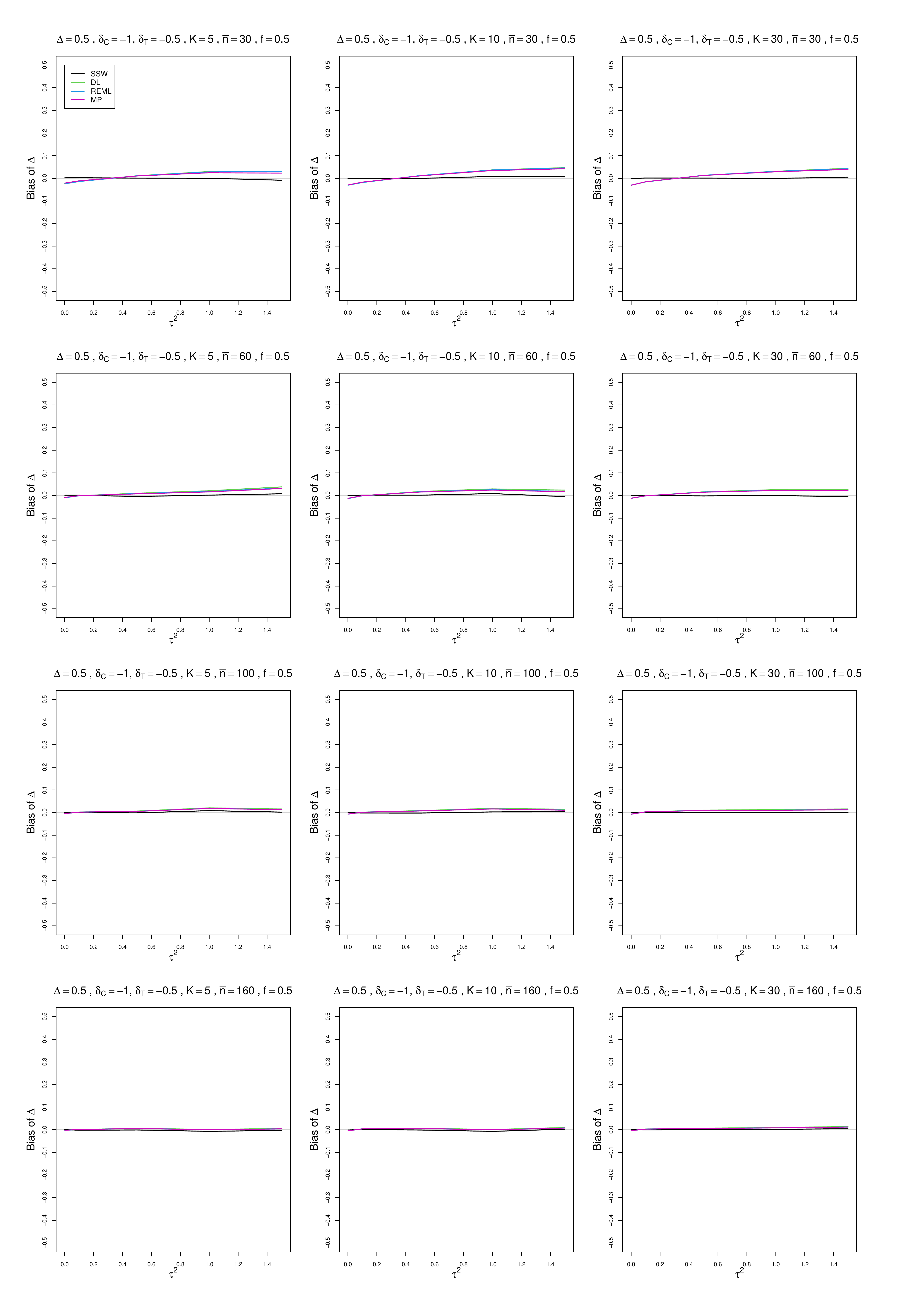}
	\caption{Bias  of estimators of overall effect measure $\Delta$ (DL, REML, MP and SSW) vs $\tau^2$, for unequal sample sizes $\bar{n}=30,\;60,\;100$ and $160$, $\delta_{iC} = -1$, $\Delta=0.5$ and  $f = 0.5$.   }
	\label{PlotBiasOfDelta_deltaC_-1deltaT=-0,5_DSM_unequal_sample_sizes.pdf}
\end{figure}

\begin{figure}[ht]
	\centering
	\includegraphics[scale=0.33]{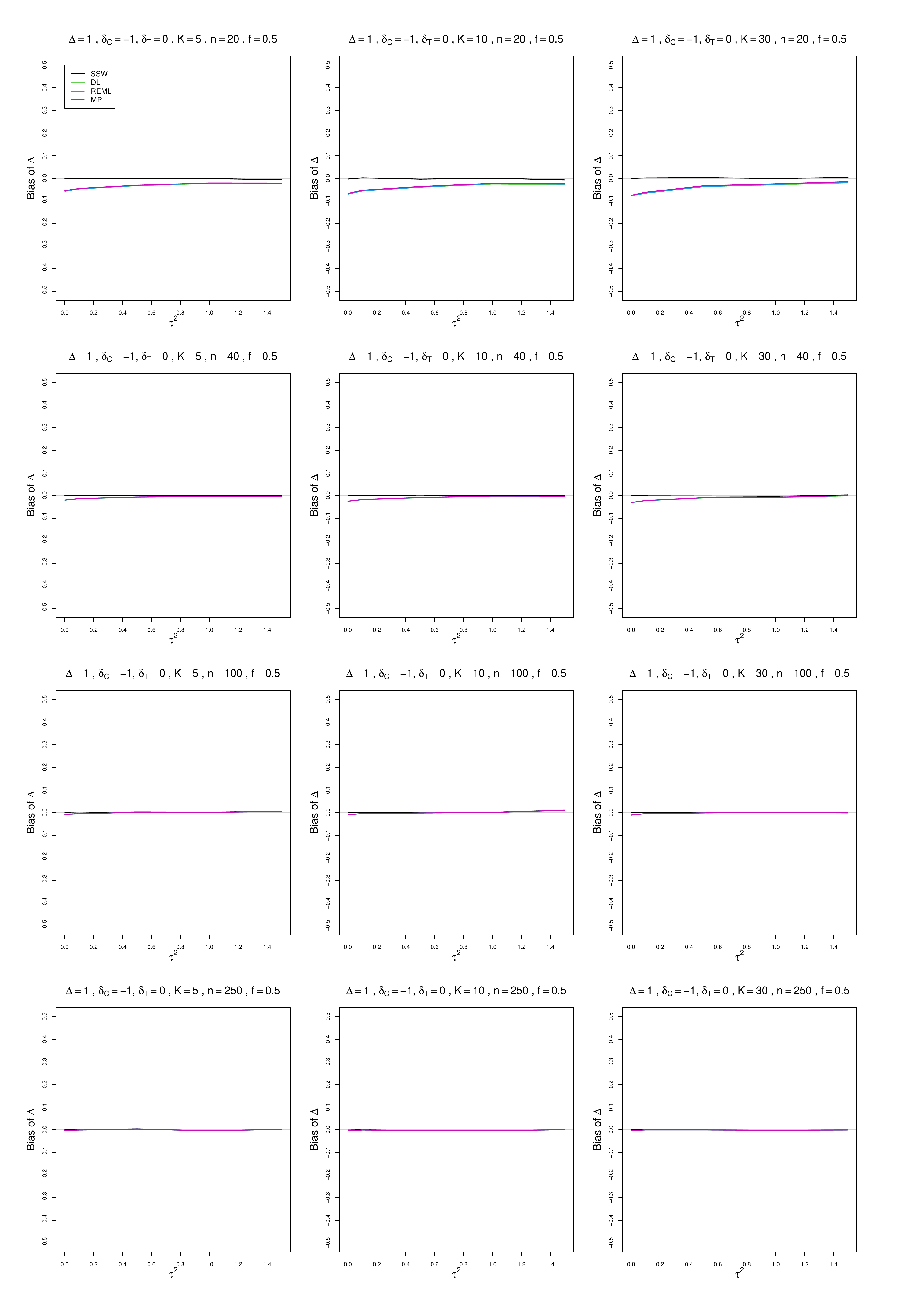}
	\caption{Bias  of estimators of overall effect measure $\Delta$ (DL, REML, MP and SSW) vs $\tau^2$, for equal sample sizes $n=20,\;40,\;100$ and $250$, $\delta_{iC} = -1$, $\Delta=1$ and  $f = 0.5$.   }
	\label{PlotBiasOfDelta_deltaC_--1deltaT=0_DSM_equal_sample_sizes.pdf}
\end{figure}

\begin{figure}[ht]
	\centering
	\includegraphics[scale=0.33]{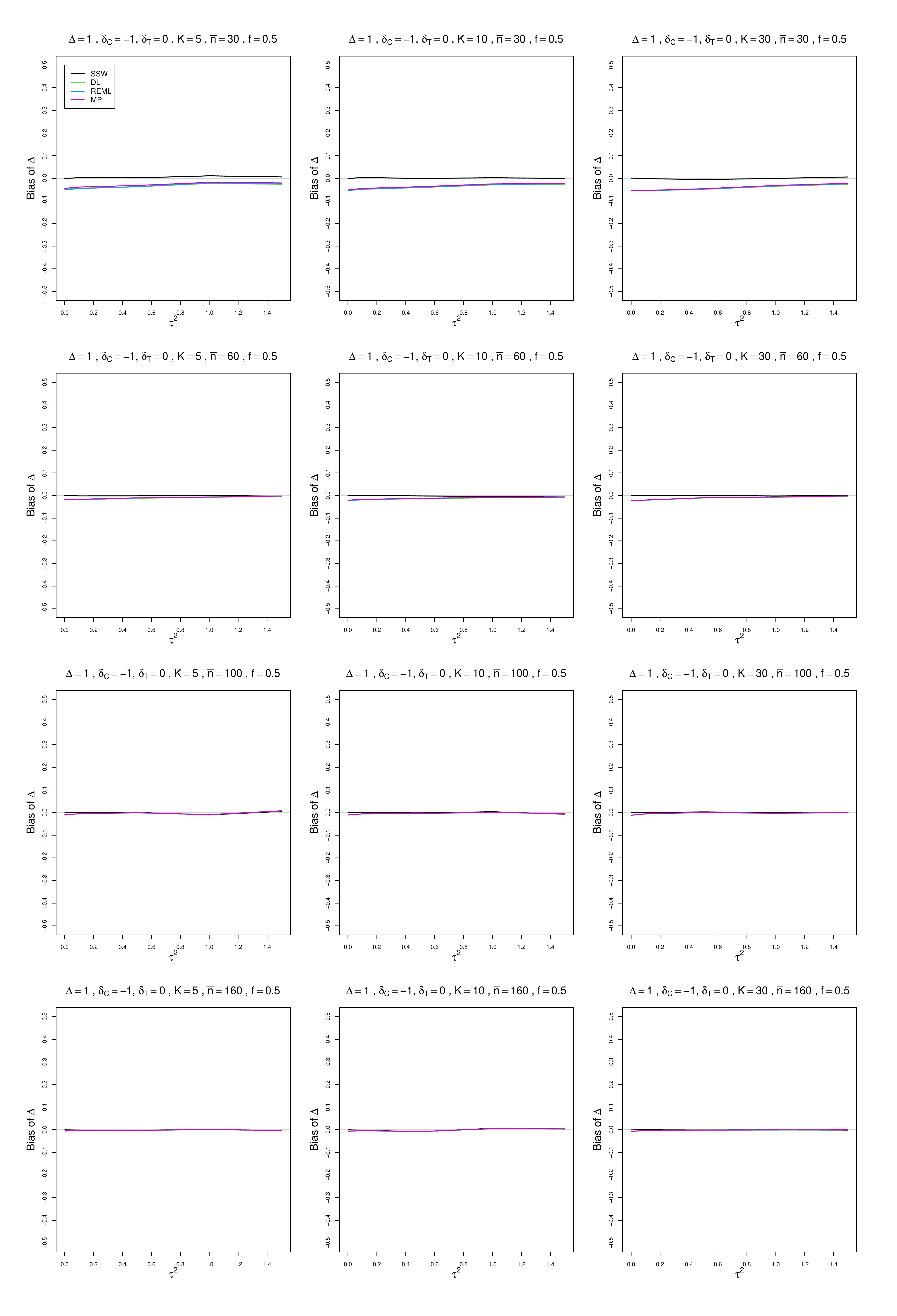}
	\caption{Bias  of estimators of overall effect measure $\Delta$ (DL, REML, MP and SSW) vs $\tau^2$, for unequal sample sizes $\bar{n}=30,\;60,\;100$ and $160$, $\delta_{iC} = -1$, $\Delta=1$ and  $f = 0.5$.   }
	\label{PlotBiasOfDelta_deltaC_-1deltaT=0_DSM_unequal_sample_sizes.pdf}
\end{figure}

\begin{figure}[ht]
	\centering
	\includegraphics[scale=0.33]{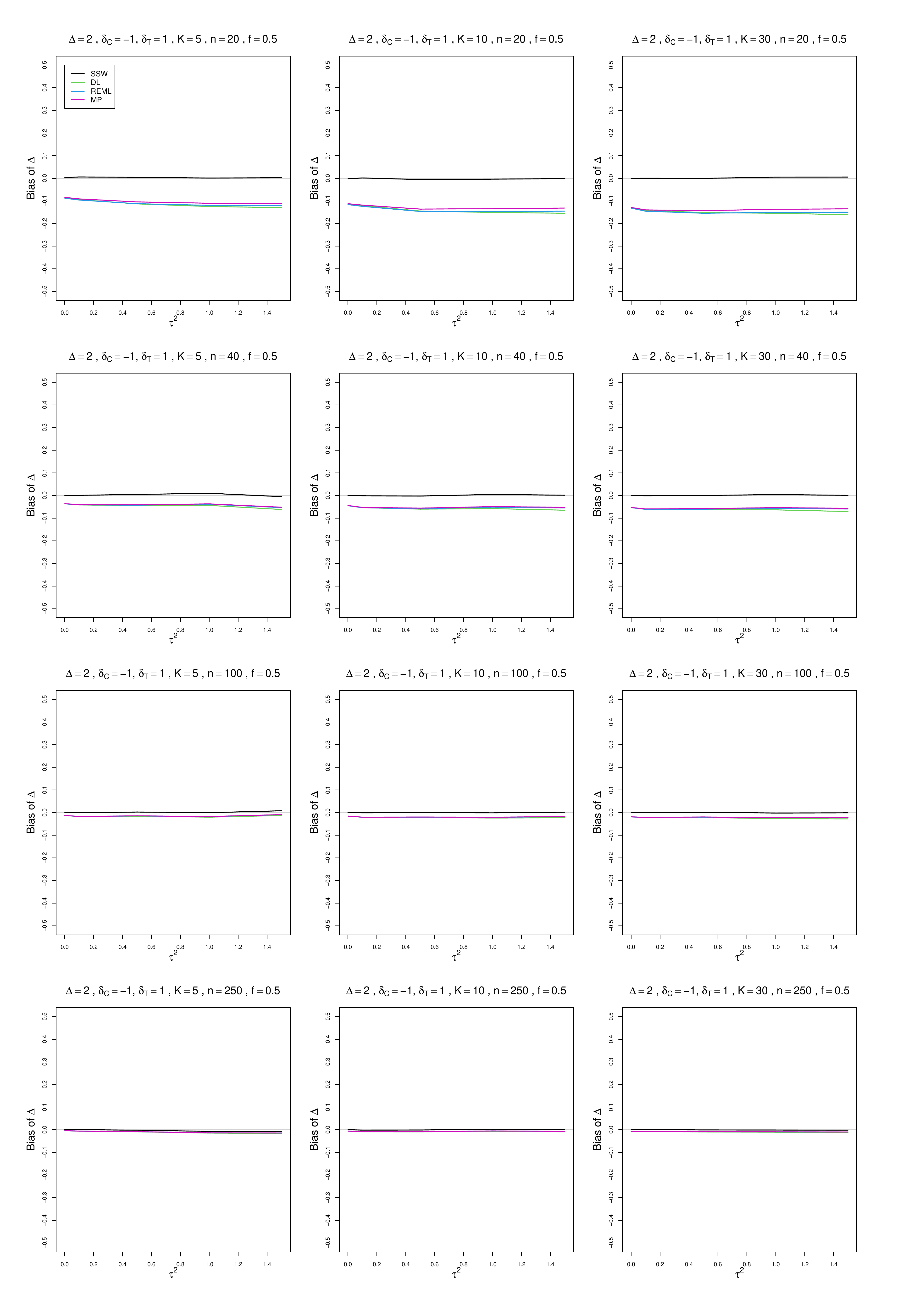}
	\caption{Bias  of estimators of overall effect measure $\Delta$ (DL, REML, MP and SSW) vs $\tau^2$, for equal sample sizes $n=20,\;40,\;100$ and $250$, $\delta_{iC} = -1$, $\Delta=2$ and  $f = 0.5$.   }
	\label{PlotBiasOfDelta_deltaC_-1deltaT=1_DSM_equal_sample_sizes.pdf}
\end{figure}

\begin{figure}[ht]
	\centering
	\includegraphics[scale=0.33]{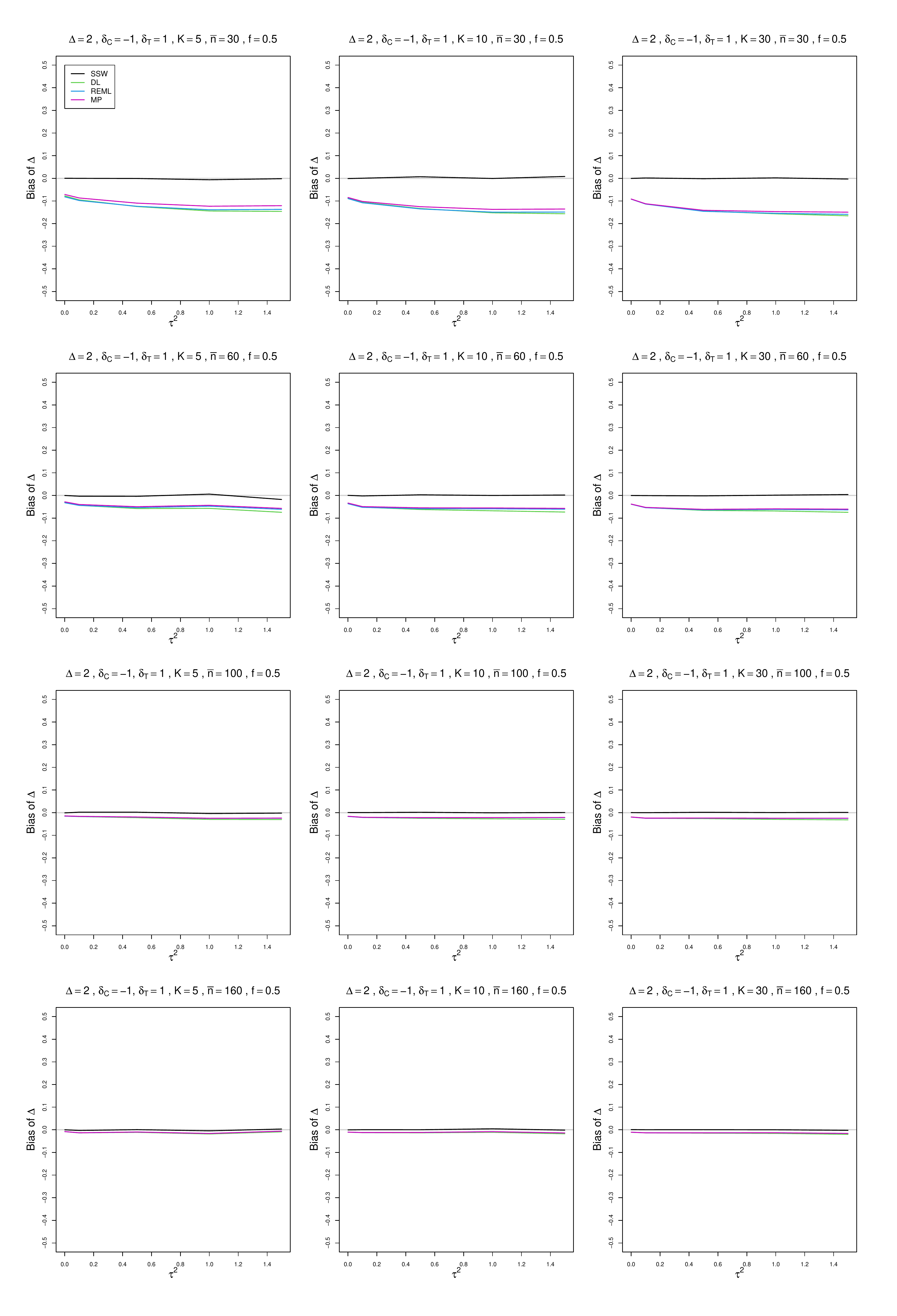}
	\caption{Bias  of estimators of overall effect measure $\Delta$ (DL, REML, MP and SSW) vs $\tau^2$, for unequal sample sizes $\bar{n}=30,\;60,\;100$ and $160$, $\delta_{iC} = -1$, $\Delta=2$ and  $f = 0.5$.   }
	\label{PlotBiasOfDelta_deltaC_-1deltaT=1_DSM_unequal_sample_sizes.pdf}
\end{figure}


\begin{figure}[ht]
	\centering
	\includegraphics[scale=0.33]{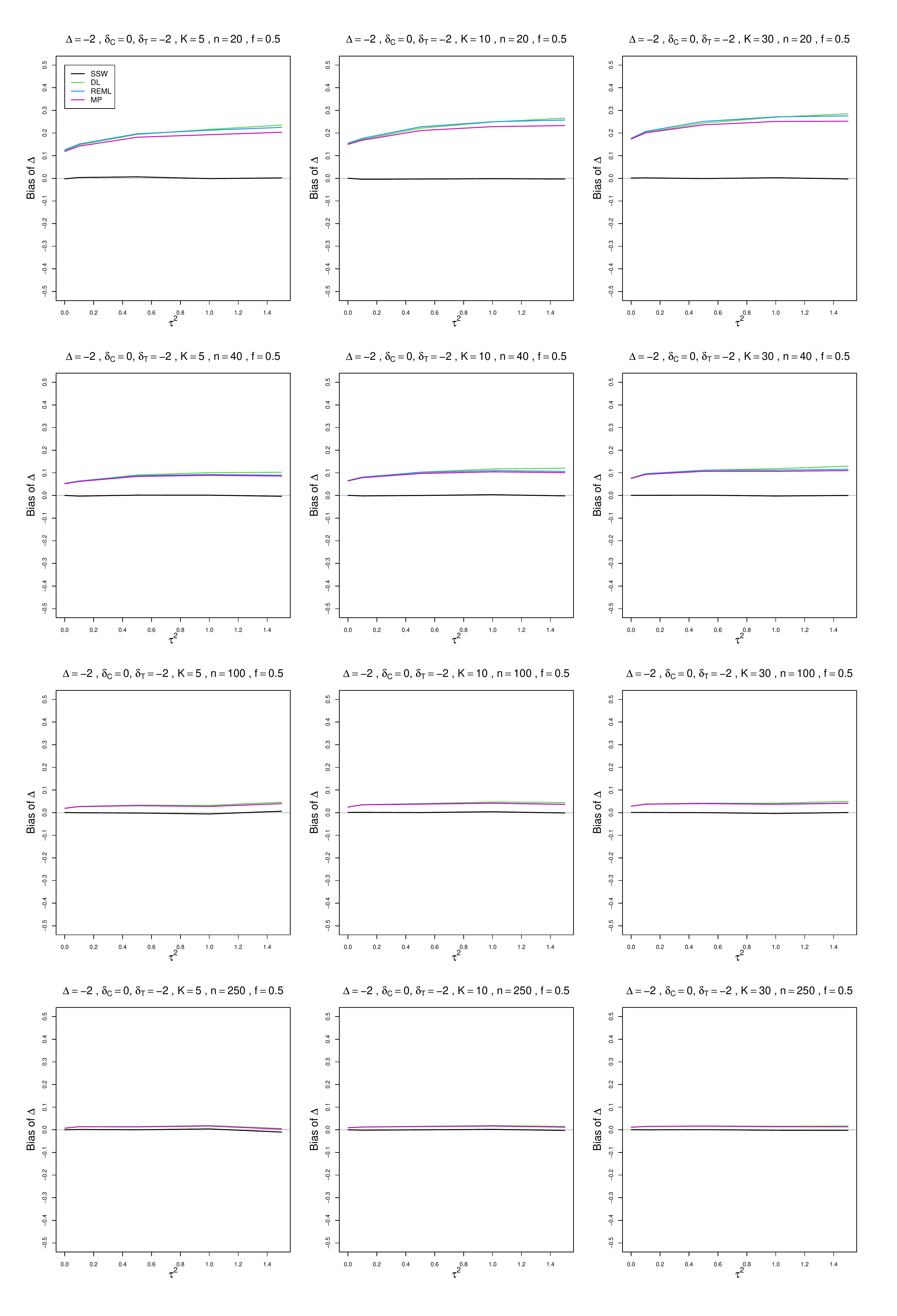}
	\caption{Bias  of estimators of overall effect measure $\Delta$ (DL, REML, MP and SSW) vs $\tau^2$, for equal sample sizes $n=20,\;40,\;100$ and $250$, $\delta_{iC} = 0$, $\Delta=-2$ and  $f = 0.5$.   }
	\label{PlotBiasOfDelta_deltaC_0deltaT=-2_DSM_equal_sample_sizes.pdf}
\end{figure}

\begin{figure}[ht]
	\centering
	\includegraphics[scale=0.33]{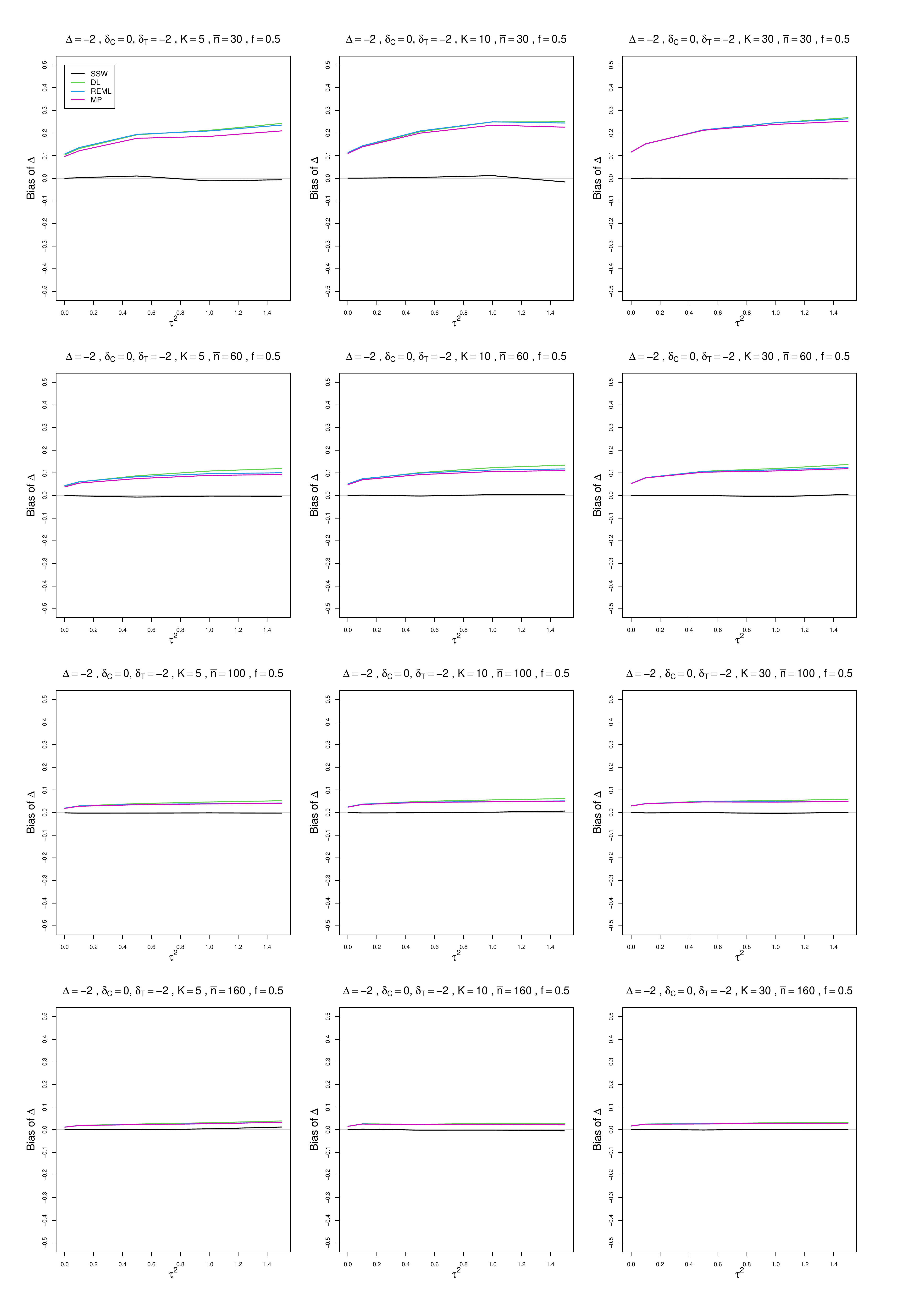}
	\caption{Bias  of estimators of overall effect measure $\Delta$ (DL, REML, MP and SSW ) vs $\tau^2$, for unequal sample sizes $\bar{n}=30,\;60,\;100$ and $160$, $\delta_{iC} = 0$, $\Delta=-2$ and  $f = 0.5$.   }
	\label{PlotBiasOfDelta_deltaC_-1deltaT=-3_DSM_unequal_sample_sizes.pdf}
\end{figure}

\begin{figure}[ht]
	\centering
	\includegraphics[scale=0.33]{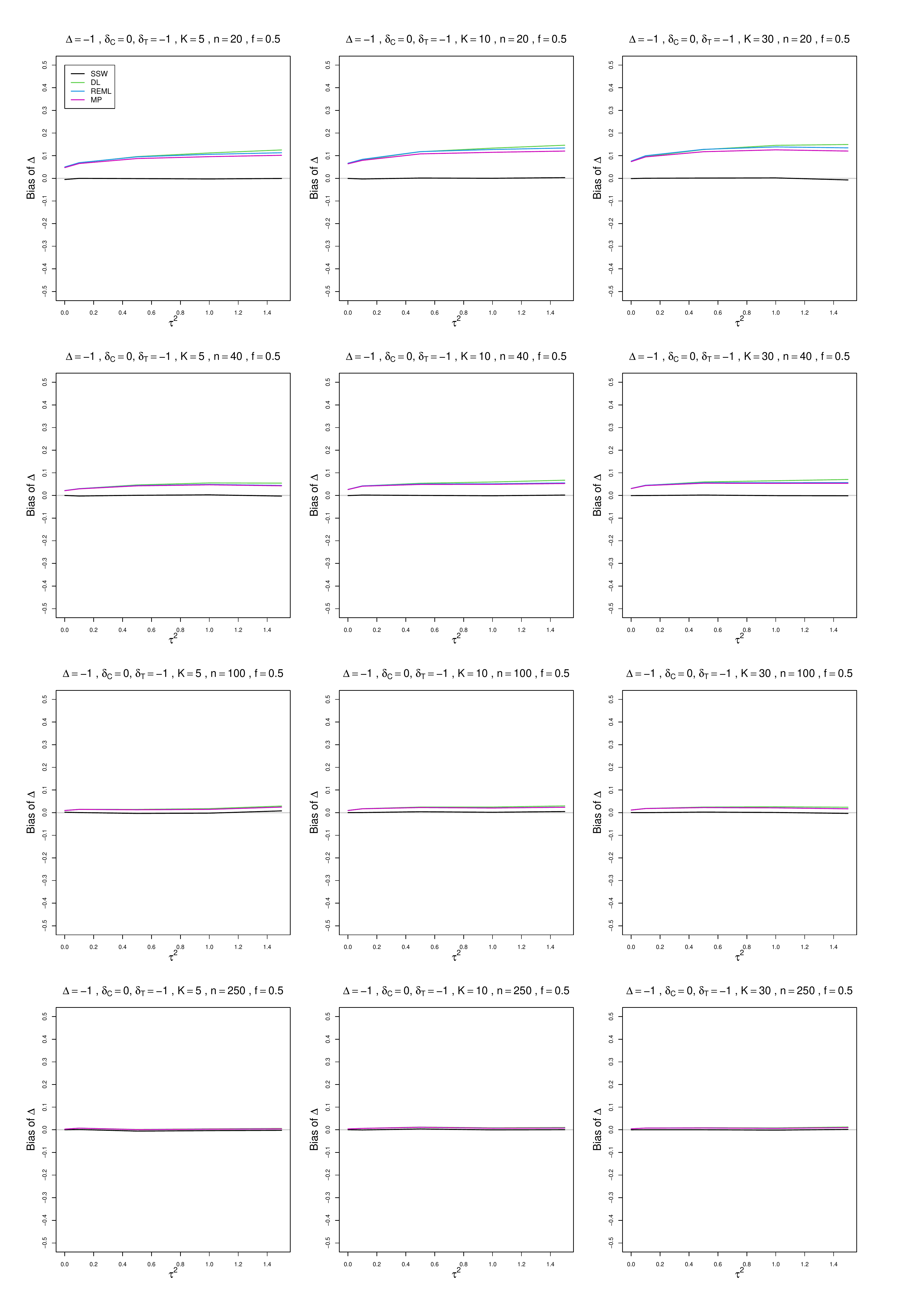}
	\caption{Bias  of estimators of overall effect measure $\Delta$ (DL, REML, MP and SSW) vs $\tau^2$, for equal sample sizes $n=20,\;40,\;100$ and $250$, $\delta_{iC} = 0$, $\Delta=-1$ and  $f = 0.5$.   }
	\label{PlotBiasOfDelta_deltaC_-0deltaT=-1_DSM_equal_sample_sizes.pdf}
\end{figure}

\begin{figure}[ht]
	\centering
	\includegraphics[scale=0.33]{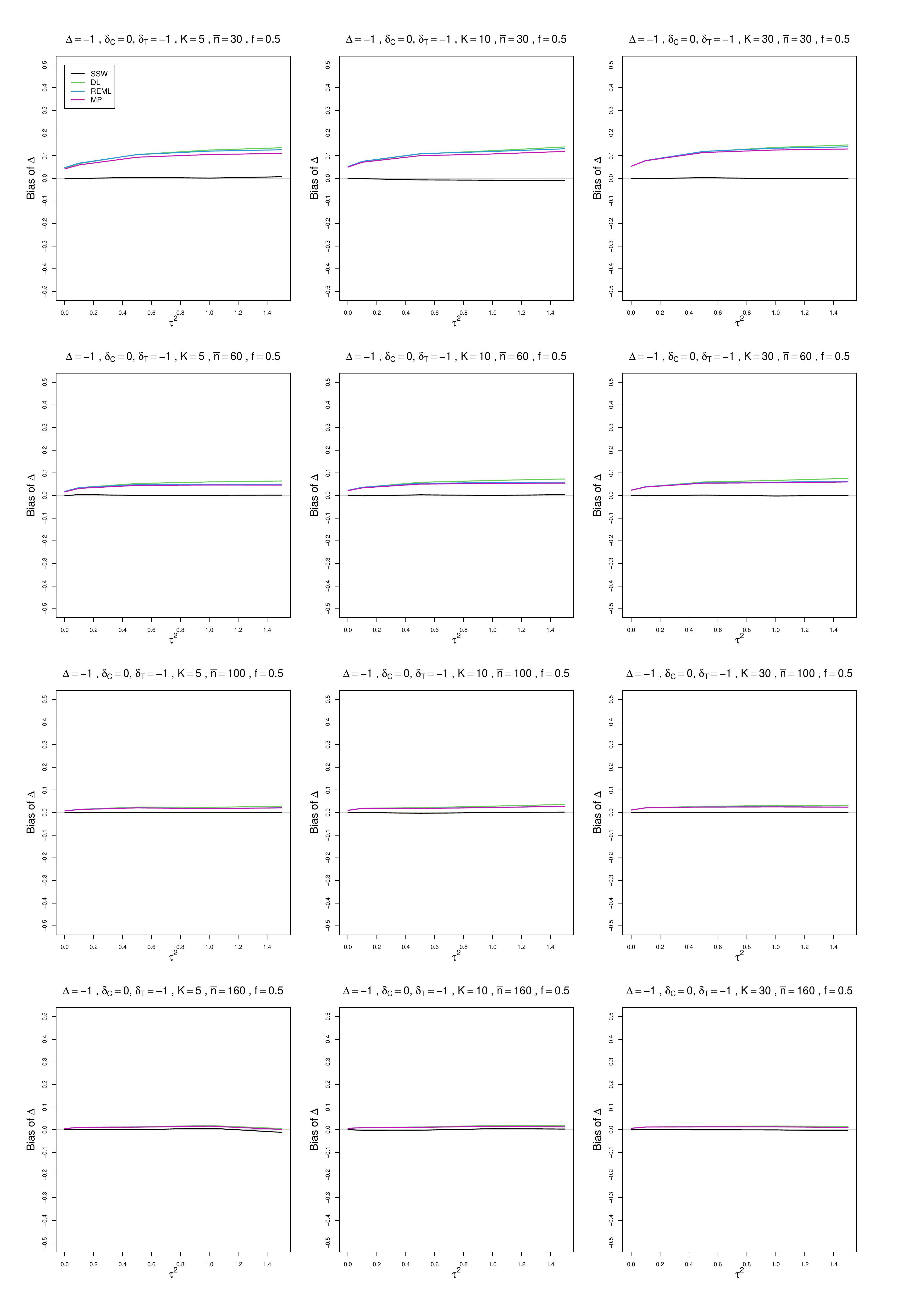}
	\caption{Bias  of estimators of overall effect measure $\Delta$ (DL, REML, MP and SSW) vs $\tau^2$, for unequal sample sizes $\bar{n}=30,\;60,\;100$ and $160$, $\delta_{iC} = 0$, $\Delta=-1$ and  $f = 0.5$.   }
	\label{PlotBiasOfDelta_deltaC_0deltaT=-1_DSM_unequal_sample_sizes.pdf}
\end{figure}

\begin{figure}[ht]
	\centering
	\includegraphics[scale=0.33]{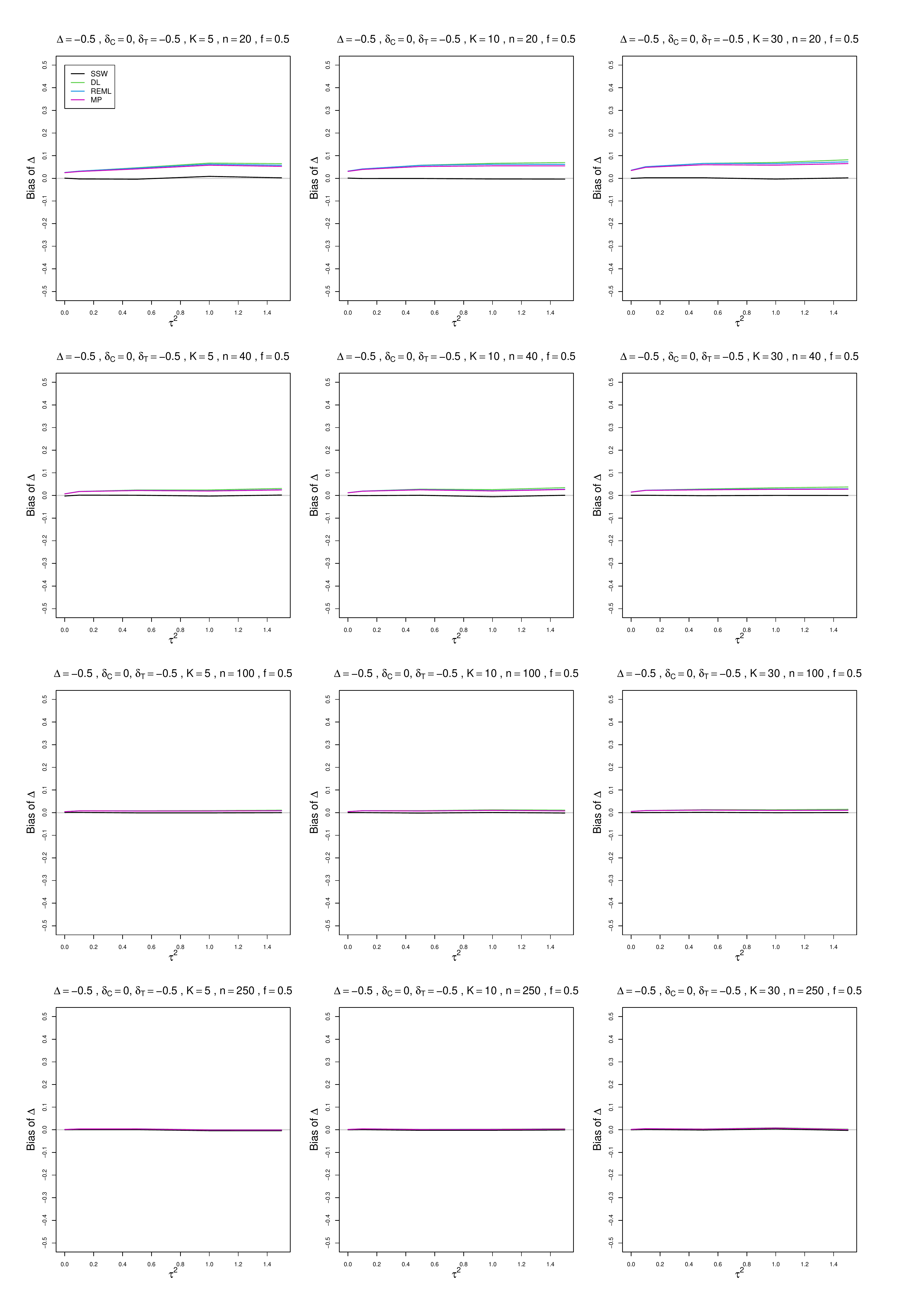}
	\caption{Bias  of estimators of overall effect measure $\Delta$ (DL, REML, MP and SSW) vs $\tau^2$, for equal sample sizes $n=20,\;40,\;100$ and $250$, $\delta_{iC} = 0$, $\Delta=-0.5$ and  $f = 0.5$.   }
	\label{PlotBiasOfDelta_deltaC_-0deltaT=-0.5_DSM_equal_sample_sizes.pdf}
\end{figure}

\begin{figure}[ht]
	\centering
	\includegraphics[scale=0.33]{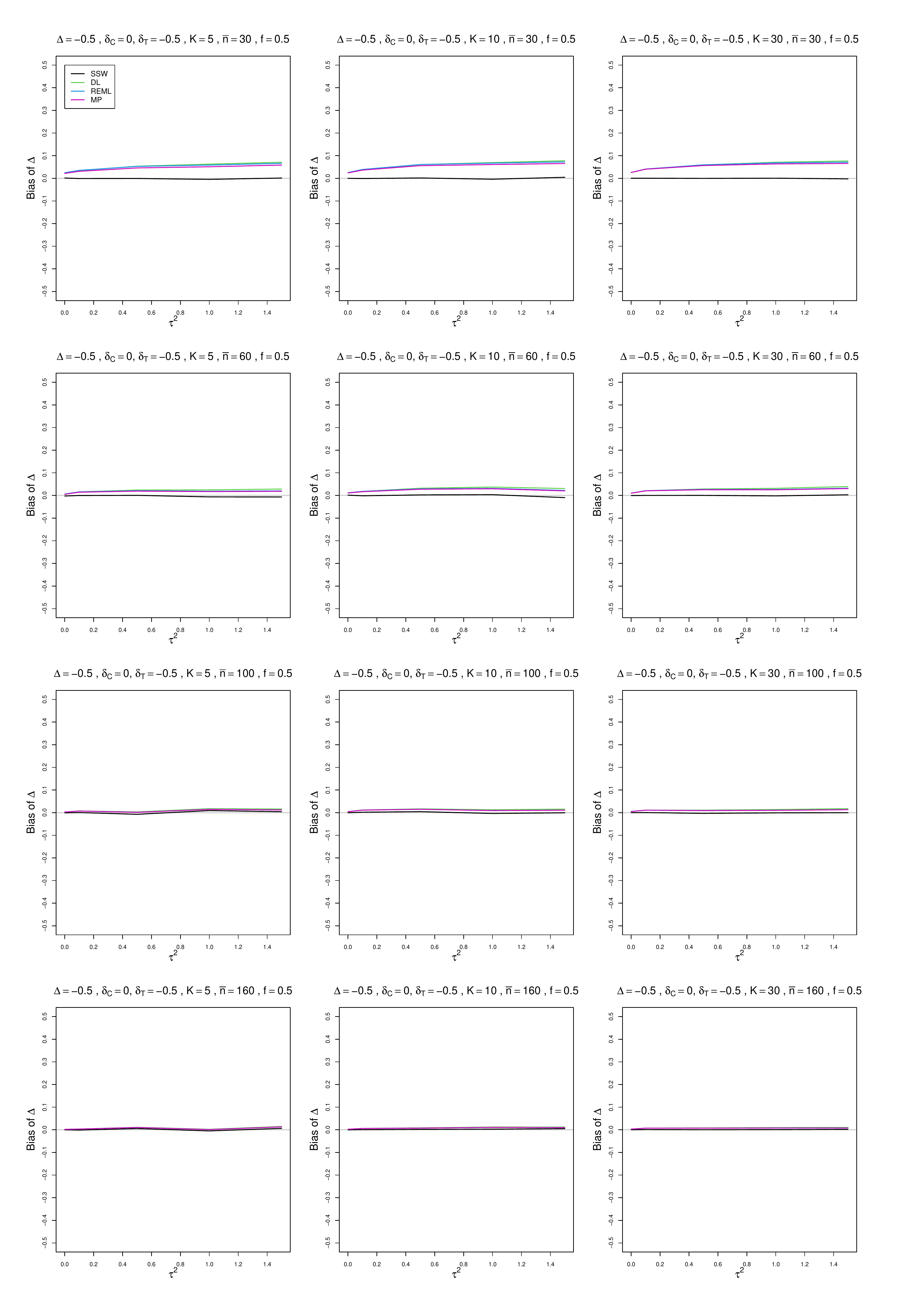}
	\caption{Bias  of estimators of overall effect measure $\Delta$ (DL, REML, MP and SSW) vs $\tau^2$, for unequal sample sizes $\bar{n}=30,\;60,\;100$ and $160$, $\delta_{iC} = 0$, $\Delta=-0.5$ and  $f = 0.5$.   }
	\label{PlotBiasOfDelta_deltaC_0deltaT=-0.5_DSM_unequal_sample_sizes.pdf}
\end{figure}

\begin{figure}[ht]
	\centering
	\includegraphics[scale=0.33]{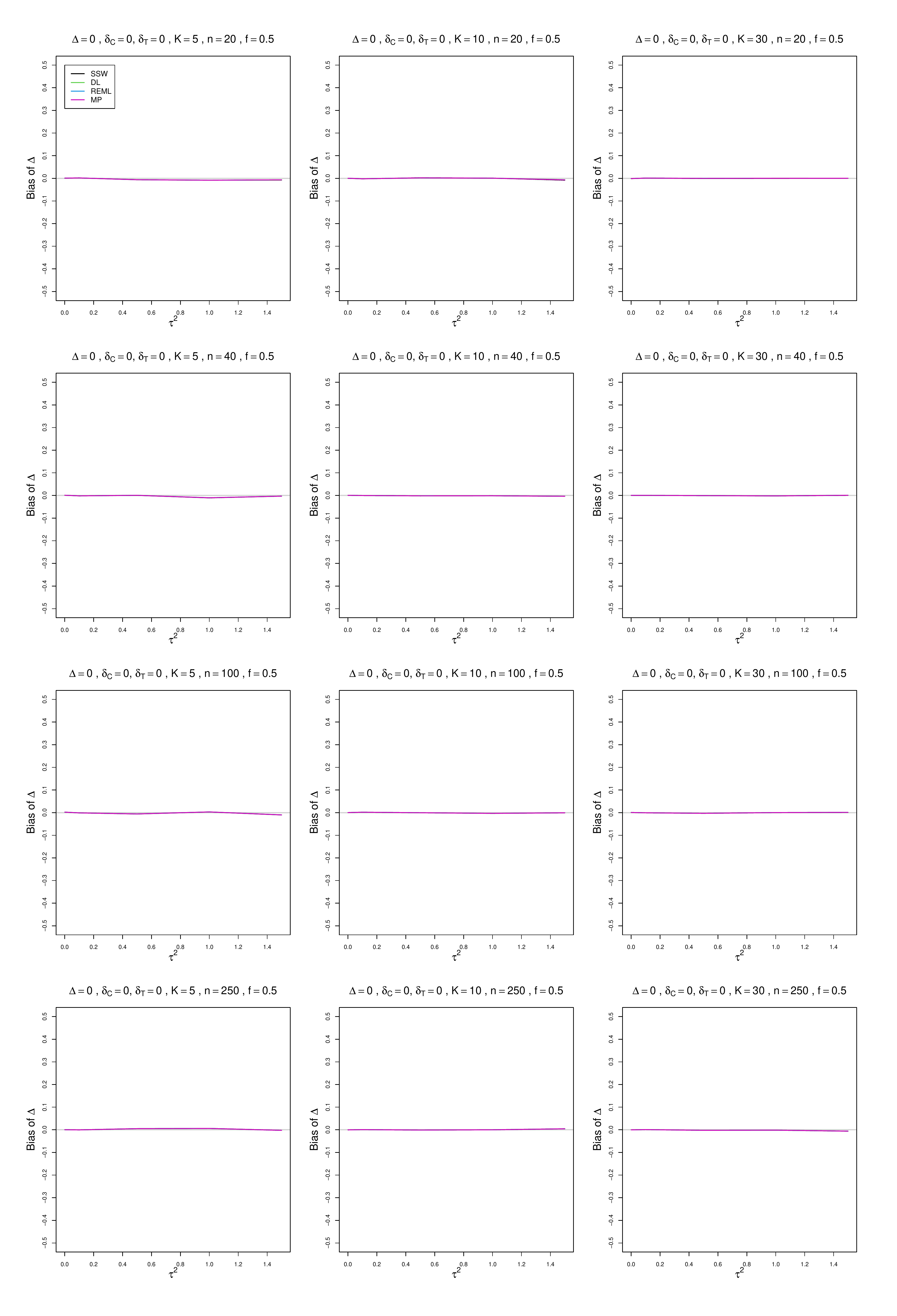}
	\caption{Bias  of estimators of overall effect measure $\Delta$ (DL, REML, MP and SSW) vs $\tau^2$, for equal sample sizes $n=20,\;40,\;100$ and $250$, $\delta_{iC} = 0$, $\Delta=0$ and  $f = 0.5$.   }
	\label{PlotBiasOfDelta_deltaC_0deltaT=0_DSM_equal_sample_sizes.pdf}
\end{figure}

\begin{figure}[ht]
	\centering
	\includegraphics[scale=0.33]{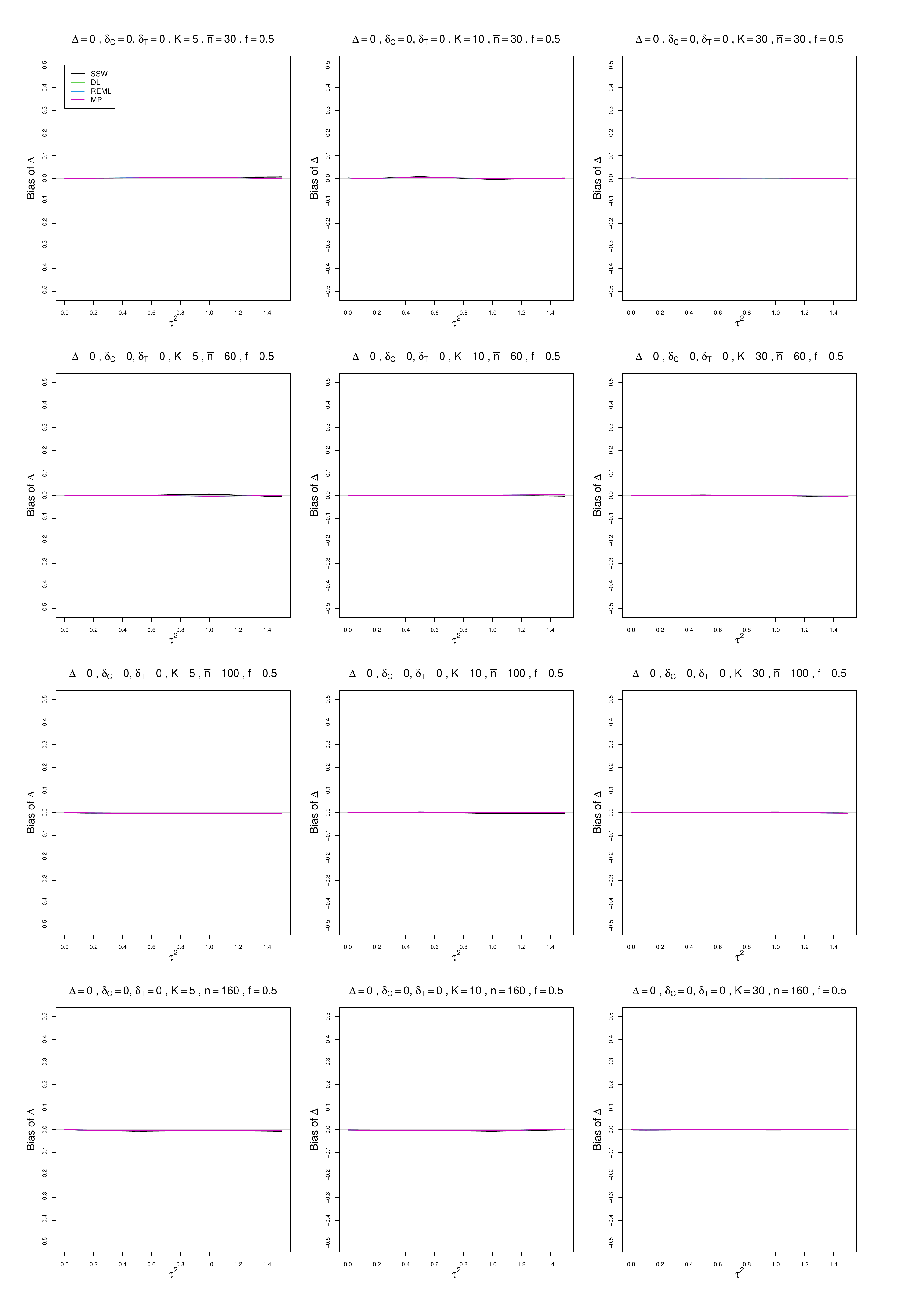}
	\caption{Bias  of estimators of overall effect measure $\Delta$ (DL, REML, MP and SSW) vs $\tau^2$, for unequal sample sizes $\bar{n}=30,\;60,\;100$ and $160$, $\delta_{iC} = 0$, $\Delta=0$ and  $f = 0.5$.   }
	\label{PlotBiasOfDelta_deltaC_0deltaT=0_DSM_unequal_sample_sizes.pdf}
\end{figure}

\begin{figure}[ht]
	\centering
	\includegraphics[scale=0.33]{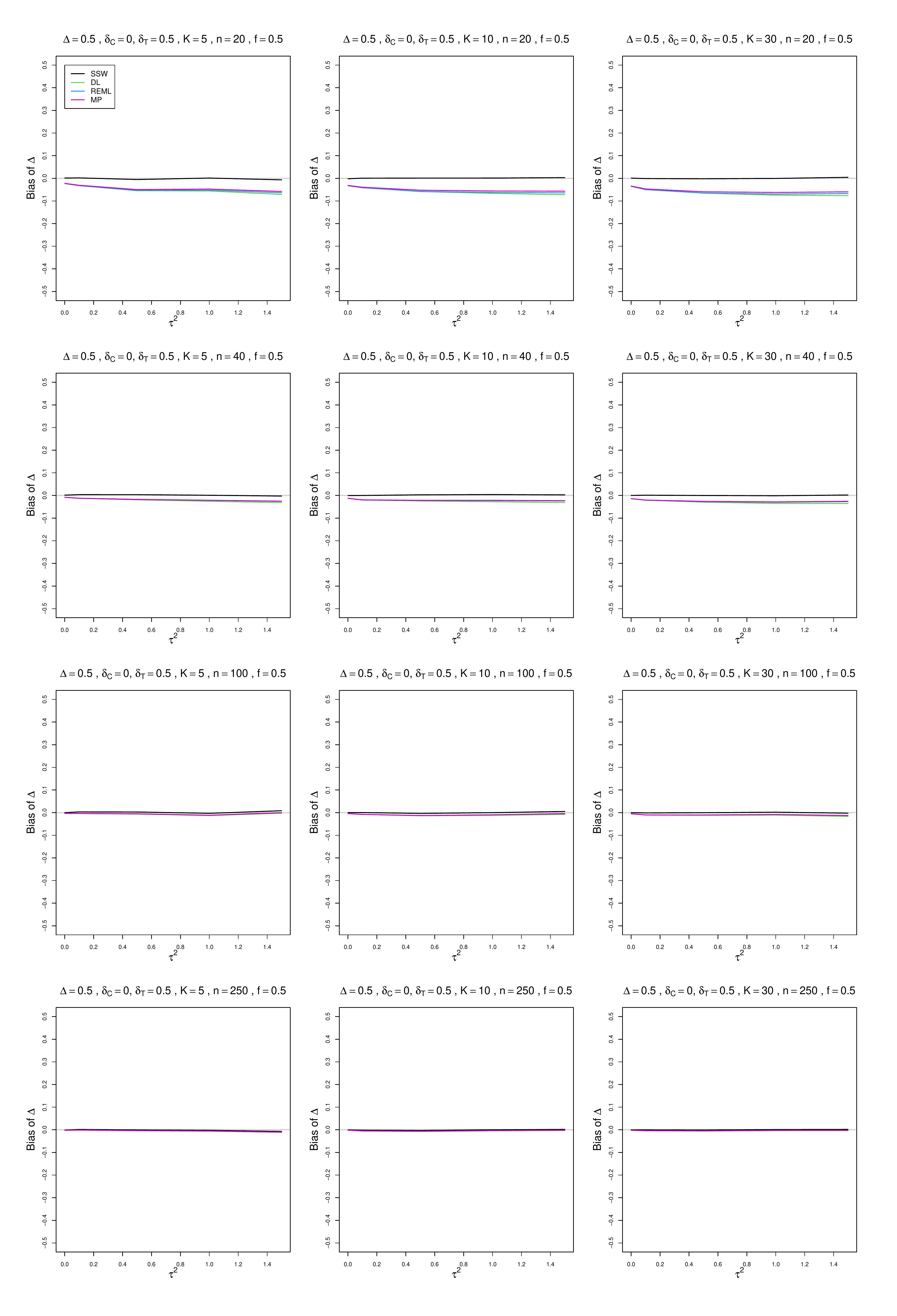}
	\caption{Bias  of estimators of overall effect measure $\Delta$ (DL, REML, MP and SSW ) vs $\tau^2$, for equal sample sizes $n=20,\;40,\;100$ and $250$, $\delta_{iC} = 0$, $\Delta=0.5$ and  $f = 0.5$.   }
	\label{PlotBiasOfDelta_deltaC_0deltaT=0.5_DSM_equal_sample_sizes.pdf}
\end{figure}

\begin{figure}[ht]
	\centering
	\includegraphics[scale=0.33]{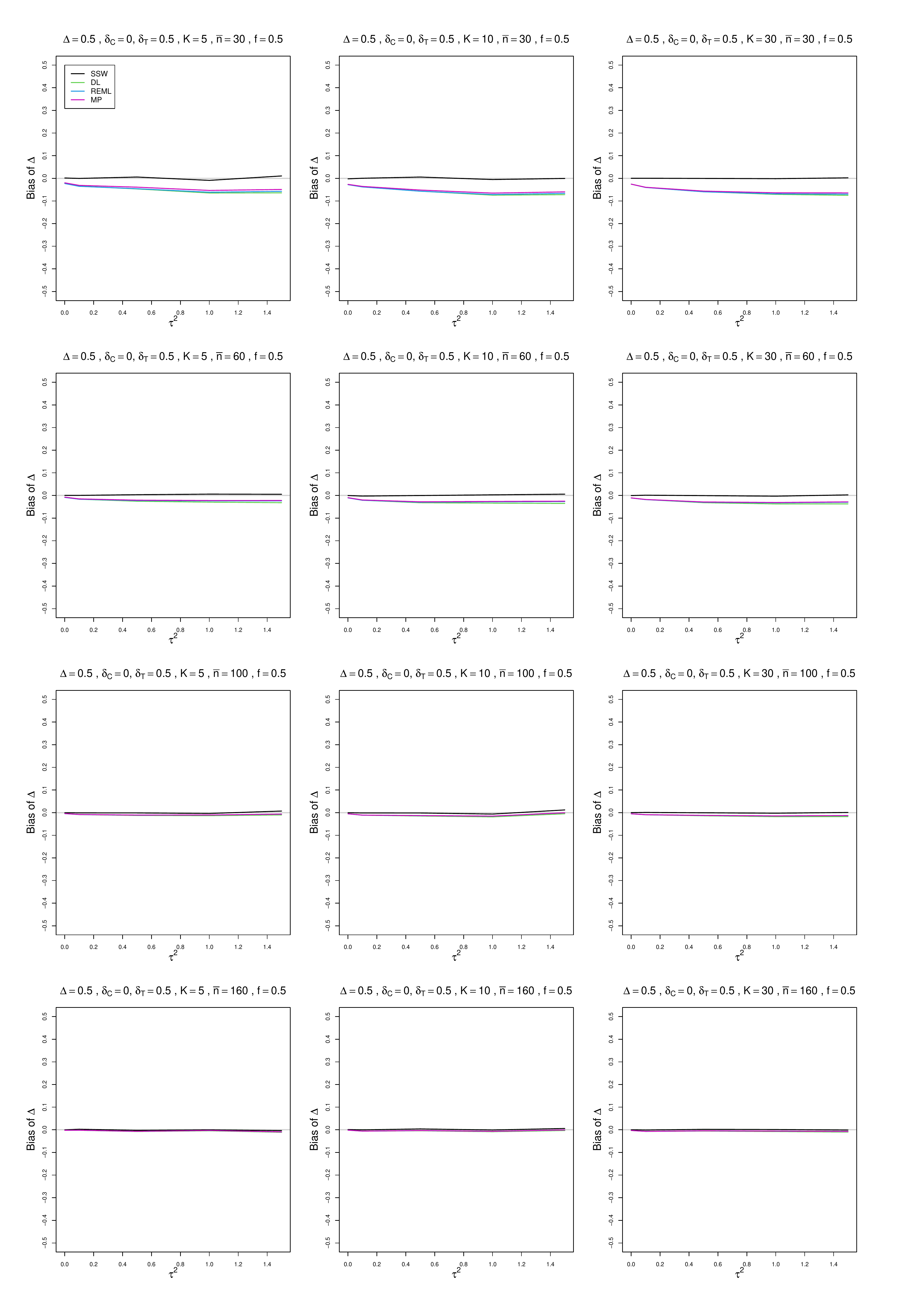}
	\caption{Bias  of estimators of overall effect measure $\Delta$ (DL, REML, MP and SSW) vs $\tau^2$, for unequal sample sizes $\bar{n}=30,\;60,\;100$ and $160$, $\delta_{iC} = 0$, $\Delta=0.5$ and  $f = 0.5$.   }
	\label{PlotBiasOfDelta_deltaC_0deltaT=-0.5_DSM_unequal_sample_sizes.pdf}
\end{figure}

\begin{figure}[ht]
	\centering
	\includegraphics[scale=0.33]{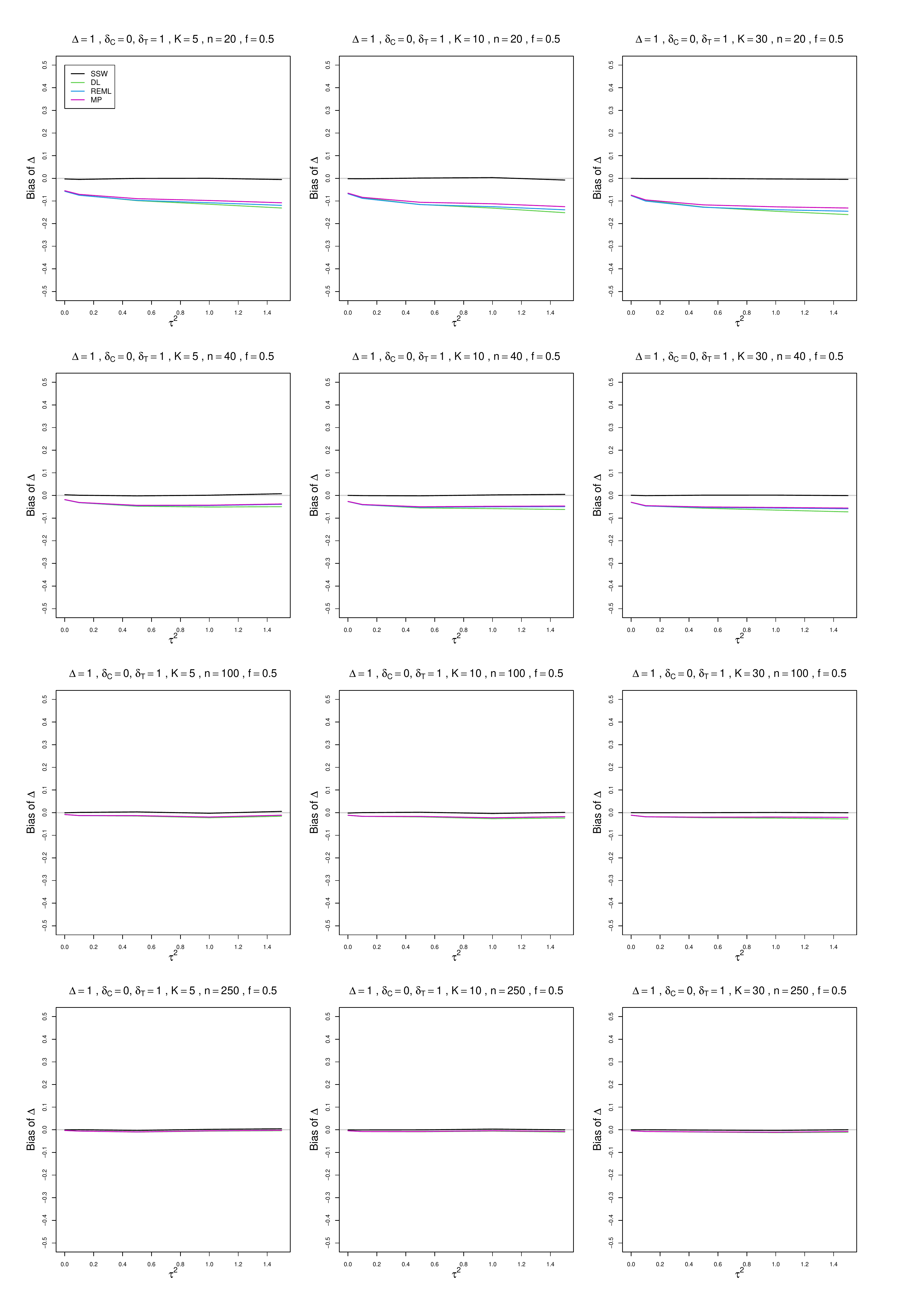}
	\caption{Bias  of estimators of overall effect measure $\Delta$ (DL, REML, MP and SSW) vs $\tau^2$, for equal sample sizes $n=20,\;40,\;100$ and $250$, $\delta_{iC} = 0$, $\Delta=1$ and  $f = 0.5$.   }
	\label{PlotBiasOfDelta_deltaC_=0deltaT=1_DSM_equal_sample_sizes.pdf}
\end{figure}

\begin{figure}[ht]
	\centering
	\includegraphics[scale=0.33]{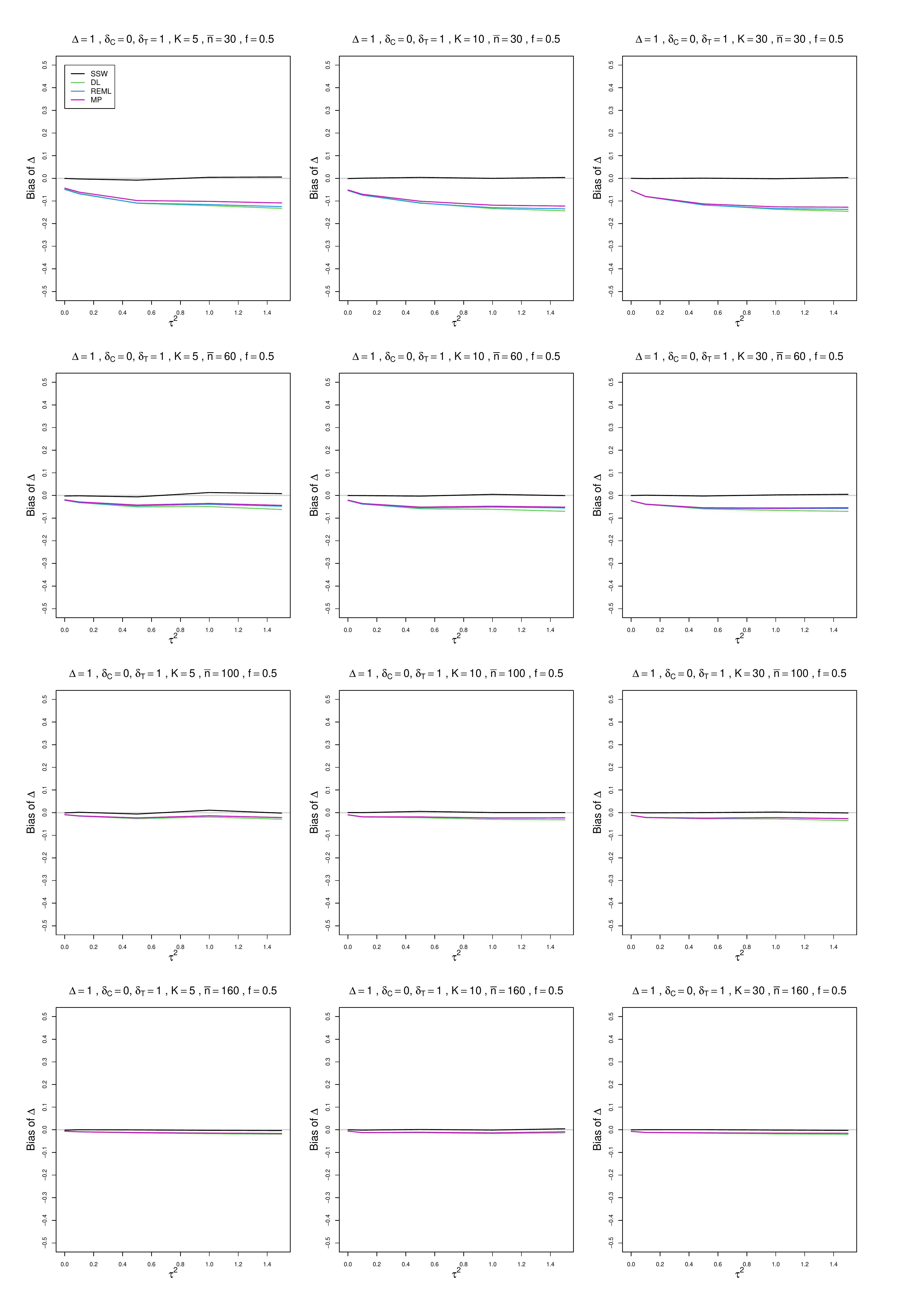}
	\caption{Bias  of estimators of overall effect measure $\Delta$ (DL, REML, MP and SSW) vs $\tau^2$, for unequal sample sizes $\bar{n}=30,\;60,\;100$ and $160$, $\delta_{iC} = 0$, $\Delta=1$ and  $f = 0.5$.   }
	\label{PlotBiasOfDelta_deltaC_0deltaT=1_DSM_unequal_sample_sizes.pdf}
\end{figure}

\begin{figure}[ht]
	\centering
	\includegraphics[scale=0.33]{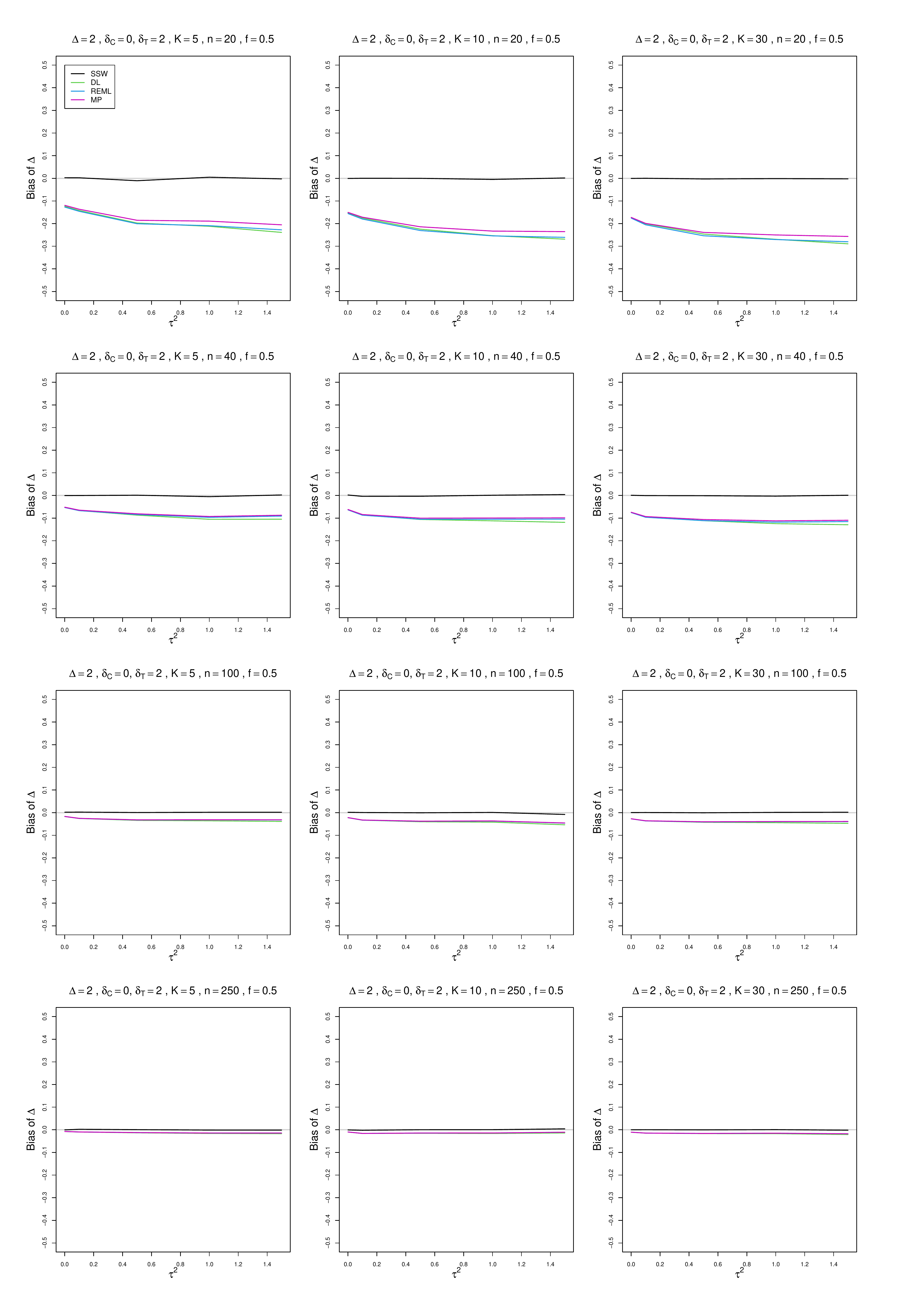}
	\caption{Bias  of estimators of overall effect measure $\Delta$ (DL, REML, MP and SSW) vs $\tau^2$, for equal sample sizes $n=20,\;40,\;100$ and $250$, $\delta_{iC} = 0$, $\Delta=2$ and  $f = 0.5$.   }
	\label{PlotBiasOfDelta_deltaC_0deltaT=2_DSM_equal_sample_sizes.pdf}
\end{figure}

\begin{figure}[ht]
	\centering
	\includegraphics[scale=0.33]{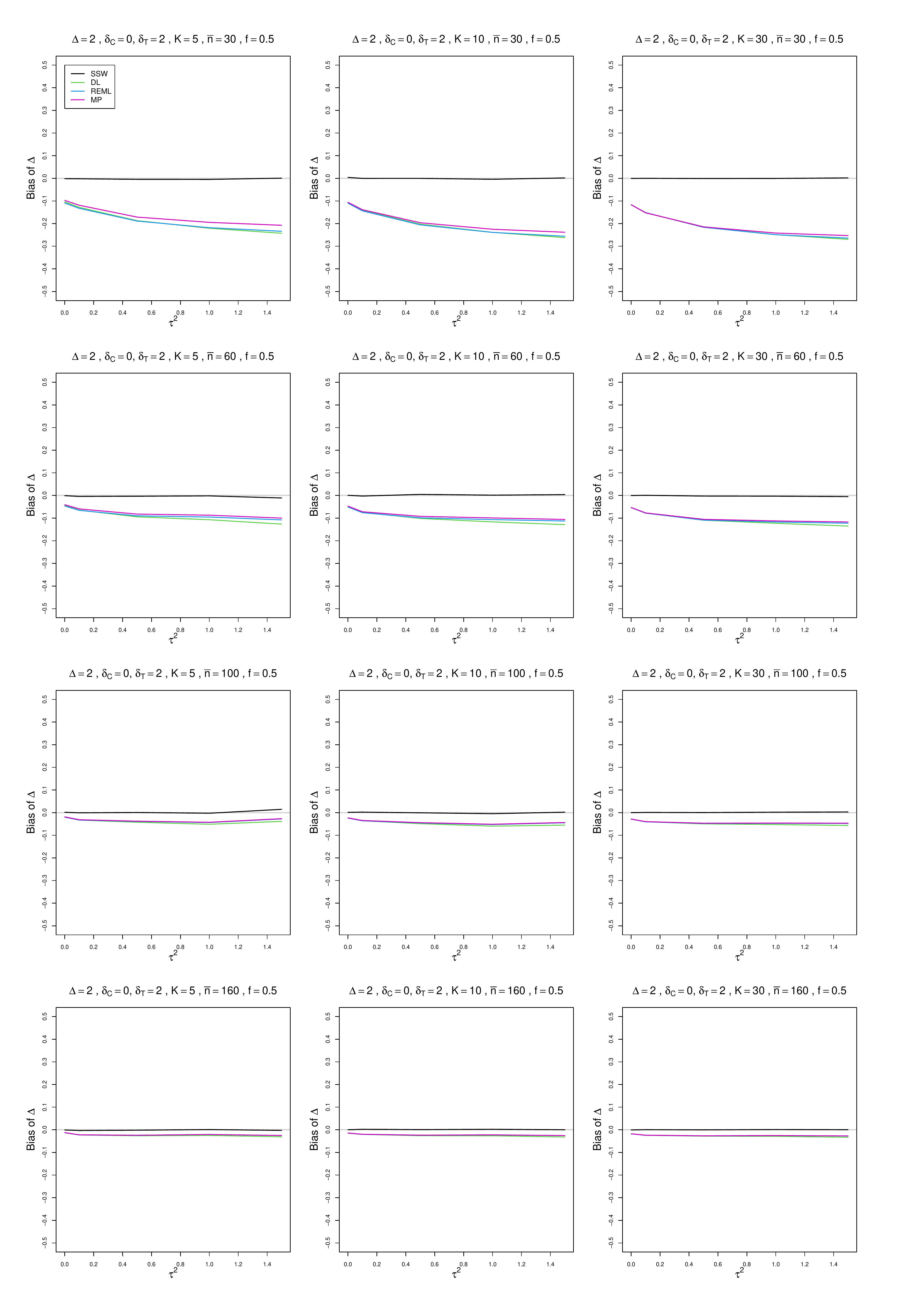}
	\caption{Bias  of estimators of overall effect measure $\Delta$ (DL, REML, MP and SSW) vs $\tau^2$, for unequal sample sizes $\bar{n}=30,\;60,\;100$ and $160$, $\delta_{iC} = 0$, $\Delta=2$ and  $f = 0.5$.   }
	\label{PlotBiasOfDelta_deltaC_0deltaT=2_DSM_unequal_sample_sizes.pdf}
\end{figure}

\clearpage
\begin{figure}[ht]
	\centering
	\includegraphics[scale=0.33]{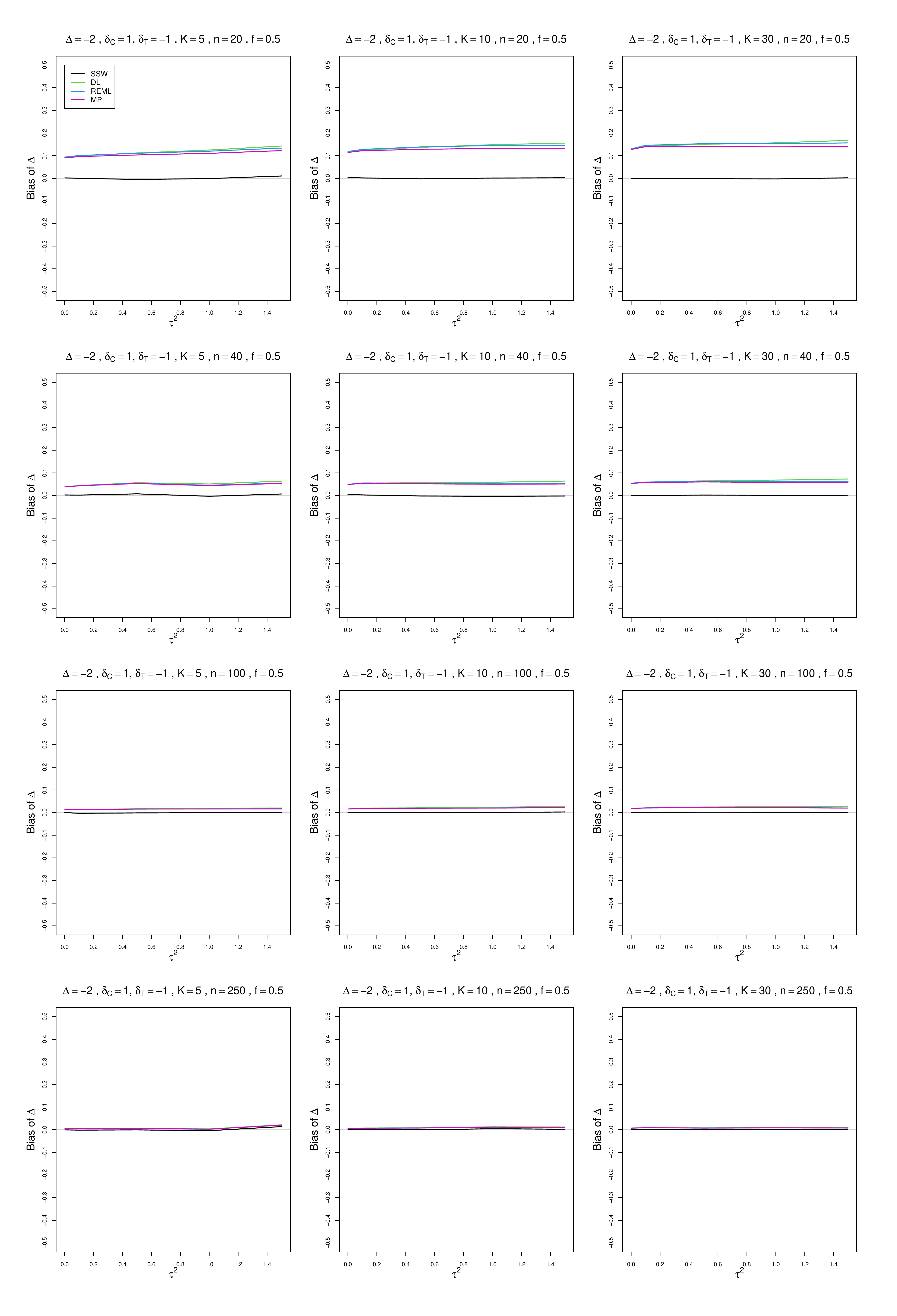}
	\caption{Bias  of estimators of overall effect measure $\Delta$ (DL, REML, MP and SSW) vs $\tau^2$, for equal sample sizes $n=20,\;40,\;100$ and $250$, $\delta_{iC} = -1$, $\Delta=-2$ and  $f = 0.5$.   }
	\label{PlotBiasOfDelta_deltaC_1deltaT=-1_DSM_equal_sample_sizes.pdf}
\end{figure}

\begin{figure}[ht]
	\centering
	\includegraphics[scale=0.33]{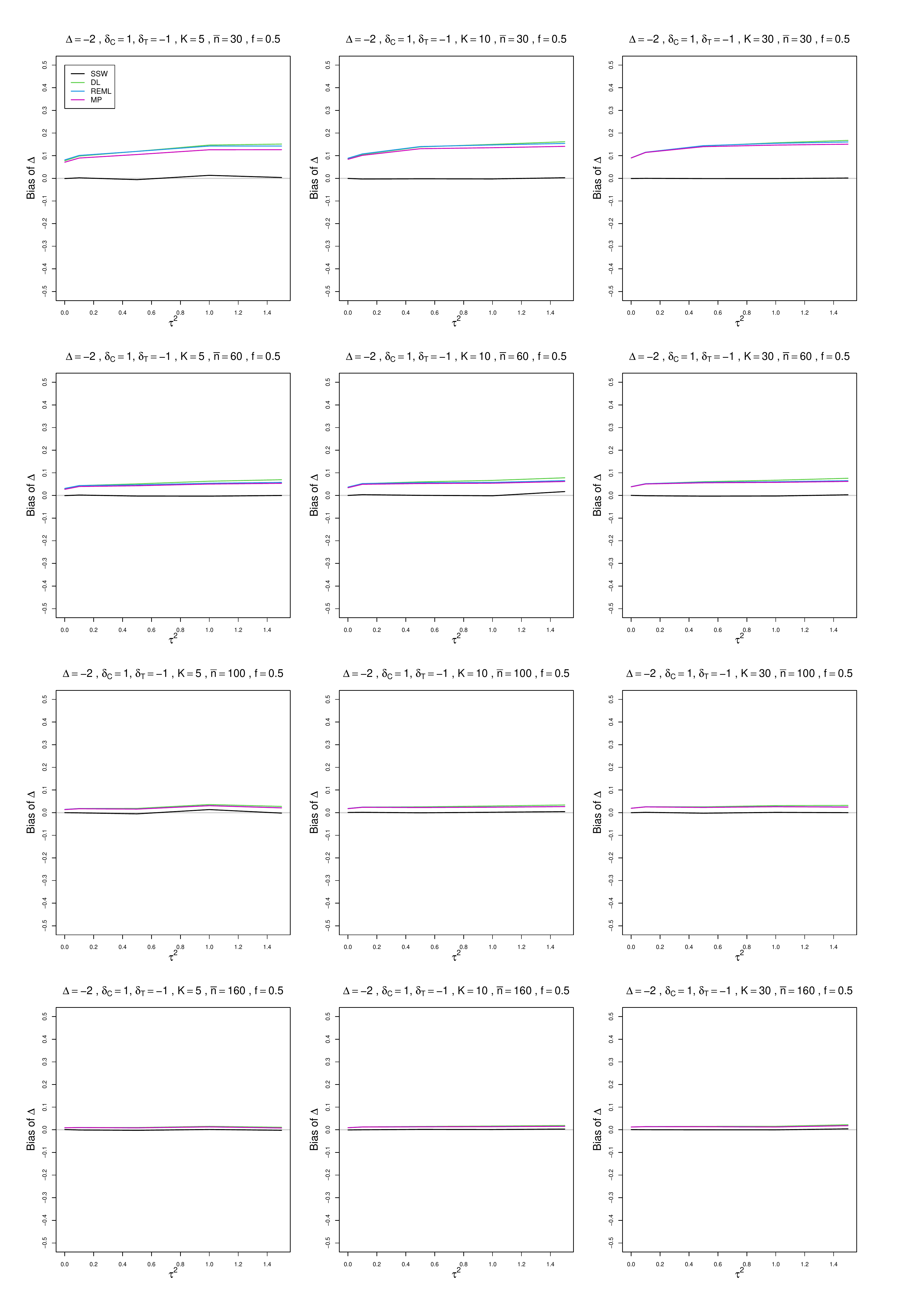}
	\caption{Bias  of estimators of overall effect measure $\Delta$ (DL, REML, MP and SSW ) vs $\tau^2$, for unequal sample sizes $\bar{n}=30,\;60,\;100$ and $160$, $\delta_{iC} = 1$, $\Delta=-2$ and  $f = 0.5$.   }
	\label{PlotBiasOfDelta_deltaC_1deltaT=-1_DSM_unequal_sample_sizes.pdf}
\end{figure}

\begin{figure}[ht]
	\centering
	\includegraphics[scale=0.33]{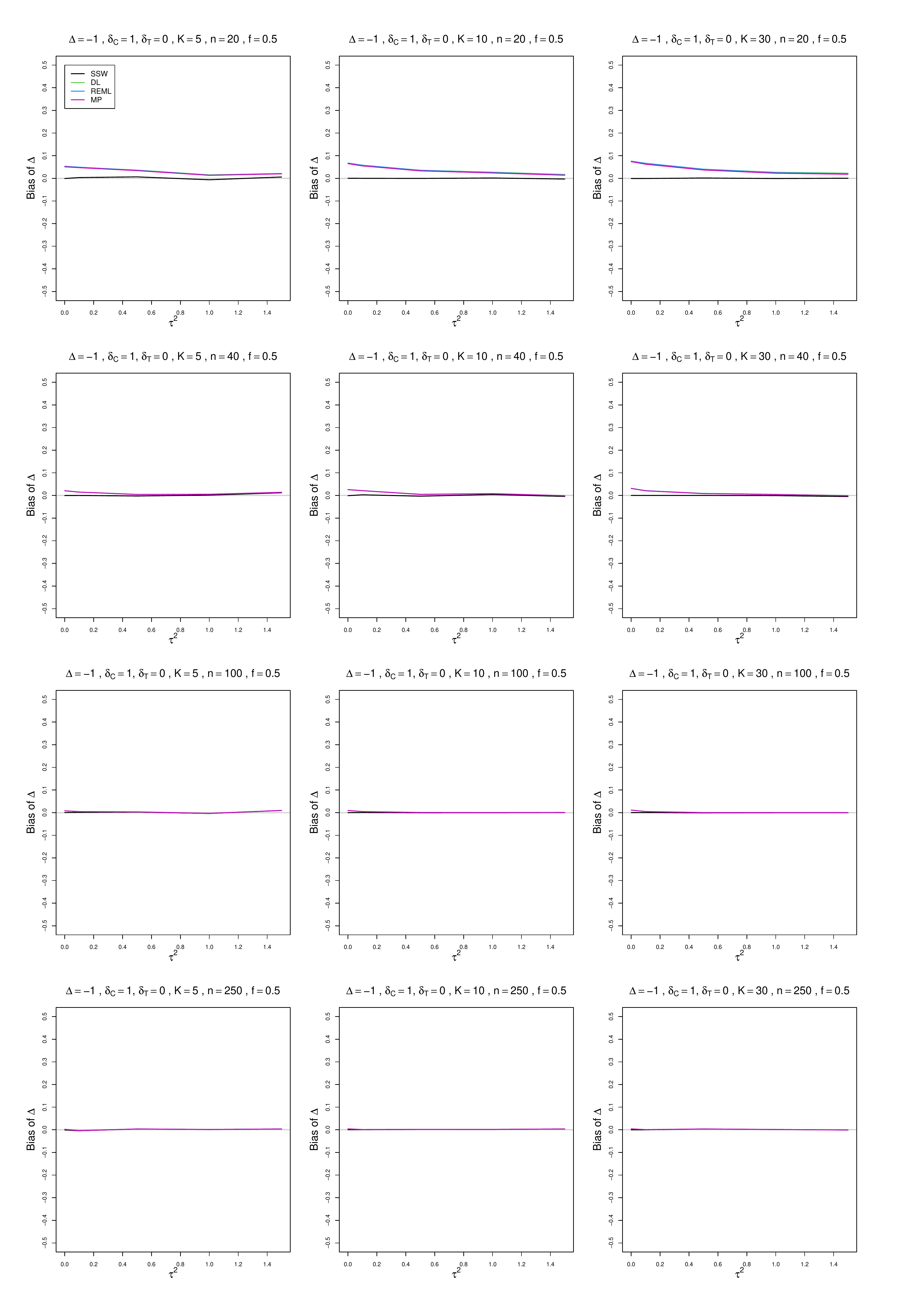}
	\caption{Bias  of estimators of overall effect measure $\Delta$ (DL, REML, MP and SSW) vs $\tau^2$, for equal sample sizes $n=20,\;40,\;100$ and $250$, $\delta_{iC} = 1$, $\Delta=-1$ and  $f = 0.5$.   }
	\label{PlotBiasOfDelta_deltaC_1deltaT=0_DSM_equal_sample_sizes.pdf}
\end{figure}

\begin{figure}[ht]
	\centering
	\includegraphics[scale=0.33]{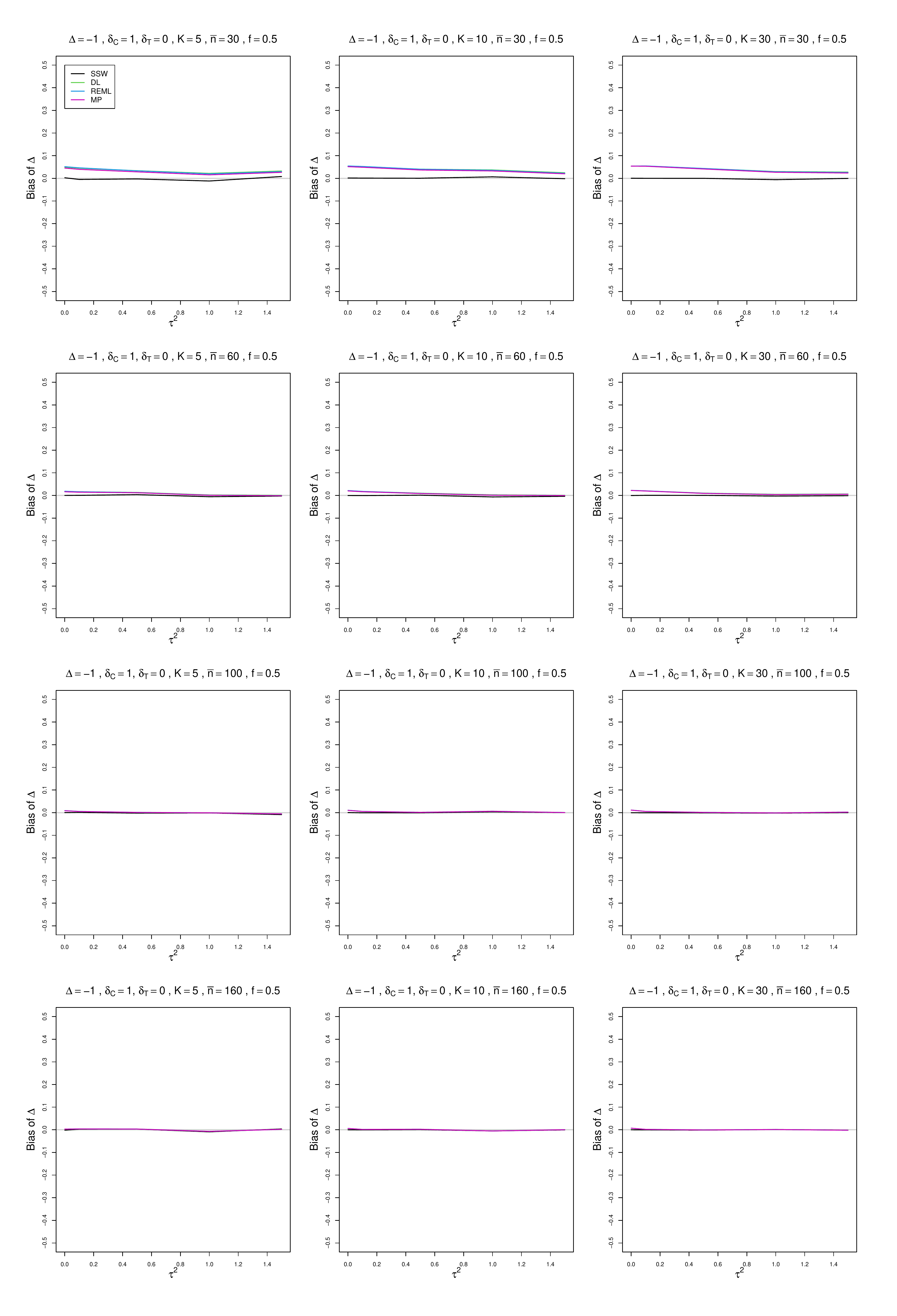}
	\caption{Bias  of estimators of overall effect measure $\Delta$ (DL, REML, MP and SSW) vs $\tau^2$, for unequal sample sizes $\bar{n}=30,\;60,\;100$ and $160$, $\delta_{iC} = 1$, $\Delta=-1$ and  $f = 0.5$.   }
	\label{PlotBiasOfDelta_deltaC_1deltaT=0_DSM_unequal_sample_sizes.pdf}
\end{figure}

\begin{figure}[ht]
	\centering
	\includegraphics[scale=0.33]{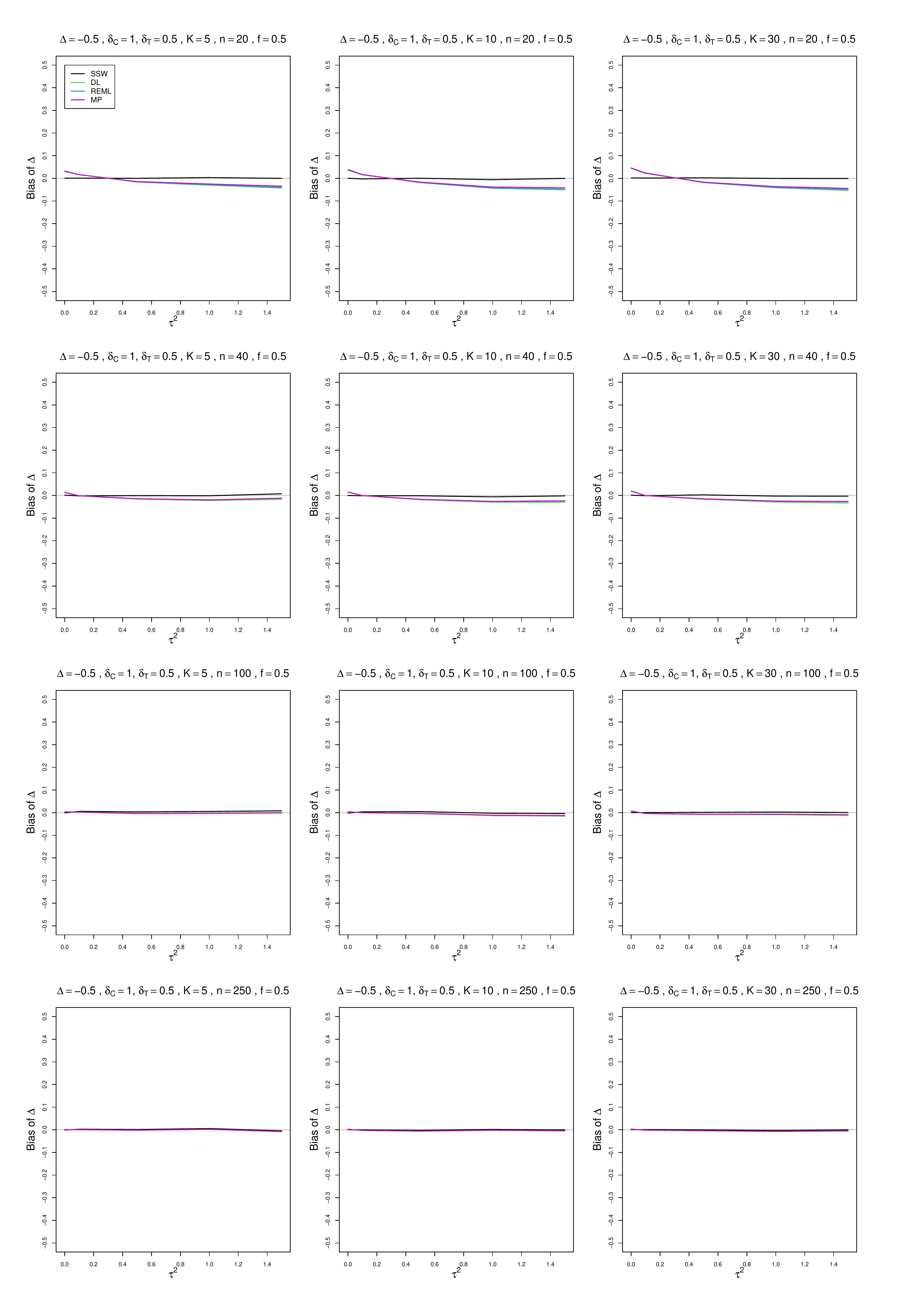}
	\caption{Bias  of estimators of overall effect measure $\Delta$ (DL, REML, MP and SSW) vs $\tau^2$, for equal sample sizes $n=20,\;40,\;100$ and $250$, $\delta_{iC} = 1$, $\Delta=-0.5$ and  $f = 0.5$.   }
	\label{PlotBiasOfDelta_deltaC_1deltaT=0.5_DSM_equal_sample_sizes.pdf}
\end{figure}

\begin{figure}[ht]
	\centering
	\includegraphics[scale=0.33]{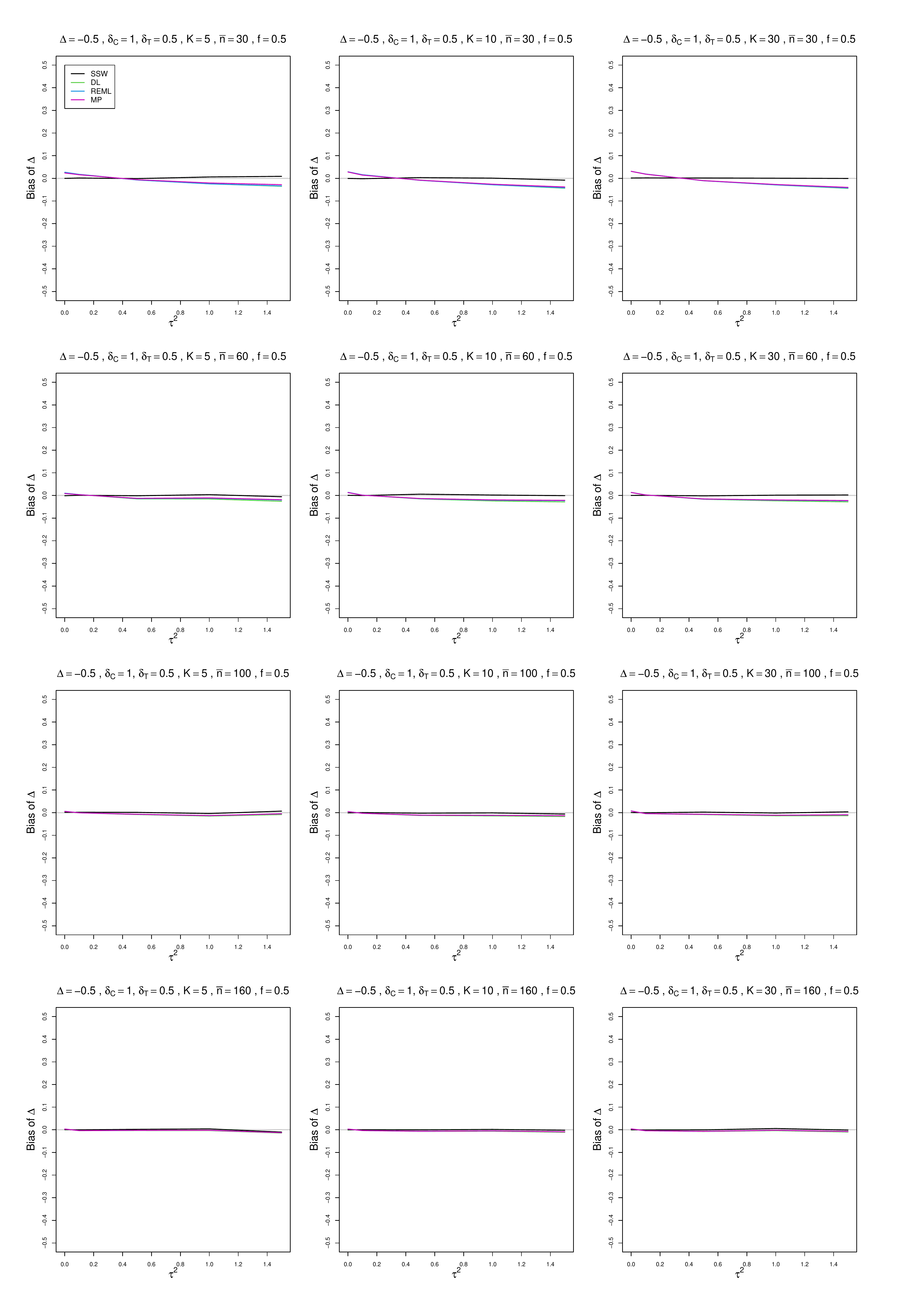}
	\caption{Bias  of estimators of overall effect measure $\Delta$ (DL, REML, MP and SSW) vs $\tau^2$, for unequal sample sizes $\bar{n}=30,\;60,\;100$ and $160$, $\delta_{iC} = 1$, $\Delta=-0.5$ and  $f = 0.5$.   }
	\label{PlotBiasOfDelta_deltaC_1deltaT=0.5_DSM_unequal_sample_sizes.pdf}
\end{figure}

\begin{figure}[ht]
	\centering
	\includegraphics[scale=0.33]{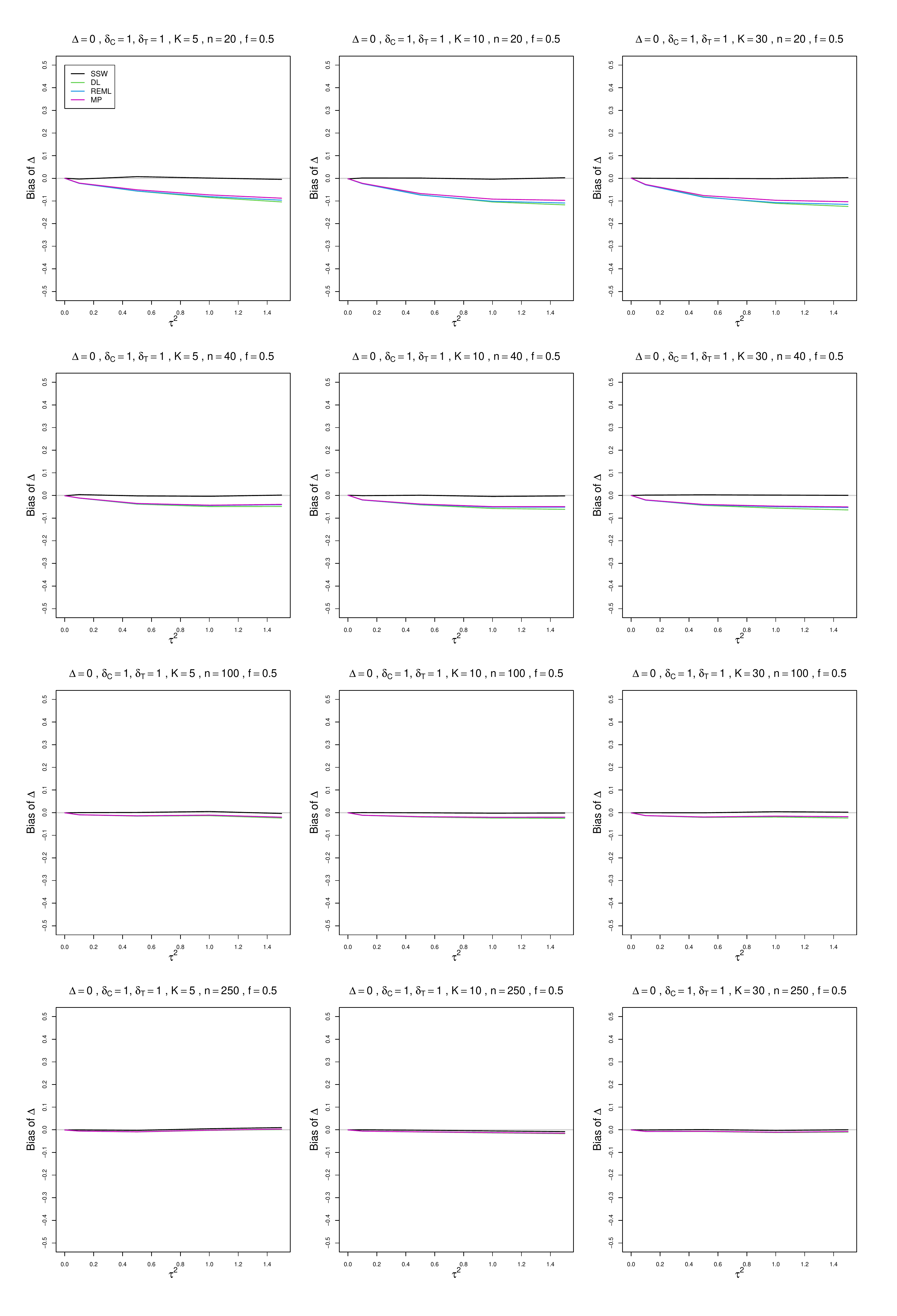}
	\caption{Bias  of estimators of overall effect measure $\Delta$ (DL, REML, MP and SSW) vs $\tau^2$, for equal sample sizes $n=20,\;40,\;100$ and $250$, $\delta_{iC} = 1$, $\Delta=0$ and  $f = 0.5$.   }
	\label{PlotBiasOfDelta_deltaC_1deltaT=1_DSM_equal_sample_sizes.pdf}
\end{figure}

\begin{figure}[ht]
	\centering
	\includegraphics[scale=0.33]{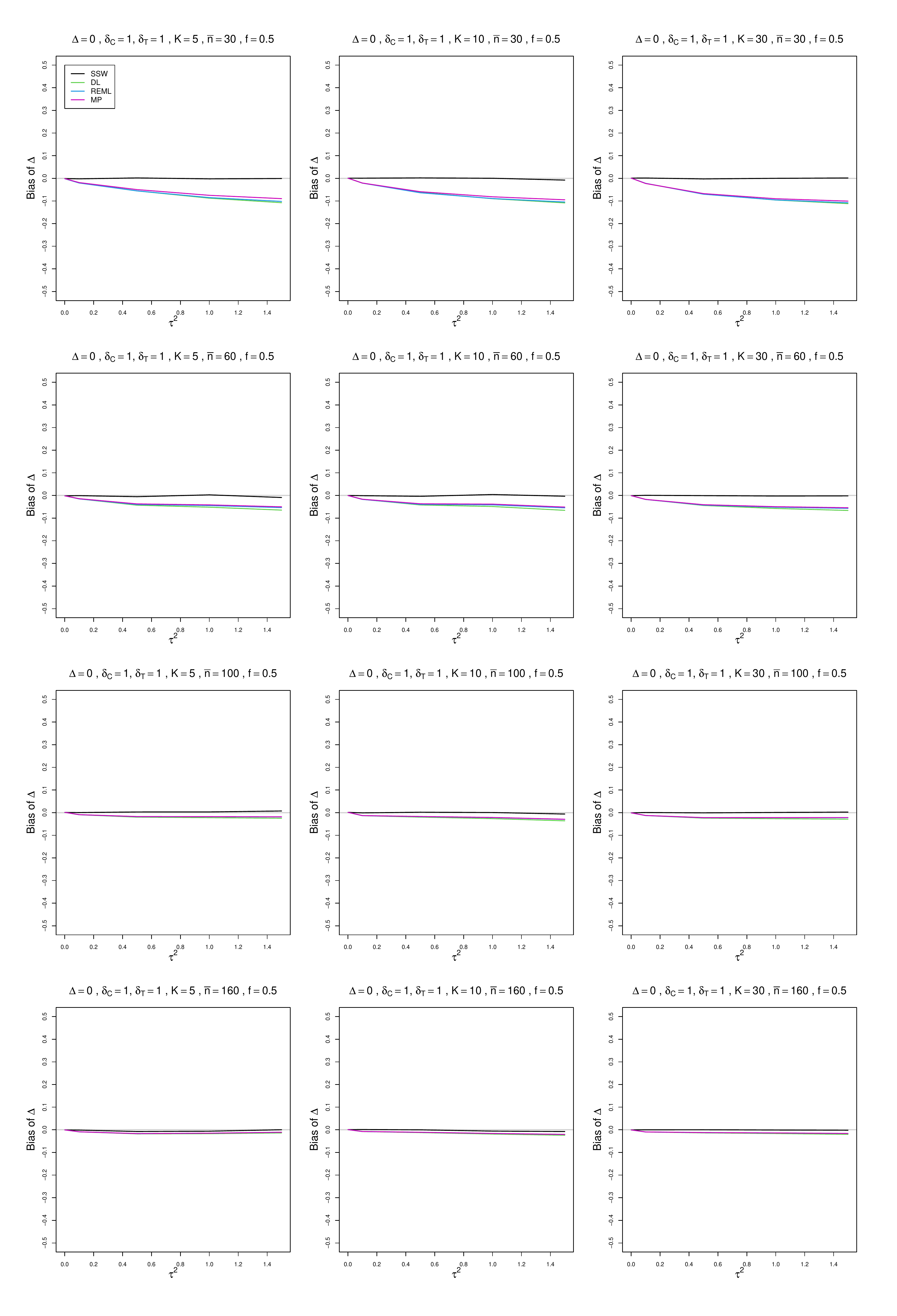}
	\caption{Bias  of estimators of overall effect measure $\Delta$ (DL, REML, MP and SSW) vs $\tau^2$, for unequal sample sizes $\bar{n}=30,\;60,\;100$ and $160$, $\delta_{iC} = 1$, $\Delta=0$ and  $f = 0.5$.   }
	\label{PlotBiasOfDelta_deltaC_1deltaT=1_DSM_unequal_sample_sizes.pdf}
\end{figure}

\begin{figure}[ht]
	\centering
	\includegraphics[scale=0.33]{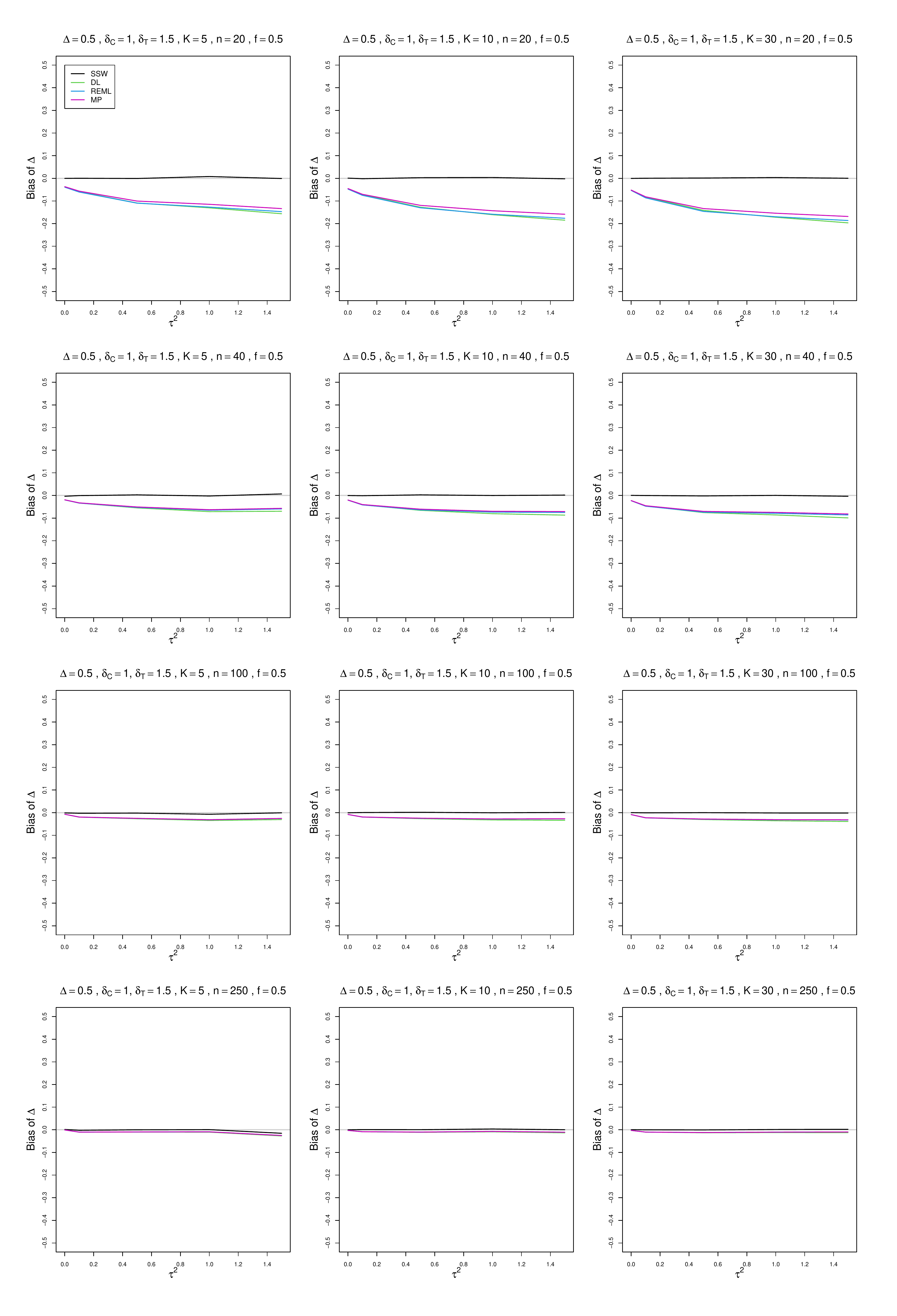}
	\caption{Bias  of estimators of overall effect measure $\Delta$ (DL, REML, MP and SSW) vs $\tau^2$, for equal sample sizes $n=20,\;40,\;100$ and $250$, $\delta_{iC} = 1$, $\Delta=0.5$ and  $f = 0.5$.   }
	\label{PlotBiasOfDelta_deltaC_1deltaT=1.5_DSM_equal_sample_sizes.pdf}
\end{figure}

\begin{figure}[ht]
	\centering
	\includegraphics[scale=0.33]{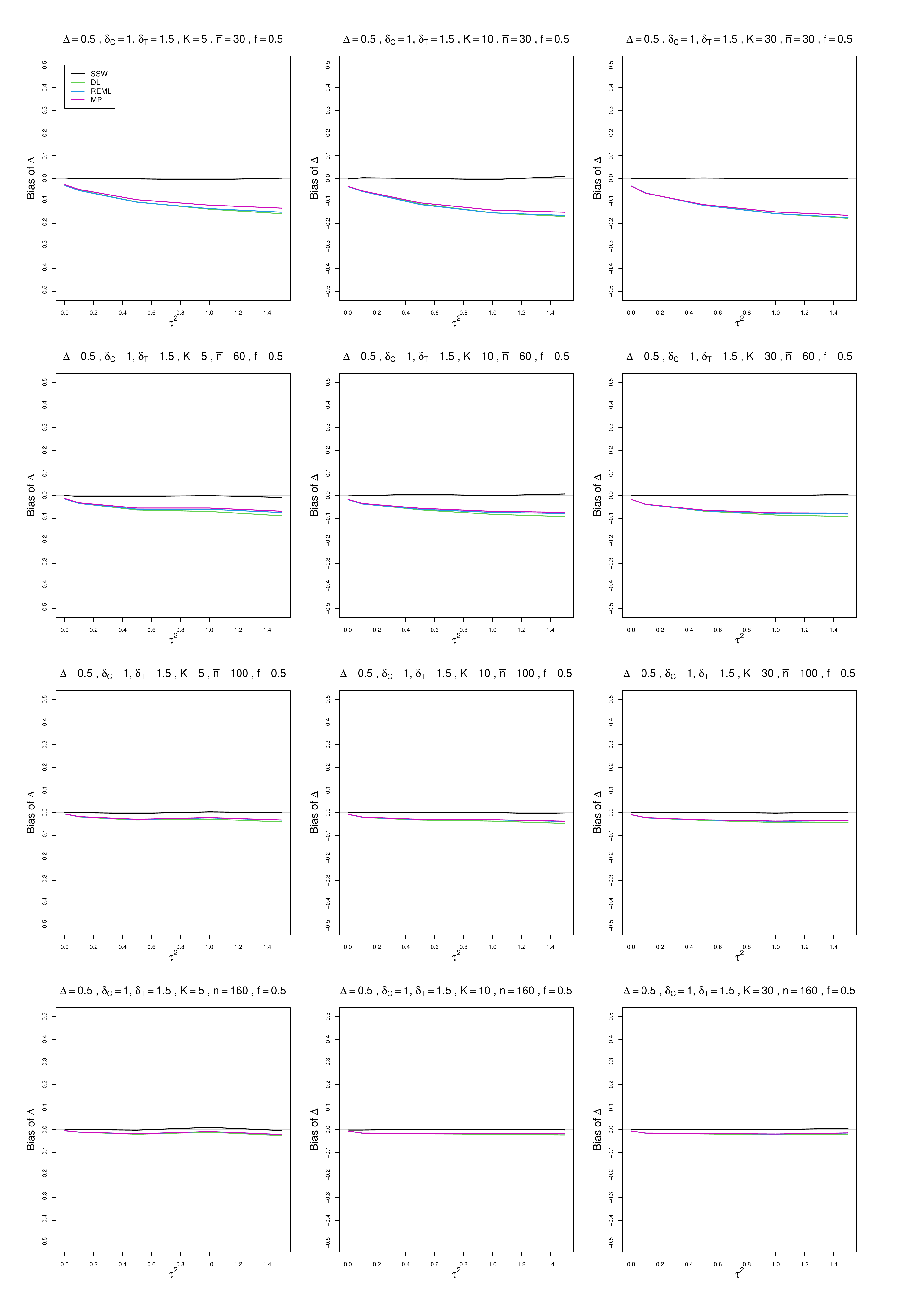}
	\caption{Bias  of estimators of overall effect measure $\Delta$ (DL, REML, MP and SSW) vs $\tau^2$, for unequal sample sizes $\bar{n}=30,\;60,\;100$ and $160$, $\delta_{iC} = 1$, $\Delta=0.5$ and  $f = 0.5$.   }
	\label{PlotBiasOfDelta_deltaC_1deltaT=1.5_DSM_unequal_sample_sizes.pdf}
\end{figure}

\begin{figure}[ht]
	\centering
	\includegraphics[scale=0.33]{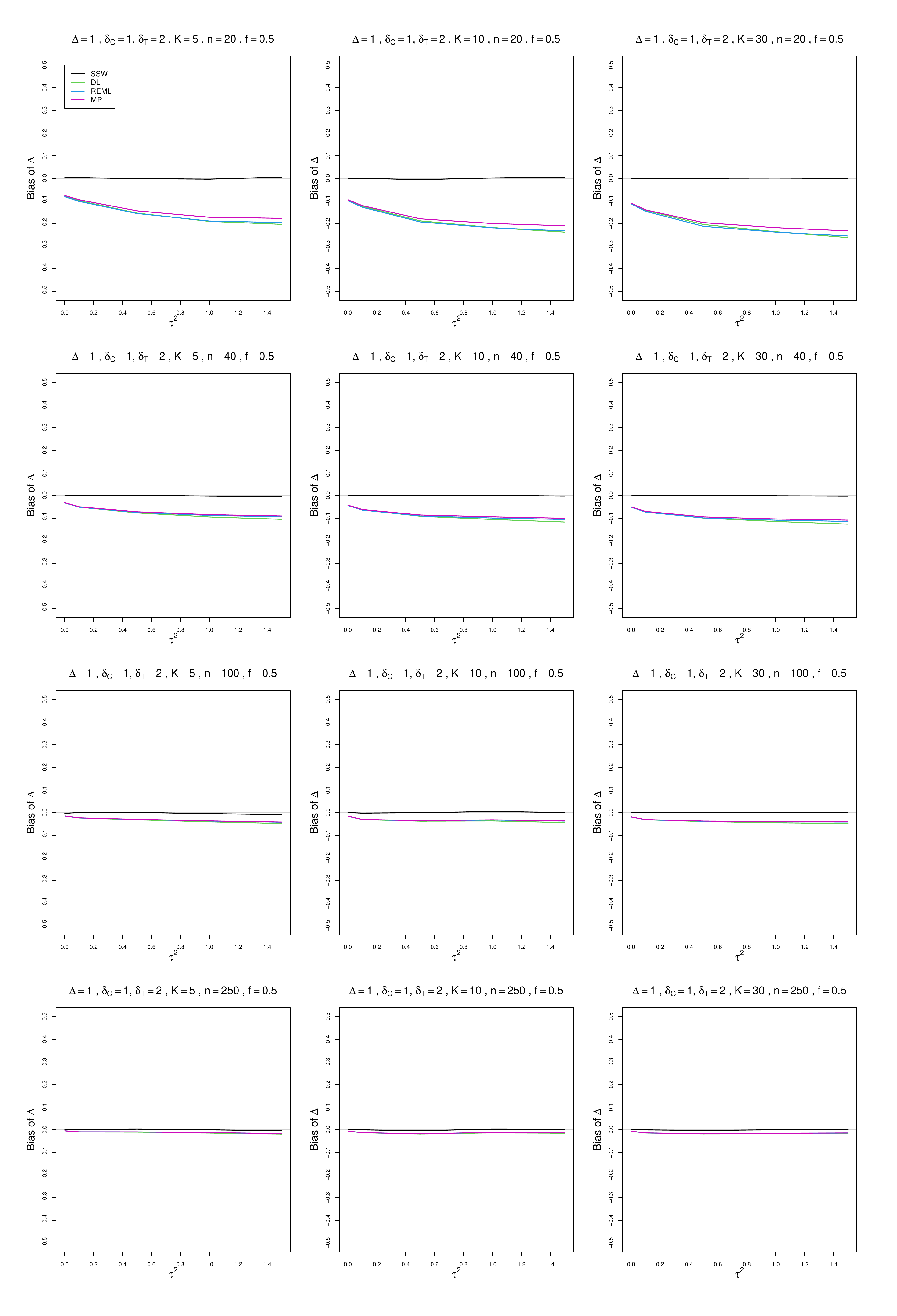}
	\caption{Bias  of estimators of overall effect measure $\Delta$ (DL, REML, MP and SSW) vs $\tau^2$, for equal sample sizes $n=20,\;40,\;100$ and $250$, $\delta_{iC} = 1$, $\Delta=1$ and  $f = 0.5$.   }
	\label{PlotBiasOfDelta_deltaC_=1deltaT=2_DSM_equal_sample_sizes.pdf}
\end{figure}

\begin{figure}[ht]
	\centering
	\includegraphics[scale=0.33]{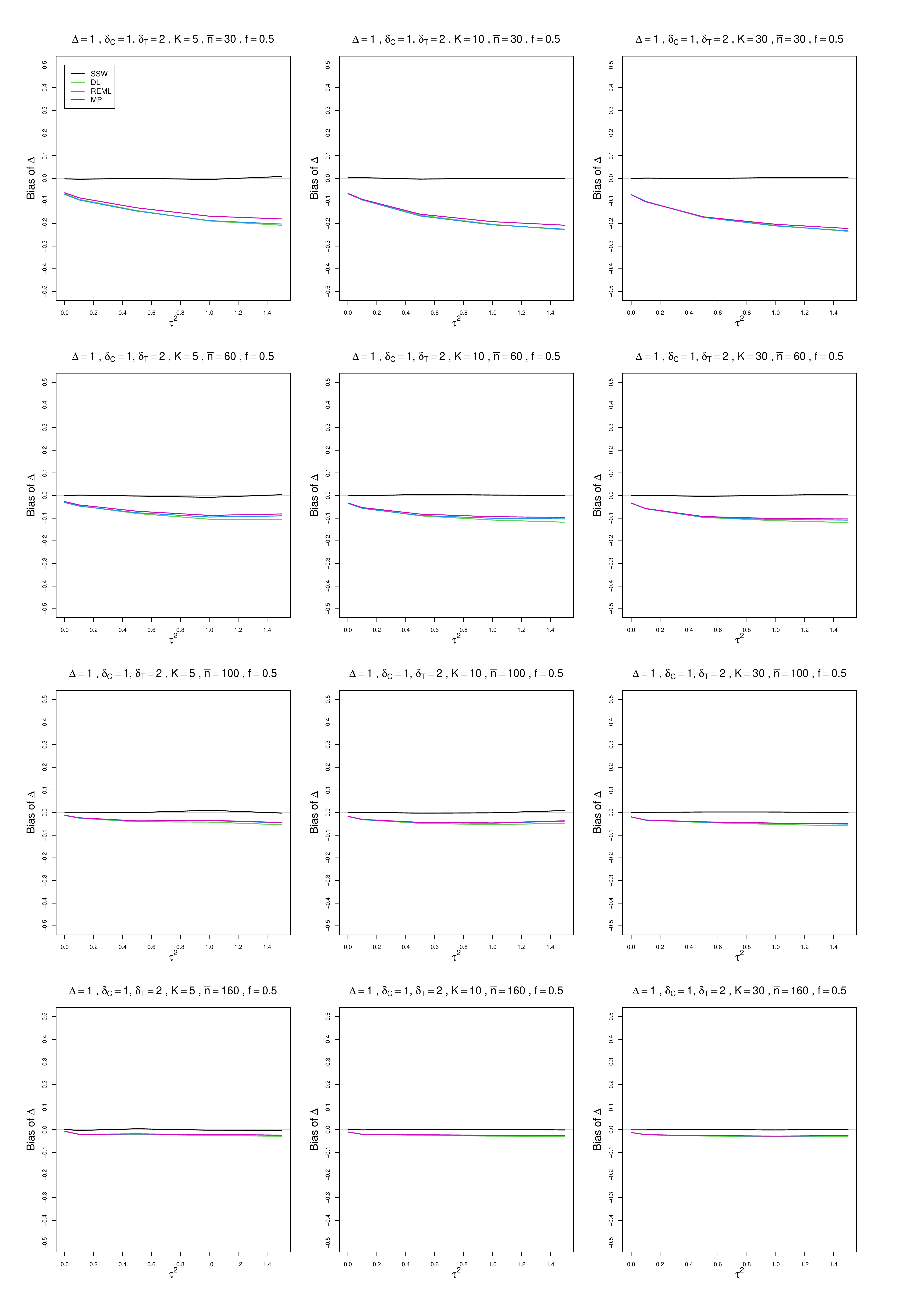}
	\caption{Bias  of estimators of overall effect measure $\Delta$ (DL, REML, MP and SSW) vs $\tau^2$, for unequal sample sizes $\bar{n}=30,\;60,\;100$ and $160$, $\delta_{iC} = 1$, $\Delta=1$ and  $f = 0.5$.   }
	\label{PlotBiasOfDelta_deltaC_1deltaT=2_DSM_unequal_sample_sizes.pdf}
\end{figure}

\begin{figure}[ht]
	\centering
	\includegraphics[scale=0.33]{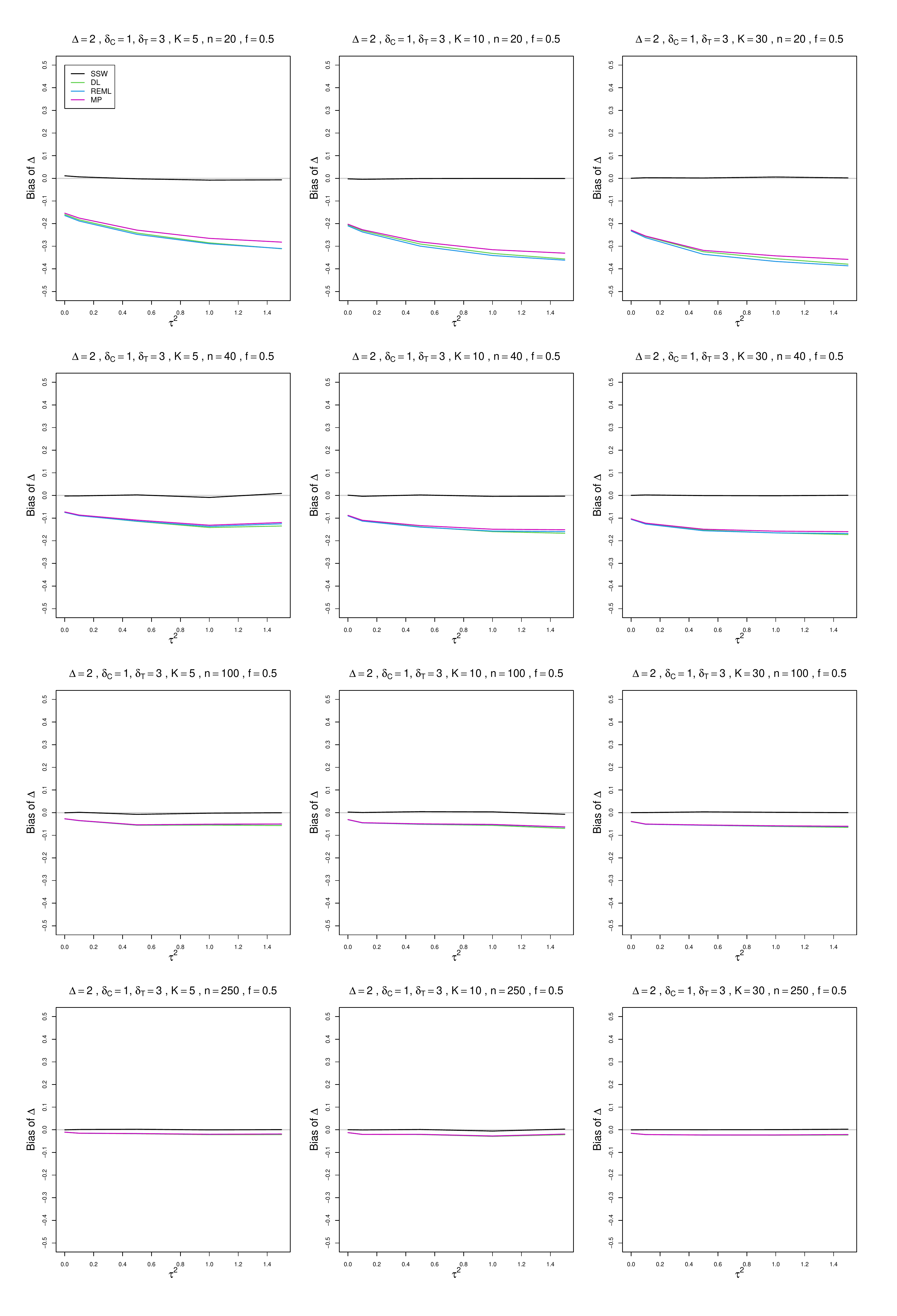}
	\caption{Bias  of estimators of overall effect measure $\Delta$ (DL, REML, MP and SSW) vs $\tau^2$, for equal sample sizes $n=20,\;40,\;100$ and $250$, $\delta_{iC} = 1$, $\Delta=2$ and  $f = 0.5$.   }
	\label{PlotBiasOfDelta_deltaC_1deltaT=3_DSM_equal_sample_sizes.pdf}
\end{figure}

\begin{figure}[ht]
	\centering
	\includegraphics[scale=0.33]{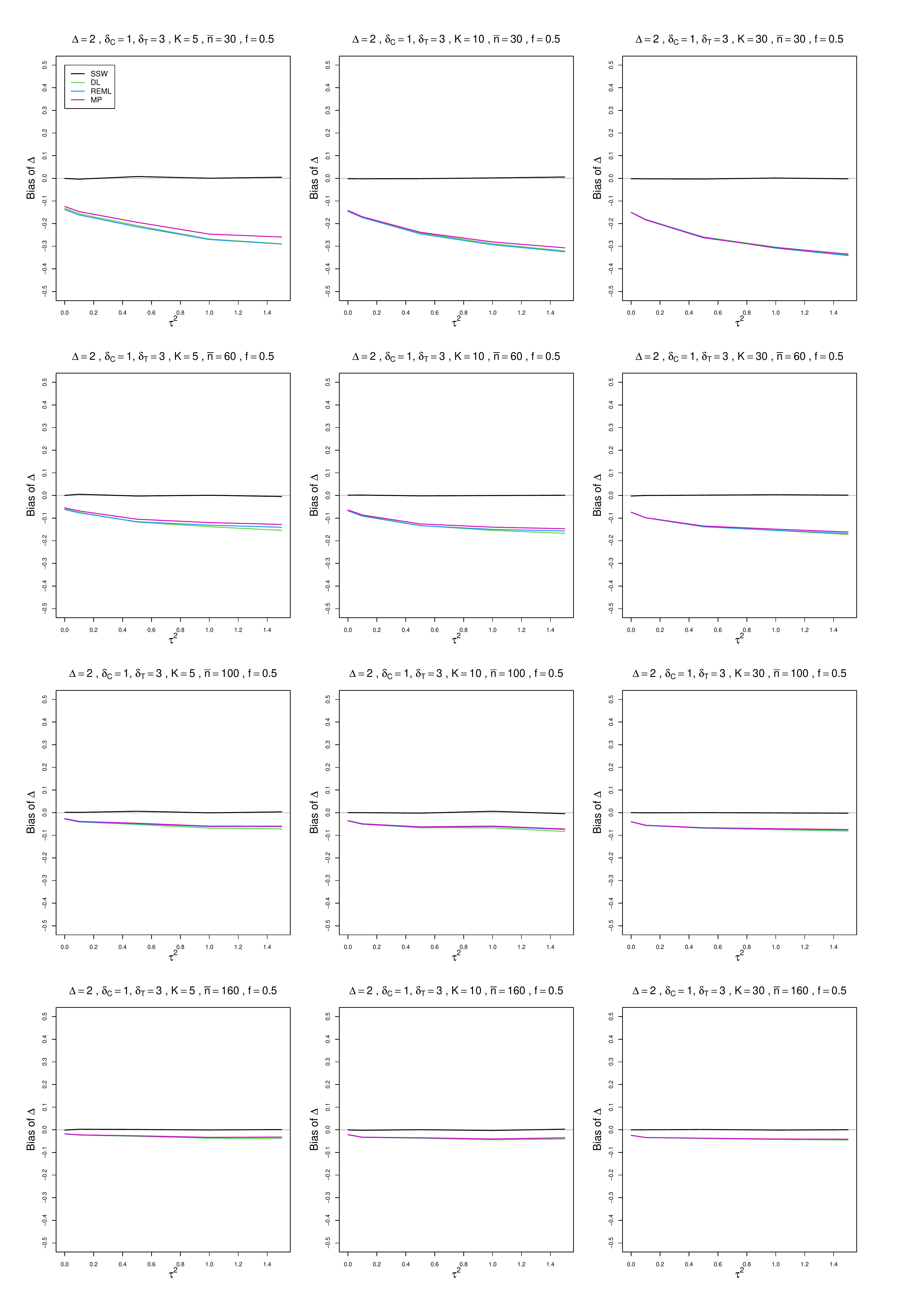}
	\caption{Bias  of estimators of overall effect measure $\Delta$ (DL, REML, MP and SSW) vs $\tau^2$, for unequal sample sizes $\bar{n}=30,\;60,\;100$ and $160$, $\delta_{iC} = 1$, $\Delta=2$ and  $f = 0.5$.   }
	\label{PlotBiasOfDelta_deltaC_1deltaT=3_DSM_unequal_sample_sizes.pdf}
\end{figure}


\begin{figure}[ht]
	\centering
	\includegraphics[scale=0.33]{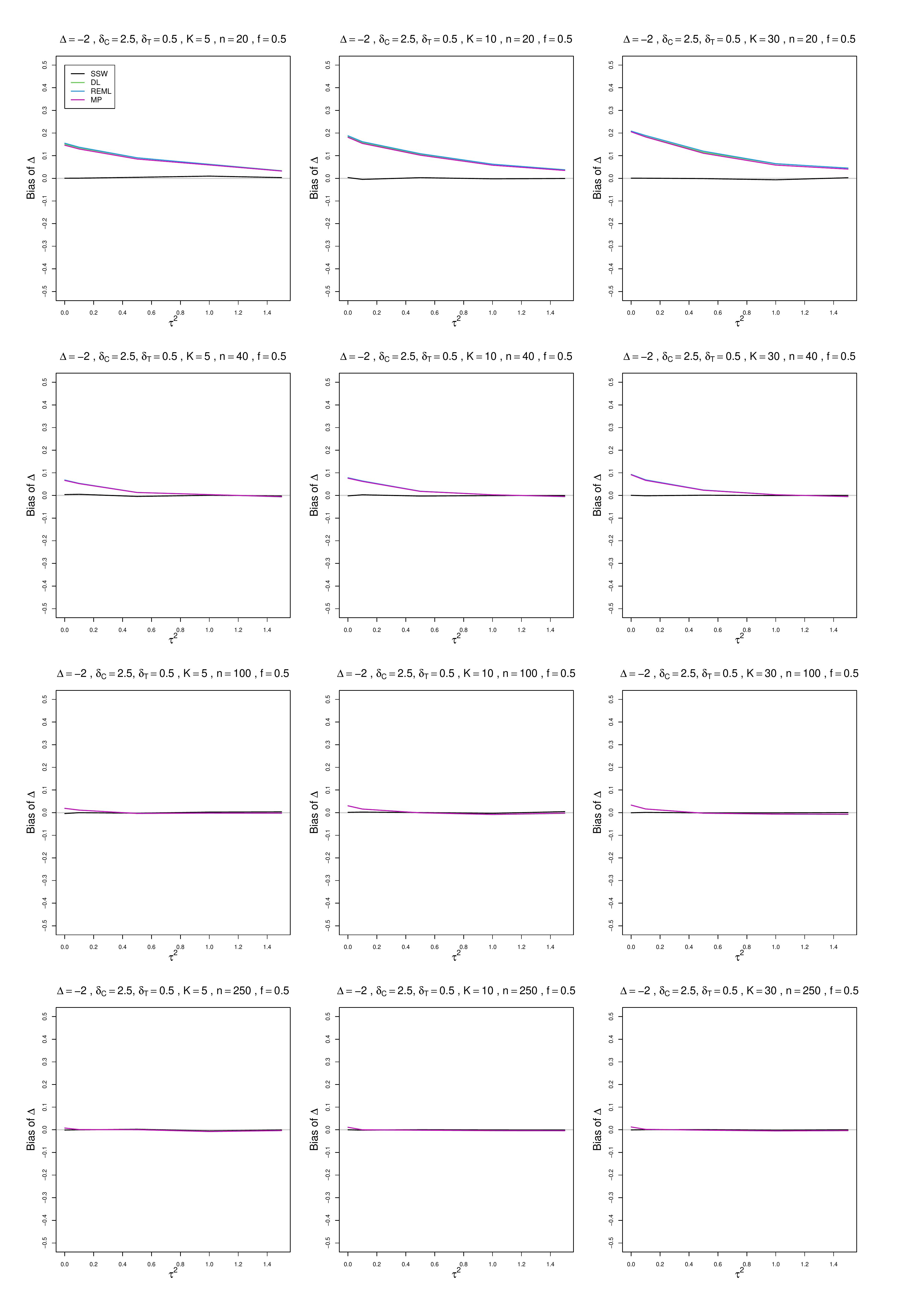}
	\caption{Bias  of estimators of overall effect measure $\Delta$ (DL, REML, MP and SSW ) vs $\tau^2$, for equal sample sizes $n=20,\;40,\;100$ and $250$, $\delta_{iC} = 2.5$, $\Delta=-2$ and  $f = 0.5$.   }
	\label{PlotBiasOfDelta_deltaC_2.5deltaT=0.5_DSM_equal_sample_sizes.pdf}
\end{figure}

\begin{figure}[ht]
	\centering
	\includegraphics[scale=0.33]{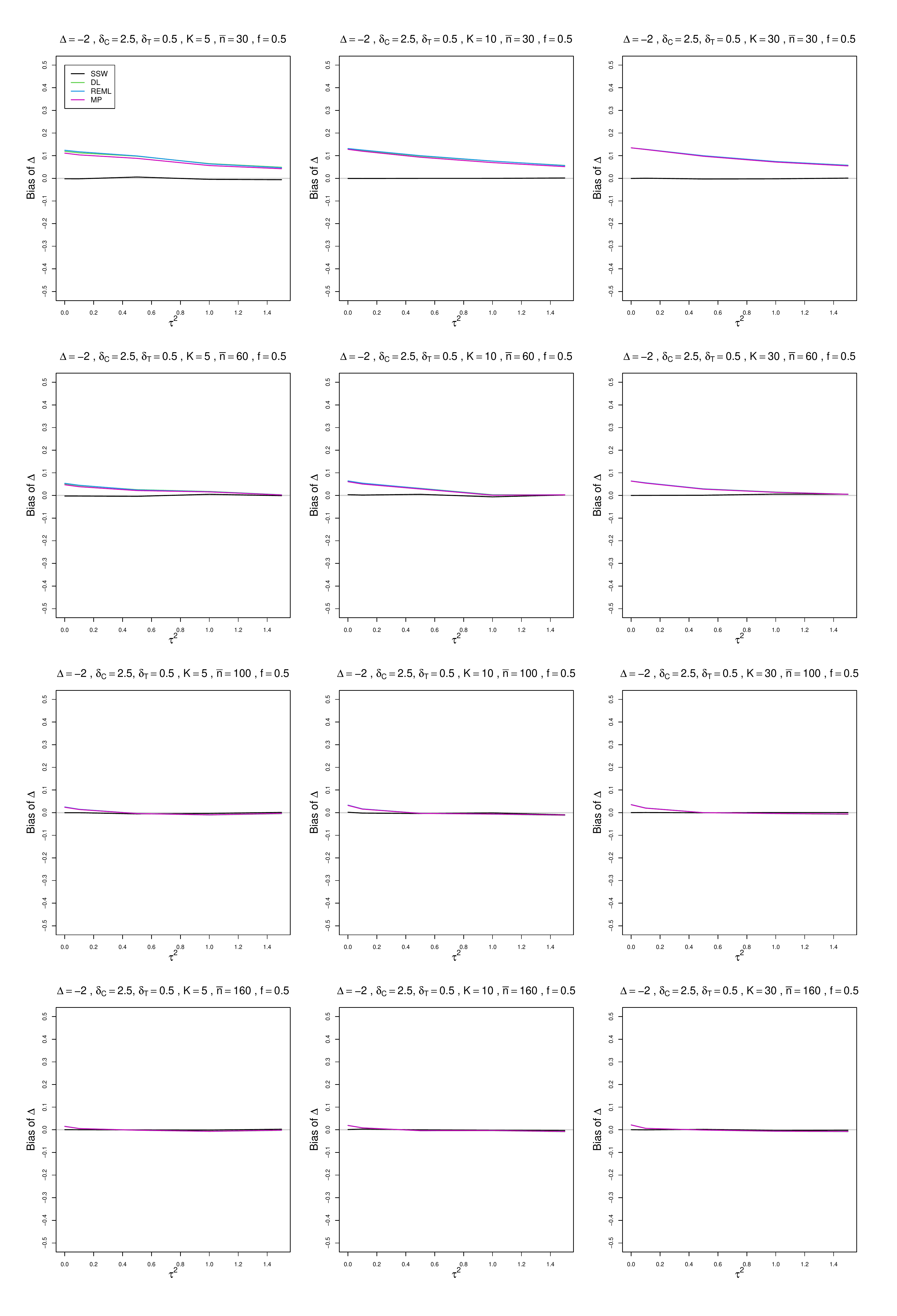}
	\caption{Bias  of estimators of overall effect measure $\Delta$ (DL, REML, MP and SSW) vs $\tau^2$, for unequal sample sizes $\bar{n}=30,\;60,\;100$ and $160$, $\delta_{iC} = 2.5$, $\Delta=-2$ and  $f = 0.5$.   }
	\label{PlotBiasOfDelta_deltaC_2.5deltaT=0.5_DSM_unequal_sample_sizes.pdf}
\end{figure}

\begin{figure}[ht]
	\centering
	\includegraphics[scale=0.33]{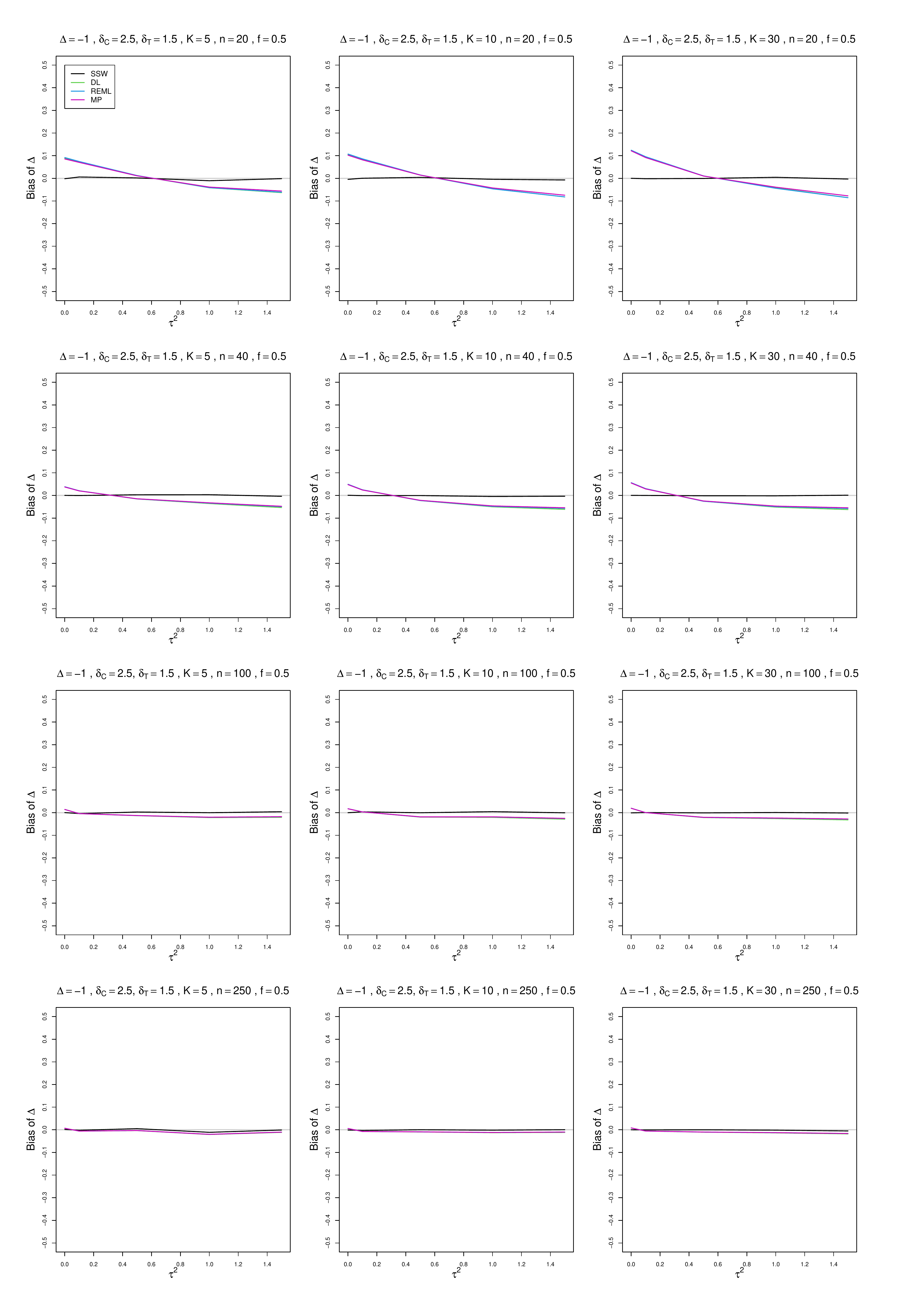}
	\caption{Bias  of estimators of overall effect measure $\Delta$ (DL, REML, MP and SSW) vs $\tau^2$, for equal sample sizes $n=20,\;40,\;100$ and $250$, $\delta_{iC} = 2.5$, $\Delta=-1$ and  $f = 0.5$.   }
	\label{PlotBiasOfDelta_deltaC_2.5deltaT=1.5_DSM_equal_sample_sizes.pdf}
\end{figure}

\begin{figure}[ht]
	\centering
	\includegraphics[scale=0.33]{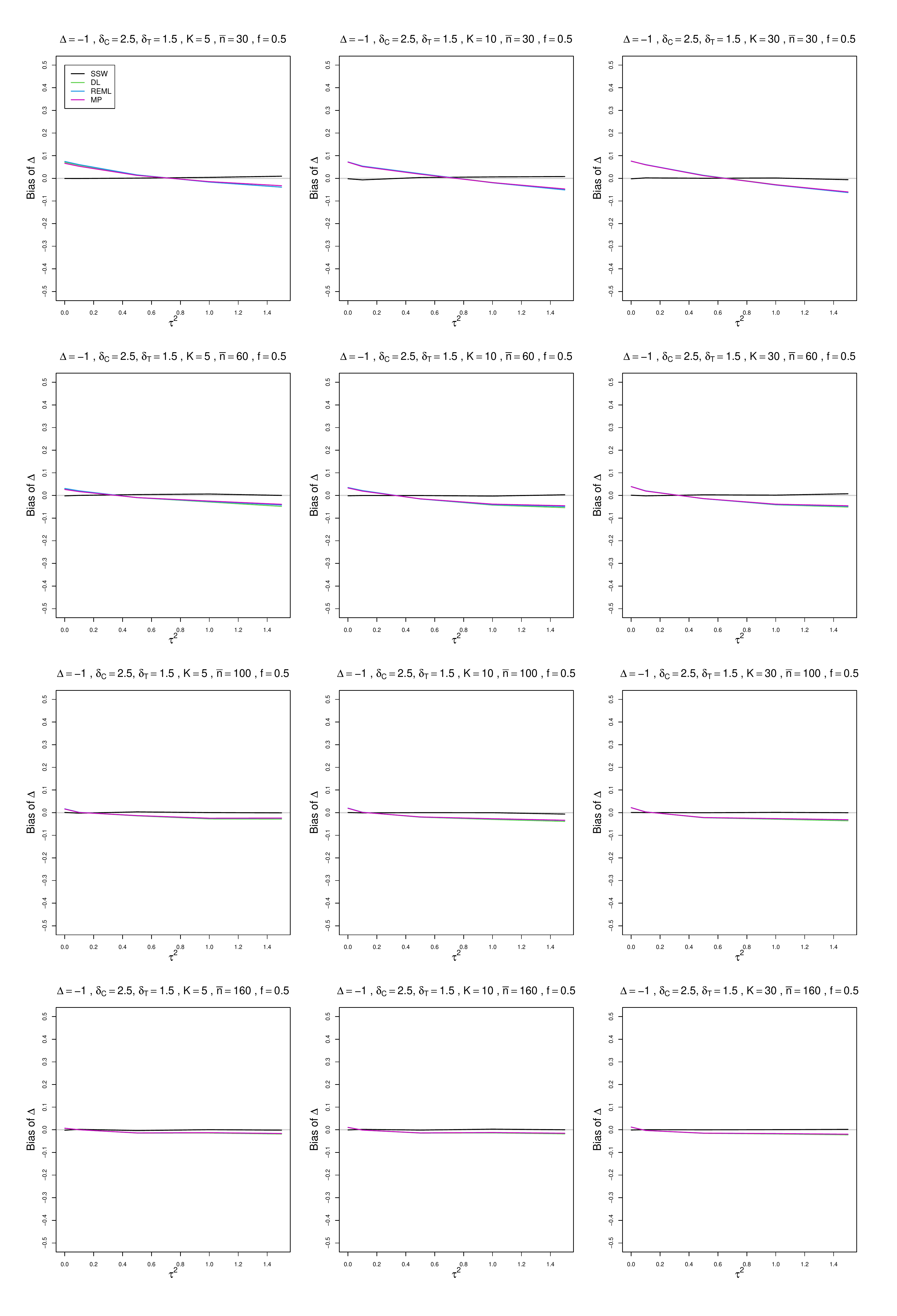}
	\caption{Bias  of estimators of overall effect measure $\Delta$ (DL, REML, MP and SSW) vs $\tau^2$, for unequal sample sizes $\bar{n}=30,\;60,\;100$ and $160$, $\delta_{iC} = 2.5$, $\Delta=-1$ and  $f = 0.5$.   }
	\label{PlotBiasOfDelta_deltaC_2.5deltaT=1.5_DSM_unequal_sample_sizes.pdf}
\end{figure}

\begin{figure}[ht]
	\centering
	\includegraphics[scale=0.33]{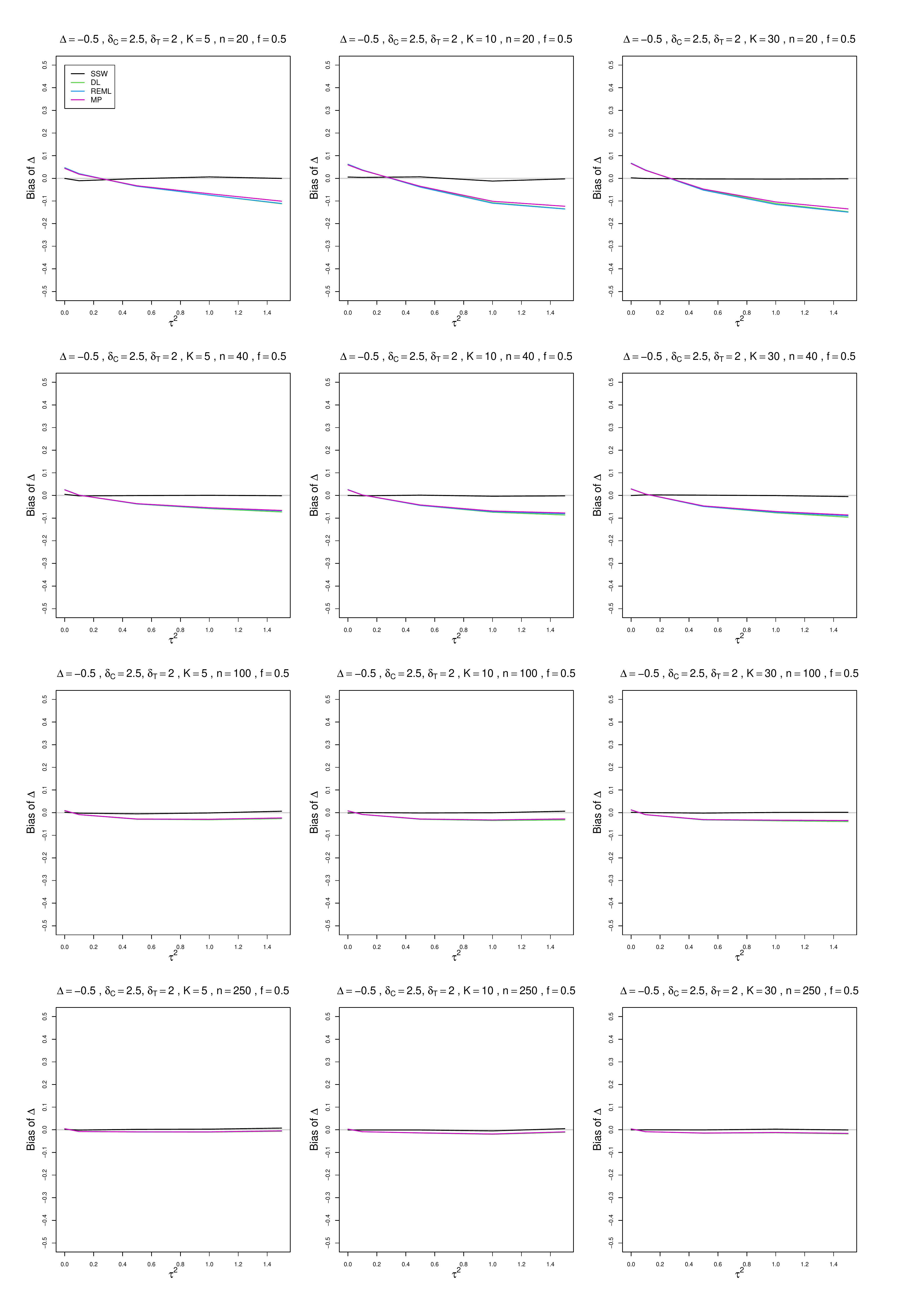}
	\caption{Bias  of estimators of overall effect measure $\Delta$ (DL, REML, MP and SSW) vs $\tau^2$, for equal sample sizes $n=20,\;40,\;100$ and $250$, $\delta_{iC} = 2.5$, $\Delta=-0.5$ and  $f = 0.5$.   }
	\label{PlotBiasOfDelta_deltaC_2.5deltaT=2_DSM_equal_sample_sizes.pdf}
\end{figure}

\begin{figure}[ht]
	\centering
	\includegraphics[scale=0.33]{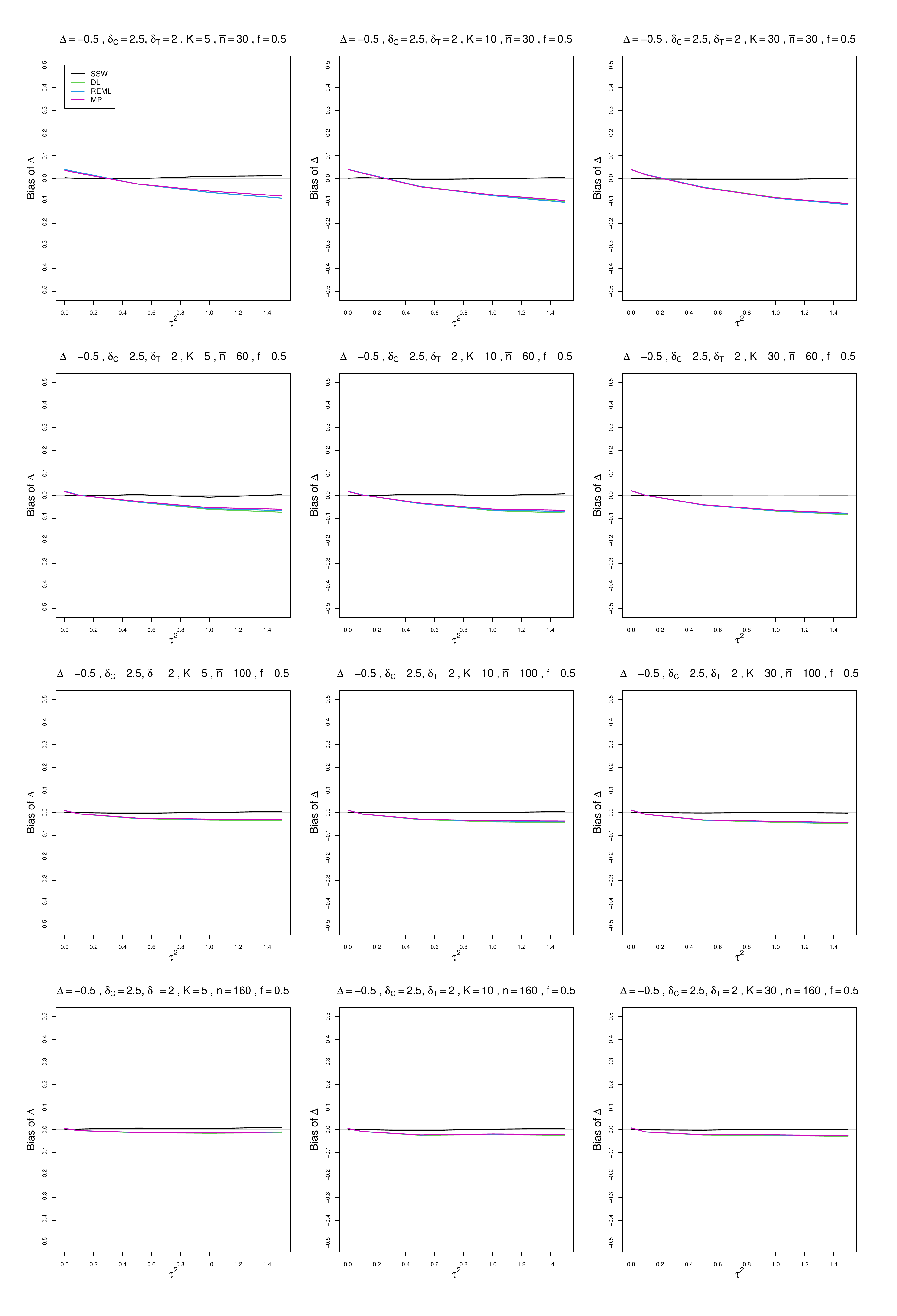}
	\caption{Bias  of estimators of overall effect measure $\Delta$ (DL, REML, MP and SSW) vs $\tau^2$, for unequal sample sizes $\bar{n}=30,\;60,\;100$ and $160$, $\delta_{iC} = 2.5$, $\Delta=-0.5$ and  $f = 0.5$.   }
	\label{PlotBiasOfDelta_deltaC_2.5deltaT=2_DSM_unequal_sample_sizes.pdf}
\end{figure}

\begin{figure}[ht]
	\centering
	\includegraphics[scale=0.33]{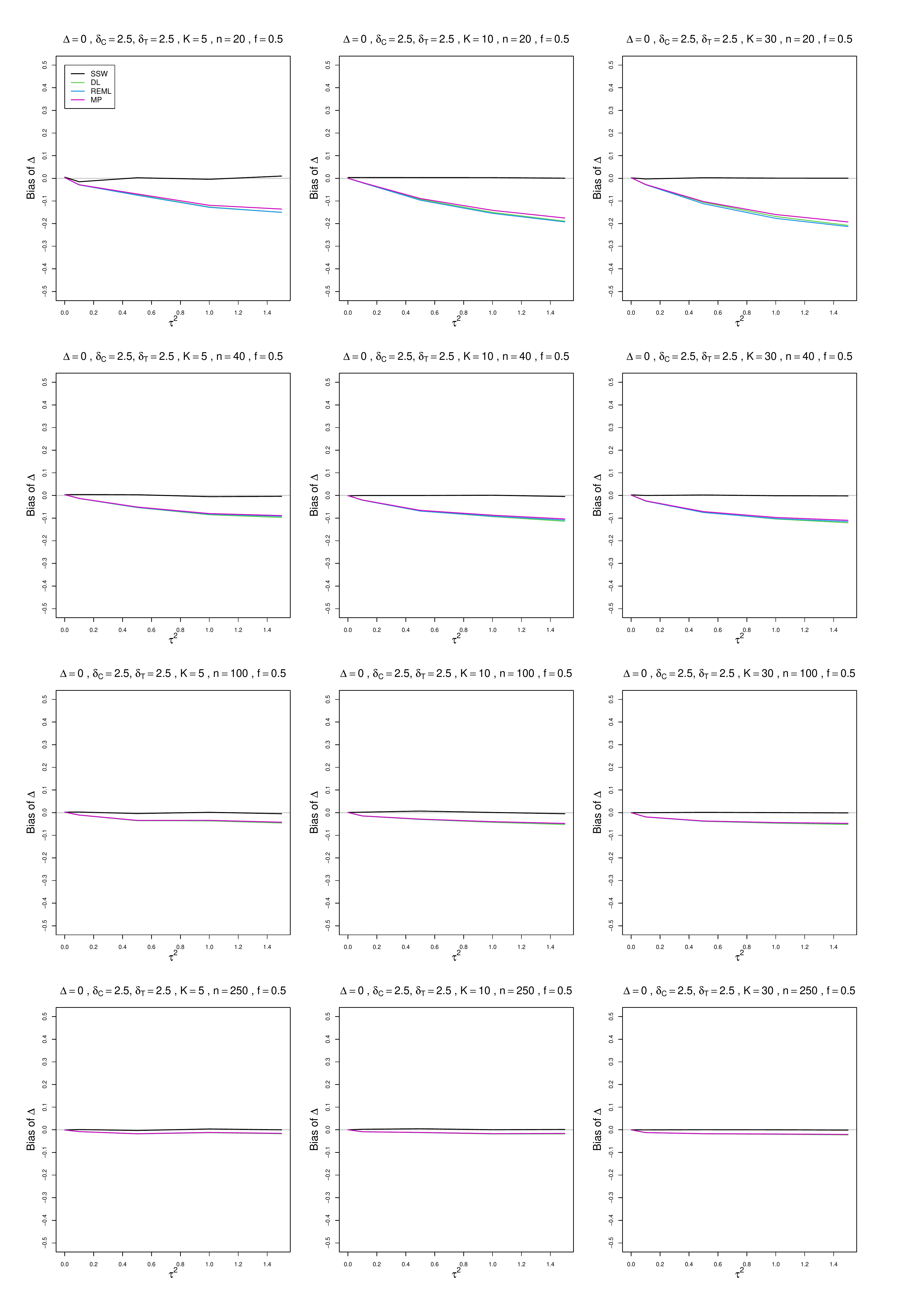}
	\caption{Bias  of estimators of overall effect measure $\Delta$ (DL, REML, MP and SSW) vs $\tau^2$, for equal sample sizes $n=20,\;40,\;100$ and $250$, $\delta_{iC} = 2.5$, $\Delta=0$ and  $f = 0.5$.   }
	\label{PlotBiasOfDelta_deltaC_2.5deltaT=2.5_DSM_equal_sample_sizes.pdf}
\end{figure}

\begin{figure}[ht]
	\centering
	\includegraphics[scale=0.33]{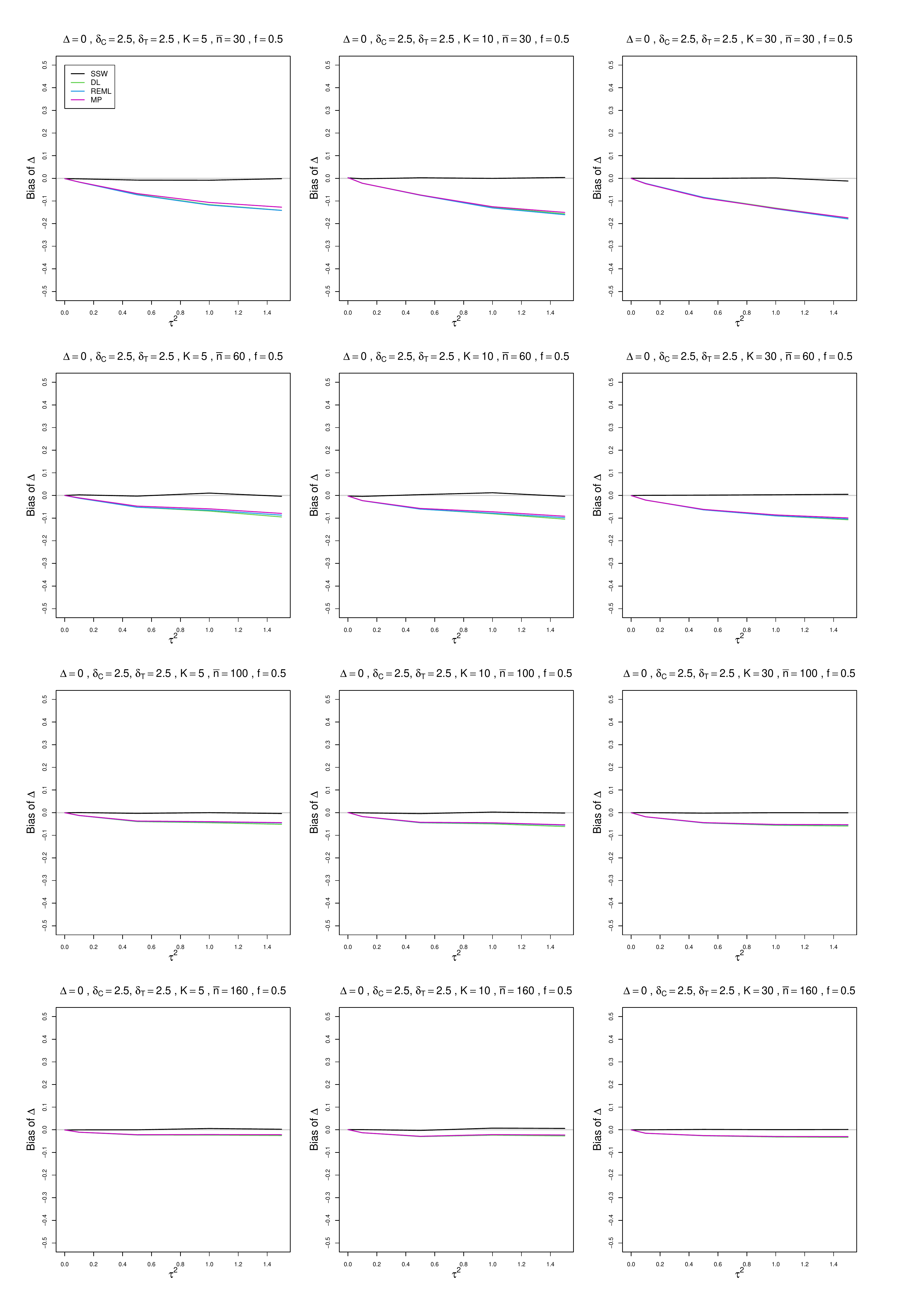}
	\caption{Bias  of estimators of overall effect measure $\Delta$ (DL, REML, MP and SSW ) vs $\tau^2$, for unequal sample sizes $\bar{n}=30,\;60,\;100$ and $160$, $\delta_{iC} = 2.5$, $\Delta=0$ and  $f = 0.5$.   }
	\label{PlotBiasOfDelta_deltaC_2.5deltaT=2.5_DSM_unequal_sample_sizes.pdf}
\end{figure}

\begin{figure}[ht]
	\centering
	\includegraphics[scale=0.33]{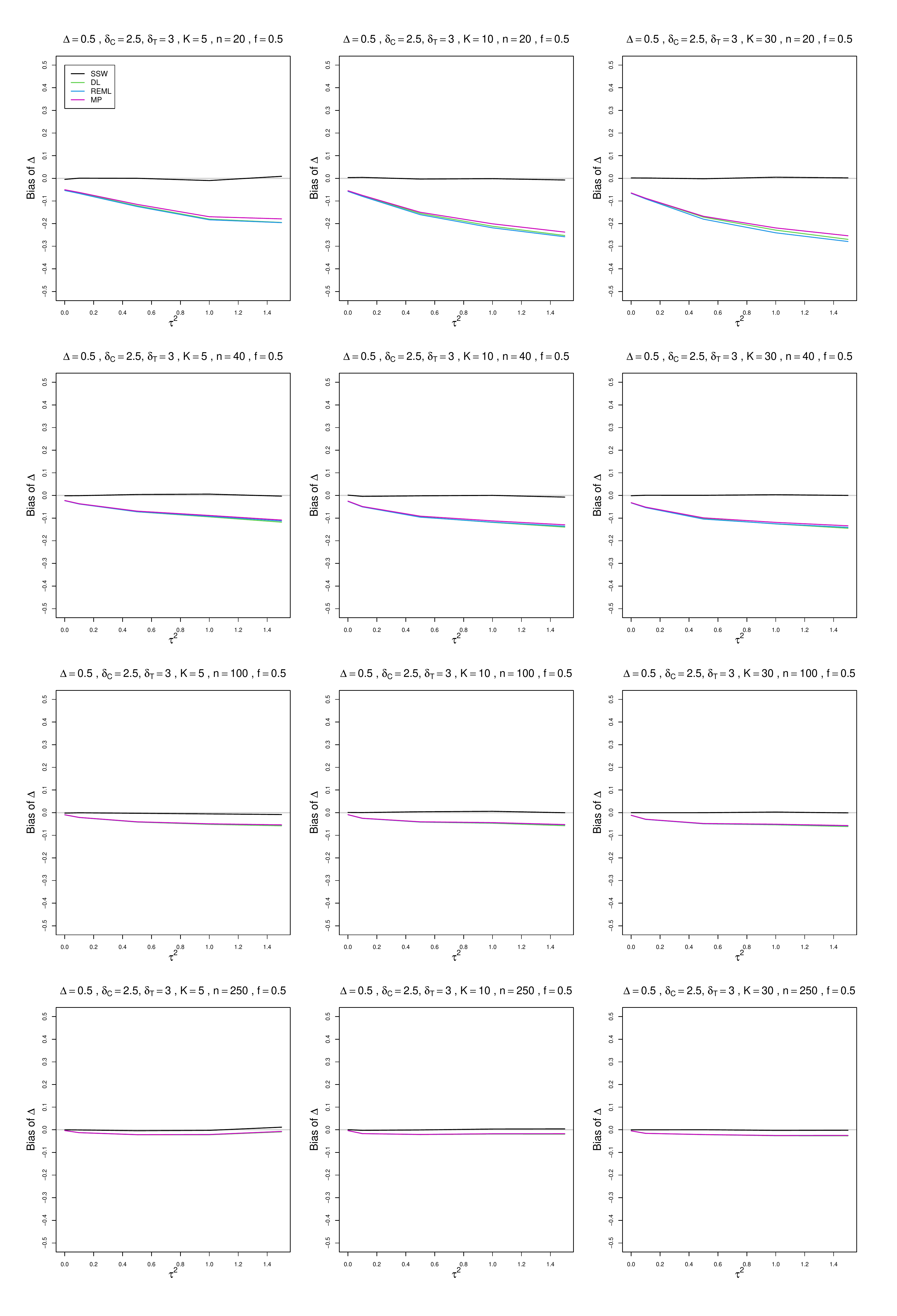}
	\caption{Bias  of estimators of overall effect measure $\Delta$ (DL, REML, MP and SSW) vs $\tau^2$, for equal sample sizes $n=20,\;40,\;100$ and $250$, $\delta_{iC} = 2.5$, $\Delta=0.5$ and  $f = 0.5$.   }
	\label{PlotBiasOfDelta_deltaC_2.5deltaT=3_DSM_equal_sample_sizes.pdf}
\end{figure}

\begin{figure}[ht]
	\centering
	\includegraphics[scale=0.33]{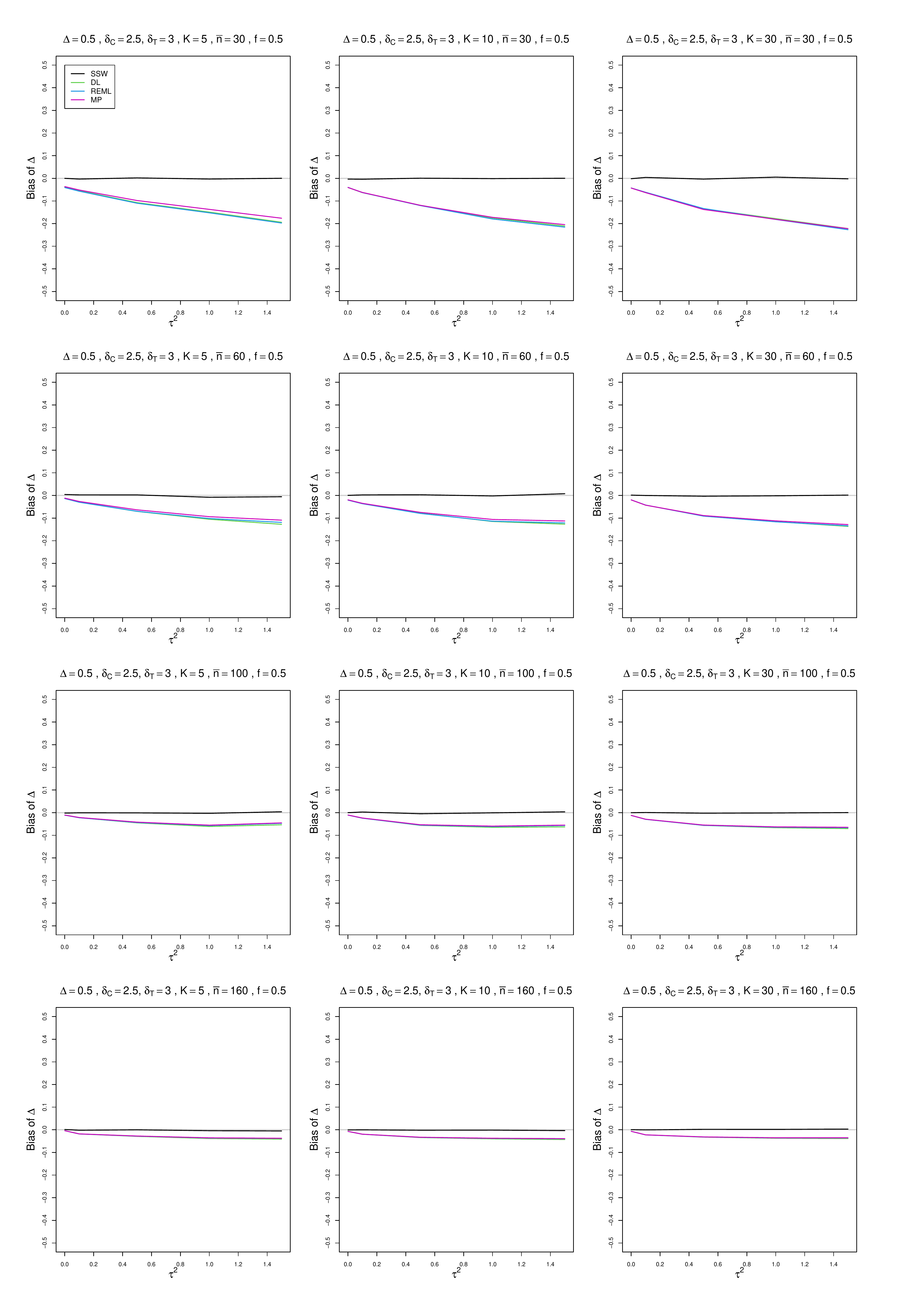}
	\caption{Bias  of estimators of overall effect measure $\Delta$ (DL, REML, MP and SSW) vs $\tau^2$, for unequal sample sizes $\bar{n}=30,\;60,\;100$ and $160$, $\delta_{iC} = 2.5$, $\Delta=0.5$ and  $f = 0.5$.   }
	\label{PlotBiasOfDelta_deltaC_2.5deltaT=3_DSM_unequal_sample_sizes.pdf}
\end{figure}

\begin{figure}[ht]
	\centering
	\includegraphics[scale=0.33]{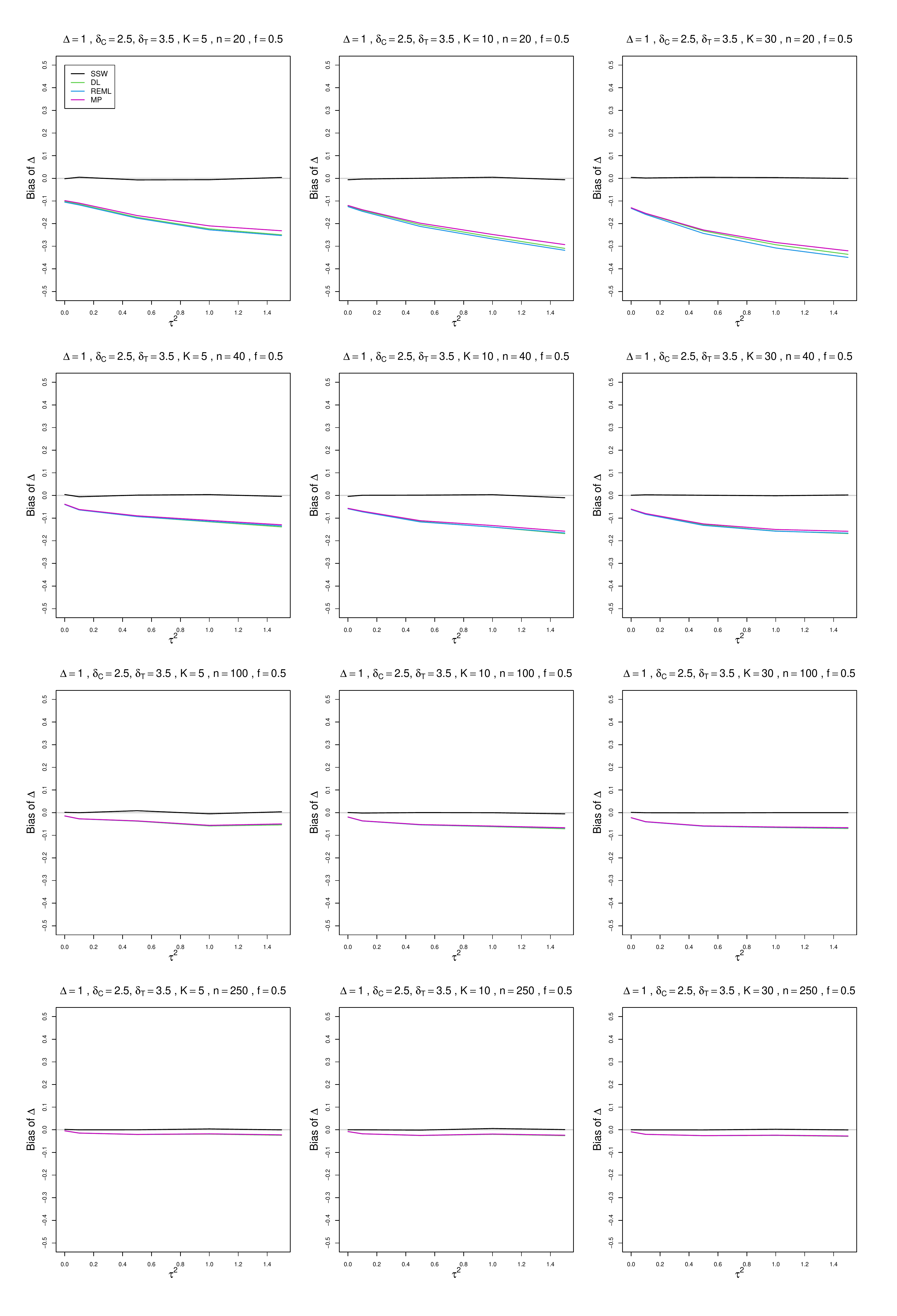}
	\caption{Bias  of estimators of overall effect measure $\Delta$ (DL, REML, MP and SSW) vs $\tau^2$, for equal sample sizes $n=20,\;40,\;100$ and $250$, $\delta_{iC} = 2.5$, $\Delta=1$ and  $f = 0.5$.   }
	\label{PlotBiasOfDelta_deltaC_2.5deltaT=3.5_DSM_equal_sample_sizes.pdf}
\end{figure}

\begin{figure}[ht]
	\centering
	\includegraphics[scale=0.33]{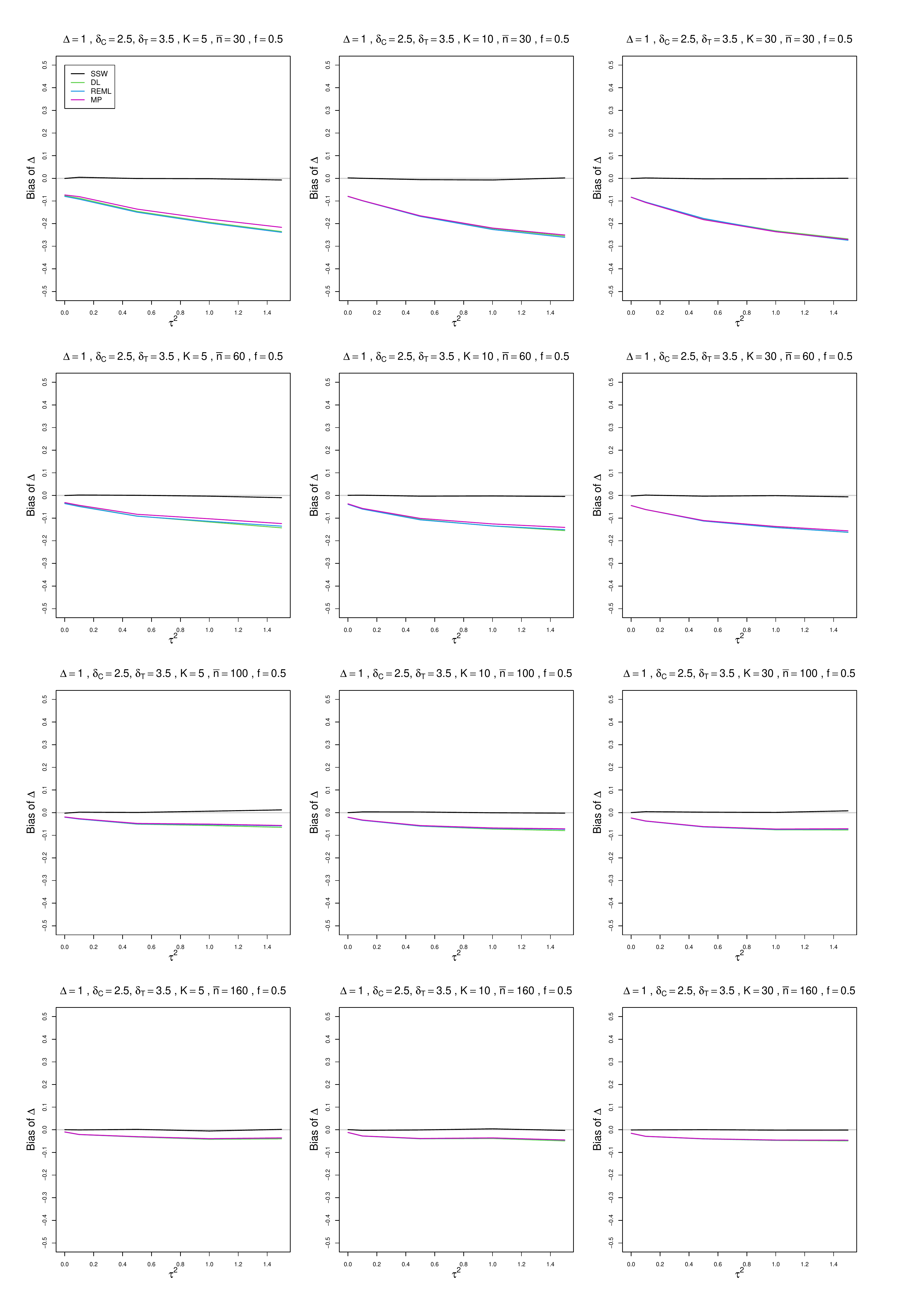}
	\caption{Bias  of estimators of overall effect measure $\Delta$ (DL, REML, MP and SSW) vs $\tau^2$, for unequal sample sizes $\bar{n}=30,\;60,\;100$ and $160$, $\delta_{iC} = 2.5$, $\Delta=1$ and  $f = 0.5$.   }
	\label{PlotBiasOfDelta_deltaC_2.5deltaT=3.5_DSM_unequal_sample_sizes.pdf}
\end{figure}

\begin{figure}[ht]
	\centering
	\includegraphics[scale=0.33]{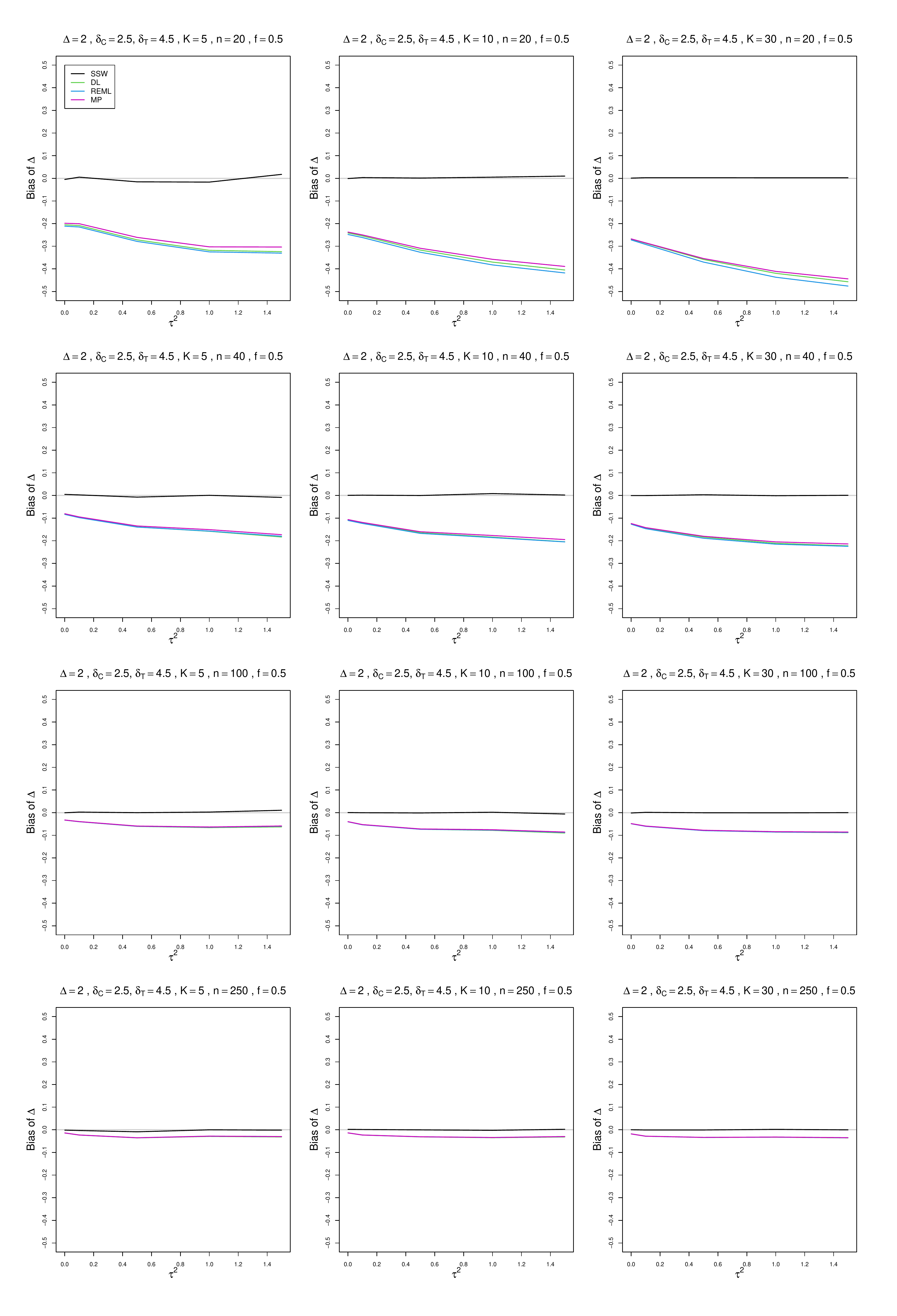}
	\caption{Bias  of estimators of overall effect measure $\Delta$ (DL, REML, MP and SSW) vs $\tau^2$, for equal sample sizes $n=20,\;40,\;100$ and $250$, $\delta_{iC} = 2.5$, $\Delta=2$ and  $f = 0.5$.   }
	\label{PlotBiasOfDelta_deltaC_2.5deltaT=2.5_DSM_equal_sample_sizes.pdf}
\end{figure}

\begin{figure}[ht]
	\centering
	\includegraphics[scale=0.33]{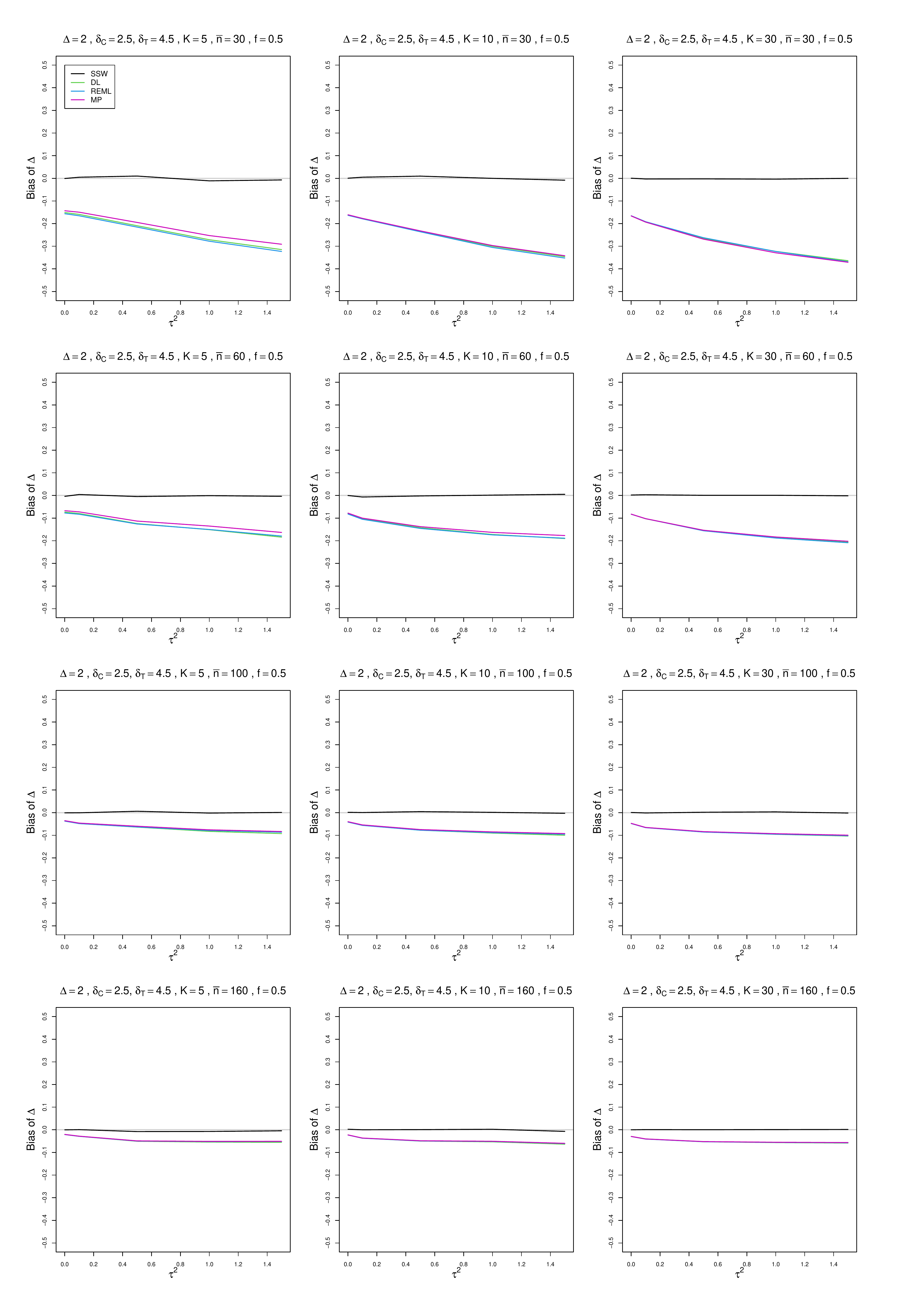}
	\caption{Bias  of estimators of overall effect measure $\Delta$ (DL, REML, MP and SSW) vs $\tau^2$, for unequal sample sizes $\bar{n}=30,\;60,\;100$ and $160$, $\delta_{iC} = 2.5$, $\Delta=2$ and  $f = 0.5$.   }
	\label{PlotBiasOfDelta_deltaC_2.5deltaT=4.5_DSM_unequal_sample_sizes.pdf}
\end{figure}


\clearpage

\section*{Appendix F: Coverage of 95\% confidence intervals for $\Delta$}


Each figure corresponds to a value of the standardized mean  in the Control arm $\delta_{C}$  (= $-2.5$, $-1$, 0, 1, 2.5)  and a value of the overall DSM $\Delta$ (= $-2$, $-1$, $-0.5$, 0, 0.5, 1, 2) . \\
The fraction of each study's sample size in the Control arm ($f$) is held constant at 0.5.

For each combination of a value of $n$ (= 20, 40, 100, 250) or  $\bar{n}$ (= 30, 60, 100, 160) and a value of $K$ (= 5, 10, 30), a panel plots coverage of 95\% confidence intervals for $\Delta$ versus $\tau^2$ (= 0, 0.1, 0.5, 1, 1.5).\\
The interval estimators of $\Delta$ are
\begin{itemize}
\item DL (DerSimonian-Laird) method, inverse-variance weights
\item REML method, inverse-variance weights
\item MP (Mandel-Paule) method, inverse-variance weights
\item HKSJ (Hartung-Knapp-Sidik-Jonkman) method centered at DL estimator of $\Delta$
\item SMC method, centered at SSW estimator of $\Delta$,  effective-sample-size weights
\item SSC method, centered at SSW estimator of $\Delta$,  effective-sample-size weights
\end{itemize}

\setcounter{figure}{0}

\clearpage
\renewcommand{\thefigure}{F.\arabic{figure}}


\begin{figure}[ht]
	\centering
	\includegraphics[scale=0.33]{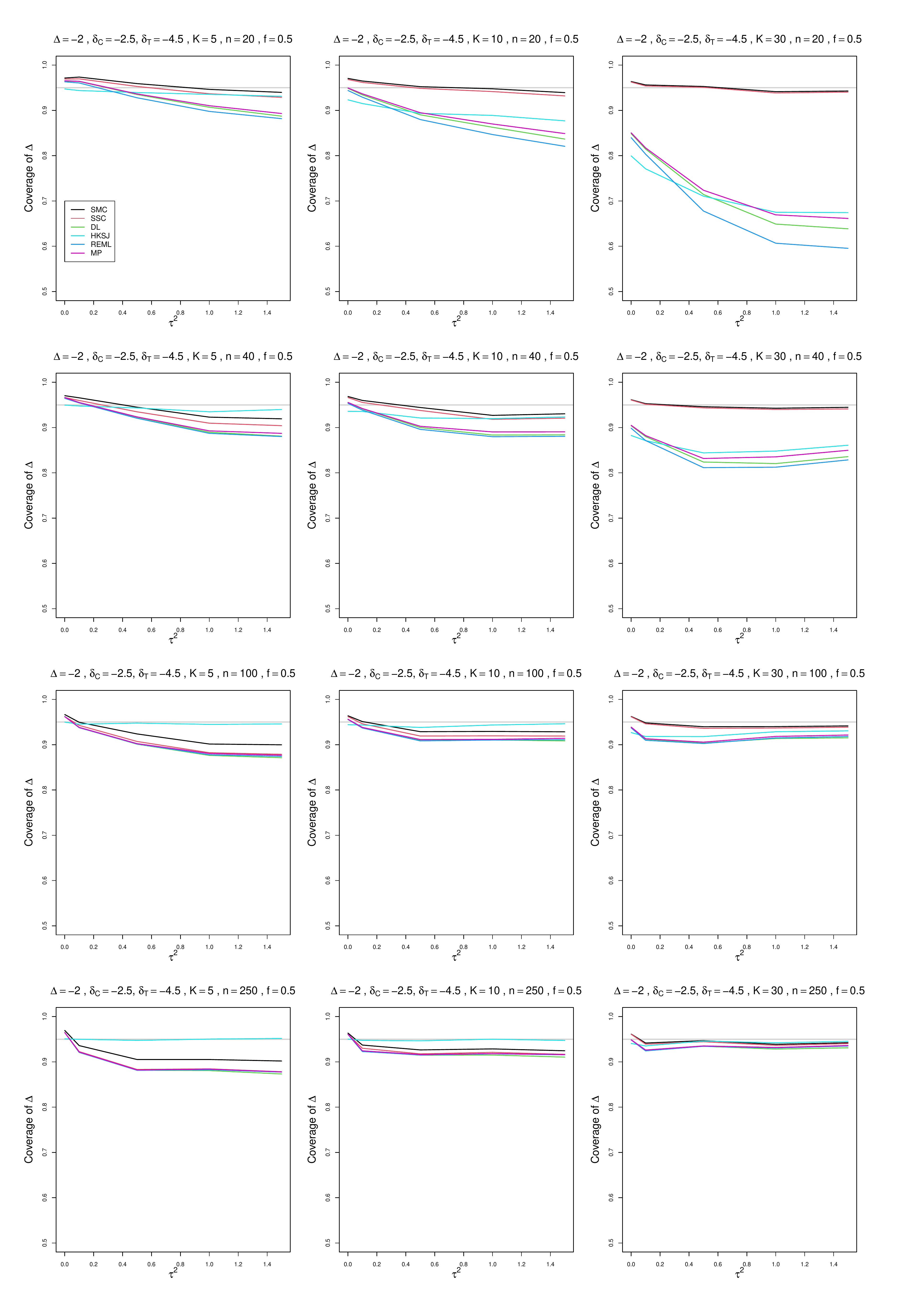}
	\caption{Coverage of 95\% confidence intervals of DSM (DL, REML, MP, HKSJ (DL), SMC and SSC intervals)  vs $\tau^2$, for equal sample sizes $n=20,\;40,\;100$ and $250$, $\delta_{iC} = -2.5$, $\Delta=-2$ and  $f = 0.5$.   }
	\label{PlotCoverageOfDelta_deltaC_-25deltaT=-4.5_DSM_equal_sample_sizes.pdf}
\end{figure}

\begin{figure}[ht]
	\centering
	\includegraphics[scale=0.33]{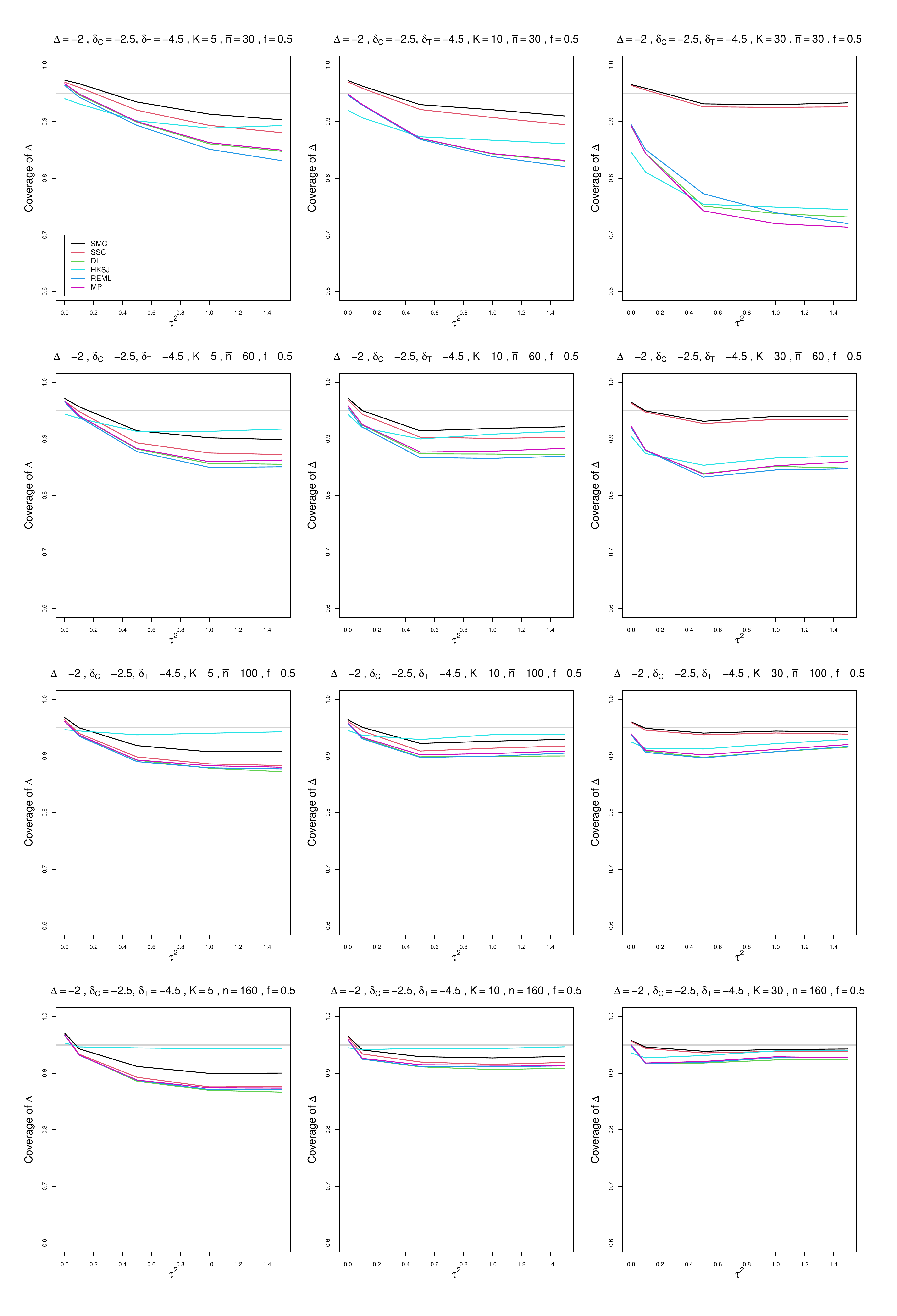}
	\caption{Coverage of 95\% confidence intervals for DSM (DL, REML, MP, HKSJ (DL), SMC and SSC intervals)  vs $\tau^2$, for unequal sample sizes $\bar{n}=30,\;60,\;100$ and $160$, $\delta_{iC} = -2.5$, $\Delta=-2$ and  $f = 0.5$.   }
	\label{PlotCoverageOfDelta_deltaC_-25deltaT=-4.5_DSM_unequal_sample_sizes.pdf}
\end{figure}
\clearpage

\begin{figure}[ht]
	\centering
	\includegraphics[scale=0.33]{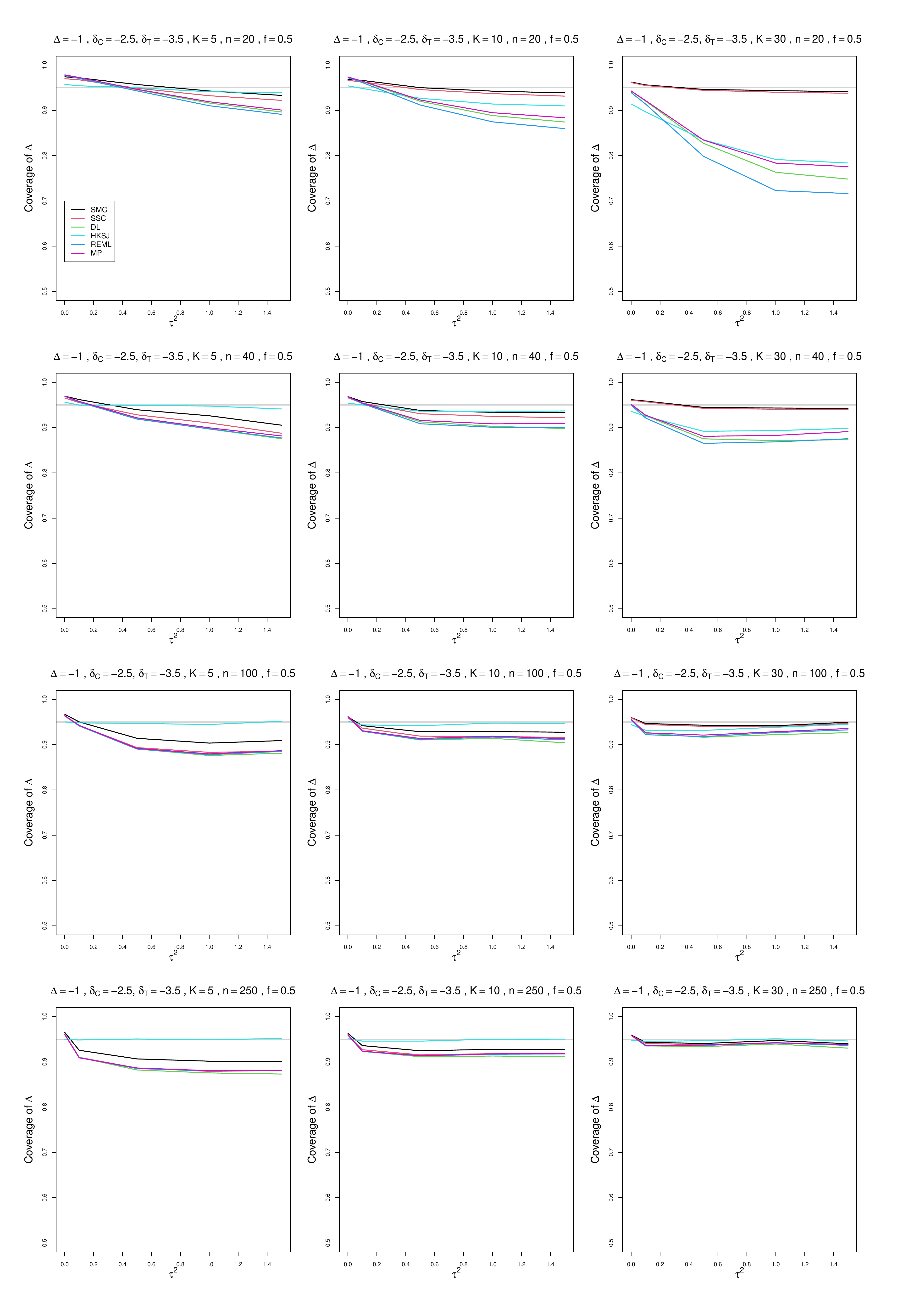}
	\caption{Coverage of 95\% confidence intervals for DSM (DL, REML, MP, HKSJ (DL), SMC and SSC intervals)  vs $\tau^2$, for equal sample sizes $n=20,\;40,\;100$ and $250$, $\delta_{iC} = -2.5$, $\Delta=-1$ and  $f = 0.5$.   }
	\label{PlotCoverageOfDelta_deltaC_-25deltaT=-3.5_DSM_equal_sample_sizes.pdf}
\end{figure}

\begin{figure}[ht]
	\centering
	\includegraphics[scale=0.33]{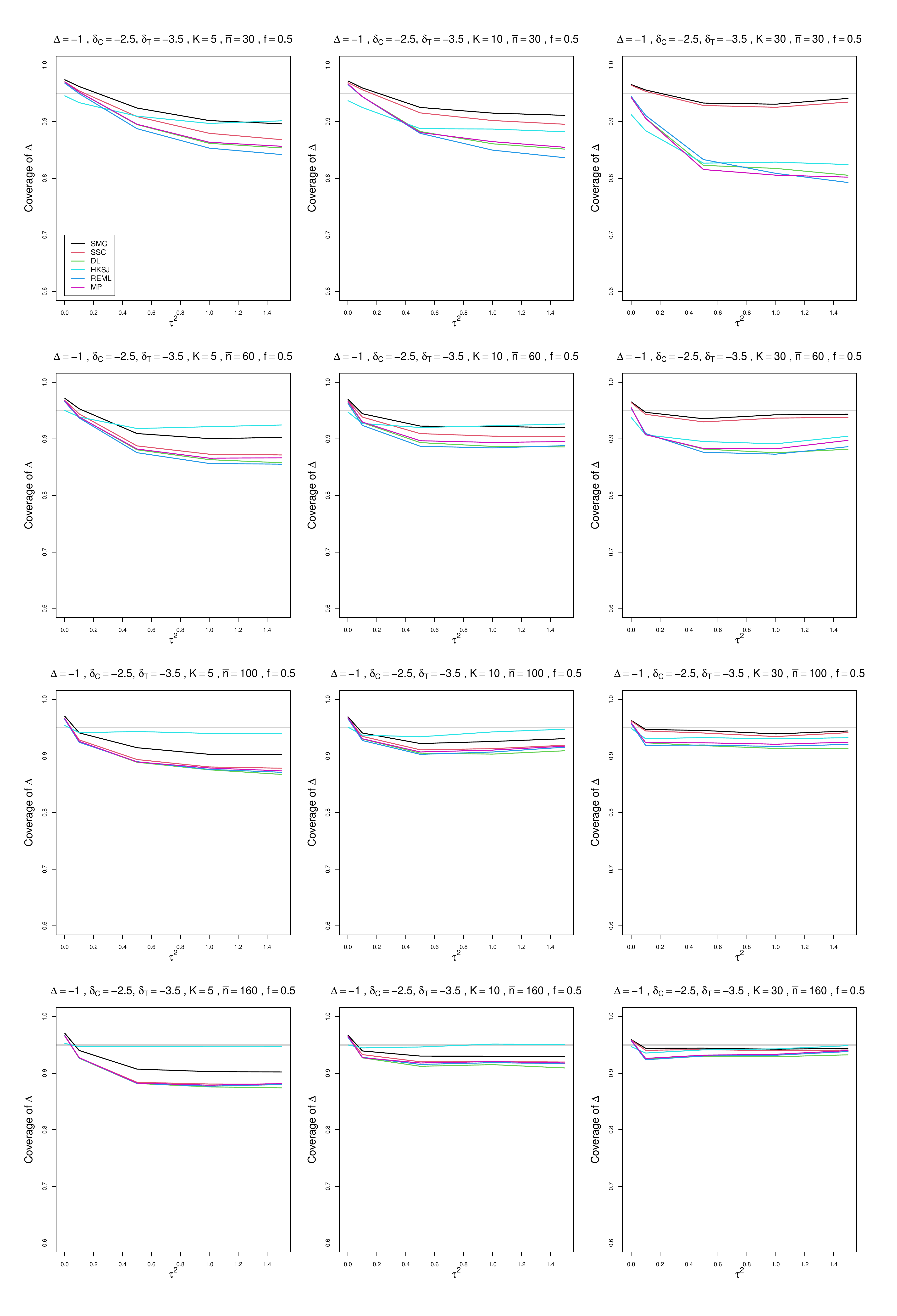}
	\caption{Coverage of 95\% confidence intervals for DSM (DL, REML, MP, HKSJ (DL), SMC and SSC intervals)  vs $\tau^2$, for unequal sample sizes $\bar{n}=30,\;60,\;100$ and $160$, $\delta_{iC} = -2.5$, $\Delta=-1$ and  $f = 0.5$.   }
	\label{PlotCoverageOfDelta_deltaC_-25deltaT=-3.5_DSM_unequal_sample_sizes.pdf}
\end{figure}

\begin{figure}[ht]
	\centering
	\includegraphics[scale=0.33]{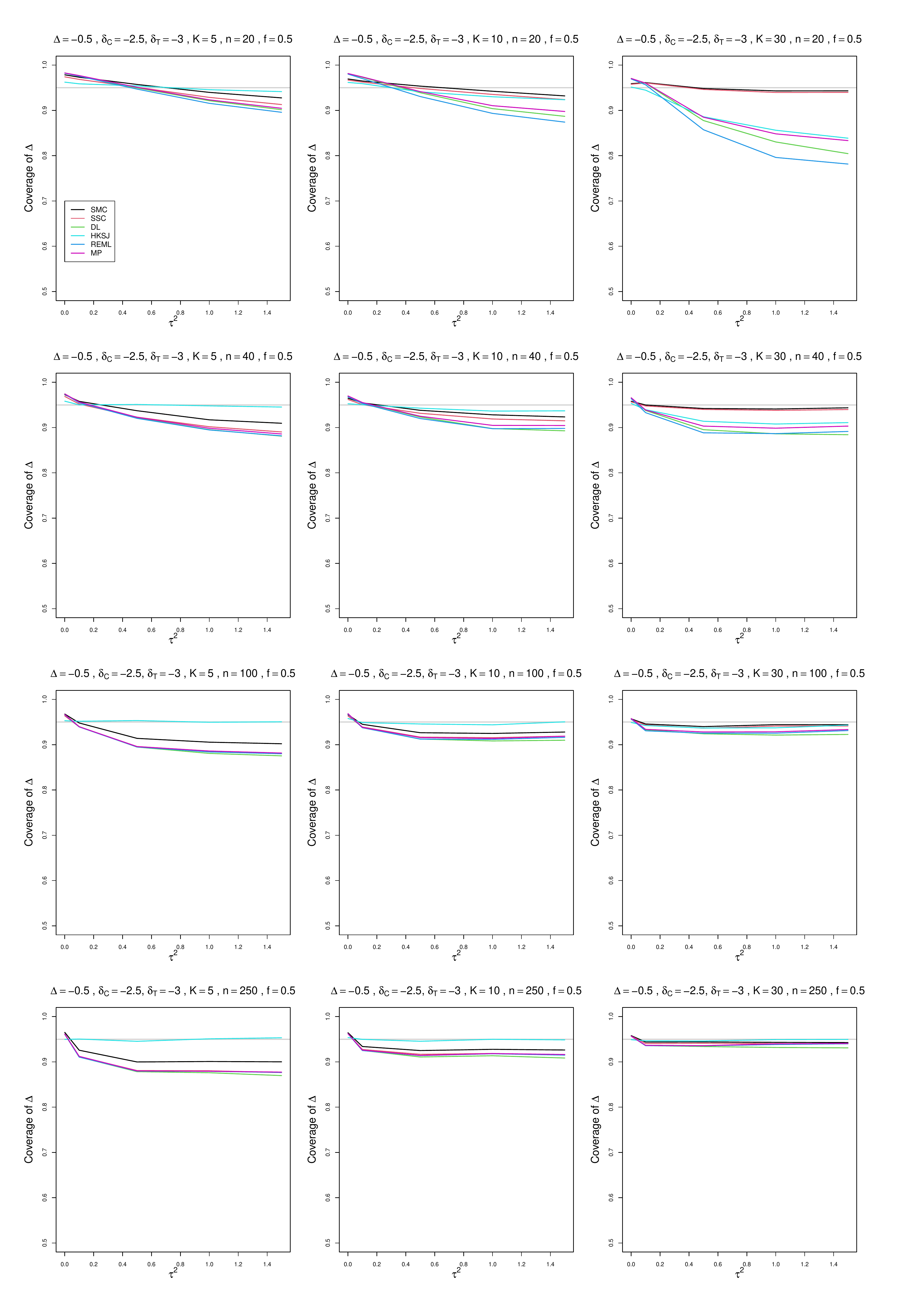}
	\caption{Coverage of 95\% confidence intervals for DSM (DL, REML, MP, HKSJ (DL), SMC and SSC intervals)  vs $\tau^2$, for equal sample sizes $n=20,\;40,\;100$ and $250$, $\delta_{iC} = -2.5$, $\Delta=-0.5$ and  $f = 0.5$.   }
	\label{PlotCoverageOfDelta_deltaC_-25deltaT=-3_DSM_equal_sample_sizes.pdf}
\end{figure}

\begin{figure}[ht]
	\centering
	\includegraphics[scale=0.33]{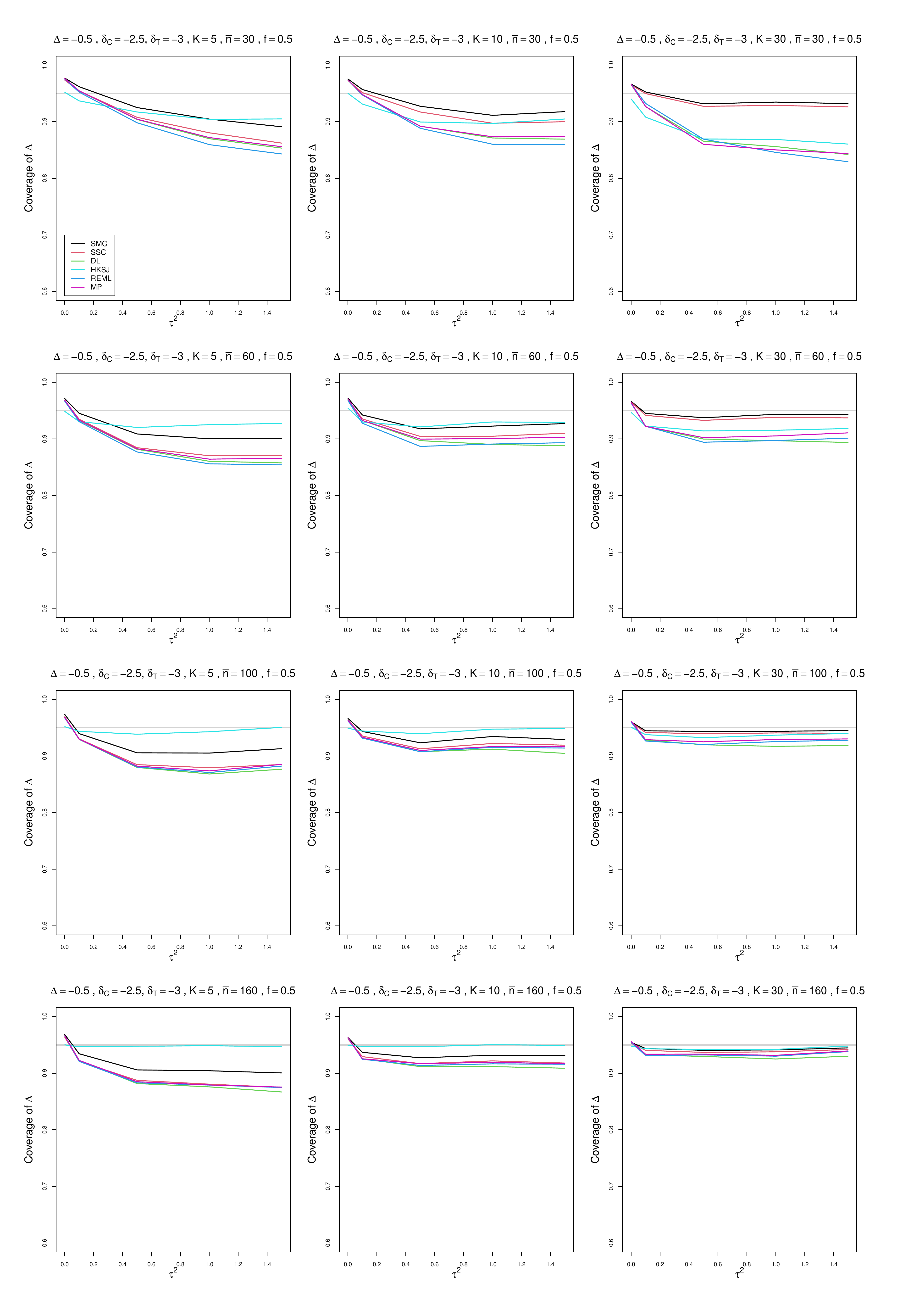}
	\caption{Coverage of 95\% confidence intervals for DSM (DL, REML, MP, HKSJ (DL), SMC and SSC intervals)  vs $\tau^2$, for unequal sample sizes $\bar{n}=30,\;60,\;100$ and $160$, $\delta_{iC} = -2.5$, $\Delta=-0.5$ and  $f = 0.5$.   }
	\label{PlotCoverageOfDelta_deltaC_-25deltaT=-3_DSM_unequal_sample_sizes.pdf}
\end{figure}

\begin{figure}[ht]
	\centering
	\includegraphics[scale=0.33]{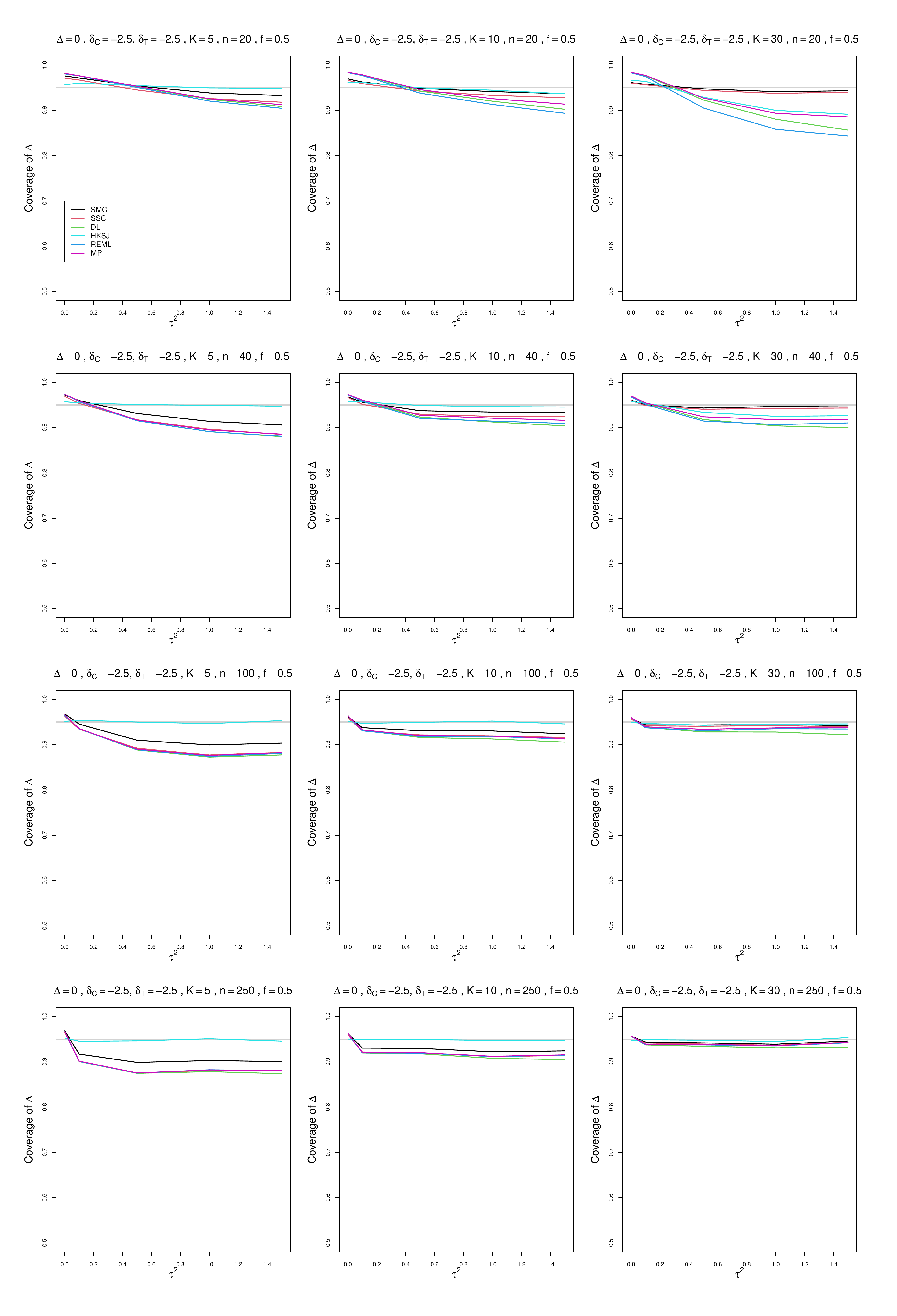}
	\caption{Coverage of 95\% confidence intervals for DSM (DL, REML, MP, HKSJ (DL), SMC and SSC intervals)  vs $\tau^2$, for equal sample sizes $n=20,\;40,\;100$ and $250$, $\delta_{iC} = -2.5$, $\Delta=0$ and  $f = 0.5$.   }
	\label{PlotCoverageOfDelta_deltaC_-25deltaT=-2.5_DSM_equal_sample_sizes.pdf}
\end{figure}

\begin{figure}[ht]
	\centering
	\includegraphics[scale=0.33]{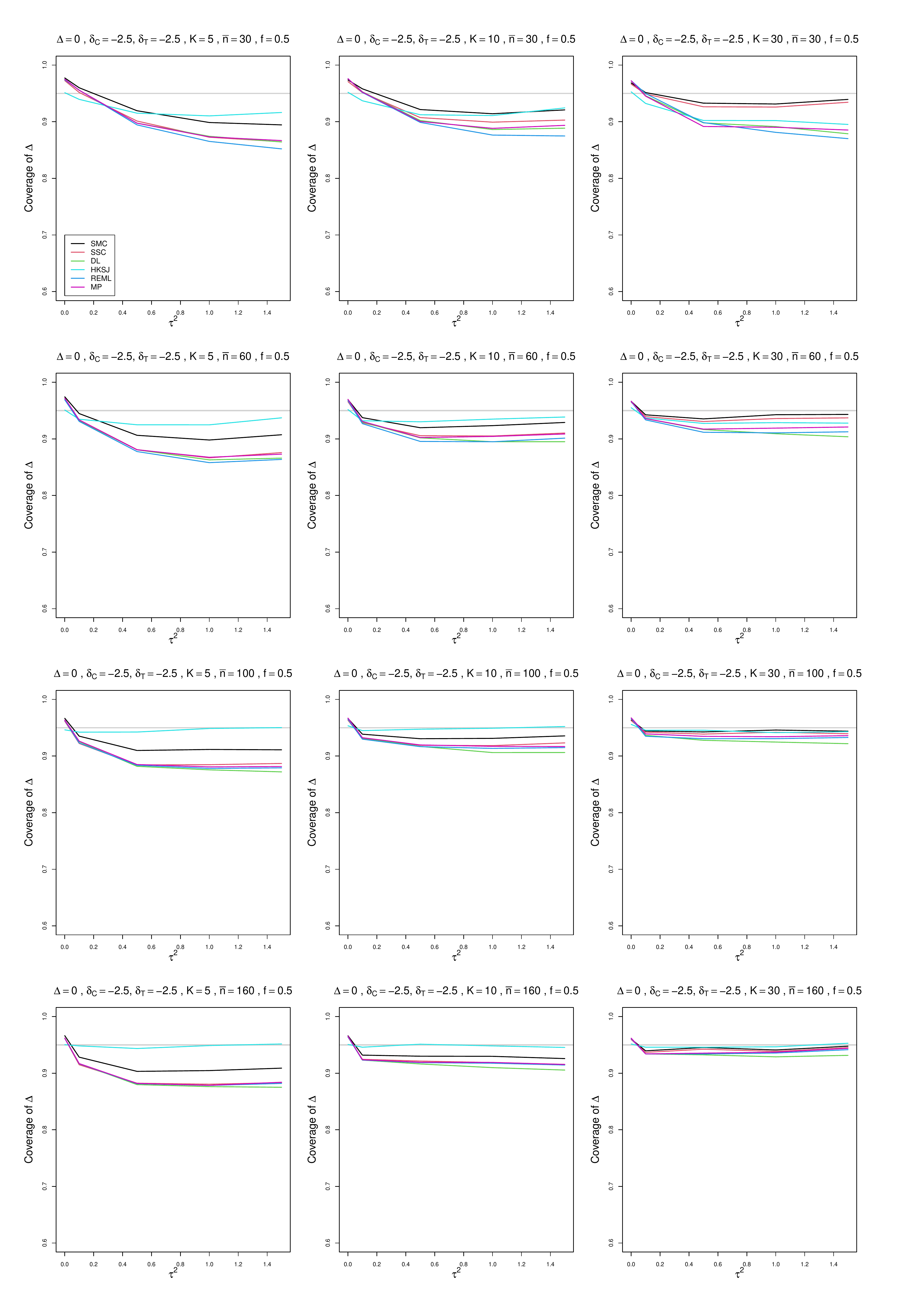}
	\caption{Coverage of 95\% confidence intervals for DSM (DL, REML, MP, HKSJ (DL), SMC and SSC intervals)  vs $\tau^2$, for unequal sample sizes $\bar{n}=30,\;60,\;100$ and $160$, $\delta_{iC} = -2.5$, $\Delta=0$ and  $f = 0.5$.   }
	\label{PlotCoverageOfDelta_deltaC_-25deltaT=-2.5_DSM_unequal_sample_sizes.pdf}
\end{figure}

\begin{figure}[ht]
	\centering
	\includegraphics[scale=0.33]{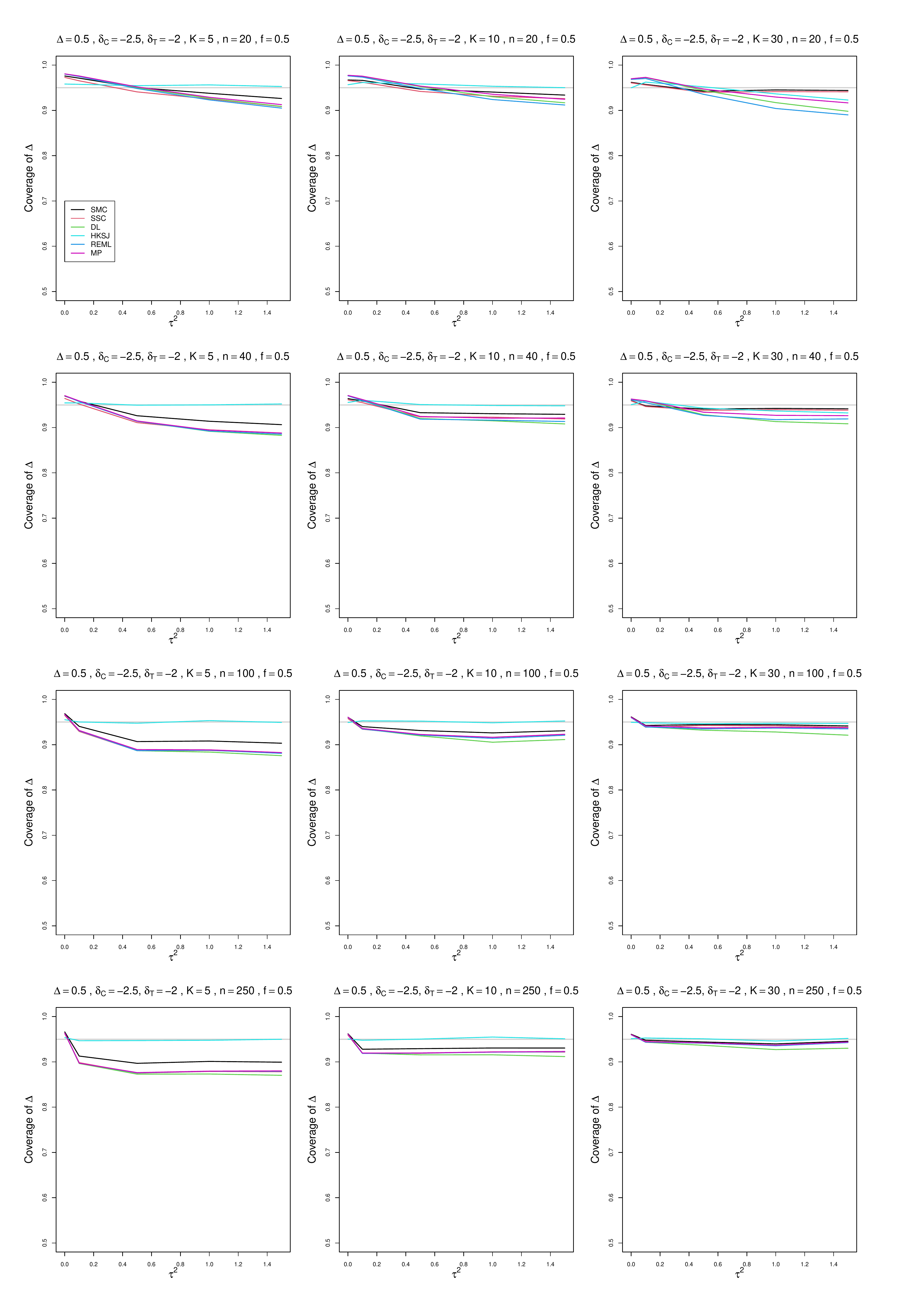}
	\caption{Coverage of 95\% confidence intervals for DSM (DL, REML, MP, HKSJ (DL), SMC and SSC intervals)  vs $\tau^2$, for equal sample sizes $n=20,\;40,\;100$ and $250$, $\delta_{iC} = -2.5$, $\Delta=0.5$ and  $f = 0.5$.   }
	\label{PlotCoverageOfDelta_deltaC_-25deltaT=-2_DSM_equal_sample_sizes.pdf}
\end{figure}

\begin{figure}[ht]
	\centering
	\includegraphics[scale=0.33]{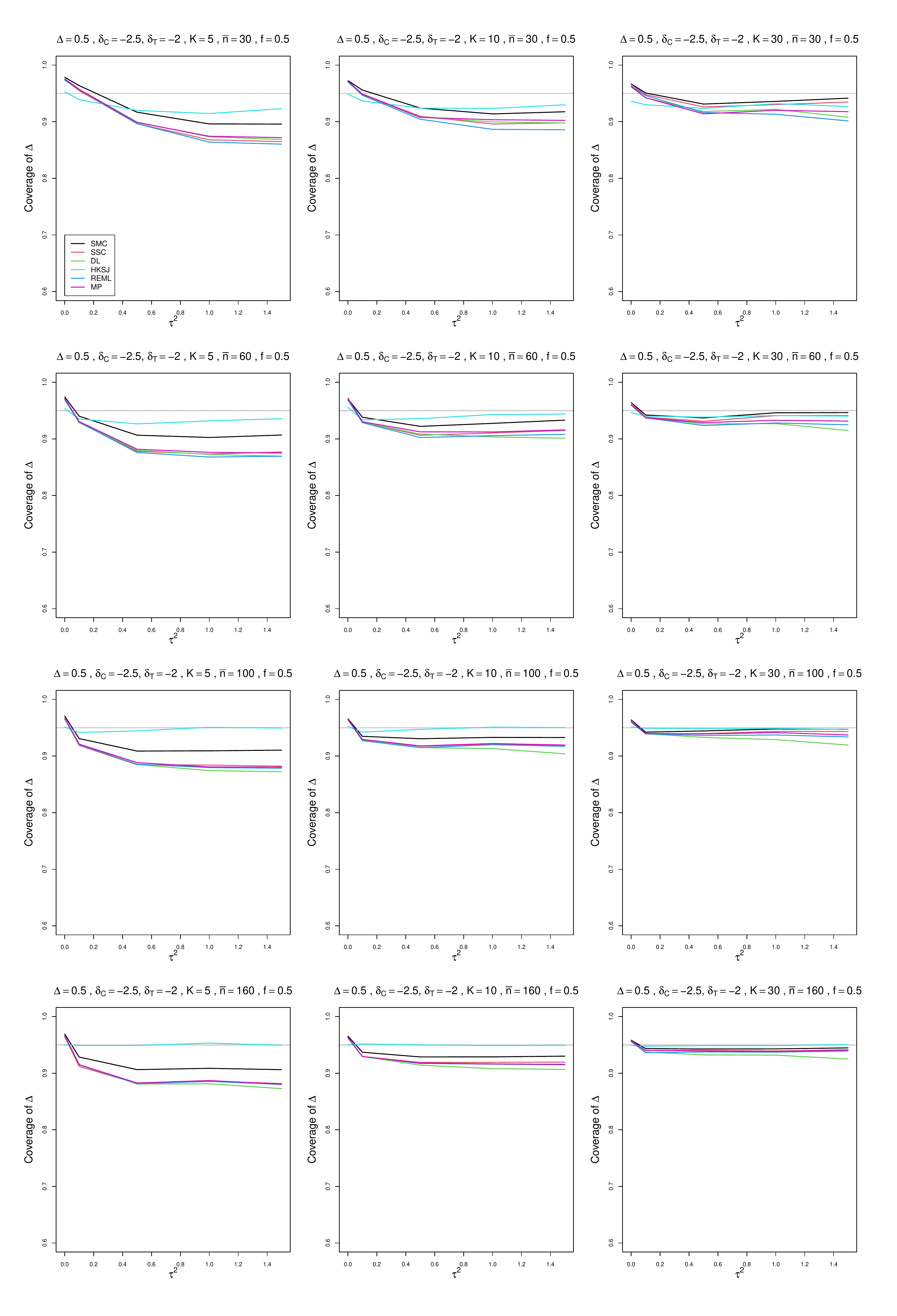}
	\caption{Coverage of 95\% confidence intervals for DSM (DL, REML, MP, HKSJ (DL), SMC and SSC intervals)  vs $\tau^2$, for unequal sample sizes $\bar{n}=30,\;60,\;100$ and $160$, $\delta_{iC} = -2.5$, $\Delta=0.5$ and  $f = 0.5$.   }
	\label{PlotCoverageOfDelta_deltaC_-25deltaT=-2_DSM_unequal_sample_sizes.pdf}
\end{figure}

\begin{figure}[ht]
	\centering
	\includegraphics[scale=0.33]{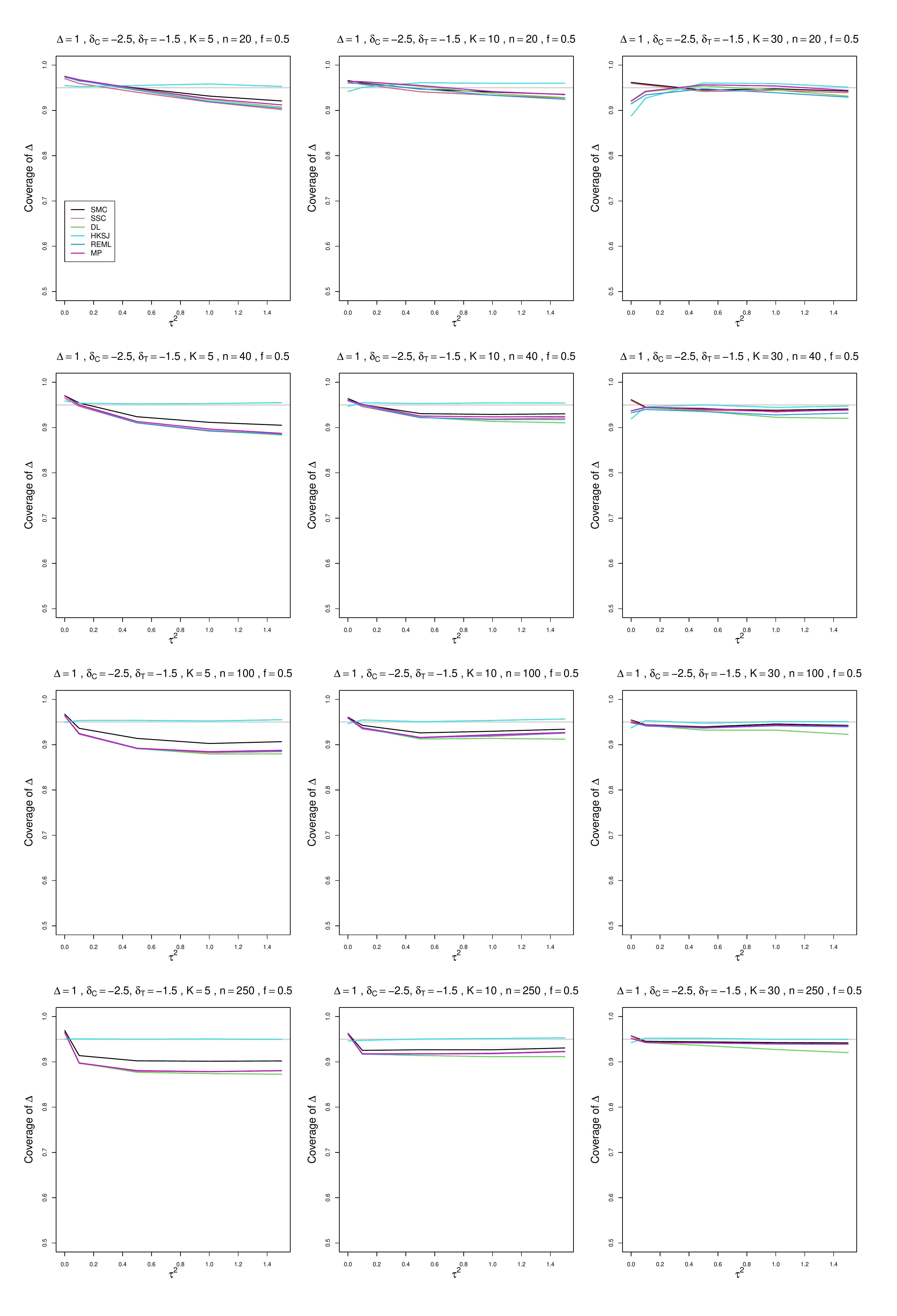}
	\caption{Coverage of 95\% confidence intervals for DSM (DL, REML, MP, HKSJ (DL), SMC and SSC intervals)  vs $\tau^2$, for equal sample sizes $n=20,\;40,\;100$ and $250$, $\delta_{iC} = -2.5$, $\Delta=1$ and  $f = 0.5$.   }
	\label{PlotCoverageOfDelta_deltaC_-25deltaT=-1.5_DSM_equal_sample_sizes.pdf}
\end{figure}

\begin{figure}[ht]
	\centering
	\includegraphics[scale=0.33]{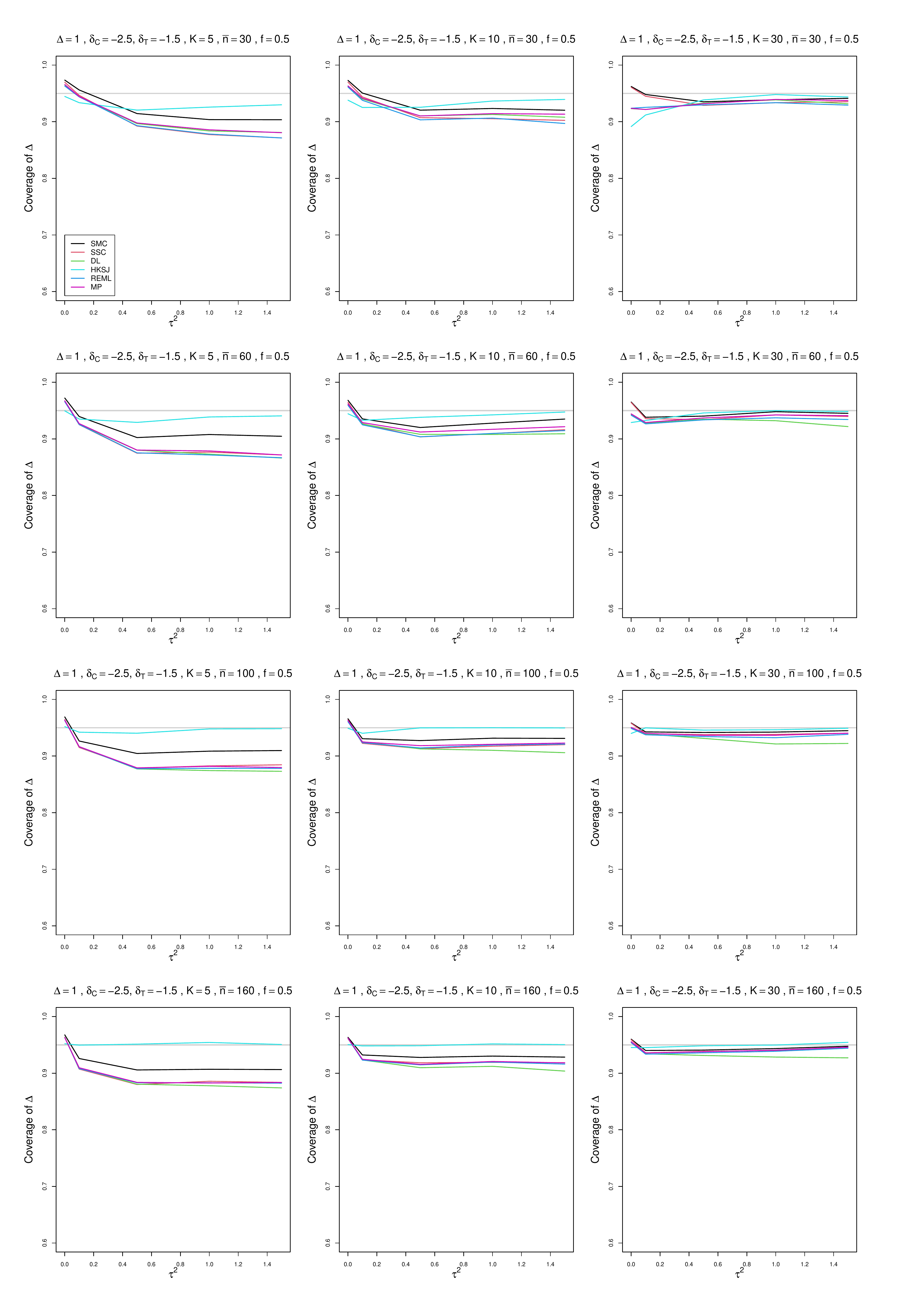}
	\caption{Coverage of 95\% confidence intervals for DSM (DL, REML, MP, HKSJ (DL), SMC and SSC intervals)  vs $\tau^2$, for unequal sample sizes $\bar{n}=30,\;60,\;100$ and $160$, $\delta_{iC} = -2.5$, $\Delta=1$ and  $f = 0.5$.   }
	\label{PlotCoverageOfDelta_deltaC_-25deltaT=-1.5_DSM_unequal_sample_sizes.pdf}
\end{figure}

\begin{figure}[ht]
	\centering
	\includegraphics[scale=0.33]{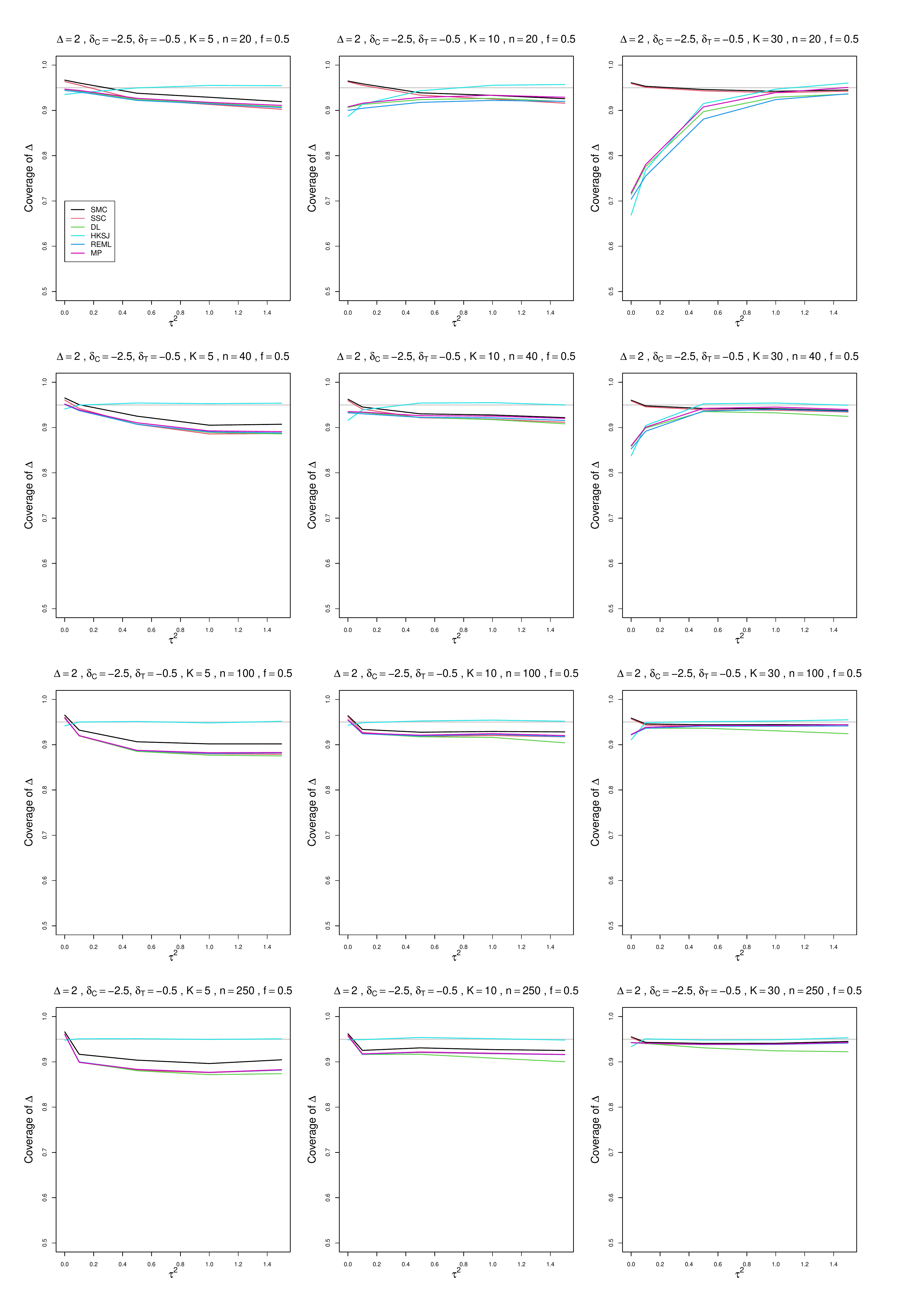}
	\caption{Coverage of 95\% confidence intervals for DSM (DL, REML, MP, HKSJ (DL), SMC and SSC intervals)  vs $\tau^2$, for equal sample sizes $n=20,\;40,\;100$ and $250$, $\delta_{iC} = -2.5$, $\Delta=2$ and  $f = 0.5$.   }
	\label{PlotCoverageOfDelta_deltaC_-25deltaT=-0.5_DSM_equal_sample_sizes.pdf}
\end{figure}

\begin{figure}[ht]
	\centering
	\includegraphics[scale=0.33]{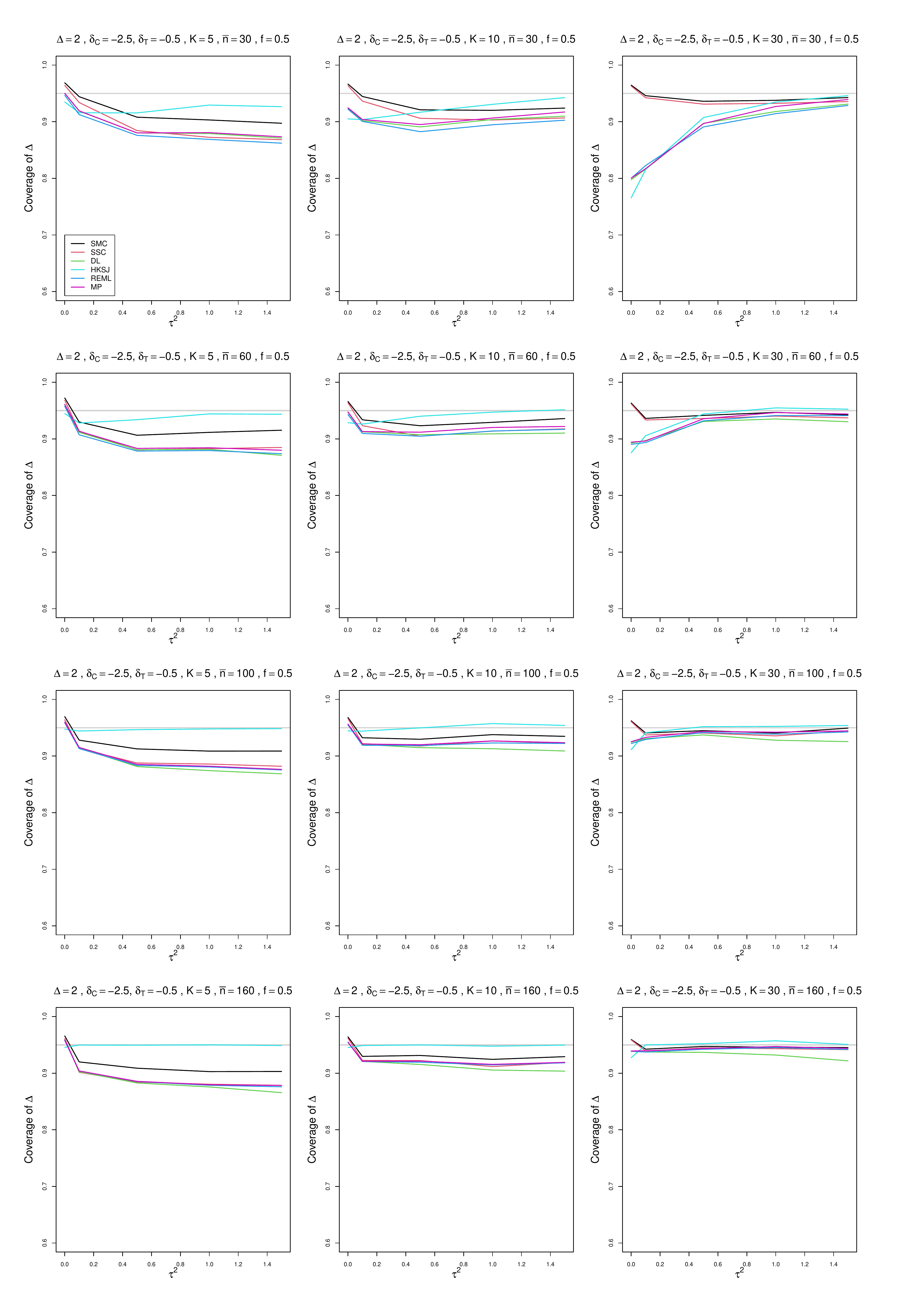}
	\caption{Coverage of 95\% confidence intervals for DSM (DL, REML, MP, HKSJ (DL), SMC and SSC intervals)  vs $\tau^2$, for unequal sample sizes $\bar{n}=30,\;60,\;100$ and $160$, $\delta_{iC} = -2.5$, $\Delta=2$ and  $f = 0.5$.   }
	\label{PlotCoverageOfDelta_deltaC_-25deltaT=-0.5_DSM_unequal_sample_sizes.pdf}
\end{figure}
\begin{figure}[ht]
	\centering
	\includegraphics[scale=0.33]{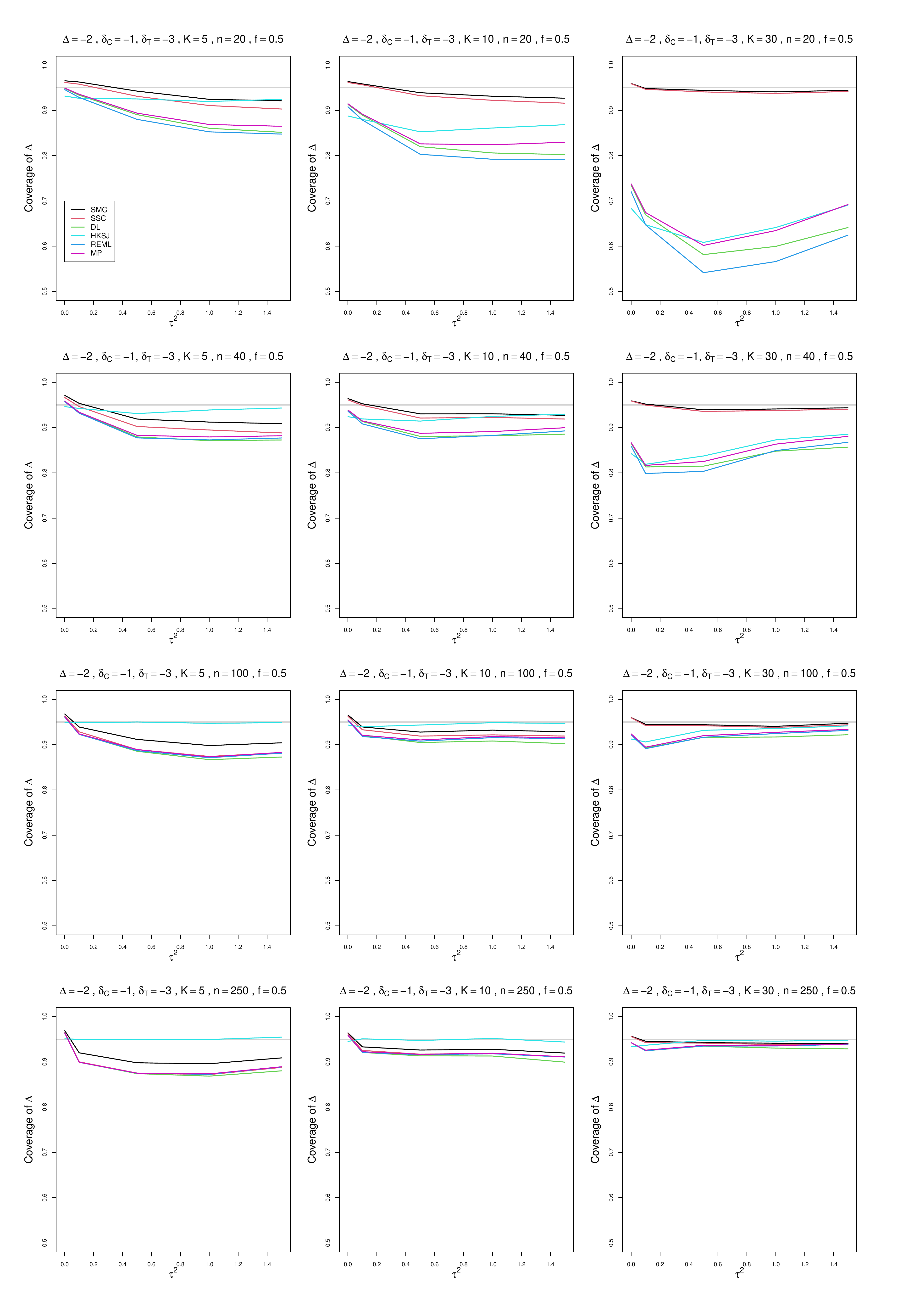}
	\caption{Coverage of 95\% confidence intervals for DSM (DL, REML, MP, HKSJ (DL), SMC and SSC intervals)  vs $\tau^2$, for equal sample sizes $n=20,\;40,\;100$ and $250$, $\delta_{iC} = -2.5$, $\Delta=-2$ and  $f = 0.5$.   }
	\label{PlotCoverageOfDelta_deltaC_-1deltaT=-3_DSM_equal_sample_sizes.pdf}
\end{figure}

\begin{figure}[ht]
	\centering
	\includegraphics[scale=0.33]{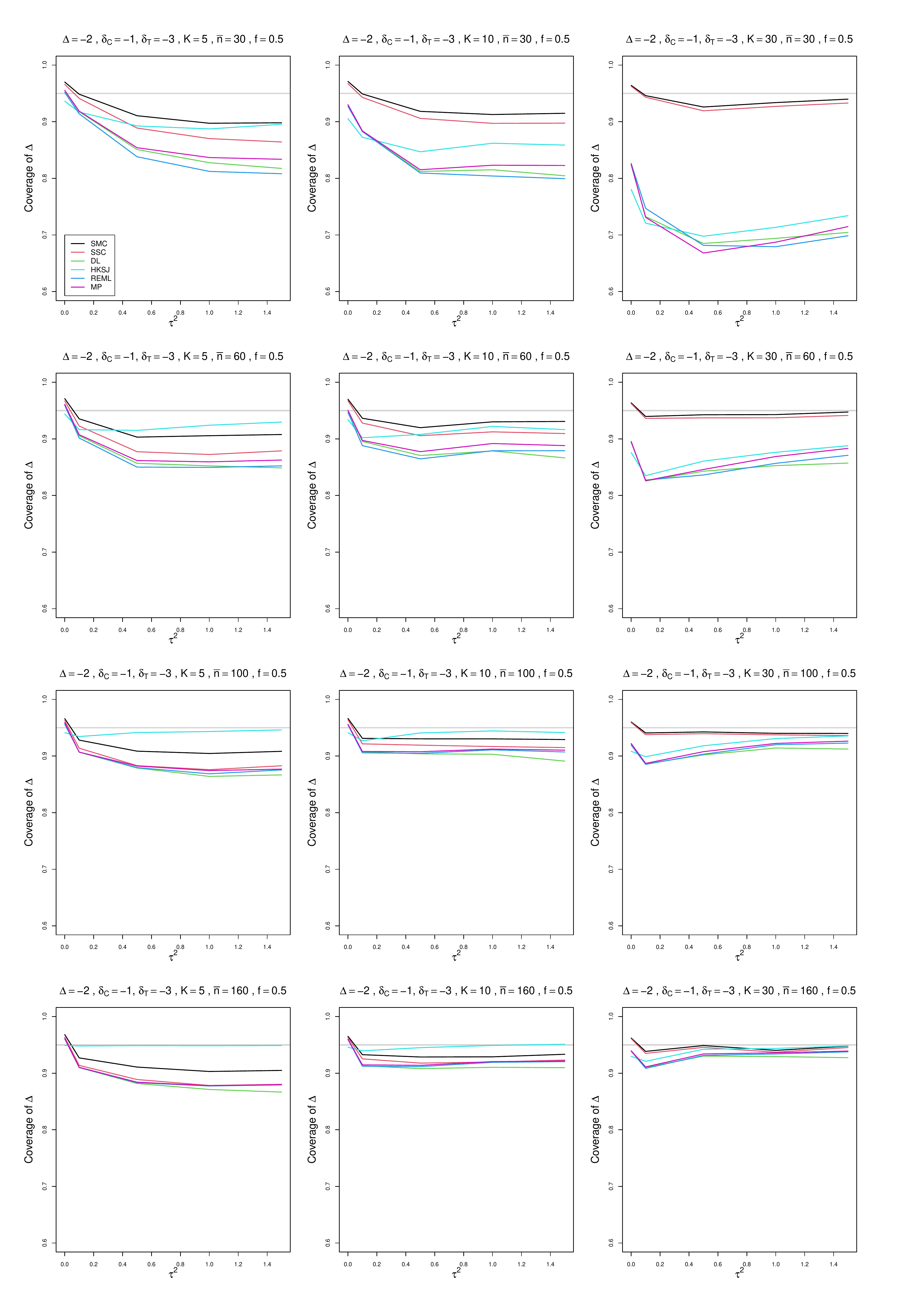}
	\caption{Coverage of 95\% confidence intervals for DSM (DL, REML, MP, HKSJ (DL), SMC and SSC intervals)  vs $\tau^2$, for unequal sample sizes $\bar{n}=30,\;60,\;100$ and $160$, $\delta_{iC} = -1$, $\Delta=-2$ and  $f = 0.5$.   }
	\label{PlotCoverageOfDelta_deltaC_-1deltaT=-3_DSM_unequal_sample_sizes.pdf}
\end{figure}

\begin{figure}[ht]
	\centering
	\includegraphics[scale=0.33]{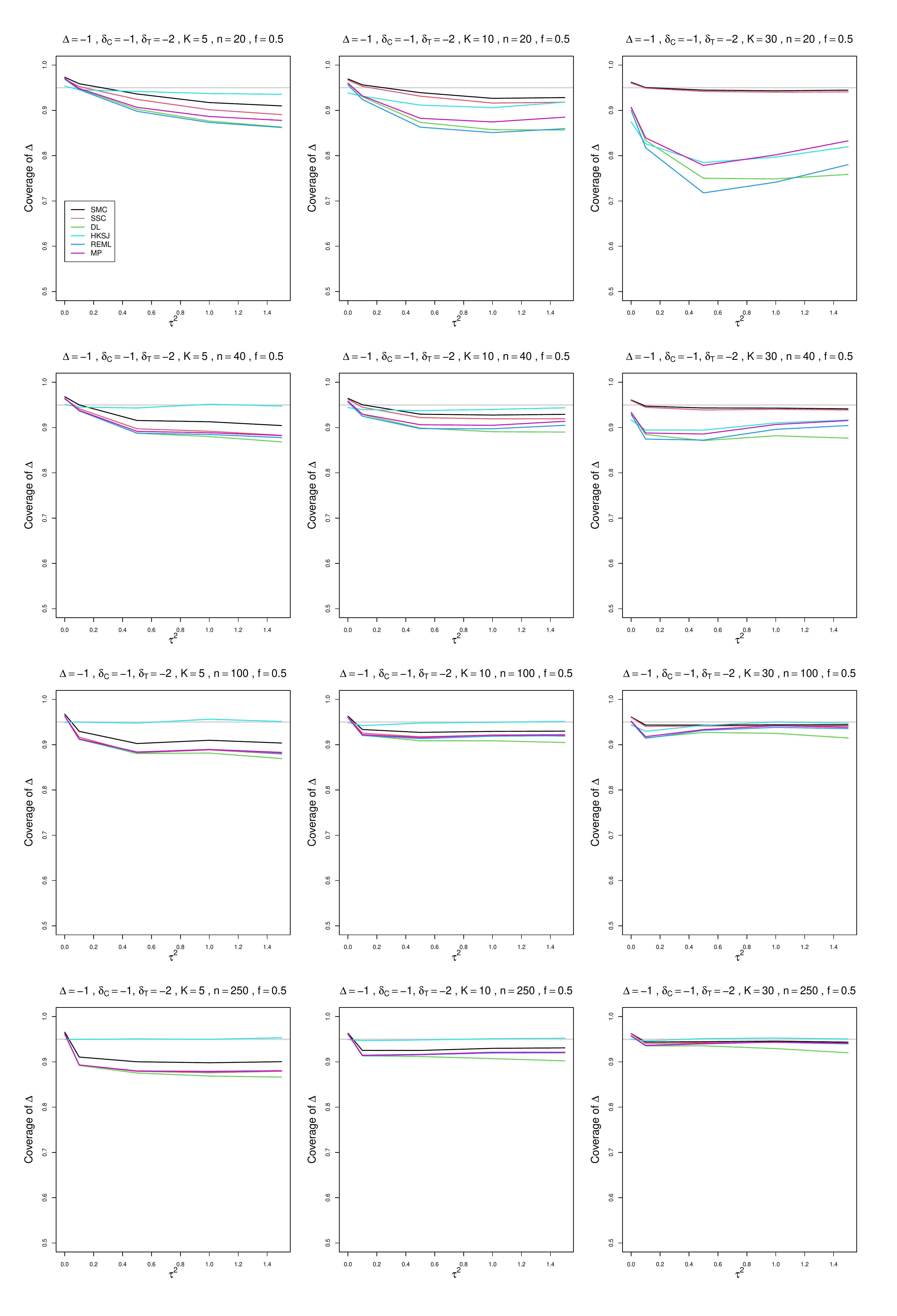}
	\caption{Coverage of 95\% confidence intervals for DSM (DL, REML, MP, HKSJ (DL), SMC and SSC intervals)  vs $\tau^2$, for equal sample sizes $n=20,\;40,\;100$ and $250$, $\delta_{iC} = -1$, $\Delta=-1$ and  $f = 0.5$.   }
	\label{PlotCoverageOfDelta_deltaC_-1deltaT=-2_DSM_equal_sample_sizes.pdf}
\end{figure}

\begin{figure}[ht]
	\centering
	\includegraphics[scale=0.33]{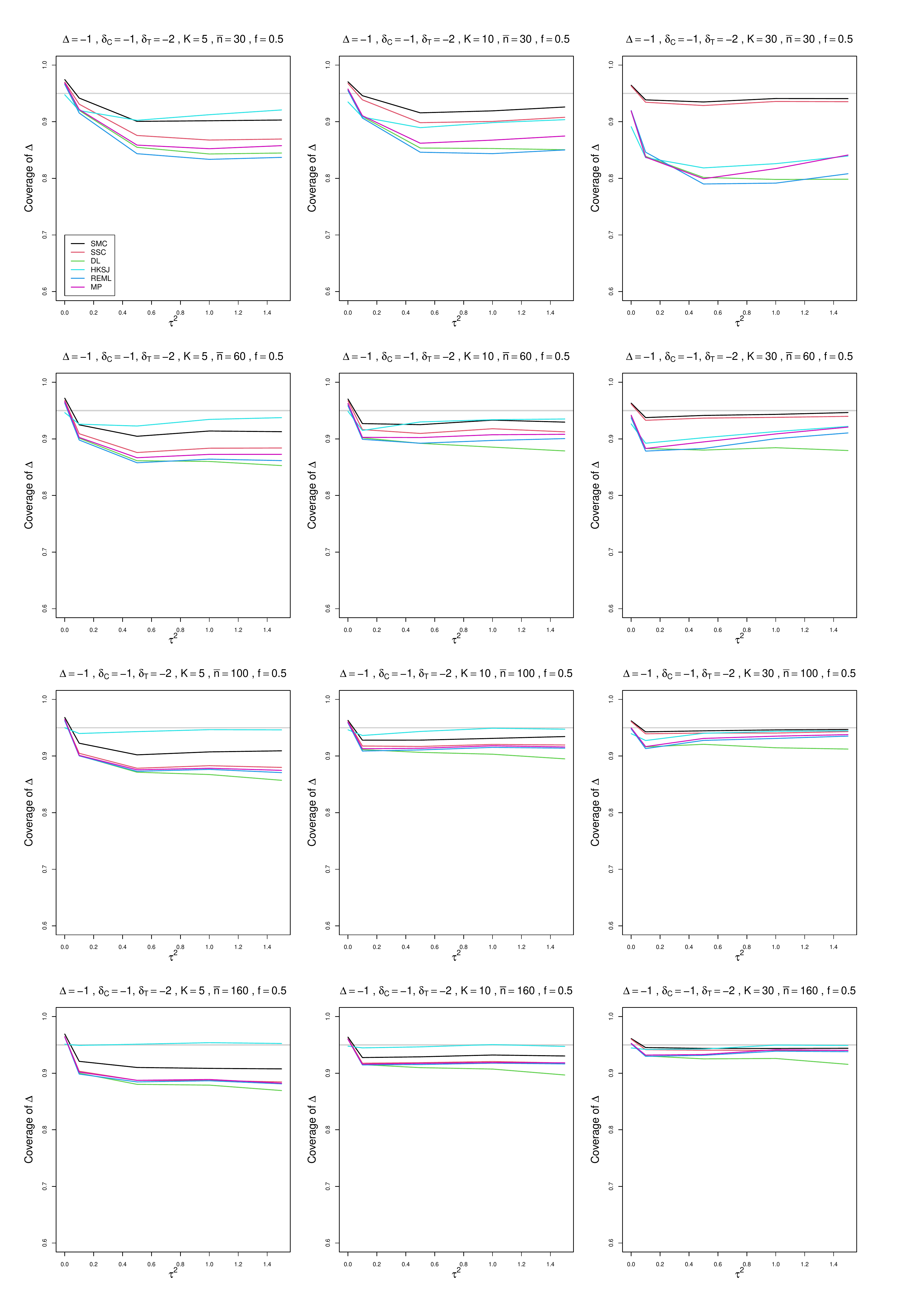}
	\caption{Coverage of 95\% confidence intervals for DSM (DL, REML, MP, HKSJ (DL), SMC and SSC intervals)  vs $\tau^2$, for unequal sample sizes $\bar{n}=30,\;60,\;100$ and $160$, $\delta_{iC} = -1$, $\Delta=-1$ and  $f = 0.5$.   }
	\label{PlotCoverageOfDelta_deltaC_-1deltaT=-2_DSM_unequal_sample_sizes.pdf}
\end{figure}

\begin{figure}[ht]
	\centering
	\includegraphics[scale=0.33]{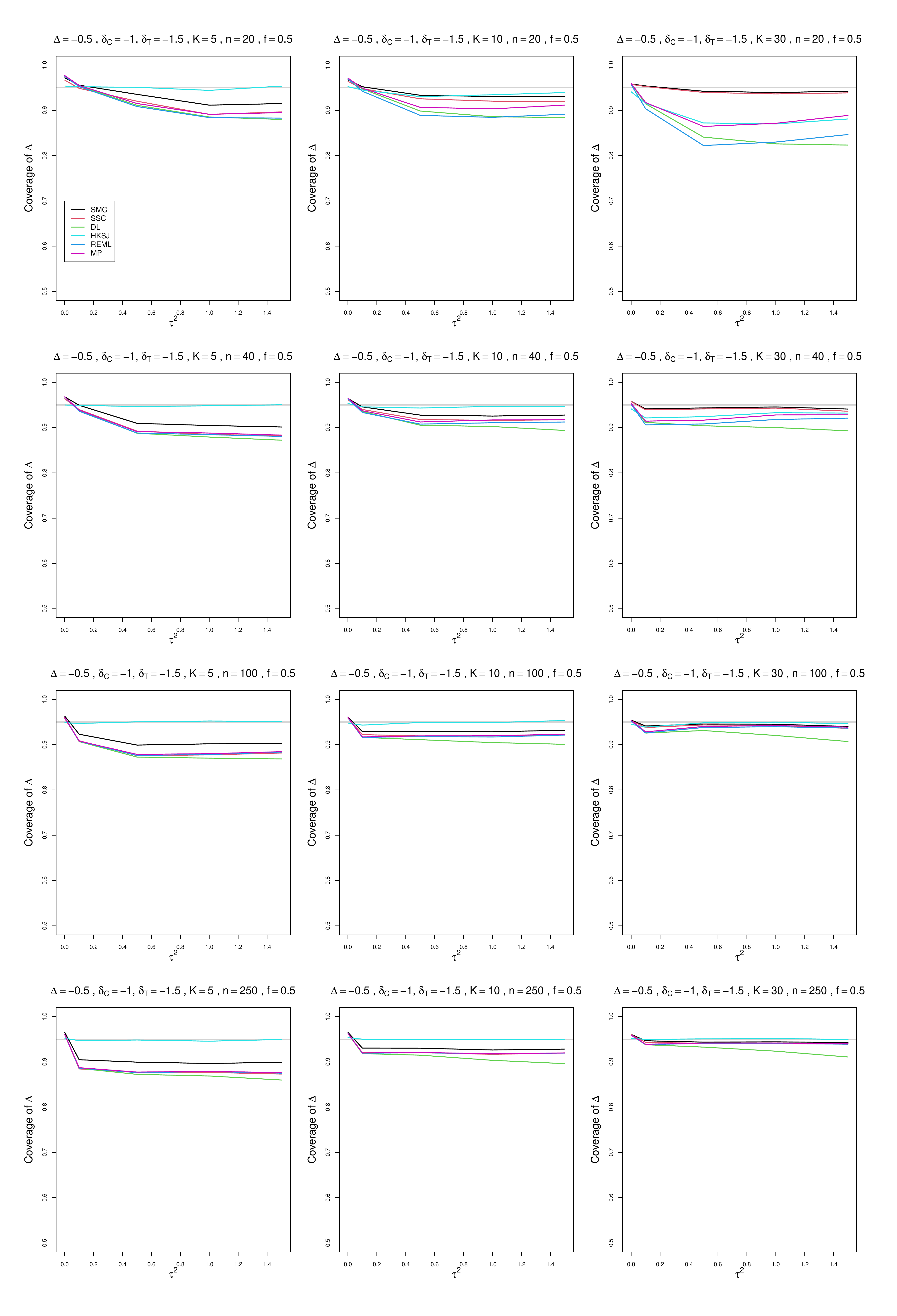}
	\caption{Coverage of 95\% confidence intervals for DSM (DL, REML, MP, HKSJ (DL), SMC and SSC intervals)  vs $\tau^2$, for equal sample sizes $n=20,\;40,\;100$ and $250$, $\delta_{iC} = -1$, $\Delta=-0.5$ and  $f = 0.5$.   }
	\label{PlotCoverageOfDelta_deltaC_-1deltaT=-1.5_DSM_equal_sample_sizes.pdf}
\end{figure}

\begin{figure}[ht]
	\centering
	\includegraphics[scale=0.33]{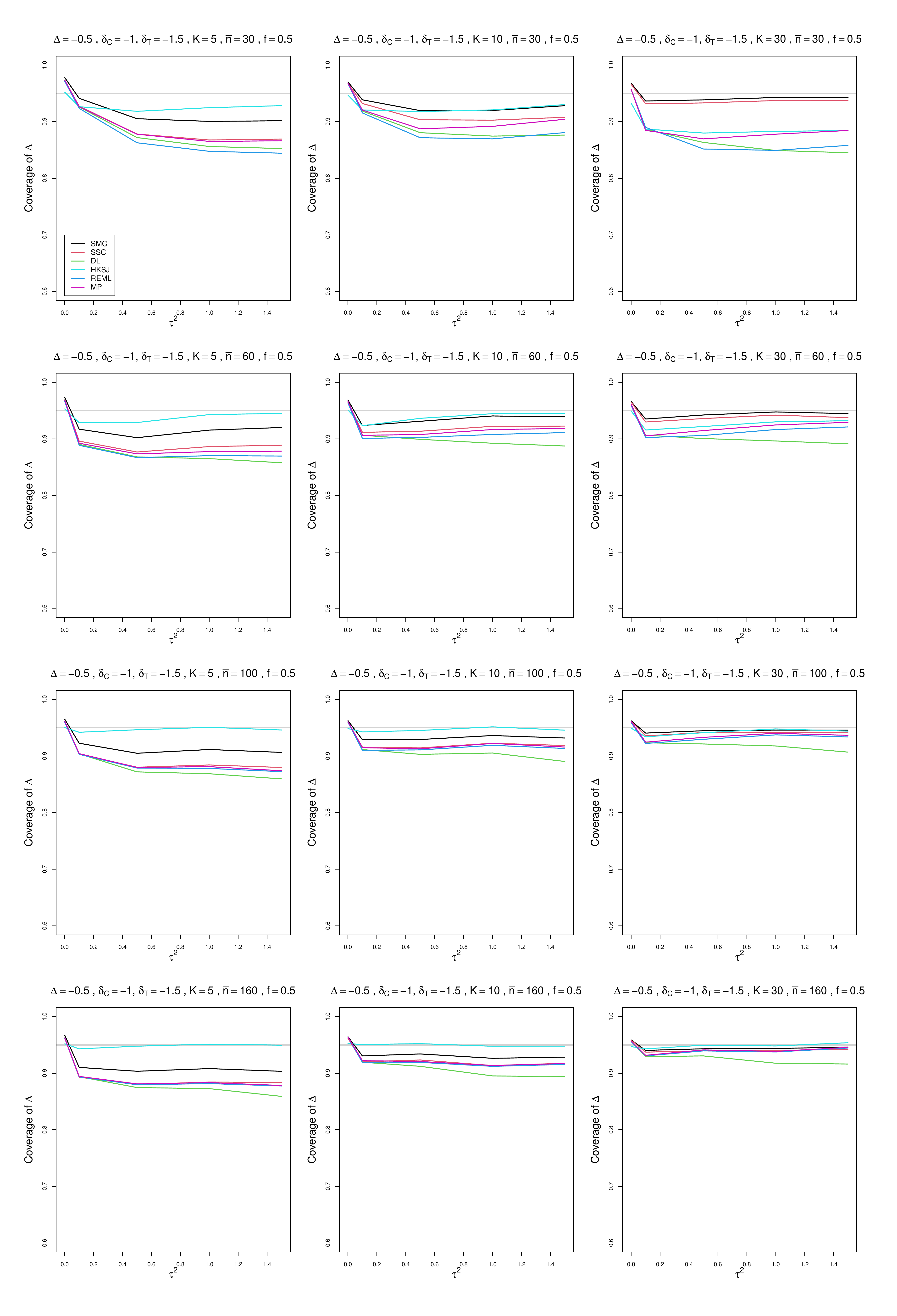}
	\caption{Coverage of 95\% confidence intervals for DSM (DL, REML, MP, HKSJ (DL), SMC and SSC intervals)  vs $\tau^2$, for unequal sample sizes $\bar{n}=30,\;60,\;100$ and $160$, $\delta_{iC} = -1$, $\Delta=-0.5$ and  $f = 0.5$.   }
	\label{PlotCoverageOfDelta_deltaC_-1deltaT=-1,5_DSM_unequal_sample_sizes.pdf}
\end{figure}

\begin{figure}[ht]
	\centering
	\includegraphics[scale=0.33]{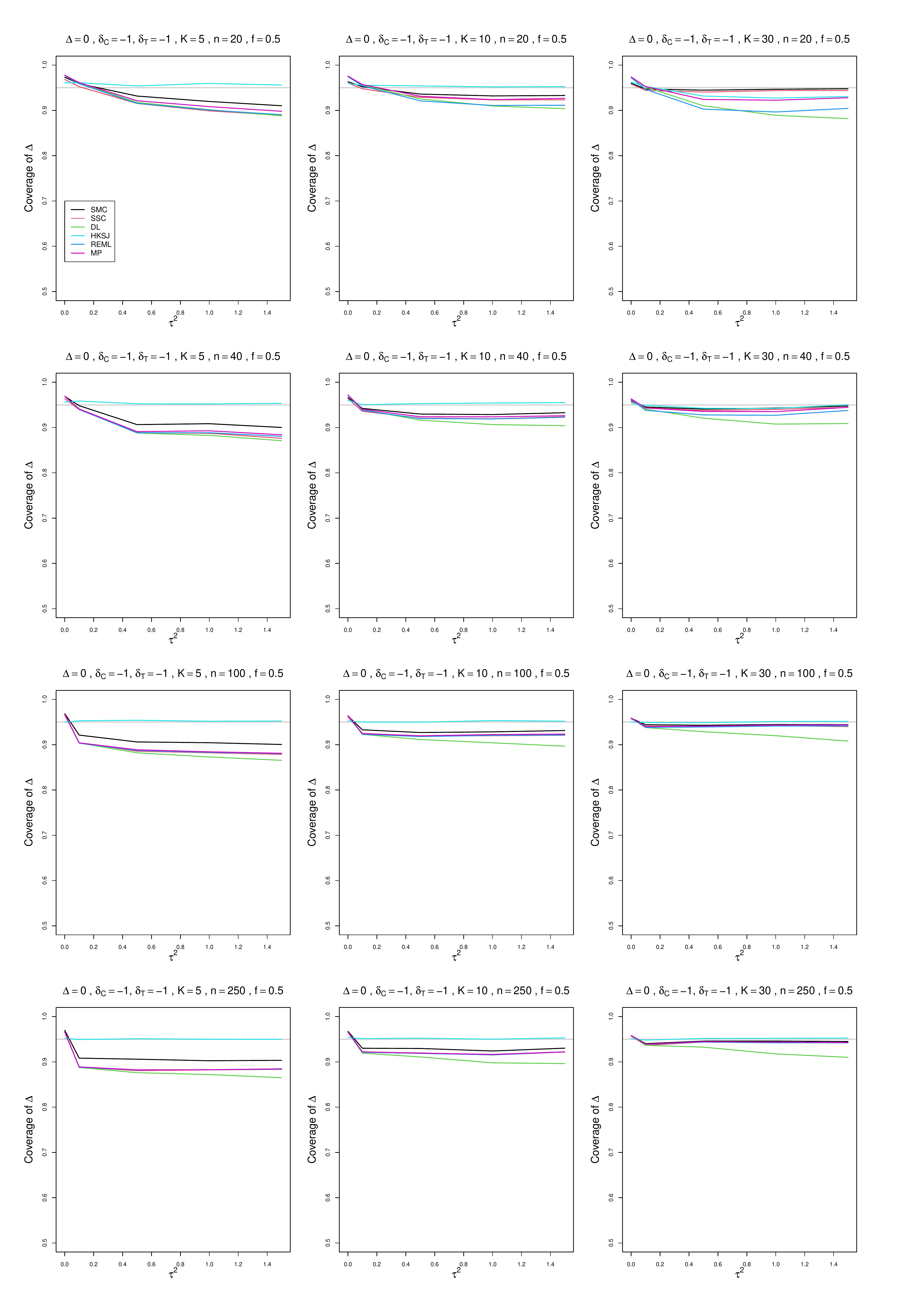}
	\caption{Coverage of 95\% confidence intervals for DSM (DL, REML, MP, HKSJ (DL), SMC and SSC intervals)  vs $\tau^2$, for equal sample sizes $n=20,\;40,\;100$ and $250$, $\delta_{iC} = -1$, $\Delta=0$ and  $f = 0.5$.   }
	\label{PlotCoverageOfDelta_deltaC_-1deltaT=-1_DSM_equal_sample_sizes.pdf}
\end{figure}

\begin{figure}[ht]
	\centering
	\includegraphics[scale=0.33]{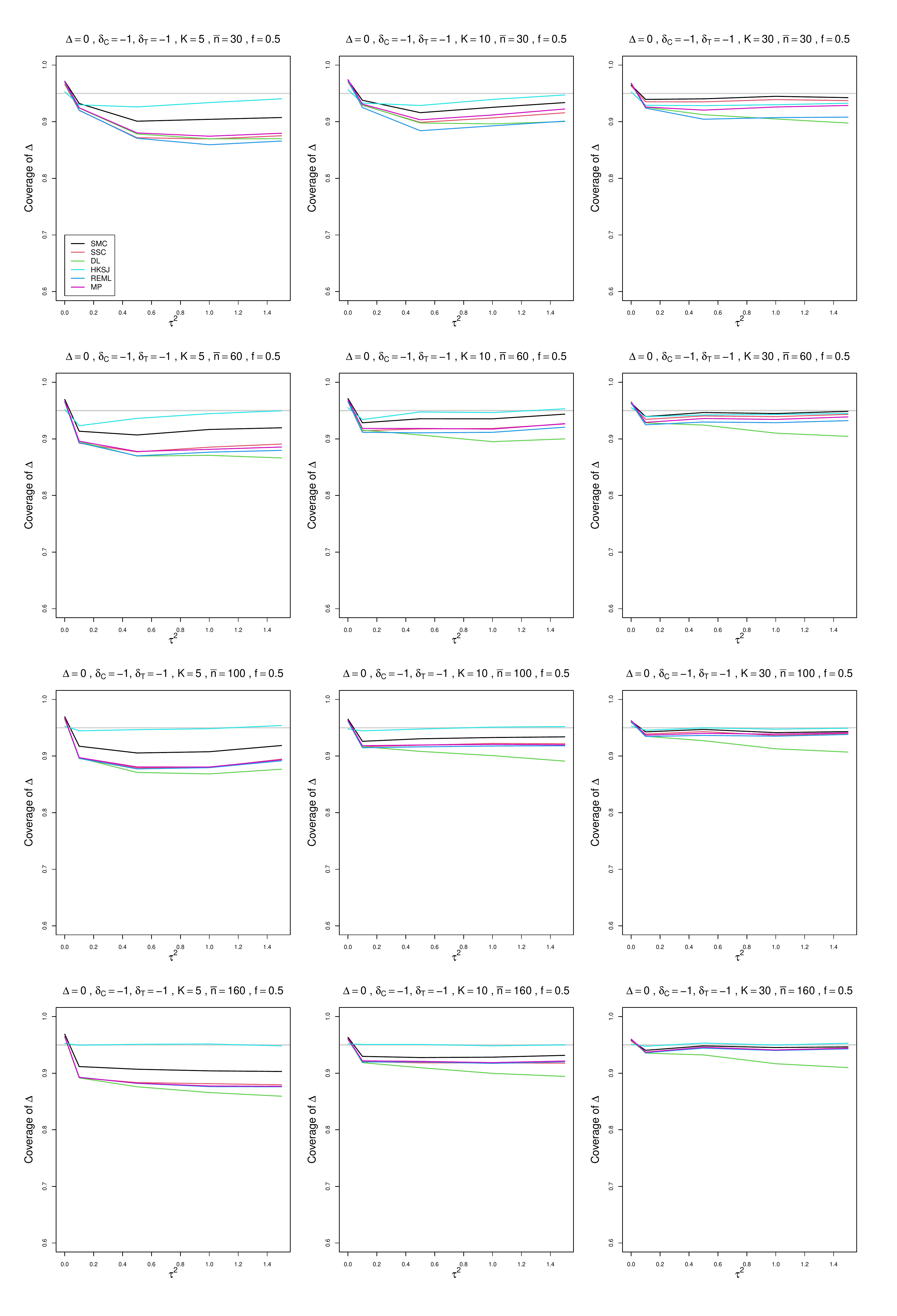}
	\caption{Coverage of 95\% confidence intervals for DSM (DL, REML, MP, HKSJ (DL), SMC and SSC intervals)  vs $\tau^2$, for unequal sample sizes $\bar{n}=30,\;60,\;100$ and $160$, $\delta_{iC} = -1$, $\Delta=0$ and  $f = 0.5$.   }
	\label{PlotCoverageOfDelta_deltaC_-1deltaT=-1_DSM_unequal_sample_sizes.pdf}
\end{figure}

\begin{figure}[ht]
	\centering
	\includegraphics[scale=0.33]{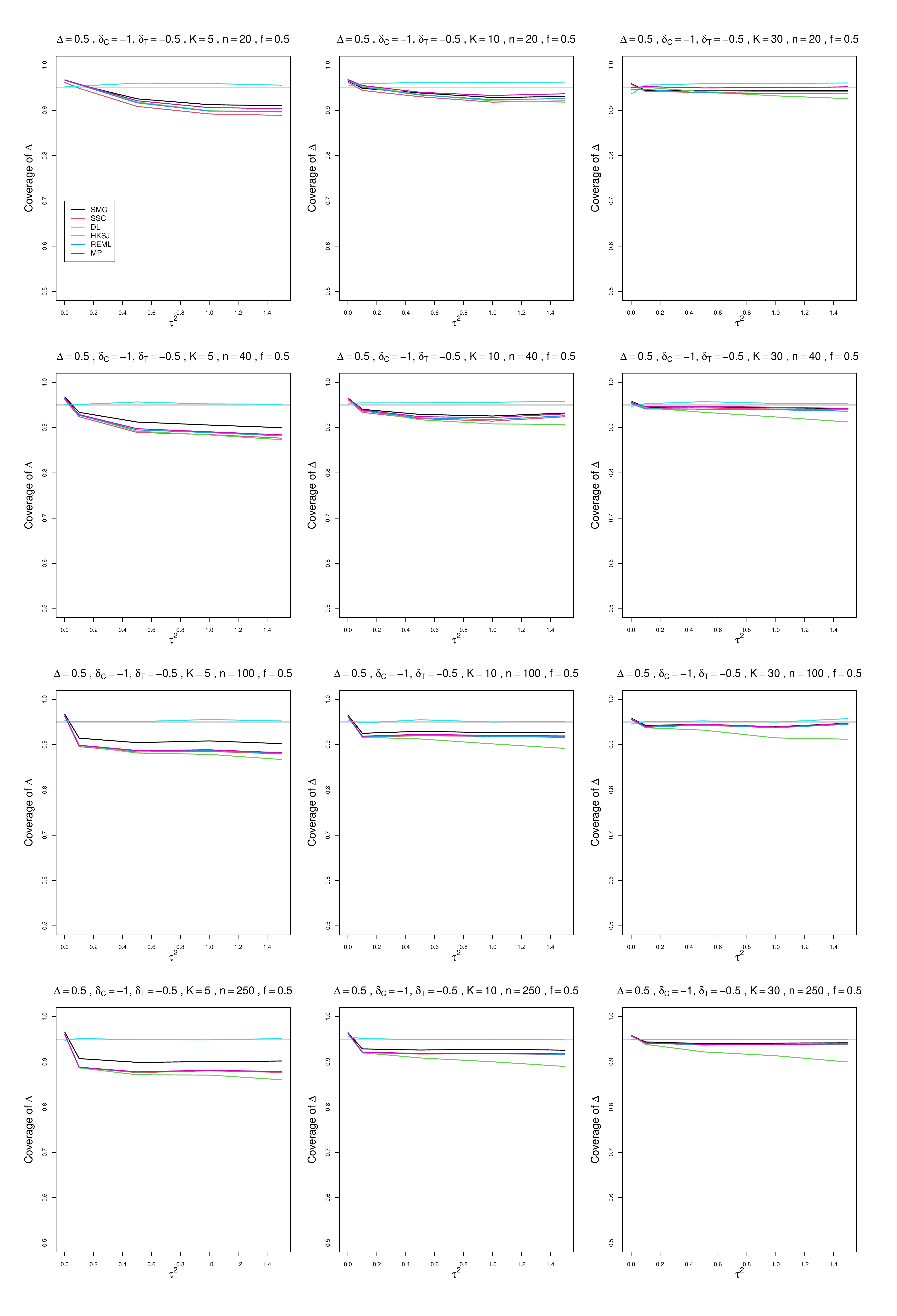}
	\caption{Coverage of 95\% confidence intervals for DSM (DL, REML, MP, HKSJ (DL), SMC and SSC intervals)  vs $\tau^2$, for equal sample sizes $n=20,\;40,\;100$ and $250$, $\delta_{iC} = -1$, $\Delta=0.5$ and  $f = 0.5$.   }
	\label{PlotCoverageOfDelta_deltaC_-1deltaT=-0.5_DSM_equal_sample_sizes.pdf}
\end{figure}

\begin{figure}[ht]
	\centering
	\includegraphics[scale=0.33]{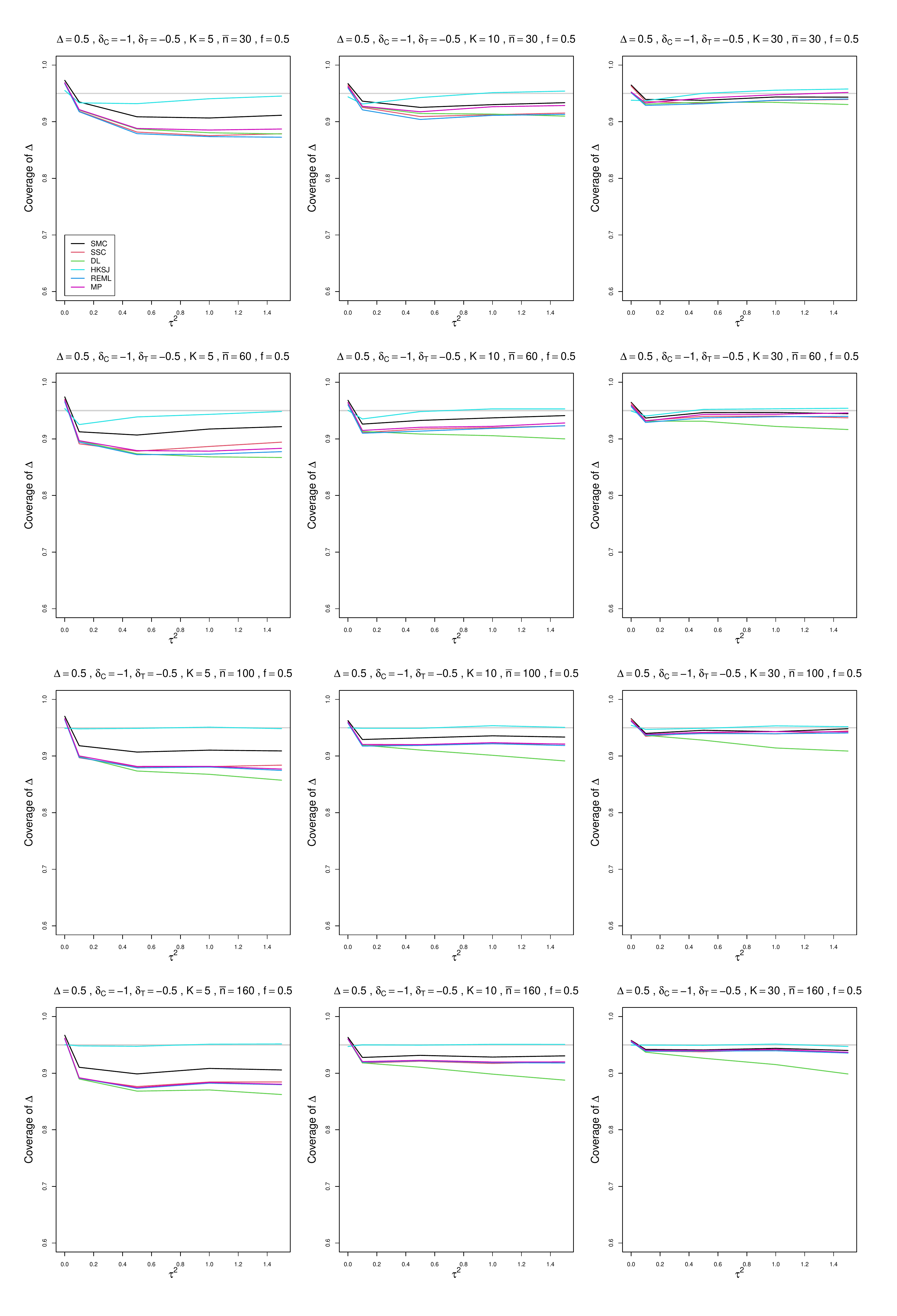}
	\caption{Coverage of 95\% confidence intervals for DSM (DL, REML, MP, HKSJ (DL), SMC and SSC intervals)  vs $\tau^2$, for unequal sample sizes $\bar{n}=30,\;60,\;100$ and $160$, $\delta_{iC} = -1$, $\Delta=0.5$ and  $f = 0.5$.   }
	\label{PlotCoverageOfDelta_deltaC_-1deltaT=-0,5_DSM_unequal_sample_sizes.pdf}
\end{figure}

\begin{figure}[ht]
	\centering
	\includegraphics[scale=0.33]{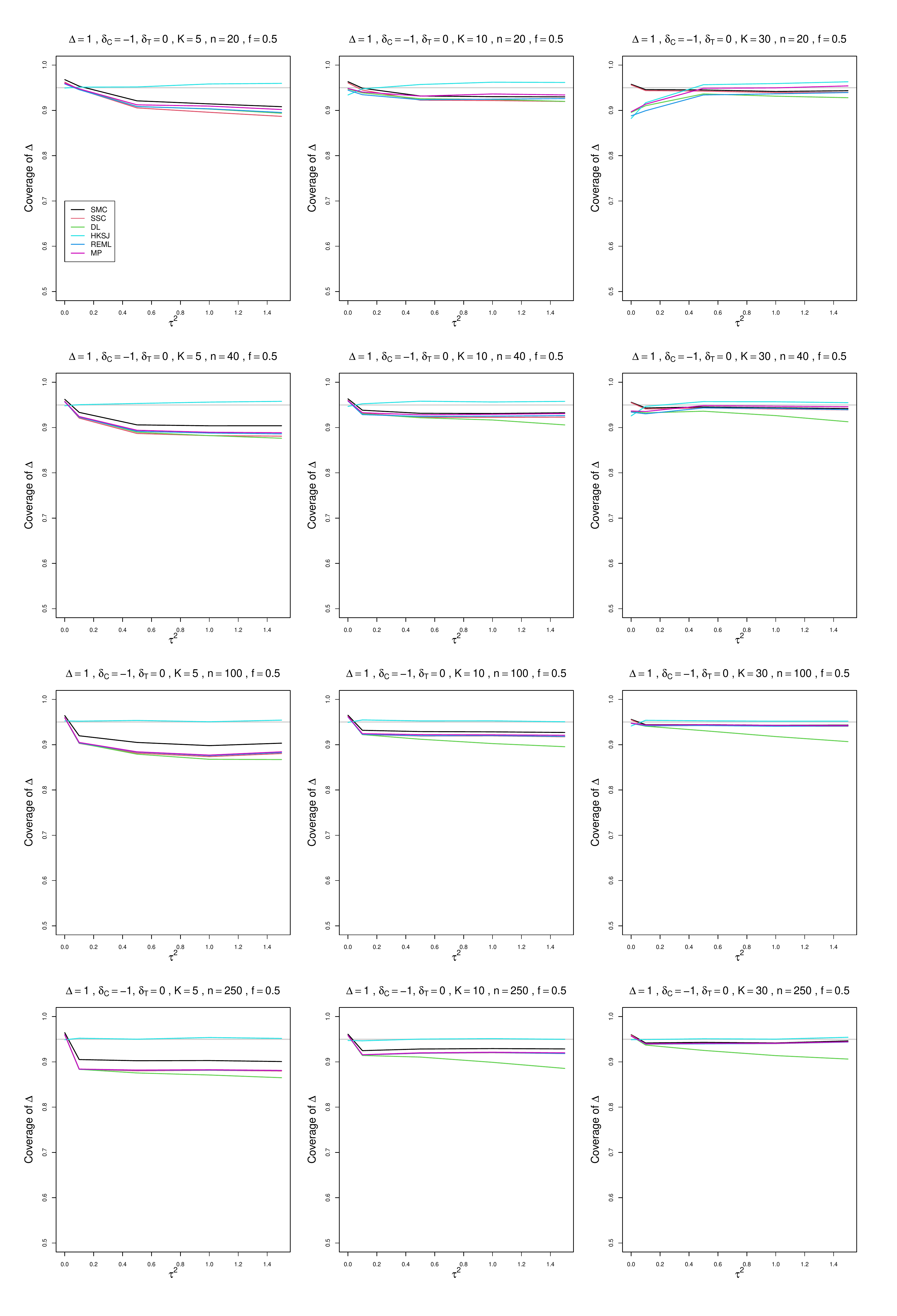}
	\caption{Coverage of 95\% confidence intervals for DSM (DL, REML, MP, HKSJ (DL), SMC and SSC intervals)  vs $\tau^2$, for equal sample sizes $n=20,\;40,\;100$ and $250$, $\delta_{iC} = -1$, $\Delta=1$ and  $f = 0.5$.   }
	\label{PlotCoverageOfDelta_deltaC_--1deltaT=0_DSM_equal_sample_sizes.pdf}
\end{figure}

\begin{figure}[ht]
	\centering
	\includegraphics[scale=0.33]{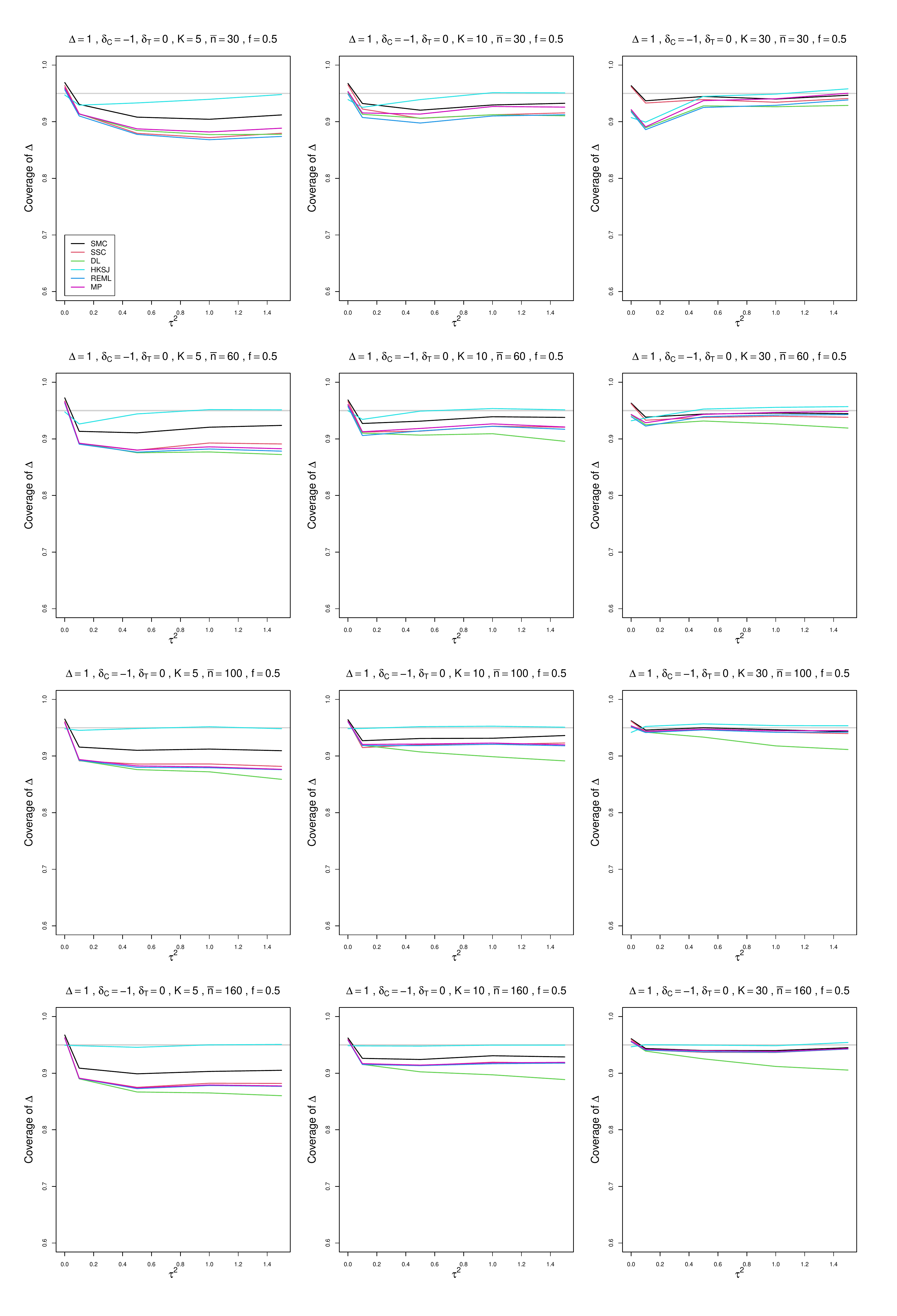}
	\caption{Coverage of 95\% confidence intervals for DSM (DL, REML, MP, HKSJ (DL), SMC and SSC intervals)  vs $\tau^2$, for unequal sample sizes $\bar{n}=30,\;60,\;100$ and $160$, $\delta_{iC} = -1$, $\Delta=1$ and  $f = 0.5$.   }
	\label{PlotCoverageOfDelta_deltaC_-1deltaT=0_DSM_unequal_sample_sizes.pdf}
\end{figure}

\begin{figure}[ht]
	\centering
	\includegraphics[scale=0.33]{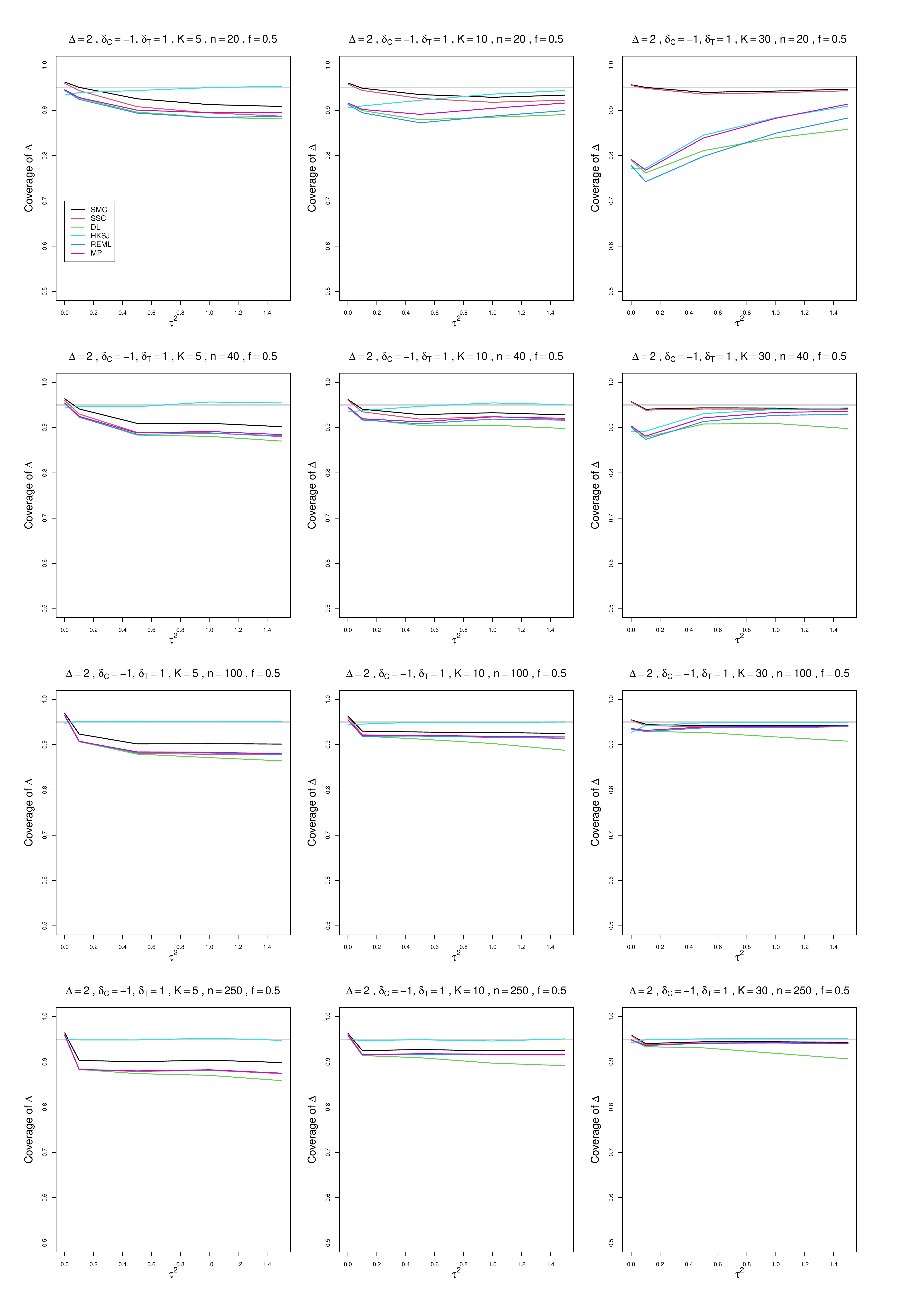}
	\caption{Coverage of 95\% confidence intervals for DSM (DL, REML, MP, HKSJ (DL), SMC and SSC intervals)  vs $\tau^2$, for equal sample sizes $n=20,\;40,\;100$ and $250$, $\delta_{iC} = -1$, $\Delta=2$ and  $f = 0.5$.   }
	\label{PlotCoverageOfDelta_deltaC_-1deltaT=1_DSM_equal_sample_sizes.pdf}
\end{figure}

\begin{figure}[ht]
	\centering
	\includegraphics[scale=0.33]{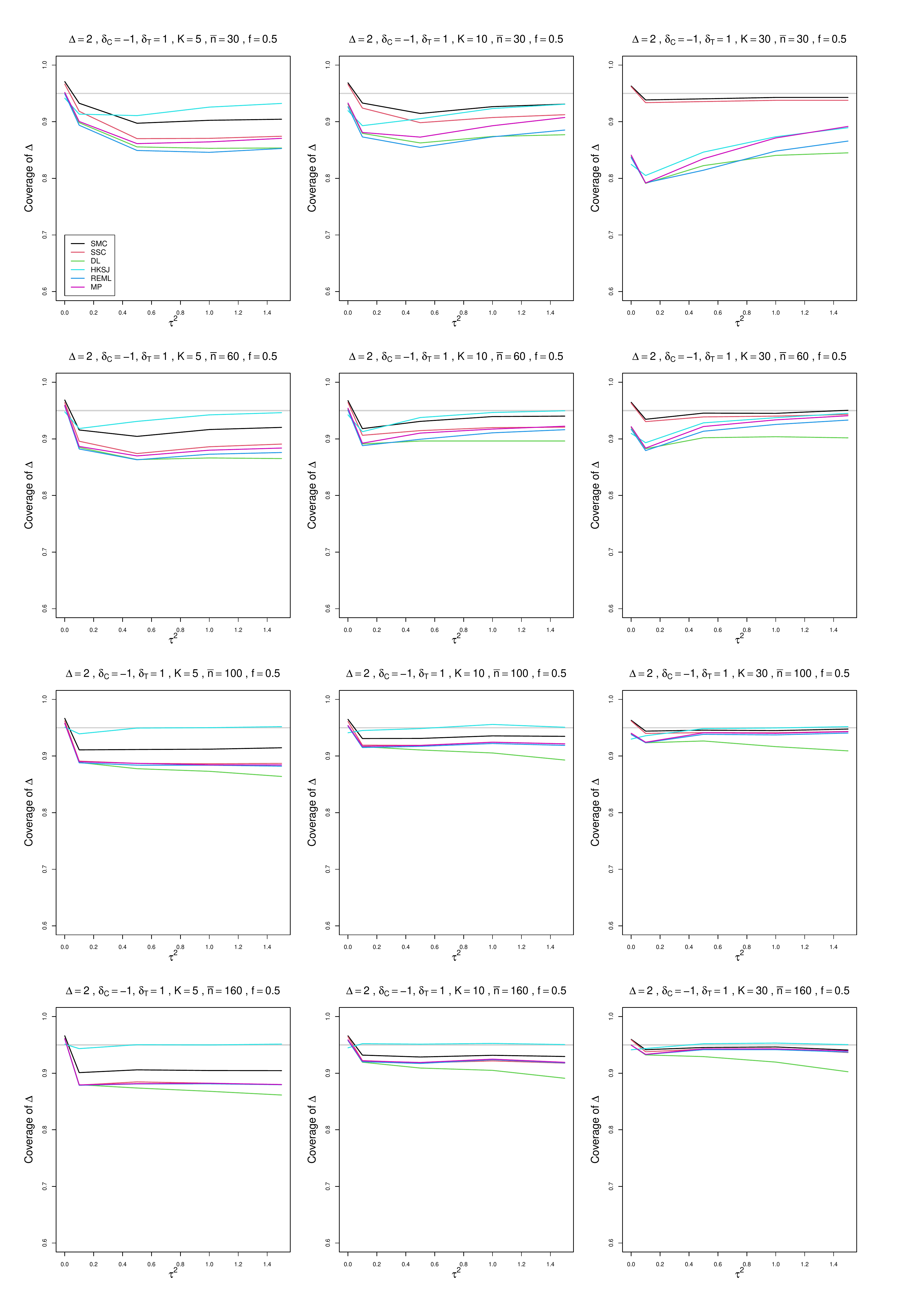}
	\caption{Coverage of 95\% confidence intervals for DSM (DL, REML, MP, HKSJ (DL), SMC and SSC intervals)  vs $\tau^2$, for unequal sample sizes $\bar{n}=30,\;60,\;100$ and $160$, $\delta_{iC} = -1$, $\Delta=2$ and  $f = 0.5$.   }
	\label{PlotCoverageOfDelta_deltaC_-1deltaT=1_DSM_unequal_sample_sizes.pdf}
\end{figure}


\begin{figure}[ht]
	\centering
	\includegraphics[scale=0.33]{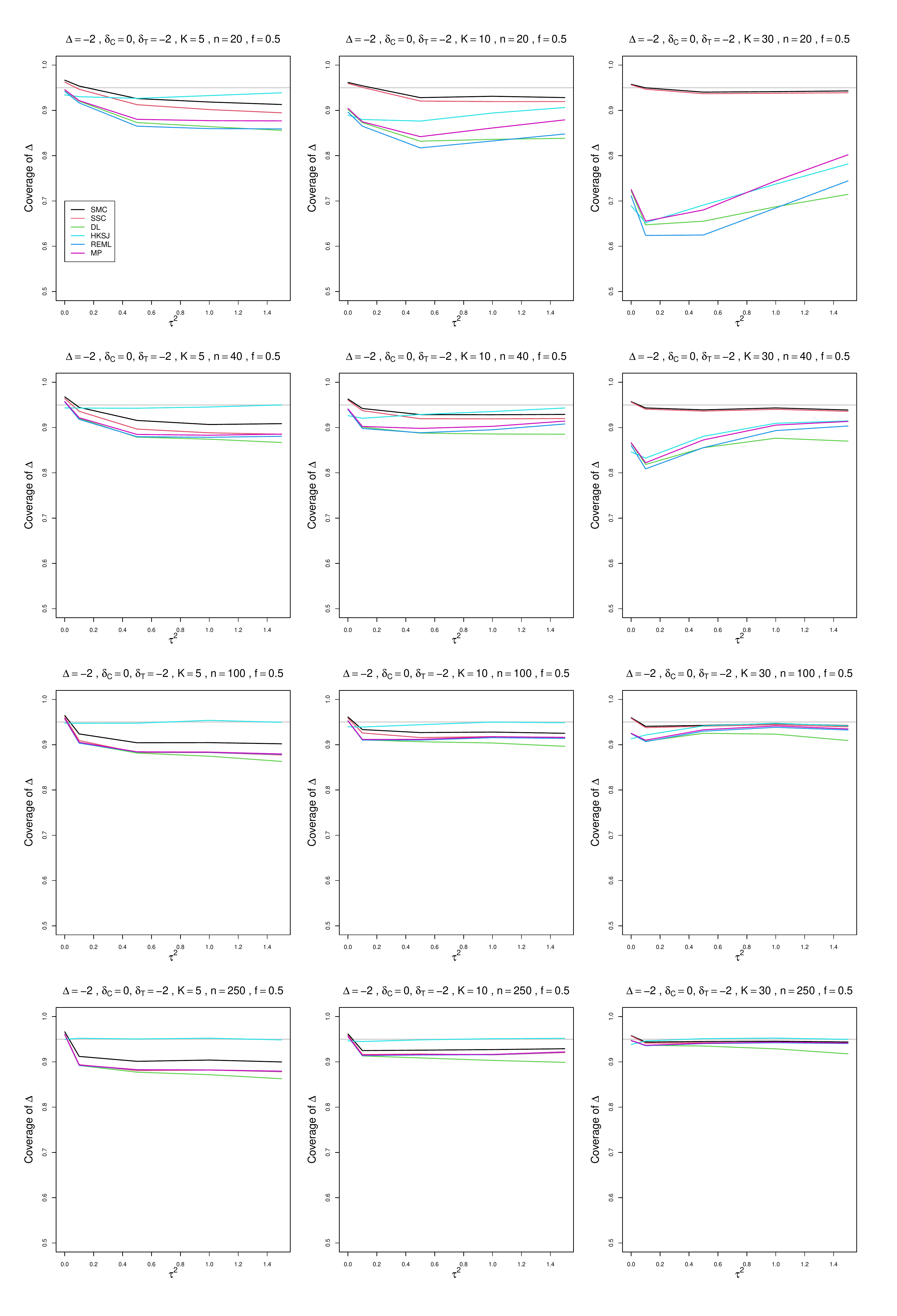}
	\caption{Coverage of 95\% confidence intervals for DSM (DL, REML, MP, HKSJ (DL), SMC and SSC intervals)  vs $\tau^2$, for equal sample sizes $n=20,\;40,\;100$ and $250$, $\delta_{iC} = 0$, $\Delta=-2$ and  $f = 0.5$.   }
	\label{PlotCoverageOfDelta_deltaC_0deltaT=-2_DSM_equal_sample_sizes.pdf}
\end{figure}

\begin{figure}[ht]
	\centering
	\includegraphics[scale=0.33]{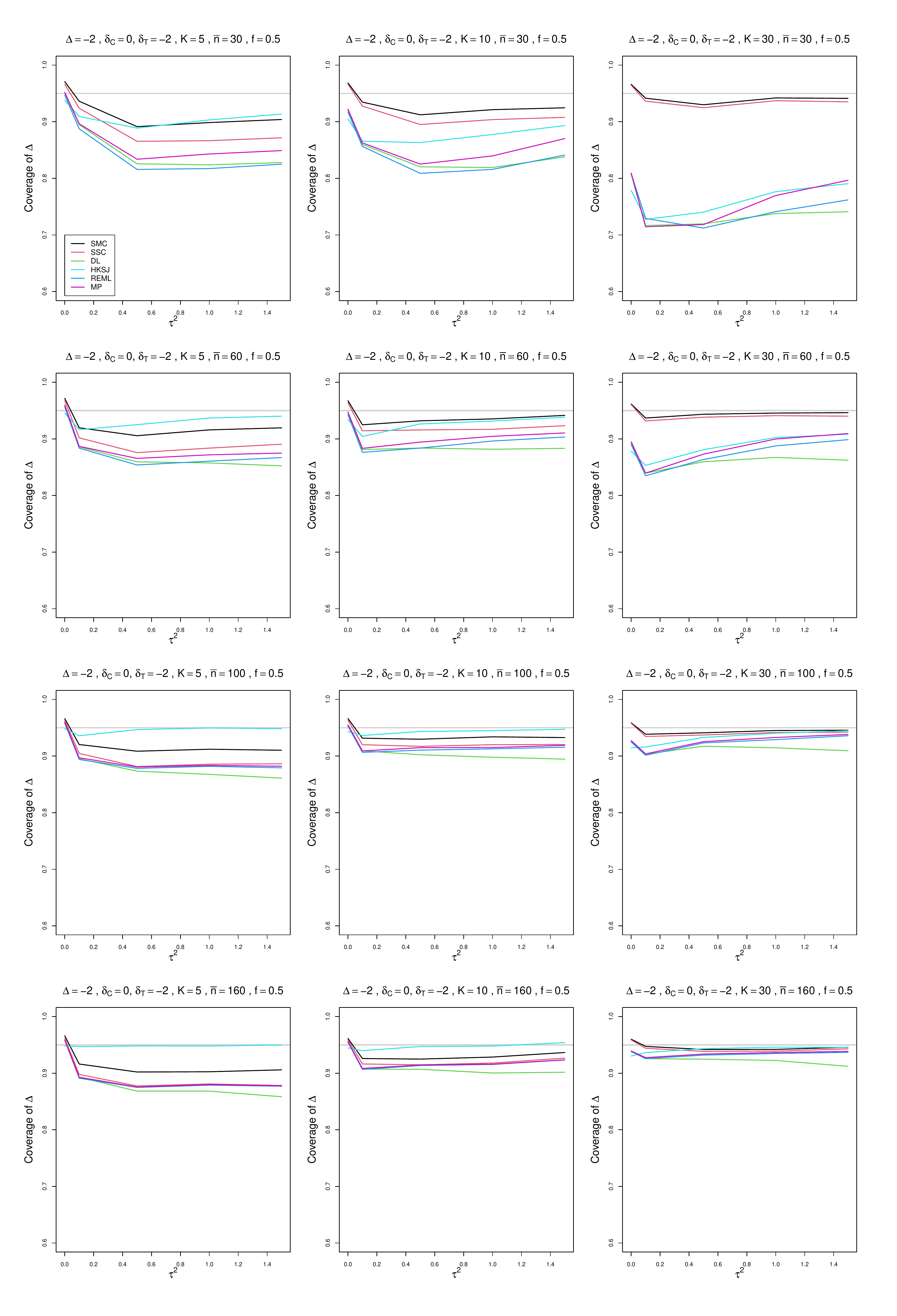}
	\caption{Coverage of 95\% confidence intervals for DSM (DL, REML, MP, HKSJ (DL), SMC and SSC intervals)  vs $\tau^2$, for unequal sample sizes $\bar{n}=30,\;60,\;100$ and $160$, $\delta_{iC} = 0$, $\Delta=-2$ and  $f = 0.5$.   }
	\label{PlotCoverageOfDelta_deltaC_-1deltaT=-3_DSM_unequal_sample_sizes.pdf}
\end{figure}

\begin{figure}[ht]
	\centering
	\includegraphics[scale=0.33]{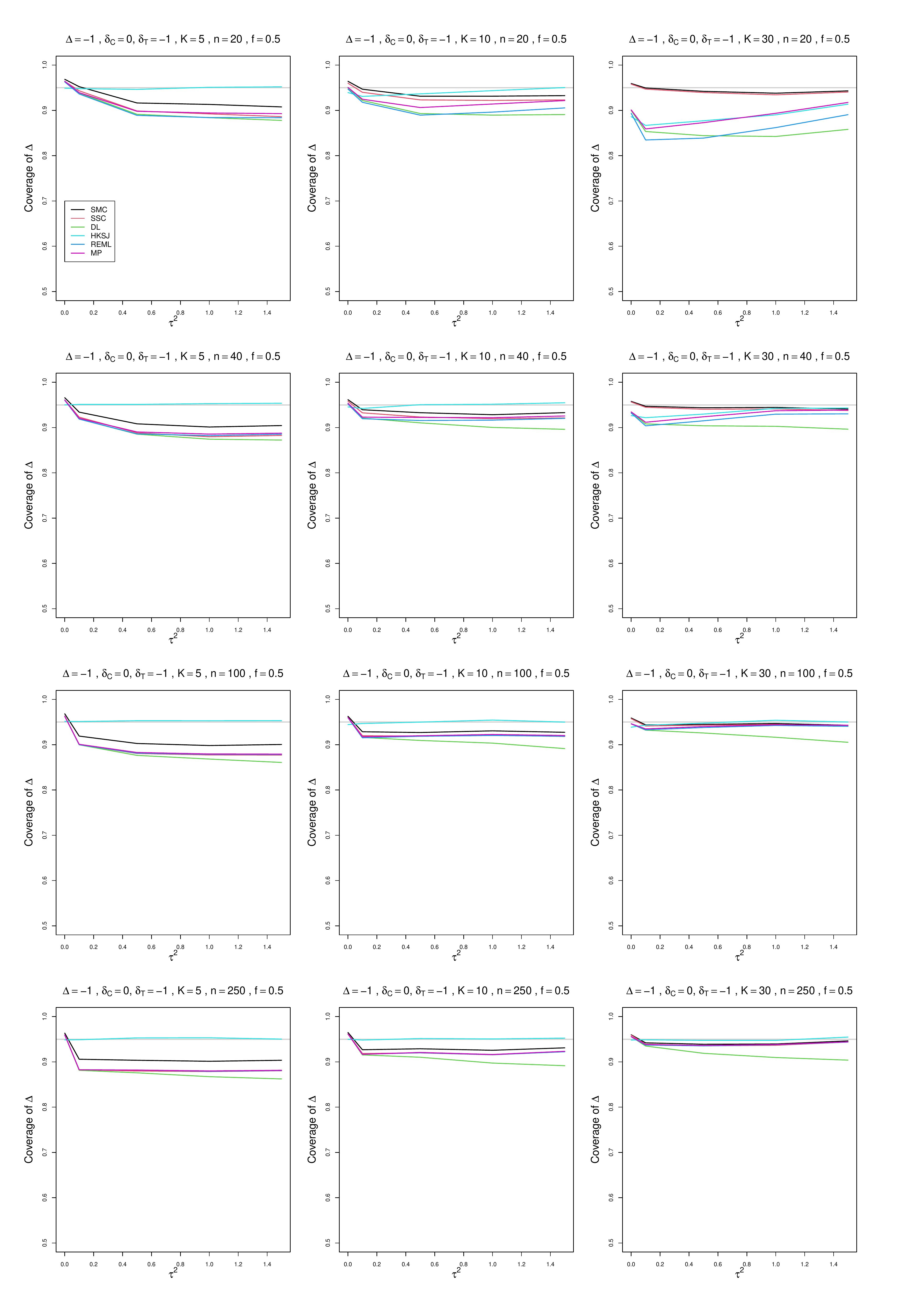}
	\caption{Coverage of 95\% confidence intervals for DSM (DL, REML, MP, HKSJ (DL), SMC and SSC intervals)  vs $\tau^2$, for equal sample sizes $n=20,\;40,\;100$ and $250$, $\delta_{iC} = 0$, $\Delta=-1$ and  $f = 0.5$.   }
	\label{PlotCoverageOfDelta_deltaC_-0deltaT=-1_DSM_equal_sample_sizes.pdf}
\end{figure}

\begin{figure}[ht]
	\centering
	\includegraphics[scale=0.33]{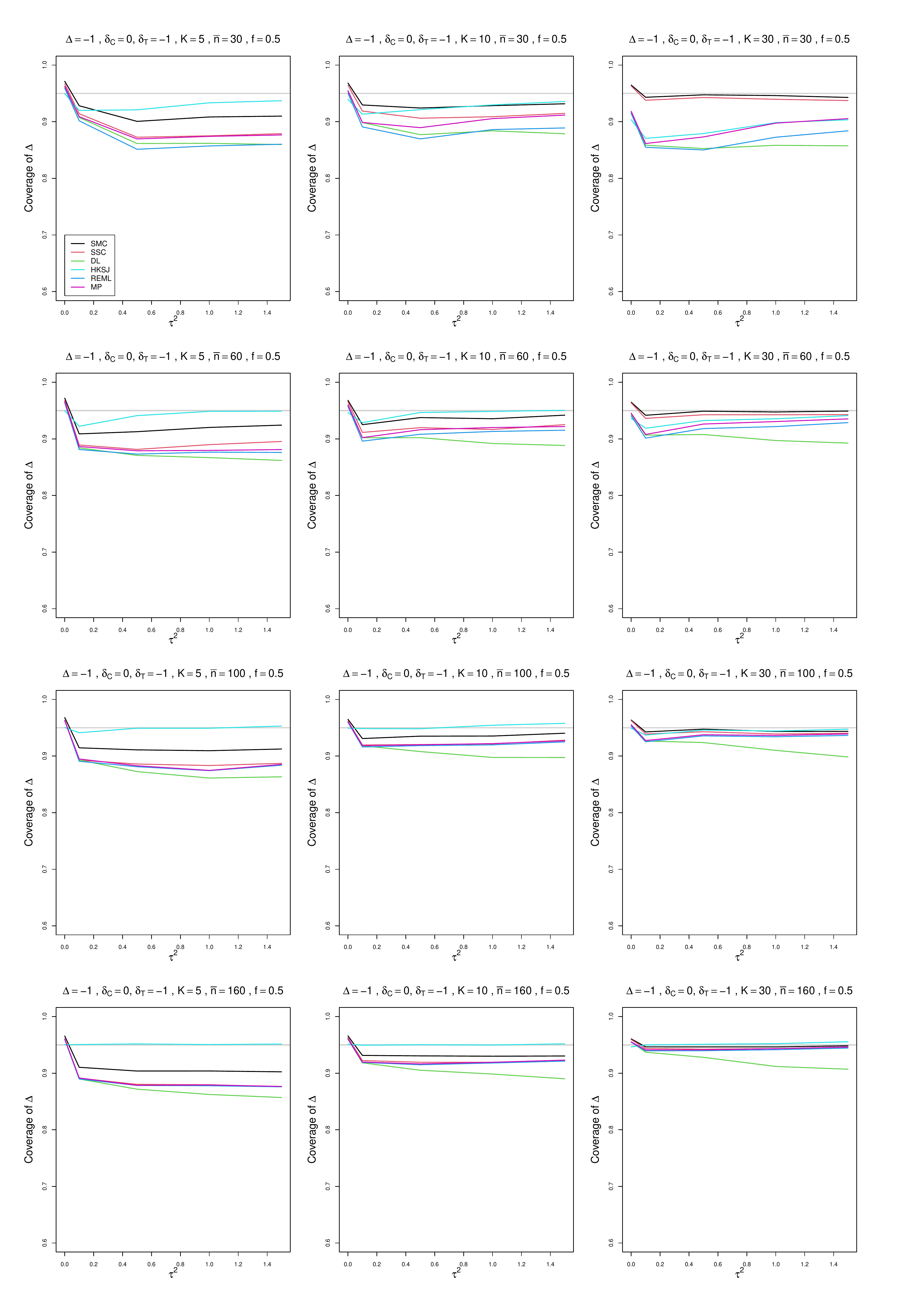}
	\caption{Coverage of 95\% confidence intervals for DSM (DL, REML, MP, HKSJ (DL), SMC and SSC intervals)  vs $\tau^2$, for unequal sample sizes $\bar{n}=30,\;60,\;100$ and $160$, $\delta_{iC} = 0$, $\Delta=-1$ and  $f = 0.5$.   }
	\label{PlotCoverageOfDelta_deltaC_0deltaT=-1_DSM_unequal_sample_sizes.pdf}
\end{figure}

\begin{figure}[ht]
	\centering
	\includegraphics[scale=0.33]{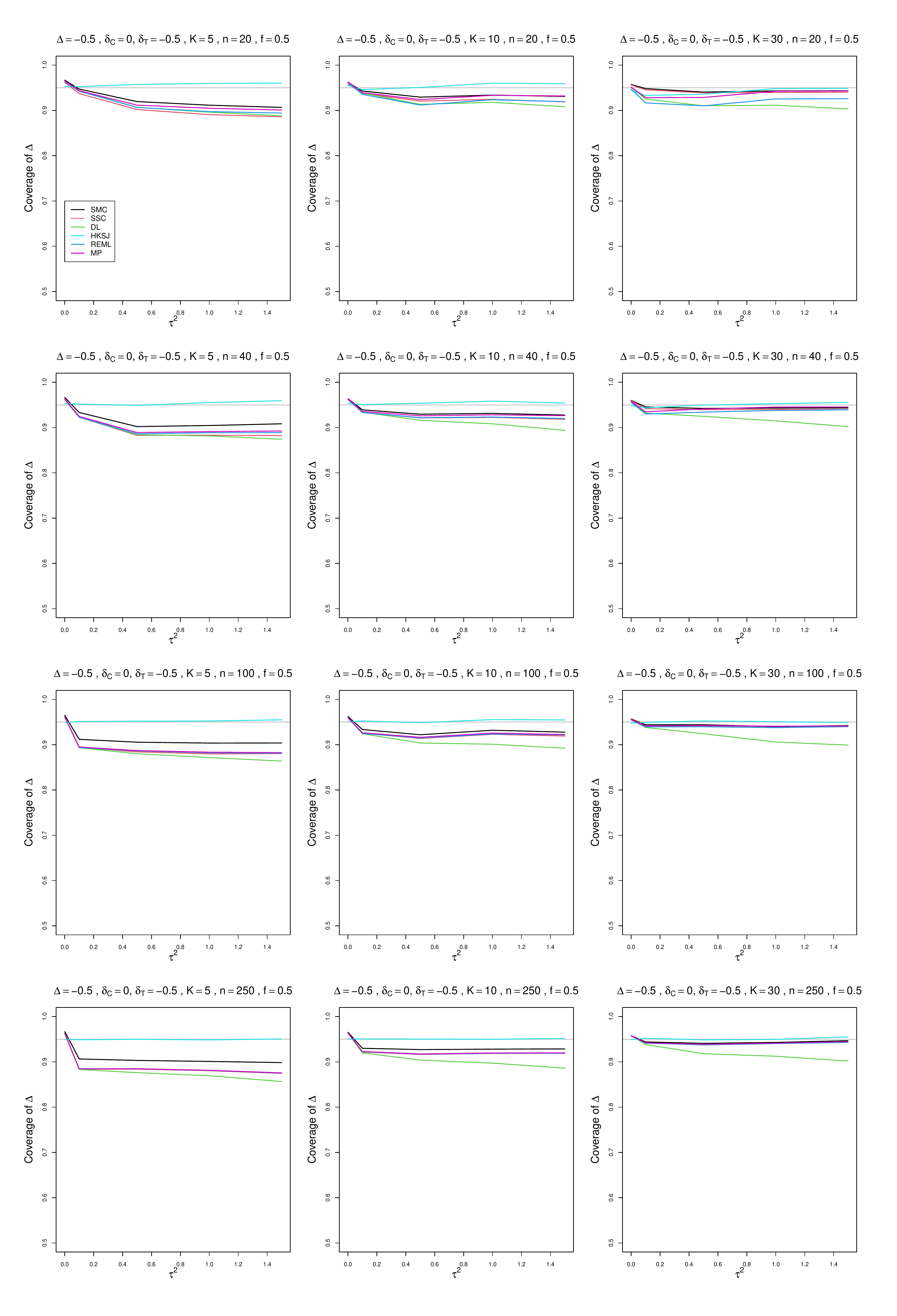}
	\caption{Coverage of 95\% confidence intervals for DSM (DL, REML, MP, HKSJ (DL), SMC and SSC intervals)  vs $\tau^2$, for equal sample sizes $n=20,\;40,\;100$ and $250$, $\delta_{iC} = 0$, $\Delta=-0.5$ and  $f = 0.5$.   }
	\label{PlotCoverageOfDelta_deltaC_-0deltaT=-0.5_DSM_equal_sample_sizes.pdf}
\end{figure}

\begin{figure}[ht]
	\centering
	\includegraphics[scale=0.33]{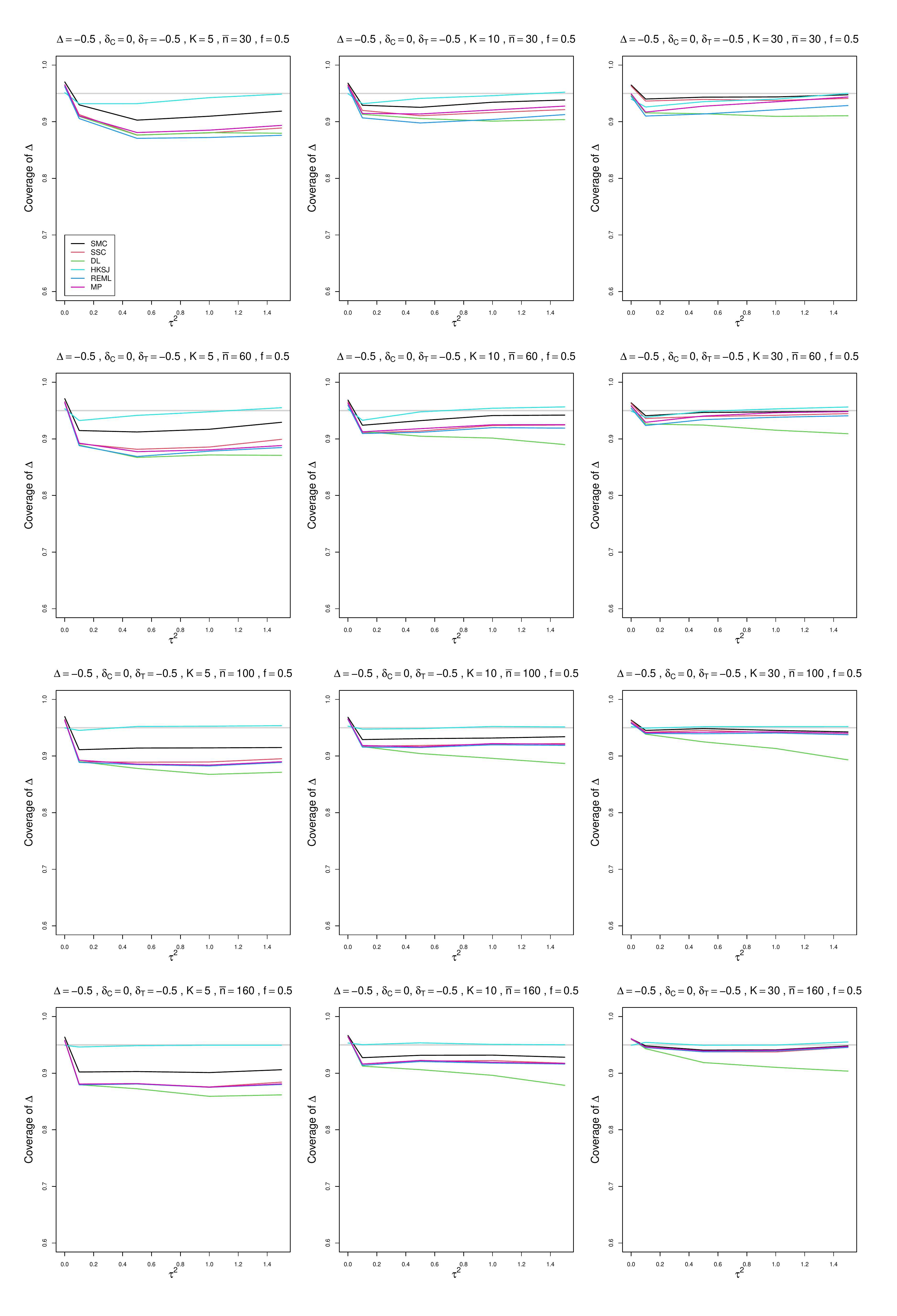}
	\caption{Coverage of 95\% confidence intervals for DSM (DL, REML, MP, HKSJ (DL), SMC and SSC intervals)  vs $\tau^2$, for unequal sample sizes $\bar{n}=30,\;60,\;100$ and $160$, $\delta_{iC} = 0$, $\Delta=-0.5$ and  $f = 0.5$.   }
	\label{PlotCoverageOfDelta_deltaC_0deltaT=-0.5_DSM_unequal_sample_sizes.pdf}
\end{figure}

\begin{figure}[ht]
	\centering
	\includegraphics[scale=0.33]{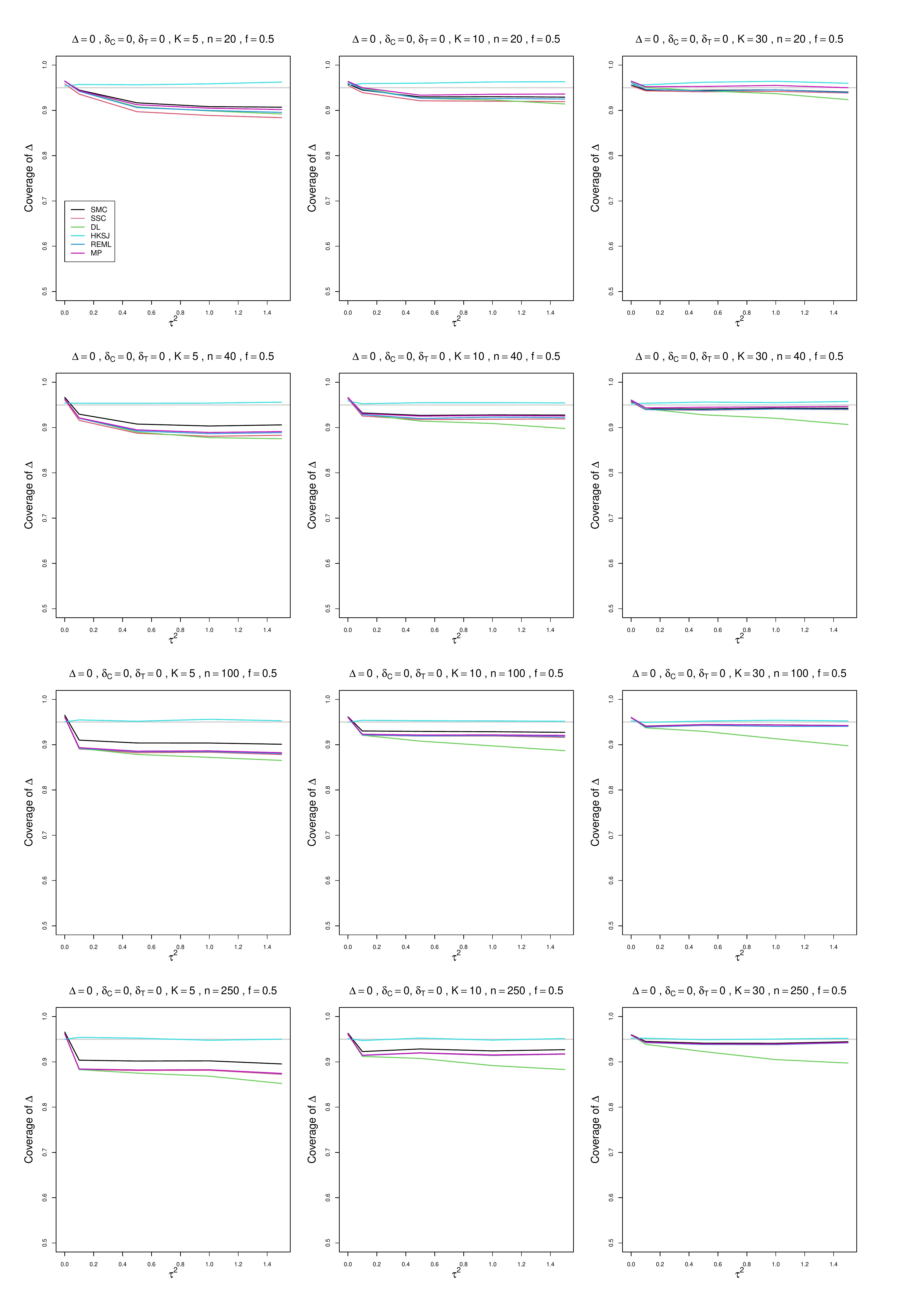}
	\caption{Coverage of 95\% confidence intervals for DSM (DL, REML, MP, HKSJ (DL), SMC and SSC intervals)  vs $\tau^2$, for equal sample sizes $n=20,\;40,\;100$ and $250$, $\delta_{iC} = 0$, $\Delta=0$ and  $f = 0.5$.   }
	\label{PlotCoverageOfDelta_deltaC_0deltaT=0_DSM_equal_sample_sizes.pdf}
\end{figure}

\begin{figure}[ht]
	\centering
	\includegraphics[scale=0.33]{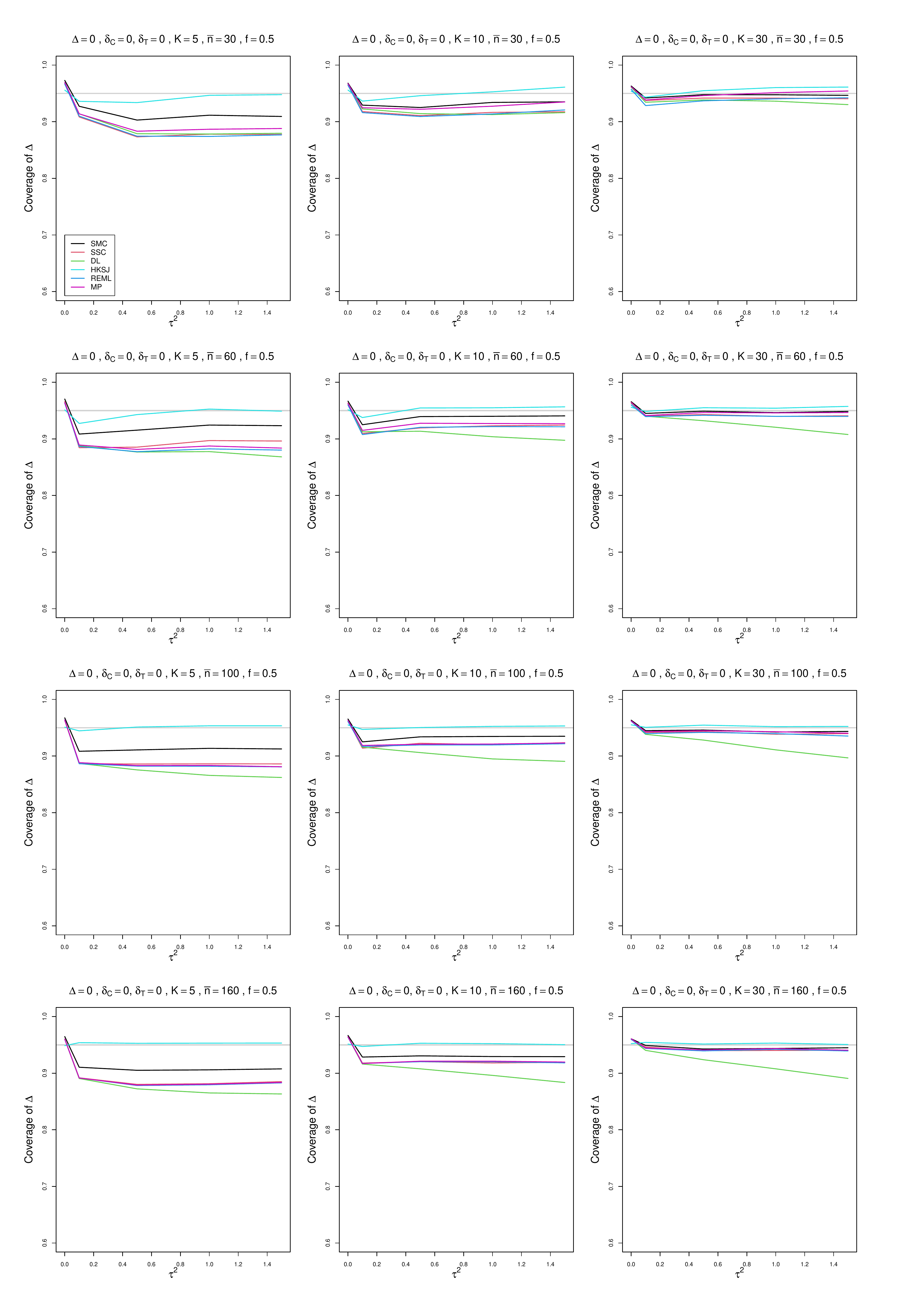}
	\caption{Coverage of 95\% confidence intervals for DSM (DL, REML, MP, HKSJ (DL), SMC and SSC intervals)  vs $\tau^2$, for unequal sample sizes $\bar{n}=30,\;60,\;100$ and $160$, $\delta_{iC} = 0$, $\Delta=0$ and  $f = 0.5$.   }
	\label{PlotCoverageOfDelta_deltaC_0deltaT=0_DSM_unequal_sample_sizes.pdf}
\end{figure}

\begin{figure}[ht]
	\centering
	\includegraphics[scale=0.33]{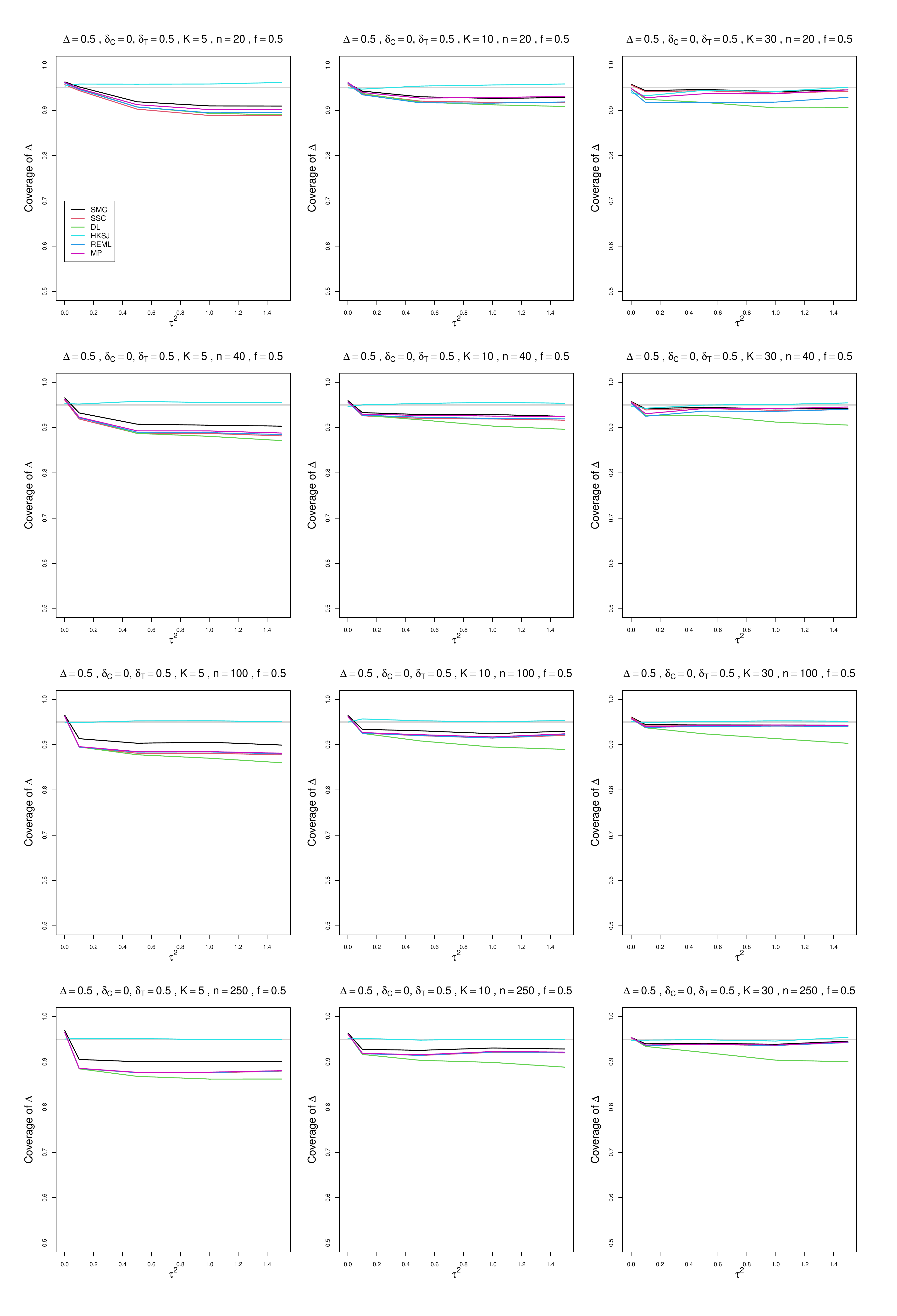}
	\caption{Coverage of 95\% confidence intervals for DSM (DL, REML, MP, HKSJ (DL), SMC and SSC intervals)  vs $\tau^2$, for equal sample sizes $n=20,\;40,\;100$ and $250$, $\delta_{iC} = 0$, $\Delta=0.5$ and  $f = 0.5$.   }
	\label{PlotCoverageOfDelta_deltaC_0deltaT=0.5_DSM_equal_sample_sizes.pdf}
\end{figure}

\begin{figure}[ht]
	\centering
	\includegraphics[scale=0.33]{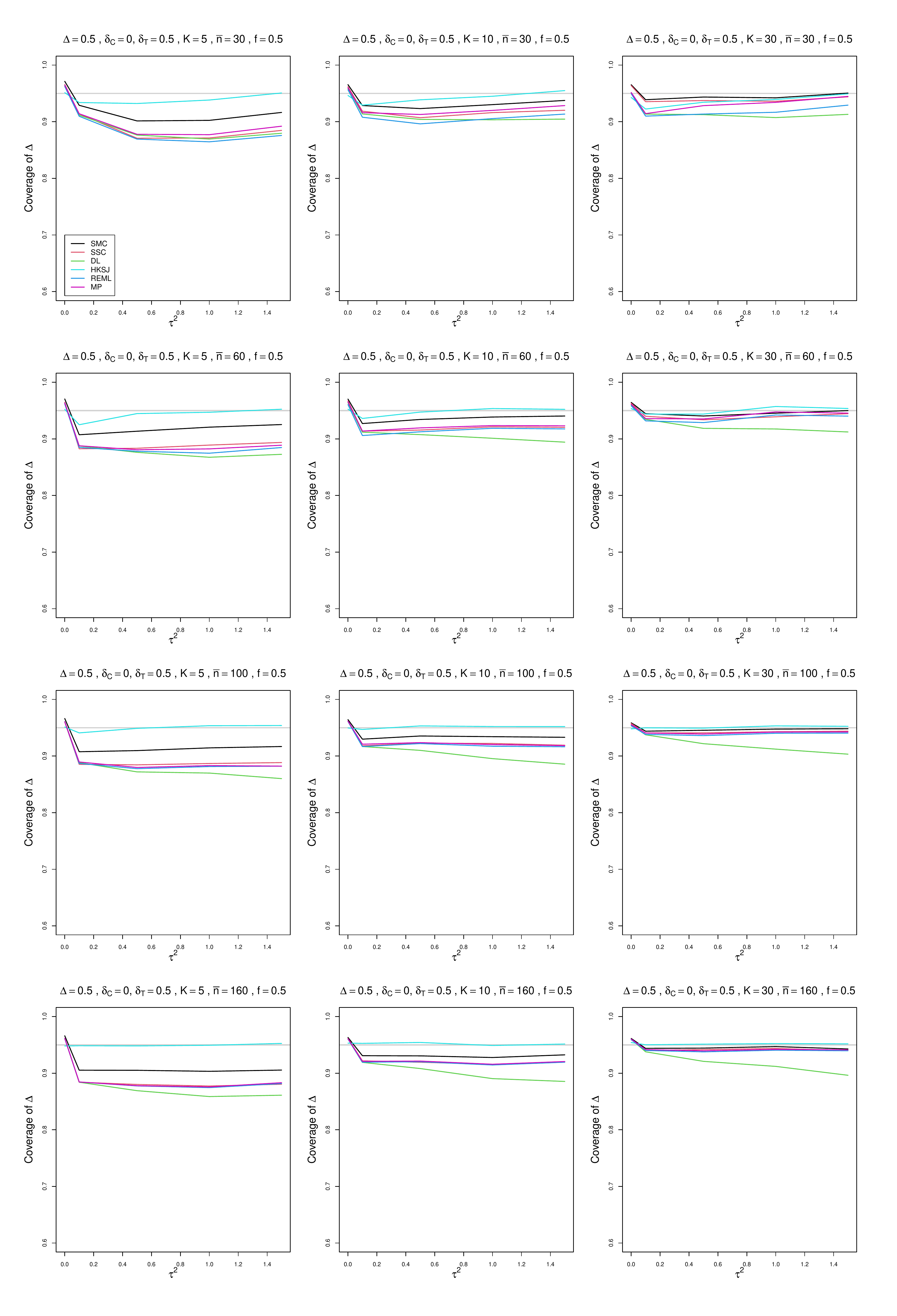}
	\caption{Coverage of 95\% confidence intervals for DSM (DL, REML, MP, HKSJ (DL), SMC and SSC intervals)  vs $\tau^2$, for unequal sample sizes $\bar{n}=30,\;60,\;100$ and $160$, $\delta_{iC} = 0$, $\Delta=0.5$ and  $f = 0.5$.   }
	\label{PlotCoverageOfDelta_deltaC_0deltaT=-0.5_DSM_unequal_sample_sizes.pdf}
\end{figure}

\begin{figure}[ht]
	\centering
	\includegraphics[scale=0.33]{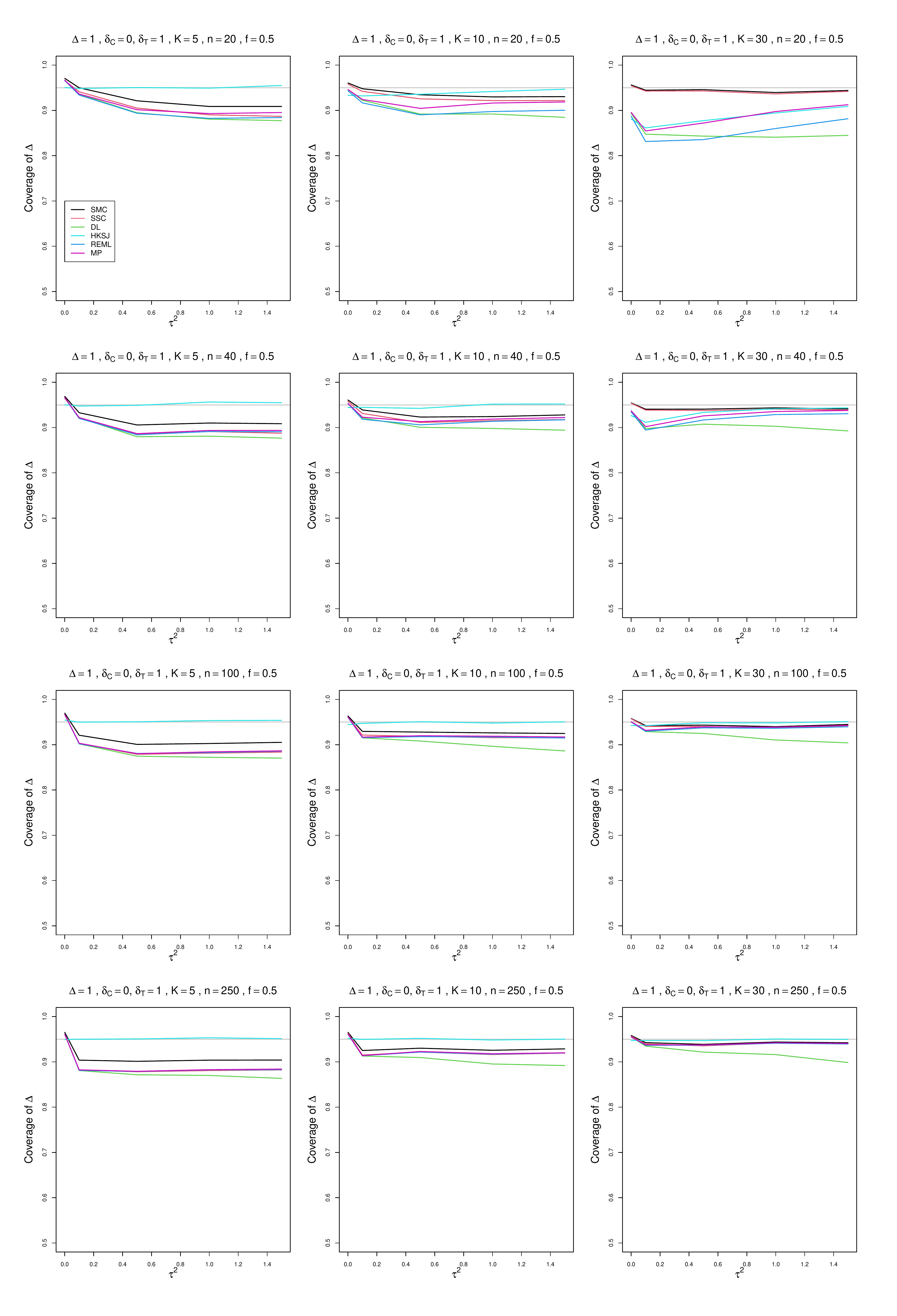}
	\caption{Coverage of 95\% confidence intervals for DSM (DL, REML, MP, HKSJ (DL), SMC and SSC intervals)  vs $\tau^2$, for equal sample sizes $n=20,\;40,\;100$ and $250$, $\delta_{iC} = 0$, $\Delta=1$ and  $f = 0.5$.   }
	\label{PlotCoverageOfDelta_deltaC_=0deltaT=1_DSM_equal_sample_sizes.pdf}
\end{figure}

\begin{figure}[ht]
	\centering
	\includegraphics[scale=0.33]{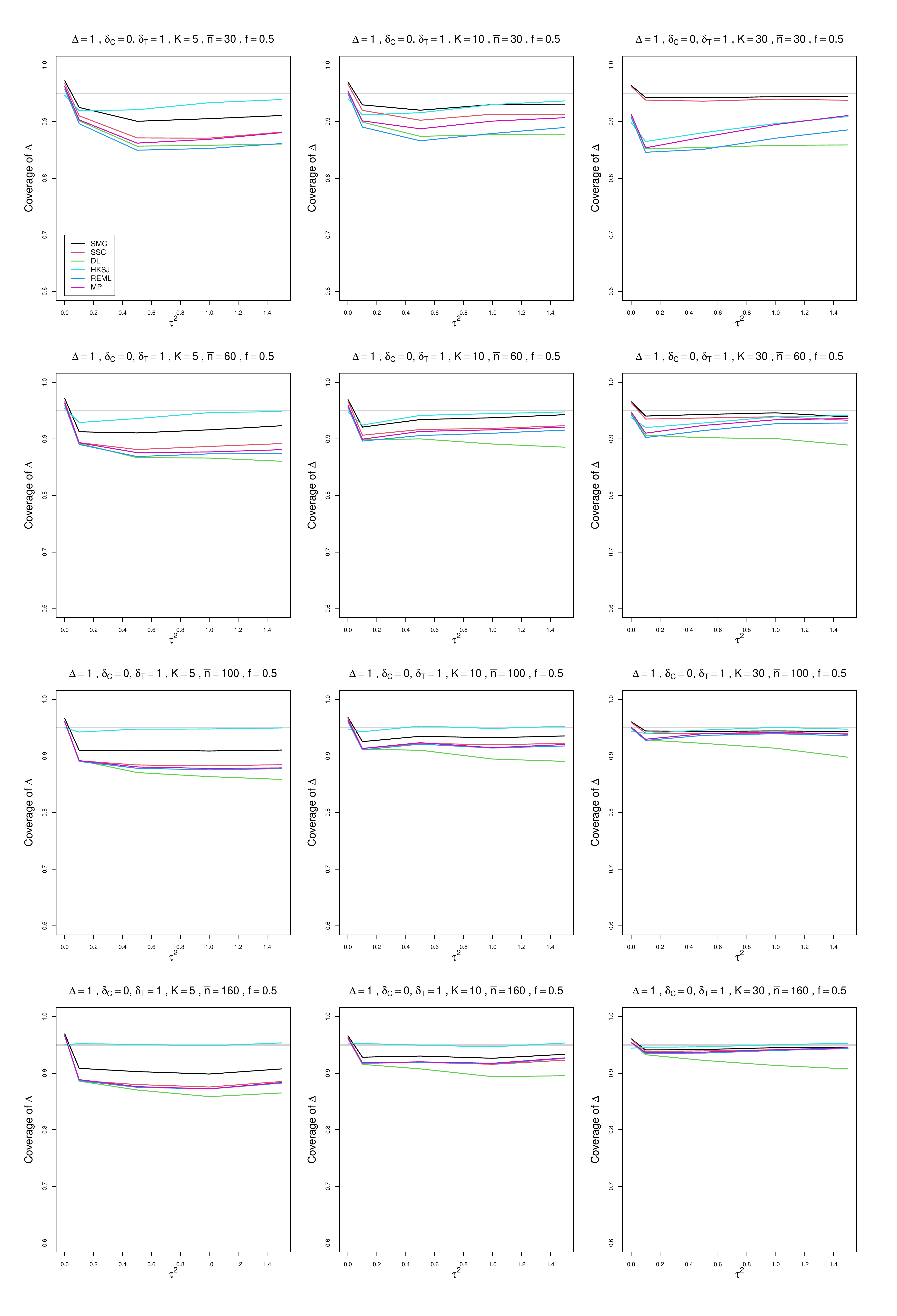}
	\caption{Coverage of 95\% confidence intervals for DSM (DL, REML, MP, HKSJ (DL), SMC and SSC intervals)  vs $\tau^2$, for unequal sample sizes $\bar{n}=30,\;60,\;100$ and $160$, $\delta_{iC} = 0$, $\Delta=1$ and  $f = 0.5$.   }
	\label{PlotCoverageOfDelta_deltaC_0deltaT=1_DSM_unequal_sample_sizes.pdf}
\end{figure}

\begin{figure}[ht]
	\centering
	\includegraphics[scale=0.33]{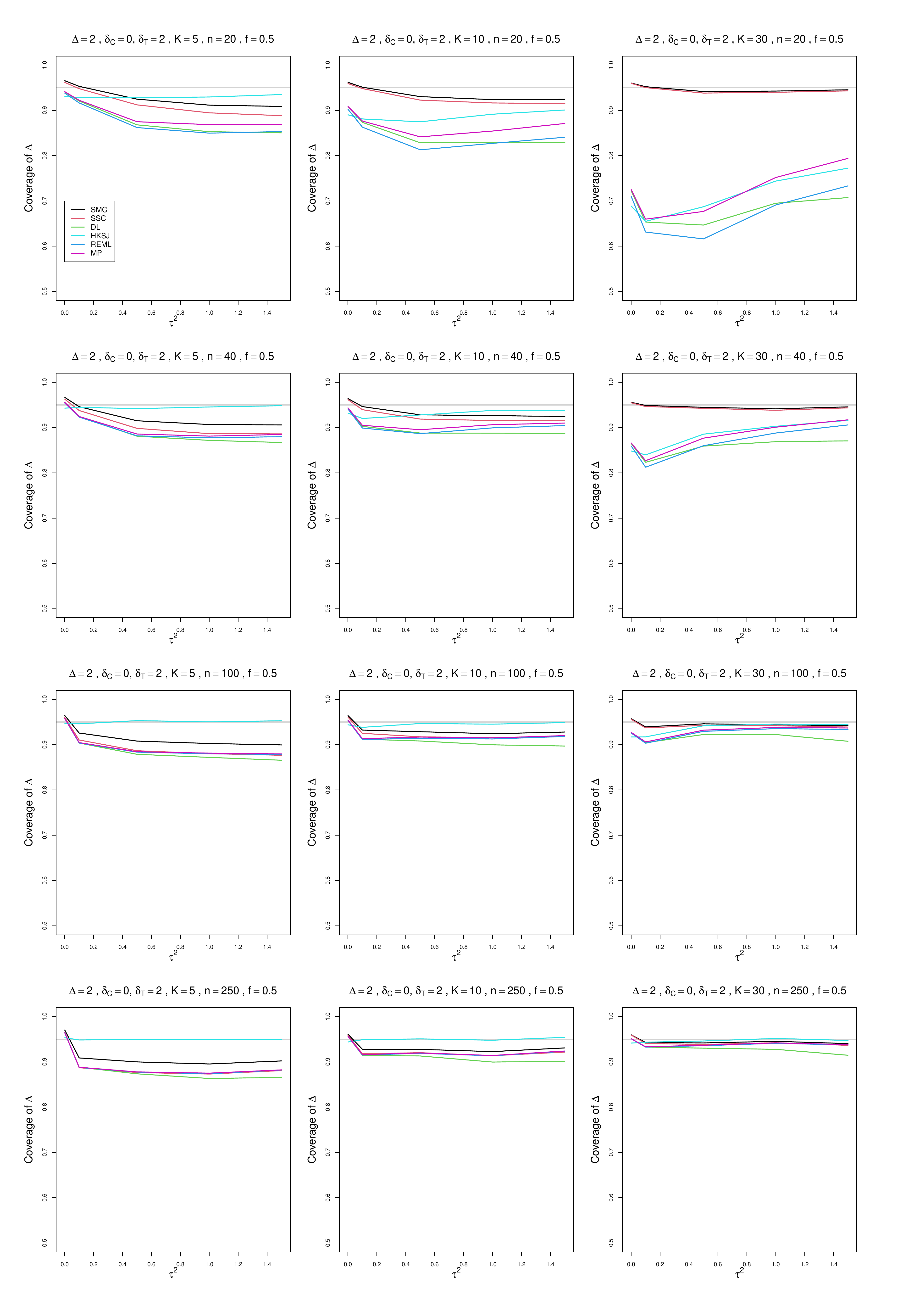}
	\caption{Coverage of 95\% confidence intervals for DSM (DL, REML, MP, HKSJ (DL), SMC and SSC intervals)  vs $\tau^2$, for equal sample sizes $n=20,\;40,\;100$ and $250$, $\delta_{iC} = 0$, $\Delta=2$ and  $f = 0.5$.   }
	\label{PlotCoverageOfDelta_deltaC_0deltaT=2_DSM_equal_sample_sizes.pdf}
\end{figure}

\begin{figure}[ht]
	\centering
	\includegraphics[scale=0.33]{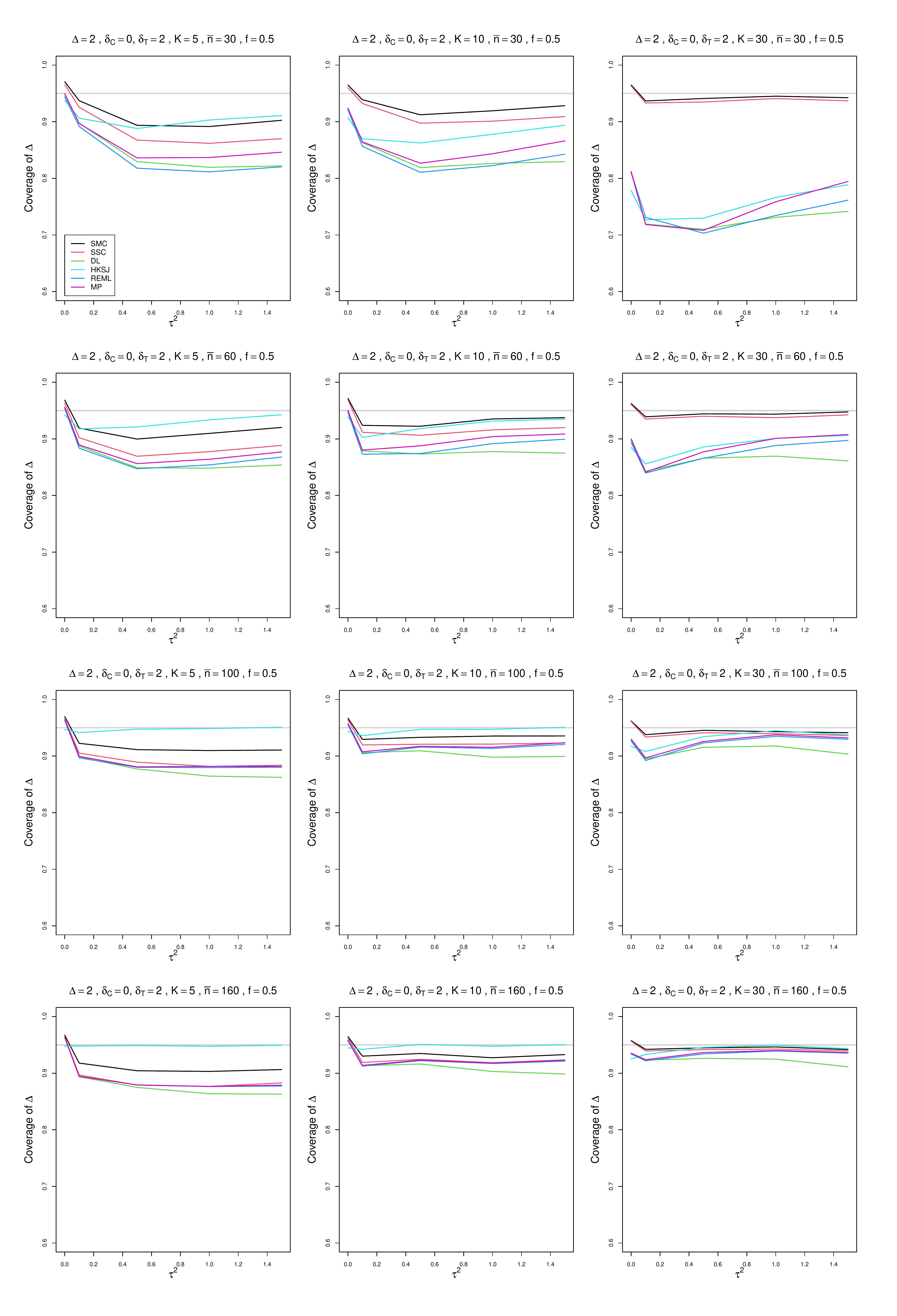}
	\caption{Coverage of 95\% confidence intervals for DSM (DL, REML, MP, HKSJ (DL), SMC and SSC intervals)  vs $\tau^2$, for unequal sample sizes $\bar{n}=30,\;60,\;100$ and $160$, $\delta_{iC} = 0$, $\Delta=2$ and  $f = 0.5$.   }
	\label{PlotCoverageOfDelta_deltaC_0deltaT=2_DSM_unequal_sample_sizes.pdf}
\end{figure}

\clearpage
\begin{figure}[ht]
	\centering
	\includegraphics[scale=0.33]{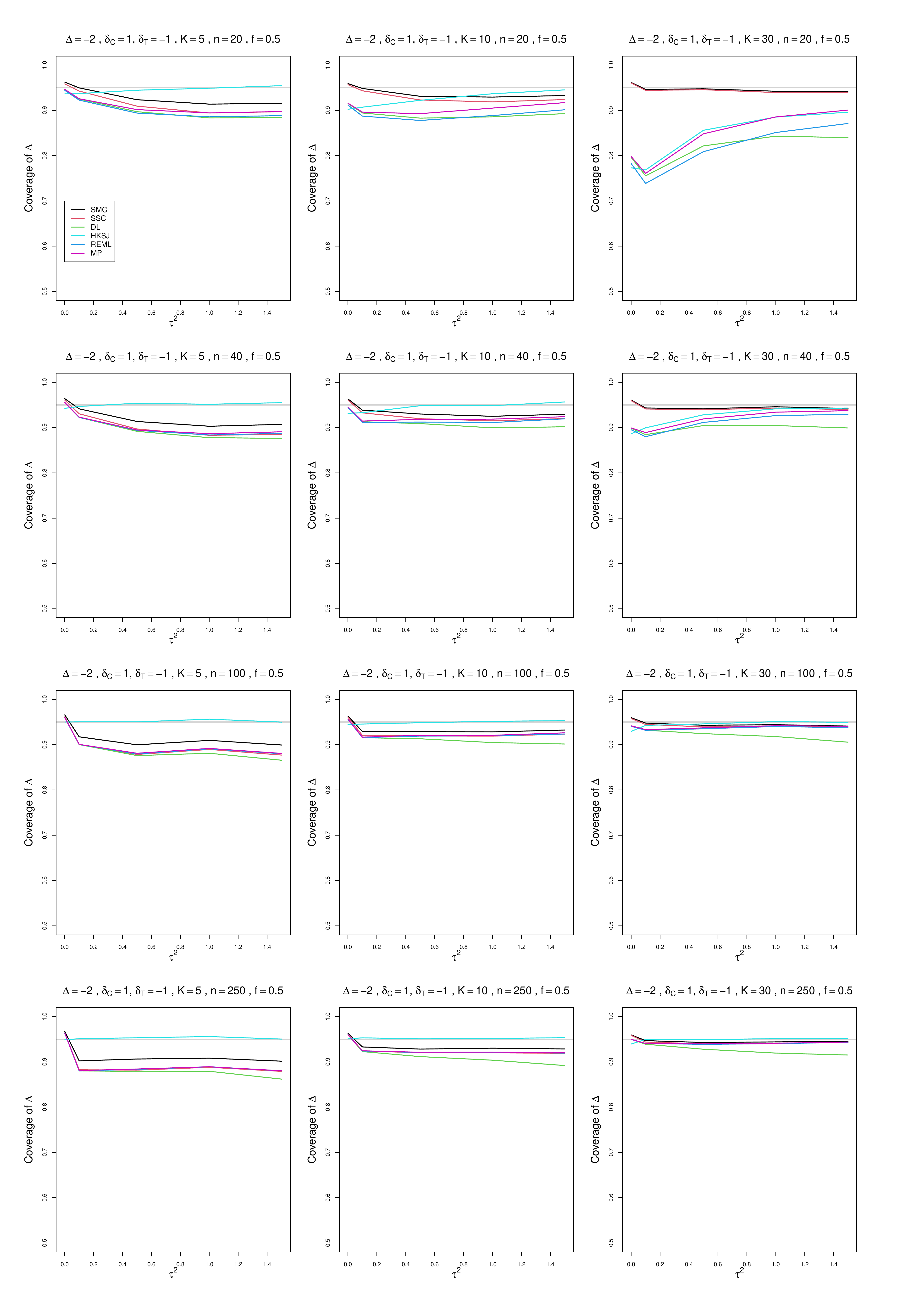}
	\caption{Coverage of 95\% confidence intervals for DSM (DL, REML, MP, HKSJ (DL), SMC and SSC intervals)  vs $\tau^2$, for equal sample sizes $n=20,\;40,\;100$ and $250$, $\delta_{iC} = -1$, $\Delta=-2$ and  $f = 0.5$.   }
	\label{PlotCoverageOfDelta_deltaC_1deltaT=-1_DSM_equal_sample_sizes.pdf}
\end{figure}

\begin{figure}[ht]
	\centering
	\includegraphics[scale=0.33]{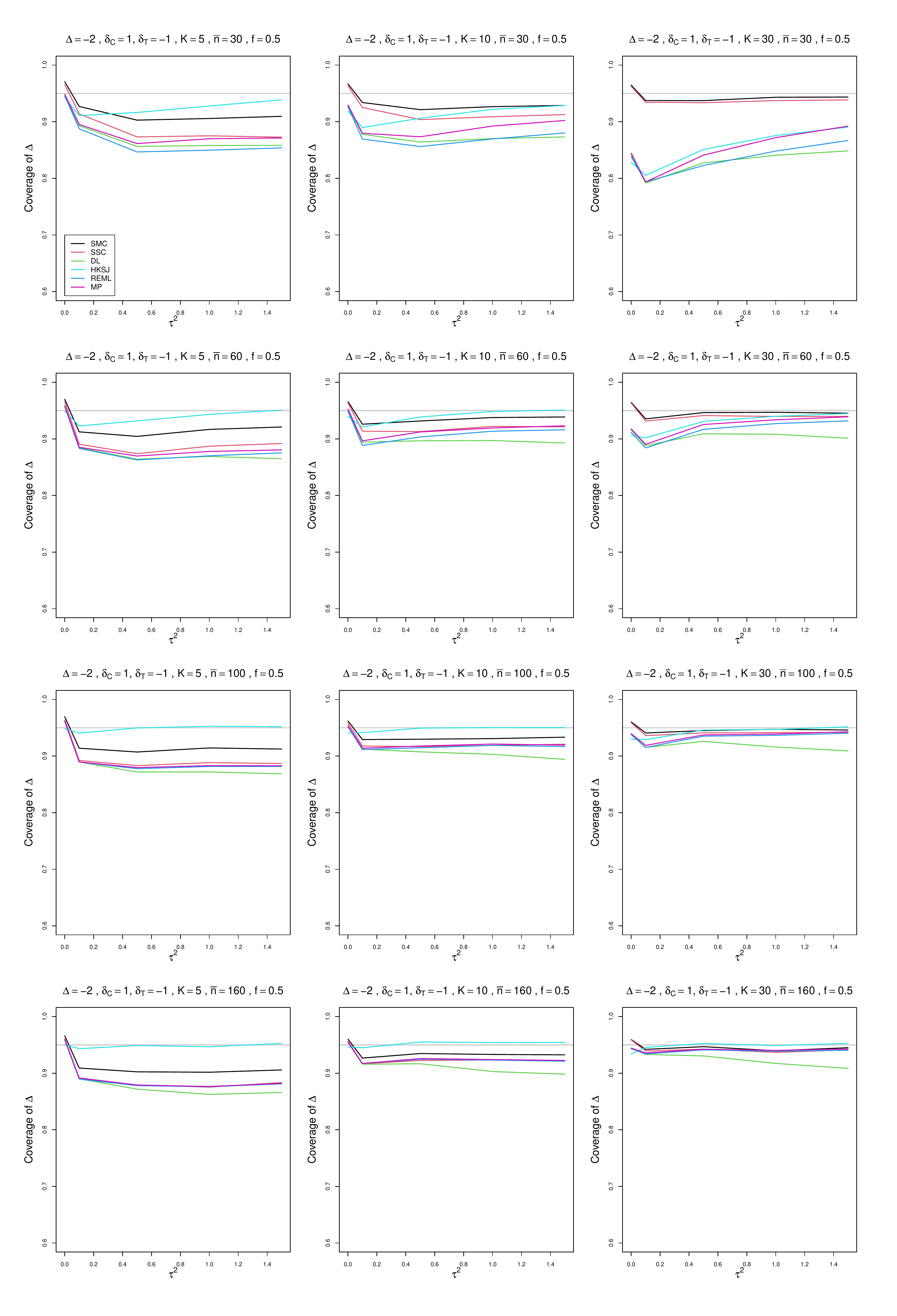}
	\caption{Coverage of 95\% confidence intervals for DSM (DL, REML, MP, HKSJ (DL), SMC and SSC intervals)  vs $\tau^2$, for unequal sample sizes $\bar{n}=30,\;60,\;100$ and $160$, $\delta_{iC} = 1$, $\Delta=-2$ and  $f = 0.5$.   }
	\label{PlotCoverageOfDelta_deltaC_1deltaT=-1_DSM_unequal_sample_sizes.pdf}
\end{figure}

\begin{figure}[ht]
	\centering
	\includegraphics[scale=0.33]{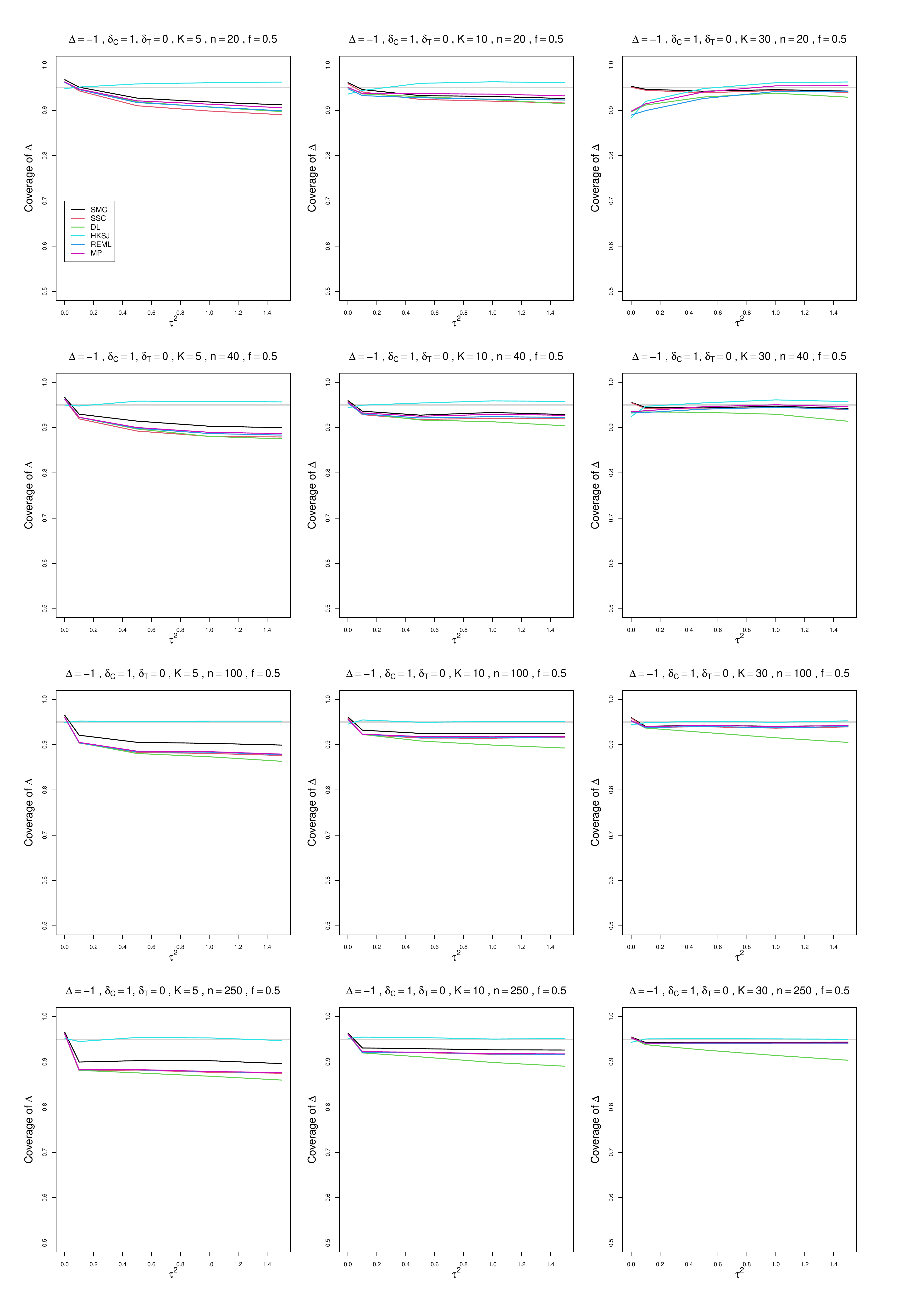}
	\caption{Coverage of 95\% confidence intervals for DSM (DL, REML, MP, HKSJ (DL), SMC and SSC intervals)  vs $\tau^2$, for equal sample sizes $n=20,\;40,\;100$ and $250$, $\delta_{iC} = 1$, $\Delta=-1$ and  $f = 0.5$.   }
	\label{PlotCoverageOfDelta_deltaC_1deltaT=0_DSM_equal_sample_sizes.pdf}
\end{figure}

\begin{figure}[ht]
	\centering
	\includegraphics[scale=0.33]{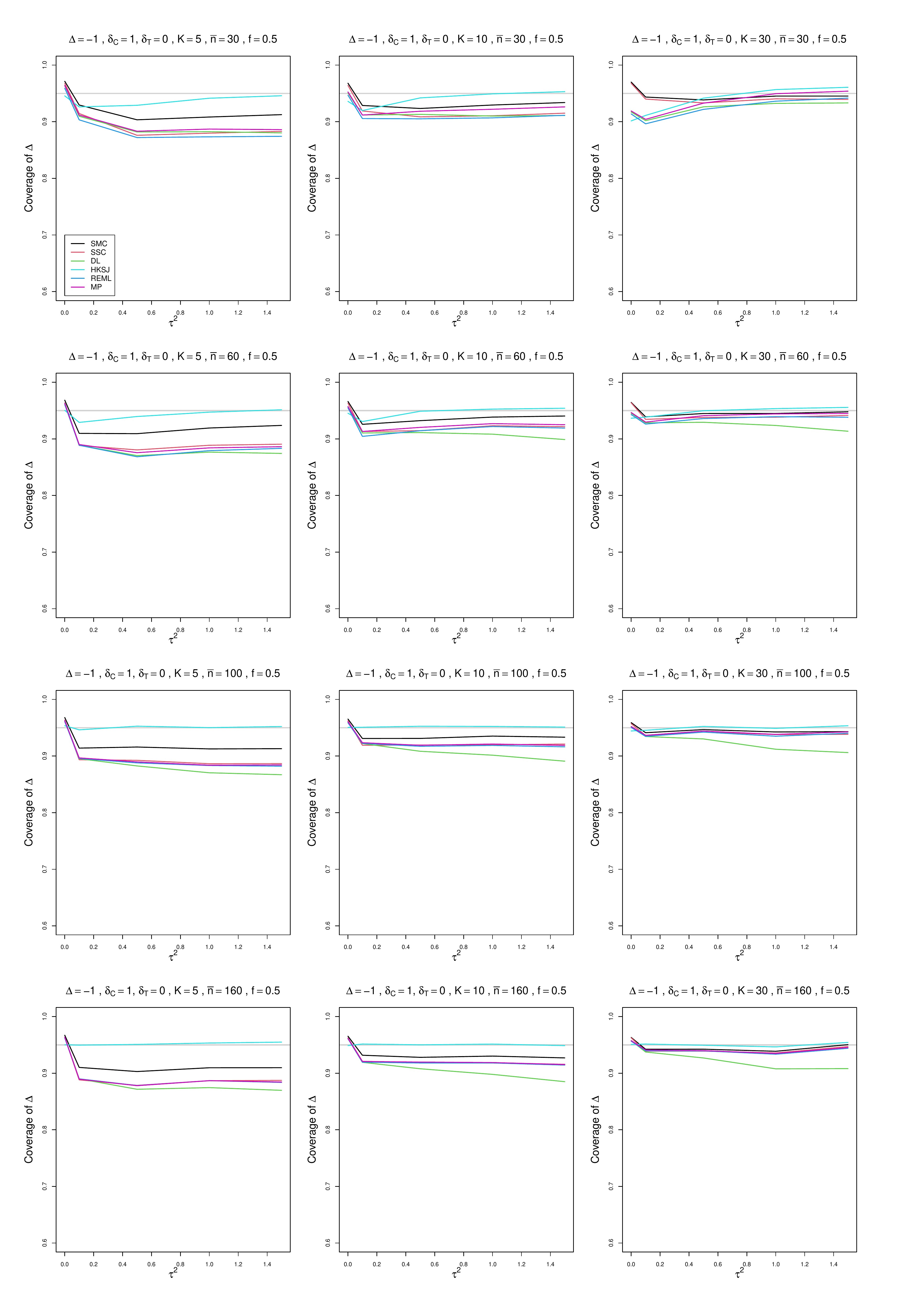}
	\caption{Coverage of 95\% confidence intervals for DSM (DL, REML, MP, HKSJ (DL), SMC and SSC intervals)  vs $\tau^2$, for unequal sample sizes $\bar{n}=30,\;60,\;100$ and $160$, $\delta_{iC} = 1$, $\Delta=-1$ and  $f = 0.5$.   }
	\label{PlotCoverageOfDelta_deltaC_1deltaT=0_DSM_unequal_sample_sizes.pdf}
\end{figure}

\begin{figure}[ht]
	\centering
	\includegraphics[scale=0.33]{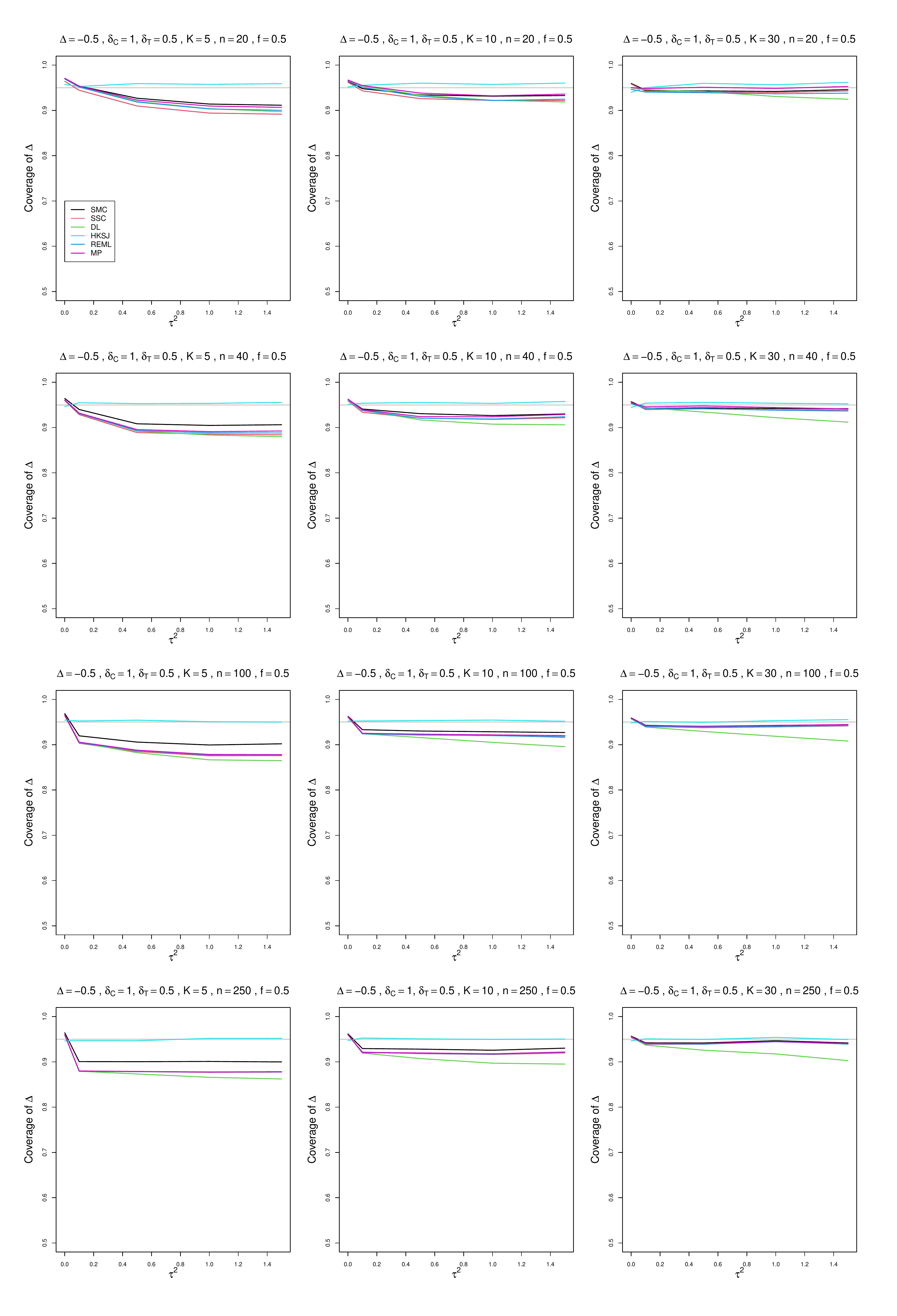}
	\caption{Coverage of 95\% confidence intervals for DSM (DL, REML, MP, HKSJ (DL), SMC and SSC intervals)  vs $\tau^2$, for equal sample sizes $n=20,\;40,\;100$ and $250$, $\delta_{iC} = 1$, $\Delta=-0.5$ and  $f = 0.5$.   }
	\label{PlotCoverageOfDelta_deltaC_1deltaT=0.5_DSM_equal_sample_sizes.pdf}
\end{figure}

\begin{figure}[ht]
	\centering
	\includegraphics[scale=0.33]{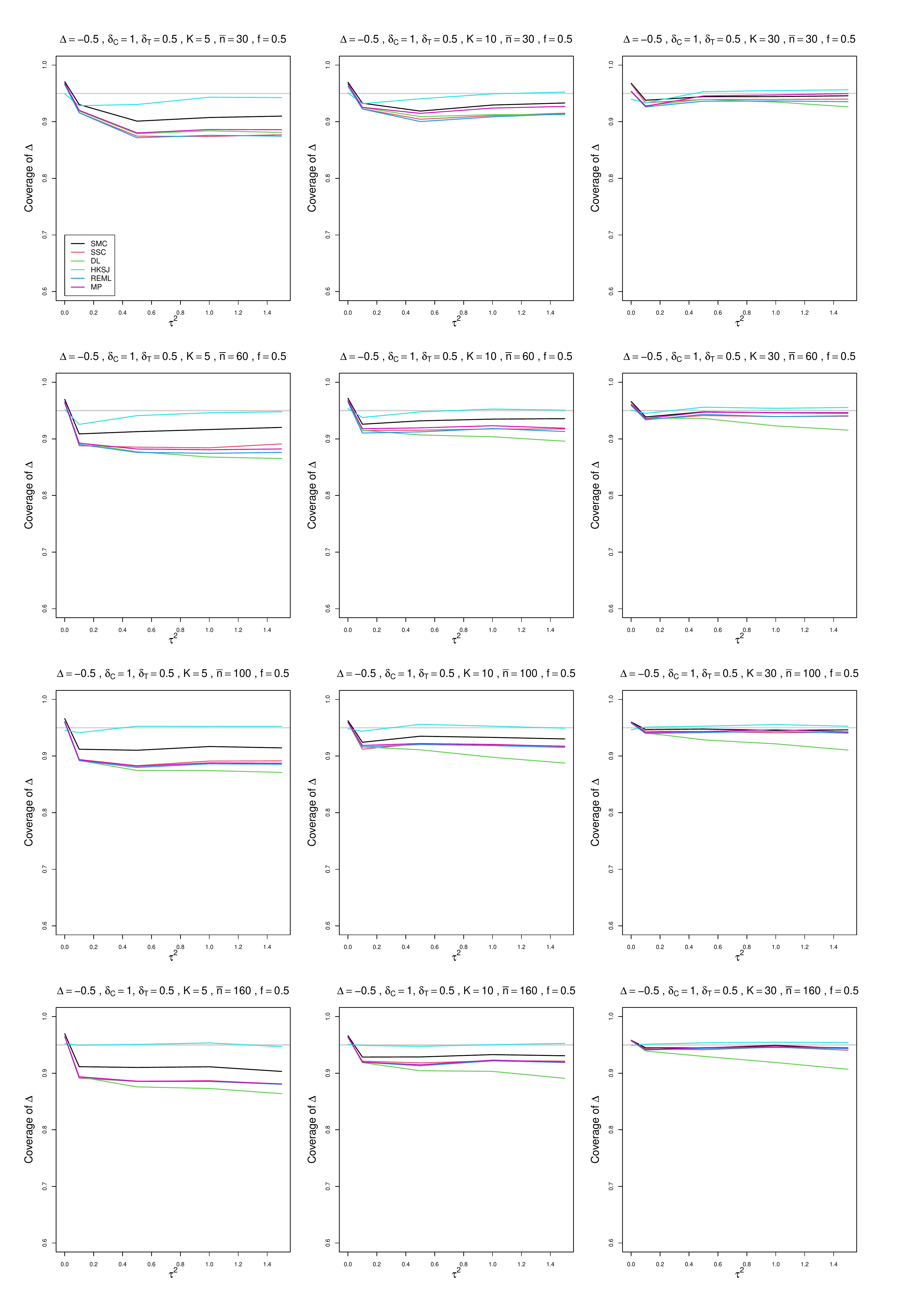}
	\caption{Coverage of 95\% confidence intervals for DSM (DL, REML, MP, HKSJ (DL), SMC and SSC intervals)  vs $\tau^2$, for unequal sample sizes $\bar{n}=30,\;60,\;100$ and $160$, $\delta_{iC} = 1$, $\Delta=-0.5$ and  $f = 0.5$.   }
	\label{PlotCoverageOfDelta_deltaC_1deltaT=0.5_DSM_unequal_sample_sizes.pdf}
\end{figure}

\begin{figure}[ht]
	\centering
	\includegraphics[scale=0.33]{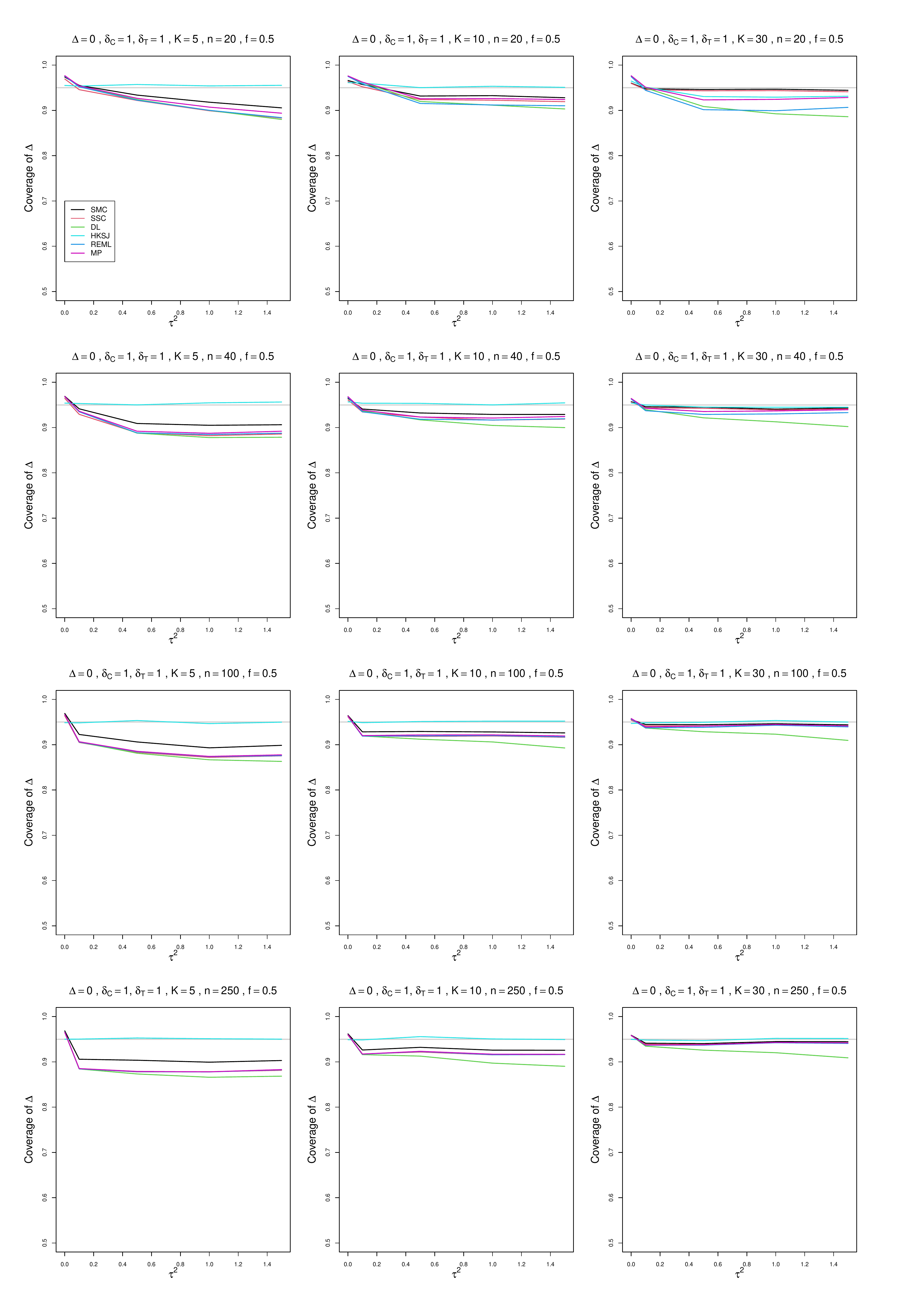}
	\caption{Coverage of 95\% confidence intervals for DSM (DL, REML, MP, HKSJ (DL), SMC and SSC intervals)  vs $\tau^2$, for equal sample sizes $n=20,\;40,\;100$ and $250$, $\delta_{iC} = 1$, $\Delta=0$ and  $f = 0.5$.   }
	\label{PlotCoverageOfDelta_deltaC_1deltaT=1_DSM_equal_sample_sizes.pdf}
\end{figure}

\begin{figure}[ht]
	\centering
	\includegraphics[scale=0.33]{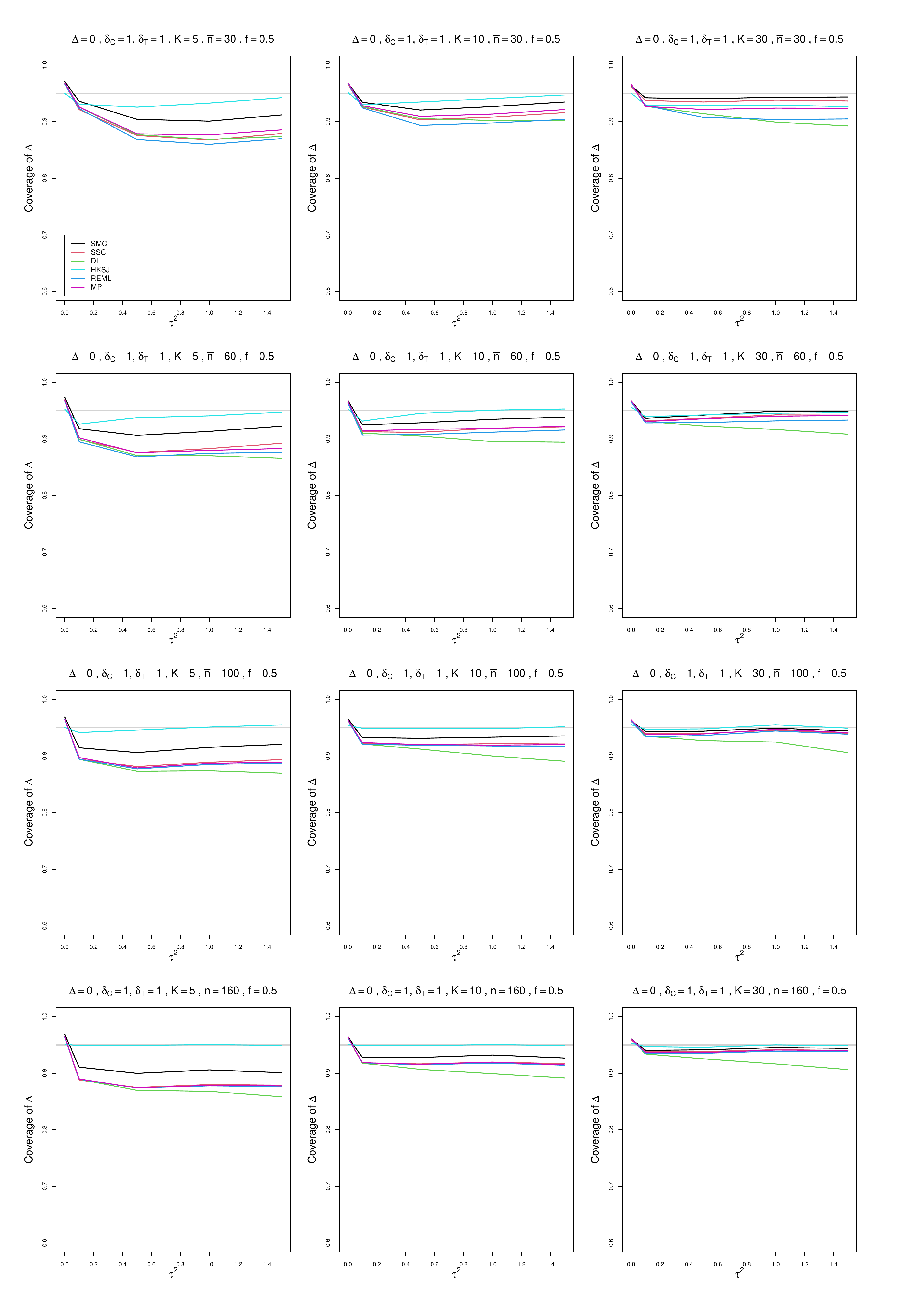}
	\caption{Coverage of 95\% confidence intervals for DSM (DL, REML, MP, HKSJ (DL), SMC and SSC intervals)  vs $\tau^2$, for unequal sample sizes $\bar{n}=30,\;60,\;100$ and $160$, $\delta_{iC} = 1$, $\Delta=0$ and  $f = 0.5$.   }
	\label{PlotCoverageOfDelta_deltaC_1deltaT=1_DSM_unequal_sample_sizes.pdf}
\end{figure}

\begin{figure}[ht]
	\centering
	\includegraphics[scale=0.33]{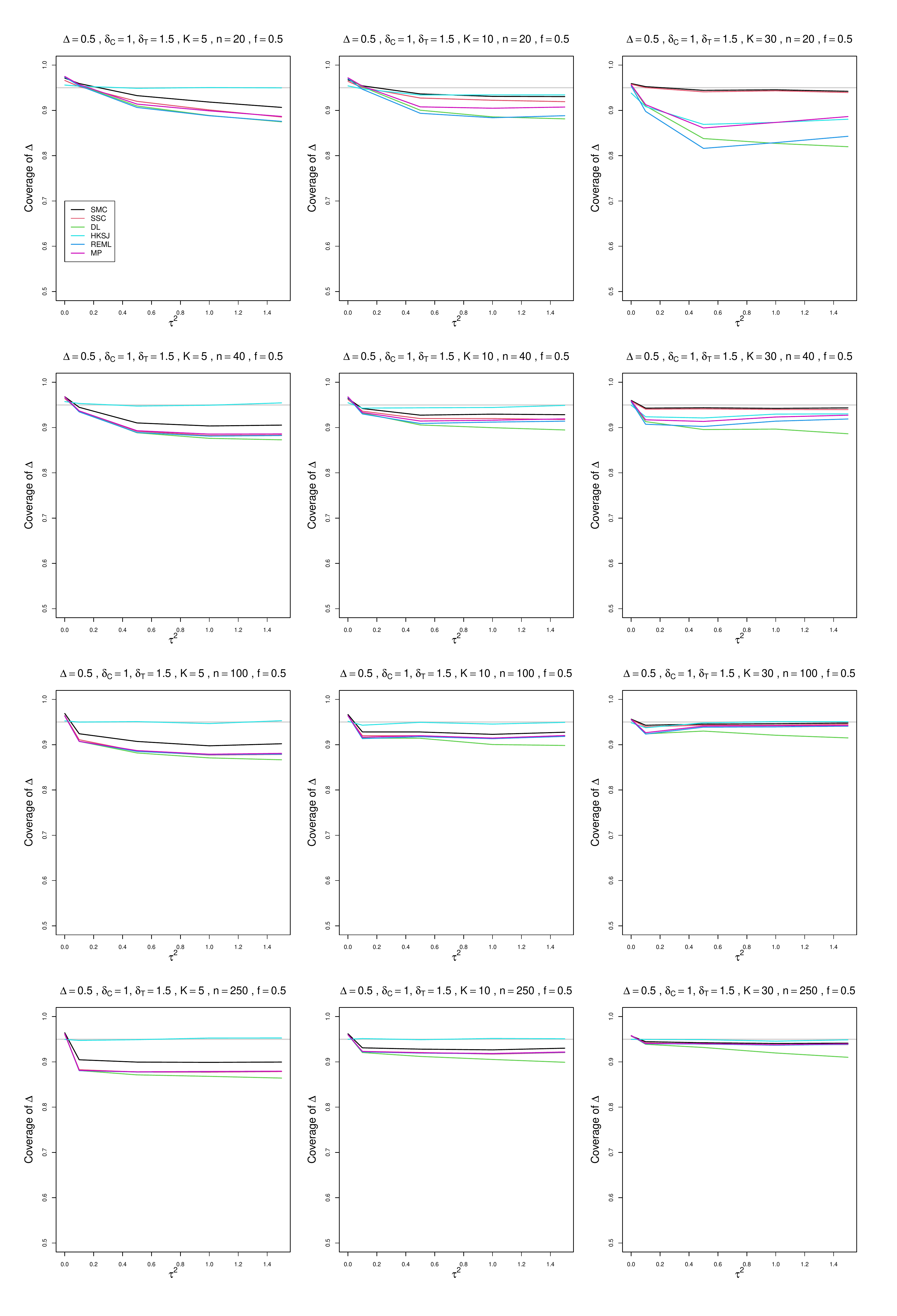}
	\caption{Coverage of 95\% confidence intervals for DSM (DL, REML, MP, HKSJ (DL), SMC and SSC intervals)  vs $\tau^2$, for equal sample sizes $n=20,\;40,\;100$ and $250$, $\delta_{iC} = 1$, $\Delta=0.5$ and  $f = 0.5$.   }
	\label{PlotCoverageOfDelta_deltaC_1deltaT=1.5_DSM_equal_sample_sizes.pdf}
\end{figure}

\begin{figure}[ht]
	\centering
	\includegraphics[scale=0.33]{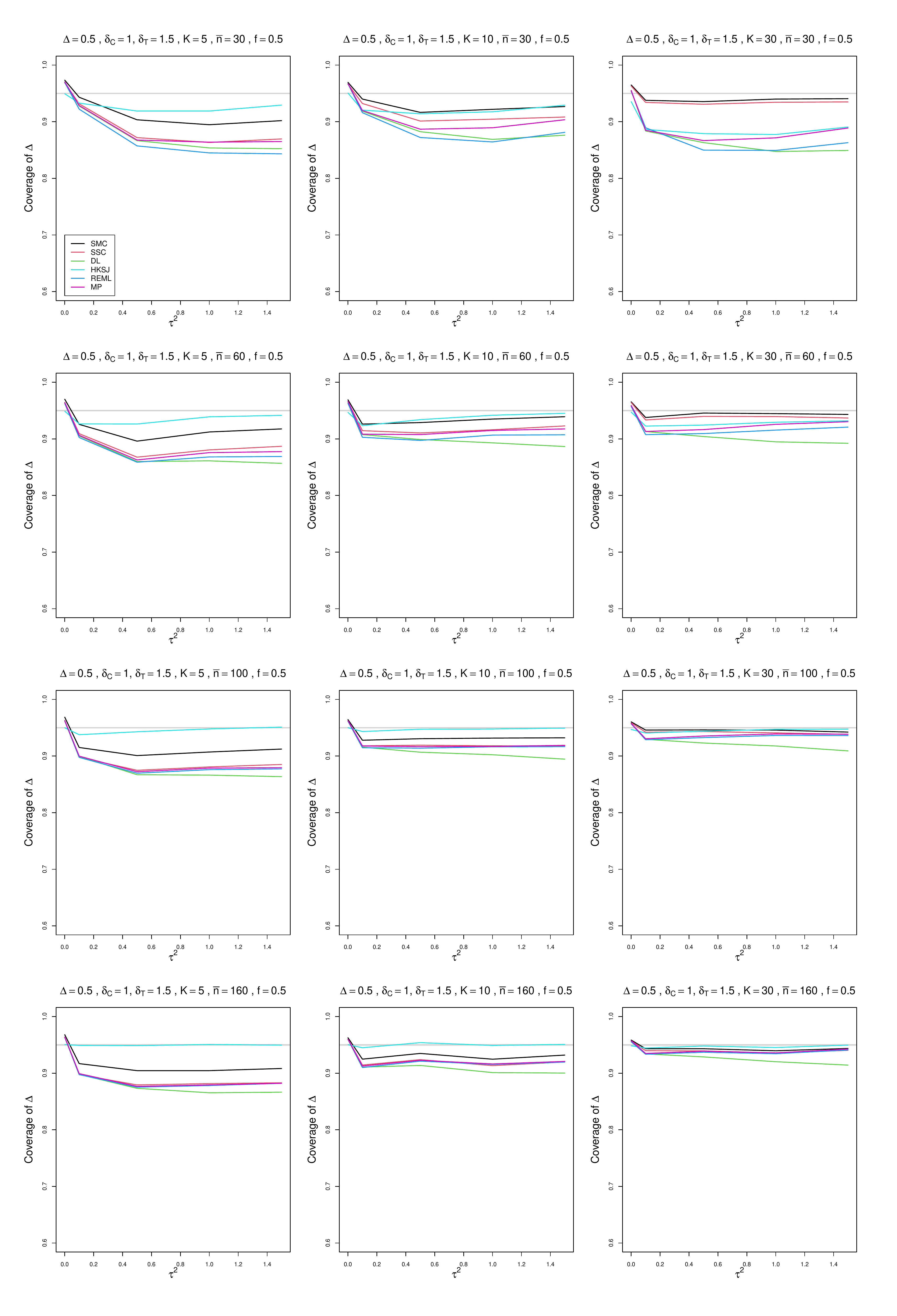}
	\caption{Coverage of 95\% confidence intervals for DSM (DL, REML, MP, HKSJ (DL), SMC and SSC intervals)  vs $\tau^2$, for unequal sample sizes $\bar{n}=30,\;60,\;100$ and $160$, $\delta_{iC} = 1$, $\Delta=0.5$ and  $f = 0.5$.   }
	\label{PlotCoverageOfDelta_deltaC_1deltaT=1.5_DSM_unequal_sample_sizes.pdf}
\end{figure}

\begin{figure}[ht]
	\centering
	\includegraphics[scale=0.33]{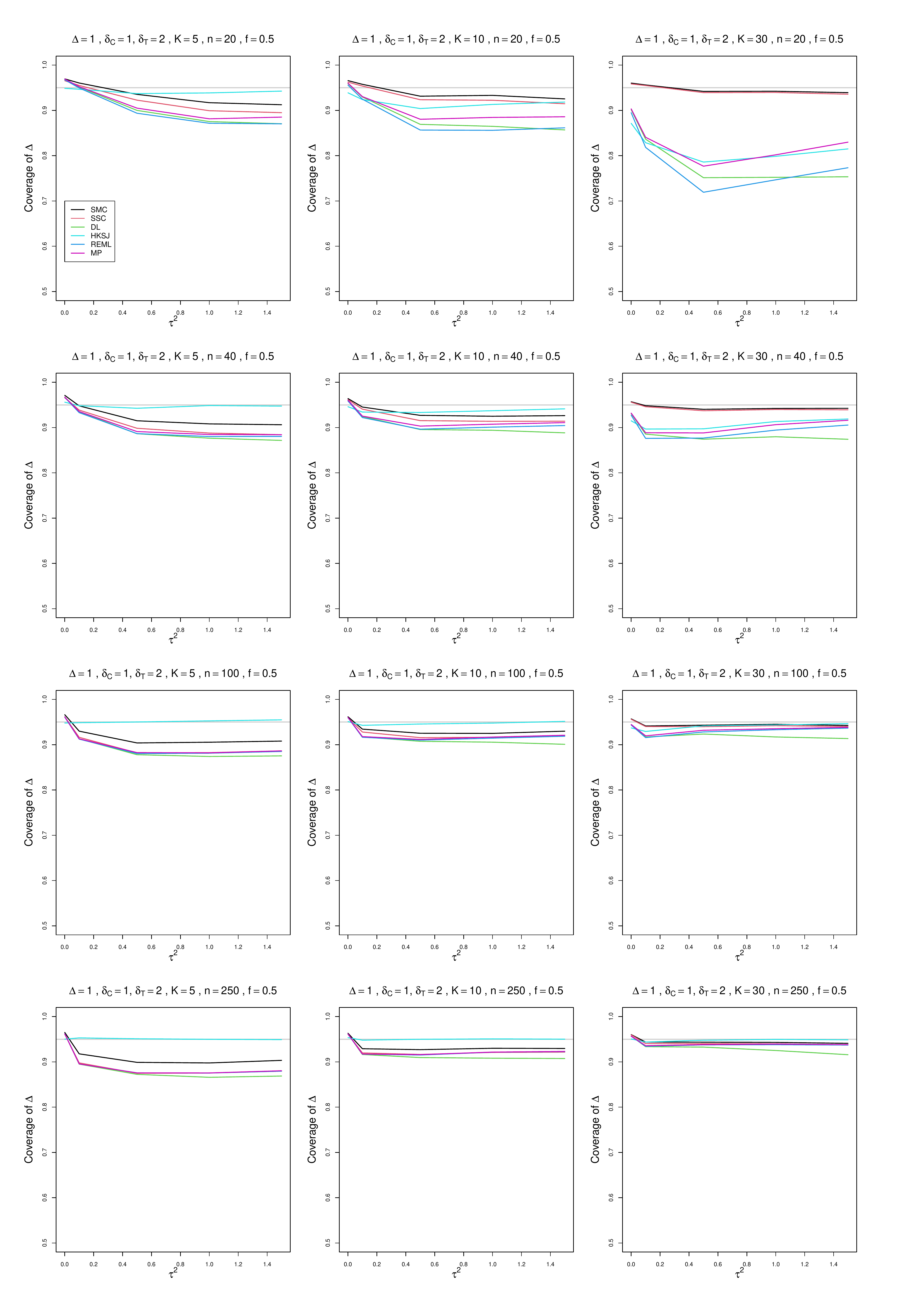}
	\caption{Coverage of 95\% confidence intervals for DSM (DL, REML, MP, HKSJ (DL), SMC and SSC intervals)  vs $\tau^2$, for equal sample sizes $n=20,\;40,\;100$ and $250$, $\delta_{iC} = 1$, $\Delta=1$ and  $f = 0.5$.   }
	\label{PlotCoverageOfDelta_deltaC_=1deltaT=2_DSM_equal_sample_sizes.pdf}
\end{figure}

\begin{figure}[ht]
	\centering
	\includegraphics[scale=0.33]{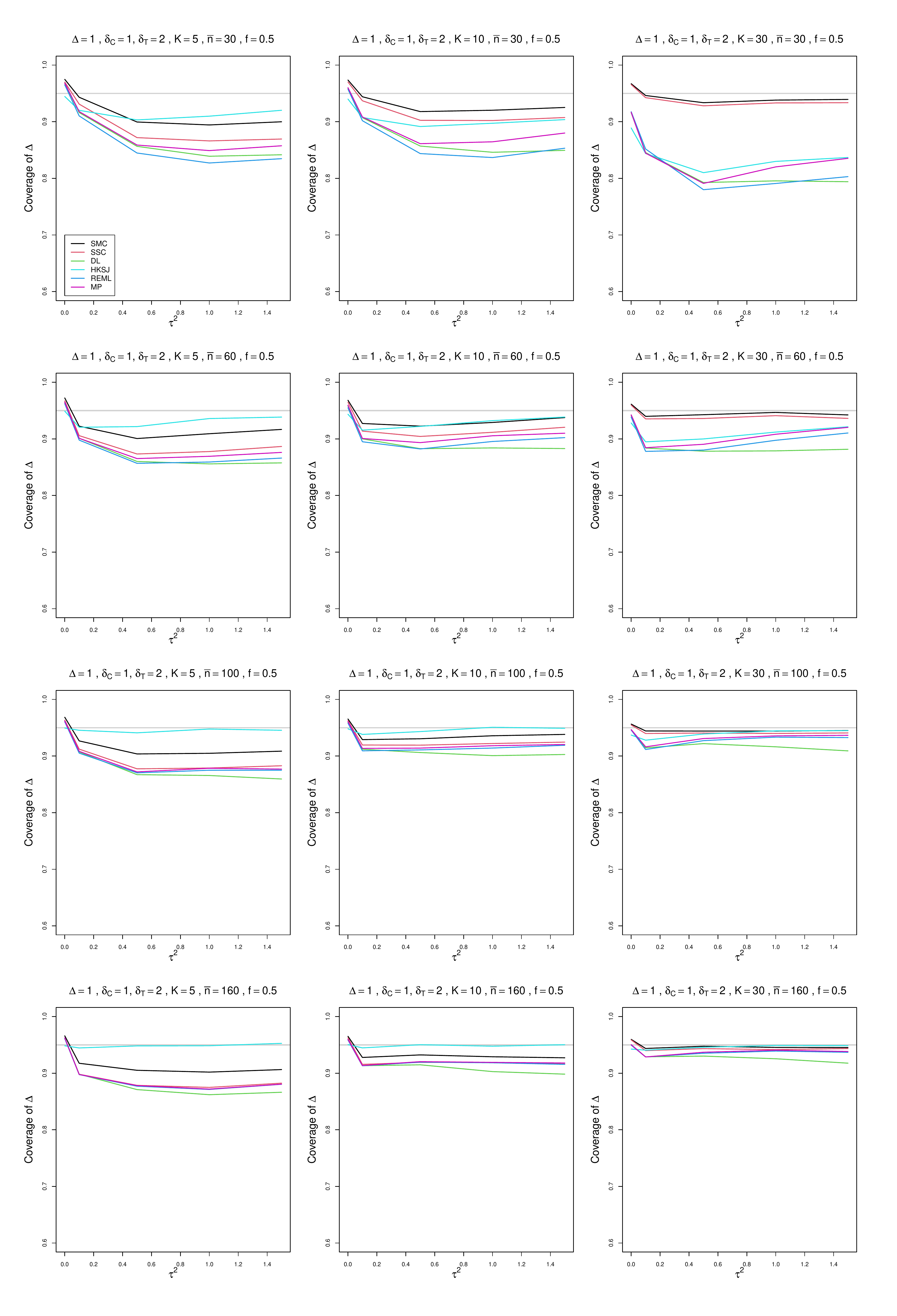}
	\caption{Coverage of 95\% confidence intervals for DSM (DL, REML, MP, HKSJ (DL), SMC and SSC intervals)  vs $\tau^2$, for unequal sample sizes $\bar{n}=30,\;60,\;100$ and $160$, $\delta_{iC} = 1$, $\Delta=1$ and  $f = 0.5$.   }
	\label{PlotCoverageOfDelta_deltaC_1deltaT=2_DSM_unequal_sample_sizes.pdf}
\end{figure}

\begin{figure}[ht]
	\centering
	\includegraphics[scale=0.33]{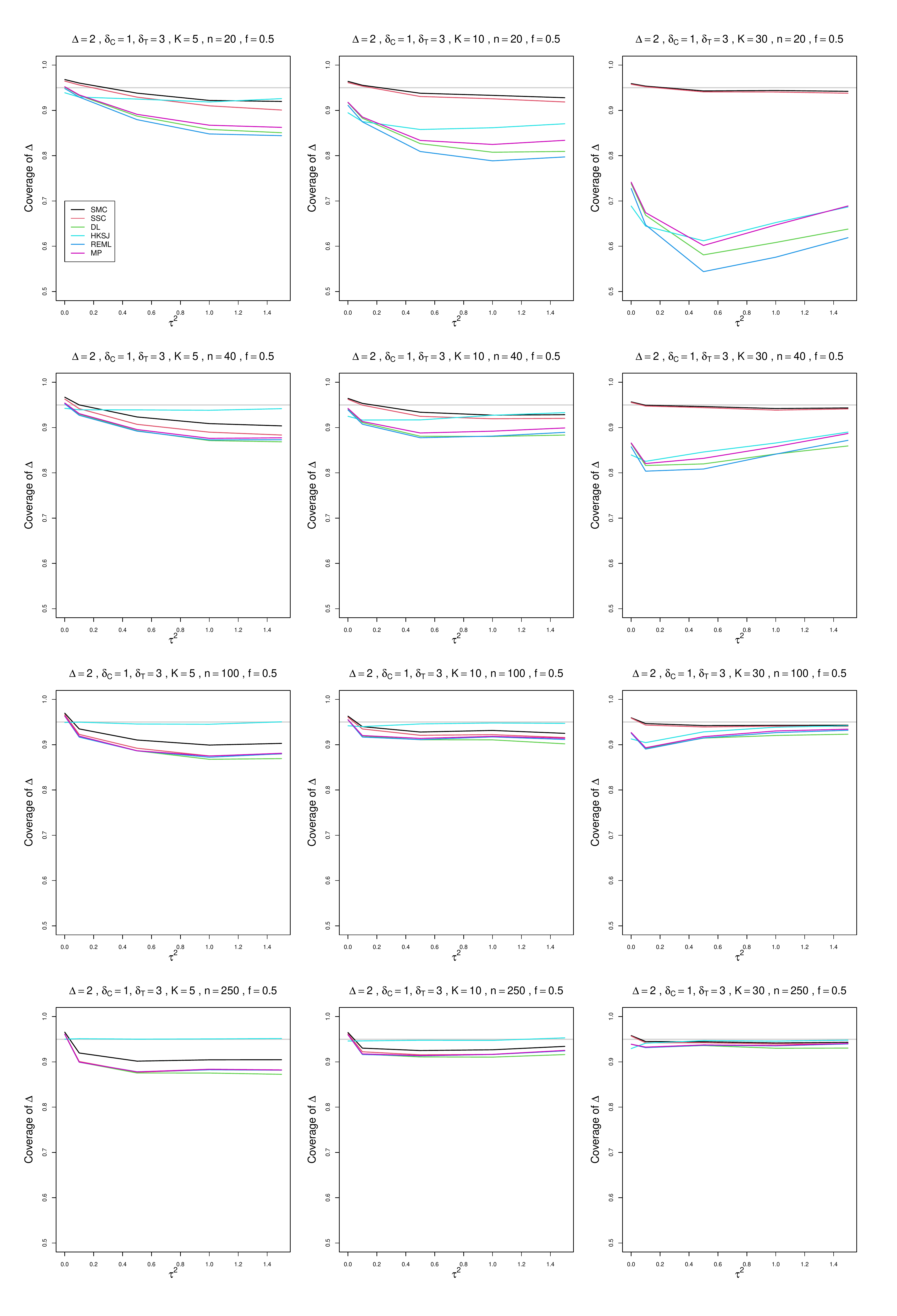}
	\caption{Coverage of 95\% confidence intervals for DSM (DL, REML, MP, HKSJ (DL), SMC and SSC intervals)  vs $\tau^2$, for equal sample sizes $n=20,\;40,\;100$ and $250$, $\delta_{iC} = 1$, $\Delta=2$ and  $f = 0.5$.   }
	\label{PlotCoverageOfDelta_deltaC_1deltaT=3_DSM_equal_sample_sizes.pdf}
\end{figure}

\begin{figure}[ht]
	\centering
	\includegraphics[scale=0.33]{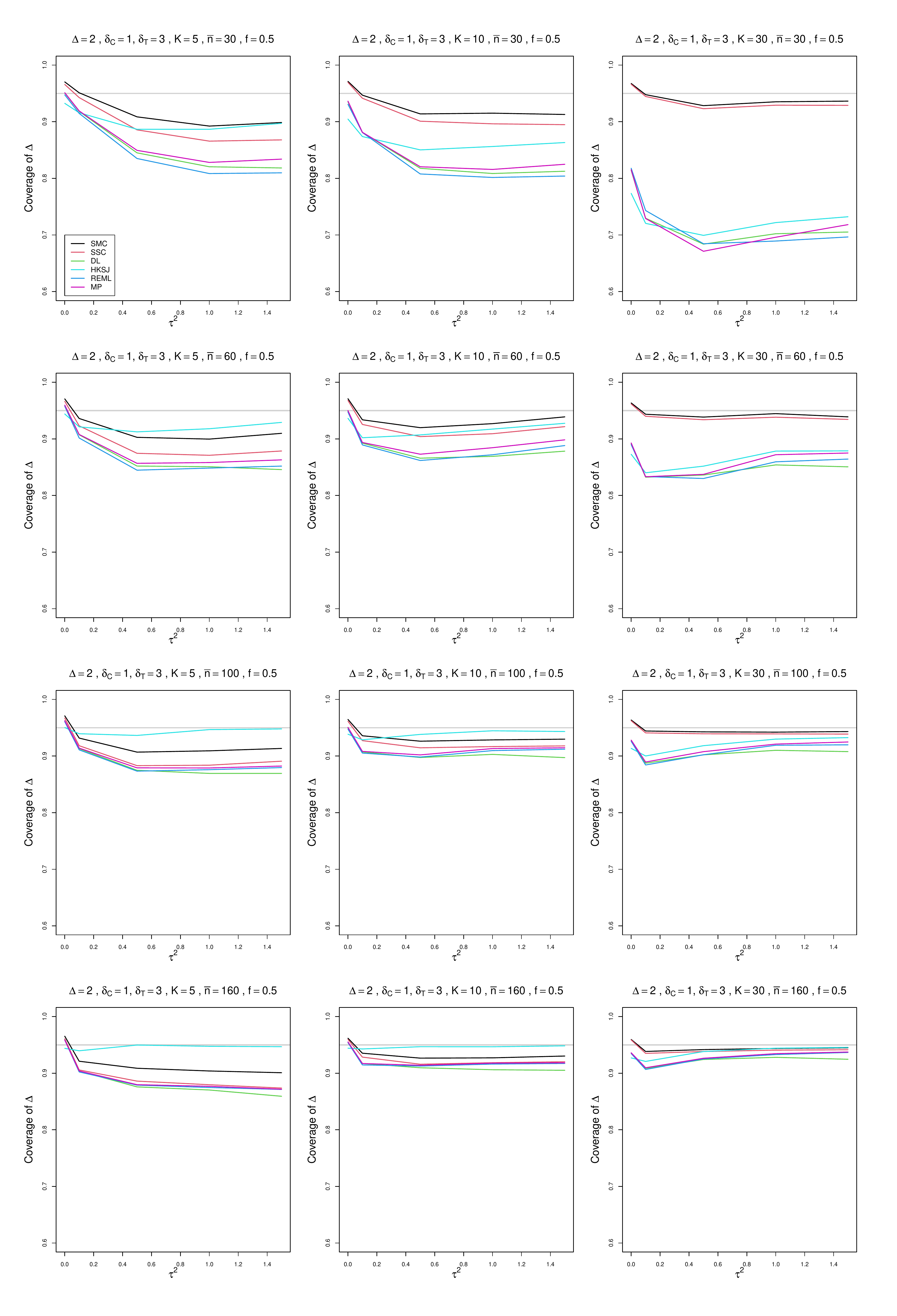}
	\caption{Coverage of 95\% confidence intervals for DSM (DL, REML, MP, HKSJ (DL), SMC and SSC intervals)  vs $\tau^2$, for unequal sample sizes $\bar{n}=30,\;60,\;100$ and $160$, $\delta_{iC} = 1$, $\Delta=2$ and  $f = 0.5$.   }
	\label{PlotCoverageOfDelta_deltaC_1deltaT=3_DSM_unequal_sample_sizes.pdf}
\end{figure}


\begin{figure}[ht]
	\centering
	\includegraphics[scale=0.33]{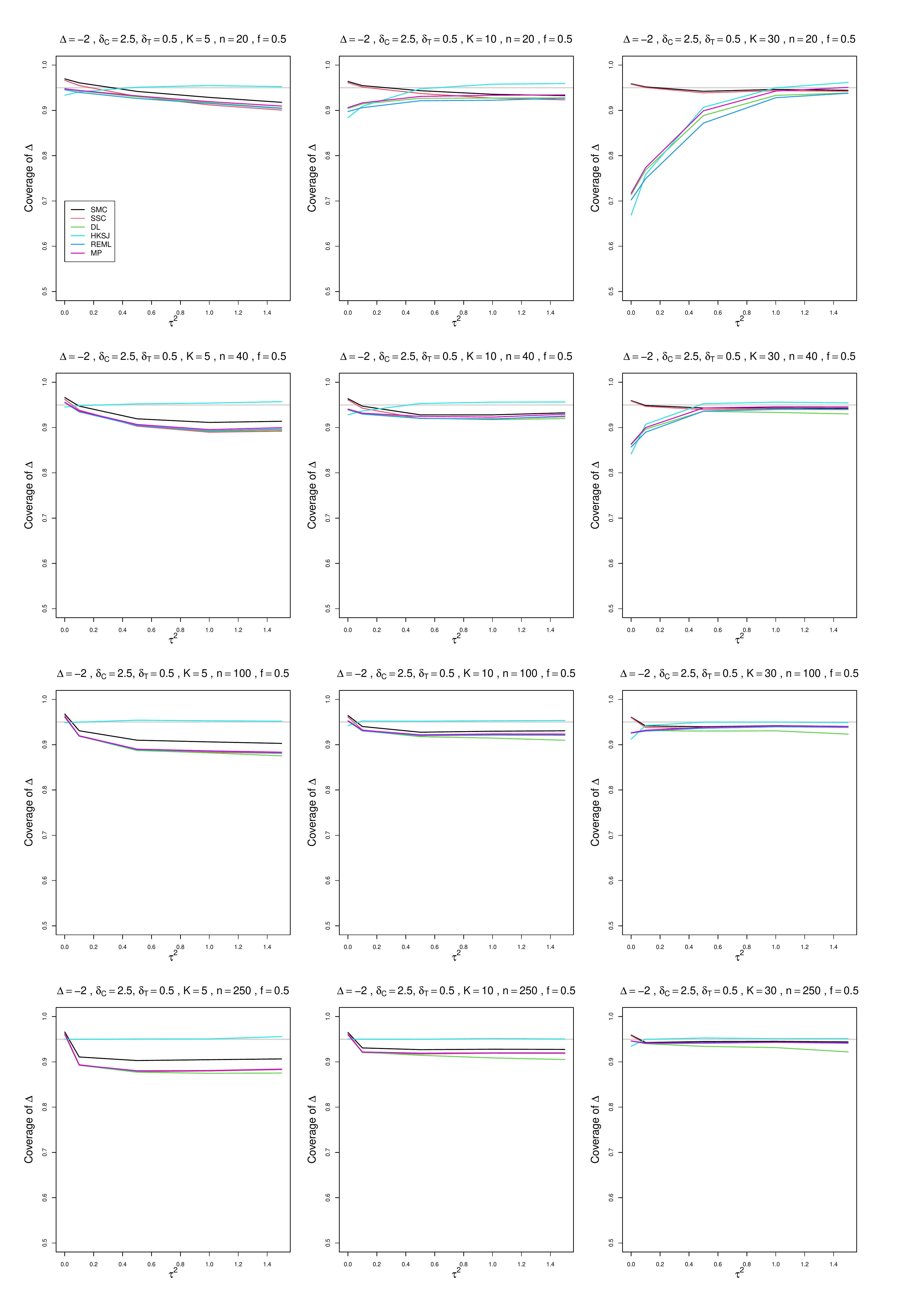}
	\caption{Coverage of 95\% confidence intervals for DSM (DL, REML, MP, HKSJ (DL), SMC and SSC intervals)  vs $\tau^2$, for equal sample sizes $n=20,\;40,\;100$ and $250$, $\delta_{iC} = 2.5$, $\Delta=-2$ and  $f = 0.5$.   }
	\label{PlotCoverageOfDelta_deltaC_2.5deltaT=0.5_DSM_equal_sample_sizes.pdf}
\end{figure}

\begin{figure}[ht]
	\centering
	\includegraphics[scale=0.33]{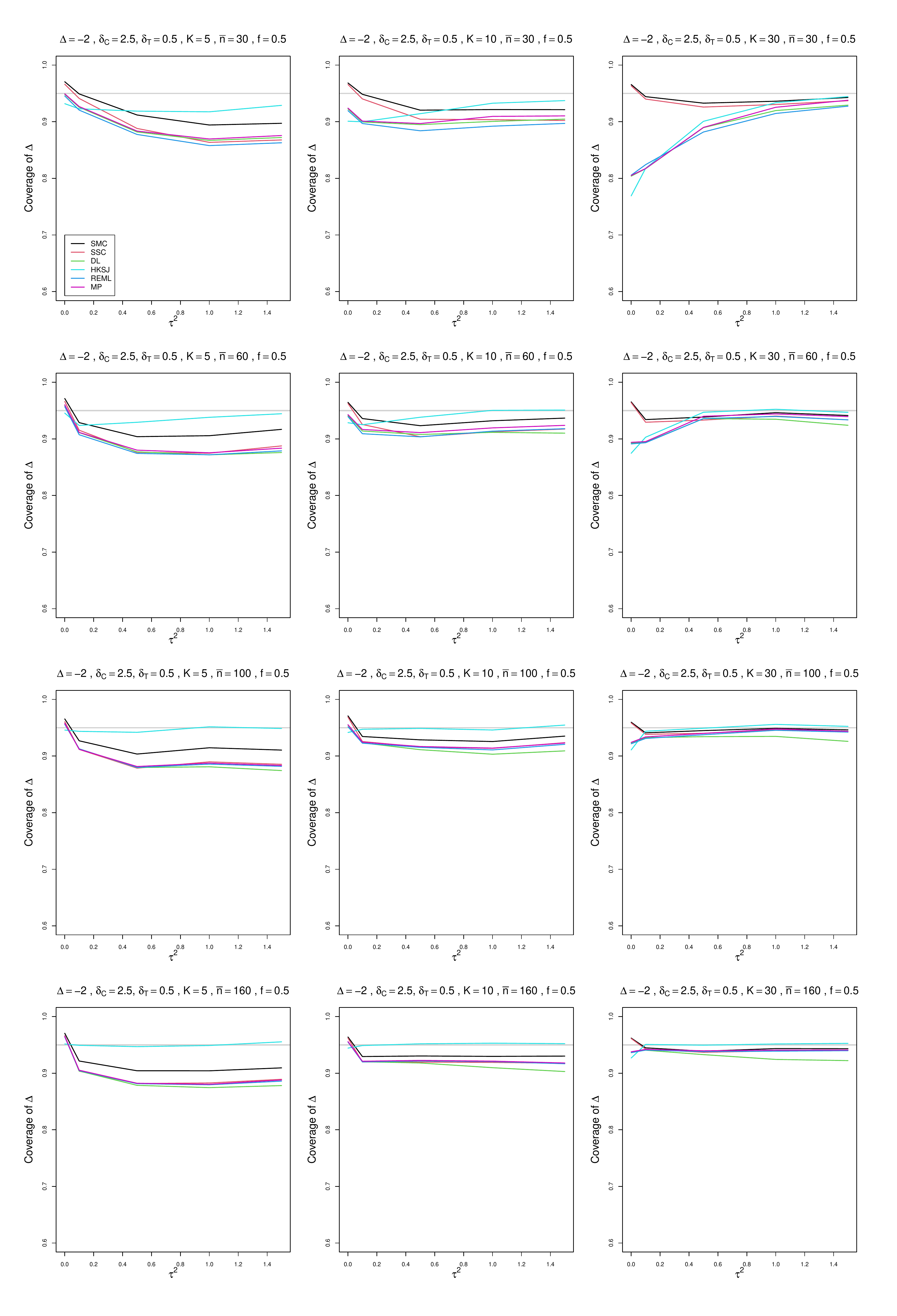}
	\caption{Coverage of 95\% confidence intervals for DSM (DL, REML, MP, HKSJ (DL), SMC and SSC intervals)  vs $\tau^2$, for unequal sample sizes $\bar{n}=30,\;60,\;100$ and $160$, $\delta_{iC} = 2.5$, $\Delta=-2$ and  $f = 0.5$.   }
	\label{PlotCoverageOfDelta_deltaC_2.5deltaT=0.5_DSM_unequal_sample_sizes.pdf}
\end{figure}

\begin{figure}[ht]
	\centering
	\includegraphics[scale=0.33]{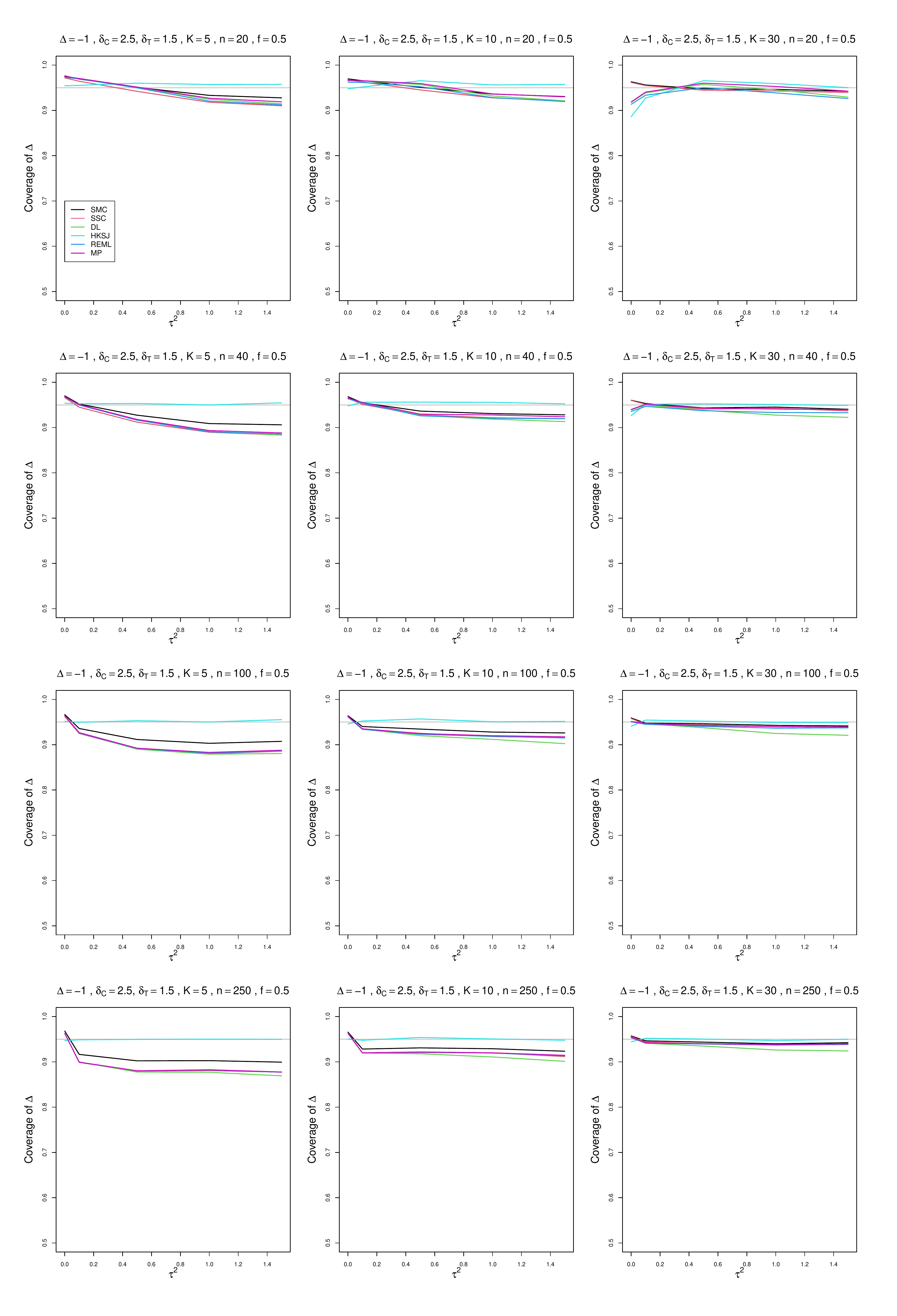}
	\caption{Coverage of 95\% confidence intervals for DSM (DL, REML, MP, HKSJ (DL), SMC and SSC intervals)  vs $\tau^2$, for equal sample sizes $n=20,\;40,\;100$ and $250$, $\delta_{iC} = 2.5$, $\Delta=-1$ and  $f = 0.5$.   }
	\label{PlotCoverageOfDelta_deltaC_2.5deltaT=1.5_DSM_equal_sample_sizes.pdf}
\end{figure}

\begin{figure}[ht]
	\centering
	\includegraphics[scale=0.33]{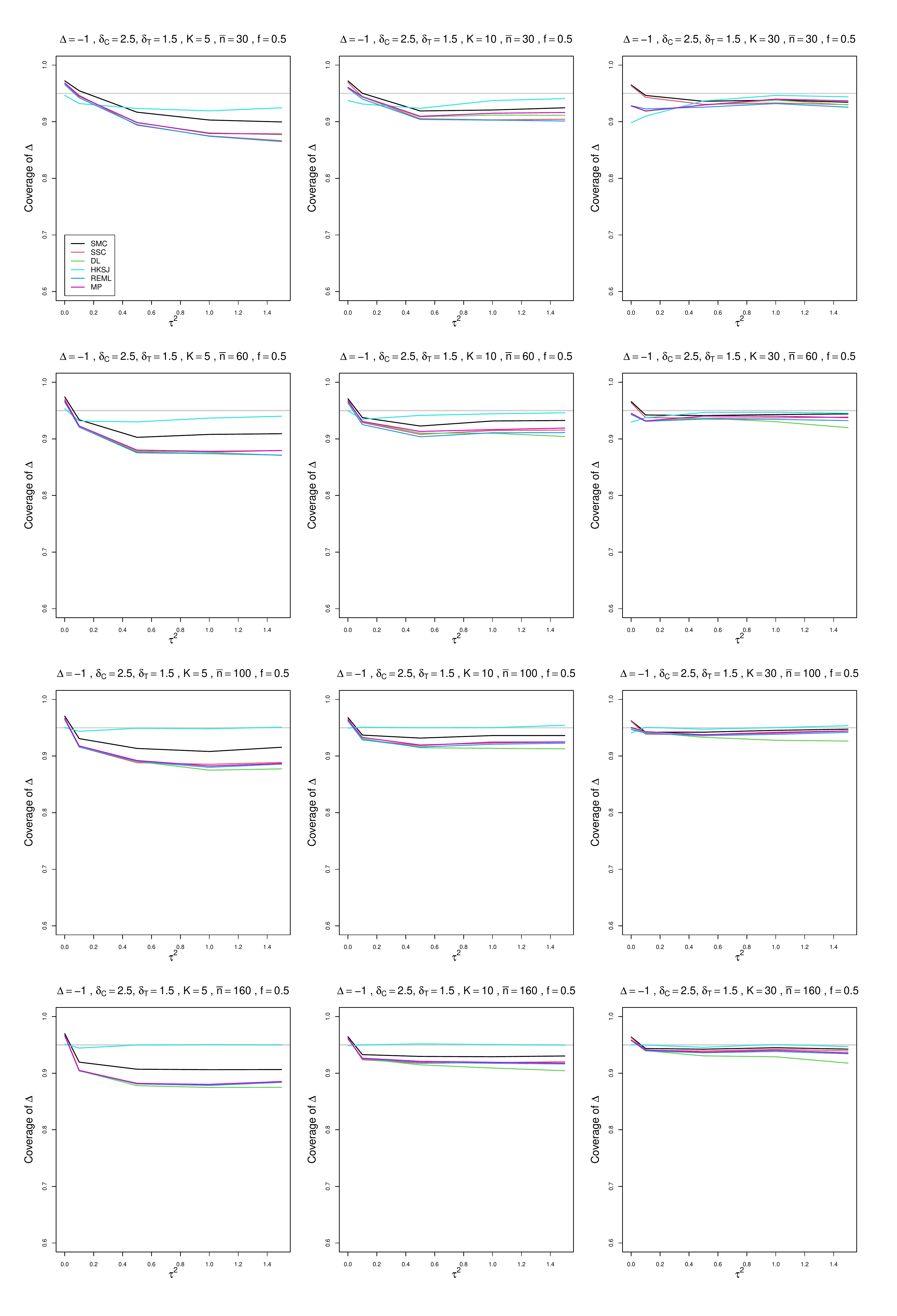}
	\caption{Coverage of 95\% confidence intervals for DSM (DL, REML, MP, HKSJ (DL), SMC and SSC intervals)  vs $\tau^2$, for unequal sample sizes $\bar{n}=30,\;60,\;100$ and $160$, $\delta_{iC} = 2.5$, $\Delta=-1$ and  $f = 0.5$.   }
	\label{PlotCoverageOfDelta_deltaC_2.5deltaT=1.5_DSM_unequal_sample_sizes.pdf}
\end{figure}

\begin{figure}[ht]
	\centering
	\includegraphics[scale=0.33]{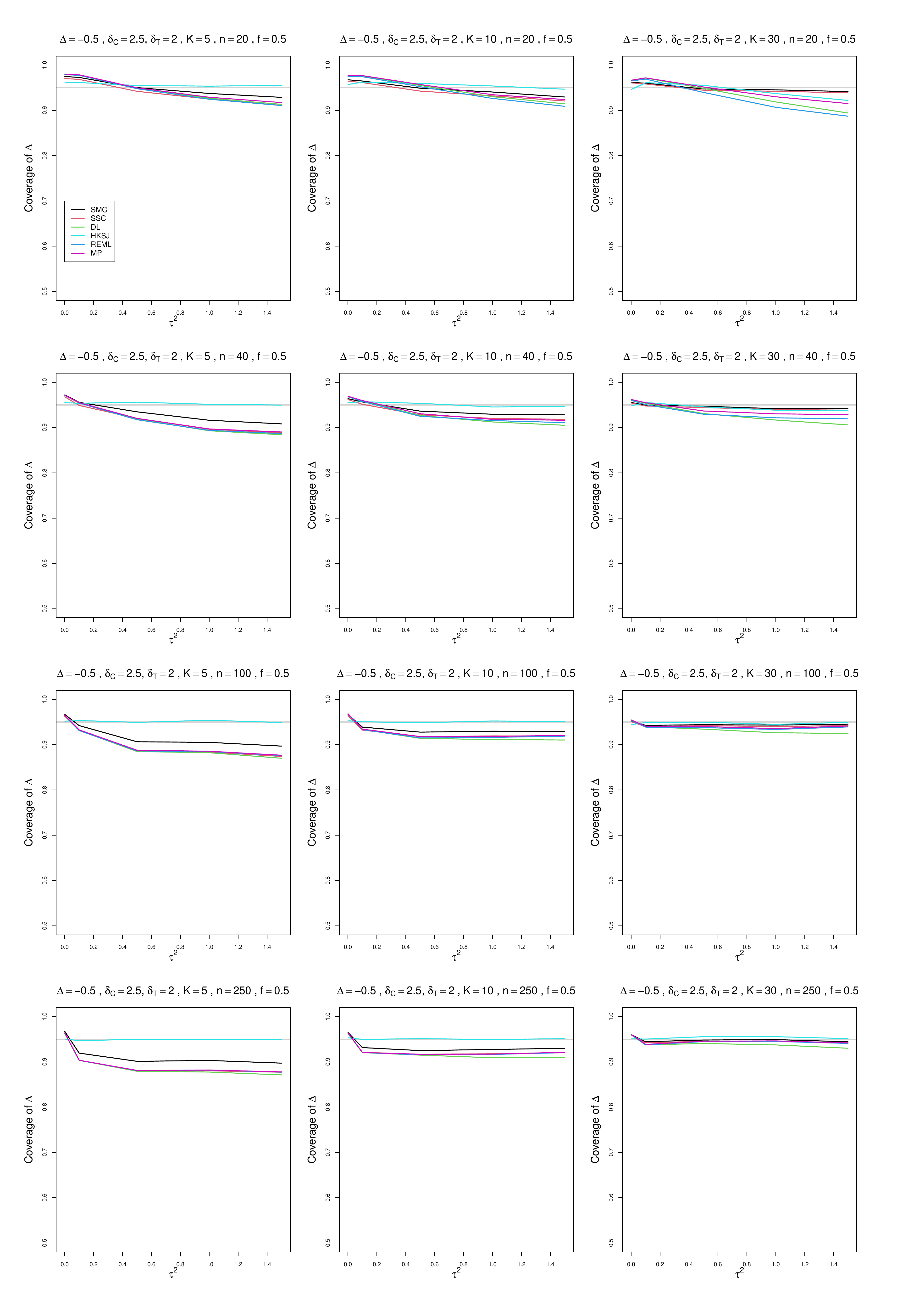}
	\caption{Coverage of 95\% confidence intervals for DSM (DL, REML, MP, HKSJ (DL), SMC and SSC intervals)  vs $\tau^2$, for equal sample sizes $n=20,\;40,\;100$ and $250$, $\delta_{iC} = 2.5$, $\Delta=-0.5$ and  $f = 0.5$.   }
	\label{PlotCoverageOfDelta_deltaC_2.5deltaT=2_DSM_equal_sample_sizes.pdf}
\end{figure}

\begin{figure}[ht]
	\centering
	\includegraphics[scale=0.33]{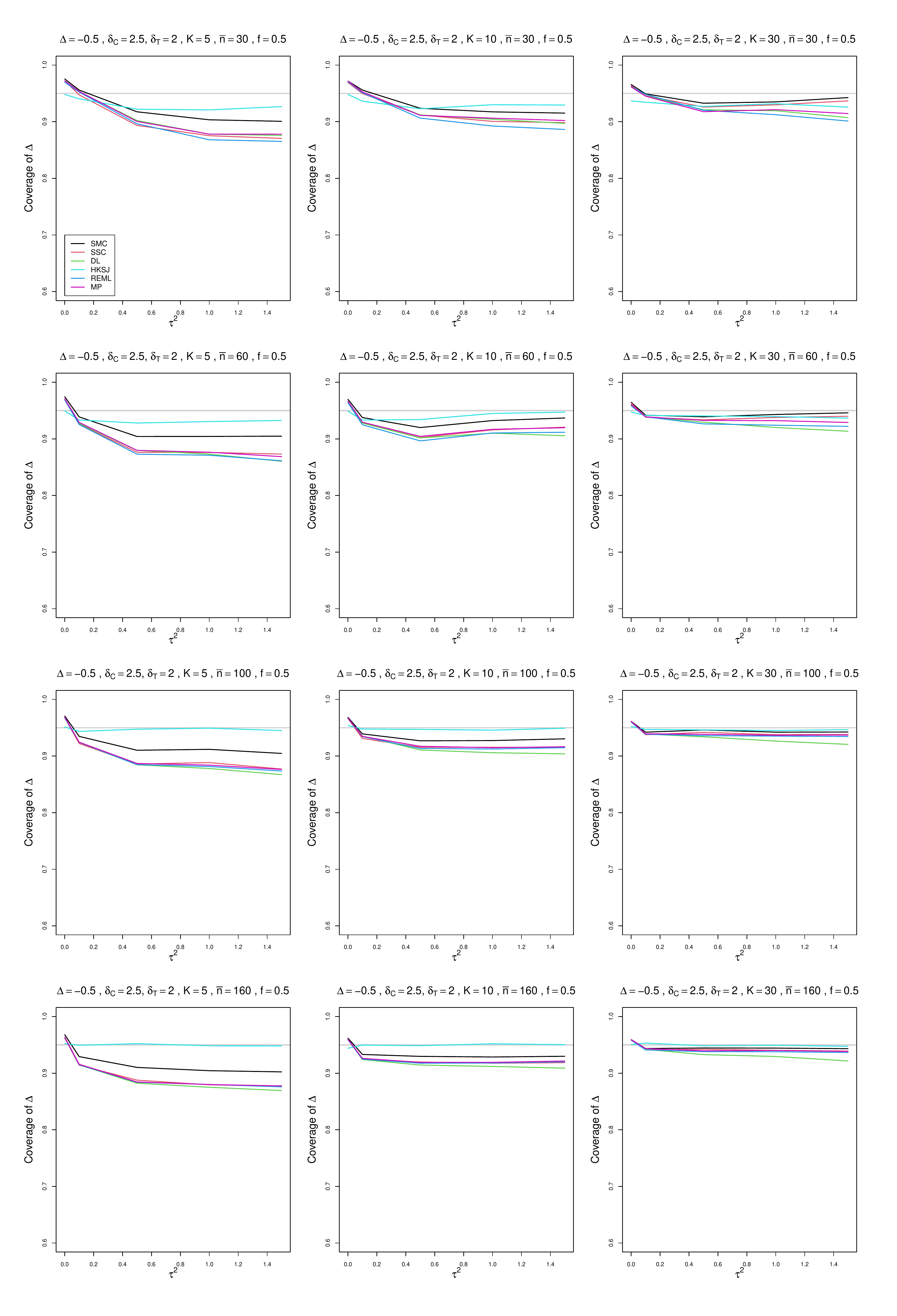}
	\caption{Coverage of 95\% confidence intervals for DSM (DL, REML, MP, HKSJ (DL), SMC and SSC intervals)  vs $\tau^2$, for unequal sample sizes $\bar{n}=30,\;60,\;100$ and $160$, $\delta_{iC} = 2.5$, $\Delta=-0.5$ and  $f = 0.5$.   }
	\label{PlotCoverageOfDelta_deltaC_2.5deltaT=2_DSM_unequal_sample_sizes.pdf}
\end{figure}

\begin{figure}[ht]
	\centering
	\includegraphics[scale=0.33]{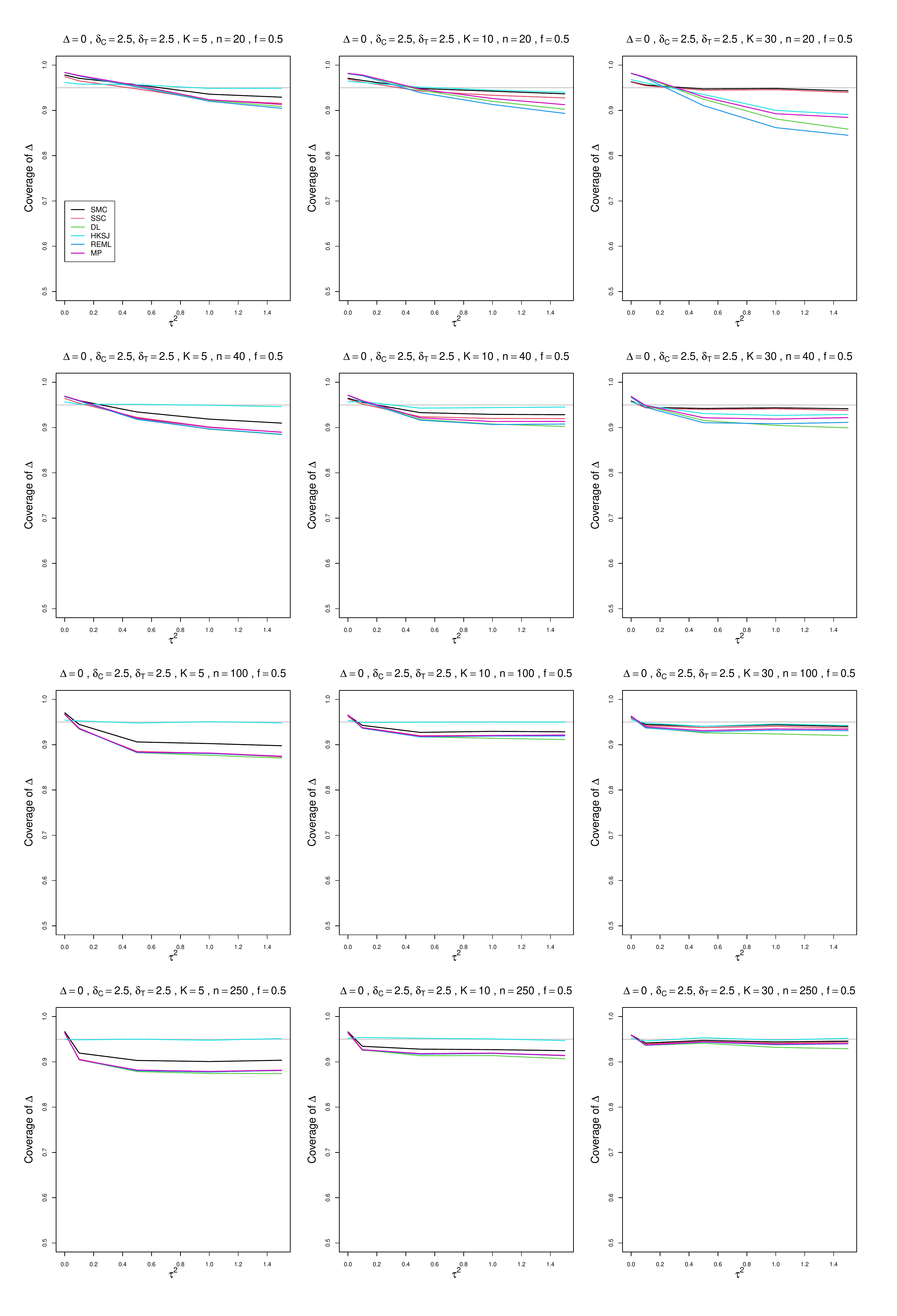}
	\caption{Coverage of 95\% confidence intervals for DSM (DL, REML, MP, HKSJ (DL), SMC and SSC intervals)  vs $\tau^2$, for equal sample sizes $n=20,\;40,\;100$ and $250$, $\delta_{iC} = 2.5$, $\Delta=0$ and  $f = 0.5$.   }
	\label{PlotCoverageOfDelta_deltaC_2.5deltaT=2.5_DSM_equal_sample_sizes.pdf}
\end{figure}

\begin{figure}[ht]
	\centering
	\includegraphics[scale=0.33]{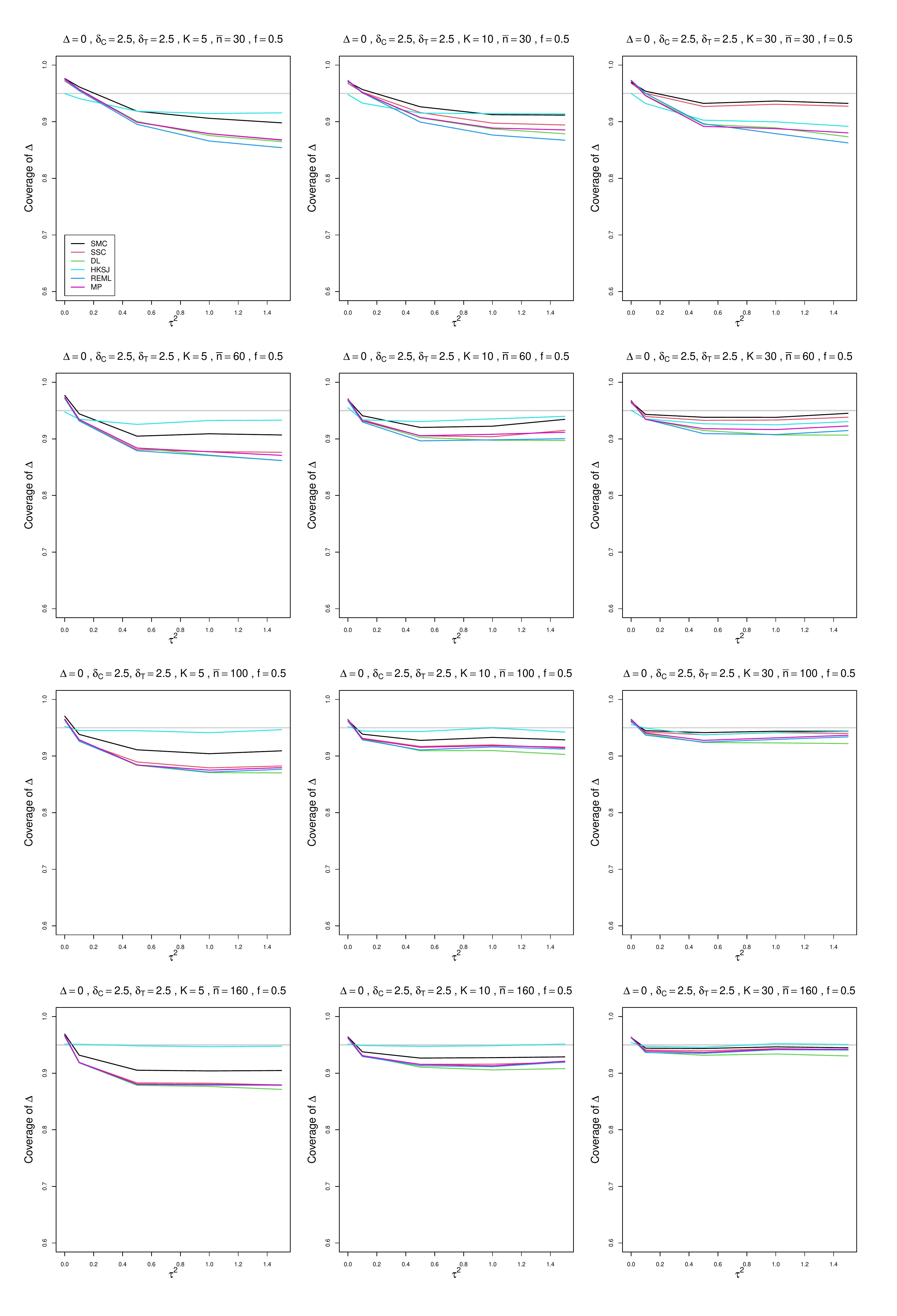}
	\caption{Coverage of 95\% confidence intervals for DSM (DL, REML, MP, HKSJ (DL), SMC and SSC intervals)  vs $\tau^2$, for unequal sample sizes $\bar{n}=30,\;60,\;100$ and $160$, $\delta_{iC} = 2.5$, $\Delta=0$ and  $f = 0.5$.   }
	\label{PlotCoverageOfDelta_deltaC_2.5deltaT=2.5_DSM_unequal_sample_sizes.pdf}
\end{figure}

\begin{figure}[ht]
	\centering
	\includegraphics[scale=0.33]{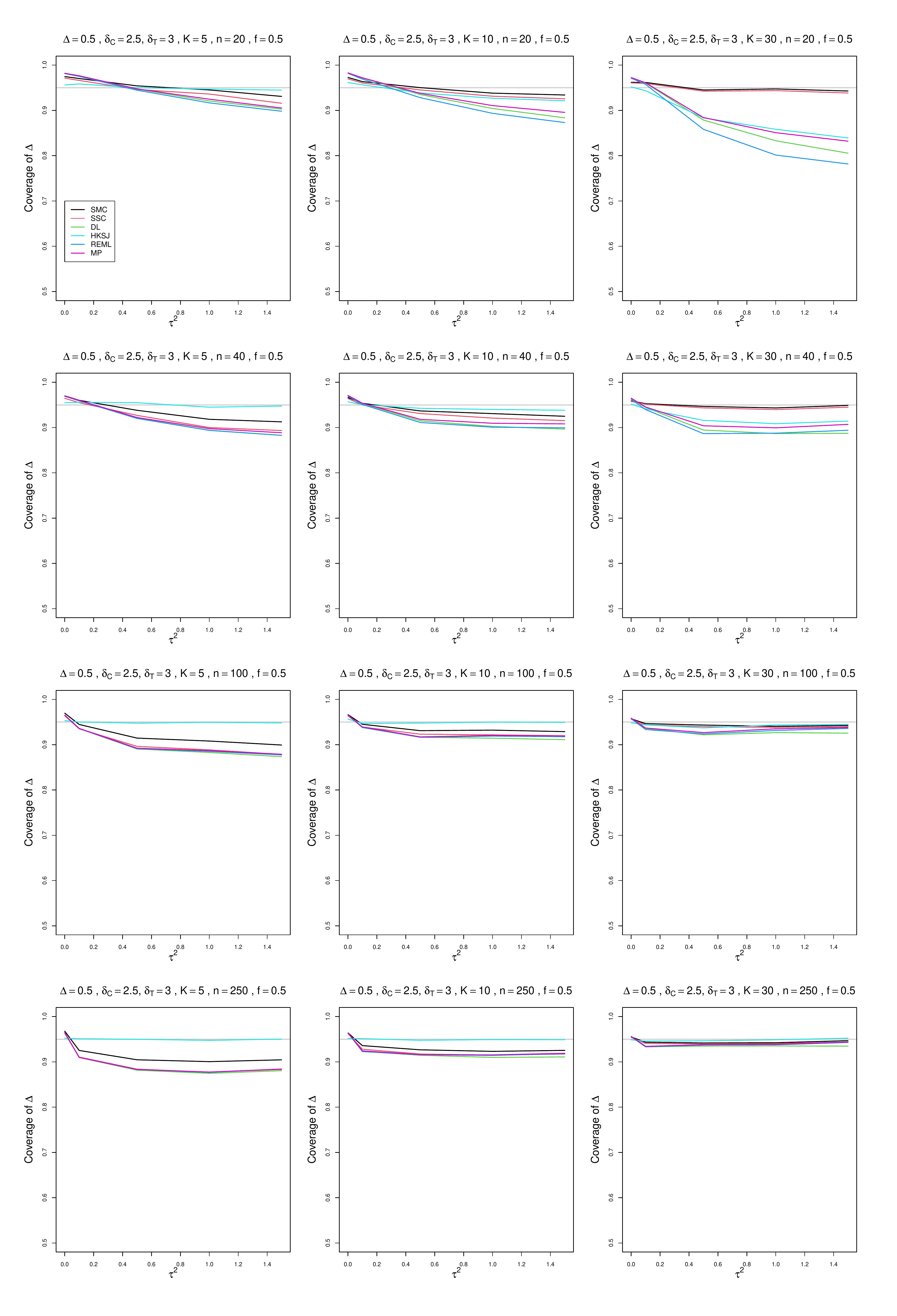}
	\caption{Coverage of 95\% confidence intervals for DSM (DL, REML, MP, HKSJ (DL), SMC and SSC intervals)  vs $\tau^2$, for equal sample sizes $n=20,\;40,\;100$ and $250$, $\delta_{iC} = 2.5$, $\Delta=0.5$ and  $f = 0.5$.   }
	\label{PlotCoverageOfDelta_deltaC_2.5deltaT=3_DSM_equal_sample_sizes.pdf}
\end{figure}

\begin{figure}[ht]
	\centering
	\includegraphics[scale=0.33]{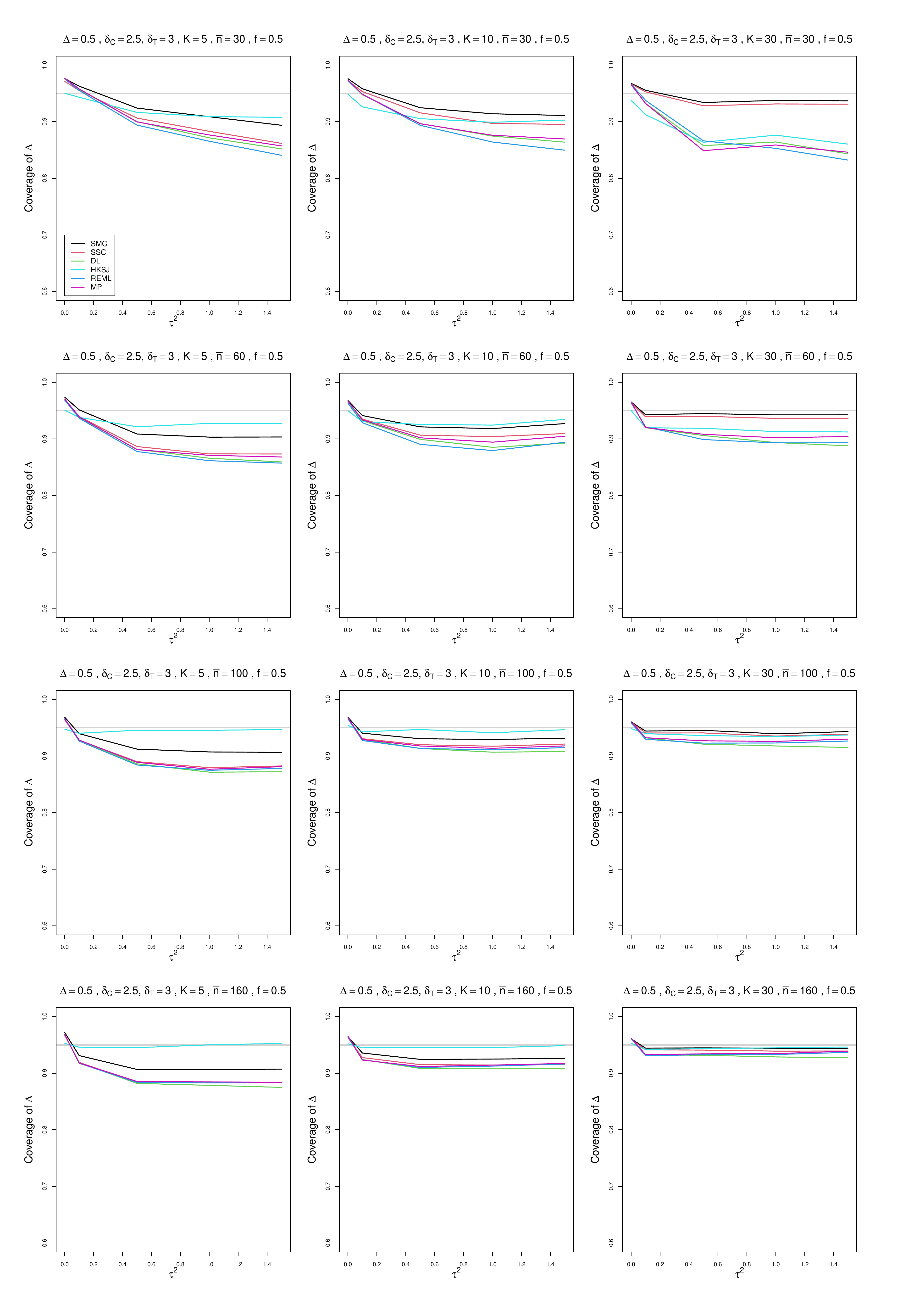}
	\caption{Coverage of 95\% confidence intervals for DSM (DL, REML, MP, HKSJ (DL), SMC and SSC intervals)  vs $\tau^2$, for unequal sample sizes $\bar{n}=30,\;60,\;100$ and $160$, $\delta_{iC} = 2.5$, $\Delta=0.5$ and  $f = 0.5$.   }
	\label{PlotCoverageOfDelta_deltaC_2.5deltaT=3_DSM_unequal_sample_sizes.pdf}
\end{figure}

\begin{figure}[ht]
	\centering
	\includegraphics[scale=0.33]{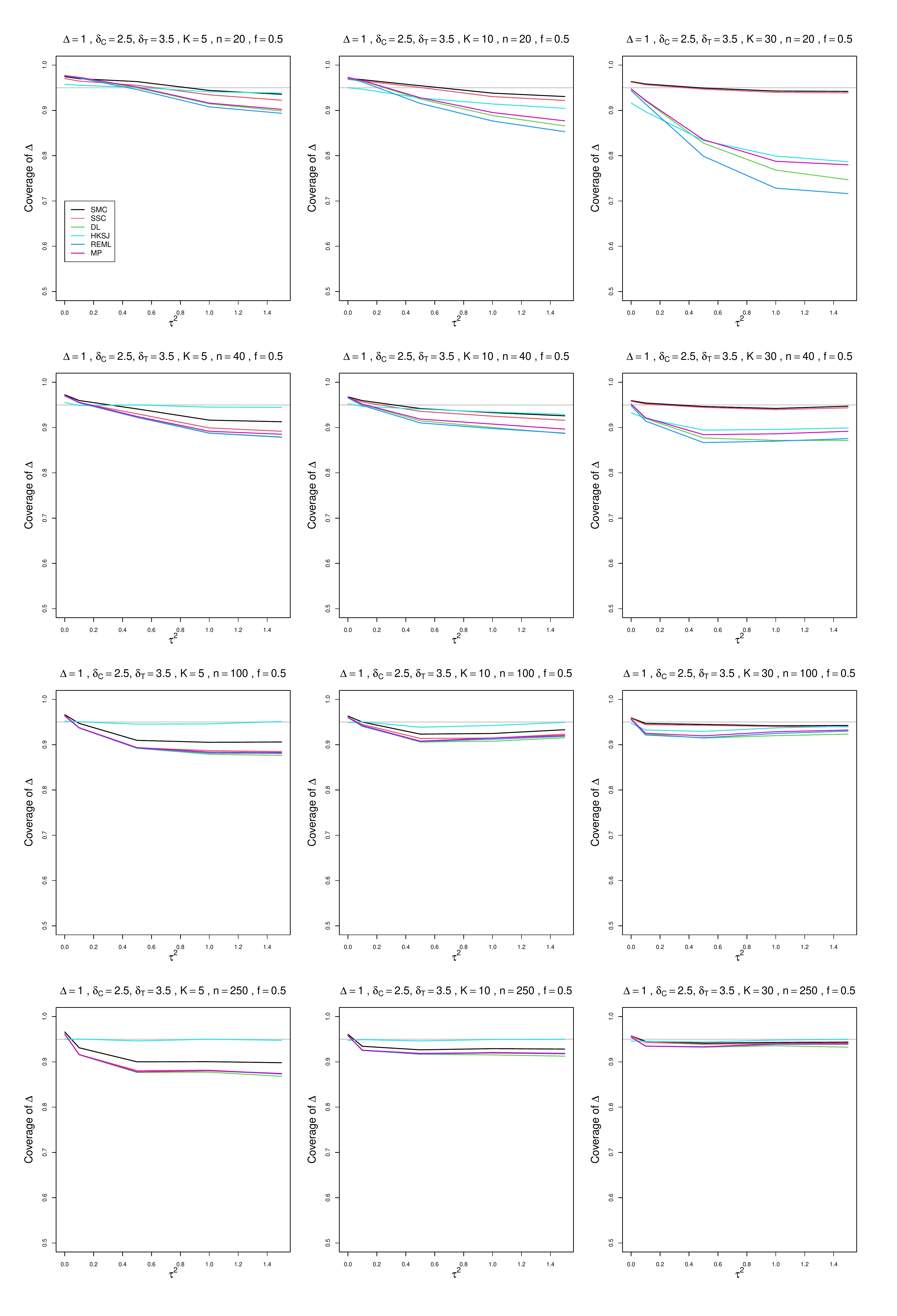}
	\caption{Coverage of 95\% confidence intervals for DSM (DL, REML, MP, HKSJ (DL), SMC and SSC intervals)  vs $\tau^2$, for equal sample sizes $n=20,\;40,\;100$ and $250$, $\delta_{iC} = 2.5$, $\Delta=1$ and  $f = 0.5$.   }
	\label{PlotCoverageOfDelta_deltaC_2.5deltaT=3.5_DSM_equal_sample_sizes.pdf}
\end{figure}

\begin{figure}[ht]
	\centering
	\includegraphics[scale=0.33]{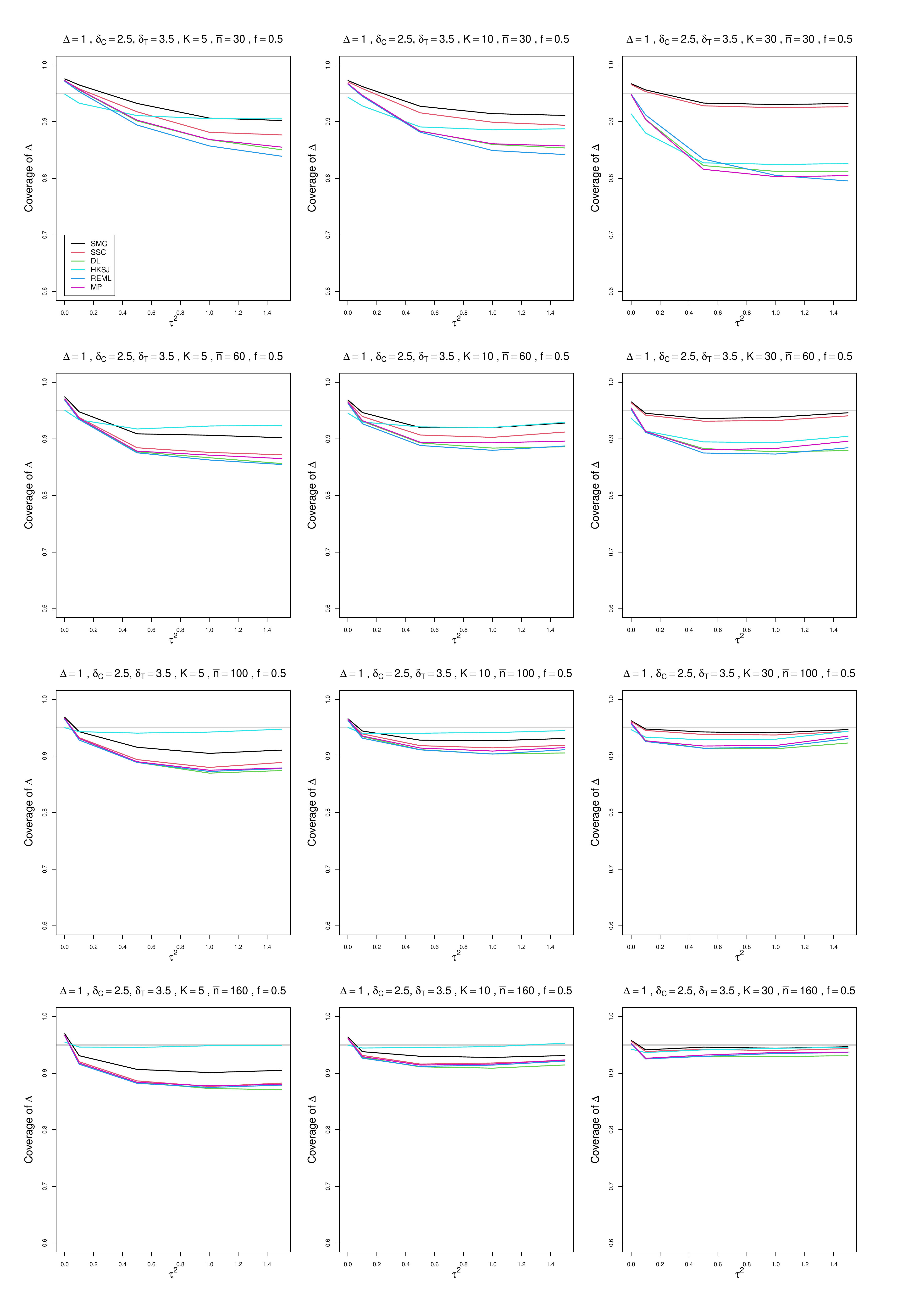}
	\caption{Coverage of 95\% confidence intervals for DSM (DL, REML, MP, HKSJ (DL), SMC and SSC intervals)  vs $\tau^2$, for unequal sample sizes $\bar{n}=30,\;60,\;100$ and $160$, $\delta_{iC} = 2.5$, $\Delta=1$ and  $f = 0.5$.   }
	\label{PlotCoverageOfDelta_deltaC_2.5deltaT=3.5_DSM_unequal_sample_sizes.pdf}
\end{figure}

\begin{figure}[ht]
	\centering
	\includegraphics[scale=0.33]{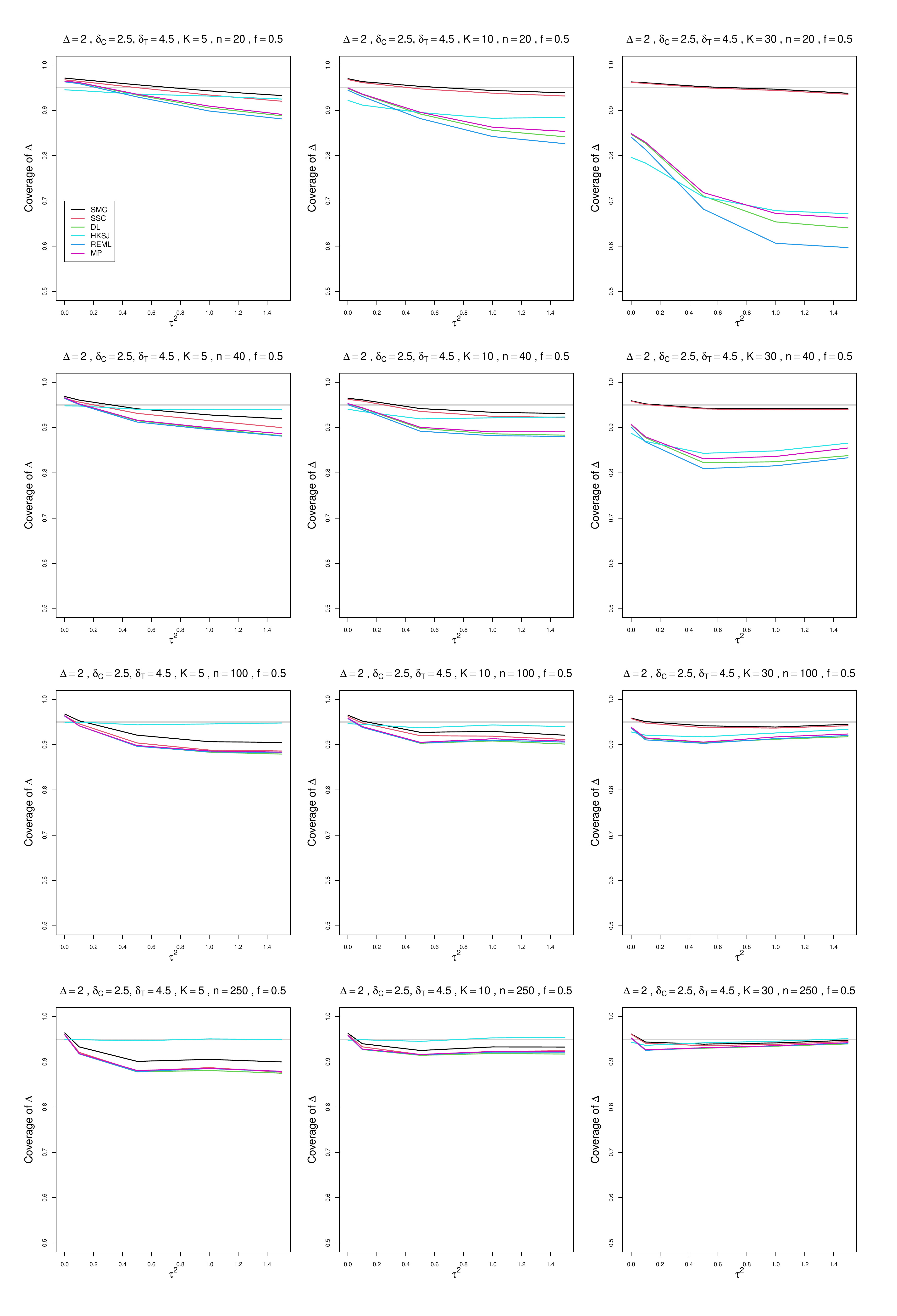}
	\caption{Coverage of 95\% confidence intervals for DSM (DL, REML, MP, HKSJ (DL), SMC and SSC intervals)  vs $\tau^2$, for equal sample sizes $n=20,\;40,\;100$ and $250$, $\delta_{iC} = 2.5$, $\Delta=2$ and  $f = 0.5$.   }
	\label{PlotCoverageOfDelta_deltaC_2.5deltaT=2.5_DSM_equal_sample_sizes.pdf}
\end{figure}

\begin{figure}[ht]
	\centering
	\includegraphics[scale=0.33]{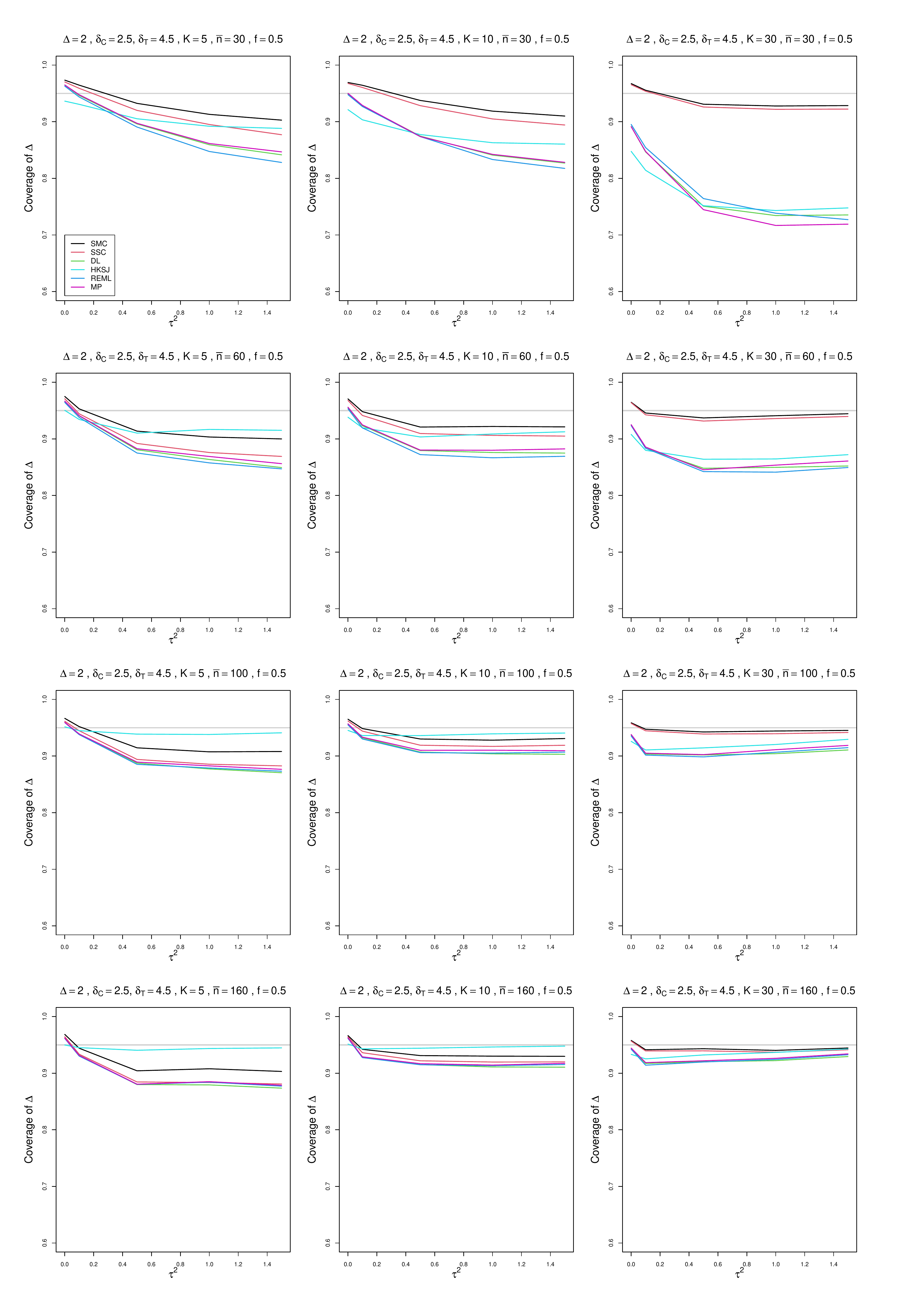}
	\caption{Coverage of 95\% confidence intervals for DSM (DL, REML, MP, HKSJ (DL), SMC and SSC intervals)  vs $\tau^2$, for unequal sample sizes $\bar{n}=30,\;60,\;100$ and $160$, $\delta_{iC} = 2.5$, $\Delta=2$ and  $f = 0.5$.   }
	\label{PlotCoverageOfDelta_deltaC_2.5deltaT=4.5_DSM_unequal_sample_sizes.pdf}
\end{figure}

\clearpage

\end{document}